# A Complete Equational Theory for Real-Clifford+CH Quantum Circuits

Alexandre Clément

Université Paris-Saclay, ENS Paris-Saclay, CNRS, Inria, LMF,
91190, Gif-sur-Yvette, France
alexandre.clement@inria.fr

## Abstract

We introduce a complete equational theory for the fragment of quantum circuits generated by the real Clifford gates plus the two-qubit controlled-Hadamard gate. That is, we give a simple set of equalities between circuits of this fragment, and prove that any other true equation can be derived from these. This is the first such completeness result for a finitely-generated, universal fragment of quantum circuits, with no parameterized gates and no need for ancillas.

## 1 Introduction

Quantum circuits form the basic language which is essentially at the core of quantum computing. The role that they play in practice is sometimes compared to that of an assembly language in classical computing. But formally, they are in fact the quantum equivalent of *boolean circuits*. Like boolean circuits, they consist of a sequence of *gates*, which are applied to some quantum data which is most often in the form of *qubits*, the quantum equivalent of bits.

Quantum computing can *a priori* be realized on a variety of physical platforms with various characteristics, and for now it is essentially impossible to predict what physical implementation will turn out to be the best, neither is it clear whether one technology will completely supplant the others; therefore, various different classes of quantum circuits are *a priori* relevant to formalize and study. Even staying inside the most standard model of quantum computing, where the data is encoded in qubits, several *fragments* of the language of quantum circuits, corresponding to particular sets of elementary gates with which to build the circuits, are relevant to consider. Some of them have received particular attention, such as the well-known Clifford+T and Toffoli-Hadamard fragments. An essential characteristic for a gate set is to be *computationally universal*, that is, to make it possible in principle to implement any arbitrary quantum program, up to arbitrarily small error. One reason that makes Clifford+T and Toffoli-Hadamard interesting is that they are universal while being generated by a very small and simple set of gates – in addition to being possibly relevant for promising implementations such as superconducting circuits.

A central task in quantum computing is to manipulate quantum circuits, and in particular to transform circuits into equivalent ones that are better according to some criterion: this has various practical applications such as optimization, satisfaction of hardware constraints, error correction, or verification. A promising approach towards manipulation of quantum circuits is via graphical rewriting. In the recent years, several *graphical languages* for formal manipulation of quantum circuits (and of other aspects of quantum processes) have been introduced, such as the ZX-, ZW-, ZH-, and ZXW-calculi, among others [2, 12, 16, 19, 29]. A central question in the formalization of such languages is the completeness of their equational theory, that is, the fact that the set of rewriting rules with which they are equipped be powerful enough so that any semantically meaningful transformation can be done. A related important question is to make the complete equational theory as simple as possible, and if possible, to be finally able to prove that it is minimal in the sense that all redundancies have been simplified out. There are now various complete equational theories for graphical languages [20, 23–25, 35], some of which being minimal [14] or near-minimal [3, 4, 32–35].

The language of quantum circuits itself can be viewed as a graphical language, and finding complete equational theories for it has its own added value compared to relying on external auxiliary languages like the ZX-calculus. For instance, applying arbitrary ZX transformations to a quantum circuit can produce a ZX-diagram from which it may be hard to get back a quantum circuit [13, 16]. Although the question of finding a complete equational theory for quantum circuits had been formulated in the early 2000's [22, 26], the first complete equational theory was found only a few years ago [11], followed by simplifications and a proof of minimality [9, 10]. Those results are for the full language of quantum circuits in the qubit setting. That is, the gate set is very simple but still contains a parameterized gate, and the minimal equational theory is relatively simple but involves non-trivial computations on the parameters.

In this paper, we consider a finitely generated fragment of quantum circuits, without parameterized gates, namely the fragment generated by real Clifford gates – the Hadamard gate $H$, the Pauli $Z$ gate, and the controlled-$Z$ gate $CZ$ – together with the controlled Hadamard gate $CH$. We call this fragment *Real-Clifford+CH*. This is a sub-fragment of Clifford+T, which contains Toffoli-Hadamard[1] and therefore is universal. To our knowledge, this is the first completeness result for a fragment of quantum circuits generated by a finite gate set, without parameterized gates.

Similarly to [11], our result is based on a back-and-forth translation into an auxiliary language for which a completeness result is known, namely a presentation of the matrix group $U_n(\mathbb{Z}[\frac{1}{\sqrt 2}])$ generated by one- and two-level matrices [17]. Those one- and two-level matrices have a similar structure to the optical circuits used in [11].

*Related works.* In addition to the graphical languages mentioned, other languages for quantum computing are equipped with complete equational theories [15, 17, 21]. The main difference is that those languages are equipped with a control constructor (or an equivalent construction) allowing one to express arbitrary multi-controlled gates, which means that they are not finitely generated from a quantum circuit point of view.

---

[1] Except in the case of 3-qubit circuits without ancillas.

The paper is organized as follows: In Section 2.1, we formally introduce Real-Clifford+CH quantum circuits and define some useful macros. In Section 3, we introduce the equational theory that we will prove to be complete. In Section 4, we state the completeness result upon which our result relies [17], and we adapt it so as to be able to lift the completeness to quantum circuits. In Sections 5 and 6, we give some additional insight about the proof of completeness. In Section 7, we prove the completeness of our equational theory for circuits on 1 and 2 qubits. Finally, in Section 8, we prove the completeness in the general case.

## 2 Real-Clifford+CH Quantum Circuits

### 2.1 The PROP of Real-Clifford+CH Quantum Circuits

We consider the *real-Clifford+CH* fragment of quantum circuits, without ancillas, generated by the following gate set:

$$\left\{ \boxed{H}, \bullet\!\!-, \, \underset{\cdot}{\overset{\cdot}{\vdash}}, \, \boxed{H} \right\} \tag{A}$$

Following [9–11], we give a rigorous formalization to Real-Clifford+CH quantum circuits by using the structure of PROP, which originates from category theory [28, 36] but can be viewed purely as an algebraic structure [8]. We start by defining *raw circuits* [11], which are the terms freely built from the generators of (A), the empty circuit ⸬, the one-qubit identity ——, and the swap ⤬, by means of the sequential composition ∘ and the parallel composition ⊗:

**Definition 2.1.** *The set $\mathbf{QC_{raw}}$ of raw Real-Clifford+CH circuits is inductively defined as follows:*

$$⸬ : 0 \to 0 \qquad \text{——} : 1 \to 1 \qquad ⤬ : 2 \to 2$$

$$\boxed{H} : 1 \to 1 \qquad \bullet\!\!- : 1 \to 1 \qquad \underset{\cdot}{\overset{\cdot}{\vdash}} : 2 \to 2 \qquad \boxed{H} : 2 \to 2$$

$$\frac{C_1 : n \to n \quad C_2 : n \to n}{C_2 \circ C_1 : n \to n} \qquad \frac{C_1 : n_1 \to n_1 \quad C_2 : n_2 \to n_2}{C_1 \otimes C_2 : n_1 + n_2 \to n_1 + n_2}$$

**Definition 2.2.** *The PROP of Real-Clifford+CH circuits, denoted by $\mathbf{QC}$, is obtained by taking the quotient of $\mathbf{QC_{raw}}$ by the axioms of PROP, shown in Figure 1.*

Roughly speaking, Equations (a) to (e) are the axioms needed to guarantee that two circuits with the same graphical representation are actually equal, and that the length of the wires does not matter (for instance, $\boxed{H} = \boxed{H} = \boxed{H}$). Equations (f) and (g) additionally give us the property that the circuits are defined up to deformation.

**Remark 1.** *The structure obtained by quotienting $\mathbf{QC_{raw}}$ only by Equations (a) to (e) is called a PRO. Depending on the context, it may be more relevant to model quantum circuits as a PRO rather than a PROP (for instance when considering implementations where the swap gate is costly), or vice-versa. Note that our completeness result can be straightforwardly adapted to the PRO setup by simply adding Equations (f) and (g) to the complete equational theory of Figure 4. Moreover, Equation (f) actually does not need to be parameterized* by an arbitrary circuit C: it is sufficient to only assume it for the generators, that is, Equation (f) can be replaced by the equations of Figure 2, since it can be recovered from them in the presence of the other axioms of PROP. Thus, there are only four equations to add to the complete equational theory in order to adapt it to the PRO setting.

**Remark 2.** *The other standard real Clifford gates can be recovered from those of (A) (see Figure 3). Note also that the swap gate can be expressed from the generators of (A) (see Equation (8)).*

**Definition 2.3 (Semantics).** *For any $n$-qubit circuit $C$, let $[\![C]\!] : \mathbb{C}^{\{0,1\}^n} \to \mathbb{C}^{\{0,1\}^n}$ be the semantics of $C$ inductively defined as*

$$[\![C_2 \circ C_1]\!] = [\![C_2]\!] \circ [\![C_1]\!]$$
$$[\![C_1 \otimes C_2]\!] = [\![C_1]\!] \otimes [\![C_2]\!]$$
$$[\![⸬]\!] = 1 \mapsto 1$$
$$[\![\text{——}]\!] = |x\rangle \mapsto |x\rangle$$
$$[\![⤬]\!] = |x, y\rangle \mapsto |y, x\rangle$$
$$[\![\bullet\!\!-]\!] = |x\rangle \mapsto (-1)^x |x\rangle$$
$$[\![\boxed{H}]\!] = |x\rangle \mapsto \frac{|0\rangle + (-1)^x |1\rangle}{\sqrt{2}}$$
$$[\![\underset{\cdot}{\overset{\cdot}{\vdash}}]\!] = |x, y\rangle \mapsto (-1)^{xy} |x, y\rangle$$
$$[\![\boxed{H}]\!] = |x, y\rangle \mapsto \begin{cases} |0, y\rangle \mapsto |0, y\rangle \\ |1, y\rangle \mapsto \frac{|1,0\rangle + (-1)^y |1,1\rangle}{\sqrt{2}} \end{cases}$$

**Remark 3.** *Strictly speaking, $[\![\cdot]\!]$ is defined on raw circuits. However, one can check that it is preserved by the equations of Figure 1, so that it is well-defined on (non-raw) circuits.*

### 2.2 Shortcuts

We now define several shortcuts that will be useful in the rest of the paper, some of which corresponding to standard gates. Note that from a formal point of view, all circuits drawn throughout the paper are still considered as being composed only of gates from (A), the shortcuts being only a convenient way to represent particular circuits.

First, we will use CNOTs, as well as standard variants of the gates, which are defined in Figure 3.

Second, it will be particularly useful to consider a change-of-basis matrix which diagonalizes $H$ into $Z$: we will use the following one:[2]

$$P := R_Y(\tfrac{\pi}{4})Z = e^{-i\frac{\pi}{8}} SHTHS = \begin{pmatrix} \cos(\frac{\pi}{8}) & \sin(\frac{\pi}{8}) \\ \sin(\frac{\pi}{8}) & -\cos(\frac{\pi}{8}) \end{pmatrix}.$$

One can check that $P^2 = I$ and $PZP = H$. Note that $P$ cannot be represented by a real-Clifford+CH circuit, since it is not in the group

---

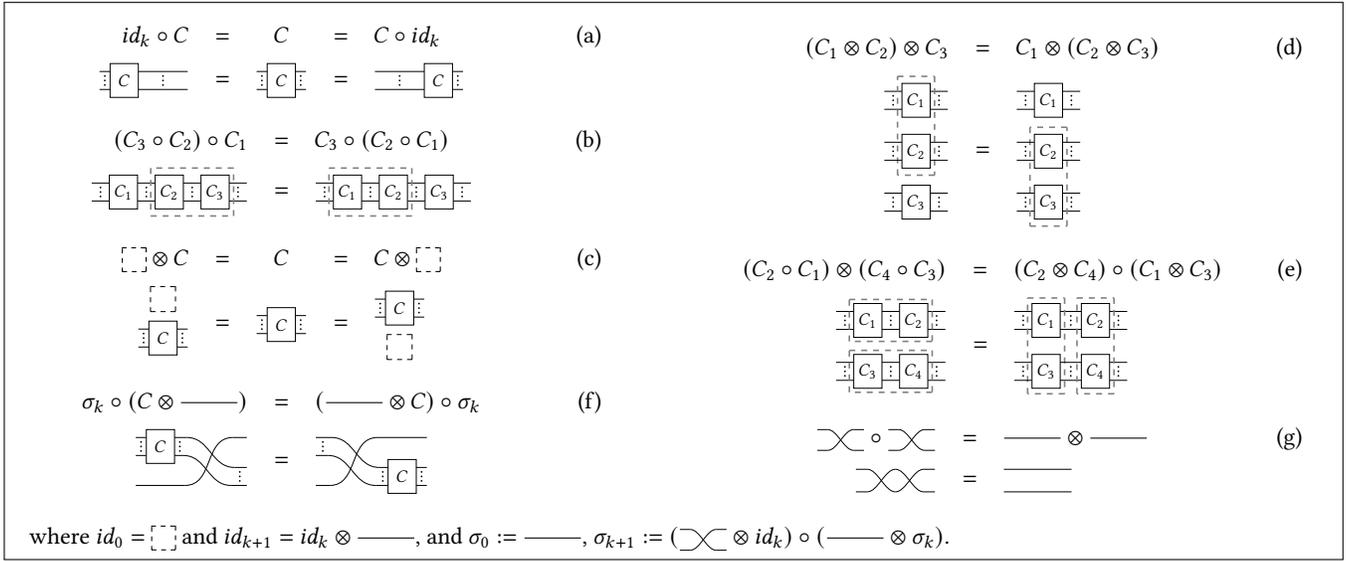

**Figure 1: Axioms of PROP.**

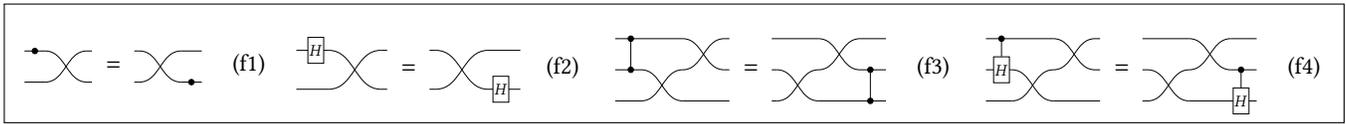

**Figure 2: Alternative equations to Equation (f).**

$U_n(\mathbb{Z}[\frac{1}{\sqrt{2}}])$. However, one has

$$P \otimes P = \frac{1}{2\sqrt{2}} \begin{pmatrix} \sqrt{2}+1 & 1 & 1 & \sqrt{2}-1 \\ 1 & -\sqrt{2}-1 & \sqrt{2}-1 & -1 \\ 1 & \sqrt{2}-1 & -\sqrt{2}-1 & -1 \\ \sqrt{2}-1 & -1 & -1 & \sqrt{2}+1 \end{pmatrix}$$

which is in that group, and can be represented by a real-Clifford+CH circuit. We draw this shortcut with a slight dashed line in order to keep track of the fact that it is actually defined using two-qubit gates:

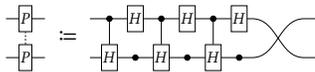

It will also be useful to represent a pair of $P$ gates acting on non-adjacent qubits:

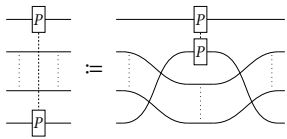

Finally, similarly to [11], another particularly useful class of shortcuts is *multi-controlled gates*.

On the one hand, we will use multi-controlled $Z$ and $H$ gates. Note that for $n \geq 3$, one cannot represent an $n$-qubit multi-controlled $Z$ or $H$ gate as an $n$-qubit real-Clifford+CH circuit.[3] However, one can represent an $(n{-}1)$-qubit multi-controlled $Z$ or $H$ gate tensored with the 1-qubit identity.

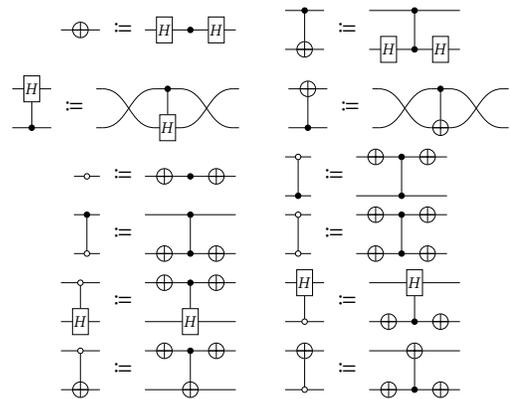

**Figure 3: Standard one- and two-qubit gates, defined as short-cuts.**

On the other hand, we will also use multi-controlled $ZX$ and $XZ$ gates (where $ZX$ and $XZ$ are just the product of the $Z$ and $X$ gates in both order), which do not have such a restriction.

---

[3]This is due to the fact that it has determinant $-1$ and that we are only working with real matrices. See also Section 4.

The definition is by mutual induction as follows. Note that we draw multi-controlled $Z$ and $H$ gates using a small dashed box, in order to keep track of the fact that the shortcut actually uses one more qubit than the multi-controlled gate itself.[4]

**Definition 2.4 (Multi-Controlled $Z$, $H$, $ZX$ and $XZ$ Gates).** *The base cases are as follows:*

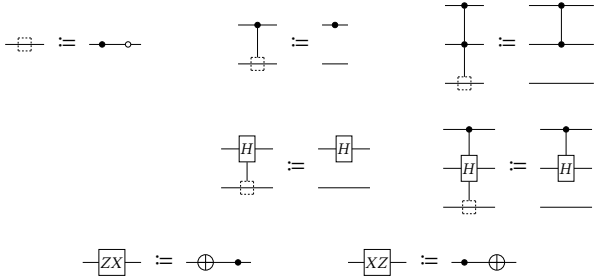

*and the recursive steps are as follows:*

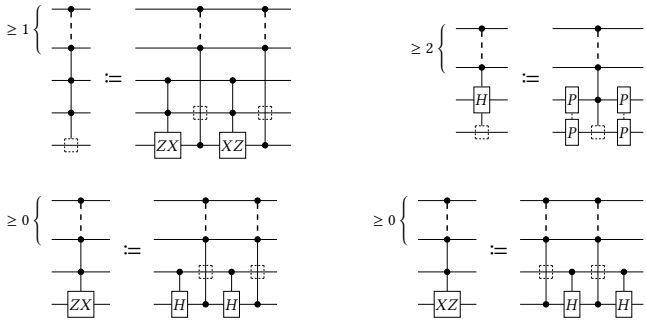

*Note that semantically, $XZ = -ZX$. This motivates the following notations: for $a \in \{0, 1\}$,*

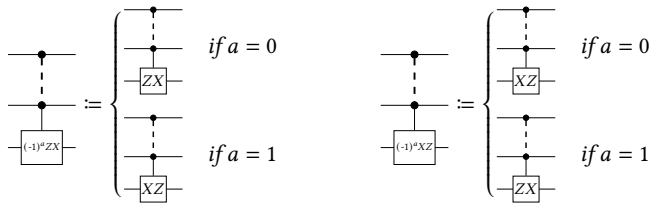

*Additionally, it will be useful to consider multi-controlled gates acting on non-adjacent qubits or where the target qubits are not at the end (note that we have already used such notations in the inductive definition above). Similarly to the $P \otimes P$ gate, these are defined by conjugating with swaps. For instance:*

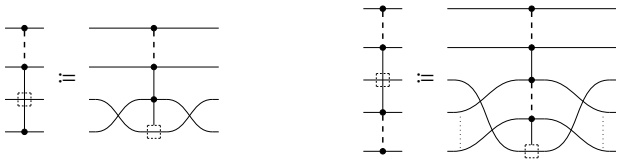

---

[4]For the multi-controlled $Z$ gate, one can interpret the dashed box as representing a 1-qubit $-I$ gate, and our drawing as representing a multi-controlled $-I$ gate. Similarly, one can see the multi-controlled $H$ tensored by the identity as being a multi-controlled $H \otimes I$ gate.

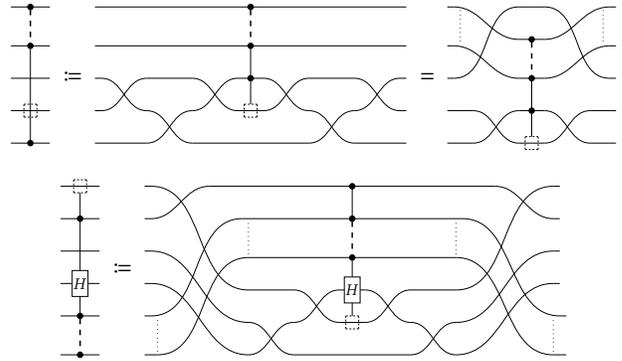

*Note that since circuits are defined up to the topological rules, we do not need to specify the exact network of swaps to put on each side of the multi-controllled gate, but only the overall permutation that they realize. The only possible ambiguity could be in the ordering of the controls: by convention, we always take the permutation that preserves the vertical order of the controls.*

*Moreover, we will also consider multi-controlled gates with negative controls (that is, with respect to $|0\rangle$ instead of $|1\rangle$), which we denote using the usual notation of white controls. These are defined by conjugating with the $X$ gate:*

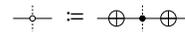

*Going further, we also define parameterized controls, whose color depends on a boolean parameter:*

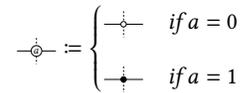

*Finally, we denote series of parametrized controls as follows, mimicking the notation used in [11]: for any $x = x_1...x_k \in \{0, 1\}^k$,*

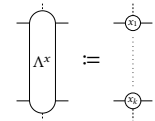

## 3 Equational Theory

The set QC of equations that we claim to be complete is given in Figure 4. We write $\text{QC} \vdash C_1 = C_2$ when $C_1$ and $C_2$ can be transformed into one another by means of these equations.[5]

Equations (1) to (10) and (12) to (15) are relatively standard. Equations (16) and (17) also express a well-known property, namely that a positively-controlled circuit commutes with a negatively-controlled one. Equation (11) essentially also expresses this property, but from the point of view of the eigenbasis of $H$: reasoning from a purely semantic point of view, one could *informally* write which makes the commutation clearly appear. Finally, Equation (18) says that the top part of ⎡P⎤/⎣P⎦ indeed behaves as a $P$ gate (see Figure 5), and Equation (19) partly witnesses the fact that the shortcut behaves as if it did not touch the bottom qubit.

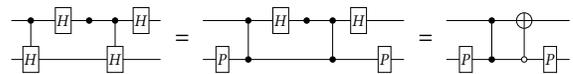

---

[5]That is, $\text{QC} \vdash \cdot = \cdot$ is the smallest congruence on circuits which satisfies the equations of Figure 5.

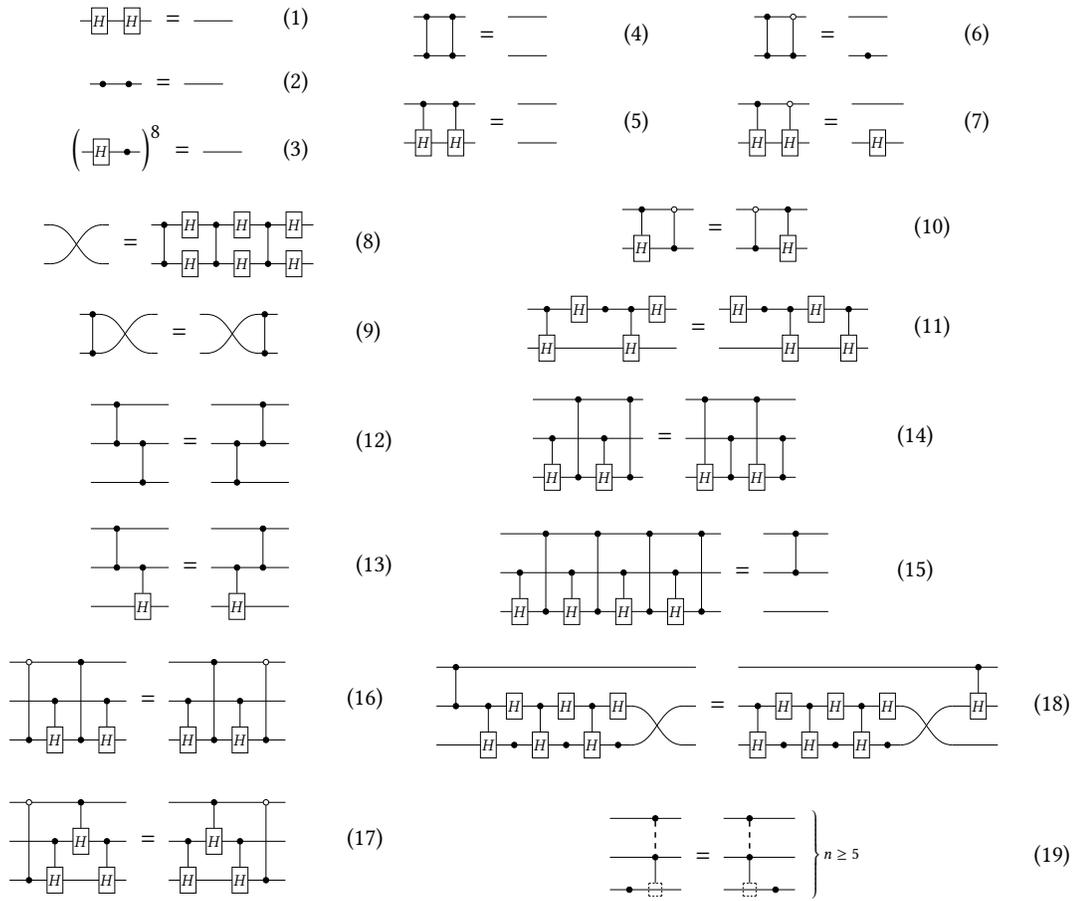

Figure 4: Complete equational theory for real-Clifford+CH quantum circuits. Equation (19) is an equation schema, consisting of an equation on $n$ qubits for each $n \geq 5$. Equations (6), (7), (14) to (17), and (19) use shortcuts, defined in Section 2.2.

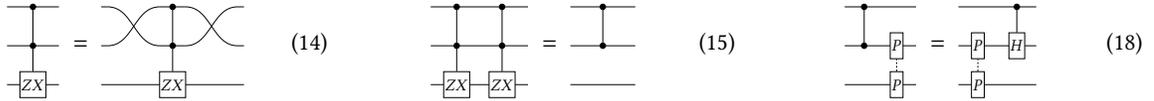

Figure 5: Some equations of Figure 4, written with more shortcuts.

**Proposition 3.1 (Soundness).** *For any circuits* $C_1, C_2 \in \mathbf{QC}$,

$$if\ \mathrm{QC} \vdash C_1 = C_2\ then\ [\![C_1]\!] = [\![C_2]\!].$$

PROOF. The semantics is compositional, therefore it suffices to check for each equation that both sides have the same semantics. This can be done by direct computation for Equations (1) to (18), and by a straightforward induction for Equation (19). □

## 4 Auxiliary Language

This paper relies essentially on a result of [17] which gives a presentation (that is, a set of generators together with a complete equational theory) of the group $U_N(\mathbb{Z}[\frac{1}{\sqrt{2}}])$ of $N$-dimensional unitary matrices with entries in $\mathbb{Z}[\frac{1}{\sqrt{2}}]$. Similarly to [11], we will turn that complete equational theory into one for quantum circuits by means of a back-and-forth encoding. However, contrary to the situation in [11], the set of matrices represented by real-Clifford+CH circuits is not exactly $U_N(\mathbb{Z}[\frac{1}{\sqrt{2}}])$, but a subgroup of it that we need to characterize first.

$$(-1)_{[a]}^2 \approx \varepsilon \tag{a1}$$

$$X_{[a,b]}^2 \approx \varepsilon \tag{a2}$$

$$(-1)_{[a]}(-1)_{[b]} \approx (-1)_{[b]}(-1)_{[a]} \tag{b1}$$

$$(-1)_{[a]}X_{[b,c]} \approx X_{[b,c]}(-1)_{[a]} \tag{b2}$$

$$X_{[a,b]}X_{[c,d]} \approx X_{[c,d]}X_{[a,b]} \tag{b3}$$

$$(-1)_{[a]}X_{[a,b]} \approx X_{[a,b]}(-1)_{[b]} \tag{c1}$$

$$X_{[b,c]}X_{[a,b]} \approx X_{[a,b]}X_{[a,c]} \tag{c2}$$

$$X_{[a,c]}X_{[b,c]} \approx X_{[b,c]}X_{[a,b]} \tag{c3}$$

$$H_{[a,b]}^2 \approx \varepsilon \tag{a3}$$

$$(-1)_{[a]}H_{[b,c]} \approx H_{[b,c]}(-1)_{[a]} \tag{b4}$$

$$X_{[a,b]}H_{[c,d]} \approx H_{[c,d]}X_{[a,b]} \tag{b5}$$

$$H_{[a,b]}H_{[c,d]} \approx H_{[c,d]}H_{[a,b]} \tag{b6}$$

$$H_{[b,c]}X_{[a,b]} \approx X_{[a,b]}H_{[a,c]} \tag{c4}$$

$$H_{[a,c]}X_{[b,c]} \approx X_{[b,c]}H_{[a,b]} \tag{c5}$$

$$(-1)_{[a]}(-1)_{[b]}H_{[a,b]} \approx H_{[a,b]}(-1)_{[a]}(-1)_{[b]} \tag{d1}$$

$$(-1)_{[b]}H_{[a,b]} \approx H_{[a,b]}X_{[a,b]} \tag{d2}$$

$$(H_{[c,d]}H_{[a,c]}H_{[b,d]})^4 \approx H_{[a,b]}H_{[c,d]} \tag{d3}$$

$$(H_{[a,c]}H_{[b,d]}H_{[a,b]}H_{[a,c]}H_{[b,d]}X_{[c,e]}X_{[d,f]})^3 \approx H_{[c,e]}H_{[d,f]}H_{[e,f]}H_{[c,e]}H_{[d,f]}X_{[c,e]}X_{[d,f]} \tag{d4}$$

$$X_{[c,b]} \approx X_{[b,c]} \tag{e1}$$

$$H_{[c,b]} \approx X_{[b,c]}H_{[b,c]}X_{[b,c]}. \tag{e2}$$

**Figure 6: The complete equational theory given in [17]. In each equation, all the indices must be distinct. In Equations (e1) and (e2), we require $b < c$, and in all other equations, the two indices of each $H$ or $X$ must be in increasing order.**

$$(-1)_{[0]}^2 \approx \varepsilon \tag{a1*}$$

$$X_{[a,b]}^2 \approx \varepsilon \tag{a2}$$

$$(-1)_{[0]}(-1)_{[1]} \approx (-1)_{[1]}(-1)_{[0]} \tag{b1*}$$

$$(-1)_{[a]}X_{[a,b]} \approx X_{[a,b]}(-1)_{[b]} \tag{c1}$$

$$H_{[a,b]}^2 \approx \varepsilon \tag{a3}$$

$$(-1)_{[1]}H_{[0,2]} \approx H_{[0,2]}(-1)_{[1]} \tag{b4*}$$

$$H_{[0,1]}H_{[2,3]} \approx H_{[2,3]}H_{[0,1]} \tag{b6*}$$

$$H_{[a,c]}X_{[b,c]} \approx X_{[b,c]}H_{[a,b]} \tag{c5}$$

$$(-1)_{[b]}H_{[a,b]} \approx H_{[a,b]}X_{[a,b]} \tag{d2}$$

$$\begin{aligned} H_{[0,1]}H_{[0,2]}H_{[1,3]}H_{[0,1]}(-1)_{[0]}(-1)_{[1]}H_{[0,2]}H_{[1,3]} \approx \\ H_{[0,2]}H_{[1,3]}H_{[0,1]}(-1)_{[0]}(-1)_{[1]}H_{[0,2]}H_{[1,3]}H_{[0,1]} \end{aligned} \tag{d3*}$$

$$\begin{aligned} H_{[0,1]}H_{[0,2]}H_{[1,3]}H_{[0,4]}H_{[1,5]}H_{[0,1]}(-1)_{[0]}(-1)_{[1]}H_{[0,4]}H_{[1,5]}H_{[0,2]}H_{[1,3]} \approx \\ H_{[0,2]}H_{[1,3]}H_{[0,4]}H_{[1,5]}H_{[0,1]}(-1)_{[0]}(-1)_{[1]}H_{[0,4]}H_{[1,5]}H_{[0,2]}H_{[1,3]}H_{[0,1]} \end{aligned} \tag{d4*}$$

$$H_{[c,b]}X_{[b,c]} \approx X_{[b,c]}H_{[b,c]} \tag{e2*}$$

**Figure 7: Simpler equational theory, equivalent to that of Figure 6. In each equation, all the indices must be distinct but are not required to be in a particular order.**

## 4.1 Expressiveness of Real-Clifford+CH Circuits

First, recall the relevant generating set for $U_N(\mathbb{Z}[\frac{1}{\sqrt{2}}])$, made of what are called *1- and 2-level* matrices:

**Definition 4.1 ([1, 17]).** *Let $M$ be an $m \times m$ matrix, let $m < N$, and let $0 \le a_1, ..., a_m \le N - 1$. An $m$-level matrix of type $M$ is the $N \times N$ matrix $M_{[a_1,...,a_m]}$ defined by*

$$(M_{[a_1,...,a_m]})_{i,j} = \begin{cases} M_{i',j'} & \text{if } i = a_{i'} \text{ and } j = a_{j'} \\ \delta_{i,j} & \text{otherwise.} \end{cases}$$

**Proposition 4.2 ([1]).** *For $N \ge 1$, a matrix belongs to the group $U_N(\mathbb{Z}[\frac{1}{\sqrt{2}}])$ if and only if it can be expressed as a product of generators from*

$$\mathcal{G}_N := \left\{ (-1)_{[a]}, X_{[a,b]}, H_{[a,b]} \mid a, b \in \{0, ..., N-1\}, \ a \ne b \right\}.$$

We can now state the completeness result of [17]:

**Definition 4.3.** *The free monoid generated by $\mathcal{G}_N$ (i.e. the monoid of words over $\mathcal{G}_N$) is denoted by $\mathcal{G}_N^*$. The empty word is denoted by $\epsilon$. The matrix associated with a word $w = G_1 G_2 ... G_k \in \mathcal{G}_N^*$ is called its semantics, and denoted by $[\![w]\!] := G_1 \cdot G_2 \cdot ... \cdot G_k$.*

*We denote by $\approx$ the smallest congruence satisfying the equations of Figure 6. That is, for any two words $w_1, w_2 \in \mathcal{G}_N^*$, one has $w_1 \approx w_2$ iff $w_1$ and $w_2$ can be transformed into one another using the equations of Figure 6.*

**Theorem 4.4 (Soundness and Completeness of $\approx$ for $\mathcal{G}_N^*$ [17]).**
*Let $w_1, w_2 \in \mathcal{G}_N^*$. Then $[\![w_1]\!] = [\![w_2]\!]$ if and only if $w_1 \approx w_2$.*

Note that contrary to [17], we have not *a priori* required that $a < b$ in $X_{[a,b]}$ or $H_{[a,b]}$, hence the additional conditions in the caption of Figure 6, and the fact that Equations (e1) and (e2) are here considered as axioms, whereas in [17] they were just shortcut definitions.

We can now express the parity conditions needed to characterize the matrix group spanned by real-Clifford+CH circuits:

**Lemma 4.5.** *Given any $A \in U_N(\mathbb{Z}[\frac{1}{\sqrt{2}}])$, the parity of the number of generators of the form $H_{[a,b]}$ (resp. of the form $(-1)_{[a]}$ or $X_{[a,b]}$) in an arbitrary decomposition of $A$ as a product of generators from $\mathcal{G}_N$ only depends on $A$.*

PROOF. By Theorem 4.4, any two decompositions of $A$ as a product of generators from $\mathcal{G}_N$ can be transformed into one another using the equations of Figure 6. One can check that all those equations preserve both the parity of the number of generators of the form $H_{[a,b]}$, and the parity of the number of generators of the form $(-1)_{[a]}$ or $X_{[a,b]}$. □

In the following, the parity of the number of $H_{[a,b]}$ in an arbitrary decomposition of $A$ will be called the *$H$-parity* of $A$, and the parity of the total number of $(-1)_{[a]}$ and $X_{[a,b]}$ in such a decomposition will be called the *$ZX$-parity* of $A$.

**Proposition 4.6.** *For $N \geq 1$, a matrix belongs to the group $U_N(\mathbb{Z}[\frac{1}{\sqrt{2}}])$ and has even $ZX$- and $H$-parities if and only if it can be expressed as a product of generators from*

$$\mathcal{P}_N := \left\{ \left((-1)_{[a]}(-1)_{[c]}\right), \left((-1)_{[a]}X_{[c,d]}\right), \left(X_{[a,b]}X_{[c,d]}\right), \left(H_{[a,b]}H_{[c,d]}\right) \right.$$
$$\left. \Big| \, a, b, c, d \in \{0, \dots, n-1\}, \, a \neq b \wedge c \neq d \right\}.$$

PROOF. The algorithm given in [17] for decomposing a given element of $U_N(\mathbb{Z}[\frac{1}{\sqrt{2}}])$ into a product of generators of $\mathcal{P}_N$ can be slightly modified so that its output just consists of a sequence of $H_{[a,b]}$ followed by a sequence of $(-1)_{[a]}$ and $X_{[a,b]}$ (the modified algorithm is described in Appendix A.1). By Lemma 4.5, these two sequences are of even length, so we can group the generators by pairs in each. Finally, it remains to get rid of the pairs of the form $\left(X_{[c,d]}(-1)_{[a]}\right)$ by noticing that this is semantically equivalent to $\left((-1)_{[a]}X_{[c,d]}\right)$ if $a \notin \{c, d\}$, to $\left((-1)_{[d]}X_{[c,d]}\right)$ if $a = c$, and to $\left((-1)_{[c]}X_{[c,d]}\right)$ if $a = d$. □

**Lemma 4.7.** *For $n = 2$, the set of matrices that can be represented by an $n$-qubit real-Clifford+CH circuit is exactly $U_{2^n}(\mathbb{Z}[\frac{1}{\sqrt{2}}])$. For $n \geq 3$, it is the subset of $U_{2^n}(\mathbb{Z}[\frac{1}{\sqrt{2}}])$ consisting of the matrices with even $ZX$- and $H$-parities.*

PROOF. First, it is clear from the definition of the semantics that any real-Clifford+CH circuit represents an element of $U_{2^n}(\mathbb{Z}[\frac{1}{\sqrt{2}}])$.

For $n = 2$, we need to show that any element of $U_4(\mathbb{Z}[\frac{1}{\sqrt{2}}])$ can be represented by a real-Clifford+CH circuit. To do so, it suffices to give for every generator of $\mathcal{G}_N$ a circuit with corresponding semantics. This is done in Definition 7.3.

For $n \geq 3$, similarly, it suffices to express each generator of $\mathcal{P}_N$ as a real-Clifford+CH circuit. This is done in Definition 8.3. □

## 4.2 A Complete Equational Theory for $\mathcal{P}_N^*$

Thus, for $n \geq 3$, the matrices represented by $n$-qubit real-Clifford+CH circuits form a subgroup of $U_{2^n}(\mathbb{Z}[\frac{1}{\sqrt{2}}])$ of index 4.

Although we are talking about groups, note that in the following we will essentially only work with the monoid structure. Indeed, this does not fundamentally change the presentations since all inverses are easy to express from the original generators, and makes more sense from a quantum circuit point of view.

There is a standard technique for obtaining a complete equational theory for a subgroup when one is known for the whole group, based on the *Reidemeister-Schreier theorem* [30, 31]. This theorem has been adapted to monoids in [6] and used for obtaining complete equational theories for 2-qubit Clifford+T [6] and 3-qubit Clifford+CS [7] quantum circuits. One drawback of this technique is that it often leads to a combinatorial explosion that makes it computationally costly, even in relatively simple cases. This is the reason that the completeness result of [7] is limited to 3 qubit, whereas the presentation of the underlying matrix group is *a priori* valid in arbitrary dimension.

Rather than trying to obtain a complete equational theory for real-Clifford+CH quantum circuits directly by applying the Reidemeister-Schreier theorem to the equational theory of [17], we adopt an alternative approach in two steps:

- first, we use the Reidemeister-Schreier theorem to obtain a presentation of the submonoid of $U_N(\mathbb{Z}[\frac{1}{\sqrt{2}}])$ generated by $\mathcal{P}_N$
- then we use a back-and-forth encoding similar to that of [11] in order to turn this presentation into a complete equational theory for real-Clifford+CH circuits.

Additionally, we perform some simplifications of the equational theories before and after each step. In particular, we start by slightly simplifying the equational theory of [17]; the simplified complete equational theory is shown in Figure 7.

**Proposition 4.8 (Soundness and Completeness of the Equational Theory of Figure 7 w.r.t. $\approx$).** *For any $w_1, w_2 \in \mathcal{G}_N^*$, one has $w_1 \approx w_2$ if and only if $w_1$ and $w_2$ can be transformed into one another using the equations of Figure 7. In other words, $\approx$ is the smallest congruence satisfying the relations of Figure 7.*

PROOF. One can check that the equations of Figure 7 are semantically sound. Hence, by Theorem 4.4, they can be derived from those of Figure 6. Conversely, it remains to prove that the equations of Figure 6 can be derived from those of Figure 7. This is done in Appendix A.2. □

Details about how we use the Reidemeister-Schreier theorem are given in Appendix A.3.1. The raw complete equational theory that we obtain by applying the Reidemeister-Schreier theorem is shown in Appendix A.3.2. Note that it contains many equations, which are not necessarily well-suited for the next step which is to translate them using the decoding function defined in Section 8; hence, before proceeding to the back-and-forth encoding, we simplify this equational theory into a smaller one, shown in Figure 8. The equational theory of Figure 8 is actually not the simplest possible from the point of view of a human reader (although much simpler than that of Appendix A.3.2), but is optimized so as to ease the decoding step.

This gives us the following completeness result:

**Definition 4.9.** *The free monoid generated by $\mathcal{P}_N$ (i.e. the monoid of words over $\mathcal{P}_N$) is denoted by $\mathcal{P}_N^*$. The empty word is denoted by $\epsilon$. The matrix associated with a word $w = P_1 P_2 ... P_k \in \mathcal{P}_N^*$ is called its semantics, and denoted by $\llbracket w \rrbracket \coloneqq P_1 \cdot P_2 \cdot ... \cdot P_k$.*

*We denote by $\approx$ the smallest congruence satisfying the equations of Figure 8. That is, for any two words $w_1, w_2 \in \mathcal{P}_N^*$, one has $w_1 \approx w_2$ iff $w_1$ and $w_2$ can be transformed into one another using the equations of Figure 8.*

**Theorem 4.10 (Soundness and Completeness of $\approx$ for $\mathcal{P}_N^*$).** *Let $w_1, w_2 \in \mathcal{P}_N^*$ with $N \geq 8$. Then $\llbracket w_1 \rrbracket = \llbracket w_2 \rrbracket$ if and only if $w_1 \approx w_2$.*

PROOF. To prove the soundness, it suffices to check that all the equations of Figure 8 preserve the semantics. The proof of completeness is given in Appendices A.3 and A.4.  □

## 5 Auxiliary Equations/Properties of Multi-Controlled Gates

Similarly to [11], the complete equational theory for Real-Clifford+CH circuits that we would obtain naively by the back-and-forth encoding technique described in Sections 7 and 8 consists mainly of equations involving a large (possibly unbounded) number of multi-controlled gates. One of the main contributions of this paper is to derive those equations from the much simpler equational theory shown in Figure 4. To do so, we will rely on auxiliary equations, that we derive from those of Figure 4 in Appendices B and D. Note that like in [11], a key role is played by basic properties of multi-controlled gates.

One can in particular recognize the main auxiliary properties used in [11], namely that the different controls of a single multi-controlled gate play the same role and therefore can be swapped (Equations (307), (313) and (353)); that a positively-controlled gate can be merged with the negatively-controlled version of the same gate (Equations (309), (310) and (354)); and that a positively-controlled gate always commutes with a gate which is negatively-controlled by the same qubit (Equations (335) to (352)). Equations (311), (312), (317) and (355) can be seen as the analogues of the additivity and periodicity properties of [11].

Other notable auxiliary equations, more specific to the real-Clifford+CH fragment, include in particular the fact that the dashed box of the multi-controlled $Z$ and $H$ gates indeed behaves as just an identity wire (Equations (274) and (277)), together with Lemmas D.9 and D.11 and their various particular cases proven beforehand in Lemma D.8.

**Remark 4.** *One can show that Equation (19) can be derived in particular from Equation (272) or from Equation (274). Thus, Equations (272) and (274) are two alternative axiom schemas that could each have* been put instead of Equation (19) in the equational theory of Figure 4, without altering its completeness.

**Remark 5.** *In particular, the 4-qubit case of Equation (272) is trivial since it follows from the topological rules. Hence, the 4-qubit case of Equation (19) can be derived from the other equations (see Equation (154) in Appendix D.3). This partly explains the maybe surprising fact that the equational theory of Figure 4 does not contain any axiom on 4 qubits.*

Note that in the proof of completeness, we will essentially prove the completeness for $n$-qubit circuits, for each $n$ separately. Therefore, we can instead reason by induction on $n$ (the base cases being given by Lemmas 7.1 and 7.6), which does not change the way that the proof of completeness itself is done, but is useful for proving auxiliary equations. This can be reformulated as the fact that we can use the following lemma:

**Lemma 5.1 (Induction Hypothesis for the Proof of Completeness).** *For any real-Clifford+CH circuits $C_1, C_2$, if $\llbracket C_1 \rrbracket = \llbracket C_2 \rrbracket$ then $\mathrm{QC} \vdash (\text{——} \otimes C_1) = (\text{——} \otimes C_2)$ and $\mathrm{QC} \vdash (C_1 \otimes \text{——}) = (C_2 \otimes \text{——})$.*

## 6 The Category of Sequentialized Circuits

As we have said, the proof of completeness will be based on a back-and-forth translation, similar to that of [11], between the PROP of real-Clifford+CH circuits and the monoids $\mathcal{G}_N^*$ and $\mathcal{P}_N^*$. However, real-Clifford+CH circuits have a richer structure: they form not only a monoid (for every particular number of qubits) – which corresponds to Equations (a) and (b) of Figure 1 – but also satisfy the other coherence relations of a PROP. Hence, we cannot directly encode circuits into words since those relations would not be preserved and therefore the encoding would be ill-defined. Instead, we will use what we call *sequentialized circuits*, that is, elements of the category $\mathrm{QC_{seq}}$ of raw circuits quotiented only by Equations (a) and (b).

The PROP $\mathrm{QC}$ of usual real-Clifford+CH circuits is recovered by quotienting by the remaining topological rules, namely Equations (c) to (g).[6] Given a sequentialized circuit $C \in \mathrm{QC_{seq}}$, we denote by $|C|$ its image by the canonical projection onto $\mathrm{QC}$. Conversely, any circuit $C \in \mathrm{QC}$ has several representants $C'$ in $\mathrm{QC_{seq}}$, satisfying $|C'| = C$.

We are now ready to prove the main result of the paper. We proceed as follows: first, we prove the completeness for 1-qubit circuits, which is easy to do by using a normal form, in Section 7.1; then we prove the completeness for 2-qubit circuits by means of a back-and-forth encoding into $\mathcal{G}_4^*$, in Section 7.2. Finally, we prove the completeness for circuits on $n \geq 3$ qubits by a back-and-forth encoding into $\mathcal{P}_{2n}^*$, in Section 8. Note that while the encoding function needs to be defined on sequentialized circuits, the decoding function can be defined directly into the PROP $\mathrm{QC}$.

---

[6] More precisely, by quotienting by the variants of Equations (c) to (g) obtained by projecting the two sides of each equation according to the canonical projection $\mathrm{QC_{raw}} \to \mathrm{QC_{raw}}/\{(a),(b)\}$.

$$((-1)_{[a]}(-1)_{[a]}) \approx \epsilon \tag{20}$$

$$((-1)_{[a]}(-1)_{[b]}) \approx ((-1)_{[b]}(-1)_{[a]}) \tag{21}$$

$$((-1)_{[a]}(-1)_{[b]})\,((-1)_{[b]}(-1)_{[c]}) \approx ((-1)_{[a]}(-1)_{[c]}) \tag{22}$$

$$((-1)_{[a]}(-1)_{[a+1]}) \approx ((-1)_{[a]}X_{[a,a+1]})\,((-1)_{[a]}X_{[a,a+1]}) \tag{23}$$

$$((-1)_{[a]}X_{[a,a+1]})\,((-1)_{[a+1]}X_{[a+1,a+2]})\,((-1)_{[a]}X_{[a,a+1]}) \approx ((-1)_{[a+1]}X_{[a+1,a+2]})\,((-1)_{[a]}X_{[a,a+1]})\,((-1)_{[a+1]}X_{[a+1,a+2]}) \tag{24}$$

$$((-1)_{[a]}X_{[a,c]}) \approx ((-1)_{[a+1]}X_{[a,a+1]})\,((-1)_{[a+1]}X_{[a+1,c]})\,((-1)_{[a]}X_{[a,a+1]}) \qquad a+1<c, \quad c-a \text{ odd} \tag{25}$$

$$((-1)_{[c]}X_{[a,c]}) \approx ((-1)_{[a+1]}X_{[a,a+1]})\,((-1)_{[c]}X_{[a+1,c]})\,((-1)_{[a]}X_{[a,a+1]}) \qquad a+1<c, \quad c-a \text{ odd} \tag{26}$$

$$((-1)_{[a]}X_{[a,c]}) \approx ((-1)_{[c-1]}X_{[c-1,c]})\,((-1)_{[a]}X_{[a,c-1]})\,((-1)_{[c]}X_{[c-1,c]}) \qquad a+1<c, \quad c-a \text{ even} \tag{27}$$

$$((-1)_{[c]}X_{[a,c]}) \approx ((-1)_{[c-1]}X_{[c-1,c]})\,((-1)_{[c-1]}X_{[a,c-1]})\,((-1)_{[c]}X_{[c-1,c]}) \qquad a+1<c, \quad c-a \text{ even} \tag{28}$$

$$((-1)_{[a]}X_{[a,a+1]})\,((-1)_{[a+1]}X_{[a,a+1]}) \approx \epsilon \tag{29}$$

$$((-1)_{[c]}X_{[a,b]}) \approx ((-1)_{[c]}(-1)_{[a]})\,((-1)_{[a]}X_{[a,b]}) \qquad c \neq a,b \tag{30}$$

$$((-1)_{[a+1]}(-1)_{[c]})\,((-1)_{[a]}X_{[a,a+1]}) \approx ((-1)_{[a]}X_{[a,a+1]})\,((-1)_{[a]}(-1)_{[c]}) \qquad c \neq a,a+1 \tag{31}$$

$$((-1)_{[a]}X_{[a,a+1]})\,((-1)_{[b]}X_{[b,b+1]}) \approx ((-1)_{[b]}X_{[b,b+1]})\,((-1)_{[a]}X_{[a,a+1]}) \qquad a+1<b \tag{32}$$

$$((-1)_{[b]}X_{[a,b]}) \approx ((-1)_{[b]}X_{[b,a]}) \tag{33}$$

$$(X_{[a,b]}X_{[c,d]}) \approx ((-1)_{[a]}X_{[a,b]})\,((-1)_{[b]}X_{[c,d]}) \tag{34}$$

$$(H_{[a,b]}H_{[c,d]}) \approx \Sigma_{a,b,c,d} \circ (H_{[0,1]}H_{[3,2]}) \circ \Sigma'_{a,b,c,d} \qquad \text{see caption} \tag{35}$$

$$(H_{[a,b]}H_{[c,d]}) \approx \mathcal{W}_1 \qquad \text{see caption} \tag{36}$$

$$(H_{[a,b]}H_{[c,d]}) \approx \mathcal{W}_2 \qquad \text{see caption} \tag{37}$$

$$((-1)_{[a]}X_{[a,a+1]})\,(H_{[0,1]}H_{[3,2]}) \approx (H_{[0,1]}H_{[3,2]})\,((-1)_{[a]}X_{[a,a+1]}) \qquad a \geq 4 \tag{38}$$

$$((-1)_{[0]}(-1)_{[1]})\,(H_{[0,1]}H_{[3,2]}) \approx (H_{[0,1]}H_{[3,2]})\,((-1)_{[0]}(-1)_{[1]}) \tag{39}$$

$$((-1)_{[3]}(-1)_{[4]})\,(H_{[0,1]}H_{[3,2]}) \approx (H_{[0,1]}H_{[3,2]})\,((-1)_{[3]}(-1)_{[4]})\,((-1)_{[2]}X_{[2,3]}) \tag{40}$$

$$(H_{[0,1]}H_{[3,2]})\,(H_{[0,1]}H_{[3,2]}) \approx \epsilon \tag{41}$$

$$(H_{[a,b]}H_{[e,f]})\,(H_{[e,f]}H_{[c,d]}) \approx (H_{[a,b]}H_{[c,d]}) \qquad \text{see caption} \tag{42}$$

$$(H_{[0,1]}H_{[3,2]})\,(H_{[3,2]}H_{[4,5]}) \approx (X_{[0,3]}X_{[1,2]})\,(H_{[3,2]}H_{[4,5]})\,(X_{[0,3]}X_{[1,2]}) \tag{43}$$

$$(X_{[0,3]}X_{[1,2]})\,(H_{[0,1]}H_{[3,2]}) \approx (H_{[0,1]}H_{[3,2]})\,(X_{[0,3]}X_{[1,2]}) \tag{44}$$

$$\begin{aligned}(H_{[0,1]}H_{[7,6]})\,(H_{[0,3]}H_{[1,2]})\,(H_{[0,1]}H_{[7,6]})\,((-1)_{[0]}(-1)_{[1]})\,(H_{[0,3]}H_{[1,2]}) &\approx \\ (H_{[0,3]}H_{[1,2]})\,((-1)_{[0]}(-1)_{[1]})\,(H_{[0,1]}H_{[7,6]})\,(H_{[0,3]}H_{[1,2]})\,(H_{[0,1]}H_{[7,6]})& \end{aligned} \tag{45}$$

$$\begin{aligned}(H_{[0,1]}H_{[7,6]})\,(H_{[0,3]}H_{[1,2]})\,(H_{[3,4]}H_{[2,5]})\,(H_{[3,2]}H_{[4,5]})\,(H_{[7,6]}H_{[4,5]})\,((-1)_{[2]}(-1)_{[3]})\,(H_{[3,4]}H_{[2,5]})\,(H_{[0,3]}H_{[1,2]}) &\approx \\ (H_{[0,3]}H_{[1,2]})\,(H_{[3,4]}H_{[2,5]})\,((-1)_{[2]}(-1)_{[3]})\,(H_{[7,6]}H_{[4,5]})\,(H_{[3,2]}H_{[4,5]})\,(H_{[3,4]}H_{[2,5]})\,(H_{[0,3]}H_{[1,2]})\,(H_{[0,1]}H_{[7,6]})& \end{aligned} \tag{46}$$

**Figure 8: Simpler equational theory, equivalent to that of Appendix A.3.2 (and therefore complete) when the dimension is at least 8.** In Equation (36), we require that the tuple $(G_n(a), G_n(b), G_n(c), G_n(d))$ be of the form $(x0y0z, x1y0z, x0y1z, x1y1z)$ for some bitstrings $x, y, z$; in Equation (37), we require that it be of the form $(x0y0z, x0y1z, x1y0z, x1y1z)$; and in Equation (35), we require that it be in neither of these two forms. $\Sigma_{a,b,c,d}$ and $\Sigma'_{a,b,c,d}$ are defined in Definition 8.3. In Equation (42), we require that $|\{a, b, c, d\}| \leq 3$ and that $e < f$ be the two smallest elements of $\{0, ..., 2^n - 1\} \setminus \{a, b, c, d\}$. The right-hand sides of Equations (36) and (37) are words specifically tailored to behave well with respect to decoding (see Definitions 8.3 and 2.4 and Appendix E.5); they are defined in Appendix A.4.1.

## 7 One- and Two-Qubit Completeness

### 7.1 One-Qubit Completeness

**Lemma 7.1 (1-Qubit Completeness).** *For any 1-qubit quantum circuits $C_1, C_2$ over $\{-\boxed{H}-, -\bullet-\}$, if $[\![C_1]\!] = [\![C_2]\!]$ then $\mathrm{QC} \vdash C_1 = C_2$.*

**Proof.** We say that a 1-qubit circuit is in *normal form* if it is of the form

$$\left(-\boxed{H}-\bullet-\right)^k \qquad \text{or} \qquad -\boxed{H}- \circ \left(-\boxed{H}-\bullet-\right)^k$$

where $k \in \{0, ..., 7\}$.

One can check by direct computation that if two circuits in normal form have the same semantics then they are equal. Thus, it remains to prove that any 1-qubit circuit can be put in normal form.

Note that a 1-qubit circuit is just a sequence of $-\boxed{H}-$ and $-\bullet-$ gates. By repeatedly applying Equations (1) and (2), we can ensure that it does not contain two consecutive identical gates. That is, it alternates between $-\boxed{H}-$ and $-\bullet-$, or in other words, it is of one of the following three forms:

$$\left(-\boxed{H}-\bullet-\right)^k \qquad -\boxed{H}-\circ\left(-\boxed{H}-\bullet-\right)^k \qquad \left(-\boxed{H}-\bullet-\right)^k\circ-\bullet-$$

By applying Equation (3) $\lfloor\frac{k}{8}\rfloor$ times, we can ensure that $k\in\{0,...,7\}$. Thus, if the circuit is in one of the first two forms, then it is in normal form. If it is of the third form, then we do the following transformations:

$$
\begin{aligned}
\left(-\boxed{H}-\bullet-\right)^k\circ-\bullet- \;&\overset{(3)}{=}\; \left(-\boxed{H}-\bullet-\right)^k\circ-\bullet-\circ\left(-\boxed{H}-\bullet-\right)^8 \\
&=\; -\bullet-\circ\left(-\bullet-\boxed{H}-\right)^k\circ\left(-\boxed{H}-\bullet-\right)^8 \\
&\overset{(2)(1)}{=}\; -\bullet-\circ\left(-\boxed{H}-\bullet-\right)^{8-k} \\
&\overset{(2)}{=}\; -\boxed{H}-\circ\left(-\boxed{H}-\bullet-\right)^{8-k-1}
\end{aligned}
$$

which gives us a circuit in normal form. □

## 7.2 Two-Qubit Completeness

**Definition 7.2 (2-Qubit Encoding).** *Let $E : \mathbf{QC}^{(2)}_{\mathbf{seq}} \to \mathcal{G}^*_4$ be defined as follows: for any 2-qubit sequentialized circuit $C$, let $E(C) = E_{0,0}(C)$, where $E_{k,\ell}$ is defined on any $m$-qubit sequentialized circuit satisfying $k+m+\ell=2$, inductively as follows:*

- *$E_{k,\ell}(C_1\otimes C_2) = E_{k+n_1,\ell}(C_2)\circ E_{k,\ell+n_2}(C_1)$, where $C_1$ (resp. $C_2$) is acting on $n_1$ (resp. $n_2$) qubits;*
- *$E_{k,\ell}(C_2\circ C_1) = E_{k,\ell}(C_2)\circ E_{k,\ell}(C_1)$;*

*and for the generators:*

- *$E_{k,\ell}(\begin{smallmatrix}\phantom{.}\end{smallmatrix}) = \epsilon$,*
- *$E_{k,\ell}(\text{———}) = \epsilon$,*
- *$E_{0,0}(\smile\frown) = X_{[1,2]}$,*
- *$E_{1,0}(-\boxed{H}-) = H_{[0,1]}H_{[2,3]}$,*
- *$E_{0,1}(-\boxed{H}-) = H_{[0,2]}H_{[1,3]}$,*
- *$E_{1,0}(-\bullet-) = (-1)_{[1]}(-1)_{[3]}$,*
- *$E_{0,1}(-\bullet-) = (-1)_{[2]}(-1)_{[3]}$,*
- *$E_{0,0}(\;\top\;) = (-1)_{[3]}$,*
- *$E_{0,0}(\;\bot\;) = H_{[2,3]}$.*

**Definition 7.3 (2-Qubit Decoding).** *Let $D: \mathcal{G}^*_4 \to \mathbf{QC}^{(2)}$ be the monoid homomorphism defined as follows on the generators:*

- $D((-1)_{[0]}) =$ ⊥,
- $D((-1)_{[1]}) =$ ⊥,
- $D((-1)_{[2]}) =$ ⊥,
- $D((-1)_{[3]}) =$ ⊥,
- $D(X_{[2,3]}) = D(X_{[3,2]}) =$ ⊕,
- $D(X_{[0,1]}) = D(X_{[1,0]}) =$ ⊕,
- $D(X_{[0,2]}) = D(X_{[2,0]}) =$ ⊕,
- $D(X_{[1,3]}) = D(X_{[3,1]}) =$ ⊕,
- $D(X_{[1,2]}) = D(X_{[2,1]}) =$ ⤬,
- $D(X_{[0,3]}) = D(X_{[3,0]}) =$ ⤬⊕,
- $D(H_{[2,3]}) =$ ⊞,
- $D(H_{[0,1]}) =$ ⊞,
- $D(H_{[0,2]}) =$ ⊞,
- $D(H_{[1,3]}) =$ ⊞,
- $D(H_{[2,1]}) =$ ⊕⊞,
- $D(H_{[0,3]}) =$ ⊕⊞,
- $D(H_{[3,2]}) =$ ⊞,
- $D(H_{[1,0]}) =$ ⊞,
- $D(H_{[2,0]}) =$ ⊕⊞,
- $D(H_{[3,1]}) =$ ⊕⊞,
- $D(H_{[1,2]}) =$ ⊞,
- $D(H_{[3,0]}) =$ ⊕⊞.

**Lemma 7.4.** *For any 2-qubit circuit $C$, and any sequentialized circuit $C'$ such that $|C'| = C$,*

$$\mathbf{QC}\vdash D(E(C')) = C.$$

PROOF. The proof is given in Appendix C.1. □

**Lemma 7.5.** *For any words $w_1, w_2 \in \mathcal{G}^*_4$,*

$$\textit{if } w_1 \approx w_2 \textit{ then } \mathbf{QC}\vdash D(w_1) = D(w_2).$$

PROOF. The proof is given in Appendix C.2. □

**Lemma 7.6 (2-Qubit Completeness).** *For any 2-qubit real-Clifford+CH quantum circuits $C_1, C_2$,*

$$\textit{if } [\![C_1]\!] = [\![C_2]\!] \textit{ then } \mathbf{QC}\vdash C_1 = C_2.$$

PROOF. Let $C'_1, C'_2$ be representatives in $\mathbf{QC}^{(2)}_{\mathbf{seq}}$ of respectively $C_1$ and $C_2$, so that $|C'_1| = C_1$ and $|C'_2| = C_2$. Since $[\![C_1]\!] = [\![C_2]\!]$ and the encoding preserves the semantics, one has $[\![E(C'_1)]\!] = [\![E(C'_1)]\!]$. Hence, by Theorem 4.4, we have $E(C'_1) \approx E(C'_2)$. By Lemma 7.5, this implies that $\mathbf{QC}\vdash D(E(C'_1)) = D(E(C'_2))$. In turn, by Lemma 7.4, this implies that $\mathbf{QC}\vdash C_1 = C_2$. □

## 8 Completeness in the General Case

To define the back-and-forth encoding between $n$-qubits circuits and the elements of $\mathcal{P}_{2^n}$ which represent $2^n$-dimensional matrices, it will be handful, like in [11], to encode the integers $0,...,2^n-1$ as $n$-length bitstrings, by using not the usual binary encoding but rather the *Gray code*, which has the advantage that the encodings of two consecutive integers always differ by exactly one bit:

**Definition 8.1 (Gray Code [11]).** *The $n$-bit Gray code of an integer $k\in\{0,...,2^n-1\}$ is the word $G_n(k)\in\{0,1\}^n$, inductively defined by $G_0(0) = \epsilon$ and*

$$G_n(k) = \begin{cases} 0G_{n-1}(k) & \textit{if } k < 2^{n-1}, \\ 1G_{n-1}(2^n-1-k) & \textit{if } k \geq 2^{n-1}. \end{cases}$$

One can easily check that $G_n \colon \{0, ..., 2^n - 1\} \to \{0,1\}^n$ is a bijection.

**Definition 8.2 (Encoding).** *For any $n \geq 3$, let $E \colon \mathbf{QC_{seq}} \to \mathcal{P}_{2^n}^*$ be defined as follows: for any $n$-qubit sequentialized circuit $C$, let $E(C) = E_{0,0}(C)$, where $E_{k,\ell}$ is defined on any $m$-qubit sequentialized circuit satisfying $k + m + \ell = n$, inductively as follows (where the order of the products is left unspecified since it does not matter in the proof):*

- $E_{k,\ell}(C_1 \otimes C_2) = E_{k+n_1,\ell}(C_2) \circ E_{k,\ell+n_2}(C_1)$, *where $C_1$ (resp. $C_2$) is acting on $n_1$ (resp. $n_2$) qubits;*

- $E_{k,\ell}(C_2 \circ C_1) = E_{k,\ell}(C_2) \circ E_{k,\ell}(C_1)$;

- $E_{k,\ell}(\overset{\cdot\;\cdot}{\underset{\cdot\;\cdot}{}}) = \epsilon$;

- $E_{k,\ell}(\underline{\quad}) = \epsilon$;

*if $\ell \geq 1$, then*

- $E_{k,\ell}(\text{---•---}) = \prod_{\substack{x \in \{0,1\}^k \\ y \in \{0,1\}^{\ell-1}}} \left( (-1)_{[G_n^{-1}(x10y)]} (-1)_{[G_n^{-1}(x11y)]} \right)$;

- $E_{k,\ell}(\overset{\bullet}{\underset{\bullet}{\mid}}) = \prod_{\substack{x \in \{0,1\}^k \\ y \in \{0,1\}^{\ell-1}}} \left( (-1)_{[G_n^{-1}(x110y)]} (-1)_{[G_n^{-1}(x111y)]} \right)$;

- $E_{k,\ell}(\text{--}\boxed{H}\text{--}) = \prod_{\substack{x \in \{0,1\}^k \\ y \in \{0,1\}^{\ell-1}}} \left( H_{[G_n^{-1}(x00y),G_n^{-1}(x10y)]} H_{[G_n^{-1}(x01y),G_n^{-1}(x11y)]} \right)$;

- $E_{k,\ell}(\overset{\bullet}{\underset{\boxed{H}}{\mid}}) = \prod_{\substack{x \in \{0,1\}^k \\ y \in \{0,1\}^{\ell-1}}} \left( H_{[G_n^{-1}(x100y),G_n^{-1}(x110y)]} H_{[G_n^{-1}(x101y),G_n^{-1}(x111y)]} \right)$;

*and if $\ell = 0$ (note that because of the condition $k + m + \ell = n$, we necessarily have $k \geq 1$ in this case), then*

- $E_{k,0}(\text{---•---}) = \prod_{x \in \{0,1\}^{k-1}} \left( (-1)_{[G_n^{-1}(0x1)]} (-1)_{[G_n^{-1}(1x1)]} \right)$;

- $E_{k,0}(\overset{\bullet}{\underset{\bullet}{\mid}}) = \prod_{x \in \{0,1\}^{k-1}} \left( (-1)_{[G_n^{-1}(0x11)]} (-1)_{[G_n^{-1}(1x11)]} \right)$;

- $E_{k,0}(\text{--}\boxed{H}\text{--}) = \prod_{x \in \{0,1\}^{k-1}} \left( H_{[G_n^{-1}(0x0),G_n^{-1}(0x1)]} H_{[G_n^{-1}(1x0),G_n^{-1}(1x1)]} \right)$;

- $E_{k,0}(\overset{\bullet}{\underset{\boxed{H}}{\mid}}) = \prod_{x \in \{0,1\}^{k-1}} \left( H_{[G_n^{-1}(0x10),G_n^{-1}(0x11)]} H_{[G_n^{-1}(1x10),G_n^{-1}(1x11)]} \right)$;

*finally, for any $\ell \geq 0$,*

- $E_{k,\ell}(\overset{\smile}{\frown}) = E_{k,\ell}(\overset{\bullet}{\underset{\bullet}{\mid}}) \circ$

$$\prod_{\substack{x \in \{0,1\}^k \\ y \in \{0,1\}^{\ell}}} \left( (-1)_{[G_n^{-1}(x11y)]} X_{[G_n^{-1}(x10y),G_n^{-1}(x11y)]} \right) \circ$$

$$\prod_{\substack{x \in \{0,1\}^k \\ y \in \{0,1\}^{\ell}}} \left( (-1)_{[G_n^{-1}(x01y)]} X_{[G_n^{-1}(x01y),G_n^{-1}(x11y)]} \right) \circ$$

$$\prod_{\substack{x \in \{0,1\}^k \\ y \in \{0,1\}^{\ell}}} \left( (-1)_{[G_n^{-1}(x11y)]} X_{[G_n^{-1}(x10y),G_n^{-1}(x11y)]} \right).$$

**Definition 8.3 (Decoding).** *For any $n \geq 3$, let $D \colon \mathcal{P}_{2^n}^* \to \mathbf{QC}$ be the monoid homomorphism defined as follows on the generators:*

**Case of** $((-1)_{[a]}(-1)_{[b]})$.

- *For any $a \in \{0, ..., 2^n - 1\}$, we define*

$$D((-1)_{[a]}(-1)_{[a]}) := (\text{---------})^{\otimes n}.$$

- *For any $a \in \{0, ..., 2^n - 2\}$, we define*

$$D\big((-1)_{[a]}(-1)_{[a+1]}\big) := \boxed{\text{circuit diagram with } \Lambda^x \text{ and } \Lambda^y}$$

*where $\beta \in \{0,1\}$, $x \in \{0,1\}^k$, and $y \in \{0,1\}^{n-k-1}$ are such that $G_n(a) = x\beta y$ and $G_n(a+1) = x\bar{\beta}y$.[7]*

- *If $a \leq b$, then we define[8]*

$$D\big((-1)_{[a]}(-1)_{[b]}\big) := \prod_{k=a}^{b-1} D\big((-1)_{[k]}(-1)_{[k+1]}\big)$$

*where we use the notations $\prod_{k=n}^{m} C_k := C_m \circ \cdots \circ C_n$ and $\prod_{k=n}^{m} C_k := C_n \circ \cdots \circ C_m$, with by convention the products being the identity if they are empty (i.e. if $m < n$).*

- *Finally, if $a > b$, we define $D\big((-1)_{[a]}(-1)_{[b]}\big) := D\big((-1)_{[b]}(-1)_{[a]}\big)$.*

**Case of** $((-1)_{[c]} X_{[a,b]})$.

- *For any $a \in \{0, ..., 2^n - 2\}$, we define*

$$D\big((-1)_{[a]} X_{[a,a+1]}\big) := \boxed{\text{circuit}} \quad , \quad D\big((-1)_{[a+1]} X_{[a,a+1]}\big) := \boxed{\text{circuit}}$$

*where $\beta \in \{0,1\}$, $x \in \{0,1\}^k$, and $y \in \{0,1\}^{n-k-1}$ are such that $G_n(a) = x\beta y$ and $G_n(a+1) = x\bar{\beta}y$.*

- *If $a < b$, then we define inductively on $b - a$ (using the previous case as base case):*

– *if $b - a$ is odd,* $D\big((-1)_{[a]} X_{[a,b]}\big) :=$

$D\big((-1)_{[a+1]} X_{[a,a+1]}\big) \circ D\big((-1)_{[a+1]} X_{[a+1,b]}\big) \circ D\big((-1)_{[a]} X_{[a,a+1]}\big)$

*and* $D\big((-1)_{[b]} X_{[a,b]}\big) :=$

$D\big((-1)_{[a+1]} X_{[a,a+1]}\big) \circ D\big((-1)_{[b]} X_{[a+1,b]}\big) \circ D\big((-1)_{[a]} X_{[a,a+1]}\big)$;

– *if $b - a$ is even,* $D\big((-1)_{[a]} X_{[a,b]}\big) :=$

$D\big((-1)_{[b-1]} X_{[b-1,b]}\big) \circ D\big((-1)_{[a]} X_{[a,b-1]}\big) \circ D\big((-1)_{[b]} X_{[b-1,b]}\big)$

*and* $D\big((-1)_{[b]} X_{[a,b]}\big) :=$

$D\big((-1)_{[b-1]} X_{[b-1,b]}\big) \circ D\big((-1)_{[b-1]} X_{[a,b-1]}\big) \circ D\big((-1)_{[b]} X_{[b-1,b]}\big).$

- *If $a > b$, then we define*

$$D\big((-1)_{[a]} X_{[a,b]}\big) := D\big((-1)_{[a]} X_{[b,a]}\big),$$

$$D\big((-1)_{[b]} X_{[a,b]}\big) := D\big((-1)_{[b]} X_{[b,a]}\big).$$

---

[7] $\beta, x, y$ necessarily exist by the properties of the Gray code.
[8] Note that the definition is consistent in the case where $b = a + 1$.

- Finally, if $c \neq a, b$, then we define[9]

$$D\left((-1)_{[c]}X_{[a,b]}\right) := D\left((-1)_{[c]}(-1)_{[a]}\right) \circ D\left((-1)_{[a]}X_{[a,b]}\right).$$

**Case of** $\left(X_{[a,b]}X_{[c,d]}\right)$**.** For any $a, b, c, d$ with $a \neq b$ and $c \neq d$, we define

$$D\left(X_{[a,b]}X_{[c,d]}\right) := D\left((-1)_{[a]}X_{[a,b]}\right) \circ D\left((-1)_{[b]}X_{[c,d]}\right).$$

**Case of** $\left(H_{[a,b]}H_{[c,d]}\right)$**.**

- If there exist $x, y, z$ such that $G_n(a) = x0y0z$, $G_n(b) = x1y0z$, $G_n(c) = x0y1z$, and $G_n(d) = x1y1z$, then we define

$$D\left(H_{[a,b]}H_{[c,d]}\right) := \left[\begin{array}{c} \Lambda^x \\ \hline H \\ \hline \Lambda^y \\ \hline \Lambda^z \end{array}\right].$$

- If there exist $x, y, z$ such that $G_n(a) = x0y0z$, $G_n(b) = x0y1z$, $G_n(c) = x1y0z$, and $G_n(d) = x1y1z$, then we define

$$D\left(H_{[a,b]}H_{[c,d]}\right) := \left[\begin{array}{c} \Lambda^x \\ \hline \Lambda^y \\ \hline H \\ \hline \Lambda^z \end{array}\right].$$

- In the other cases where $|\{a, b, c, d\}| = 4$, we define:

$$D\left(H_{[a,b]}H_{[c,d]}\right) := D(\Sigma_{a,b,c,d}) \circ \left[\begin{array}{c} \vdots \\ \hline H \end{array}\right] \circ D(\Sigma'_{a,b,c,d})$$

where $\Sigma_{a,b,c,d}$ and $\Sigma'_{a,b,c,d}$ are words over $\mathcal{P}_N$, not containing any generators of the form $\left(H_{[e,f]}H_{[g,h]}\right)$, such that $\left[\!\left[\Sigma'_{a,b,c,d}\right]\!\right] = \left[\!\left[\Sigma_{a,b,c,d}\right]\!\right]^{-1}$, and either $\left[\!\left[\Sigma_{a,b,c,d}\right]\!\right] = \sigma$ or $\left[\!\left[\Sigma_{a,b,c,d}\right]\!\right] = \sigma \circ \left[\!\left[(-1)_{[\mu]}\right]\!\right]$, where $\mu \notin \{0, 1, 3, 2\}$ and $\sigma$ is a permutation matrix mapping respectively $|0\rangle, |1\rangle, |3\rangle, |2\rangle$ to $|a\rangle, |b\rangle, |c\rangle, |d\rangle$.

Note that the precise definitions of $\Sigma_{a,b,c,d}$ and $\Sigma'_{a,b,c,d}$ do not matter in the proof. Nonetheless, an example of possible definitions is given in Appendix E.1.

- Finally, if $|\{a, b, c, d\}| < 4$, then we define $D\left(H_{[a,b]}H_{[c,d]}\right) := D\left(H_{[a,b]}H_{[e,f]}\right) \circ D\left(H_{[e,f]}H_{[c,d]}\right)$, where $e, f$ with $e < f$ are the two smallest elements of $\{0, ..., 2^n - 1\} \setminus \{a, b, c, d\}$.

---

[9]Note that since $D\left((-1)_{[a]}(-1)_{[a]}\right)$ is defined as the identity, the equality is still true if $a = c$ or $a = d$.

**Lemma 8.4.** For any bitstrings $x \in \{0,1\}^k$ and $y \in \{0,1\}^\ell$ with $k, \ell \geq 0$ and $k + \ell = n - 1$,

$$QC \vdash D\left((-1)_{[G_n^{-1}(x0y)]}(-1)_{[G_n^{-1}(x1y)]}\right) = \left[\begin{array}{c} \Lambda^x \\ \hline \\ \hline \Lambda^y \end{array}\right]$$

**Proof.** The proof is given in Appendix E.2. □

**Lemma 8.5.** For any bitstrings $x \in \{0,1\}^k$ and $y \in \{0,1\}^\ell$ with $k, \ell \geq 0$ and $k + \ell = n - 1$,

$$QC \vdash D\left((-1)_{[G_n^{-1}(x0y)]}X_{[G_n^{-1}(x0y), G_n^{-1}(x1y)]}\right) = \left[\begin{array}{c} \Lambda^x \\ \hline XZ \\ \hline \Lambda^y \end{array}\right]$$

$$QC \vdash D\left((-1)_{[G_n^{-1}(x1y)]}X_{[G_n^{-1}(x0y), G_n^{-1}(x1y)]}\right) = \left[\begin{array}{c} \Lambda^x \\ \hline ZX \\ \hline \Lambda^y \end{array}\right]$$

**Proof.** The proof is given in Appendix E.3. □

**Lemma 8.6.** For any $n$-qubit circuit $C \in \mathbf{QC}$, any $C' \in \mathbf{QC_{seq}}$ such that $|C'| = C$, and any $k, \ell$ such that $k + n + \ell \geq 3$,

$$QC \vdash D(E_{k,\ell}(C')) = id_k \otimes C \otimes id_\ell.$$

**Proof.** The proof is given in Appendix E.4. □

**Lemma 8.7.** For any $n$-qubit circuit $C \in \mathbf{QC}$ with $n \geq 3$, and any $C' \in \mathbf{QC_{seq}}$ such that $|C'| = C$,

$$QC \vdash D(E(C')) = C.$$

**Proof.** This follows directly from Lemma 8.6. □

**Lemma 8.8.** For any $w_1, w_2 \in \mathcal{P}^*_{2^n}$ with $n \geq 3$,

$$\text{if } w_1 \approx w_2 \text{ then } QC \vdash D(w_1) = D(w_2).$$

**Proof.** The proof is given in Appendix E.5. □

**Theorem 8.9 (Completeness).** For any circuits $C_1, C_2 \in \mathbf{QC}$,

$$\text{if } \left[\!\left[C_1\right]\!\right] = \left[\!\left[C_2\right]\!\right] \text{ then } QC \vdash C_1 = C_2.$$

**Proof.** Let $C_1, C_2$ be $n$-qubit circuits such that $\left[\!\left[C_1\right]\!\right] = \left[\!\left[C_2\right]\!\right]$.

If $n = 0$, then the result is trivial since $\left[\begin{array}{c}\vdots\end{array}\right]$ is the only 0-qubit circuit. If $n = 1$ or $n = 2$, then the result is given by Lemmas 7.1 and 7.6 respectively.

If $n \geq 3$, let $C'_1, C'_2$ be representatives in $\mathbf{QC_{seq}}$ of respectively $C_1$ and $C_2$, so that $|C'_1| = C_1$ and $|C'_2| = C_2$. Since $\left[\!\left[C_1\right]\!\right] = \left[\!\left[C_2\right]\!\right]$ and the encoding preserves the semantics, one has $\left[\!\left[E(C'_1)\right]\!\right] = \left[\!\left[E(C'_2)\right]\!\right]$. Hence, by Theorem 4.10, we have $E(C'_1) \approx E(C'_2)$. By Lemma 8.8, this implies that $QC \vdash D(E(C'_1)) = D(E(C'_2))$. In turn, by Lemma 8.7, this implies that $QC \vdash C_1 = C_2$. □

## 9 Conclusion and Perspectives

We have introduced a complete equational theory for unitary quantum circuits over the gate set $\left\{ \begin{array}{c} \text{—}\square\text{—}, \text{—}\bullet\text{—}, \begin{array}{c}\bullet\\\square\end{array}, \begin{array}{c}\bullet\\\boxed{H}\end{array} \right\}$. To our knowledge, this is the second universal fragment of unitary quantum circuits to be given a complete axiomatization, and the first which is finitely generated. One can note the relative similarity between our complete equational theory and that of [9], which is proved to be minimal: both equational theories are composed of a few simple equations on at most 3 qubits, plus an equation schema consisting of one equation for each number of qubits; in both cases, the equation schema has a trivial semantic interpretation, and its meaning comes from the fact that the multi-controlled gate is expressed as a macro built from 1- and 2- qubit gates. A difference between our equational theory and that of [9] is that since we are working in a discrete setting, ours does not require working with continuous parameters.

An obvious follow-up question for future work is whether the unbounded equation schema of Equation (19) is actually necessary, or can be derived from a finite number of equations. In the second case, one would have to find a derivation of it; in the first case, one would have to find an argument to show its necessity. Note that the necessity argument used in [9] does not apply to our case.

Our result is for unitary circuits, without ancillas. One could proceed in the same way as in [10] to extend the equational theory and its completeness to circuits with qubit initializations, ancillas, and qubit discarding (i.e. trace-out). If Equation (19) turned out to be necessary in the unitary setting, then another open question would be whether it can be derived in the presence of ancillas, similarly to what is done in [9, 10].

Another open problem is to find complete equational theories for other relevant fragments of quantum circuits. In particular, one could a priori proceed in the same way as we do in this paper in order to obtain complete equational theories for the Toffoli-Hadamard and Clifford+CS fragments, relying on the corresponding matrix group presentations given respectively in [27] and [5]. Doing the same for Clifford+T would require first to extend the presentation of [18] to arbitrary dimension.

### Acknowledgments

This work has been partially funded by the French National Research Agency (ANR) within the framework of "Plan France 2030", under the research projects EPIQ ANR-22-PETQ-0007, HQI-Acquisition ANR-22-PNCQ-0001 and HQI-R&D ANR-22-PNCQ-0002.

## Appendix Contents

## A  Auxiliary Language

### A.1  Modified Exact Synthesis Algorithm

### A.2  Slight Simplification of the Complete Equational Theory of [17]

**Lemma A.1.** *The equations of Figure 6 can be derived from those of Figure 7.*

PROOF. Strictly speaking, except for Equations (e1) and (e2), we only need to consider the cases where the two indices of each generator are in increasing order. However, for more simplicity, we prove each equation in full generality without this restriction (note that the two indices of each generator still need to be distinct).

*Proof of Equation* (a1):

- If $a = 0$, then Equation (a1) is just Equation (a1*).
- If $a \neq 0$:

$$
\begin{aligned}
(-1)^2_{[a]} \;&\overset{(a2)}{\approx}\; X^2_{[0,a]}(-1)^2_{[a]} \\
&\overset{(c1)}{\approx}\; X_{[0,a]}(-1)^2_{[0]}X_{[0,a]} \\
&\overset{(a1^*)}{\approx}\; X^2_{[0,a]} \\
&\overset{(a2)}{\approx}\; \epsilon
\end{aligned}
$$

**Algorithm 1** Exact synthesis for $U_N(\mathbb{Z}[\frac{1}{\sqrt{2}}])$ (cf. [17]).

---

**Input:** An element $A$ of $U_N(\mathbb{Z}[\frac{1}{\sqrt{2}}])$
**Output:** A sequence $\mathbf{W}_1, \ldots, \mathbf{W}_\ell$ of words over $\mathcal{G}_N$ such that $\mathbf{W}_\ell \cdots \mathbf{W}_1 A = I$

1:  $B \leftarrow A$
2:  **for** $j \leftarrow N-1, N-2, \ldots, 0$ **do**
3:      $k \leftarrow \text{lde}(Be_j)$                                                   ▷ See [17] for the definition of lde
4:      **while** $k > 0$ **do**
5:          Let $i_1$ be the least integer such that $\sqrt{2}^k B[i_1, j] \equiv 1 \pmod 2$ or $\equiv 1 + \sqrt{2} \pmod 2$
6:          Let $i_2$ be the least integer $> i_1$ such that $\sqrt{2}^k B[i_2, j] \equiv \sqrt{2}^k B[i_1, j] \pmod 2$    ▷ See [1] for the existence of $i_1$ and $i_2$
7:          $\mathbf{W} \leftarrow H_{[i_1, i_2]}$
8:          Output $\mathbf{W}$
9:          $B \leftarrow \mathbf{W}B, k \leftarrow \text{lde}(Be_j)$
10:      **end while**
11:  **end for**                                                                      ▷ $\text{lde}(Be_j) = 0$ for all $j$, therefore $B$ is now a permutation matrix
12:  **for** $j \leftarrow N-1, N-2, \ldots, 0$ **do**
13:      $v \leftarrow Be_j$
14:      **if** $v \neq e_j$ **then**
15:          There are $a \leq j$ and $\tau \in \{0, 1\}$ such that $v = (-1)^\tau e_a$.
16:          **if** $a = j$ **then** $\mathbf{W} \leftarrow (-1)^\tau_{[a]}$
17:          **if** $a < j$ **then** $\mathbf{W} \leftarrow X_{[a,j]}(-1)^\tau_{[a]}$
18:          Output $\mathbf{W}$
19:          $B \leftarrow \mathbf{W}B$
20:      **end if**
21:  **end for**

---

*Proof of Equation* (e1):

$$
\begin{aligned}
X_{[b,c]} &\overset{\text{(a2)}}{\approx} X_{[b,c]} X_{[b,c]} X_{[b,c]} \\
&\overset{\text{(a3)}}{\approx} X_{[b,c]} H_{[b,c]} H_{[b,c]} X_{[b,c]} X_{[b,c]} \\
&\overset{\text{(d2)}}{\approx} X_{[b,c]} H_{[b,c]} (-1)_{[c]} H_{[b,c]} X_{[b,c]} \\
&\overset{\text{(e2*)}}{\approx} H_{[c,b]} X_{[b,c]} (-1)_{[c]} H_{[b,c]} X_{[b,c]} \\
&\overset{\text{(c1)}}{\approx} H_{[c,b]} (-1)_{[b]} X_{[b,c]} H_{[b,c]} X_{[b,c]} \\
&\overset{\text{(e2*)}}{\approx} H_{[c,b]} (-1)_{[b]} H_{[c,b]} X_{[b,c]} X_{[b,c]} \\
&\overset{\text{(a2)}}{\approx} H_{[c,b]} (-1)_{[b]} H_{[c,b]} \\
&\overset{\text{(d2)}}{\approx} H_{[c,b]} H_{[c,b]} X_{[c,b]} \\
&\overset{\text{(a3)}}{\approx} X_{[c,b]}
\end{aligned}
$$

*Proof of Equation* (e2):

$$
\begin{aligned}
H_{[c,b]} &\overset{\text{(a2)}}{\approx} H_{[c,b]} X_{[b,c]} X_{[b,c]} \\
&\overset{\text{(e2*)}}{\approx} X_{[b,c]} H_{[b,c]} X_{[b,c]}
\end{aligned}
$$

*Proof of Equation* (c3):

$$
\begin{aligned}
X_{[b,c]} X_{[a,b]} &\overset{\text{(a3)}}{\approx} X_{[b,c]} H_{[a,b]} H_{[a,b]} X_{[a,b]} \\
&\overset{\text{(d2)}}{\approx} X_{[b,c]} H_{[a,b]} (-1)_{[b]} H_{[a,b]}
\end{aligned}
$$

$$\stackrel{\text{(c5)}}{\approx} H_{[a,c]} X_{[b,c]} (-1)_{[b]} H_{[a,b]}$$

$$\stackrel{\text{(a2)}}{\approx} H_{[a,c]} X_{[b,c]} (-1)_{[b]} X_{[b,c]} X_{[b,c]} H_{[a,b]}$$

$$\stackrel{\text{(c1)}}{\approx} H_{[a,c]} X_{[b,c]} X_{[b,c]} (-1)_{[c]} X_{[b,c]} H_{[a,b]}$$

$$\stackrel{\text{(a2)}}{\approx} H_{[a,c]} (-1)_{[c]} X_{[b,c]} H_{[a,b]}$$

$$\stackrel{\text{(c5)}}{\approx} H_{[a,c]} (-1)_{[c]} H_{[a,c]} X_{[b,c]}$$

$$\stackrel{\text{(d2)}}{\approx} H_{[a,c]} H_{[a,c]} X_{[a,c]} X_{[b,c]}$$

$$\stackrel{\text{(a3)}}{\approx} X_{[a,c]} X_{[b,c]}$$

*Proof of Equation* (c2):

$$X_{[a,b]} X_{[a,c]} \stackrel{\text{(e1)}}{\approx} X_{[a,b]} X_{[c,a]}$$

$$\stackrel{\text{(c3)}}{\approx} X_{[c,b]} X_{[a,b]}$$

$$\stackrel{\text{(e1)}}{\approx} X_{[b,c]} X_{[a,b]}$$

*Proof of Equation* (c4):

$$X_{[a,b]} H_{[a,c]} \stackrel{\text{(a2)}}{\approx} X_{[a,b]} H_{[a,c]} X_{[c,a]} X_{[c,a]}$$

$$\stackrel{\text{(e2*)}}{\approx} X_{[a,b]} X_{[c,a]} H_{[c,a]} X_{[c,a]}$$

$$\stackrel{\text{(c3)}}{\approx} X_{[c,b]} X_{[a,b]} H_{[c,a]} X_{[c,a]}$$

$$\stackrel{\text{(c5)}}{\approx} X_{[c,b]} H_{[c,b]} X_{[a,b]} X_{[c,a]}$$

$$\stackrel{\text{(c3)}}{\approx} X_{[c,b]} H_{[c,b]} X_{[c,b]} X_{[a,b]}$$

$$\stackrel{\text{(e2*)}}{\approx} H_{[b,c]} X_{[c,b]} X_{[c,b]} X_{[a,b]}$$

$$\stackrel{\text{(a2)}}{\approx} H_{[b,c]} X_{[a,b]}$$

The following variants of Equations (c2), (c4), (c5) and (d2) will be useful to prove the remaining equations:

$$X_{[a,b]} X_{[b,c]} \approx X_{[a,c]} X_{[a,b]} \tag{c2$^{\dagger}$}$$

$$X_{[a,b]} H_{[b,c]} \approx H_{[a,c]} X_{[a,b]} \tag{c4$^{\dagger}$}$$

$$X_{[b,c]} H_{[a,c]} \approx H_{[a,b]} X_{[b,c]} \tag{c5$^{\dagger}$}$$

$$H_{[a,b]} (-1)_{[b]} \approx X_{[a,b]} H_{[a,b]} \tag{d2$^{\dagger}$}$$

*Proof of Equation* (c2$^{\dagger}$):

$$X_{[a,c]} X_{[a,b]} \stackrel{\text{(a2)}}{\approx} X_{[a,b]} X_{[a,b]} X_{[a,c]} X_{[a,b]}$$

$$\stackrel{\text{(c2)}}{\approx} X_{[a,b]} X_{[b,c]} X_{[a,b]} X_{[a,b]}$$

$$\stackrel{\text{(a2)}}{\approx} X_{[a,b]} X_{[b,c]}$$

*Proof of Equation* (c4$^{\dagger}$):

$$H_{[a,c]} X_{[a,b]} \stackrel{\text{(a2)}}{\approx} X_{[a,b]} X_{[a,b]} H_{[a,c]} X_{[a,b]}$$

$$\stackrel{\text{(c4)}}{\approx} X_{[a,b]} H_{[b,c]} X_{[a,b]} X_{[a,b]}$$

$$\stackrel{\text{(a2)}}{\approx} X_{[a,b]} H_{[b,c]}$$

*Proof of Equation* (c5†):

$$H_{[a,b]}X_{[b,c]} \overset{\text{(a2)}}{\approx} X_{[b,c]}X_{[b,c]}H_{[a,b]}X_{[b,c]}$$
$$\overset{\text{(c5)}}{\approx} X_{[b,c]}H_{[a,b]}X_{[b,c]}X_{[b,c]}$$
$$\overset{\text{(a2)}}{\approx} X_{[b,c]}H_{[a,c]}$$

*Proof of Equation* (d2†):

$$H_{[a,b]}(-1)_{[b]} \overset{\text{(a3)(a2)}}{\approx} H_{[a,b]}(-1)_{[b]}H_{[a,b]}X_{[a,b]}X_{[a,b]}H_{[a,b]}$$
$$\overset{\text{(d2)}}{\approx} H_{[a,b]}(-1)_{[b]}(-1)_{[b]}H_{[a,b]}X_{[a,b]}H_{[a,b]}$$
$$\overset{\text{(a1)(a3)}}{\approx} X_{[a,b]}H_{[a,b]}$$

*Proof of Equation* (b2):[10]

$$(-1)_{[c]}X_{[a,b]} \overset{\text{(a2)}}{\approx} X_{[a,c]}X_{[a,c]}(-1)_{[c]}X_{[a,b]}$$
$$\overset{\text{(c1)}}{\approx} X_{[a,c]}(-1)_{[a]}X_{[a,c]}X_{[a,b]}$$
$$\overset{\text{(c2†)}}{\approx} X_{[a,c]}(-1)_{[a]}X_{[a,b]}X_{[b,c]}$$
$$\overset{\text{(c1)}}{\approx} X_{[a,c]}X_{[a,b]}(-1)_{[b]}X_{[b,c]}$$
$$\overset{\text{(c1)}}{\approx} X_{[a,c]}X_{[a,b]}X_{[b,c]}(-1)_{[c]}$$
$$\overset{\text{(c2†)}}{\approx} X_{[a,c]}X_{[a,c]}X_{[a,b]}(-1)_{[c]}$$
$$\overset{\text{(a2)}}{\approx} X_{[a,b]}(-1)_{[c]}$$

*Proof of Equation* (b5):

$$H_{[c,d]}X_{[a,b]} \overset{\text{(a2)}}{\approx} X_{[a,c]}X_{[a,c]}H_{[c,d]}X_{[a,b]}$$
$$\overset{\text{(c4†)}}{\approx} X_{[a,c]}H_{[a,d]}X_{[a,c]}X_{[a,b]}$$
$$\overset{\text{(c2†)}}{\approx} X_{[a,c]}H_{[a,d]}X_{[a,b]}X_{[b,c]}$$
$$\overset{\text{(c4†)}}{\approx} X_{[a,c]}X_{[a,b]}H_{[b,d]}X_{[b,c]}$$
$$\overset{\text{(c4†)}}{\approx} X_{[a,c]}X_{[a,b]}X_{[b,c]}H_{[c,d]}$$
$$\overset{\text{(c2†)}}{\approx} X_{[a,c]}X_{[a,c]}X_{[a,b]}H_{[c,d]}$$
$$\overset{\text{(a2)}}{\approx} X_{[a,b]}H_{[c,d]}$$

*Proof of Equation* (b3):[11]

$$X_{[c,d]}X_{[a,b]} \overset{\text{(a2)}}{\approx} X_{[a,c]}X_{[a,c]}X_{[c,d]}X_{[a,b]}$$
$$\overset{\text{(c2†)}}{\approx} X_{[a,c]}X_{[a,d]}X_{[a,c]}X_{[a,b]}$$
$$\overset{\text{(c2†)}}{\approx} X_{[a,c]}X_{[a,d]}X_{[a,b]}X_{[b,c]}$$

---

[10]We have renamed the indices in order to emphasize the similarity with the proofs of Equations (b5) and (b3).

[11]Note that the proof is identical to that of Equation (b5), with $H$ replaced by $X$ and Equation (c4†) replaced by Equation (c2†).

$$\overset{(c2^\dagger)}{\approx} X_{[a,c]}X_{[a,b]}X_{[b,d]}X_{[b,c]}$$

$$\overset{(c2^\dagger)}{\approx} X_{[a,c]}X_{[a,b]}X_{[b,c]}X_{[c,d]}$$

$$\overset{(c2^\dagger)}{\approx} X_{[a,c]}X_{[a,c]}X_{[a,b]}X_{[c,d]}$$

$$\overset{(a2)}{\approx} X_{[a,b]}X_{[c,d]}$$

*Proof of Equation* (b1):

- If $\{a,b\} = \{0,1\}$, then Equation (b1) is just Equation (b1*).
- If $a = 0$ and $b \neq 0, 1$:

$$(-1)_{[0]}(-1)_{[b]} \overset{(a2)}{\approx} X_{[1,b]}X_{[1,b]}(-1)_{[0]}(-1)_{[b]}$$
$$\overset{(b2)}{\approx} X_{[1,b]}(-1)_{[0]}X_{[1,b]}(-1)_{[b]}$$
$$\overset{(c1)}{\approx} X_{[1,b]}(-1)_{[0]}(-1)_{[1]}X_{[1,b]}$$
$$\overset{(b1^*)}{\approx} X_{[1,b]}(-1)_{[1]}(-1)_{[0]}X_{[1,b]}$$
$$\overset{(b2)}{\approx} X_{[1,b]}(-1)_{[1]}X_{[1,b]}(-1)_{[0]}$$
$$\overset{(c1)}{\approx} X_{[1,b]}X_{[1,b]}(-1)_{[b]}(-1)_{[0]}$$
$$\overset{(a2)}{\approx} (-1)_{[b]}(-1)_{[0]}$$

- If $b = 0$ and $a \neq 0, 1$, or if $a = 1$ and $b \neq 0, 1$, or if $b = 1$ and $a \neq 0, 1$, then the proof is identical, with $a$ replaced by $b$ and/or 0 replaced by 1.
- If $\{a,b\} \cap \{0,1\} = \emptyset$:

$$(-1)_{[a]}(-1)_{[b]} \overset{(a2)}{\approx} X_{[0,a]}X_{[0,a]}(-1)_{[a]}(-1)_{[b]}$$
$$\overset{(a2)}{\approx} X_{[0,a]}X_{[1,b]}X_{[1,b]}X_{[0,a]}(-1)_{[a]}(-1)_{[b]}$$
$$\overset{(c1)}{\approx} X_{[0,a]}X_{[1,b]}X_{[1,b]}(-1)_{[0]}X_{[0,a]}(-1)_{[b]}$$
$$\overset{(b2)}{\approx} X_{[0,a]}X_{[1,b]}X_{[1,b]}(-1)_{[0]}(-1)_{[b]}X_{[0,a]}$$
$$\overset{(b2)}{\approx} X_{[0,a]}X_{[1,b]}(-1)_{[0]}X_{[1,b]}(-1)_{[b]}X_{[0,a]}$$
$$\overset{(c1)}{\approx} X_{[0,a]}X_{[1,b]}(-1)_{[0]}(-1)_{[1]}X_{[1,b]}X_{[0,a]}$$
$$\overset{(b1^*)}{\approx} X_{[0,a]}X_{[1,b]}(-1)_{[1]}(-1)_{[0]}X_{[1,b]}X_{[0,a]}$$
$$\overset{(b2)}{\approx} X_{[0,a]}X_{[1,b]}(-1)_{[1]}X_{[1,b]}(-1)_{[0]}X_{[0,a]}$$
$$\overset{(c1)}{\approx} X_{[0,a]}X_{[1,b]}(-1)_{[1]}X_{[1,b]}X_{[0,a]}(-1)_{[a]}$$
$$\overset{(c1)}{\approx} X_{[0,a]}X_{[1,b]}X_{[1,b]}(-1)_{[b]}X_{[0,a]}(-1)_{[a]}$$
$$\overset{(b2)}{\approx} X_{[0,a]}X_{[1,b]}X_{[1,b]}X_{[0,a]}(-1)_{[b]}(-1)_{[a]}$$
$$\overset{(a2)}{\approx} X_{[0,a]}X_{[0,a]}(-1)_{[b]}(-1)_{[a]}$$
$$\overset{(a2)}{\approx} (-1)_{[b]}(-1)_{[a]}$$

The proofs of Equations (b4) and (b6) follow the same principle as the proof of Equation (b1): using Equation (a2) followed by Equations (c1), (b2), (c4), (c5) and (b5) (and maybe (e1)), we can arbitrarily rename the indices up to conjugating by some $X$ generators, so that we can conclude using Equation (b4*) (resp. Equation (b6*)).

*Proof of Equation* (d1):

$$
\begin{aligned}
&(-1)_{[a]}(-1)_{[b]}H_{[a,b]}\\
\overset{(d2)}{\approx}\ &(-1)_{[a]}H_{[a,b]}X_{[a,b]}\\
\overset{(e1)}{\approx}\ &(-1)_{[a]}H_{[a,b]}X_{[b,a]}\\
\overset{(e2^*)}{\approx}\ &(-1)_{[a]}X_{[b,a]}H_{[b,a]}\\
\overset{(a3)}{\approx}\ &(-1)_{[a]}H_{[b,a]}H_{[b,a]}X_{[b,a]}H_{[b,a]}\\
\overset{(d2)}{\approx}\ &H_{[b,a]}X_{[b,a]}H_{[b,a]}X_{[b,a]}H_{[b,a]}\\
\overset{(d2)}{\approx}\ &H_{[b,a]}X_{[b,a]}(-1)_{[a]}H_{[b,a]}H_{[b,a]}\\
\overset{(a3)}{\approx}\ &H_{[b,a]}X_{[b,a]}(-1)_{[a]}\\
\overset{(e1)}{\approx}\ &H_{[b,a]}X_{[a,b]}(-1)_{[a]}\\
\overset{(e2^*)}{\approx}\ &X_{[a,b]}H_{[a,b]}(-1)_{[a]}\\
\overset{(a3)}{\approx}\ &H_{[a,b]}H_{[a,b]}X_{[a,b]}H_{[a,b]}(-1)_{[a]}\\
\overset{(d2)}{\approx}\ &H_{[a,b]}(-1)_{[b]}H_{[a,b]}H_{[a,b]}(-1)_{[a]}\\
\overset{(a3)}{\approx}\ &H_{[a,b]}(-1)_{[b]}(-1)_{[a]}\\
\overset{(b1)}{\approx}\ &H_{[a,b]}(-1)_{[a]}(-1)_{[b]}
\end{aligned}
$$

□

*Proof of Equation* (d3): By proceeding similarly to the proofs of Equations (b1) to (b6), we can rename the indices arbitrarily, therefore it suffices to prove the following equation:

$$(H_{[2,3]}H_{[0,2]}H_{[1,3]})^4 \approx H_{[0,1]}H_{[2,3]}$$

which we do as follows:

$$
\begin{aligned}
&(H_{[2,3]}H_{[0,2]}H_{[1,3]})^4\\
=\ &H_{[2,3]}H_{[0,2]}H_{[1,3]}H_{[2,3]}H_{[0,2]}H_{[1,3]}H_{[2,3]}H_{[0,2]}H_{[1,3]}H_{[2,3]}H_{[0,2]}H_{[1,3]}\\
\overset{(a1)(b1)}{\approx}\ &H_{[2,3]}(-1)_{[2]}(-1)_{[3]}(-1)_{[2]}(-1)_{[3]}H_{[0,2]}H_{[1,3]}H_{[2,3]}H_{[0,2]}H_{[1,3]}H_{[2,3]}H_{[0,2]}H_{[1,3]}H_{[2,3]}H_{[0,2]}H_{[1,3]}\\
\overset{(b4)(d2)(b5)}{\approx}\ &H_{[2,3]}(-1)_{[2]}(-1)_{[3]}H_{[0,2]}H_{[1,3]}X_{[0,2]}X_{[1,3]}H_{[2,3]}H_{[0,2]}H_{[1,3]}H_{[2,3]}H_{[0,2]}H_{[1,3]}\\
\overset{(c5^\dagger)(c4^\dagger)}{\approx}\ &H_{[2,3]}(-1)_{[2]}(-1)_{[3]}H_{[0,2]}H_{[1,3]}H_{[0,1]}X_{[0,2]}X_{[1,3]}H_{[0,2]}H_{[1,3]}H_{[2,3]}H_{[0,2]}H_{[1,3]}H_{[2,3]}H_{[0,2]}H_{[1,3]}\\
\overset{(b5)(d2^\dagger)(b4)}{\approx}\ &H_{[2,3]}(-1)_{[2]}(-1)_{[3]}H_{[0,2]}H_{[1,3]}H_{[0,1]}H_{[0,2]}H_{[1,3]}(-1)_{[2]}(-1)_{[3]}H_{[2,3]}H_{[0,2]}H_{[1,3]}H_{[2,3]}H_{[0,2]}H_{[1,3]}\\
\overset{(d1)}{\approx}\ &H_{[2,3]}(-1)_{[2]}(-1)_{[3]}H_{[0,2]}H_{[1,3]}H_{[0,1]}H_{[0,2]}H_{[1,3]}H_{[2,3]}(-1)_{[2]}(-1)_{[3]}H_{[0,2]}H_{[1,3]}H_{[2,3]}H_{[0,2]}H_{[1,3]}\\
\overset{(b4)(d2)(b5)}{\approx}\ &H_{[2,3]}(-1)_{[2]}(-1)_{[3]}H_{[0,2]}H_{[1,3]}H_{[0,1]}H_{[0,2]}H_{[1,3]}H_{[2,3]}H_{[0,2]}H_{[1,3]}X_{[0,2]}X_{[1,3]}H_{[2,3]}H_{[0,2]}H_{[1,3]}\\
\overset{(c5^\dagger)(c4^\dagger)}{\approx}\ &H_{[2,3]}(-1)_{[2]}(-1)_{[3]}H_{[0,2]}H_{[1,3]}H_{[0,1]}H_{[0,2]}H_{[1,3]}H_{[2,3]}H_{[0,2]}H_{[1,3]}H_{[0,1]}X_{[0,2]}X_{[1,3]}H_{[0,2]}H_{[1,3]}\\
\overset{(b5)(d2^\dagger)(b4)}{\approx}\ &H_{[2,3]}(-1)_{[2]}(-1)_{[3]}H_{[0,2]}H_{[1,3]}H_{[0,1]}H_{[0,2]}H_{[1,3]}H_{[2,3]}H_{[0,2]}H_{[1,3]}H_{[0,1]}H_{[0,2]}H_{[1,3]}(-1)_{[2]}(-1)_{[3]}\\
\overset{(a1)(b1)}{\approx}\ &H_{[2,3]}(-1)_{[2]}(-1)_{[3]}H_{[0,2]}H_{[1,3]}H_{[0,1]}(-1)_{[0]}(-1)_{[1]}(-1)_{[2]}(-1)_{[3]}H_{[0,2]}H_{[1,3]}H_{[2,3]}H_{[0,2]}H_{[1,3]}H_{[0,2]}H_{[1,3]}(-1)_{[2]}(-1)_{[3]}\\
\overset{(b1)(b4)(d1)}{\approx}\ &H_{[2,3]}(-1)_{[2]}(-1)_{[3]}H_{[0,2]}H_{[1,3]}H_{[0,1]}(-1)_{[0]}(-1)_{[1]}(-1)_{[2]}(-1)_{[3]}H_{[0,2]}H_{[1,3]}H_{[2,3]}H_{[0,2]}H_{[1,3]}(-1)_{[2]}(-1)_{[3]}(-1)_{[0]}(-1)_{[1]}H_{[0,1]}H_{[0,2]}H_{[1,3]}(-1)_{[2]}(-1)_{[3]}\\
\overset{(b4)(d2)(b5)}{\approx}\ &H_{[2,3]}(-1)_{[2]}(-1)_{[3]}H_{[0,2]}H_{[1,3]}H_{[0,1]}(-1)_{[0]}(-1)_{[1]}H_{[0,2]}X_{[0,2]}X_{[1,3]}H_{[2,3]}H_{[0,2]}H_{[1,3]}(-1)_{[2]}(-1)_{[3]}(-1)_{[0]}(-1)_{[1]}H_{[0,1]}H_{[0,2]}H_{[1,3]}(-1)_{[2]}(-1)_{[3]}\\
\overset{(c5^\dagger)(c4^\dagger)}{\approx}\ &H_{[2,3]}(-1)_{[2]}(-1)_{[3]}H_{[0,2]}H_{[1,3]}H_{[0,1]}(-1)_{[0]}(-1)_{[1]}H_{[0,2]}H_{[1,3]}H_{[0,1]}X_{[0,2]}X_{[1,3]}H_{[0,2]}H_{[1,3]}(-1)_{[2]}(-1)_{[3]}(-1)_{[0]}(-1)_{[1]}H_{[0,1]}H_{[0,2]}H_{[1,3]}(-1)_{[2]}(-1)_{[3]}\\
\overset{(b5)(d2^\dagger)(b4)}{\approx}\ &H_{[2,3]}(-1)_{[2]}(-1)_{[3]}H_{[0,2]}H_{[1,3]}H_{[0,1]}(-1)_{[0]}(-1)_{[1]}H_{[0,2]}H_{[1,3]}H_{[0,1]}H_{[0,2]}H_{[1,3]}(-1)_{[2]}(-1)_{[3]}(-1)_{[0]}(-1)_{[1]}H_{[0,1]}H_{[0,2]}H_{[1,3]}(-1)_{[2]}(-1)_{[3]}\\
\overset{(b1)(a1)}{\approx}\ &H_{[2,3]}(-1)_{[2]}(-1)_{[3]}H_{[0,2]}H_{[1,3]}H_{[0,1]}(-1)_{[0]}(-1)_{[1]}H_{[0,2]}H_{[1,3]}H_{[0,1]}H_{[0,2]}H_{[1,3]}(-1)_{[0]}(-1)_{[1]}H_{[0,1]}H_{[0,2]}H_{[1,3]}(-1)_{[2]}(-1)_{[3]}\\
\overset{(d3^*)}{\approx}\ &H_{[2,3]}(-1)_{[2]}(-1)_{[3]}H_{[0,1]}H_{[0,2]}H_{[1,3]}H_{[0,1]}(-1)_{[0]}(-1)_{[1]}H_{[0,2]}H_{[1,3]}H_{[0,1]}H_{[0,2]}H_{[1,3]}(-1)_{[0]}(-1)_{[1]}H_{[0,1]}H_{[0,2]}H_{[1,3]}(-1)_{[2]}(-1)_{[3]}
\end{aligned}
$$

$$\overset{\text{(b1)(b6)(a3)(a1)}}{\approx} H_{[2,3]}(-1)_{[2]}(-1)_{[3]}H_{[0,1]}(-1)_{[2]}(-1)_{[3]}$$

$$\overset{\text{(b4)(b6)(b1)(a1)}}{\approx} H_{[0,1]}H_{[2,3]}$$

*Proof of Equation* (d4)*:* By proceeding similarly to the proofs of Equations (b1) to (b6), we can rename the indices arbitrarily, therefore it suffices to prove the following equation:

$$(H_{[0,4]}H_{[1,5]}H_{[0,1]}H_{[0,4]}H_{[1,5]}X_{[4,2]}X_{[5,3]})^3 \approx H_{[4,2]}H_{[5,3]}H_{[2,3]}H_{[4,2]}H_{[5,3]}X_{[4,2]}X_{[5,3]}$$

which we do as follows:

$$(H_{[0,4]}H_{[1,5]}H_{[0,1]}H_{[0,4]}H_{[1,5]}X_{[4,2]}X_{[5,3]})^3$$

$$= H_{[0,4]}H_{[1,5]}H_{[0,1]}H_{[0,4]}H_{[1,5]}X_{[4,2]}X_{[5,3]}H_{[0,4]}H_{[1,5]}H_{[0,1]}H_{[0,4]}H_{[1,5]}X_{[4,2]}X_{[5,3]}H_{[0,4]}H_{[1,5]}H_{[0,1]}H_{[0,4]}H_{[1,5]}X_{[4,2]}X_{[5,3]}$$

$$\overset{\text{(b5)(c5)}}{\approx} H_{[0,4]}H_{[1,5]}H_{[0,1]}H_{[0,4]}H_{[1,5]}H_{[0,2]}H_{[1,3]}H_{[0,1]}H_{[0,2]}H_{[1,3]}X_{[4,2]}X_{[5,3]}X_{[4,2]}X_{[5,3]}H_{[0,4]}H_{[1,5]}H_{[0,1]}H_{[0,4]}H_{[1,5]}X_{[4,2]}X_{[5,3]}$$

$$\overset{\text{(b3)(a2)}}{\approx} H_{[0,4]}H_{[1,5]}H_{[0,1]}H_{[0,4]}H_{[1,5]}H_{[0,2]}H_{[1,3]}H_{[0,1]}H_{[0,2]}H_{[1,3]}H_{[0,4]}H_{[1,5]}H_{[0,1]}H_{[0,4]}H_{[1,5]}X_{[4,2]}X_{[5,3]}$$

$$\overset{\text{(a1)(b1)}}{\approx} H_{[0,4]}H_{[1,5]}H_{[0,1]}(-1)_{[0]}(-1)_{[1]}(-1)_{[4]}(-1)_{[5]}(-1)_{[4]}(-1)_{[5]}(-1)_{[0]}(-1)_{[1]}H_{[0,4]}H_{[1,5]}H_{[0,2]}H_{[1,3]}H_{[0,1]}H_{[0,2]}H_{[1,3]}H_{[0,4]}H_{[1,5]}H_{[0,1]}H_{[0,4]}H_{[1,5]}X_{[4,2]}X_{[5,3]}$$

$$\overset{\text{(d1)(b4)(b1)}}{\approx} (-1)_{[0]}(-1)_{[1]}(-1)_{[4]}(-1)_{[5]}H_{[0,4]}H_{[1,5]}(-1)_{[4]}(-1)_{[5]}H_{[0,1]}(-1)_{[0]}(-1)_{[1]}H_{[0,4]}H_{[1,5]}H_{[0,2]}H_{[1,3]}H_{[0,1]}H_{[0,2]}H_{[1,3]}H_{[0,4]}H_{[1,5]}H_{[0,1]}H_{[0,4]}H_{[1,5]}X_{[4,2]}X_{[5,3]}$$

$$\overset{\text{(b4)(d2$^\dagger$)(b5)}}{\approx} (-1)_{[0]}(-1)_{[1]}(-1)_{[4]}(-1)_{[5]}H_{[0,4]}H_{[1,5]}(-1)_{[4]}(-1)_{[5]}H_{[0,1]}(-1)_{[0]}(-1)_{[1]}H_{[0,4]}H_{[1,5]}H_{[0,2]}H_{[1,3]}H_{[0,1]}H_{[0,2]}H_{[1,3]}H_{[0,4]}H_{[1,5]}H_{[0,1]}H_{[0,4]}H_{[1,5]}X_{[4,2]}X_{[5,3]}$$

$$\overset{\text{(a3)(b6)}}{\approx} (-1)_{[0]}(-1)_{[1]}(-1)_{[4]}(-1)_{[5]}X_{[0,4]}X_{[1,5]}H_{[0,2]}H_{[1,3]}H_{[0,2]}H_{[1,3]}H_{[0,4]}H_{[1,5]}H_{[0,2]}H_{[1,3]}H_{[0,2]}H_{[1,3]}H_{[0,4]}H_{[1,5]}H_{[0,4]}H_{[1,5]}X_{[4,2]}X_{[5,3]}$$

$$\overset{\text{(d4$^*$)}}{\approx} (-1)_{[0]}(-1)_{[1]}(-1)_{[4]}(-1)_{[5]}X_{[0,4]}X_{[1,5]}H_{[0,2]}H_{[1,3]}H_{[0,2]}H_{[1,3]}H_{[0,4]}(-1)_{[0]}(-1)_{[1]}H_{[1,5]}H_{[0,4]}H_{[0,2]}H_{[1,3]}H_{[0,2]}H_{[1,3]}H_{[0,4]}H_{[1,5]}H_{[0,4]}H_{[1,5]}X_{[4,2]}X_{[5,3]}$$

$$\overset{\text{(b6)(a3)}}{\approx} (-1)_{[0]}(-1)_{[1]}(-1)_{[4]}(-1)_{[5]}X_{[0,4]}X_{[1,5]}H_{[0,2]}H_{[1,3]}H_{[0,1]}H_{[0,2]}H_{[1,3]}H_{[0,4]}H_{[1,5]}H_{[0,1]}(-1)_{[0]}(-1)_{[1]}H_{[0,4]}H_{[1,5]}X_{[4,2]}X_{[5,3]}$$

$$\overset{\text{(d1)(a3)}}{\approx} (-1)_{[0]}(-1)_{[1]}(-1)_{[4]}(-1)_{[5]}X_{[0,4]}X_{[1,5]}H_{[0,2]}H_{[1,3]}H_{[0,1]}H_{[0,2]}H_{[1,3]}H_{[0,4]}H_{[1,5]}(-1)_{[0]}(-1)_{[1]}H_{[0,4]}X_{[4,2]}X_{[5,3]}$$

$$\overset{\text{(c4)(c5)}}{\approx} (-1)_{[0]}(-1)_{[1]}(-1)_{[4]}(-1)_{[5]}H_{[4,2]}H_{[5,3]}H_{[4,2]}H_{[5,3]}H_{[0,4]}X_{[1,5]}H_{[0,4]}H_{[1,5]}(-1)_{[0]}(-1)_{[1]}H_{[0,4]}X_{[4,2]}X_{[5,3]}$$

$$\overset{\text{(b5)(d2$^\dagger$)(b4)}}{\approx} (-1)_{[0]}(-1)_{[1]}(-1)_{[4]}(-1)_{[5]}H_{[4,2]}H_{[5,3]}H_{[4,2]}H_{[5,3]}H_{[0,4]}H_{[1,5]}(-1)_{[4]}(-1)_{[5]}(-1)_{[0]}(-1)_{[1]}H_{[0,4]}H_{[1,5]}X_{[4,2]}X_{[5,3]}$$

$$\overset{\text{(b1)(b4)(d1)}}{\approx} (-1)_{[0]}(-1)_{[1]}(-1)_{[4]}(-1)_{[5]}H_{[4,2]}H_{[5,3]}H_{[4,5]}H_{[4,2]}H_{[5,3]}(-1)_{[4]}(-1)_{[5]}(-1)_{[0]}(-1)_{[1]}H_{[0,4]}H_{[1,5]}X_{[4,2]}X_{[5,3]}$$

$$\overset{\text{(b6)(a3)}}{\approx} (-1)_{[0]}(-1)_{[1]}(-1)_{[4]}(-1)_{[5]}H_{[4,2]}H_{[5,3]}H_{[4,5]}H_{[4,2]}H_{[5,3]}(-1)_{[4]}(-1)_{[5]}(-1)_{[0]}(-1)_{[1]}X_{[4,2]}X_{[5,3]}$$

$$\overset{\text{(b1)(b4)(a1)}}{\approx} (-1)_{[4]}(-1)_{[5]}H_{[4,2]}H_{[5,3]}H_{[4,5]}H_{[4,2]}H_{[5,3]}(-1)_{[4]}(-1)_{[5]}X_{[4,2]}X_{[5,3]}$$

$$\overset{\text{(a1)(b1)}}{\approx} (-1)_{[4]}(-1)_{[5]}(-1)_{[2]}(-1)_{[3]}(-1)_{[2]}(-1)_{[3]}H_{[4,2]}H_{[5,3]}H_{[4,5]}H_{[4,2]}H_{[5,3]}(-1)_{[4]}(-1)_{[5]}X_{[4,2]}X_{[5,3]}$$

$$\overset{\text{(b4)(d2)(b5)}}{\approx} (-1)_{[4]}(-1)_{[5]}(-1)_{[2]}(-1)_{[3]}H_{[4,2]}H_{[5,3]}X_{[4,2]}X_{[5,3]}H_{[4,5]}H_{[4,2]}H_{[5,3]}(-1)_{[4]}(-1)_{[5]}X_{[4,2]}X_{[5,3]}$$

$$\overset{\text{(c4)(c5)}}{\approx} (-1)_{[4]}(-1)_{[5]}(-1)_{[2]}(-1)_{[3]}H_{[4,2]}H_{[5,3]}H_{[2,3]}X_{[4,2]}X_{[5,3]}H_{[4,2]}H_{[5,3]}(-1)_{[4]}(-1)_{[5]}X_{[4,2]}X_{[5,3]}$$

$$\overset{\text{(b5)(d2$^\dagger$)(b4)}}{\approx} (-1)_{[4]}(-1)_{[5]}(-1)_{[2]}(-1)_{[3]}H_{[4,2]}H_{[5,3]}H_{[2,3]}H_{[4,2]}H_{[5,3]}(-1)_{[2]}(-1)_{[3]}(-1)_{[4]}(-1)_{[5]}X_{[4,2]}X_{[5,3]}$$

$$\overset{\text{(b1)(b4)(d1)(a1)}}{\approx} H_{[4,2]}H_{[5,3]}H_{[2,3]}H_{[4,2]}H_{[5,3]}X_{[4,2]}X_{[5,3]}$$

## A.3 Applying the Reidemeister-Schreier Theorem

*A.3.1 Elements of the Reidemeister-Schreier Theorem for Monoids.* We denote by $U_N^{\text{even}}(\mathbb{Z}[\frac{1}{\sqrt{2}}])$ the submonoid (or subgroup) of $U_N(\mathbb{Z}[\frac{1}{\sqrt{2}}])$ consisting of the matrices with even $H$- and $ZX$-parities.

In order to apply the Reidemeister-Schreier theorem for monoids described in [6] (Section 4.2), it suffices to provide the following elements:

- a set $C$, consisting of a representative of each of the four cosets of $U_N^{\text{even}}(\mathbb{Z}[\frac{1}{\sqrt{2}}])$ as a subgroup of $U_N(\mathbb{Z}[\frac{1}{\sqrt{2}}])$, the representative of $U_N^{\text{even}}(\mathbb{Z}[\frac{1}{\sqrt{2}}])$ itself being necessarily the identity $I$
- a function $f: \mathcal{P}_N \to \mathcal{G}_N^*$ (where $\mathcal{G}_N^*$ denotes the free monoid generated by $\mathcal{G}_N$) preserving the semantics
- a function $h: C \times \mathcal{G}_N \to \mathcal{P}_N^* \times C$ such that for any $(c, g)$, if $(p, c') = h(c, g)$ then $[\![cg]\!] = [\![pc']\!]$.

Then, let $h^{**}: C \times \mathcal{G}_N^* \to \mathcal{P}_N^* \times C$ be the function defined by $h^{**}(c_0, x_1 \ldots x_k) = (w_1 \cdot \ldots \cdot w_k, c_k)$, where $h(c_{i-1}, x_i) = (w_i, c_i)$ for all $i = 1, \ldots, k$. The Reidemeister-Schreier theorem for monoids stated in [6] tells us that the equational theory consisting of the following equations is complete:

(a) for every $x \in \mathcal{P}_N$, an equation $x \approx v$, where $v$ is such that $h^{**}(I, f(x)) = (v, c)$ (note that in this case, the preservation of the semantics implies that $c = I$)

(b) for every equation of Figure 7, of the form $w \approx w'$, and every $c \in C$, an equation $v \approx v'$, where $v, v'$ are such that $h^{**}(c, w) = (v, c')$ and $h^{**}(c, w') = (v', c'')$ (note that in this case, the preservation of the semantics implies that $c' = c''$).

The set $C$ and the functions $f$ and $h$ are defined in Definition A.2 below (by case distinction on the output of $f$ and $h$). The equational theory built as just described is given in Appendix A.3.2.

**Definition A.2 (Elements of the Reidemeister-Schreier theorem for monoids [6]).**

$$C = \{\epsilon, (-1)_{[1]}, H_{[0,1]}, H_{[0,1]}(-1)_{[1]}\}$$

$$
\begin{aligned}
f \colon ((-1)_{[a]}(-1)_{[b]}) &\mapsto (-1)_{[a]}(-1)_{[b]} \\
((-1)_{[a]}X_{[c,d]}) &\mapsto (-1)_{[a]}X_{[c,d]} \\
(X_{[a,b]}X_{[c,d]}) &\mapsto X_{[a,b]}X_{[c,d]} \\
(H_{[a,b]}H_{[c,d]}) &\mapsto H_{[a,b]}H_{[c,d]}
\end{aligned}
$$

$$
\begin{aligned}
h \colon (\epsilon, (-1)_{[a]}) &\mapsto \left(((-1)_{[1]}(-1)_{[a]}), (-1)_{[1]}\right) \\[4pt]
(\epsilon, X_{[a,b]}) &\mapsto
\begin{cases}
\left(((-1)_{[ab]}X_{[a,b]}), (-1)_{[1]}\right) & \text{if } a = 1 \text{ or } b = 1 \\
\left(((-1)_{[1]}X_{[a,b]}), (-1)_{[1]}\right) & \text{if } a, b \neq 1
\end{cases} \\[4pt]
(\epsilon, H_{[a,b]}) &\mapsto \left((H_{[a,b]}H_{[0,1]}), H_{[0,1]}\right) \\[8pt]
((-1)_{[1]}, (-1)_{[a]}) &\mapsto \left(((-1)_{[1]}(-1)_{[a]}), \epsilon\right) \\[6pt]
((-1)_{[1]}, X_{[a,b]}) &\mapsto \left(((-1)_{[1]}X_{[a,b]}), \epsilon\right) \\[6pt]
((-1)_{[1]}, H_{[a,b]}) &\mapsto
\begin{cases}
\left(((-1)_{[b]}X_{[a,b]})(H_{[a,b]}H_{[0,1]}), H_{[0,1]}(-1)_{[1]}\right) & \text{if } a = 1 \text{ or } b = 1 \\
\left((H_{[a,b]}H_{[0,1]}), H_{[0,1]}(-1)_{[1]}\right) & \text{if } a, b \neq 1
\end{cases} \\[8pt]
(H_{[0,1]}, (-1)_{[a]}) &\mapsto
\begin{cases}
\left(((-1)_{[1]}(-1)_{[a]}), H_{[0,1]}(-1)_{[1]}\right) & \text{if } a \in \{0, 1\} \\
\left(((-1)_{[a]}X_{[0,1]}), H_{[0,1]}(-1)_{[1]}\right) & \text{if } a \notin \{0, 1\}
\end{cases} \\[8pt]
(H_{[0,1]}, X_{[a,b]}) &\mapsto
\begin{cases}
\left(((-1)_{[1]}X_{[0,1]}), H_{[0,1]}(-1)_{[1]}\right) & \text{if } \{a, b\} = \{0, 1\} \\
\left((X_{[0,1]}X_{[a,b]})(H_{[a+b,1]}H_{[0,1]}), H_{[0,1]}(-1)_{[1]}\right) & \text{if } \{a, b\} \cap \{0, 1\} = \{0\} \\
\left(((-1)_{[ab]}X_{[a,b]})(H_{[0,ab]}H_{[0,1]}), H_{[0,1]}(-1)_{[1]}\right) & \text{if } \{a, b\} \cap \{0, 1\} = \{1\} \\
\left((X_{[a,b]}X_{[0,1]}), H_{[0,1]}(-1)_{[1]}\right) & \text{if } \{a, b\} \cap \{0, 1\} = \emptyset
\end{cases} \\[8pt]
(H_{[0,1]}, H_{[a,b]}) &\mapsto \left((H_{[0,1]}H_{[a,b]}), \epsilon\right) \\[8pt]
(H_{[0,1]}(-1)_{[1]}, (-1)_{[a]}) &\mapsto
\begin{cases}
\left(((-1)_{[1]}(-1)_{[a]}), H_{[0,1]}\right) & \text{if } a \in \{0, 1\} \\
\left(((-1)_{[a]}X_{[0,1]}), H_{[0,1]}\right) & \text{if } a \notin \{0, 1\}
\end{cases}
\end{aligned}
$$

$$\left(H_{[0,1]}(-1)_{[1]}, X_{[a,b]}\right) \quad \mapsto \quad \begin{cases} \left(\left((-1)_{[0]}X_{[0,1]}\right), H_{[0,1]}\right) & \text{if } \{a,b\} = \{0,1\} \\ \left(\left(X_{[0,1]}X_{[a,b]}\right)\left(H_{[a+b,1]}H_{[0,1]}\right), H_{[0,1]}\right) & \text{if } \{a,b\} \cap \{0,1\} = \{0\} \\ \left(\left(X_{[0,1]}X_{[a,b]}\right)\left(H_{[0,ab]}H_{[0,1]}\right), H_{[0,1]}\right) & \text{if } \{a,b\} \cap \{0,1\} = \{1\} \\ \left(\left(X_{[a,b]}X_{[0,1]}\right), H_{[0,1]}\right) & \text{if } \{a,b\} \cap \{0,1\} = \emptyset \end{cases}$$

$$\left(H_{[0,1]}(-1)_{[1]}, H_{[a,b]}\right) \quad \mapsto \quad \begin{cases} \left(\left(H_{[0,1]}H_{[a,b]}\right)\left((-1)_{[a]}X_{[a,b]}\right), (-1)_{[1]}\right) & \text{if } a = 1 \text{ or } b = 1 \\ \left(\left(H_{[0,1]}H_{[a,b]}\right), (-1)_{[1]}\right) & \text{if } a, b \neq 1 \end{cases}$$

### A.3.2 Raw Output of the Reidemeister-Schreier Theorem.

We follow the following naming convention for the equations:

- The equations coming from Item (*a*) are called (DE-**) or (DE-**-*k*), where ** is either ZZ, ZX, XX, or HH depending on whether $x$ is $\left((-1)_{[a]}(-1)_{[b]}\right)$, $\left((-1)_{[a]}X_{[c,d]}\right)$, $\left(X_{[a,b]}X_{[c,d]}\right)$, or $\left(H_{[a,b]}H_{[c,d]}\right)$ respectively.
- The equations coming from Item (*b*) are called (**-††) or (**-††-*k*), where ** is either $\epsilon$, Z, H, or HZ depending on whether $c$ is $\epsilon$, $(-1)_{[1]}$, $H_{[0,1]}$, or $H_{[0,1]}(-1)_{[1]}$ respectively, and †† is the label of the original equation in Figure 7.

In both cases, the additional index $k$ is used when we need to split cases depending on the values of the parameters.

The equational theory is the following:

$$\left((-1)_{[a]}(-1)_{[b]}\right) \approx \left((-1)_{[1]}(-1)_{[a]}\right)\left((-1)_{[1]}(-1)_{[b]}\right) \tag{DE-ZZ}$$

$$\left((-1)_{[a]}X_{[c,d]}\right) \approx \left((-1)_{[1]}(-1)_{[a]}\right)\left((-1)_{[1]}X_{[c,d]}\right) \tag{DE-ZX}$$

$$\left(X_{[a,b]}X_{[c,d]}\right) \approx \left((-1)_{[ab]}X_{[a,b]}\right)\left((-1)_{[1]}X_{[c,d]}\right) \qquad (a = 1 \text{ or } b = 1) \tag{DE-XX-1}$$

$$\left(X_{[a,b]}X_{[c,d]}\right) \approx \left((-1)_{[1]}X_{[a,b]}\right)\left((-1)_{[1]}X_{[c,d]}\right) \qquad (a, b \neq 1) \tag{DE-XX-2}$$

$$\left(H_{[a,b]}H_{[c,d]}\right) \approx \left(H_{[a,b]}H_{[0,1]}\right)\left(H_{[0,1]}H_{[c,d]}\right) \tag{DE-HH}$$

$$\left((-1)_{[1]}(-1)_{[0]}\right)^2 \approx \epsilon \tag{$\epsilon$-a1}$$

$$\left((-1)_{[1]}(-1)_{[0]}\right)^2 \approx \epsilon \tag{Z-a1}$$

$$\left((-1)_{[1]}(-1)_{[0]}\right)^2 \approx \epsilon \tag{H-a1}$$

$$\left((-1)_{[1]}(-1)_{[0]}\right)^2 \approx \epsilon \tag{HZ-a1}$$

$$\left(H_{[j,k]}H_{[0,1]}\right)\left(H_{[0,1]}H_{[j,k]}\right) \approx \epsilon \tag{$\epsilon$-a3}$$

$$\left((-1)_{[k]}X_{[j,k]}\right)\left(H_{[j,k]}H_{[0,1]}\right)\left(H_{[0,1]}H_{[j,k]}\right)\left((-1)_{[j]}X_{[j,k]}\right) \approx \epsilon \qquad j = 1 \text{ or } k = 1 \tag{Z-a3-1}$$

$$\left(H_{[j,k]}H_{[0,1]}\right)\left(H_{[0,1]}H_{[j,k]}\right) \approx \epsilon \qquad j, k \neq 1 \tag{Z-a3-2}$$

$$\left(H_{[0,1]}H_{[j,k]}\right)\left(H_{[j,k]}H_{[0,1]}\right) \approx \epsilon \tag{H-a3}$$

$$\left(H_{[0,1]}H_{[j,k]}\right)\left((-1)_{[j]}X_{[j,k]}\right)\left((-1)_{[k]}X_{[j,k]}\right)\left(H_{[j,k]}H_{[0,1]}\right) \approx \epsilon \qquad j = 1 \text{ or } k = 1 \tag{HZ-a3-1}$$

$$\left(H_{[0,1]}H_{[j,k]}\right)\left(H_{[j,k]}H_{[0,1]}\right) \approx \epsilon \qquad j, k \neq 1 \tag{HZ-a3-2}$$

$$\left((-1)_{[j,k]}X_{[j,k]}\right)\left((-1)_{[1]}X_{[j,k]}\right) \approx \epsilon \qquad j = 1 \text{ or } k = 1 \tag{$\epsilon$-a2-1}$$

$$\left((-1)_{[1]}X_{[j,k]}\right)\left((-1)_{[1]}X_{[j,k]}\right) \approx \epsilon \qquad j, k \neq 1 \tag{$\epsilon$-a2-2}$$

$$\left((-1)_{[1]}X_{[j,k]}\right)\left((-1)_{[j,k]}X_{[j,k]}\right) \approx \epsilon \qquad j = 1 \text{ or } k = 1 \tag{Z-a2-1}$$

$$\left((-1)_{[1]}X_{[j,k]}\right)\left((-1)_{[1]}X_{[j,k]}\right) \approx \epsilon \qquad j, k \neq 1 \tag{Z-a2-2}$$

$$\left((-1)_{[1]}X_{[0,1]}\right)\left((-1)_{[0]}X_{[0,1]}\right) \approx \epsilon \qquad \{j,k\} = \{0,1\} \tag{H-a2-1}$$

$$\left(X_{[0,1]}X_{[j,k]}\right)\left(H_{[j+k,1]}H_{[0,1]}\right)\left(X_{[0,1]}X_{[j,k]}\right)\left(H_{[j+k,1]}H_{[0,1]}\right) \approx \epsilon \qquad \{j,k\} \cap \{0,1\} = \{0\} \tag{H-a2-2}$$

$$\left((-1)_{[j,k]}X_{[j,k]}\right)\left(H_{[0,jk]}H_{[0,1]}\right)\left(X_{[0,1]}X_{[j,k]}\right)\left(H_{[0,jk]}H_{[0,1]}\right) \approx \epsilon \qquad \{j,k\} \cap \{0,1\} = \{1\} \tag{H-a2-3}$$

$$\left(X_{[j,k]}X_{[0,1]}\right)\left(X_{[j,k]}X_{[0,1]}\right) \approx \epsilon \qquad \{j,k\} \cap \{0,1\} = \emptyset \tag{H-a2-4}$$

$$((-1)_{[0]} X_{[0,1]}) \; ((-1)_{[1]} X_{[0,1]}) \;\approx\; \epsilon \qquad\qquad \{j,k\} = \{0,1\} \qquad\qquad \text{(HZ-a2-1)}$$

$$(X_{[0,1]} X_{[j,k]}) \; (H_{[j+k,1]} H_{[0,1]}) \; (X_{[0,1]} X_{[j,k]}) \; (H_{[j+k,1]} H_{[0,1]}) \;\approx\; \epsilon \qquad \{j,k\} \cap \{0,1\} = \{0\} \qquad \text{(HZ-a2-2)}$$

$$(X_{[0,1]} X_{[j,k]}) \; (H_{[0,jk]} H_{[0,1]}) \; ((-1)_{[jk]} X_{[j,k]}) \; (H_{[0,jk]} H_{[0,1]}) \;\approx\; \epsilon \qquad \{j,k\} \cap \{0,1\} = \{1\} \qquad \text{(HZ-a2-3)}$$

$$(X_{[j,k]} X_{[0,1]}) \; (X_{[j,k]} X_{[0,1]}) \;\approx\; \epsilon \qquad\qquad \{j,k\} \cap \{0,1\} = \emptyset \qquad\qquad \text{(HZ-a2-4)}$$

$$((-1)_{[1]}(-1)_{[0]}) \; ((-1)_{[1]}(-1)_{[1]}) \;\approx\; ((-1)_{[1]}(-1)_{[1]}) \; ((-1)_{[1]}(-1)_{[0]}) \qquad\qquad (\epsilon\text{-b1}^*)$$

$$((-1)_{[1]}(-1)_{[0]}) \; ((-1)_{[1]}(-1)_{[1]}) \;\approx\; ((-1)_{[1]}(-1)_{[1]}) \; ((-1)_{[1]}(-1)_{[0]}) \qquad\qquad (\text{Z-b1}^*)$$

$$((-1)_{[1]}(-1)_{[0]}) \; ((-1)_{[1]}(-1)_{[1]}) \;\approx\; ((-1)_{[1]}(-1)_{[1]}) \; ((-1)_{[1]}(-1)_{[0]}) \qquad\qquad (\text{H-b1}^*)$$

$$((-1)_{[1]}(-1)_{[0]}) \; ((-1)_{[1]}(-1)_{[1]}) \;\approx\; ((-1)_{[1]}(-1)_{[1]}) \; ((-1)_{[1]}(-1)_{[0]}) \qquad\qquad (\text{HZ-b1}^*)$$

$$((-1)_{[1]}(-1)_{[1]}) \; (H_{[0,2]} H_{[0,1]}) \;\approx\; (H_{[0,2]} H_{[0,1]}) \; ((-1)_{[1]}(-1)_{[1]}) \qquad\qquad (\epsilon\text{-b4}^*)$$

$$((-1)_{[1]}(-1)_{[1]}) \; (H_{[0,2]} H_{[0,1]}) \;\approx\; (H_{[0,2]} H_{[0,1]}) \; ((-1)_{[1]}(-1)_{[1]}) \qquad\qquad (\text{Z-b4}^*)$$

$$((-1)_{[1]}(-1)_{[1]}) \; (H_{[0,1]} H_{[0,2]}) \;\approx\; (H_{[0,1]} H_{[0,2]}) \; ((-1)_{[1]}(-1)_{[1]}) \qquad\qquad (\text{H-b4}^*)$$

$$((-1)_{[1]}(-1)_{[1]}) \; (H_{[0,1]} H_{[0,2]}) \;\approx\; (H_{[0,1]} H_{[0,2]}) \; ((-1)_{[1]}(-1)_{[1]}) \qquad\qquad (\text{HZ-b4}^*)$$

$$(H_{[0,1]} H_{[0,1]}) \; (H_{[0,1]} H_{[2,3]}) \;\approx\; (H_{[2,3]} H_{[0,1]}) \; (H_{[0,1]} H_{[0,1]}) \qquad\qquad (\epsilon\text{-b6}^*)$$

$$((-1)_{[1]} X_{[0,1]}) \; (H_{[0,1]} H_{[0,1]}) \; (H_{[0,1]} H_{[2,3]}) \;\approx\; (H_{[2,3]} H_{[0,1]}) \; (H_{[0,1]} H_{[0,1]}) \; ((-1)_{[0]} X_{[0,1]}) \quad (\text{Z-b6}^*)$$

$$(H_{[0,1]} H_{[0,1]}) \; (H_{[2,3]} H_{[0,1]}) \;\approx\; (H_{[0,1]} H_{[2,3]}) \; (H_{[0,1]} H_{[0,1]}) \qquad\qquad (\text{H-b6}^*)$$

$$(H_{[0,1]} H_{[0,1]}) \; ((-1)_{[0]} X_{[0,1]}) \; (H_{[2,3]} H_{[0,1]}) \;\approx\; (H_{[0,1]} H_{[2,3]}) \; ((-1)_{[1]} X_{[0,1]}) \; (H_{[0,1]} H_{[0,1]}) \quad (\text{HZ-b6}^*)$$

$$((-1)_{[jk]} X_{[j,k]}) \; ((-1)_{[1]}(-1)_{[k]}) \;\approx\; ((-1)_{[1]}(-1)_{[j]}) \; ((-1)_{[1]} X_{[j,k]}) \qquad (j=1 \text{ or } k=1) \qquad (\epsilon\text{-c1-1})$$

$$((-1)_{[1]} X_{[j,k]}) \; ((-1)_{[1]}(-1)_{[k]}) \;\approx\; ((-1)_{[1]}(-1)_{[j]}) \; ((-1)_{[1]} X_{[j,k]}) \qquad j,k \neq 1 \qquad (\epsilon\text{-c1-2})$$

$$((-1)_{[1]} X_{[j,k]}) \; ((-1)_{[1]}(-1)_{[k]}) \;\approx\; ((-1)_{[1]}(-1)_{[j]}) \; ((-1)_{[jk]} X_{[j,k]}) \qquad (j=1 \text{ or } k=1) \qquad (\text{Z-c1-1})$$

$$((-1)_{[1]} X_{[j,k]}) \; ((-1)_{[1]}(-1)_{[k]}) \;\approx\; ((-1)_{[1]}(-1)_{[j]}) \; ((-1)_{[1]} X_{[j,k]}) \qquad j,k \neq 1 \qquad (\text{Z-c1-2})$$

$$((-1)_{[1]} X_{[0,1]}) \; ((-1)_{[1]}(-1)_{[k]}) \;\approx\; ((-1)_{[1]}(-1)_{[j]}) \; ((-1)_{[0]} X_{[0,1]}) \qquad \{j,k\} = \{0,1\} \qquad (\text{H-c1-1})$$

$$(X_{[0,1]} X_{[0,k]}) \; (H_{[k,1]} H_{[0,1]}) \; ((-1)_{[k]} X_{[0,1]}) \;\approx\; ((-1)_{[1]}(-1)_{[0]}) \; (X_{[0,1]} X_{[0,k]}) \; (H_{[k,1]} H_{[0,1]}) \qquad j=0, k \notin \{0,1\} \qquad (\text{H-c1-2})$$

$$((-1)_{[k]} X_{[1,k]}) \; (H_{[0,k]} H_{[0,1]}) \; ((-1)_{[k]} X_{[0,1]}) \;\approx\; ((-1)_{[1]}(-1)_{[1]}) \; (X_{[0,1]} X_{[1,k]}) \; (H_{[0,k]} H_{[0,1]}) \qquad j=1, k \notin \{0,1\} \qquad (\text{H-c1-3})$$

$$(X_{[0,1]} X_{[j,0]}) \; (H_{[j,1]} H_{[0,1]}) \; ((-1)_{[1]}(-1)_{[0]}) \;\approx\; ((-1)_{[j]} X_{[0,1]}) \; (X_{[0,1]} X_{[j,1]}) \; (H_{[j,1]} H_{[0,1]}) \qquad k=0, j \notin \{0,1\} \qquad (\text{H-c1-4})$$

$$((-1)_{[j]} X_{[j,1]}) \; (H_{[0,j]} H_{[0,1]}) \; ((-1)_{[1]}(-1)_{[1]}) \;\approx\; ((-1)_{[j]} X_{[0,1]}) \; (X_{[0,1]} X_{[j,1]}) \; (H_{[0,j]} H_{[0,1]}) \qquad k=1, j \notin \{0,1\} \qquad (\text{H-c1-5})$$

$$(X_{[j,k]} X_{[0,1]}) \; ((-1)_{[k]} X_{[0,1]}) \;\approx\; ((-1)_{[j]} X_{[0,1]}) \; (X_{[j,k]} X_{[0,1]}) \qquad \{j,k\} \cap \{0,1\} = \emptyset \qquad (\text{H-c1-6})$$

$$((-1)_{[0]} X_{[0,1]}) \; ((-1)_{[1]}(-1)_{[k]}) \;\approx\; ((-1)_{[1]}(-1)_{[j]}) \; ((-1)_{[1]} X_{[0,1]}) \qquad \{j,k\} = \{0,1\} \qquad (\text{HZ-c1-1})$$

$$(X_{[0,1]} X_{[0,k]}) \; (H_{[k,1]} H_{[0,1]}) \; ((-1)_{[k]} X_{[0,1]}) \;\approx\; ((-1)_{[1]}(-1)_{[0]}) \; (X_{[0,1]} X_{[0,k]}) \; (H_{[k,1]} H_{[0,1]}) \qquad j=0, k \notin \{0,1\} \qquad (\text{HZ-c1-2})$$

$$(X_{[0,1]} X_{[1,k]}) \; (H_{[0,k]} H_{[0,1]}) \; ((-1)_{[k]} X_{[0,1]}) \;\approx\; ((-1)_{[1]}(-1)_{[1]}) \; ((-1)_{[k]} X_{[1,k]}) \; (H_{[0,k]} H_{[0,1]}) \qquad j=1, k \notin \{0,1\} \qquad (\text{HZ-c1-3})$$

$$(X_{[0,1]} X_{[j,0]}) \; (H_{[j,1]} H_{[0,1]}) \; ((-1)_{[1]}(-1)_{[0]}) \;\approx\; ((-1)_{[j]} X_{[0,1]}) \; (X_{[0,1]} X_{[j,1]}) \; (H_{[j,1]} H_{[0,1]}) \qquad k=0, j \notin \{0,1\} \qquad (\text{HZ-c1-4})$$

$$(X_{[0,1]} X_{[j,1]}) \; (H_{[0,j]} H_{[0,1]}) \; ((-1)_{[1]}(-1)_{[1]}) \;\approx\; ((-1)_{[j]} X_{[0,1]}) \; ((-1)_{[j]} X_{[j,1]}) \; (H_{[0,j]} H_{[0,1]}) \qquad k=1, j \notin \{0,1\} \qquad (\text{HZ-c1-5})$$

$$(X_{[j,k]} X_{[0,1]}) \; ((-1)_{[k]} X_{[0,1]}) \;\approx\; ((-1)_{[j]} X_{[0,1]}) \; (X_{[j,k]} X_{[0,1]}) \qquad \{j,k\} \cap \{0,1\} = \emptyset \qquad (\text{HZ-c1-6})$$

$$((-1)_{[1]} X_{[j,k]}) \; ((-1)_{[j]} X_{[1,j]}) \; (H_{[1,j]} H_{[0,1]}) \;\approx\; (H_{[1,k]} H_{[0,1]}) \; (X_{[0,1]} X_{[j,k]}) \; (H_{[j+k,1]} H_{[0,1]}) \qquad l=1, \; (j=0 \text{ or } k=0) \qquad (\epsilon\text{-c5-1})$$

$$((-1)_{[1]} X_{[j,k]}) \; ((-1)_{[j]} X_{[1,j]}) \; (H_{[1,j]} H_{[0,1]}) \;\approx\; (H_{[1,k]} H_{[0,1]}) \; (X_{[j,k]} X_{[0,1]}) \qquad l=1, \; j,k \notin \{0,1\} \qquad (\epsilon\text{-c5-2})$$

$$((-1)_{[0]} X_{[1,0]}) \; ((-1)_{[1]} X_{[l,1]}) \; (H_{[l,1]} H_{[0,1]}) \;\approx\; (H_{[l,0]} H_{[0,1]}) \; ((-1)_{[1]} X_{[0,1]}) \qquad j=1, k=0 \qquad (\epsilon\text{-c5-3})$$

$$((-1)_{[k]} X_{[1,k]}) \; ((-1)_{[1]} X_{[l,1]}) \; (H_{[l,1]} H_{[0,1]}) \;\approx\; (H_{[l,k]} H_{[0,1]}) \; ((-1)_{[k]} X_{[1,k]}) \; (H_{[0,k]} H_{[0,1]}) \qquad j=1, k \notin \{0,1\} \qquad (\epsilon\text{-c5-4})$$

$$((-1)_{[0]}X_{[0,1]})\,(H_{[l,0]}H_{[0,1]}) \;\approx\; (H_{[l,1]}H_{[0,1]})\,((-1)_{[1]}X_{[0,1]}) \qquad\qquad k=1,\,j=0 \qquad (\epsilon\text{-c5-5})$$

$$((-1)_{[j]}X_{[j,1]})\,(H_{[l,j]}H_{[0,1]}) \;\approx\; (H_{[l,1]}H_{[0,1]})\,((-1)_{[j]}X_{[j,1]})\,(H_{[0,j]}H_{[0,1]}) \qquad k=1,\,j\notin\{0,1\} \qquad (\epsilon\text{-c5-6})$$

$$((-1)_{[1]}X_{[j,k]})\,(H_{[l,j]}H_{[0,1]}) \;\approx\; (H_{[l,k]}H_{[0,1]})\,(X_{[0,1]}X_{[j,k]})\,(H_{[j+k,1]}H_{[0,1]}) \qquad l\neq 1,\,\{j,k\}\cap\{0,1\}=\{0\} \qquad (\epsilon\text{-c5-7})$$

$$((-1)_{[1]}X_{[j,k]})\,(H_{[l,j]}H_{[0,1]}) \;\approx\; (H_{[l,k]}H_{[0,1]})\,(X_{[j,k]}X_{[0,1]}) \qquad l\neq 1,\,\{j,k\}\cap\{0,1\}=\emptyset \qquad (\epsilon\text{-c5-8})$$

$$((-1)_{[1]}X_{[j,k]})\,(H_{[1,j]}H_{[0,1]}) \;\approx\; {}_{((-1)_{[k]}X_{[1,k]})\,(H_{[1,k]}H_{[0,1]})\,(X_{[0,1]}X_{[j,k]})\,(H_{[j+k,1]}H_{[0,1]})} \qquad l=1,\,(j=0\text{ or }k=0) \qquad (Z\text{-c5-1})$$

$$((-1)_{[1]}X_{[j,k]})\,(H_{[1,j]}H_{[0,1]}) \;\approx\; ((-1)_{[k]}X_{[1,k]})\,(H_{[1,k]}H_{[0,1]})\,(X_{[j,k]}X_{[0,1]}) \qquad l=1,\,j,k\notin\{0,1\} \qquad (Z\text{-c5-2})$$

$$((-1)_{[1]}X_{[0,1]})\,(H_{[l,0]}H_{[0,1]}) \;\approx\; ((-1)_{[1]}X_{[l,1]})\,(H_{[l,1]}H_{[0,1]})\,((-1)_{[0]}X_{[0,1]}) \qquad k=1,\,j=0 \qquad (Z\text{-c5-3})$$

$$((-1)_{[1]}X_{[j,1]})\,(H_{[l,j]}H_{[0,1]}) \;\approx\; {}_{((-1)_{[1]}X_{[l,1]})\,(H_{[l,1]}H_{[0,1]})\,(X_{[0,1]}X_{[j,1]})\,(H_{[0,j]}H_{[0,1]})} \qquad k=1,\,j\notin\{0,1\} \qquad (Z\text{-c5-4})$$

$$((-1)_{[1]}X_{[1,0]})\,(H_{[1,1]}H_{[0,1]}) \;\approx\; (H_{[l,0]}H_{[0,1]})\,((-1)_{[0]}X_{[0,1]}) \qquad j=1,\,k=0 \qquad (Z\text{-c5-5})$$

$$((-1)_{[1]}X_{[1,k]})\,(H_{[l,1]}H_{[0,1]}) \;\approx\; (H_{[l,k]}H_{[0,1]})\,(X_{[0,1]}X_{[1,k]})\,(H_{[0,k]}H_{[0,1]}) \qquad j=1,\,k\notin\{0,1\} \qquad (Z\text{-c5-6})$$

$$((-1)_{[1]}X_{[j,k]})\,(H_{[l,j]}H_{[0,1]}) \;\approx\; (H_{[l,k]}H_{[0,1]})\,(X_{[0,1]}X_{[j,k]})\,(H_{[j+k,1]}H_{[0,1]}) \qquad l\neq 1,\,\{j,k\}\cap\{0,1\}=\{0\} \qquad (Z\text{-c5-7})$$

$$((-1)_{[1]}X_{[j,k]})\,(H_{[l,j]}H_{[0,1]}) \;\approx\; (H_{[l,k]}H_{[0,1]})\,(X_{[j,k]}X_{[0,1]}) \qquad l\neq 1,\,\{j,k\}\cap\{0,1\}=\emptyset \qquad (Z\text{-c5-8})$$

$$(X_{[0,1]}X_{[j,k]})\,(H_{[j+k,1]}H_{[0,1]})\,(H_{[0,1]}H_{[1,j]})\,((-1)_{[1]}X_{[1,j]}) \;\approx\; (H_{[0,1]}H_{[1,k]})\,((-1)_{[1]}X_{[j,k]}) \qquad l=1,\,(j=0\text{ or }k=0) \qquad (H\text{-c5-1})$$

$$(X_{[j,k]}X_{[0,1]})\,(H_{[0,1]}H_{[1,j]})\,((-1)_{[1]}X_{[1,j]}) \;\approx\; (H_{[0,1]}H_{[1,k]})\,((-1)_{[1]}X_{[j,k]}) \qquad l=1,\,j,k\notin\{0,1\} \qquad (H\text{-c5-2})$$

$$((-1)_{[1]}X_{[0,1]})\,(H_{[0,1]}H_{[l,1]})\,((-1)_{[l]}X_{[l,1]}) \;\approx\; (H_{[0,1]}H_{[l,0]})\,((-1)_{[0]}X_{[1,0]}) \qquad j=1,\,k=0 \qquad (H\text{-c5-3})$$

$$((-1)_{[k]}X_{[1,k]})\,(H_{[0,k]}H_{[0,1]})\,(H_{[0,1]}H_{[l,1]})\,((-1)_{[l]}X_{[l,1]}) \;\approx\; (H_{[0,1]}H_{[l,k]})\,((-1)_{[k]}X_{[1,k]}) \qquad j=1,\,k\notin\{0,1\} \qquad (H\text{-c5-4})$$

$$((-1)_{[1]}X_{[0,1]})\,(H_{[0,1]}H_{[l,0]}) \;\approx\; (H_{[0,1]}H_{[l,1]})\,((-1)_{[0]}X_{[0,1]}) \qquad k=1,\,j=0 \qquad (H\text{-c5-5})$$

$$((-1)_{[j]}X_{[j,1]})\,(H_{[0,j]}H_{[0,1]})\,(H_{[0,1]}H_{[l,j]}) \;\approx\; (H_{[0,1]}H_{[l,1]})\,((-1)_{[j]}X_{[j,1]}) \qquad k=1,\,j\notin\{0,1\} \qquad (H\text{-c5-6})$$

$$(X_{[0,1]}X_{[j,k]})\,(H_{[j+k,1]}H_{[0,1]})\,(H_{[0,1]}H_{[l,j]}) \;\approx\; (H_{[0,1]}H_{[l,k]})\,((-1)_{[1]}X_{[j,k]}) \qquad l\neq 1,\,\{j,k\}\cap\{0,1\}=\{0\} \qquad (H\text{-c5-7})$$

$$(X_{[j,k]}X_{[0,1]})\,(H_{[0,1]}H_{[l,j]}) \;\approx\; (H_{[0,1]}H_{[l,k]})\,((-1)_{[1]}X_{[j,k]}) \qquad l\neq 1,\,\{j,k\}\cap\{0,1\}=\emptyset \qquad (H\text{-c5-8})$$

$$(X_{[0,1]}X_{[j,k]})\,(H_{[j+k,1]}H_{[0,1]})\,(H_{[0,1]}H_{[1,j]}) \;\approx\; (H_{[0,1]}H_{[1,k]})\,((-1)_{[1]}X_{[1,k]})\,((-1)_{[1]}X_{[j,k]}) \qquad l=1,\,(j=0\text{ or }k=0) \qquad (HZ\text{-c5-1})$$

$$(X_{[j,k]}X_{[0,1]})\,(H_{[0,1]}H_{[1,j]}) \;\approx\; (H_{[0,1]}H_{[1,k]})\,((-1)_{[1]}X_{[1,k]})\,((-1)_{[1]}X_{[j,k]}) \qquad l=1,\,j,k\notin\{0,1\} \qquad (HZ\text{-c5-2})$$

$$((-1)_{[0]}X_{[0,1]})\,(H_{[0,1]}H_{[l,0]}) \;\approx\; (H_{[0,1]}H_{[l,1]})\,((-1)_{[1]}X_{[l,1]})\,((-1)_{[1]}X_{[1,1]}) \qquad k=1,\,j=0 \qquad (HZ\text{-c5-3})$$

$$(X_{[0,1]}X_{[j,1]})\,(H_{[0,j]}H_{[0,1]})\,(H_{[0,1]}H_{[l,j]}) \;\approx\; (H_{[0,1]}H_{[l,1]})\,((-1)_{[l]}X_{[l,1]})\,((-1)_{[1]}X_{[j,1]}) \qquad k=1,\,j\notin\{0,1\} \qquad (HZ\text{-c5-4})$$

$$((-1)_{[0]}X_{[0,1]})\,(H_{[0,1]}H_{[l,1]}) \;\approx\; (H_{[0,1]}H_{[l,0]})\,((-1)_{[1]}X_{[1,0]}) \qquad j=1,\,k=0 \qquad (HZ\text{-c5-5})$$

$$(X_{[0,1]}X_{[1,k]})\,(H_{[0,k]}H_{[0,1]})\,(H_{[0,1]}H_{[l,1]}) \;\approx\; (H_{[0,1]}H_{[l,k]})\,((-1)_{[1]}X_{[1,k]}) \qquad j=1,\,k\notin\{0,1\} \qquad (HZ\text{-c5-6})$$

$$(X_{[0,1]}X_{[j,k]})\,(H_{[j+k,1]}H_{[0,1]})\,(H_{[0,1]}H_{[l,j]}) \;\approx\; (H_{[0,1]}H_{[l,k]})\,((-1)_{[1]}X_{[j,k]}) \qquad l\neq 1,\,\{j,k\}\cap\{0,1\}=\{0\} \qquad (HZ\text{-c5-7})$$

$$(X_{[j,k]}X_{[0,1]})\,(H_{[0,1]}H_{[l,j]}) \;\approx\; (H_{[0,1]}H_{[l,k]})\,((-1)_{[1]}X_{[j,k]}) \qquad l\neq 1,\,\{j,k\}\cap\{0,1\}=\emptyset \qquad (HZ\text{-c5-8})$$

$$((-1)_{[0]}X_{[j,k]})\,((-1)_{[k]}X_{[j,k]})\,(H_{[j,k]}H_{[0,1]}) \;\approx\; (H_{[k,j]}H_{[0,1]})\,((-1)_{[1]}X_{[0,1]}) \qquad \{j,k\}=\{0,1\} \qquad (\epsilon\text{-e2}^*\text{-1})$$

$$((-1)_{[jk]}X_{[j,k]})\,((-1)_{[k]}X_{[j,k]})\,(H_{[j,k]}H_{[0,1]}) \;\approx\; (H_{[k,j]}H_{[0,1]})\,((-1)_{[jk]}X_{[j,k]})\,(H_{[0,jk]}H_{[0,1]}) \qquad \{j,k\}\cap\{0,1\}=\{1\} \qquad (\epsilon\text{-e2}^*\text{-2})$$

$$((-1)_{[1]}X_{[j,k]})\,(H_{[j,k]}H_{[0,1]}) \;\approx\; (H_{[k,j]}H_{[0,1]})\,(X_{[0,1]}X_{[j,k]})\,(H_{[j+k,1]}H_{[0,1]}) \qquad \{j,k\}\cap\{0,1\}=\{0\} \qquad (\epsilon\text{-e2}^*\text{-3})$$

$$((-1)_{[1]}X_{[j,k]})\,(H_{[j,k]}H_{[0,1]}) \;\approx\; (H_{[k,j]}H_{[0,1]})\,(X_{[j,k]}X_{[0,1]}) \qquad \{j,k\}\cap\{0,1\}=\emptyset \qquad (\epsilon\text{-e2}^*\text{-4})$$

$$((-1)_{[1]}X_{[j,k]})\,(H_{[j,k]}H_{[0,1]}) \;\approx\; ((-1)_{[j]}X_{[k,j]})\,(H_{[k,j]}H_{[0,1]})\,((-1)_{[0]}X_{[0,1]}) \qquad \{j,k\}=\{0,1\} \qquad (Z\text{-e2}^*\text{-1})$$

$$((-1)_{[1]}X_{[j,k]})\,(H_{[j,k]}H_{[0,1]}) \;\approx\; {}_{((-1)_{[j]}X_{[k,j]})\,(H_{[k,j]}H_{[0,1]})\,(X_{[0,1]}X_{[j,k]})\,(H_{[0,jk]}H_{[0,1]})} \qquad \{j,k\}\cap\{0,1\}=\{1\} \qquad (Z\text{-e2}^*\text{-2})$$

$$((-1)_{[1]}X_{[j,k]})\,(H_{[j,k]}H_{[0,1]}) \;\approx\; (H_{[k,j]}H_{[0,1]})\,(X_{[0,1]}X_{[j,k]})\,(H_{[j+k,1]}H_{[0,1]}) \qquad \{j,k\}\cap\{0,1\}=\{0\} \qquad (Z\text{-e2}^*\text{-3})$$

$$((-1)_{[1]}X_{[j,k]})\,(H_{[j,k]}H_{[0,1]}) \;\approx\; (H_{[k,j]}H_{[0,1]})\,(X_{[j,k]}X_{[0,1]}) \qquad \{j,k\}\cap\{0,1\}=\emptyset \qquad (Z\text{-e2}^*\text{-4})$$

$$((-1)_{[1]}X_{[0,1]})\,(H_{[0,1]}H_{[j]})\,((-1)_{[j]}X_{[j,k]}) \;\approx\; (H_{[0,1]}H_{[k,j]})\,((-1)_{[0]}X_{[j,k]}) \qquad \{j,k\}=\{0,1\} \qquad (H\text{-e2}^*\text{-1})$$

$$((-1)_{[jk]}X_{[j,k]})\,(H_{[0,jk]}H_{[0,1]})\,(H_{[0,1]}H_{[j,k]})\,((-1)_{[j]}X_{[j,k]}) \;\approx\; (H_{[0,1]}H_{[k,j]})\,((-1)_{[jk]}X_{[j,k]}) \qquad \{j,k\}\cap\{0,1\}=\{1\} \qquad (H\text{-e2}^*\text{-2})$$

$$(X_{[0,1]}X_{[j,k]})\,(H_{[j+k,1]}H_{[0,1]})\,(H_{[0,1]}H_{[j,k]}) \;\approx\; (H_{[0,1]}H_{[k,j]})\,((-1)_{[1]}X_{[j,k]}) \qquad \{j,k\}\cap\{0,1\}=\{0\} \qquad (H\text{-e2}^*\text{-3})$$

$$(X_{[j,k]}X_{[0,1]})\,(H_{[0,1]}H_{[j,k]}) \;\approx\; (H_{[0,1]}H_{[k,j]})\,((-1)_{[1]}X_{[j,k]}) \qquad \{j,k\}\cap\{0,1\}=\emptyset \qquad (H\text{-e2}^*\text{-4})$$

$$((-1)_{[0]}X_{[0,1]})\,(H_{[0,1]}H_{[j,k]}) \;\approx\; (H_{[0,1]}H_{[k,j]})\,((-1)_{[k]}X_{[k,j]})\,((-1)_{[1]}X_{[j,k]}) \qquad \{j,k\}=\{0,1\} \qquad (HZ\text{-e2}^*\text{-1})$$

$$\left(X_{[0,1]}X_{[j,k]}\right)\left(H_{[0,jk]}H_{[0,1]}\right)\left(H_{[0,1]}H_{[j,k]}\right) \approx \left(H_{[0,1]}H_{[k,j]}\right)\left((-1)_{[k]}X_{[k,j]}\right)\left((-1)_{[1]}X_{[j,k]}\right) \qquad \{j,k\}\cap\{0,1\}=\{1\} \qquad \text{(HZ-e2*-2)}$$

$$\left(X_{[0,1]}X_{[j,k]}\right)\left(H_{[j+k,1]}H_{[0,1]}\right)\left(H_{[0,1]}H_{[j,k]}\right) \approx \left(H_{[0,1]}H_{[k,j]}\right)\left((-1)_{[1]}X_{[j,k]}\right) \qquad \{j,k\}\cap\{0,1\}=\{0\} \qquad \text{(HZ-e2*-3)}$$

$$\left(X_{[j,k]}X_{[0,1]}\right)\left(H_{[0,1]}H_{[j,k]}\right) \approx \left(H_{[0,1]}H_{[k,j]}\right)\left((-1)_{[1]}X_{[j,k]}\right) \qquad \{j,k\}\cap\{0,1\}=\emptyset \qquad \text{(HZ-e2*-4)}$$

$$\left(H_{[j,k]}H_{[0,1]}\right)\left((-1)_{[1]}X_{[0,1]}\right) \approx (-1)_{[1]}(-1)_{[k]}\left((-1)_{[k]}X_{[j,k]}\right)\left(H_{[j,k]}H_{[0,1]}\right) \qquad \{j,k\}=\{0,1\} \qquad \text{($\epsilon$-d2-1)}$$

$$\left(H_{[j,k]}H_{[0,1]}\right)\left((-1)_{[jk]}X_{[j,k]}\right)\left(H_{[0,jk]}H_{[0,1]}\right) \approx (-1)_{[1]}(-1)_{[k]}\left((-1)_{[k]}X_{[j,k]}\right)\left(H_{[j,k]}H_{[0,1]}\right) \qquad \{j,k\}\cap\{0,1\}=\{1\} \qquad \text{($\epsilon$-d2-2)}$$

$$\left(H_{[j,k]}H_{[0,1]}\right)\left(X_{[0,1]}X_{[j,k]}\right)\left(H_{[j+k,1]}H_{[0,1]}\right) \approx (-1)_{[1]}(-1)_{[k]}\left(H_{[j,k]}H_{[0,1]}\right) \qquad \{j,k\}\cap\{0,1\}=\{0\} \qquad \text{($\epsilon$-d2-3)}$$

$$\left(H_{[j,k]}H_{[0,1]}\right)\left(X_{[j,k]}X_{[0,1]}\right) \approx (-1)_{[1]}(-1)_{[k]}\left(H_{[j,k]}H_{[0,1]}\right) \qquad \{j,k\}\cap\{0,1\}=\emptyset \qquad \text{($\epsilon$-d2-4)}$$

$$\left((-1)_{[k]}X_{[j,k]}\right)\left(H_{[j,k]}H_{[0,1]}\right)\left((-1)_{[0]}X_{[0,1]}\right) \approx (-1)_{[1]}(-1)_{[k]}\left(H_{[j,k]}H_{[0,1]}\right) \qquad \{j,k\}=\{0,1\} \qquad \text{(Z-d2-1)}$$

$$\left((-1)_{[k]}X_{[j,k]}\right)\left(H_{[j,k]}H_{[0,1]}\right)\left(X_{[0,1]}X_{[j,k]}\right)\left(H_{[0,jk]}H_{[0,1]}\right) \approx (-1)_{[1]}(-1)_{[k]}\left(H_{[j,k]}H_{[0,1]}\right) \qquad \{j,k\}\cap\{0,1\}=\{1\} \qquad \text{(Z-d2-2)}$$

$$\left(H_{[j,k]}H_{[0,1]}\right)\left(X_{[0,1]}X_{[j,k]}\right)\left(H_{[j+k,1]}H_{[0,1]}\right) \approx (-1)_{[1]}(-1)_{[k]}\left(H_{[j,k]}H_{[0,1]}\right) \qquad \{j,k\}\cap\{0,1\}=\{0\} \qquad \text{(Z-d2-3)}$$

$$\left(H_{[j,k]}H_{[0,1]}\right)\left(X_{[j,k]}X_{[0,1]}\right) \approx (-1)_{[1]}(-1)_{[k]}\left(H_{[j,k]}H_{[0,1]}\right) \qquad \{j,k\}\cap\{0,1\}=\emptyset \qquad \text{(Z-d2-4)}$$

$$\left(H_{[0,1]}H_{[j,1]}\right)\left((-1)_{[1]}\right)\left((-1)_{[j]}X_{[j,1]}\right) \approx (-1)_{[1]}(-1)_{[j]}\left(H_{[0,1]}H_{[j,1]}\right) \qquad k=1 \qquad \text{(H-d2-1)}$$

$$\left(H_{[0,1]}H_{[1,0]}\right)\left((-1)_{[0]}X_{[1,0]}\right) \approx (-1)_{[1]}(-1)_{[0]}\left(H_{[0,1]}H_{[1,0]}\right)\left((-1)_{[1]}X_{[1,0]}\right) \qquad j=1,k=0 \qquad \text{(H-d2-2)}$$

$$\left(H_{[0,1]}H_{[1,k]}\right)\left((-1)_{[k]}X_{[1,k]}\right) \approx (-1)_{[k]}X_{[0,1]}\left(H_{[0,1]}H_{[1,k]}\right)\left((-1)_{[1]}X_{[1,k]}\right) \qquad j=1,k\notin\{0,1\} \qquad \text{(H-d2-3)}$$

$$\left(H_{[0,1]}H_{[j,0]}\right)\left((-1)_{[1]}X_{[j,0]}\right) \approx (-1)_{[1]}(-1)_{[0]}\left(H_{[0,1]}H_{[j,0]}\right) \qquad j\neq 1,k=0 \qquad \text{(H-d2-4)}$$

$$\left(H_{[0,1]}H_{[j,k]}\right)\left((-1)_{[1]}X_{[j,k]}\right) \approx (-1)_{[k]}X_{[0,1]}\left(H_{[0,1]}H_{[j,k]}\right) \qquad j\neq 1,k\notin\{0,1\} \qquad \text{(H-d2-5)}$$

$$\left(H_{[0,1]}H_{[j,1]}\right)\left((-1)_{[j]}X_{[j,1]}\right)\left((-1)_{[1]}X_{[j,1]}\right) \approx (-1)_{[1]}(-1)_{[j]}\left(H_{[0,1]}H_{[j,1]}\right) \qquad k=1 \qquad \text{(HZ-d2-1)}$$

$$\left(H_{[0,1]}H_{[1,0]}\right)\left((-1)_{[1]}X_{[1,0]}\right)\left((-1)_{[1]}X_{[1,0]}\right) \approx (-1)_{[1]}(-1)_{[0]}\left(H_{[0,1]}H_{[1,0]}\right) \qquad j=1,k=0 \qquad \text{(HZ-d2-2)}$$

$$\left(H_{[0,1]}H_{[1,k]}\right)\left((-1)_{[1]}X_{[1,k]}\right)\left((-1)_{[1]}X_{[1,k]}\right) \approx (-1)_{[k]}X_{[0,1]}\left(H_{[0,1]}H_{[1,k]}\right) \qquad j=1,k\notin\{0,1\} \qquad \text{(HZ-d2-3)}$$

$$\left(H_{[0,1]}H_{[j,0]}\right)\left((-1)_{[1]}X_{[j,0]}\right) \approx (-1)_{[1]}(-1)_{[0]}\left(H_{[0,1]}H_{[j,0]}\right) \qquad j\neq 1,k=0 \qquad \text{(HZ-d2-4)}$$

$$\left(H_{[0,1]}H_{[j,k]}\right)\left((-1)_{[1]}X_{[j,k]}\right) \approx (-1)_{[k]}X_{[0,1]}\left(H_{[0,1]}H_{[j,k]}\right) \qquad j\neq 1,k\notin\{0,1\} \qquad \text{(HZ-d2-5)}$$

$$\begin{aligned}
&\left(H_{[0,1]}H_{[0,1]}\right)\left(H_{[0,1]}H_{[0,2]}\right)\left(H_{[1,3]}H_{[0,1]}\right)\left(H_{[0,1]}H_{[0,1]}\right)\\
&\left((-1)_{[1]}(-1)_{[0]}\right)\left((-1)_{[1]}(-1)_{[1]}\right)\left(H_{[0,2]}H_{[0,1]}\right)\left(H_{[0,1]}H_{[1,3]}\right) \approx\\
&\left(H_{[0,2]}H_{[0,1]}\right)\left(H_{[0,1]}H_{[1,3]}\right)\left(H_{[0,1]}H_{[0,1]}\right)\left((-1)_{[1]}(-1)_{[0]}\right)\left((-1)_{[1]}(-1)_{[1]}\right)\\
&\left(H_{[0,1]}H_{[0,2]}\right)\left(H_{[1,3]}H_{[0,1]}\right)\left(H_{[0,1]}H_{[0,1]}\right)
\end{aligned} \qquad \text{($\epsilon$-d3*)}$$

$$\begin{aligned}
&\left((-1)_{[1]}X_{[0,1]}\right)\left(H_{[0,1]}H_{[0,1]}\right)\left(H_{[0,1]}H_{[0,2]}\right)\left((-1)_{[3]}X_{[1,3]}\right)\left(H_{[1,3]}H_{[0,1]}\right)\left(H_{[0,1]}H_{[0,1]}\right)\left((-1)_{[0]}X_{[0,1]}\right)\\
&\left((-1)_{[1]}(-1)_{[0]}\right)\left((-1)_{[1]}(-1)_{[1]}\right)\left(H_{[0,2]}H_{[0,1]}\right)\left(H_{[0,1]}H_{[1,3]}\right)\left((-1)_{[1]}X_{[1,3]}\right) \approx\\
&\left(H_{[0,2]}H_{[0,1]}\right)\left(H_{[0,1]}H_{[1,3]}\right)\left((-1)_{[1]}X_{[1,3]}\right)\left((-1)_{[1]}X_{[0,1]}\right)\left(H_{[0,1]}H_{[0,1]}\right)\left((-1)_{[1]}(-1)_{[0]}\right)\left((-1)_{[1]}(-1)_{[1]}\right)\\
&\left(H_{[0,1]}H_{[0,2]}\right)\left((-1)_{[3]}X_{[1,3]}\right)\left(H_{[1,3]}H_{[0,1]}\right)\left(H_{[0,1]}H_{[0,1]}\right)\left((-1)_{[0]}X_{[0,1]}\right)
\end{aligned} \qquad \text{(Z-d3*)}$$

$$\begin{aligned}
&\left(H_{[0,1]}H_{[0,1]}\right)\left(H_{[0,2]}H_{[0,1]}\right)\left(H_{[0,1]}H_{[1,3]}\right)\left(H_{[0,1]}H_{[0,1]}\right)\\
&\left((-1)_{[1]}(-1)_{[0]}\right)\left((-1)_{[1]}(-1)_{[1]}\right)\left(H_{[0,1]}H_{[0,2]}\right)\left(H_{[1,3]}H_{[0,1]}\right) \approx\\
&\left(H_{[0,1]}H_{[0,2]}\right)\left(H_{[1,3]}H_{[0,1]}\right)\left(H_{[0,1]}H_{[0,1]}\right)\left((-1)_{[1]}(-1)_{[0]}\right)\left((-1)_{[1]}(-1)_{[1]}\right)\\
&\left(H_{[0,2]}H_{[0,1]}\right)\left(H_{[0,1]}H_{[1,3]}\right)\left(H_{[0,1]}H_{[0,1]}\right)
\end{aligned} \qquad \text{(H-d3*)}$$

$$\begin{aligned}
&\left(H_{[0,1]}H_{[0,1]}\right)\left((-1)_{[0]}X_{[0,1]}\right)\left(H_{[0,2]}H_{[0,1]}\right)\left(H_{[0,1]}H_{[1,3]}\right)\left((-1)_{[1]}X_{[1,3]}\right)\left((-1)_{[0]}X_{[0,1]}\right)\left(H_{[0,1]}H_{[0,1]}\right)\\
&\left((-1)_{[1]}(-1)_{[0]}\right)\left((-1)_{[1]}(-1)_{[1]}\right)\left(H_{[0,1]}H_{[0,2]}\right)\left((-1)_{[3]}X_{[1,3]}\right)\left(H_{[1,3]}H_{[0,1]}\right) \approx\\
&\left(H_{[0,2]}H_{[0,1]}\right)\left((-1)_{[3]}X_{[1,3]}\right)\left(H_{[1,3]}H_{[0,1]}\right)\left(H_{[0,1]}H_{[0,1]}\right)\left((-1)_{[0]}X_{[0,1]}\right)\left((-1)_{[1]}(-1)_{[0]}\right)\left((-1)_{[1]}(-1)_{[1]}\right)\\
&\left(H_{[0,2]}H_{[0,1]}\right)\left(H_{[0,1]}H_{[1,3]}\right)\left((-1)_{[1]}X_{[1,3]}\right)\left((-1)_{[1]}X_{[0,1]}\right)\left(H_{[0,1]}H_{[0,1]}\right)
\end{aligned} \qquad \text{(HZ-d3*)}$$

$$
\begin{aligned}
(H_{[0,1]}H_{[0,1]}) \; (H_{[0,1]}H_{[0,2]}) \; (H_{[1,3]}H_{[0,1]}) \; (H_{[0,1]}H_{[0,4]}) \; (H_{[1,5]}H_{[0,1]}) \; (H_{[0,1]}H_{[0,1]}) \\
((-1)_{[1]}(-1)_{[0]}) \; ((-1)_{[1]}(-1)_{[1]}) \; (H_{[0,4]}H_{[0,1]}) \; (H_{[0,1]}H_{[1,5]}) \; (H_{[0,2]}H_{[0,1]}) \; (H_{[0,1]}H_{[1,3]}) \quad \approx \\
(H_{[0,2]}H_{[0,1]}) \; (H_{[0,1]}H_{[1,3]}) \; (H_{[0,4]}H_{[0,1]}) \; (H_{[0,1]}H_{[1,5]}) \; (H_{[0,1]}H_{[0,1]}) \; ((-1)_{[1]}(-1)_{[0]}) \; ((-1)_{[1]}(-1)_{[1]}) \\
(H_{[0,1]}H_{[0,4]}) \; (H_{[1,5]}H_{[0,1]}) \; (H_{[0,1]}H_{[0,2]}) \; (H_{[1,3]}H_{[0,1]}) \; (H_{[0,1]}H_{[0,1]})
\end{aligned}
\tag{$\epsilon$-d4*}
$$

$$
\begin{aligned}
((-1)_{[1]}X_{[0,1]}) \; (H_{[0,1]}H_{[0,1]}) \; (H_{[0,1]}H_{[0,2]}) \; ((-1)_{[3]}X_{[1,3]}) \; (H_{[1,3]}H_{[0,1]}) \; (H_{[0,1]}H_{[0,4]}) \; ((-1)_{[5]}X_{[1,5]}) \; (H_{[1,5]}H_{[0,1]}) \; (H_{[0,1]}H_{[0,1]}) \; ((-1)_{[0]}X_{[0,1]}) \\
((-1)_{[1]}(-1)_{[0]}) \; ((-1)_{[1]}(-1)_{[1]}) \; (H_{[0,4]}H_{[0,1]}) \; (H_{[0,1]}H_{[1,5]}) \; ((-1)_{[1]}X_{[1,5]}) \; (H_{[0,2]}H_{[0,1]}) \; ((-1)_{[1]}(-1)_{[0]}) \; ((-1)_{[1]}(-1)_{[1]}) \quad \approx \\
(H_{[0,2]}H_{[0,1]}) \; (H_{[0,1]}H_{[1,3]}) \; ((-1)_{[1]}X_{[1,3]}) \; (H_{[0,4]}H_{[0,1]}) \; (H_{[0,1]}H_{[1,5]}) \; ((-1)_{[1]}X_{[1,5]}) \; ((-1)_{[1]}X_{[0,1]}) \; ((-1)_{[1]}(-1)_{[0]}) \; ((-1)_{[1]}(-1)_{[1]}) \\
(H_{[0,1]}H_{[0,4]}) \; ((-1)_{[5]}X_{[1,5]}) \; (H_{[1,5]}H_{[0,1]}) \; (H_{[0,1]}H_{[0,2]}) \; ((-1)_{[3]}X_{[1,3]}) \; (H_{[1,3]}H_{[0,1]}) \; (H_{[0,1]}H_{[0,1]}) \; ((-1)_{[0]}X_{[0,1]})
\end{aligned}
\tag{Z-d4*}
$$

$$
\begin{aligned}
(H_{[0,1]}H_{[0,1]}) \; (H_{[0,2]}H_{[0,1]}) \; (H_{[0,1]}H_{[1,3]}) \; (H_{[0,4]}H_{[0,1]}) \; (H_{[0,1]}H_{[1,5]}) \; (H_{[0,1]}H_{[0,1]}) \\
((-1)_{[1]}(-1)_{[0]}) \; ((-1)_{[1]}(-1)_{[1]}) \; (H_{[0,1]}H_{[0,4]}) \; (H_{[1,5]}H_{[0,1]}) \; (H_{[0,1]}H_{[0,2]}) \; (H_{[1,3]}H_{[0,1]}) \quad \approx \\
(H_{[0,1]}H_{[0,2]}) \; (H_{[1,3]}H_{[0,1]}) \; (H_{[0,1]}H_{[0,4]}) \; (H_{[1,5]}H_{[0,1]}) \; (H_{[0,1]}H_{[0,1]}) \; ((-1)_{[1]}(-1)_{[0]}) \; ((-1)_{[1]}(-1)_{[1]}) \\
(H_{[0,4]}H_{[0,1]}) \; (H_{[0,1]}H_{[1,5]}) \; (H_{[0,2]}H_{[0,1]}) \; (H_{[0,1]}H_{[1,3]}) \; (H_{[0,1]}H_{[0,1]})
\end{aligned}
\tag{H-d4*}
$$

$$
\begin{aligned}
(H_{[0,1]}H_{[0,1]}) \; ((-1)_{[0]}X_{[0,1]}) \; (H_{[0,2]}H_{[0,1]}) \; (H_{[0,1]}H_{[1,3]}) \; ((-1)_{[1]}X_{[1,3]}) \; (H_{[0,4]}H_{[0,1]}) \; (H_{[0,1]}H_{[1,5]}) \; ((-1)_{[1]}X_{[1,5]}) \; ((-1)_{[1]}X_{[0,1]}) \; (H_{[0,1]}H_{[0,1]}) \\
((-1)_{[1]}(-1)_{[0]}) \; ((-1)_{[1]}(-1)_{[1]}) \; (H_{[0,1]}H_{[0,4]}) \; ((-1)_{[5]}X_{[1,5]}) \; (H_{[1,5]}H_{[0,1]}) \; (H_{[0,1]}H_{[0,2]}) \; ((-1)_{[3]}X_{[1,3]}) \; (H_{[1,3]}H_{[0,1]}) \quad \approx \\
(H_{[0,1]}H_{[0,2]}) \; ((-1)_{[3]}X_{[1,3]}) \; (H_{[1,3]}H_{[0,1]}) \; (H_{[0,1]}H_{[0,1]}) \; ((-1)_{[5]}X_{[1,5]}) \; (H_{[1,5]}H_{[0,1]}) \; (H_{[0,1]}H_{[0,1]}) \; ((-1)_{[0]}X_{[0,1]}) \; ((-1)_{[1]}(-1)_{[0]}) \; ((-1)_{[1]}(-1)_{[1]}) \\
(H_{[0,4]}H_{[0,1]}) \; (H_{[0,1]}H_{[1,5]}) \; ((-1)_{[1]}X_{[1,5]}) \; (H_{[0,2]}H_{[0,1]}) \; (H_{[0,1]}H_{[1,3]}) \; ((-1)_{[1]}X_{[1,3]}) \; ((-1)_{[1]}X_{[0,1]}) \; (H_{[0,1]}H_{[0,1]})
\end{aligned}
\tag{HZ-d4*}
$$

## A.4 Simplifying the Output of the Reidemeister-Schreier Theorem

*A.4.1 Details about Equations* (36) *and* (37). The right-hand sides of Equations (36) and (37) are defined respectively as follows:

$$
\begin{aligned}
\mathcal{W}_1 :=\; & \left( \overleftarrow{\prod_{j=m+1}^{m+\ell}} E_{n-j-2,j}(\,\smile\!\frown\,) \right) \circ \left( \overleftarrow{\prod_{j=0}^{m-1}} E_{n-j-2,j}(\,\smile\!\frown\,) \circ E_{n-j-3,j+1}(\,\smile\!\frown\,) \right) \circ E_{n-2,0}(\,\smile\!\frown\,) \circ \\[6pt]
& \left( \overleftarrow{\prod_{j=k+\ell}^{n-3}} \mathcal{E}_{j,n-j-2}^{1-z_j}(\,-\!\oplus\!-\,) \right) \circ \left( \overleftarrow{\prod_{j=k}^{k+\ell-1}} \mathcal{E}_{j,n-j-2}^{1-y_j}(\,-\!\oplus\!-\,) \right) \circ \left( \overleftarrow{\prod_{j=0}^{k-1}} \mathcal{E}_{j,n-j-2}^{1-x_j}(\,-\!\oplus\!-\,) \right) \circ \\[6pt]
& (H_{[0,1]}H_{[3,2]}) \circ \\[6pt]
& \left( \overrightarrow{\prod_{j=0}^{k-1}} \mathcal{E}_{j,n-j-2}^{1-x_j}(\,-\!\oplus\!-\,) \right) \circ \left( \overrightarrow{\prod_{j=k}^{k+\ell-1}} \mathcal{E}_{j,n-j-2}^{1-y_j}(\,-\!\oplus\!-\,) \right) \circ \left( \overrightarrow{\prod_{j=k+\ell}^{n-3}} \mathcal{E}_{j,n-j-2}^{1-z_j}(\,-\!\oplus\!-\,) \right) \circ \\[6pt]
& E_{n-2,0}(\,\smile\!\frown\,) \circ \left( \overrightarrow{\prod_{j=0}^{m-1}} E_{n-j-3,j+1}(\,\smile\!\frown\,) \circ E_{n-j-2,j}(\,\smile\!\frown\,) \right) \circ \left( \overrightarrow{\prod_{j=m+1}^{m+\ell}} E_{n-j-2,j}(\,\smile\!\frown\,) \right)
\end{aligned}
$$

and

$$
\mathcal{W}_2 :=\; \left( \overleftarrow{\prod_{j=m+1}^{m+\ell}} E_{n-j-2,j}(\,\smile\!\frown\,) \right) \circ \left( \overleftarrow{\prod_{j=0}^{m-1}} E_{n-j-2,j}(\,\smile\!\frown\,) \circ E_{n-j-3,j+1}(\,\smile\!\frown\,) \right) \circ
$$

$$\left(\overleftarrow{\prod_{j=k+\ell}^{n-3}} \mathcal{E}_{j,n-j-2}^{1-z_j}(-\oplus-)\right) \circ \left(\overleftarrow{\prod_{j=k}^{k+\ell-1}} \mathcal{E}_{j,n-j-2}^{1-y_j}(-\oplus-)\right) \circ \left(\overleftarrow{\prod_{j=0}^{k-1}} \mathcal{E}_{j,n-j-2}^{1-x_j}(-\oplus-)\right) \circ$$

$$(H_{[0,1]} H_{[3,2]}) \circ$$

$$\left(\overrightarrow{\prod_{j=0}^{k-1}} \mathcal{E}_{j,n-j-2}^{1-x_j}(-\oplus-)\right) \circ \left(\overrightarrow{\prod_{j=k}^{k+\ell-1}} \mathcal{E}_{j,n-j-2}^{1-y_j}(-\oplus-)\right) \circ \left(\overrightarrow{\prod_{j=k+\ell}^{n-3}} \mathcal{E}_{j,n-j-2}^{1-z_j}(-\oplus-)\right) \circ$$

$$\left(\overrightarrow{\prod_{j=0}^{m-1}} E_{n-j-3,j+1}(\smile\frown) \circ E_{n-j-2,j}(\smile\frown)\right) \circ \left(\overrightarrow{\prod_{j=m+1}^{m+\ell}} E_{n-j-2,j}(\smile\frown)\right)$$

where the notations $\overleftarrow{\prod}$ and $\overrightarrow{\prod}$ are introduced in Definition 8.3, and $\mathcal{E}_{k,\ell}^{1-x}(-\oplus-)$ is a shortcut defined as follows (where the order of the product is left unspecified):

$$\mathcal{E}_{k,\ell}^{1-x}(-\oplus-) := \begin{cases} \epsilon & \text{if } x = 1 \\ E_{k,\ell}(-\bullet-) \circ \displaystyle\prod_{\substack{x \in \{0,1\}^k \\ y \in \{0,1\}^\ell}} \left((-1)_{\lfloor G_n^{-1}(x1y)\rfloor} X_{[G_n^{-1}(x0y),G_n^{-1}(x1y)]}\right) & \text{if } x = 0. \end{cases}$$

*A.4.2 Proof of Completeness of $\approx$ for $\mathcal{P}_N^*$.* The goal of this appendix is to prove the completeness part of Theorem 4.10 (stated as Corollary A.12 below). For this purpose, we prove increasingly strong partial completeness results, together with some auxiliary equations, so as to be finally able to derive the equations of Appendix A.3.2 with minimum effort.

**Lemma A.3.** *The equations of Figure 9 are consequences of the equations of Figure 8.*

PROOF. [12]

*Proof of Equation* (47):

$$\begin{aligned} \left((-1)_{[a+1]} X_{[a,a+1]}\right) \left((-1)_{[a]} X_{[a,a+1]}\right) &\overset{(20)(22)(21)}{\approx} \left((-1)_{[a]}(-1)_{[a+1]}\right)^2 \left((-1)_{[a+1]} X_{[a,a+1]}\right) \left((-1)_{[a]} X_{[a,a+1]}\right) \\ &\overset{(23)}{\approx} \left((-1)_{[a]} X_{[a,a+1]}\right)^4 \left((-1)_{[a+1]} X_{[a,a+1]}\right) \left((-1)_{[a]} X_{[a,a+1]}\right) \\ &\overset{(29)}{\approx} \left((-1)_{[a]} X_{[a,a+1]}\right)^4 \\ &\overset{(23)(21)(22)(20)}{\approx} \epsilon \end{aligned}$$

*Proof of Equation* (48):

$$\begin{aligned} \left((-1)_{[a+1]} X_{[a,a+1]}\right) &\overset{(20)(22)(21)}{\approx} \left((-1)_{[a]}(-1)_{[a+1]}\right)^2 \left((-1)_{[a+1]} X_{[a,a+1]}\right) \\ &\overset{(23)}{\approx} \left((-1)_{[a]} X_{[a,a+1]}\right)^4 \left((-1)_{[a+1]} X_{[a,a+1]}\right) \\ &\overset{(29)}{\approx} \left((-1)_{[a]} X_{[a,a+1]}\right)^3 \end{aligned}$$

*Proof of Equation* (49): If $c - a$ is odd, then Equation (49) is just Equation (25), therefore we only have to consider the case where $c - a$ is even. We proceed by induction on $c - a$. If $c - a = 2$, then we have

$$\begin{aligned} \left((-1)_{[a]} X_{[a,a+2]}\right) &\overset{(27)}{\approx} \left((-1)_{[a+1]} X_{[a+1,a+2]}\right) \left((-1)_{[a]} X_{[a,a+1]}\right) \left((-1)_{[a+2]} X_{[a+1,a+2]}\right) \\ &\overset{(48)}{\approx} \left((-1)_{[a+1]} X_{[a+1,a+2]}\right) \left((-1)_{[a]} X_{[a,a+1]}\right) \left((-1)_{[a+1]} X_{[a+1,a+2]}\right)^3 \\ &\overset{(24)}{\approx} \left((-1)_{[a]} X_{[a,a+1]}\right)^3 \left((-1)_{[a+1]} X_{[a+1,a+2]}\right) \left((-1)_{[a]} X_{[a,a+1]}\right) \end{aligned}$$

---

[12]Note that throughout all the paper, whenever we write a derivation to prove an equation to which parameter conditions are attached, we implicitly assume the conditions of the equation that we are proving.

$$\left((-1)_{[a+1]}X_{[a,a+1]}\right)\left((-1)_{[a]}X_{[a,a+1]}\right) \approx \epsilon \tag{47}$$

$$\left((-1)_{[a+1]}X_{[a,a+1]}\right) \approx \left((-1)_{[a]}X_{[a,a+1]}\right)^3 \tag{48}$$

$$\left((-1)_{[a]}X_{[a,c]}\right) \approx \left((-1)_{[a+1]}X_{[a,a+1]}\right)\left((-1)_{[a+1]}X_{[a+1,c]}\right)\left((-1)_{[a]}X_{[a,a+1]}\right) \qquad a+1<c \tag{49}$$

$$\left((-1)_{[c]}X_{[a,c]}\right) \approx \left((-1)_{[a+1]}X_{[a,a+1]}\right)\left((-1)_{[c]}X_{[a+1,c]}\right)\left((-1)_{[a]}X_{[a,a+1]}\right) \qquad a+1<c \tag{50}$$

$$\left((-1)_{[a]}X_{[a,b]}\right)\left((-1)_{[b]}X_{[a,b]}\right) \approx \epsilon \tag{51}$$

$$\left((-1)_{[a]}(-1)_{[b]}\right) \approx \left((-1)_{[a]}X_{[a,b]}\right)\left((-1)_{[a]}X_{[a,b]}\right) \qquad a\neq b \tag{52}$$

$$\left((-1)_{[c]}X_{[a,b]}\right) \approx \left((-1)_{[c]}(-1)_{[a]}\right)\left((-1)_{[a]}X_{[a,b]}\right) \tag{53}$$

$$\left((-1)_{[a]}X_{[a,c]}\right) \approx \left((-1)_{[a]}X_{[a,a+1]}\right)^3\left((-1)_{[a+1]}X_{[a+1,c]}\right)\left((-1)_{[a]}X_{[a,a+1]}\right) \qquad a+1<c \tag{54}$$

$$\left((-1)_{[c]}(-1)_{[d]}\right)\left((-1)_{[a]}X_{[a,c]}\right) \approx \left((-1)_{[a]}X_{[a,b]}\right)\left((-1)_{[c]}(-1)_{[d]}\right) \qquad \{a,a+1\}\cap\{c,d\}=\emptyset \tag{55}$$

$$\left((-1)_{[a]}(-1)_{[b]}\right)\left((-1)_{[a]}X_{[a,b]}\right) \approx \left((-1)_{[a]}X_{[a,b]}\right)\left((-1)_{[a]}(-1)_{[b]}\right) \tag{56}$$

$$\left((-1)_{[b]}(-1)_{[a]}\right)\left((-1)_{[a]}X_{[a,b]}\right) \approx \left((-1)_{[a]}X_{[a,b]}\right)\left((-1)_{[b]}(-1)_{[a]}\right) \tag{57}$$

$$\left((-1)_{[a]}(-1)_{[c]}\right)\left((-1)_{[a]}X_{[a,b]}\right) \approx \left((-1)_{[a]}X_{[a,b]}\right)\left((-1)_{[b]}(-1)_{[c]}\right) \qquad c\neq a,b \tag{58}$$

$$\left((-1)_{[c]}(-1)_{[a]}\right)\left((-1)_{[a]}X_{[a,b]}\right) \approx \left((-1)_{[a]}X_{[a,b]}\right)\left((-1)_{[c]}(-1)_{[b]}\right) \qquad c\neq a,b \tag{59}$$

$$\left((-1)_{[b]}(-1)_{[c]}\right)\left((-1)_{[a]}X_{[a,b]}\right) \approx \left((-1)_{[a]}X_{[a,b]}\right)\left((-1)_{[a]}(-1)_{[c]}\right) \qquad c\neq a,b \tag{60}$$

$$\left((-1)_{[c]}(-1)_{[b]}\right)\left((-1)_{[a]}X_{[a,b]}\right) \approx \left((-1)_{[a]}X_{[a,b]}\right)\left((-1)_{[c]}(-1)_{[a]}\right) \qquad c\neq a,b \tag{61}$$

$$\left((-1)_{[c]}(-1)_{[d]}\right)\left((-1)_{[a]}X_{[a,b]}\right) \approx \left((-1)_{[a]}X_{[a,b]}\right)\left((-1)_{[c]}(-1)_{[d]}\right) \qquad \{a,b\}\cap\{c,d\}=\emptyset \tag{62}$$

$$\left((-1)_{[a]}(-1)_{[b]}\right)\left((-1)_{[c]}(-1)_{[d]}\right) \approx \left((-1)_{[a]}(-1)_{[c]}\right)\left((-1)_{[b]}(-1)_{[d]}\right) \tag{63}$$

$$\left((-1)_{[a]}X_{[a,b]}\right)\left((-1)_{[c]}X_{[c,d]}\right) \approx \left((-1)_{[c]}X_{[c,d]}\right)\left((-1)_{[a]}X_{[a,b]}\right) \qquad \{a,b\}\cap\{c,d\}=\emptyset \tag{64}$$

**Figure 9: Useful auxiliary equations.**

$$\overset{(48)}{\approx} \quad \left((-1)_{[a+1]}X_{[a,a+1]}\right)\left((-1)_{[a+1]}X_{[a+1,a+2]}\right)\left((-1)_{[a]}X_{[a,a+1]}\right)$$

And if $c-a\geq 4$, then we have

$$\left((-1)_{[a]}X_{[a,c]}\right)$$

$$\overset{(27)}{\approx} \quad \left((-1)_{[c-1]}X_{[c-1,c]}\right)\left((-1)_{[a]}X_{[a,c-1]}\right)\left((-1)_{[c]}X_{[c-1,c]}\right)$$

$$\overset{(25)}{\approx} \quad \left((-1)_{[c-1]}X_{[c-1,c]}\right)\left((-1)_{[a+1]}X_{[a,a+1]}\right)\left((-1)_{[a+1]}X_{[a+1,c-1]}\right)\left((-1)_{[a]}X_{[a,a+1]}\right)\left((-1)_{[c]}X_{[c-1,c]}\right)$$

$$\overset{\substack{\text{induction}\\\text{hypothesis}}}{\approx} \quad \left((-1)_{[c-1]}X_{[c-1,c]}\right)\left((-1)_{[a+1]}X_{[a,a+1]}\right)\left((-1)_{[a+2]}X_{[a+1,a+2]}\right)\left((-1)_{[a+2]}X_{[a+2,c-1]}\right)\left((-1)_{[a+1]}X_{[a+1,a+2]}\right)\left((-1)_{[a]}X_{[a,a+1]}\right)\left((-1)_{[c]}X_{[c-1,c]}\right)$$

$$\overset{(48)(32)}{\approx} \quad \left((-1)_{[a+1]}X_{[a,a+1]}\right)\left((-1)_{[a+2]}X_{[a+1,a+2]}\right)\left((-1)_{[c-1]}X_{[c-1,c]}\right)\left((-1)_{[a+2]}X_{[a+2,c-1]}\right)\left((-1)_{[c]}X_{[c-1,c]}\right)\left((-1)_{[a+1]}X_{[a+1,a+2]}\right)\left((-1)_{[a]}X_{[a,a+1]}\right)$$

$$\overset{(27)}{\approx} \quad \left((-1)_{[a+1]}X_{[a,a+1]}\right)\left((-1)_{[a+2]}X_{[a+1,a+2]}\right)\left((-1)_{[a+2]}X_{[a+2,c]}\right)\left((-1)_{[a+1]}X_{[a+1,a+2]}\right)\left((-1)_{[a]}X_{[a,a+1]}\right)$$

$$\overset{(25)}{\approx} \quad \left((-1)_{[a+1]}X_{[a,a+1]}\right)\left((-1)_{[a+1]}X_{[a+1,c]}\right)\left((-1)_{[a]}X_{[a,a+1]}\right)$$

*Proof of Equation* (50): If $c-a$ is odd, then Equation (50) is just Equation (26), therefore we only have to consider the case where $c-a$ is even. We proceed by induction on $c-a$. If $c-a=2$, then we have

$$\overset{(28)}{\approx} \quad \left((-1)_{[a+1]}X_{[a+1,a+2]}\right)\left((-1)_{[a+1]}X_{[a,a+1]}\right)\left((-1)_{[a+2]}X_{[a+1,a+2]}\right)$$

$$\overset{(47)(29)}{\approx} \quad \left((-1)_{[a+1]}X_{[a+1,a+2]}\right)\left((-1)_{[a+1]}X_{[a,a+1]}\right)\left((-1)_{[a+2]}X_{[a+1,a+2]}\right)$$
$$\left((-1)_{[a+1]}X_{[a,a+1]}\right)\left((-1)_{[a+1]}X_{[a+1,a+2]}\right)\left((-1)_{[a]}X_{[a,a+1]}\right)$$
$$\left((-1)_{[a+1]}X_{[a,a+1]}\right)\left((-1)_{[a+2]}X_{[a+1,a+2]}\right)\left((-1)_{[a]}X_{[a,a+1]}\right)$$

$$\overset{(49)}{\approx} \quad \left((-1)_{[a+1]}X_{[a+1,a+2]}\right)\left((-1)_{[a+1]}X_{[a,a+1]}\right)\left((-1)_{[a+2]}X_{[a+1,a+2]}\right)$$
$$\left((-1)_{[a]}X_{[a,a+2]}\right)$$

$$\left((-1)_{[a+1]}X_{[a,a+1]}\right)\left((-1)_{[a+2]}X_{[a+1,a+2]}\right)\left((-1)_{[a]}X_{[a,a+1]}\right)$$

$$\overset{(27)}{\approx}\ \left((-1)_{[a+1]}X_{[a+1,a+2]}\right)\left((-1)_{[a+1]}X_{[a,a+1]}\right)\left((-1)_{[a+2]}X_{[a+1,a+2]}\right)$$
$$\left((-1)_{[a+1]}X_{[a+1,a+2]}\right)\left((-1)_{[a]}X_{[a,a+1]}\right)\left((-1)_{[a+2]}X_{[a+1,a+2]}\right)$$
$$\left((-1)_{[a+1]}X_{[a,a+1]}\right)\left((-1)_{[a+2]}X_{[a+1,a+2]}\right)\left((-1)_{[a]}X_{[a,a+1]}\right)$$

$$\overset{(47)(29)}{\approx}\ \left((-1)_{[a+1]}X_{[a,a+1]}\right)\left((-1)_{[a+2]}X_{[a+1,a+2]}\right)\left((-1)_{[a]}X_{[a,a+1]}\right)$$

And if $c-a\geq 4$, then we have

$$\left((-1)_{[c]}X_{[a,c]}\right)$$

$$\overset{(28)}{\approx}\ \left((-1)_{[c-1]}X_{[c-1,c]}\right)\left((-1)_{[c-1]}X_{[a,c-1]}\right)\left((-1)_{[c]}X_{[c-1,c]}\right)$$

$$\overset{(26)}{\approx}\ \left((-1)_{[c-1]}X_{[c-1,c]}\right)\left((-1)_{[a+1]}X_{[a,a+1]}\right)\left((-1)_{[c-1]}X_{[a+1,c-1]}\right)\left((-1)_{[a]}X_{[a,a+1]}\right)\left((-1)_{[c]}X_{[c-1,c]}\right)$$

$$\overset{\text{induction}}{\underset{\text{hypothesis}}{\approx}}\ (-1)_{[c-1]}X_{[c-1,c]})((-1)_{[a+1]}X_{[a,a+1]})((-1)_{[a+2]}X_{[a+1,a+2]})((-1)_{[c-1]}X_{[a+2,c-1]})((-1)_{[a+1]}X_{[a+1,a+2]})((-1)_{[a]}X_{[a,a+1]})((-1)_{[c]}X_{[c-1,c]})$$

$$\overset{(48)(32)}{\approx}\ ((-1)_{[a+1]}X_{[a,a+1]})((-1)_{[a+2]}X_{[a+1,a+2]})((-1)_{[c-1]}X_{[c-1,c]})((-1)_{[c-1]}X_{[a+2,c-1]})((-1)_{[c]}X_{[c-1,c]})((-1)_{[a+1]}X_{[a+1,a+2]})((-1)_{[a]}X_{[a,a+1]})$$

$$\overset{(28)}{\approx}\ \left((-1)_{[a+1]}X_{[a,a+1]}\right)\left((-1)_{[a+2]}X_{[a+1,a+2]}\right)\left((-1)_{[c]}X_{[a+2,c]}\right)\left((-1)_{[a+1]}X_{[a+1,a+2]}\right)\left((-1)_{[a]}X_{[a,a+1]}\right)$$

$$\overset{(26)}{\approx}\ \left((-1)_{[a+1]}X_{[a,a+1]}\right)\left((-1)_{[c]}X_{[a+1,c]}\right)\left((-1)_{[a]}X_{[a,a+1]}\right)$$

*Proof of Equations* (51) *and* (52): First, we prove both equations in the case $a<b$ by decreasing finite induction on $a$.
If $a=b-1$, then Equation (51) is an instance of Equation (29). Assuming Equation (51) for $a+1$, we prove it for $a$, as follows:

$$\left((-1)_{[a]}X_{[a,b]}\right)\left((-1)_{[b]}X_{[a,b]}\right)$$

$$\overset{(49)(50)}{\approx}\ \left((-1)_{[a+1]}X_{[a,a+1]}\right)\left((-1)_{[a+1]}X_{[a+1,b]}\right)\left((-1)_{[a]}X_{[a,a+1]}\right)\left((-1)_{[a+1]}X_{[a,a+1]}\right)\left((-1)_{[b]}X_{[a+1,b]}\right)\left((-1)_{[a]}X_{[a,a+1]}\right)$$

$$\overset{(29)}{\approx}\ \left((-1)_{[a+1]}X_{[a,a+1]}\right)\left((-1)_{[a+1]}X_{[a+1,b]}\right)\left((-1)_{[b]}X_{[a+1,b]}\right)\left((-1)_{[a]}X_{[a,a+1]}\right)$$

$$\overset{\text{induction}}{\underset{\text{hypothesis}}{\approx}}\ \left((-1)_{[a+1]}X_{[a,a+1]}\right)\left((-1)_{[a]}X_{[a,a+1]}\right)$$

$$\overset{(20)(22)(21)}{\approx}\ \left((-1)_{[a]}(-1)_{[a+1]}\right)^2\left((-1)_{[a+1]}X_{[a,a+1]}\right)\left((-1)_{[a]}X_{[a,a+1]}\right)$$

$$\overset{(23)}{\approx}\ \left((-1)_{[a]}X_{[a,a+1]}\right)^4\left((-1)_{[a+1]}X_{[a,a+1]}\right)\left((-1)_{[a]}X_{[a,a+1]}\right)$$

$$\overset{(29)}{\approx}\ \left((-1)_{[a]}X_{[a,a+1]}\right)^4$$

$$\overset{(23)(21)(22)(20)}{\approx}\ \epsilon$$

If $a=b-1$, then Equation (52) is an instance of Equation (23). Assuming Equation (52) for $a+1$, we prove it for $a$, as follows:

$$\left((-1)_{[a]}X_{[a,b]}\right)\left((-1)_{[a]}X_{[a,b]}\right)$$

$$\overset{(49)}{\approx}\ \left((-1)_{[a+1]}X_{[a,a+1]}\right)\left((-1)_{[a+1]}X_{[a+1,b]}\right)\left((-1)_{[a]}X_{[a,a+1]}\right)\left((-1)_{[a+1]}X_{[a,a+1]}\right)\left((-1)_{[a+1]}X_{[a+1,b]}\right)\left((-1)_{[a]}X_{[a,a+1]}\right)$$

$$\overset{(29)}{\approx}\ \left((-1)_{[a+1]}X_{[a,a+1]}\right)\left((-1)_{[a+1]}X_{[a+1,b]}\right)\left((-1)_{[a+1]}X_{[a+1,b]}\right)\left((-1)_{[a]}X_{[a,a+1]}\right)$$

$$\overset{\text{induction}}{\underset{\text{hypothesis}}{\approx}}\ \left((-1)_{[a+1]}X_{[a,a+1]}\right)\left((-1)_{[a+1]}(-1)_{[b]}\right)\left((-1)_{[a]}X_{[a,a+1]}\right)$$

$$\overset{(31)}{\approx}\ \left((-1)_{[a+1]}X_{[a,a+1]}\right)\left((-1)_{[a]}X_{[a,a+1]}\right)\left((-1)_{[a]}(-1)_{[b]}\right)$$

$$\overset{(20)(22)(21)}{\approx}\ \left((-1)_{[a]}(-1)_{[a+1]}\right)^2\left((-1)_{[a+1]}X_{[a,a+1]}\right)\left((-1)_{[a]}X_{[a,a+1]}\right)\left((-1)_{[a]}(-1)_{[b]}\right)$$

$$\overset{(23)}{\approx}\ \left((-1)_{[a]}X_{[a,a+1]}\right)^4\left((-1)_{[a+1]}X_{[a,a+1]}\right)\left((-1)_{[a]}X_{[a,a+1]}\right)\left((-1)_{[a]}(-1)_{[b]}\right)$$

$$\overset{(29)}{\approx}\ \left((-1)_{[a]}X_{[a,a+1]}\right)^4\left((-1)_{[a]}(-1)_{[b]}\right)$$

$$\overset{(23)(21)(22)(20)}{\approx}\ \left((-1)_{[a]}(-1)_{[b]}\right)$$

It remains to prove Equations (51) and (52) in the case $b < a$. By Equation (33), this is equivalent to proving the following equations for $a < b$:

$$\left((-1)_{[b]}X_{[a,b]}\right)\left((-1)_{[a]}X_{[a,b]}\right) \approx \epsilon$$

$$\left((-1)_{[b]}X_{[a,b]}\right)\left((-1)_{[b]}X_{[a,b]}\right) \approx \left((-1)_{[b]}(-1)_{[a]}\right)$$

The proof of the first one is as follows:

$$\left((-1)_{[b]}X_{[a,b]}\right)\left((-1)_{[a]}X_{[a,b]}\right)$$

$$\overset{(20)(22)(21)}{\approx} \left((-1)_{[a]}(-1)_{[b]}\right)^2\left((-1)_{[b]}X_{[a,b]}\right)\left((-1)_{[a]}X_{[a,b]}\right)$$

$$\overset{\substack{\text{case } a < b \\ \text{of } (52)}}{\approx} \left((-1)_{[a]}X_{[a,b]}\right)^4\left((-1)_{[b]}X_{[a,b]}\right)\left((-1)_{[a]}X_{[a,b]}\right)$$

$$\overset{\substack{\text{case } a < b \\ \text{of } (51)}}{\approx} \left((-1)_{[a]}X_{[a,b]}\right)^3\left((-1)_{[a]}X_{[a,b]}\right)$$

$$\overset{\substack{\text{case } a < b \\ \text{of } (52)}}{\approx} \left((-1)_{[a]}(-1)_{[b]}\right)^2$$

$$\overset{(21)(22)(20)}{\approx} \epsilon$$

and the proof of the second one is as follows:

$$\left((-1)_{[b]}X_{[a,b]}\right)\left((-1)_{[b]}X_{[a,b]}\right)$$

$$\overset{(20)(22)}{\approx} \left((-1)_{[b]}(-1)_{[a]}\right)\left((-1)_{[a]}(-1)_{[b]}\right)\left((-1)_{[b]}X_{[a,b]}\right)\left((-1)_{[b]}X_{[a,b]}\right)$$

$$\overset{\substack{\text{case } a < b \\ \text{of } (52)}}{\approx} \left((-1)_{[b]}(-1)_{[a]}\right)\left((-1)_{[a]}X_{[a,b]}\right)\left((-1)_{[a]}X_{[a,b]}\right)\left((-1)_{[b]}X_{[a,b]}\right)\left((-1)_{[b]}X_{[a,b]}\right)$$

$$\overset{\substack{\text{case } a < b \\ \text{of } (51)}}{\approx} \left((-1)_{[b]}(-1)_{[a]}\right)$$

*Proof of Equation* (53): If $c \neq a, b$, then the equation is just Equation (30). If $c = a$, then the result follows from Equation (20). Therefore, it only remains to treat the case where $c = b$:

$$\left((-1)_{[b]}X_{[a,b]}\right)$$

$$\overset{(51)(33)}{\approx} \left((-1)_{[b]}X_{[a,b]}\right)\left((-1)_{[b]}X_{[a,b]}\right)\left((-1)_{[a]}X_{[a,b]}\right)$$

$$\overset{(33)(52)}{\approx} \left((-1)_{[b]}(-1)_{[a]}\right)\left((-1)_{[a]}X_{[a,b]}\right)$$

*Proof of Equation* (54):

$$\left((-1)_{[a]}X_{[a,c]}\right) \overset{(49)}{\approx} \left((-1)_{[a+1]}X_{[a,a+1]}\right)\left((-1)_{[a+1]}X_{[a+1,c]}\right)\left((-1)_{[a]}X_{[a,a+1]}\right)$$

$$\overset{(53)}{\approx} \left((-1)_{[a+1]}(-1)_{[a]}\right)\left((-1)_{[a]}X_{[a,a+1]}\right)\left((-1)_{[a+1]}X_{[a+1,c]}\right)\left((-1)_{[a]}X_{[a,a+1]}\right)$$

$$\overset{(21)}{\approx} \left((-1)_{[a]}(-1)_{[a+1]}\right)\left((-1)_{[a]}X_{[a,a+1]}\right)\left((-1)_{[a+1]}X_{[a+1,c]}\right)\left((-1)_{[a]}X_{[a,a+1]}\right)$$

$$\overset{(23)}{\approx} \left((-1)_{[a]}X_{[a,a+1]}\right)^3\left((-1)_{[a+1]}X_{[a+1,c]}\right)\left((-1)_{[a]}X_{[a,a+1]}\right)$$

*Proof of Equation* (55):

$$\left((-1)_{[c]}(-1)_{[d]}\right)\left((-1)_{[a]}X_{[a,a+1]}\right) \overset{(22)(21)}{\approx} \left((-1)_{[a+1]}(-1)_{[c]}\right)\left((-1)_{[a+1]}(-1)_{[d]}\right)\left((-1)_{[a]}X_{[a,a+1]}\right)$$

$$\overset{(31)}{\approx} \left((-1)_{[a]}X_{[a,a+1]}\right)\left((-1)_{[a]}(-1)_{[c]}\right)\left((-1)_{[a]}(-1)_{[d]}\right)$$

$$\overset{(21)(22)}{\approx} \left((-1)_{[a]}X_{[a,a+1]}\right)\left((-1)_{[c]}(-1)_{[d]}\right)$$

*Proof of Equation* (56):

$$\left((-1)_{[a]}(-1)_{[b]}\right)\left((-1)_{[a]}X_{[a,b]}\right) \overset{(52)}{\approx} \left((-1)_{[a]}(-1)_{[b]}\right)^3$$

$$\overset{(52)}{\approx} \left((-1)_{[a]}X_{[a,b]}\right)\left((-1)_{[a]}(-1)_{[b]}\right)$$

*Proof of Equation* (57):

$$
\begin{aligned}
\left((-1)_{[b]}(-1)_{[a]}\right)\left((-1)_{[a]}X_{[a,b]}\right) &\overset{(21)}{\approx} \left((-1)_{[a]}(-1)_{[b]}\right)\left((-1)_{[a]}X_{[a,b]}\right) \\
&\overset{(56)}{\approx} \left((-1)_{[a]}X_{[a,b]}\right)\left((-1)_{[a]}(-1)_{[b]}\right) \\
&\overset{(21)}{\approx} \left((-1)_{[a]}X_{[a,b]}\right)\left((-1)_{[b]}(-1)_{[a]}\right)
\end{aligned}
$$

*Proof of Equation* (58): First, we prove Equation (58) in the case where $a < b$ and $c \notin [a,b]$ (that is, either $c < a < b$ or $a < b < c$) by decreasing finite induction on $a$.

If $a = b - 1$, then we have

$$
\begin{aligned}
\left((-1)_{[a]}(-1)_{[c]}\right)\left((-1)_{[a]}X_{[a,a+1]}\right) &\overset{(22)}{\approx} \left((-1)_{[a]}(-1)_{[a+1]}\right)\left((-1)_{[a+1]}(-1)_{[c]}\right)\left((-1)_{[a]}X_{[a,a+1]}\right) \\
&\overset{(31)}{\approx} \left((-1)_{[a]}(-1)_{[a+1]}\right)\left((-1)_{[a]}X_{[a,a+1]}\right)\left((-1)_{[a]}(-1)_{[c]}\right) \\
&\overset{(23)}{\approx} \left((-1)_{[a]}X_{[a,a+1]}\right)^3\left((-1)_{[a]}(-1)_{[c]}\right) \\
&\overset{(23)}{\approx} \left((-1)_{[a]}X_{[a,a+1]}\right)\left((-1)_{[a]}(-1)_{[a+1]}\right)\left((-1)_{[a]}(-1)_{[c]}\right) \\
&\overset{(21)(22)}{\approx} \left((-1)_{[a]}X_{[a,a+1]}\right)\left((-1)_{[a+1]}(-1)_{[c]}\right)
\end{aligned}
$$

Assuming $c \notin [a,b]$ and Equation (58) for $a + 1$, we prove Equation (58) for $a$, as follows:

$$
\begin{aligned}
&\left((-1)_{[a]}(-1)_{[c]}\right)\left((-1)_{[a]}X_{[a,b]}\right) \\
\overset{(49)}{\approx}\ & \left((-1)_{[a]}(-1)_{[c]}\right)\left((-1)_{[a+1]}X_{[a,a+1]}\right)\left((-1)_{[a+1]}X_{[a+1,b]}\right)\left((-1)_{[a]}X_{[a,a+1]}\right) \\
\overset{(51)(33)}{\approx}\ & \left((-1)_{[a+1]}X_{[a,a+1]}\right)\left((-1)_{[a]}X_{[a,a+1]}\right)\left((-1)_{[a]}(-1)_{[c]}\right)\left((-1)_{[a+1]}X_{[a,a+1]}\right)\left((-1)_{[a+1]}X_{[a+1,b]}\right)\left((-1)_{[a]}X_{[a,a+1]}\right) \\
\overset{(31)}{\approx}\ & \left((-1)_{[a+1]}X_{[a,a+1]}\right)\left((-1)_{[a+1]}(-1)_{[c]}\right)\left((-1)_{[a]}X_{[a,a+1]}\right)\left((-1)_{[a+1]}X_{[a,a+1]}\right)\left((-1)_{[a+1]}X_{[a+1,b]}\right)\left((-1)_{[a]}X_{[a,a+1]}\right) \\
\overset{(29)}{\approx}\ & \left((-1)_{[a+1]}X_{[a,a+1]}\right)\left((-1)_{[a+1]}(-1)_{[c]}\right)\left((-1)_{[a+1]}X_{[a+1,b]}\right)\left((-1)_{[a]}X_{[a,a+1]}\right) \\
\overset{\substack{\text{induction}\\\text{hypothesis}}}{\approx}\ & \left((-1)_{[a+1]}X_{[a,a+1]}\right)\left((-1)_{[a+1]}X_{[a+1,b]}\right)\left((-1)_{[b]}(-1)_{[c]}\right)\left((-1)_{[a]}X_{[a,a+1]}\right) \\
\overset{(55)}{\approx}\ & \left((-1)_{[a+1]}X_{[a,a+1]}\right)\left((-1)_{[a+1]}X_{[a+1,b]}\right)\left((-1)_{[a]}X_{[a,a+1]}\right)\left((-1)_{[b]}(-1)_{[c]}\right) \\
\overset{(49)}{\approx}\ & \left((-1)_{[a]}X_{[a,b]}\right)\left((-1)_{[b]}(-1)_{[c]}\right)
\end{aligned}
$$

Then, we prove Equation (58) in the case $a < c < b$, again by decreasing finite induction on $a$.

If $a = c - 1$, then since $b \neq c$, we necessarily have $b > a + 1$. We have

$$
\begin{aligned}
&\left((-1)_{[a]}(-1)_{[a+1]}\right)\left((-1)_{[a]}X_{[a,b]}\right) \\
\overset{(49)}{\approx}\ & \left((-1)_{[a]}(-1)_{[a+1]}\right)\left((-1)_{[a+1]}X_{[a,a+1]}\right)\left((-1)_{[a+1]}X_{[a+1,b]}\right)\left((-1)_{[a]}X_{[a,a+1]}\right) \\
\overset{(21)(52)(33)}{\approx}\ & \left((-1)_{[a+1]}X_{[a,a+1]}\right)^3\left((-1)_{[a+1]}X_{[a+1,b]}\right)\left((-1)_{[a]}X_{[a,a+1]}\right) \\
\overset{(33)(52)}{\approx}\ & \left((-1)_{[a+1]}X_{[a,a+1]}\right)\left((-1)_{[a+1]}(-1)_{[a]}\right)\left((-1)_{[a+1]}X_{[a+1,b]}\right)\left((-1)_{[a]}X_{[a,a+1]}\right) \\
\overset{\substack{\text{previous}\\\text{case}}}{\approx}\ & \left((-1)_{[a+1]}X_{[a,a+1]}\right)\left((-1)_{[a+1]}X_{[a+1,b]}\right)\left((-1)_{[b]}(-1)_{[a]}\right)\left((-1)_{[a]}X_{[a,a+1]}\right) \\
\overset{(21)}{\approx}\ & \left((-1)_{[a+1]}X_{[a,a+1]}\right)\left((-1)_{[a+1]}X_{[a+1,b]}\right)\left((-1)_{[a]}(-1)_{[b]}\right)\left((-1)_{[a]}X_{[a,a+1]}\right) \\
\overset{\substack{\text{previous}\\\text{case}}}{\approx}\ & \left((-1)_{[a+1]}X_{[a,a+1]}\right)\left((-1)_{[a+1]}X_{[a+1,b]}\right)\left((-1)_{[a]}X_{[a,a+1]}\right)\left((-1)_{[a+1]}(-1)_{[b]}\right) \\
\overset{(49)}{\approx}\ & \left((-1)_{[a]}X_{[a,b]}\right)\left((-1)_{[a+1]}(-1)_{[b]}\right) \\
\overset{(21)}{\approx}\ & \left((-1)_{[a]}X_{[a,b]}\right)\left((-1)_{[b]}(-1)_{[a+1]}\right)
\end{aligned}
$$

Assuming $a + 1 < c < b$ and Equation (58) for $a + 1$, we prove Equation (58) for $a$, in the same way as in the previous case:

$$
\begin{aligned}
&\left((-1)_{[a]}(-1)_{[c]}\right)\left((-1)_{[a]}X_{[a,b]}\right) \\
\overset{(49)}{\approx}\ & \left((-1)_{[a]}(-1)_{[c]}\right)\left((-1)_{[a+1]}X_{[a,a+1]}\right)\left((-1)_{[a+1]}X_{[a+1,b]}\right)\left((-1)_{[a]}X_{[a,a+1]}\right)
\end{aligned}
$$

$$
\overset{(51)(33)}{\approx} \quad \left((-1)_{[a+1]}X_{[a,a+1]}\right)\left((-1)_{[a]}X_{[a,a+1]}\right)\left((-1)_{[a]}(-1)_{[c]}\right)\left((-1)_{[a+1]}X_{[a,a+1]}\right)\left((-1)_{[a+1]}X_{[a+1,b]}\right)\left((-1)_{[a]}X_{[a,a+1]}\right)
$$

$$
\overset{(31)}{\approx} \quad \left((-1)_{[a+1]}X_{[a,a+1]}\right)\left((-1)_{[a+1]}(-1)_{[c]}\right)\left((-1)_{[a]}X_{[a,a+1]}\right)\left((-1)_{[a+1]}X_{[a,a+1]}\right)\left((-1)_{[a+1]}X_{[a+1,b]}\right)\left((-1)_{[a]}X_{[a,a+1]}\right)
$$

$$
\overset{(29)}{\approx} \quad \left((-1)_{[a+1]}X_{[a,a+1]}\right)\left((-1)_{[a+1]}(-1)_{[c]}\right)\left((-1)_{[a+1]}X_{[a+1,b]}\right)\left((-1)_{[a]}X_{[a,a+1]}\right)
$$

$$
\overset{\text{induction}}{\underset{\text{hypothesis}}{\approx}} \quad \left((-1)_{[a+1]}X_{[a,a+1]}\right)\left((-1)_{[a+1]}X_{[a+1,b]}\right)\left((-1)_{[b]}(-1)_{[c]}\right)\left((-1)_{[a]}X_{[a,a+1]}\right)
$$

$$
\overset{(55)}{\approx} \quad \left((-1)_{[a+1]}X_{[a,a+1]}\right)\left((-1)_{[a+1]}X_{[a+1,b]}\right)\left((-1)_{[a]}X_{[a,a+1]}\right)\left((-1)_{[b]}(-1)_{[c]}\right)
$$

$$
\overset{(49)}{\approx} \quad \left((-1)_{[a]}X_{[a,b]}\right)\left((-1)_{[b]}(-1)_{[c]}\right)
$$

Finally, it remains to prove Equation (58) in the case $b < a$:

$$
\left((-1)_{[a]}(-1)_{[c]}\right)\left((-1)_{[a]}X_{[a,b]}\right)
$$

$$
\overset{(33)}{\approx} \quad \left((-1)_{[a]}(-1)_{[c]}\right)\left((-1)_{[a]}X_{[b,a]}\right)
$$

$$
\overset{(53)}{\approx} \quad \left((-1)_{[a]}(-1)_{[c]}\right)\left((-1)_{[a]}(-1)_{[b]}\right)\left((-1)_{[b]}X_{[b,a]}\right)
$$

$$
\overset{(21)(22)}{\approx} \quad \left((-1)_{[b]}(-1)_{[c]}\right)\left((-1)_{[b]}X_{[b,a]}\right)
$$

$$
\overset{\text{previous}}{\underset{\text{cases}}{\approx}} \quad \left((-1)_{[b]}X_{[b,a]}\right)\left((-1)_{[a]}(-1)_{[c]}\right)
$$

$$
\overset{(22)}{\approx} \quad \left((-1)_{[b]}X_{[b,a]}\right)\left((-1)_{[a]}(-1)_{[b]}\right)\left((-1)_{[b]}(-1)_{[c]}\right)
$$

$$
\overset{(52)}{\approx} \quad \left((-1)_{[b]}X_{[b,a]}\right)\left((-1)_{[a]}X_{[a,b]}\right)\left((-1)_{[a]}X_{[a,b]}\right)\left((-1)_{[b]}(-1)_{[c]}\right)
$$

$$
\overset{(33)(51)}{\approx} \quad \left((-1)_{[a]}X_{[a,b]}\right)\left((-1)_{[b]}(-1)_{[c]}\right)
$$

*Proof of Equation* (59):

$$
\left((-1)_{[c]}(-1)_{[a]}\right)\left((-1)_{[a]}X_{[a,b]}\right) \quad \overset{(21)}{\approx} \quad \left((-1)_{[a]}(-1)_{[c]}\right)\left((-1)_{[a]}X_{[a,b]}\right)
$$

$$
\overset{(58)}{\approx} \quad \left((-1)_{[a]}X_{[a,b]}\right)\left((-1)_{[b]}(-1)_{[c]}\right)
$$

$$
\overset{(21)}{\approx} \quad \left((-1)_{[a]}X_{[a,b]}\right)\left((-1)_{[c]}(-1)_{[b]}\right)
$$

*Proof of Equation* (60):

$$
\left((-1)_{[b]}(-1)_{[c]}\right)\left((-1)_{[a]}X_{[a,b]}\right) \quad \overset{(22)}{\approx} \quad \left((-1)_{[b]}(-1)_{[a]}\right)\left((-1)_{[a]}(-1)_{[c]}\right)\left((-1)_{[a]}X_{[a,b]}\right)
$$

$$
\overset{(58)(57)}{\approx} \quad \left((-1)_{[a]}X_{[a,b]}\right)\left((-1)_{[b]}(-1)_{[a]}\right)\left((-1)_{[b]}(-1)_{[c]}\right)
$$

$$
\overset{(21)(22)}{\approx} \quad \left((-1)_{[a]}X_{[a,b]}\right)\left((-1)_{[a]}(-1)_{[c]}\right)
$$

*Proof of Equation* (61):

$$
\left((-1)_{[c]}(-1)_{[b]}\right)\left((-1)_{[a]}X_{[a,b]}\right) \quad \overset{(21)}{\approx} \quad \left((-1)_{[b]}(-1)_{[c]}\right)\left((-1)_{[a]}X_{[a,b]}\right)
$$

$$
\overset{(60)}{\approx} \quad \left((-1)_{[a]}X_{[a,b]}\right)\left((-1)_{[a]}(-1)_{[c]}\right)
$$

$$
\overset{(21)}{\approx} \quad \left((-1)_{[a]}X_{[a,b]}\right)\left((-1)_{[c]}(-1)_{[a]}\right)
$$

*Proof of Equation* (62):

$$
\left((-1)_{[c]}(-1)_{[d]}\right)\left((-1)_{[a]}X_{[a,b]}\right) \quad \overset{(22)}{\approx} \quad \left((-1)_{[c]}(-1)_{[a]}\right)\left((-1)_{[a]}(-1)_{[d]}\right)\left((-1)_{[a]}X_{[a,b]}\right)
$$

$$
\overset{(58)(59)}{\approx} \quad \left((-1)_{[a]}X_{[a,b]}\right)\left((-1)_{[c]}(-1)_{[b]}\right)\left((-1)_{[b]}(-1)_{[d]}\right)
$$

$$
\overset{(22)}{\approx} \quad \left((-1)_{[a]}X_{[a,b]}\right)\left((-1)_{[c]}(-1)_{[d]}\right)
$$

*Proof of Equation* (63):

$$
\left((-1)_{[a]}(-1)_{[b]}\right)\left((-1)_{[c]}(-1)_{[d]}\right) \quad \overset{(20)}{\approx} \quad \left((-1)_{[a]}(-1)_{[b]}\right)\left((-1)_{[b]}(-1)_{[b]}\right)\left((-1)_{[c]}(-1)_{[d]}\right)
$$

$$
\overset{(22)}{\approx} \quad \left((-1)_{[a]}(-1)_{[b]}\right)\left((-1)_{[b]}(-1)_{[c]}\right)\left((-1)_{[c]}(-1)_{[b]}\right)\left((-1)_{[c]}(-1)_{[d]}\right)
$$

$$\overset{(21)}{\approx} \quad \left((-1)_{[a]}(-1)_{[b]}\right)\left((-1)_{[b]}(-1)_{[c]}\right)\left((-1)_{[b]}(-1)_{[c]}\right)\left((-1)_{[c]}(-1)_{[d]}\right)$$

$$\overset{(22)}{\approx} \quad \left((-1)_{[a]}(-1)_{[c]}\right)\left((-1)_{[b]}(-1)_{[d]}\right)$$

*Proof of Equation* (64): First, we consider the case where either $a < b < c < d$ or $c < d < a < b$. We proceed by induction on $b - a$. If $b - a = 1$, then we proceed by induction on $d - c$. If $d - c = 1$, then Equation (64) is just an instance of Equation (32). If $d - c > 1$, then we have

$$\left((-1)_{[a]}X_{[a,a+1]}\right)\left((-1)_{[c]}X_{[c,d]}\right)$$

$$\overset{(54)}{\approx} \quad \left((-1)_{[a]}X_{[a,a+1]}\right)\left((-1)_{[c]}X_{[c,c+1]}\right)^3\left((-1)_{[c+1]}X_{[c+1,d]}\right)\left((-1)_{[c]}X_{[c,c+1]}\right)$$

$$\overset{\text{induction}}{\underset{\text{hypothesis}}{\approx}} \quad \left((-1)_{[c]}X_{[c,c+1]}\right)^3\left((-1)_{[c+1]}X_{[c+1,d]}\right)\left((-1)_{[c]}X_{[c,c+1]}\right)\left((-1)_{[a]}X_{[a,a+1]}\right)$$

$$\overset{(54)}{\approx} \quad \left((-1)_{[c]}X_{[c,d]}\right)\left((-1)_{[a]}X_{[a,a+1]}\right)$$

Now, if $b - a > 1$, we have

$$\left((-1)_{[a]}X_{[a,b]}\right)\left((-1)_{[c]}X_{[c,d]}\right)$$

$$\overset{(54)}{\approx} \quad \left((-1)_{[a]}X_{[a,a+1]}\right)^3\left((-1)_{[a+1]}X_{[a+1,c]}\right)\left((-1)_{[a]}X_{[a,a+1]}\right)\left((-1)_{[c]}X_{[c,d]}\right)$$

$$\overset{\text{induction}}{\underset{\text{hypothesis}}{\approx}} \quad \left((-1)_{[c]}X_{[c,d]}\right)\left((-1)_{[a]}X_{[a,a+1]}\right)^3\left((-1)_{[a+1]}X_{[a+1,c]}\right)\left((-1)_{[a]}X_{[a,a+1]}\right)$$

$$\overset{(54)}{\approx} \quad \left((-1)_{[c]}X_{[c,d]}\right)\left((-1)_{[a]}X_{[a,b]}\right)$$

Then we consider the case where $c < a < b < d$. Again, we proceed by induction on $b - a$. If $b - a = 1$, then we proceed by induction on $a - c$. If $a - c = 1$, then we have

$$\left((-1)_{[c+1]}X_{[c+1,c+2]}\right)\left((-1)_{[c]}X_{[c,d]}\right)$$

$$\overset{(49)}{\approx} \quad \left((-1)_{[c+1]}X_{[c+1,c+2]}\right)\left((-1)_{[c+1]}X_{[c,c+1]}\right)\left((-1)_{[c+1]}X_{[c+1,d]}\right)\left((-1)_{[c]}X_{[c,c+1]}\right)$$

$$\overset{(49)}{\approx} \quad \left((-1)_{[c+1]}X_{[c+1,c+2]}\right)\left((-1)_{[c+1]}X_{[c,c+1]}\right)\left((-1)_{[c+2]}X_{[c+1,c+2]}\right)\left((-1)_{[c+2]}X_{[c+2,d]}\right)\left((-1)_{[c+1]}X_{[c+1,c+2]}\right)\left((-1)_{[c]}X_{[c,c+1]}\right)$$

$$\overset{(53)}{\approx} \quad \left((-1)_{[c+1]}X_{[c+1,c+2]}\right)\left((-1)_{[c+1]}(-1)_{[c]}\right)\left((-1)_{[c]}X_{[c,c+1]}\right)\left((-1)_{[c+2]}(-1)_{[c+1]}\right)\left((-1)_{[c+1]}X_{[c+1,c+2]}\right)\left((-1)_{[c+2]}X_{[c+2,d]}\right)\left((-1)_{[c+1]}X_{[c+1,c+2]}\right)\left((-1)_{[c]}X_{[c,c+1]}\right)$$

$$\overset{(59)}{\approx} \quad \left((-1)_{[c+1]}X_{[c+1,c+2]}\right)\left((-1)_{[c+1]}(-1)_{[c]}\right)\left((-1)_{[c+2]}(-1)_{[c]}\right)\left((-1)_{[c]}X_{[c,c+1]}\right)\left((-1)_{[c+1]}X_{[c+1,c+2]}\right)\left((-1)_{[c+2]}X_{[c+2,d]}\right)\left((-1)_{[c+1]}X_{[c+1,c+2]}\right)\left((-1)_{[c]}X_{[c,c+1]}\right)$$

$$\overset{(60)(58)}{\approx} \quad \left((-1)_{[c+2]}(-1)_{[c]}\right)\left((-1)_{[c+1]}(-1)_{[c]}\right)\left((-1)_{[c+1]}X_{[c+1,c+2]}\right)\left((-1)_{[c]}X_{[c,c+1]}\right)\left((-1)_{[c+1]}X_{[c+1,c+2]}\right)\left((-1)_{[c+2]}X_{[c+2,d]}\right)\left((-1)_{[c+1]}X_{[c+1,c+2]}\right)\left((-1)_{[c]}X_{[c,c+1]}\right)$$

$$\overset{(24)}{\approx} \quad \left((-1)_{[c+2]}(-1)_{[c]}\right)\left((-1)_{[c+1]}(-1)_{[c]}\right)\left((-1)_{[c]}X_{[c,c+1]}\right)\left((-1)_{[c+1]}X_{[c+1,c+2]}\right)\left((-1)_{[c]}X_{[c,c+1]}\right)\left((-1)_{[c+2]}X_{[c+2,d]}\right)\left((-1)_{[c+1]}X_{[c+1,c+2]}\right)\left((-1)_{[c]}X_{[c,c+1]}\right)$$

$$\overset{(53)}{\approx} \quad \left((-1)_{[c+2]}(-1)_{[c]}\right)\left((-1)_{[c+1]}X_{[c,c+1]}\right)\left((-1)_{[c+1]}X_{[c+1,c+2]}\right)\left((-1)_{[c]}X_{[c,c+1]}\right)\left((-1)_{[c+2]}X_{[c+2,d]}\right)\left((-1)_{[c+1]}X_{[c+1,c+2]}\right)\left((-1)_{[c]}X_{[c,c+1]}\right)$$

$$\overset{(33)(61)}{\approx} \quad \left((-1)_{[c+1]}X_{[c,c+1]}\right)\left((-1)_{[c+2]}(-1)_{[c+1]}\right)\left((-1)_{[c+1]}X_{[c+1,c+2]}\right)\left((-1)_{[c]}X_{[c,c+1]}\right)\left((-1)_{[c+2]}X_{[c+2,d]}\right)\left((-1)_{[c+1]}X_{[c+1,c+2]}\right)\left((-1)_{[c]}X_{[c,c+1]}\right)$$

$$\overset{(53)}{\approx} \quad \left((-1)_{[c+1]}X_{[c,c+1]}\right)\left((-1)_{[c+2]}X_{[c+1,c+2]}\right)\left((-1)_{[c]}X_{[c,c+1]}\right)\left((-1)_{[c+2]}X_{[c+2,d]}\right)\left((-1)_{[c+1]}X_{[c+1,c+2]}\right)\left((-1)_{[c]}X_{[c,c+1]}\right)$$

$$\overset{\text{previous case}}{\approx} \quad \left((-1)_{[c+1]}X_{[c,c+1]}\right)\left((-1)_{[c+1]}X_{[c+1,c+2]}\right)\left((-1)_{[c+2]}X_{[c+2,d]}\right)\left((-1)_{[c+1]}X_{[c+1,c+2]}\right)\left((-1)_{[c]}X_{[c,c+1]}\right)$$

$$\overset{(24)}{\approx} \quad \left((-1)_{[c+1]}X_{[c,c+1]}\right)\left((-1)_{[c+2]}X_{[c+1,c+2]}\right)\left((-1)_{[c+2]}X_{[c+2,d]}\right)\left((-1)_{[c+1]}X_{[c+1,c+2]}\right)\left((-1)_{[c]}X_{[c,c+1]}\right)\left((-1)_{[c+1]}X_{[c+1,c+2]}\right)$$

$$\overset{(49)}{\approx} \quad \left((-1)_{[c]}X_{[c,d]}\right)\left((-1)_{[c+1]}X_{[c+1,c+2]}\right)$$

And if $a - c > 1$, we have

$$\left((-1)_{[a]}X_{[a,a+1]}\right)\left((-1)_{[c]}X_{[c,d]}\right)$$

$$\overset{(54)}{\approx} \quad \left((-1)_{[a]}X_{[a,a+1]}\right)\left((-1)_{[c]}X_{[c,c+1]}\right)^3\left((-1)_{[c+1]}X_{[c+1,d]}\right)\left((-1)_{[c]}X_{[c,c+1]}\right)$$

$$\overset{(32)}{\approx} \quad \left((-1)_{[c]}X_{[c,c+1]}\right)^3\left((-1)_{[a]}X_{[a,a+1]}\right)\left((-1)_{[c+1]}X_{[c+1,d]}\right)\left((-1)_{[c]}X_{[c,c+1]}\right)$$

$$\overset{\text{induction}}{\underset{\text{hypothesis}}{\approx}} \quad \left((-1)_{[c]}X_{[c,c+1]}\right)^3\left((-1)_{[c+1]}X_{[c+1,d]}\right)\left((-1)_{[a]}X_{[a,a+1]}\right)\left((-1)_{[c]}X_{[c,c+1]}\right)$$

$$\overset{(32)}{\approx} \quad \left((-1)_{[c]}X_{[c,c+1]}\right)^3\left((-1)_{[c+1]}X_{[c+1,d]}\right)\left((-1)_{[c]}X_{[c,c+1]}\right)\left((-1)_{[a]}X_{[a,a+1]}\right)$$

$$\overset{(54)}{\approx} \quad \left((-1)_{[c]}X_{[c,d]}\right)\left((-1)_{[a]}X_{[a,a+1]}\right)$$

Now, if $b - a > 1$, we have

$$\left((-1)_{[a]}X_{[a,b]}\right)\left((-1)_{[c]}X_{[c,d]}\right)$$

$$\overset{(54)}{\approx} \quad (-1)_{[a]}X_{[a,a+1]}\big)^3 \, \big((-1)_{[a+1]}X_{[a+1,c]}\big) \, \big((-1)_{[a]}X_{[a,a+1]}\big) \, \big((-1)_{[c]}X_{[c,d]}\big)$$

$$\underset{\text{hypothesis}}{\overset{\text{induction}}{\approx}} \quad (-1)_{[c]}X_{[c,d]}\big) \, \big((-1)_{[a]}X_{[a,a+1]}\big)^3 \, \big((-1)_{[a+1]}X_{[a+1,c]}\big) \, \big((-1)_{[a]}X_{[a,a+1]}\big)$$

$$\overset{(54)}{\approx} \quad (-1)_{[c]}X_{[c,d]}\big) \, \big((-1)_{[a]}X_{[a,b]}\big)$$

Then we consider the case where $a < c < b < d$. We proceed by induction on $b - c$. If $b - c = 1$, then we proceed by induction on $c - a$. If $c - a = 1$, then we have

$$(-1)_{[a]}X_{[a,a+2]}\big) \, \big((-1)_{[a+1]}X_{[a+1,d]}\big)$$

$$\overset{(49)}{\approx} \quad (-1)_{[a+1]}X_{[a+1,a+1]}\big) \, \big((-1)_{[a+1]}X_{[a+1,a+2]}\big) \, \big((-1)_{[a]}X_{[a,a+1]}\big) \, \big((-1)_{[a+2]}X_{[a+1,a+2]}\big) \, \big((-1)_{[a+2]}X_{[a+2,d]}\big) \, \big((-1)_{[a+1]}X_{[a+1,a+2]}\big)$$

$$\overset{(53)}{\approx} \quad (-1)_{[a+1]}(-1)_{[a]}\big) \, \big((-1)_{[a]}X_{[a,a+1]}\big) \, \big((-1)_{[a+1]}X_{[a+1,a+2]}\big) \, \big((-1)_{[a]}X_{[a,a+1]}\big) \, \big((-1)_{[a+2]}X_{[a+1,a+2]}\big) \, \big((-1)_{[a+2]}X_{[a+2,d]}\big) \, \big((-1)_{[a+1]}X_{[a+1,a+2]}\big)$$

$$\overset{(24)}{\approx} \quad (-1)_{[a+1]}(-1)_{[a]}\big) \, \big((-1)_{[a+1]}X_{[a+1,a+2]}\big) \, \big((-1)_{[a]}X_{[a,a+1]}\big) \, \big((-1)_{[a+1]}X_{[a+1,a+2]}\big) \, \big((-1)_{[a+2]}X_{[a+1,a+2]}\big) \, \big((-1)_{[a+2]}X_{[a+2,d]}\big) \, \big((-1)_{[a+1]}X_{[a+1,a+2]}\big)$$

$$\overset{(29)}{\approx} \quad (-1)_{[a+1]}(-1)_{[a]}\big) \, \big((-1)_{[a+1]}X_{[a+1,a+2]}\big) \, \big((-1)_{[a]}X_{[a,a+1]}\big) \, \big((-1)_{[a+2]}X_{[a+2,d]}\big) \, \big((-1)_{[a+1]}X_{[a+1,a+2]}\big)$$

$$\underset{\text{first case}}{\approx} \quad (-1)_{[a+1]}(-1)_{[a]}\big) \, \big((-1)_{[a+1]}X_{[a+1,a+2]}\big) \, \big((-1)_{[a+2]}X_{[a+2,d]}\big) \, \big((-1)_{[a]}X_{[a,a+1]}\big) \, \big((-1)_{[a+1]}X_{[a+1,a+2]}\big)$$

$$\overset{(29)}{\approx} \quad (-1)_{[a+1]}(-1)_{[a]}\big) \, \big((-1)_{[a+1]}X_{[a+1,a+2]}\big) \, \big((-1)_{[a+2]}X_{[a+2,d]}\big) \, \big((-1)_{[a+1]}X_{[a+1,a+2]}\big) \, \big((-1)_{[a+2]}X_{[a+1,a+2]}\big) \, \big((-1)_{[a]}X_{[a,a+1]}\big) \, \big((-1)_{[a+1]}X_{[a+1,a+2]}\big)$$

$$\overset{(53)}{\approx} \quad (-1)_{[a+1]}(-1)_{[a]}\big) \, \big((-1)_{[a+1]}X_{[a+1,a+2]}\big) \, \big((-1)_{[a+2]}X_{[a+2,d]}\big) \, \big((-1)_{[a+1]}X_{[a+1,a+2]}\big) \, \big((-1)_{[a+2]}(-1)_{[a+1]}\big) \, \big((-1)_{[a+1]}X_{[a+1,a+2]}\big) \, \big((-1)_{[a]}X_{[a,a+1]}\big) \, \big((-1)_{[a+1]}X_{[a+1,a+2]}\big)$$

$$\overset{(24)}{\approx} \quad (-1)_{[a+1]}(-1)_{[a]}\big) \, \big((-1)_{[a+1]}X_{[a+1,a+2]}\big) \, \big((-1)_{[a+2]}X_{[a+2,d]}\big) \, \big((-1)_{[a+1]}X_{[a+1,a+2]}\big) \, \big((-1)_{[a+2]}(-1)_{[a+1]}\big) \, \big((-1)_{[a]}X_{[a,a+1]}\big) \, \big((-1)_{[a+1]}X_{[a+1,a+2]}\big) \, \big((-1)_{[a]}X_{[a,a+1]}\big)$$

$$\overset{(22)(21)(53)}{\approx} \quad (-1)_{[a+1]}(-1)_{[a]}\big) \, \big((-1)_{[a+1]}X_{[a+1,a+2]}\big) \, \big((-1)_{[a+2]}X_{[a+2,d]}\big) \, \big((-1)_{[a+1]}X_{[a+1,a+2]}\big) \, \big((-1)_{[a+2]}(-1)_{[a+1]}\big) \, \big((-1)_{[a]}X_{[a,a+1]}\big) \, \big((-1)_{[a+1]}X_{[a+1,a+2]}\big) \, \big((-1)_{[a]}X_{[a,a+1]}\big)$$

$$\overset{(59)(62)(60)}{\approx} \quad (-1)_{[a+1]}(-1)_{[a]}\big) \, \big((-1)_{[a+2]}(-1)_{[a]}\big) \, \big((-1)_{[a+1]}X_{[a+1,a+2]}\big) \, \big((-1)_{[a+2]}X_{[a+2,d]}\big) \, \big((-1)_{[a+1]}X_{[a+1,a+2]}\big) \, \big((-1)_{[a]}X_{[a,a+1]}\big) \, \big((-1)_{[a+1]}X_{[a+1,a+2]}\big) \, \big((-1)_{[a]}X_{[a,a+1]}\big)$$

$$\overset{(21)(22)}{\approx} \quad (-1)_{[a+2]}(-1)_{[a+1]}\big) \, \big((-1)_{[a+1]}X_{[a+1,a+2]}\big) \, \big((-1)_{[a+2]}X_{[a+2,d]}\big) \, \big((-1)_{[a+1]}X_{[a+1,a+2]}\big) \, \big((-1)_{[a]}X_{[a,a+1]}\big) \, \big((-1)_{[a+1]}X_{[a+1,a+2]}\big) \, \big((-1)_{[a]}X_{[a,a+1]}\big)$$

$$\overset{(53)}{\approx} \quad (-1)_{[a+2]}X_{[a+1,a+2]}\big) \, \big((-1)_{[a+2]}X_{[a+2,d]}\big) \, \big((-1)_{[a+1]}X_{[a+1,a+2]}\big) \, \big((-1)_{[a+1]}X_{[a,a+1]}\big) \, \big((-1)_{[a+1]}X_{[a+1,a+2]}\big) \, \big((-1)_{[a]}X_{[a,a+1]}\big)$$

$$\overset{(49)}{\approx} \quad (-1)_{[a+1]}X_{[a+1,d]}\big) \, \big((-1)_{[a]}X_{[a,a+2]}\big)$$

And if $c - a > 1$, we have

$$(-1)_{[a]}X_{[a,c+1]}\big) \, \big((-1)_{[c]}X_{[c,d]}\big)$$

$$\overset{(54)}{\approx} \quad (-1)_{[a]}X_{[a,a+1]}\big)^3 \, \big((-1)_{[a+1]}X_{[a+1,c+1]}\big) \, \big((-1)_{[a]}X_{[a,a+1]}\big) \, \big((-1)_{[c]}X_{[c,d]}\big)$$

$$\underset{\text{first case}}{\approx} \quad (-1)_{[a]}X_{[a,a+1]}\big)^3 \, \big((-1)_{[a+1]}X_{[a+1,c+1]}\big) \, \big((-1)_{[c]}X_{[c,d]}\big) \, \big((-1)_{[a]}X_{[a,a+1]}\big)$$

$$\underset{\text{hypothesis}}{\overset{\text{induction}}{\approx}} \quad (-1)_{[a]}X_{[a,a+1]}\big)^3 \, \big((-1)_{[c]}X_{[c,d]}\big) \, \big((-1)_{[a+1]}X_{[a+1,c+1]}\big) \, \big((-1)_{[a]}X_{[a,a+1]}\big)$$

$$\underset{\text{first case}}{\approx} \quad (-1)_{[c]}X_{[c,d]}\big) \, \big((-1)_{[a]}X_{[a,a+1]}\big)^3 \, \big((-1)_{[a+1]}X_{[a+1,c+1]}\big) \, \big((-1)_{[a]}X_{[a,a+1]}\big)$$

$$\overset{(54)}{\approx} \quad (-1)_{[c]}X_{[c,d]}\big) \, \big((-1)_{[a]}X_{[a,c+1]}\big)$$

Now, if $b - c > 1$, we have

$$(-1)_{[a]}X_{[a,b]}\big) \, \big((-1)_{[c]}X_{[c,d]}\big)$$

$$\overset{(54)}{\approx} \quad (-1)_{[a]}X_{[a,b]}\big) \, \big((-1)_{[c]}X_{[c,c+1]}\big)^3 \, \big((-1)_{[c+1]}X_{[c+1,d]}\big) \, \big((-1)_{[c]}X_{[c,c+1]}\big)$$

$$\underset{\text{previous case}}{\approx} \quad (-1)_{[c]}X_{[c,c+1]}\big)^3 \, \big((-1)_{[a]}X_{[a,b]}\big) \, \big((-1)_{[c+1]}X_{[c+1,d]}\big) \, \big((-1)_{[c]}X_{[c,c+1]}\big)$$

$$\underset{\text{hypothesis}}{\overset{\text{induction}}{\approx}} \quad (-1)_{[c]}X_{[c,c+1]}\big)^3 \, \big((-1)_{[c+1]}X_{[c+1,d]}\big) \, \big((-1)_{[a]}X_{[a,b]}\big) \, \big((-1)_{[c]}X_{[c,c+1]}\big)$$

$$\underset{\text{previous case}}{\approx} \quad (-1)_{[c]}X_{[c,c+1]}\big)^3 \, \big((-1)_{[c+1]}X_{[c+1,d]}\big) \, \big((-1)_{[c]}X_{[c,c+1]}\big) \, \big((-1)_{[a]}X_{[a,b]}\big)$$

$$\overset{(54)}{\approx} \quad (-1)_{[c]}X_{[c,d]}\big) \, \big((-1)_{[a]}X_{[a,b]}\big)$$

By symmetry, the case $a < c < d < b$ is similar to the case $c < a < b < d$, and the case $c < a < d < b$ is similar to the case $a < c < b < d$. Thus, we have treated all the cases where $a < b$ and $c < d$.

Finally, if $b < a$ and $c < d$, we have

$$(-1)_{[a]}X_{[a,b]}\big) \, \big((-1)_{[c]}X_{[c,d]}\big)$$

$$\overset{(51)}{\approx} \quad (-1)_{[a]}X_{[a,b]}\big) \, \big((-1)_{[c]}X_{[c,d]}\big) \, \big((-1)_{[b]}X_{[b,a]}\big) \, \big((-1)_{[a]}X_{[b,a]}\big)$$

$$\overset{\text{previous cases}}{\approx} \quad \left((-1)_{[a]}X_{[a,b]}\right)\left((-1)_{[b]}X_{[b,a]}\right)\left((-1)_{[c]}X_{[c,d]}\right)\left((-1)_{[a]}X_{[b,a]}\right)$$

$$\overset{(33)(51)}{\approx} \quad \left((-1)_{[c]}X_{[c,d]}\right)\left((-1)_{[a]}X_{[a,b]}\right)$$

The case where $a < b$ and $d < c$ follows by symmetry. Finally, the case where $b < a$ and $d < c$ follows by exactly the same derivation:

$$\left((-1)_{[a]}X_{[a,b]}\right)\left((-1)_{[c]}X_{[c,d]}\right)$$

$$\overset{(51)}{\approx} \quad \left((-1)_{[a]}X_{[a,b]}\right)\left((-1)_{[c]}X_{[c,d]}\right)\left((-1)_{[b]}X_{[b,a]}\right)\left((-1)_{[a]}X_{[b,a]}\right)$$

$$\overset{\text{previous cases}}{\approx} \quad \left((-1)_{[a]}X_{[a,b]}\right)\left((-1)_{[b]}X_{[b,a]}\right)\left((-1)_{[c]}X_{[c,d]}\right)\left((-1)_{[a]}X_{[b,a]}\right)$$

$$\overset{(33)(51)}{\approx} \quad \left((-1)_{[c]}X_{[c,d]}\right)\left((-1)_{[a]}X_{[a,b]}\right)$$

$\square$

**Lemma A.4.** *Any word over the generators of $\mathcal{P}_N$ containing no generators of the form $\left(H_{[a,b]}H_{[c,d]}\right)$ is $\approx$-equivalent to a unique word, called its* normal form, *of the form*

$$\left((-1)_{[a_1]}(-1)_{[b_1]}\right)\cdots\left((-1)_{[a_k]}(-1)_{[b_k]}\right)\prod_{i=1}^{N-1}\overrightarrow{\left(\prod_{j=d_i}^{N-i-1}\left((-1)_{[j]}X_{[j,j+1]}\right)\right)} \tag{B}$$

*with $k \geq 0$; $a_1 < b_1 < \cdots < a_k < b_k$; and for every $i \in \{1,...,N\}$, $d_i \in \{0,...,N-i\}$ (with the empty products being the identity). Moreover, two words with the same semantics have the same normal form.*

**Remark 6.** *In the following, when we talk about words of the form (B) or being put in the form (B), we implicitly also require that they satisfy the conditions on the parameters $k, a_i, b_i, d_i$ given in Lemma A.4.*

PROOF OF LEMMA A.4.

*Uniqueness given the semantics:* Given a word of the form (B), its semantics is the product of a permutation matrix with a diagonal matrix whose diagonal elements are all $\pm 1$. It is easy to see that the $d_i$'s are uniquely determined by the permutation, as the product $\prod_{i=1}^{N-1}\overrightarrow{\left(\prod_{j=d_i}^{N-i-1}X_{[j,j+1]}\right)}$ corresponds to a well-known canonical decomposition of the permutation as a product of transpositions. Then, the $d_i$'s being known, $k$ together with the $a_i$'s and the $b_i$'s are uniquely determined by the elements of the diagonal matrix.

*Existence:* First, all generators of $\mathcal{P}_N$, except those of the form $\left(H_{[a,b]}H_{[c,d]}\right)$, can be written by means of the equations of Figure 8 as a product of generators of the form $\left((-1)_{[j]}X_{[j,j+1]}\right)$. Indeed: first, by Equations (34) and (53), any of these generators can be written as a product of generators of the form $\left((-1)_{[a]}(-1)_{[b]}\right)$ and $\left((-1)_{[a]}X_{[a,b]}\right)$; then by Equations (33), (53), (20) and (21), one can ensure that $a < b$ in all generators of the form $\left((-1)_{[a]}(-1)_{[b]}\right)$ and $\left((-1)_{[a]}X_{[a,b]}\right)$; then by Equation (52), one can write $\left((-1)_{[a]}\right)$ as a product of two $\left((-1)_{[a]}X_{[a,b]}\right)$; finally, when $a < b$, Equation (54) allows us to write $\left((-1)_{[a]}X_{[a,b]}\right)$, as a product of generators of the form $\left((-1)_{[j]}X_{[j,j+1]}\right)$ by direct induction on $b - a$.

Therefore, it suffices to show that any product of $\left((-1)_{[j]}X_{[j,j+1]}\right)$ can be put in the form (B). We proceed by induction: The identity is of the form (B), namely with $k = 0$ and $d_i = N-i$ for all $i$. Then it remains to show that for any word $P$ of the form (B) and any $a \in \{0,...,N-2\}$, the word $P\left((-1)_{[a]}X_{[a,a+1]}\right)$ can be put in the form (B):

$$\left((-1)_{[a_1]}(-1)_{[b_1]}\right)\cdots\left((-1)_{[a_k]}(-1)_{[b_k]}\right)\prod_{i=1}^{N-1}\overrightarrow{\left(\prod_{j=d_i}^{N-i-1}\left((-1)_{[j]}X_{[j,j+1]}\right)\right)}\circ\left((-1)_{[a]}X_{[a,a+1]}\right)$$

$$\overset{(54)}{\approx}\left((-1)_{[a_1]}(-1)_{[b_1]}\right)\cdots\left((-1)_{[a_k]}(-1)_{[b_k]}\right)\prod_{i=1}^{N-a}\overrightarrow{\left(\prod_{j=d_i}^{N-i-1}\left((-1)_{[j]}X_{[j,j+1]}\right)\right)}\circ\left((-1)_{[a]}X_{[a,a+1]}\right)\circ\prod_{i=N-a+1}^{N-1}\overrightarrow{\left(\prod_{j=d_i}^{N-i-1}\left((-1)_{[j]}X_{[j,j+1]}\right)\right)}$$

$$=\left((-1)_{[a_1]}(-1)_{[b_1]}\right)\cdots\left((-1)_{[a_k]}(-1)_{[b_k]}\right)\prod_{i=1}^{N-a-2}\overrightarrow{\left(\prod_{j=d_i}^{N-i-1}\left((-1)_{[j]}X_{[j,j+1]}\right)\right)}\circ\prod_{j=d_{N-a-1}}^{a}\left((-1)_{[j]}X_{[j,j+1]}\right)\circ\prod_{j=d_{N-a}}^{a-1}\left((-1)_{[j]}X_{[j,j+1]}\right)\circ\left((-1)_{[a]}X_{[a,a+1]}\right)\circ\prod_{i=N-a+1}^{N-1}\overrightarrow{\left(\prod_{j=d_i}^{N-i-1}\left((-1)_{[j]}X_{[j,j+1]}\right)\right)}$$

- If $d_{N-a-1} = d_{N-a} + 1$, then by Equation (64), this can be rewritten into

$$((-1)_{[a_1]}(-1)_{[b_1]} \cdots (-1)_{[a_k]}(-1)_{[b_k]}) \prod_{i=1}^{N-a-2} \left( \prod_{j=d_i}^{N-i-1} ((-1)_{[j]} X_{[j,j+1]}) \right) \circ \prod_{j=d_{N-a-1}}^{a} ((-1)_{[j]} X_{[j,j+1]}) ((-1)_{[j-1]} X_{[j-1,j]}) \circ (-1)_{[a]} X_{[a,a+1]} \circ \prod_{i=N-a+1}^{N-1} \left( \prod_{j=d_i}^{N-i-1} ((-1)_{[j]} X_{[j,j+1]}) \right)$$

$$\overset{(24)}{\approx} ((-1)_{[a_1]}(-1)_{[b_1]} \cdots (-1)_{[a_k]}(-1)_{[b_k]}) \prod_{i=1}^{N-a-2} \left( \prod_{j=d_i}^{N-i-1} ((-1)_{[j]} X_{[j,j+1]}) \right) \circ (-1)_{[d_{N-a-1}-1]} X_{[d_{N-a-1}-1,d_{N-a-1}]} \circ \prod_{j=d_{N-a-1}}^{a} ((-1)_{[j]} X_{[j,j+1]}) ((-1)_{[j-1]} X_{[j-1,j]}) \circ \prod_{i=N-a+1}^{N-1} \left( \prod_{j=d_i}^{N-i-1} ((-1)_{[j]} X_{[j,j+1]}) \right)$$

$$\overset{(64)}{\approx} ((-1)_{[a_1]}(-1)_{[b_1]} \cdots (-1)_{[a_k]}(-1)_{[b_k]}) \prod_{i=1}^{N-a-2} \left( \prod_{j=d_i}^{N-i-1} ((-1)_{[j]} X_{[j,j+1]}) \right) \circ (-1)_{[d_{N-a-1}-1]} X_{[d_{N-a-1}-1,d_{N-a-1}]} \circ \prod_{j=d_{N-a-1}}^{a} ((-1)_{[j]} X_{[j,j+1]}) \circ \prod_{j=d_{N-a-1}-1}^{a-1} ((-1)_{[j]} X_{[j,j+1]}) \circ \prod_{i=N-a+1}^{N-1} \left( \prod_{j=d_i}^{N-i-1} ((-1)_{[j]} X_{[j,j+1]}) \right)$$

$$= ((-1)_{[a_1]}(-1)_{[b_1]} \cdots (-1)_{[a_k]}(-1)_{[b_k]}) \prod_{i=1}^{N-a-2} \left( \prod_{j=d_i}^{N-i-1} ((-1)_{[j]} X_{[j,j+1]}) \right) \circ \prod_{j=d_{N-a-1}-1}^{a} ((-1)_{[j]} X_{[j,j+1]}) \circ \prod_{j=d_{N-a-1}-1}^{a-1} ((-1)_{[j]} X_{[j,j+1]}) \circ \prod_{i=N-a+1}^{N-1} \left( \prod_{j=d_i}^{N-i-1} ((-1)_{[j]} X_{[j,j+1]}) \right)$$

which is in the form (B).

- If $d_{N-a-1} > d_{N-a} + 1$, then this is equal to

$$((-1)_{[a_1]}(-1)_{[b_1]} \cdots (-1)_{[a_k]}(-1)_{[b_k]}) \prod_{i=1}^{N-a-2} \left( \prod_{j=d_i}^{N-i-1} ((-1)_{[j]} X_{[j,j+1]}) \right) \circ \prod_{j=d_{N-a}}^{a} ((-1)_{[j]} X_{[j,j+1]}) \circ \prod_{j=d_{N-a-1}}^{d_{N-a-1}-2} ((-1)_{[j]} X_{[j,j+1]}) \circ \prod_{j=d_{N-a-1}-1}^{a-1} ((-1)_{[j]} X_{[j,j+1]}) \circ ((-1)_{[a]} X_{[a,a+1]}) \circ \prod_{i=N-a+1}^{N-1} \left( \prod_{j=d_i}^{N-i-1} ((-1)_{[j]} X_{[j,j+1]}) \right)$$

$$\overset{(64)}{\approx} ((-1)_{[a_1]}(-1)_{[b_1]} \cdots (-1)_{[a_k]}(-1)_{[b_k]}) \prod_{i=1}^{N-a-2} \left( \prod_{j=d_i}^{N-i-1} ((-1)_{[j]} X_{[j,j+1]}) \right) \circ \prod_{j=d_{N-a}}^{d_{N-a-1}-2} ((-1)_{[j]} X_{[j,j+1]}) \circ \prod_{j=d_{N-a-1}}^{a} ((-1)_{[j]} X_{[j,j+1]}) \circ \prod_{j=d_{N-a-1}-1}^{a-1} ((-1)_{[j]} X_{[j,j+1]}) \circ ((-1)_{[a]} X_{[a,a+1]}) \circ \prod_{i=N-a+1}^{N-1} \left( \prod_{j=d_i}^{N-i-1} ((-1)_{[j]} X_{[j,j+1]}) \right)$$

$$\overset{(64)}{\approx} ((-1)_{[a_1]}(-1)_{[b_1]} \cdots (-1)_{[a_k]}(-1)_{[b_k]}) \prod_{i=1}^{N-a-2} \left( \prod_{j=d_i}^{N-i-1} ((-1)_{[j]} X_{[j,j+1]}) \right) \circ \prod_{j=d_{N-a}}^{d_{N-a-1}-2} ((-1)_{[j]} X_{[j,j+1]}) \circ \prod_{j=d_{N-a-1}}^{a} ((-1)_{[j]} X_{[j,j+1]}) ((-1)_{[j-1]} X_{[j-1,j]}) \circ ((-1)_{[a]} X_{[a,a+1]}) \circ \prod_{i=N-a+1}^{N-1} \left( \prod_{j=d_i}^{N-i-1} ((-1)_{[j]} X_{[j,j+1]}) \right)$$

$$\overset{(24)}{\approx} ((-1)_{[a_1]}(-1)_{[b_1]} \cdots (-1)_{[a_k]}(-1)_{[b_k]}) \prod_{i=1}^{N-a-2} \left( \prod_{j=d_i}^{N-i-1} ((-1)_{[j]} X_{[j,j+1]}) \right) \circ \prod_{j=d_{N-a}}^{d_{N-a-1}-2} ((-1)_{[j]} X_{[j,j+1]}) \circ ((-1)_{[d_{N-a-1}-1]} X_{[d_{N-a-1}-1,d_{N-a-1}]}) \circ \prod_{j=d_{N-a-1}}^{a} ((-1)_{[j]} X_{[j,j+1]}) ((-1)_{[j-1]} X_{[j-1,j]}) \circ \prod_{i=N-a+1}^{N-1} \left( \prod_{j=d_i}^{N-i-1} ((-1)_{[j]} X_{[j,j+1]}) \right)$$

$$= ((-1)_{[a_1]}(-1)_{[b_1]} \cdots (-1)_{[a_k]}(-1)_{[b_k]}) \prod_{i=1}^{N-a-2} \left( \prod_{j=d_i}^{N-i-1} ((-1)_{[j]} X_{[j,j+1]}) \right) \circ \prod_{j=d_{N-a}}^{a} ((-1)_{[j]} X_{[j,j+1]}) \circ \prod_{j=d_{N-a-1}}^{a} ((-1)_{[j]} X_{[j,j+1]}) ((-1)_{[j-1]} X_{[j-1,j]}) \circ \prod_{i=N-a+1}^{N-1} \left( \prod_{j=d_i}^{N-i-1} ((-1)_{[j]} X_{[j,j+1]}) \right)$$

$$\overset{(64)}{\approx} ((-1)_{[a_1]}(-1)_{[b_1]} \cdots (-1)_{[a_k]}(-1)_{[b_k]}) \prod_{i=1}^{N-a-2} \left( \prod_{j=d_i}^{N-i-1} ((-1)_{[j]} X_{[j,j+1]}) \right) \circ \prod_{j=d_{N-a}}^{d_{N-a-1}-1} ((-1)_{[j]} X_{[j,j+1]}) \circ \prod_{j=d_{N-a-1}}^{a} ((-1)_{[j]} X_{[j,j+1]}) \circ \prod_{j=d_{N-a-1}-1}^{a-1} ((-1)_{[j]} X_{[j,j+1]}) \circ \prod_{i=N-a+1}^{N-1} \left( \prod_{j=d_i}^{N-i-1} ((-1)_{[j]} X_{[j,j+1]}) \right)$$

$$= ((-1)_{[a_1]}(-1)_{[b_1]} \cdots (-1)_{[a_k]}(-1)_{[b_k]}) \prod_{i=1}^{N-a-2} \left( \prod_{j=d_i}^{N-i-1} ((-1)_{[j]} X_{[j,j+1]}) \right) \circ \prod_{j=d_{N-a}}^{a} ((-1)_{[j]} X_{[j,j+1]}) \circ \prod_{j=d_{N-a-1}-1}^{a-1} ((-1)_{[j]} X_{[j,j+1]}) \circ \prod_{i=N-a+1}^{N-1} \left( \prod_{j=d_i}^{N-i-1} ((-1)_{[j]} X_{[j,j+1]}) \right)$$

which is in the form (B).

- And if $d_{N-a-1} < d_{N-a} + 1$, then this is equal to

$$\left((-1)_{[a_1]}(-1)_{[b_1]}\cdots(-1)_{[a_k]}(-1)_{[b_k]}\right)\prod_{i=1}^{N-a-2}\left(\prod_{j=d_i}^{N-i-1}((-1)_{[j]}X_{[j,j+1]})\right)\circ\prod_{j=d_{N-a-1}}^{d_{N-a}-1}(-1)_{[j]}X_{[j,j+1]})\circ\prod_{j=d_{N-a}+1}^{a}(-1)_{[j]}X_{[j,j+1]})\circ(-1)_{[a]}X_{[a,a+1]})\circ\prod_{i=N-a+1}^{N-1}\left(\prod_{j=d_i}^{N-i-1}((-1)_{[j]}X_{[j,j+1]})\right)$$

$$\stackrel{(64)}{\approx}\left((-1)_{[a_1]}(-1)_{[b_1]}\cdots(-1)_{[a_k]}(-1)_{[b_k]}\right)\prod_{i=1}^{N-a-2}\left(\prod_{j=d_i}^{N-i-1}((-1)_{[j]}X_{[j,j+1]})\right)\circ\prod_{j=d_{N-a}}^{d_{N-a}}(-1)_{[j]}X_{[j,j+1]})\circ\prod_{j=d_{N-a}+1}^{a}(-1)_{[j]}X_{[j,j+1]})(-1)_{[j-1]}X_{[j-1,j]})\circ(-1)_{[a]}X_{[a,a+1]})\circ\prod_{i=N-a+1}^{N-1}\left(\prod_{j=d_i}^{N-i-1}((-1)_{[j]}X_{[j,j+1]})\right)$$

$$\stackrel{(23)}{\approx}\left((-1)_{[a_1]}(-1)_{[b_1]}\cdots(-1)_{[a_k]}(-1)_{[b_k]}\right)\prod_{i=1}^{N-a-2}\left(\prod_{j=d_i}^{N-i-1}((-1)_{[j]}X_{[j,j+1]})\right)\circ\prod_{j=d_{N-a}}^{d_{N-a}}(-1)_{[j]}X_{[j,j+1]})\circ(-1)_{[d_{N-a}]}X_{[d_{N-a},d_{N-a}+1]})\circ\prod_{j=d_{N-a}+1}^{a}(-1)_{[j]}X_{[j,j+1]})(-1)_{[j-1]}X_{[j-1,j]})\circ\prod_{i=N-a+1}^{N-1}\left(\prod_{j=d_i}^{N-i-1}((-1)_{[j]}X_{[j,j+1]})\right)$$

$$\stackrel{(56)\text{ to }(62)}{\approx}\left((-1)_{[a_1]}(-1)_{[b_1]}\cdots(-1)_{[a_k]}(-1)_{[b_k]}(-1)_{[a]}(-1)_{[\beta]}\right)\prod_{i=1}^{N-a-2}\left(\prod_{j=d_i}^{N-i-1}((-1)_{[j]}X_{[j,j+1]})\right)\circ\prod_{j=d_{N-a}}^{d_{N-a}-1}(-1)_{[j]}X_{[j,j+1]})\circ\prod_{j=d_{N-a}+1}^{a}(-1)_{[j]}X_{[j,j+1]})(-1)_{[j-1]}X_{[j-1,j]})\circ\prod_{i=N-a+1}^{N-1}\left(\prod_{j=d_i}^{N-i-1}((-1)_{[j]}X_{[j,j+1]})\right)\quad\text{for some }\alpha,\beta\in\{0,\ldots,N-1\}$$

$$\stackrel{(64)}{\approx}\left((-1)_{[a_1]}(-1)_{[b_1]}\cdots(-1)_{[a_k]}(-1)_{[b_k]}(-1)_{[a]}(-1)_{[\beta]}\right)\prod_{i=1}^{N-a-2}\left(\prod_{j=d_i}^{N-i-1}((-1)_{[j]}X_{[j,j+1]})\right)\circ\prod_{j=d_{N-a}}^{d_{N-a}-1}(-1)_{[j]}X_{[j,j+1]})\circ\prod_{j=d_{N-a}+1}^{a-1}(-1)_{[j]}X_{[j,j+1]})\circ\prod_{i=N-a+1}^{N-1}\left(\prod_{j=d_i}^{N-i-1}((-1)_{[j]}X_{[j,j+1]})\right)$$

$$=\left((-1)_{[a_1]}(-1)_{[b_1]}\cdots(-1)_{[a_k]}(-1)_{[b_k]}(-1)_{[a]}(-1)_{[\beta]}\right)\prod_{i=1}^{N-a-2}\left(\prod_{j=d_i}^{N-i-1}((-1)_{[j]}X_{[j,j+1]})\right)\circ\prod_{j=d_{N-a}}^{a}(-1)_{[j]}X_{[j,j+1]})\circ\prod_{j=d_{N-a}+1}^{a-1}(-1)_{[j]}X_{[j,j+1]})\circ\prod_{i=N-a+1}^{N-1}\left(\prod_{j=d_i}^{N-i-1}((-1)_{[j]}X_{[j,j+1]})\right)$$

finally, to put this in the form (B), it suffices to put the left part composed of $(-1)$'s in the appropriate form, which can be done using Equations (20) to (22) and (63). □

**Corollary A.5.** *Given any two words $w, w'$ over the generators of $\mathcal{P}_N$ containing no generators of the form $(H_{[a,b]}H_{[c,d]})$, if $[\![w]\!] = [\![w']\!]$ then $w \approx w'$.*

Proof. Since $w$ and $w'$ are semantically equivalent, by Lemma A.4 they are $\approx$-equivalent to the same normal form (of the form (B)), which implies that they are $\approx$-equivalent to each other. □

**Lemma A.6.** *For any word $w$ over the generators of $\mathcal{P}_N$ containing no generators of the form $(H_{[a,b]}H_{[c,d]})$, if $[\![w(H_{[0,1]}H_{[3,2]})]\!] = [\![(H_{[0,1]}H_{[3,2]})w]\!]$, then $w$ is $\approx$-equivalent to a word of the form*

$$((-1)_{[0]}(-1)_{[1]})^\alpha\,((-1)_{[3]}(-1)_{[2]})^\beta\,(X_{[0,3]}X_{[1,2]})^\gamma\,Q$$

*where $Q$ is a product of generators of the form $(-1)_{[a]}X_{[a,a+1]}$ with $a \geq 4$, and $\alpha, \beta, \gamma \in \{0, 1\}$.*

Proof. By analyzing the effect of $[\![w(H_{[0,1]}H_{[3,2]})]\!]$ and $[\![(H_{[0,1]}H_{[3,2]})w]\!]$ on the vectors of the canonical basis of $\mathbb{C}^N$, noting in particular that $[\![w]\!]$ is a just permutation matrix up to the signs of its entries, one can show that $w$ must be semantically equivalent to a word of the indicated form. Then the result follows by Corollary A.5. □

**Corollary A.7.** *For any words $A, A', B, B'$ over the generators of $\mathcal{P}_N$ containing no generators of the form $(H_{[a,b]}H_{[c,d]})$, if $[\![AA']\!] = [\![BB']\!] = I$ and $[\![A(H_{[0,1]}H_{[3,2]})A']\!] = [\![B(H_{[0,1]}H_{[3,2]})B']\!]$, then $A(H_{[0,1]}H_{[3,2]})A' \approx B(H_{[0,1]}H_{[3,2]})B'$.*

Proof. One has

$$\begin{aligned}
[\![B'A(H_{[0,1]}H_{[3,2]})]\!] &= [\![B'A(H_{[0,1]}H_{[3,2]})A'A]\!]\\
&= [\![B'B(H_{[0,1]}H_{[3,2]})B'A]\!]\\
&= [\![(H_{[0,1]}H_{[3,2]})B'A]\!].
\end{aligned}$$

Therefore, by Lemma A.6, $B'A$ is $\approx$-equivalent to a word of the form

$$((-1)_{[0]}(-1)_{[1]})^\alpha\,((-1)_{[3]}(-1)_{[2]})^\beta\,(X_{[0,3]}X_{[1,2]})^\gamma\,Q$$

where $Q$ is a product of generators of the form $(-1)_{[a]}X_{[a,a+1]}$ with $a \geq 4$, and $\alpha, \beta, \gamma \in \{0, 1\}$. Then one has

$$\begin{aligned}
A(H_{[0,1]}H_{[3,2]})A' &\stackrel{\text{Corollary A.5}}{\approx} BB'A(H_{[0,1]}H_{[3,2]})A'\\
&\approx B((-1)_{[0]}(-1)_{[1]})^\alpha\,((-1)_{[3]}(-1)_{[2]})^\beta\,(X_{[0,3]}X_{[1,2]})^\gamma\,Q(H_{[0,1]}H_{[3,2]})A'\\
&\stackrel{(38)}{\approx} B((-1)_{[0]}(-1)_{[1]})^\alpha\,((-1)_{[3]}(-1)_{[2]})^\beta\,(X_{[0,3]}X_{[1,2]})^\gamma\,(H_{[0,1]}H_{[3,2]})QA'
\end{aligned}$$

$$\left(H_{[a,b]}H_{[c,d]}\right) \approx \Sigma_{a,b,c,d} \circ \left(H_{[0,1]}H_{[3,2]}\right) \circ \Sigma'_{a,b,c,d} \qquad \{a,b\} \cap \{c,d\} = \emptyset \tag{65}$$

$$\left(H_{[a,b]}H_{[c,d]}\right) \approx \left(H_{[c,d]}H_{[a,b]}\right) \qquad \{a,b\} \cap \{c,d\} = \emptyset \tag{66}$$

$$\left(H_{[b,a]}H_{[c,d]}\right) \approx \left(H_{[a,b]}H_{[c,d]}\right)\left((-1)_{[b]}X_{[a,b]}\right) \qquad \{a,b\} \cap \{c,d\} = \emptyset \tag{67}$$

$$\left((-1)_{[a]}X_{[a,a+1]}\right)\left(H_{[c,d]}\right) \approx \left(H_{[a+1,b]}H_{[c,d]}\right)\left((-1)_{[a]}X_{[a,a+1]}\right) \qquad |\{a,a+1,b,c,d\}| = 5 \tag{68}$$

$$\left((-1)_{[a]}X_{[a,a+1]}\right)\left(H_{[a+1,b]}H_{[c,d]}\right) \approx \left(H_{[a,b]}H_{[c,d]}\right)\left((-1)_{[a]}X_{[a,b]}\right)\left((-1)_{[a]}X_{[a,a+1]}\right) \qquad |\{a,a+1,b,c,d\}| = 5 \tag{69}$$

$$\left(H_{[a,b]}H_{[a,b]}\right) \approx \epsilon \tag{70}$$

$$\left(H_{[a,b]}H_{[c,d]}\right)H_{[e,f]} \approx \left(H_{[a,b]}H_{[e,f]}\right) \tag{71}$$

$$\left(H_{[a,b]}H_{[b,a]}\right) \approx \left((-1)_{[b]}X_{[a,b]}\right) \tag{72}$$

$$\left(H_{[a,b]}H_{[c,d]}\right)\left(H_{[e,f]}H_{[g,h]}\right) \approx \left(H_{[a,b]}H_{[e,f]}\right)\left(H_{[c,d]}H_{[g,h]}\right) \qquad \{c,d\} \cap \{e,f\} = \emptyset \tag{73}$$

$$\left((-1)_{[a]}X_{[a,b]}\right)\left(H_{[c,d]}H_{[e,f]}\right) \approx \left(H_{[c,d]}H_{[e,f]}\right)\left((-1)_{[a]}X_{[a,b]}\right) \qquad \{a,b\} \cap \{c,d,e,f\} = \emptyset \tag{74}$$

$$\left((-1)_{[a]}X_{[a,b]}\right)\left(H_{[a,b]}H_{[c,d]}\right) \approx \left(H_{[a,b]}H_{[c,d]}\right)\left((-1)_{[b]}X_{[a,b]}\right) \qquad \{a,b\} \cap \{c,d\} = \emptyset \tag{75}$$

$$\left((-1)_{[a]}X_{[a,e]}\right)\left(H_{[a,b]}H_{[c,d]}\right) \approx \left(H_{[e,b]}H_{[c,d]}\right)\left((-1)_{[a]}X_{[a,e]}\right) \qquad |\{a,e,b,c,d\}| = 5 \tag{76}$$

$$\left((-1)_{[a]}X_{[a,e]}\right)\left(H_{[a,b]}H_{[c,d]}\right) \approx \left(H_{[a,b]}H_{[c,d]}\right)\left((-1)_{[a]}X_{[a,b]}\right)\left((-1)_{[a]}X_{[a,e]}\right) \qquad |\{a,e,b,c,d\}| = 5 \tag{77}$$

**Figure 10: More useful auxiliary equations. Equation (65) is the same as Equation (35) but without the restriction on the tuple** $(G_n(a), G_n(b), G_n(c), G_n(d))$**.**

$$
\begin{aligned}
&\overset{\text{Corollary A.5}}{\approx} && B\left((-1)_{[0]}(-1)_{[1]}\right)^{\alpha}\left(X_{[0,3]}X_{[1,2]}\right)^{\gamma}\left((-1)_{[0]}(-1)_{[1]}\right)^{\beta}\left(H_{[0,1]}H_{[3,2]}\right)QA' \\
&\overset{(39)}{\approx} && B\left((-1)_{[0]}(-1)_{[1]}\right)^{\alpha}\left(X_{[0,3]}X_{[1,2]}\right)^{\gamma}\left(H_{[0,1]}H_{[3,2]}\right)\left((-1)_{[0]}(-1)_{[1]}\right)^{\beta}QA' \\
&\overset{(44)}{\approx} && B\left((-1)_{[0]}(-1)_{[1]}\right)^{\alpha}\left(H_{[0,1]}H_{[3,2]}\right)\left(X_{[0,3]}X_{[1,2]}\right)^{\gamma}\left((-1)_{[0]}(-1)_{[1]}\right)^{\beta}QA' \\
&\overset{(39)}{\approx} && B\left(H_{[0,1]}H_{[3,2]}\right)\left((-1)_{[0]}(-1)_{[1]}\right)^{\alpha}\left(X_{[0,3]}X_{[1,2]}\right)^{\gamma}\left((-1)_{[0]}(-1)_{[1]}\right)^{\beta}QA' \\
&\overset{\text{Corollary A.5}}{\approx} && B\left(H_{[0,1]}H_{[3,2]}\right)\left((-1)_{[0]}(-1)_{[1]}\right)^{\alpha}\left((-1)_{[3]}(-1)_{[2]}\right)^{\beta}\left(X_{[0,3]}X_{[1,2]}\right)^{\gamma}QA' \\
&\approx && B\left(H_{[0,1]}H_{[3,2]}\right)B'AA' \\
&\overset{\text{Corollary A.5}}{\approx} && B\left(H_{[0,1]}H_{[3,2]}\right)B'.
\end{aligned}
$$

$\square$

**Corollary A.8.** *The equations of Figure 10 are consequences of the equations of Figure 8.*

PROOF.

*Proof of Equation* (65): Given Equations (35), (36) and (37) and their soundness, and the fact that Equation (65) is sound for all values of $a, b, c, d$, the result follows directly from Corollary A.7.

*Proof of Equation* (66):

$$
\begin{aligned}
\left(H_{[a,b]}H_{[c,d]}\right) &\overset{(65)}{\approx} \Sigma_{a,b,c,d} \circ \left(H_{[0,1]}H_{[3,2]}\right) \circ \Sigma'_{a,b,c,d} \\
&\overset{\text{Corollary A.7}}{\approx} \Sigma_{c,d,a,b} \circ \left(H_{[0,1]}H_{[3,2]}\right) \circ \Sigma'_{c,d,a,b} \\
&\overset{(65)}{\approx} \left(H_{[c,d]}H_{[a,b]}\right)
\end{aligned}
$$

*Proof of Equation* (67):

$$
\begin{aligned}
\left(H_{[b,a]}H_{[c,d]}\right) &\overset{(65)}{\approx} \Sigma_{b,a,c,d} \circ \left(H_{[0,1]}H_{[3,2]}\right) \circ \Sigma'_{b,a,c,d} \\
&\overset{\text{Corollary A.7}}{\approx} \Sigma_{c,d,a,b} \circ \left((-1)_{[3]}(-1)_{[4]}\right)\left(H_{[0,1]}H_{[3,2]}\right)\left((-1)_{[2]}(-1)_{[4]}\right) \circ \Sigma'_{c,d,a,b} \\
&\overset{(40)}{\approx} \Sigma_{c,d,a,b} \circ \left(H_{[0,1]}H_{[3,2]}\right)\left((-1)_{[3]}(-1)_{[4]}\right)\left((-1)_{[2]}X_{[2,3]}\right)\left((-1)_{[2]}(-1)_{[4]}\right) \circ \Sigma'_{c,d,a,b} \\
&\overset{\text{Corollary A.5}}{\approx} \Sigma_{c,d,a,b} \circ \left(H_{[0,1]}H_{[3,2]}\right) \circ \Sigma'_{c,d,a,b} \circ \left((-1)_{[b]}X_{[a,b]}\right)
\end{aligned}
$$

$$\overset{\text{Corollary A.7}}{\approx} \Sigma_{a,b,c,d} \circ \left(H_{[0,1]}H_{[3,2]}\right) \circ \Sigma'_{a,b,c,d} \circ \left((-1)_{[b]}X_{[a,b]}\right)$$

$$\overset{(65)}{\approx} \left(H_{[a,b]}H_{[c,d]}\right) \left((-1)_{[b]}X_{[a,b]}\right)$$

*Proof of Equation* (68):

$$\left((-1)_{[a]}X_{[a,a+1]}\right) \left(H_{[a,b]}H_{[c,d]}\right)$$

$$\overset{\text{Corollary A.5}}{\approx} \left((-1)_{[a]}X_{[a,a+1]}\right) \left(H_{[a,b]}H_{[c,d]}\right) \left((-1)_{[a+1]}X_{[a,a+1]}\right) \left((-1)_{[a]}X_{[a,a+1]}\right)$$

$$\overset{(65)}{\approx} \left((-1)_{[a]}X_{[a,a+1]}\right) \circ \Sigma_{a,b,c,d} \circ \left(H_{[0,1]}H_{[3,2]}\right) \circ \Sigma'_{a,b,c,d} \circ \left((-1)_{[a+1]}X_{[a,a+1]}\right) \left((-1)_{[a]}X_{[a,a+1]}\right)$$

$$\overset{\text{Corollary A.7}}{\approx} \Sigma_{a+1,b,c,d} \circ \left(H_{[0,1]}H_{[3,2]}\right) \circ \Sigma'_{a+1,b,c,d} \circ \left((-1)_{[a]}X_{[a,a+1]}\right)$$

$$\overset{(65)}{\approx} \left(H_{[a+1,b]}H_{[c,d]}\right) \left((-1)_{[a]}X_{[a,a+1]}\right)$$

*Proof of Equation* (69):

$$\left((-1)_{[a]}X_{[a,a+1]}\right) \left(H_{[a+1,b]}H_{[c,d]}\right)$$

$$\overset{\text{Corollary A.5}}{\approx} \left((-1)_{[a]}X_{[a,a+1]}\right) \left(H_{[a+1,b]}H_{[c,d]}\right) \left((-1)_{[a+1]}X_{[a,a+1]}\right) \left((-1)_{[a]}X_{[a,a+1]}\right)$$

$$\overset{(65)}{\approx} \left((-1)_{[a]}X_{[a,a+1]}\right) \circ \Sigma_{a+1,b,c,d} \circ \left(H_{[0,1]}H_{[3,2]}\right) \circ \Sigma'_{a+1,b,c,d} \circ \left((-1)_{[a+1]}X_{[a,a+1]}\right) \left((-1)_{[a]}X_{[a,a+1]}\right)$$

$$\overset{\text{Corollary A.7}}{\approx} \Sigma_{c,d,a,b} \circ \left((-1)_{[3]}(-1)_{[4]}\right) \left(H_{[0,1]}H_{[3,2]}\right) \left((-1)_{[3]}(-1)_{[4]}\right) \circ \Sigma'_{c,d,a,b} \circ \left((-1)_{[a]}X_{[a,a+1]}\right)$$

$$\overset{(40)}{\approx} \Sigma_{c,d,a,b} \circ \left(H_{[0,1]}H_{[3,2]}\right) \left((-1)_{[3]}(-1)_{[4]}\right) \left((-1)_{[2]}X_{[2,3]}\right) \left((-1)_{[3]}(-1)_{[4]}\right) \circ \Sigma'_{c,d,a,b} \circ \left((-1)_{[a]}X_{[a,a+1]}\right)$$

$$\overset{\text{Corollary A.5}}{\approx} \Sigma_{c,d,a,b} \circ \left(H_{[0,1]}H_{[3,2]}\right) \circ \Sigma'_{c,d,a,b} \circ \left((-1)_{[a]}X_{[a,b]}\right) \circ \left((-1)_{[a]}X_{[a,a+1]}\right)$$

$$\overset{\text{Corollary A.7}}{\approx} \Sigma_{a,b,c,d} \circ \left(H_{[0,1]}H_{[3,2]}\right) \circ \Sigma'_{a,b,c,d} \circ \left((-1)_{[a]}X_{[a,b]}\right) \circ \left((-1)_{[a]}X_{[a,a+1]}\right)$$

$$\overset{(65)}{\approx} \left(H_{[a,b]}H_{[c,d]}\right) \left((-1)_{[a]}X_{[a,b]}\right) \circ \left((-1)_{[a]}X_{[a,a+1]}\right)$$

*Proof of Equation* (70):

$$\left(H_{[a,b]}H_{[a,b]}\right) \overset{(42)}{\approx} \left(H_{[a,b]}H_{[e,f]}\right) \left(H_{[e,f]}H_{[a,b]}\right)$$

$$\overset{(66)}{\approx} \left(H_{[a,b]}H_{[e,f]}\right) \left(H_{[a,b]}H_{[e,f]}\right)$$

$$\overset{(65)}{\approx} \Sigma_{a,b,e,f} \circ \left(H_{[0,1]}H_{[3,2]}\right) \circ \Sigma'_{a,b,e,f} \circ \Sigma_{a,b,e,f} \circ \left(H_{[0,1]}H_{[3,2]}\right) \circ \Sigma'_{a,b,e,f}$$

$$\overset{\text{Corollary A.5}}{\approx} \Sigma_{a,b,e,f} \circ \left(H_{[0,1]}H_{[3,2]}\right) \left(H_{[0,1]}H_{[3,2]}\right) \circ \Sigma'_{a,b,e,f}$$

$$\overset{(41)}{\approx} \Sigma_{a,b,e,f} \circ \Sigma'_{a,b,e,f}$$

$$\overset{\text{Corollary A.5}}{\approx} \epsilon$$

where $e, f$ are defined in Figure 8.

*Proof of Equation* (71): First, we consider the case where $a, b, c, d, e, f$ are all distinct. Let $\mathcal{T}_{a,b,c,d,e,f}$ and $\mathcal{T}'_{a,b,c,d,e,f}$ be words over the generators of $\mathcal{P}_N$ containing no generators of the form $\left(H_{[g,h]}H_{[i,j]}\right)$, such that $\left[\!\left[\mathcal{T}'_{a,b,c,d,e,f}\right]\!\right] = \left[\!\left[\mathcal{T}_{a,b,c,d,e,f}\right]\!\right]^{-1}$, and either $\left[\!\left[\mathcal{T}_{a,b,c,d,e,f}\right]\!\right] = \tau$ or $\left[\!\left[\mathcal{T}_{a,b,c,d,e,f}\right]\!\right] = \tau \circ \left[\!\left[(-1)_{[v]}\right]\!\right]$, where $v \notin \{0, 1, 3, 2, 4, 5\}$ and $\tau$ is a permutation matrix mapping respectively $|0\rangle, |1\rangle, |3\rangle, |2\rangle, |4\rangle, |5\rangle$ to $|a\rangle, |b\rangle, |c\rangle, |d\rangle, |e\rangle, |f\rangle$.[13] We have

$$\left(H_{[a,b]}H_{[c,d]}\right) \left(H_{[c,d]}H_{[e,f]}\right)$$

$$\overset{(65)}{\approx} \Sigma_{a,b,c,d} \circ \left(H_{[0,1]}H_{[3,2]}\right) \circ \Sigma'_{a,b,c,d} \circ \Sigma_{c,d,e,f} \circ \left(H_{[0,1]}H_{[3,2]}\right) \circ \Sigma'_{c,d,e,f}$$

$$\overset{\text{Corollary A.7}}{\approx} \mathcal{T}_{a,b,c,d,e,f} \circ \left(H_{[0,1]}H_{[3,2]}\right) \circ \mathcal{T}'_{a,b,c,d,e,f} \circ \mathcal{T}_{a,b,c,d,e,f} \circ \Sigma_{3,2,4,5} \circ \left(H_{[0,1]}H_{[3,2]}\right) \circ \Sigma'_{3,2,4,5} \circ \mathcal{T}'_{a,b,c,d,e,f}$$

$$\overset{(65)}{\approx} \mathcal{T}_{a,b,c,d,e,f} \circ \left(H_{[0,1]}H_{[3,2]}\right) \circ \mathcal{T}'_{a,b,c,d,e,f} \circ \mathcal{T}_{a,b,c,d,e,f} \circ \left(H_{[3,2]}H_{[4,5]}\right) \circ \mathcal{T}'_{a,b,c,d,e,f}$$

---

[13] $\mathcal{T}_{a,b,c,d,e,f}$ and $\mathcal{T}'_{a,b,c,d,e,f}$ (as well as $\Xi^{(a,b,e,f)}_{i,j \,|\, c,d}$ and $\Xi^{(a,b,e,f)'}_{i,j \,|\, c,d}$ below) can be constructed in a similar way to $\Sigma_{a,b,c,d}$ and $\Sigma_{a,b,c,d}$ (see Appendix E.1).

$$\overset{\text{Corollary A.5}}{\approx} \mathcal{T}_{a,b,c,d,e,f} \circ \left( H_{[0,1]} H_{[3,2]} \right) \left( H_{[3,2]} H_{[4,5]} \right) \circ \mathcal{T}'_{a,b,c,d,e,f}$$

$$\overset{(43)}{\approx} \mathcal{T}_{a,b,c,d,e,f} \circ \left( X_{[0,3]} X_{[1,2]} \right) \left( H_{[3,2]} H_{[4,5]} \right) \left( X_{[0,3]} X_{[1,2]} \right) \circ \mathcal{T}'_{a,b,c,d,e,f}$$

$$\overset{(65)}{\approx} \mathcal{T}_{a,b,c,d,e,f} \circ \left( X_{[0,3]} X_{[1,2]} \right) \circ \Sigma_{3,2,4,5} \circ \left( H_{[0,1]} H_{[3,2]} \right) \circ \Sigma'_{3,2,4,5} \circ \left( X_{[0,3]} X_{[1,2]} \right) \circ \mathcal{T}'_{a,b,c,d,e,f}$$

$$\overset{\text{Corollary A.7}}{\approx} \Sigma_{a,b,e,f} \circ \left( H_{[0,1]} H_{[3,2]} \right) \circ \Sigma'_{a,b,e,f}$$

$$\overset{(65)}{\approx} \left( H_{[a,b]} H_{[e,f]} \right)$$

Then, we consider the case where $|\{a, b, e, f\}| \leq 3$ and $\{c, d\} \cap \{a, b, e, f\} = \emptyset$. Let $i, j$, with $i < j$, be the two smallest elements of $\{0, ..., N-1\} \setminus \{a, b, e, f\}$. Let $\Xi^{(a,b,e,f)}_{i,j \,|\, c,d}$ and $\Xi^{(a,b,e,f)'}_{i,j \,|\, c,d}$ be words over the generators of $\mathcal{P}_N$ containing no generators of the form $\left( H_{[g,h]} H_{[k,\ell]} \right)$, such that $\left[\!\left[ \Xi^{(a,b,e,f)'}_{i,j \,|\, c,d} \right]\!\right] = \left[\!\left[ \Xi^{(a,b,e,f)}_{i,j \,|\, c,d} \right]\!\right]^{-1}$, and either $\left[\!\left[ \Xi^{(a,b,e,f)}_{i,j \,|\, c,d} \right]\!\right] = \xi$ or $\left[\!\left[ \Xi^{(a,b,e,f)}_{i,j \,|\, c,d} \right]\!\right] = \xi \circ \left[\!\left[ (-1)_{[\rho]} \right]\!\right]$, where[14] $\rho \notin \{a, b, c, d, e, f, i, j\}$ and $\xi$ is a permutation matrix mapping respectively $|a\rangle, |b\rangle, |e\rangle, |f\rangle, |i\rangle, |j\rangle$ to $|a\rangle, |b\rangle, |e\rangle, |f\rangle, |c\rangle, |d\rangle$. We have

$$\left( H_{[a,b]} H_{[c,d]} \right) \left( H_{[c,d]} H_{[e,f]} \right)$$

$$\overset{(65)}{\approx} \Sigma_{a,b,c,d} \circ \left( H_{[0,1]} H_{[3,2]} \right) \circ \Sigma'_{a,b,c,d} \circ \Sigma_{c,d,e,f} \circ \left( H_{[0,1]} H_{[3,2]} \right) \circ \Sigma'_{c,d,e,f}$$

$$\overset{\text{Corollary A.7}}{\approx} \Xi^{(a,b,e,f)}_{i,j \,|\, c,d} \circ \Sigma_{a,b,i,j} \circ \left( H_{[0,1]} H_{[3,2]} \right) \circ \Sigma'_{a,b,i,j} \circ \Xi^{(a,b,e,f)'}_{i,j \,|\, c,d} \circ \Xi^{(a,b,e,f)}_{i,j \,|\, c,d} \circ \Sigma_{i,j,e,f} \circ \left( H_{[0,1]} H_{[3,2]} \right) \circ \Sigma'_{i,j,e,f} \circ \Xi^{(a,b,e,f)'}_{i,j \,|\, c,d}$$

$$\overset{(65)}{\approx} \Xi^{(a,b,e,f)}_{i,j \,|\, c,d} \circ \left( H_{[a,b]} H_{[i,j]} \right) \circ \Xi^{(a,b,e,f)'}_{i,j \,|\, c,d} \circ \Xi^{(a,b,e,f)}_{i,j \,|\, c,d} \circ \left( H_{[i,j]} H_{[e,f]} \right) \circ \Xi^{(a,b,e,f)'}_{i,j \,|\, c,d}$$

$$\overset{\text{Corollary A.5}}{\approx} \Xi^{(a,b,e,f)}_{i,j \,|\, c,d} \circ \left( H_{[a,b]} H_{[i,j]} \right) \left( H_{[i,j]} H_{[e,f]} \right) \circ \Xi^{(a,b,e,f)'}_{i,j \,|\, c,d}$$

$$\overset{(42)}{\approx} \Xi^{(a,b,e,f)}_{i,j \,|\, c,d} \circ \left( H_{[a,b]} H_{[e,f]} \right) \circ \Xi^{(a,b,e,f)'}_{i,j \,|\, c,d}$$

$$\overset{(65)}{\approx} \Xi^{(a,b,e,f)}_{i,j \,|\, c,d} \circ \Sigma_{a,b,e,f} \circ \left( H_{[0,1]} H_{[3,2]} \right) \circ \Sigma'_{a,b,e,f} \circ \Xi^{(a,b,e,f)'}_{i,j \,|\, c,d}$$

$$\overset{\text{Corollary A.7}}{\approx} \Sigma_{a,b,e,f} \circ \left( H_{[0,1]} H_{[3,2]} \right) \circ \Sigma'_{a,b,e,f}$$

$$\overset{(65)}{\approx} \left( H_{[a,b]} H_{[e,f]} \right)$$

We have treated all the cases where $\{c, d\} \cap \{a, b, e, f\} = \emptyset$, therefore it remains to treat the case where $\{c, d\} \cap \{a, b, e, f\} \neq \emptyset$. In this case, let $i, j$, with $i < j$, be the two smallest elements of $\{0, ..., N-1\} \setminus \{a, b, c, d, e, f\}$. We have

$$\left( H_{[a,b]} H_{[c,d]} \right) \left( H_{[c,d]} H_{[e,f]} \right)$$

$$\overset{\text{previous cases}}{\approx} \left( H_{[a,b]} H_{[i,j]} \right) \left( H_{[i,j]} H_{[c,d]} \right) \left( H_{[c,d]} H_{[i,j]} \right) \left( H_{[i,j]} H_{[e,f]} \right)$$

$$\overset{\text{previous cases}}{\approx} \left( H_{[a,b]} H_{[i,j]} \right) \left( H_{[i,j]} H_{[i,j]} \right) \left( H_{[i,j]} H_{[e,f]} \right)$$

$$\overset{(70)}{\approx} \left( H_{[a,b]} H_{[i,j]} \right) \left( H_{[i,j]} H_{[e,f]} \right)$$

$$\overset{\text{previous cases}}{\approx} \left( H_{[a,b]} H_{[e,f]} \right)$$

*Proof of Equation* (72): Let $c, d$ be such that $c \neq d$ and $\{a, b\} \cap \{c, d\} = \emptyset$. Such indices exist since $N \geq 4$. We have

$$\left( H_{[a,b]} H_{[b,a]} \right) \overset{(71)}{\approx} \left( H_{[a,b]} H_{[c,d]} \right) \left( H_{[c,d]} H_{[b,a]} \right)$$

$$\overset{(66)}{\approx} \left( H_{[a,b]} H_{[c,d]} \right) \left( H_{[b,a]} H_{[c,d]} \right)$$

$$\overset{(67)}{\approx} \left( H_{[a,b]} H_{[c,d]} \right) \left( H_{[a,b]} H_{[c,d]} \right) \left( (-1)_{[b]} X_{[a,b]} \right)$$

$$\overset{(66)}{\approx} \left( H_{[a,b]} H_{[c,d]} \right) \left( H_{[c,d]} H_{[a,b]} \right) \left( (-1)_{[b]} X_{[a,b]} \right)$$

$$\overset{(71)}{\approx} \left( H_{[a,b]} H_{[a,b]} \right) \left( (-1)_{[b]} X_{[a,b]} \right)$$

$$\overset{(70)}{\approx} \left( (-1)_{[b]} X_{[a,b]} \right)$$

---

[14] $\rho$ exists since $N \geq 8$, and $|\{a, b, e, f\}| \leq 3$ implies $|\{a, b, c, d, e, f, i, j\}| < 8$.

*Proof of Equation* (73):

$$\begin{aligned}
\left(H_{[a,b]}H_{[c,d]}\right)\left(H_{[e,f]}H_{[g,h]}\right) &\overset{(70)}{\approx} \left(H_{[a,b]}H_{[c,d]}\right)\left(H_{[c,d]}H_{[c,d]}\right)\left(H_{[e,f]}H_{[g,h]}\right) \\
&\overset{(71)}{\approx} \left(H_{[a,b]}H_{[c,d]}\right)\left(H_{[c,d]}H_{[e,f]}\right)\left(H_{[e,f]}H_{[c,d]}\right)\left(H_{[e,f]}H_{[g,h]}\right) \\
&\overset{(66)}{\approx} \left(H_{[a,b]}H_{[c,d]}\right)\left(H_{[c,d]}H_{[e,f]}\right)\left(H_{[c,d]}H_{[e,f]}\right)\left(H_{[e,f]}H_{[g,h]}\right) \\
&\overset{(71)}{\approx} H_{[a,b]}H_{[e,f]}\right)\left(H_{[c,d]}H_{[g,h]}\right)
\end{aligned}$$

*Proof of Equation* (74):

$$\begin{aligned}
\left((-1)_{[a]}X_{[a,b]}\right)\left(H_{[c,d]}H_{[e,f]}\right) &\overset{(33)}{\approx} \left((-1)_{[a]}X_{[b,a]}\right)\left(H_{[c,d]}H_{[e,f]}\right) \\
&\overset{(72)}{\approx} \left(H_{[b,a]}H_{[a,b]}\right)\left(H_{[c,d]}H_{[e,f]}\right) \\
&\overset{(73)}{\approx} \left(H_{[b,a]}H_{[c,d]}\right)\left(H_{[a,b]}H_{[e,f]}\right) \\
&\overset{(66)}{\approx} \left(H_{[c,d]}H_{[b,a]}\right)\left(H_{[e,f]}H_{[a,b]}\right) \\
&\overset{(73)}{\approx} \left(H_{[c,d]}H_{[e,f]}\right)\left(H_{[b,a]}H_{[a,b]}\right) \\
&\overset{(72)}{\approx} \left(H_{[c,d]}H_{[e,f]}\right)\left((-1)_{[a]}X_{[b,a]}\right) \\
&\overset{(33)}{\approx} \left(H_{[c,d]}H_{[e,f]}\right)\left((-1)_{[a]}X_{[a,b]}\right)
\end{aligned}$$

*Proof of Equation* (75):

$$\begin{aligned}
\left((-1)_{[a]}X_{[a,b]}\right)\left(H_{[a,b]}H_{[c,d]}\right) &\overset{(33)(72)}{\approx} \left(H_{[b,a]}H_{[a,b]}\right)\left(H_{[a,b]}H_{[c,d]}\right) \\
&\overset{(71)}{\approx} \left(H_{[b,a]}H_{[c,d]}\right) \\
&\overset{(67)}{\approx} \left(H_{[a,b]}H_{[c,d]}\right)\left((-1)_{[b]}X_{[a,b]}\right)
\end{aligned}$$

*Proof of Equations* (76) *and* (77):

- If $e - a = 1$, then Equations (76) and (77) are just instances of Equations (68) and (69) respectively.
- If $a - e = 1$, then we proceed as follows.[15] For Equation (76):

$$\begin{aligned}
&\phantom{\approx}\left((-1)_{[a]}X_{[a,a-1]}\right)\left(H_{[a,b]}H_{[c,d]}\right) \\
&\overset{\text{Corollary A.5}}{\approx} \left((-1)_{[a-1]}X_{[a-1,a]}\right)^3\left(H_{[a,b]}H_{[c,d]}\right) \\
&\overset{(69)}{\approx} \left((-1)_{[a-1]}X_{[a-1,a]}\right)^2\left(H_{[a-1,b]}H_{[c,d]}\right)\left((-1)_{[a-1]}X_{[a-1,b]}\right)\left((-1)_{[a-1]}X_{[a-1,a]}\right) \\
&\overset{(68)}{\approx} \left((-1)_{[a-1]}X_{[a-1,a]}\right)\left(H_{[a,b]}H_{[c,d]}\right)\left((-1)_{[a-1]}X_{[a-1,a]}\right)\left((-1)_{[a-1]}X_{[a-1,b]}\right)\left((-1)_{[a-1]}X_{[a-1,a]}\right) \\
&\overset{(69)}{\approx} \left(H_{[a-1,b]}H_{[c,d]}\right)\left((-1)_{[a-1]}X_{[a-1,b]}\right)\left((-1)_{[a-1]}X_{[a-1,a]}\right)\left((-1)_{[a-1]}X_{[a-1,a]}\right)\left((-1)_{[a-1]}X_{[a-1,b]}\right)\left((-1)_{[a-1]}X_{[a-1,a]}\right) \\
&\overset{\text{Corollary A.5}}{\approx} \left(H_{[a-1,b]}H_{[c,d]}\right)\left((-1)_{[a]}X_{[a,a-1]}\right)
\end{aligned}$$

and for Equation (77):

$$\begin{aligned}
&\phantom{\approx}\left((-1)_{[a]}X_{[a,a-1]}\right)\left(H_{[a-1,b]}H_{[c,d]}\right) \\
&\overset{\text{Corollary A.5}}{\approx} \left((-1)_{[a-1]}X_{[a-1,a]}\right)^3\left(H_{[a-1,b]}H_{[c,d]}\right) \\
&\overset{(68)}{\approx} \left((-1)_{[a-1]}X_{[a-1,a]}\right)^2\left(H_{[a,b]}H_{[c,d]}\right)\left((-1)_{[a-1]}X_{[a-1,a]}\right) \\
&\overset{(69)}{\approx} \left((-1)_{[a-1]}X_{[a-1,a]}\right)\left(H_{[a-1,b]}H_{[c,d]}\right)\left((-1)_{[a-1]}X_{[a-1,b]}\right)\left((-1)_{[a-1]}X_{[a-1,a]}\right)^2 \\
&\overset{(68)}{\approx} \left(H_{[a,b]}H_{[c,d]}\right)\left((-1)_{[a-1]}X_{[a-1,a]}\right)\left((-1)_{[a-1]}X_{[a-1,b]}\right)\left((-1)_{[a-1]}X_{[a-1,a]}\right)^2 \\
&\overset{\text{Corollary A.5}}{\approx} \left(H_{[a,b]}H_{[c,d]}\right)\left((-1)_{[a]}X_{[a,b]}\right)\left((-1)_{[a]}X_{[a,a-1]}\right)
\end{aligned}$$

---

[15] Notice that the derivations are identical to the case below where $e < a$ and $\{a, ..., e\} \cap \{b, c, d\} = \emptyset$

- If $a < e$ and $\{a, ..., e\} \cap \{b, c, d\} = \emptyset$, then we proceed by finite induction on $e - a$. The base case has already been treated. The induction step for Equation (76) is as follows:

$$
\begin{aligned}
\left((-1)_{[a]} X_{[a,e]}\right) \left(H_{[a,b]} H_{[c,d]}\right) \quad &\overset{(49)}{\approx} \quad \left((-1)_{[a+1]} X_{[a,a+1]}\right) \left((-1)_{[a+1]} X_{[a+1,e]}\right) \left((-1)_{[a]} X_{[a,a+1]}\right) \left(H_{[a,b]} H_{[c,d]}\right) \\
&\overset{\text{induction}}{\underset{\text{hypothesis}}{\approx}} \quad \left((-1)_{[a+1]} X_{[a,a+1]}\right) \left((-1)_{[a+1]} X_{[a+1,e]}\right) \left(H_{[a+1,b]} H_{[c,d]}\right) \left((-1)_{[a]} X_{[a,a+1]}\right) \\
&\overset{\text{induction}}{\underset{\text{hypothesis}}{\approx}} \quad \left((-1)_{[a+1]} X_{[a,a+1]}\right) \left(H_{[e,b]} H_{[c,d]}\right) \left((-1)_{[a+1]} X_{[a+1,e]}\right) \left((-1)_{[a]} X_{[a,a+1]}\right) \\
&\overset{(74)}{\approx} \quad \left(H_{[e,b]} H_{[c,d]}\right) \left((-1)_{[a+1]} X_{[a,a+1]}\right) \left((-1)_{[a+1]} X_{[a+1,e]}\right) \left((-1)_{[a]} X_{[a,a+1]}\right) \\
&\overset{(49)}{\approx} \quad \left(H_{[e,b]} H_{[c,d]}\right) \left((-1)_{[a]} X_{[a,e]}\right)
\end{aligned}
$$

and the induction step for Equation (77) is as follows:

$$
\begin{aligned}
&\quad \left((-1)_{[a]} X_{[a,e]}\right) \left(H_{[e,b]} H_{[c,d]}\right) \\
\overset{(49)}{\approx} \quad &\left((-1)_{[a+1]} X_{[a,a+1]}\right) \left((-1)_{[a+1]} X_{[a+1,e]}\right) \left((-1)_{[a]} X_{[a,a+1]}\right) \left(H_{[e,b]} H_{[c,d]}\right) \\
\overset{(74)}{\approx} \quad &\left((-1)_{[a+1]} X_{[a,a+1]}\right) \left((-1)_{[a+1]} X_{[a+1,e]}\right) \left(H_{[e,b]} H_{[c,d]}\right) \left((-1)_{[a]} X_{[a,a+1]}\right) \\
\overset{\text{induction}}{\underset{\text{hypothesis}}{\approx}} \quad &\left((-1)_{[a+1]} X_{[a,a+1]}\right) \left(H_{[a+1,b]} H_{[c,d]}\right) \left((-1)_{[a+1]} X_{[a+1,b]}\right) \left((-1)_{[a+1]} X_{[a+1,e]}\right) \left((-1)_{[a]} X_{[a,a+1]}\right) \\
\overset{\substack{\text{case } a-e=1 \\ \text{for (76)}}}{\approx} \quad &\left(H_{[a,b]} H_{[c,d]}\right) \left((-1)_{[a+1]} X_{[a,a+1]}\right) \left((-1)_{[a+1]} X_{[a+1,b]}\right) \left((-1)_{[a+1]} X_{[a+1,e]}\right) \left((-1)_{[a]} X_{[a,a+1]}\right) \\
\overset{(51)}{\approx} \quad &{\scriptstyle \left(H_{[a,b]} H_{[c,d]}\right) \left((-1)_{[a+1]} X_{[a,a+1]}\right) \left((-1)_{[a+1]} X_{[a+1,b]}\right) \left((-1)_{[a]} X_{[a,a+1]}\right) \left((-1)_{[a+1]} X_{[a,a+1]}\right) \left((-1)_{[a+1]} X_{[a+1,e]}\right) \left((-1)_{[a]} X_{[a,a+1]}\right)} \\
\overset{(49)}{\approx} \quad &\left(H_{[e,b]} H_{[c,d]}\right) \left((-1)_{[a]} X_{[a,b]}\right) \left((-1)_{[a]} X_{[a,e]}\right)
\end{aligned}
$$

- If $e < a$ and $\{e, ..., a\} \cap \{b, c, d\} = \emptyset$, then we deduce the result from the previous case, as follows. For Equation (76):

$$
\begin{aligned}
&\quad \left((-1)_{[a]} X_{[a,e]}\right) \left(H_{[a,b]} H_{[c,d]}\right) \\
\overset{\text{Corollary A.5}}{\approx} \quad &\left((-1)_{[e]} X_{[e,a]}\right)^3 \left(H_{[a,b]} H_{[c,d]}\right) \\
\overset{\substack{\text{case } a<e \\ \text{of (77)}}}{\approx} \quad &\left((-1)_{[e]} X_{[e,a]}\right)^2 \left(H_{[e,b]} H_{[c,d]}\right) \left((-1)_{[e]} X_{[e,b]}\right) \left((-1)_{[e]} X_{[e,a]}\right) \\
\overset{\substack{\text{case } a<e \\ \text{of (76)}}}{\approx} \quad &\left((-1)_{[e]} X_{[e,a]}\right) \left(H_{[a,b]} H_{[c,d]}\right) \left((-1)_{[e]} X_{[e,a]}\right) \left((-1)_{[e]} X_{[e,b]}\right) \left((-1)_{[e]} X_{[e,a]}\right) \\
\overset{\substack{\text{case } a<e \\ \text{of (77)}}}{\approx} \quad &\left(H_{[e,b]} H_{[c,d]}\right) \left((-1)_{[e]} X_{[e,b]}\right) \left((-1)_{[e]} X_{[e,a]}\right) \left((-1)_{[e]} X_{[e,a]}\right) \left((-1)_{[e]} X_{[e,b]}\right) \left((-1)_{[e]} X_{[e,a]}\right) \\
\overset{\text{Corollary A.5}}{\approx} \quad &\left(H_{[e,b]} H_{[c,d]}\right) \left((-1)_{[a]} X_{[a,e]}\right)
\end{aligned}
$$

and for Equation (77):

$$
\begin{aligned}
&\quad \left((-1)_{[a]} X_{[a,e]}\right) \left(H_{[e,b]} H_{[c,d]}\right) \\
\overset{\text{Corollary A.5}}{\approx} \quad &\left((-1)_{[e]} X_{[e,a]}\right)^3 \left(H_{[e,b]} H_{[c,d]}\right) \\
\overset{\substack{\text{case } a<e \\ \text{of (76)}}}{\approx} \quad &\left((-1)_{[e]} X_{[e,a]}\right)^2 \left(H_{[a,b]} H_{[c,d]}\right) \left((-1)_{[e]} X_{[e,a]}\right) \\
\overset{\substack{\text{case } a<e \\ \text{of (77)}}}{\approx} \quad &\left((-1)_{[e]} X_{[e,a]}\right) \left(H_{[e,b]} H_{[c,d]}\right) \left((-1)_{[e]} X_{[e,b]}\right) \left((-1)_{[e]} X_{[e,a]}\right)^2 \\
\overset{\substack{\text{case } a<e \\ \text{of (76)}}}{\approx} \quad &\left(H_{[a,b]} H_{[c,d]}\right) \left((-1)_{[e]} X_{[e,a]}\right) \left((-1)_{[e]} X_{[e,b]}\right) \left((-1)_{[e]} X_{[e,a]}\right)^2 \\
\overset{\text{Corollary A.5}}{\approx} \quad &\left(H_{[a,b]} H_{[c,d]}\right) \left((-1)_{[a]} X_{[a,b]}\right) \left((-1)_{[a]} X_{[a,e]}\right)
\end{aligned}
$$

- To treat the general case, it will be useful to prove the following auxiliary equation:

$$
\left((-1)_{[a]} X_{[a,c]}\right) \left(H_{[a,b]} H_{[c,d]}\right) \quad \approx \quad \left(H_{[c,b]} H_{[a,d]}\right) \left((-1)_{[a]} X_{[a,d]}\right) \left((-1)_{[a]} X_{[a,c]}\right) \qquad {\scriptstyle |\{a,b,c,d\}| = 4, \ [a,c] \cap \{b,d\} = \emptyset, \ |a-c| \le N-4} \tag{78}
$$

where $[a, c]$ denotes the interval $\{a, ..., c\}$ if $a < c$, or $\{c, ..., a\}$ if $c < a$.

To prove Equation (78), let $f_b, g_b, f_d, g_d \in \{0, ..., N-1\} \setminus [a,c]$ be such that $f_b \neq g_b$, $f_d \neq g_d$, $f_b, g_b \neq d$ and $f_d, g_d \neq b$. Such indices exist thanks to the condition $|a-c| \leq N-4$. One has the following derivation (where the references to Equations (76) and (77) refer to their restricted versions proven above):

$$\left((-1)_{[a]} X_{[a,c]}\right) \left(H_{[a,b]} H_{[c,d]}\right)$$

$$\overset{(71)}{\approx} \left((-1)_{[a]} X_{[a,c]}\right) \left(H_{[a,b]} H_{[f_d,g_d]}\right) \left(H_{[f_d,g_d]} H_{[f_b,g_b]}\right) \left(H_{[f_b,g_b]} H_{[c,d]}\right)$$

$$\overset{(66)}{\approx} \left((-1)_{[a]} X_{[a,c]}\right) \left(H_{[a,b]} H_{[f_d,g_d]}\right) \left(H_{[f_d,g_d]} H_{[f_b,g_b]}\right) \left(H_{[c,d]} H_{[f_b,g_b]}\right)$$

$$\overset{(76)}{\approx} \left(H_{[c,b]} H_{[f_d,g_d]}\right) \left((-1)_{[a]} X_{[a,c]}\right) \left(H_{[f_d,g_d]} H_{[f_b,g_b]}\right) \left(H_{[c,d]} H_{[f_b,g_b]}\right)$$

$$\overset{(74)}{\approx} \left(H_{[c,b]} H_{[f_d,g_d]}\right) \left(H_{[f_d,g_d]} H_{[f_b,g_b]}\right) \left((-1)_{[a]} X_{[a,c]}\right) \left(H_{[c,d]} H_{[f_b,g_b]}\right)$$

$$\overset{(77)}{\approx} \left(H_{[c,b]} H_{[f_d,g_d]}\right) \left(H_{[f_d,g_d]} H_{[f_b,g_b]}\right) \left(H_{[a,d]} H_{[f_b,g_b]}\right) \left((-1)_{[a]} X_{[a,d]}\right) \left((-1)_{[a]} X_{[a,c]}\right)$$

$$\overset{(66)}{\approx} \left(H_{[c,b]} H_{[f_d,g_d]}\right) \left(H_{[f_d,g_d]} H_{[f_b,g_b]}\right) \left(H_{[f_b,g_b]} H_{[a,d]}\right) \left((-1)_{[a]} X_{[a,d]}\right) \left((-1)_{[a]} X_{[a,c]}\right)$$

$$\overset{(71)}{\approx} \left(H_{[c,b]} H_{[a,d]}\right) \left((-1)_{[a]} X_{[a,d]}\right) \left((-1)_{[a]} X_{[a,c]}\right)$$

Now, to prove Equations (76) and (77) in the general case, let $i_0 < \cdots < i_k$ be such that $1 \leq k \leq 4$, $\{i_0, i_k\} = \{a, e\}$ and $\{i_1, i_2, ..., i_{k-1}\} = [a, e] \cap \{b, c, d\}$ (namely, $i_1, ..., i_{k-1}$ are the indices among $b, c, d$ that are between $a$ and $e$). By Corollary A.5, one has

$$\left((-1)_{[a]} X_{[a,e]}\right) \approx \left((-1)_{[i_1]} X_{[i_0,i_1]}\right) \cdots \left((-1)_{[i_{k-1}]} X_{[i_{k-2},i_{k-1}]}\right) \circ \left((-1)_{[i_{k-1}]} X_{[i_{k-1},i_k]}\right) \left((-1)_{[i_{k-2}]} X_{[i_{k-2},i_{k-1}]}\right) \cdots \left((-1)_{[i_0]} X_{[i_0,i_1]}\right) \quad \text{(C)}$$

and for every $j \in \{0, ..., k-2\}$,

$$\left((-1)_{[i_{j+1}]} X_{[i_j,i_{j+1}]}\right) \approx \left((-1)_{[i_j]} X_{[i_j,i_{j+1}]}\right)^3 \quad \text{(D)}$$

Every pair of indices $i_j, i_{j+1}$ is such that the other three indices are outside of the interval $[i_j, i_{j+1}]$, which also implies that $|i_j - i_{j+1}| \leq N-4$. Therefore, starting from $\left((-1)_{[a]} X_{[a,e]}\right) \left(H_{[a,b]} H_{[c,d]}\right)$ (resp. $\left((-1)_{[a]} X_{[a,e]}\right) \left(H_{[e,b]} H_{[c,d]}\right)$) and after decomposing $\left((-1)_{[a]} X_{[a,e]}\right)$ according to Equations (C) and (D), we can use the already proven cases of Equations (76) and (77) together with Equation (78) and Equation (75) (all four being potentially used in combination with Equations (66) and (67)) to push all the $\left((-1)_{[i_j]} X_{[i_j,i_{j+1}]}\right)$ to the right. One can check that we get $\left(H_{[e,b]} H_{[c,d]}\right)$ (resp. $\left(H_{[a,b]} H_{[c,d]}\right)$), multiplied on its right by a product of generators of the form $\left((-1)_{[f]} X_{[f,g]}\right)$ and $\left((-1)_{[g]} X_{[f,g]}\right)$. By Corollary A.5, we can rewrite this product into $\left((-1)_{[a]} X_{[a,e]}\right)$ (resp. into $\left((-1)_{[a]} X_{[a,b]}\right) \left((-1)_{[a]} X_{[a,e]}\right)$), which finishes the proof. □

**Corollary A.9.** *Any word over the generators of $\mathcal{P}_N$ containing at most one generator of the form $\left(H_{[a,b]} H_{[c,d]}\right)$ is $\approx$-equivalent to a unique word, called its* normal form, *of the form*

$$P \quad \text{or} \quad \left(H_{[a,b]} H_{[c,d]}\right) P$$

*with $a < b$; $c < d$; $\{a, b\} \neq \{c, d\}$; if $\{a, b\} \cap \{c, d\} = \emptyset$ then $a < c$; and $P$ being of the form (B). Moreover, two words with the same semantics have the same normal form.*

Proof.

*Uniqueness given the semantics:* By analysis of the effect of $[\![P]\!]$ and $[\![\left(H_{[a,b]} H_{[c,d]}\right) P]\!]$ on the canonical basis of $\mathbb{C}^N$, noting that $[\![P]\!]$ is a permutation matrix up to the signs of its entries.

*Existence:* If the word does not contain any $\left(H_{[a,b]} H_{[c,d]}\right)$, then the result is just a consequence of Lemma A.4.

If the word contains a single $\left(H_{[a,b]} H_{[c,d]}\right)$, then it is of the form $w = p_1 \left(H_{[a,b]} H_{[c,d]}\right) p_2$ where $p_1, p_2$ are two words containing no generators of the form $\left(H_{[e,f]} H_{[g,h]}\right)$. As we have show in the proof of Lemma A.4, the word $p_1$ can be written as a product of generators of the form $\left((-1)_{[j]} X_{[j,j+1]}\right)$, therefore it suffices to prove the result for $p_1 = \left((-1)_{[j]} X_{[j,j+1]}\right)$. For practical purposes, we will prove the result in a slightly more general case, namely for $p_1 = \left((-1)_{[j]} X_{[j,k]}\right)$.

Let $e, f \in \{0, ..., N-1\}$ such that $e \neq f$ and $e, f \notin \{a, b, c, d, j, k\}$. Such indices exist since $N \geq 8$. One has

$$\left((-1)_{[j]} X_{[j,k]}\right) \left(H_{[a,b]} H_{[c,d]}\right) \overset{(71)}{\approx} \left((-1)_{[j]} X_{[j,k]}\right) \left(H_{[a,b]} H_{[e,f]}\right) \left(H_{[e,f]} H_{[c,d]}\right)$$

$$\overset{(66)}{\approx} \left((-1)_{[j]} X_{[j,k]}\right) \left(H_{[a,b]} H_{[e,f]}\right) \left(H_{[c,d]} H_{[e,f]}\right)$$

Let us first consider $\left((-1)_{[j]} X_{[j,k]}\right) \left(H_{[a,b]} H_{[e,f]}\right)$. There are seven cases:

- If $a = j$ and $b = k$, then

$$\left((-1)_{[j]} X_{[j,k]}\right) \left(H_{[a,b]} H_{[e,f]}\right) \quad \overset{(75)}{\approx} \quad \left(H_{[a,b]} H_{[e,f]}\right) \left((-1)_{[k]} X_{[j,k]}\right)$$

$$\overset{(33)}{\approx} \quad \left(H_{[a,b]} H_{[e,f]}\right) \left((-1)_{[k]} X_{[k,j]}\right)$$

- If $a = k$ and $b = j$, then

$$\left((-1)_{[j]} X_{[j,k]}\right) \left(H_{[a,b]} H_{[e,f]}\right)$$

$$= \quad \left((-1)_{[j]} X_{[j,k]}\right) \left(H_{[k,j]} H_{[e,f]}\right)$$

$$\overset{(67)}{\approx} \quad \left((-1)_{[j]} X_{[j,k]}\right) \left(H_{[j,k]} H_{[e,f]}\right) \left((-1)_{[k]} X_{[j,k]}\right)$$

$$\overset{(75)}{\approx} \quad \left(H_{[j,k]} H_{[e,f]}\right) \left((-1)_{[k]} X_{[j,k]}\right) \left((-1)_{[k]} X_{[j,k]}\right)$$

$$\overset{(53)(21)}{\approx} \quad \left(H_{[j,k]} H_{[e,f]}\right) \left((-1)_{[j]} (-1)_{[k]}\right) \left((-1)_{[j]} X_{[j,k]}\right) \left((-1)_{[k]} (-1)_{[j]}\right) \left((-1)_{[j]} X_{[j,k]}\right)$$

$$\overset{(56)}{\approx} \quad \left(H_{[j,k]} H_{[e,f]}\right) \left((-1)_{[j]} X_{[j,k]}\right) \left((-1)_{[j]} (-1)_{[k]}\right) \left((-1)_{[k]} (-1)_{[j]}\right) \left((-1)_{[j]} X_{[j,k]}\right)$$

$$\overset{(22)(20)}{\approx} \quad \left(H_{[j,k]} H_{[e,f]}\right) \left((-1)_{[j]} X_{[j,k]}\right)^2$$

- If $a = j$ and $b \neq k$, then

$$\left((-1)_{[j]} X_{[j,k]}\right) \left(H_{[a,b]} H_{[e,f]}\right) \quad = \quad \left((-1)_{[j]} X_{[j,k]}\right) \left(H_{[j,b]} H_{[e,f]}\right)$$

$$\overset{(76)}{\approx} \quad \left(H_{[k,b]} H_{[e,f]}\right) \left((-1)_{[j]} X_{[j,k]}\right)$$

- If $a = k$ and $b \neq j$, then

$$\left((-1)_{[j]} X_{[j,k]}\right) \left(H_{[a,b]} H_{[e,f]}\right) \quad = \quad \left((-1)_{[j]} X_{[j,k]}\right) \left(H_{[k,b]} H_{[e,f]}\right)$$

$$\overset{(77)}{\approx} \quad \left(H_{[j,b]} H_{[e,f]}\right) \left((-1)_{[j]} X_{[j,b]}\right) \left((-1)_{[j]} X_{[j,k]}\right)$$

- If $b = j$ and $a \neq k$, then

$$\left((-1)_{[j]} X_{[j,k]}\right) \left(H_{[a,b]} H_{[e,f]}\right)$$

$$= \quad \left((-1)_{[j]} X_{[j,k]}\right) \left(H_{[a,j]} H_{[e,f]}\right)$$

$$\overset{(67)}{\approx} \quad \left((-1)_{[j]} X_{[j,k]}\right) \left(H_{[j,a]} H_{[e,f]}\right) \left((-1)_{[a]} X_{[j,a]}\right)$$

$$\overset{(76)}{\approx} \quad \left(H_{[k,a]} H_{[e,f]}\right) \left((-1)_{[j]} X_{[j,k]}\right) \left((-1)_{[a]} X_{[j,a]}\right)$$

$$\overset{(67)}{\approx} \quad \left(H_{[a,k]} H_{[e,f]}\right) \left((-1)_{[k]} X_{[a,k]}\right) \left((-1)_{[j]} X_{[j,k]}\right) \left((-1)_{[a]} X_{[j,a]}\right)$$

$$\overset{(33)}{\approx} \quad \left(H_{[a,k]} H_{[e,f]}\right) \left((-1)_{[k]} X_{[k,a]}\right) \left((-1)_{[j]} X_{[j,k]}\right) \left((-1)_{[a]} X_{[j,a]}\right)$$

$$\overset{(51)}{\approx} \quad \left(H_{[a,k]} H_{[e,f]}\right) \left((-1)_{[j]} X_{[j,k]}\right) \left((-1)_{[k]} X_{[j,k]}\right) \left((-1)_{[k]} X_{[k,a]}\right) \left((-1)_{[j]} X_{[j,k]}\right) \left((-1)_{[a]} X_{[j,a]}\right)$$

$$\overset{\text{Corollary A.5}}{\approx} \quad \left(H_{[a,k]} H_{[e,f]}\right) \left((-1)_{[j]} X_{[j,k]}\right) \left((-1)_{[j]} X_{[j,a]}\right) \left((-1)_{[a]} X_{[j,a]}\right)$$

$$\overset{(51)}{\approx} \quad (H_{[a,k]}H_{[e,f]}) \left((-1)_{[j]}X_{[j,k]}\right)$$

- If $b = k$ and $a \neq j$, then

$$\left((-1)_{[j]}X_{[j,k]}\right)(H_{[a,b]}H_{[e,f]})$$

$$= \quad \left((-1)_{[j]}X_{[j,k]}\right)(H_{[a,k]}H_{[e,f]})$$

$$\overset{(67)}{\approx} \quad \left((-1)_{[j]}X_{[j,k]}\right)(H_{[k,a]}H_{[e,f]})\left((-1)_{[a]}X_{[k,a]}\right)$$

$$\overset{(77)}{\approx} \quad (H_{[j,a]}H_{[e,f]})\left((-1)_{[j]}X_{[j,a]}\right)\left((-1)_{[j]}X_{[j,k]}\right)\left((-1)_{[a]}X_{[k,a]}\right)$$

$$\overset{(33)}{\approx} \quad (H_{[j,a]}H_{[e,f]})\left((-1)_{[j]}X_{[j,a]}\right)\left((-1)_{[j]}X_{[j,k]}\right)\left((-1)_{[a]}X_{[a,k]}\right)$$

- If $\{a, b\} \cap \{j, k\} = \emptyset$, then

$$\left((-1)_{[j]}X_{[j,k]}\right)(H_{[a,b]}H_{[e,f]}) \quad \overset{(74)}{\approx} \quad (H_{[a,b]}H_{[e,f]})\left((-1)_{[j]}X_{[j,k]}\right)$$

In all seven cases, we have transformed $\left((-1)_{[j]}X_{[j,k]}\right)(H_{[a,b]}H_{[e,f]})$ into some $(H_{[a',b']}H_{[e,f]})\,q$, where $\{a', b'\} \cap \{e, f\} = \emptyset$, and $q$ is the product of between one and three generators of the form $\left((-1)_{[j']}X_{[j',k']}\right)$, which are such that $j', k'$ are distinct from $e, f$. Then we can apply the same reasoning to transform $q\,(H_{[c,d]}H_{[e,f]})$ into some $(H_{[c',d']}H_{[e,f]})\,p$, where $\{c', d'\} \cap \{e, f\} = \emptyset$, and $p$ is the product of between one and nine generators of the form $\left((-1)_{[j']}X_{[j',k']}\right)$.

This shows that the word $w = p_1\,(H_{[a,b]}H_{[e,f]})\,p_2$ can be put in the form $(H_{[a',b']}H_{[e,f]})\,(H_{[c',d']}H_{[e,f]})\,p_3$, where $p_3$ is a product of generators of the form $\left((-1)_{[j']}X_{[j',k']}\right)$. The next step is to ensure that the indices $a', b', c', d'$ satisfy the required conditions.

If $a' < b'$, then we proceed as follows

$$(H_{[a',b']}H_{[e,f]})\,(H_{[c',d']}H_{[e,f]})\,p_3 \quad \overset{(67)}{\approx} \quad (H_{[b',a']}H_{[e,f]})\left((-1)_{[a']}X_{[b',a']}\right)(H_{[c',d']}H_{[e,f]})\,p_3$$

$$\overset{(33)}{\approx} \quad (H_{[b',a']}H_{[e,f]})\left((-1)_{[a']}X_{[a',b']}\right)(H_{[c',d']}H_{[e,f]})\,p_3$$

and we apply the same reasoning as above to obtain $(H_{[b',a']}H_{[e,f]})\,(H_{[c'',d'']}H_{[e,f]})\,p_3'$, where $p_3'$ is a product of generators of the form $\left((-1)_{[j']}X_{[j',k']}\right)$. Thus, we have got $(H_{[a'',b'']}H_{[e,f]})\,(H_{[c'',d'']}H_{[e,f]})\,p_3'$, where $p_3'$ is a product of generators of the form $\left((-1)_{[j']}X_{[j',k']}\right)$, and $a'' < b''$. Then, if $c'' < d''$, we apply Equations (67) and (33) one more time to get $(H_{[a'',b'']}H_{[e,f]})\,(H_{[d'',c'']}H_{[e,f]})\,p_3''$, where $p_3''$ is still a product of generators of the form $\left((-1)_{[j']}X_{[j',k']}\right)$. Thus, we have got $(H_{[a'',b'']}H_{[e,f]})\,(H_{[c''',d''']}H_{[e,f]})\,p_3''$, where $p_3''$ is a product of generators of the form $\left((-1)_{[j']}X_{[j',k']}\right)$, and where $a'' < b''$ and $c''' < d'''$.

Finally, we have

$$(H_{[a'',b'']}H_{[e,f]})\,(H_{[c''',d''']}H_{[e,f]})\,p_3'' \quad \overset{(66)}{\approx} \quad (H_{[a'',b'']}H_{[e,f]})\,(H_{[e,f]}H_{[c''',d''']})\,p_3''$$

$$\overset{(71)}{\approx} \quad (H_{[a'',b'']}H_{[c''',d''']})\,p_3''$$

Up to applying Equation (70) in the case where $\{a'', b''\} = \{c''', d'''\}$, or Equation (66) in the case where $\{a'', b''\} \cap \{c''', d'''\} = \emptyset$ and $a'' > c'''$, and to applying Lemma A.4 to $p_3''$, we get the desired form. □

**Corollary A.10.** *Given any two words $w, w'$ over the generators of $\mathcal{P}_N$, such that one of the two contains at most one generator of the form $(H_{[a,b]}H_{[c,d]})$, and the other contains at most two generators of this form, if $[\![w]\!] = [\![w']\!]$ then $w \approx w'$.*

Proof. By applying Corollary A.9 as many times as the number of generators of the form $(H_{[a,b]}H_{[c,d]})$, we can transform $w$ (resp. $w'$) into a word $\tilde{w}$ (resp. $\tilde{w}'$) containing at most as many such generators as $w$ (resp. $w'$) and where all such generators are on the left. Then by analysis of the effect of $[\![\tilde{w}]\!]$ and $[\![\tilde{w}']\!]$ on the canonical basis of $\mathbb{C}^N$, one can show that if $\tilde{w}$ or $\tilde{w}'$ contains two generators of the form $(H_{[a,b]}H_{[c,d]})$, then one can necessarily make them cancel out by means of Equations (66), (70), (71) and (73). Then the result follows by applying Corollary A.9 again, thanks to the uniqueness of the normal form. □

**Corollary A.11.** *The equations of Figure 11 are consequences of the equations of Figure 8.*
Proof.

$$\left(H_{[0,1]}H_{[0,2]}\right)\left(H_{[1,3]}H_{[0,1]}\right)\left((-1)_{[1]}(-1)_{[0]}\right)\left(H_{[0,2]}H_{[1,3]}\right) \approx$$
$$\left(H_{[0,2]}H_{[1,3]}\right)\left((-1)_{[1]}(-1)_{[0]}\right)\left(H_{[0,1]}H_{[0,2]}\right)\left(H_{[1,3]}H_{[0,1]}\right) \tag{79}$$

$$\left(H_{[0,1]}H_{[0,2]}\right)\left(H_{[1,3]}H_{[0,4]}\right)\left(H_{[1,5]}H_{[0,1]}\right)\left((-1)_{[1]}(-1)_{[0]}\right)\left(H_{[0,4]}H_{[1,5]}\right)\left(H_{[0,2]}H_{[1,3]}\right) \approx$$
$$\left(H_{[0,2]}H_{[1,3]}\right)\left(H_{[0,4]}H_{[1,5]}\right)\left((-1)_{[1]}(-1)_{[0]}\right)\left(H_{[0,1]}H_{[0,4]}\right)\left(H_{[1,5]}H_{[0,2]}\right)\left(H_{[1,3]}H_{[0,1]}\right) \tag{80}$$

**Figure 11:** Simpler versions of Equations (45) and (46), more complicated to decode but simpler to use in derivations.

*Proof of Equation (79):*

$$\left(H_{[0,1]}H_{[0,2]}\right)\left(H_{[1,3]}H_{[0,1]}\right)\left((-1)_{[1]}(-1)_{[0]}\right)\left(H_{[0,2]}H_{[1,3]}\right)$$

$\overset{\text{Corollary A.5}}{\approx} \left((-1)_{[4]}X_{[2,3]}\right)\left((-1)_{[4]}X_{[2,3]}\right)\left(H_{[0,1]}H_{[0,2]}\right)\left(H_{[1,3]}H_{[0,1]}\right)\left((-1)_{[1]}(-1)_{[0]}\right)\left(H_{[0,2]}H_{[1,3]}\right)$

$\overset{\text{Corollary A.10}}{\approx} \left((-1)_{[4]}X_{[2,3]}\right)\left(H_{[0,1]}H_{[0,3]}\right)\left((-1)_{[4]}X_{[2,3]}\right)\left(H_{[1,3]}H_{[0,1]}\right)\left((-1)_{[1]}(-1)_{[0]}\right)\left(H_{[0,2]}H_{[1,3]}\right)$

$\overset{\text{Corollary A.10}}{\approx} \left((-1)_{[4]}X_{[2,3]}\right)\left(H_{[0,1]}H_{[0,3]}\right)\left(H_{[1,2]}H_{[0,1]}\right)\left((-1)_{[4]}X_{[2,3]}\right)\left((-1)_{[1]}(-1)_{[0]}\right)\left(H_{[0,2]}H_{[1,3]}\right)$

$\overset{\text{Corollary A.5}}{\approx} \left((-1)_{[4]}X_{[2,3]}\right)\left(H_{[0,1]}H_{[0,3]}\right)\left(H_{[1,2]}H_{[0,1]}\right)\left((-1)_{[1]}(-1)_{[0]}\right)\left((-1)_{[4]}X_{[2,3]}\right)\left(H_{[0,2]}H_{[1,3]}\right)$

$\overset{\text{Corollary A.10}}{\approx} \left((-1)_{[4]}X_{[2,3]}\right)\left(H_{[0,1]}H_{[0,3]}\right)\left(H_{[1,2]}H_{[0,1]}\right)\left((-1)_{[1]}(-1)_{[0]}\right)\left(H_{[0,3]}H_{[1,2]}\right)\left((-1)_{[4]}X_{[2,3]}\right)$

$\overset{(70)}{\approx} \left((-1)_{[4]}X_{[2,3]}\right)\left(H_{[0,1]}H_{[0,3]}\right)\left(H_{[1,2]}H_{[0,1]}\right)\left(H_{[7,6]}H_{[7,6]}\right)\left((-1)_{[1]}(-1)_{[0]}\right)\left(H_{[0,3]}H_{[1,2]}\right)\left((-1)_{[4]}X_{[2,3]}\right)$

$\overset{(73)}{\approx} \left((-1)_{[4]}X_{[2,3]}\right)\left(H_{[0,1]}H_{[0,3]}\right)\left(H_{[1,2]}H_{[7,6]}\right)\left(H_{[0,1]}H_{[7,6]}\right)\left((-1)_{[1]}(-1)_{[0]}\right)\left(H_{[0,3]}H_{[1,2]}\right)\left((-1)_{[4]}X_{[2,3]}\right)$

$\overset{(66)}{\approx} \left((-1)_{[4]}X_{[2,3]}\right)\left(H_{[0,1]}H_{[0,3]}\right)\left(H_{[7,6]}H_{[1,2]}\right)\left(H_{[0,1]}H_{[7,6]}\right)\left((-1)_{[1]}(-1)_{[0]}\right)\left(H_{[0,3]}H_{[1,2]}\right)\left((-1)_{[4]}X_{[2,3]}\right)$

$\overset{(73)}{\approx} \left((-1)_{[4]}X_{[2,3]}\right)\left(H_{[0,1]}H_{[7,6]}\right)\left(H_{[0,3]}H_{[1,2]}\right)\left(H_{[0,1]}H_{[7,6]}\right)\left((-1)_{[1]}(-1)_{[0]}\right)\left(H_{[0,3]}H_{[1,2]}\right)\left((-1)_{[4]}X_{[2,3]}\right)$

$\overset{(21)}{\approx} \left((-1)_{[4]}X_{[2,3]}\right)\left(H_{[0,1]}H_{[7,6]}\right)\left(H_{[0,3]}H_{[1,2]}\right)\left(H_{[0,1]}H_{[7,6]}\right)\left((-1)_{[0]}(-1)_{[1]}\right)\left(H_{[0,3]}H_{[1,2]}\right)\left((-1)_{[4]}X_{[2,3]}\right)$

$\overset{(45)}{\approx} \left((-1)_{[4]}X_{[2,3]}\right)\left(H_{[0,3]}H_{[1,2]}\right)\left((-1)_{[0]}(-1)_{[1]}\right)\left(H_{[0,1]}H_{[7,6]}\right)\left(H_{[0,3]}H_{[1,2]}\right)\left(H_{[0,1]}H_{[7,6]}\right)\left((-1)_{[4]}X_{[2,3]}\right)$

$\overset{(21)}{\approx} \left((-1)_{[4]}X_{[2,3]}\right)\left(H_{[0,3]}H_{[1,2]}\right)\left((-1)_{[1]}(-1)_{[0]}\right)\left(H_{[0,1]}H_{[7,6]}\right)\left(H_{[0,3]}H_{[1,2]}\right)\left(H_{[0,1]}H_{[7,6]}\right)\left((-1)_{[4]}X_{[2,3]}\right)$

$\overset{(73)}{\approx} \left((-1)_{[4]}X_{[2,3]}\right)\left(H_{[0,3]}H_{[1,2]}\right)\left((-1)_{[1]}(-1)_{[0]}\right)\left(H_{[0,1]}H_{[0,3]}\right)\left(H_{[7,6]}H_{[1,2]}\right)\left(H_{[0,1]}H_{[7,6]}\right)\left((-1)_{[4]}X_{[2,3]}\right)$

$\overset{(66)}{\approx} \left((-1)_{[4]}X_{[2,3]}\right)\left(H_{[0,3]}H_{[1,2]}\right)\left((-1)_{[1]}(-1)_{[0]}\right)\left(H_{[0,1]}H_{[0,3]}\right)\left(H_{[1,2]}H_{[7,6]}\right)\left(H_{[0,1]}H_{[7,6]}\right)\left((-1)_{[4]}X_{[2,3]}\right)$

$\overset{(73)}{\approx} \left((-1)_{[4]}X_{[2,3]}\right)\left(H_{[0,3]}H_{[1,2]}\right)\left((-1)_{[1]}(-1)_{[0]}\right)\left(H_{[0,1]}H_{[0,3]}\right)\left(H_{[1,2]}H_{[0,1]}\right)\left(H_{[7,6]}H_{[7,6]}\right)\left((-1)_{[4]}X_{[2,3]}\right)$

$$\overset{(70)}{\approx} \quad \left((-1)_{[4]} X_{[2,3]}\right) \left(H_{[0,3]} H_{[1,2]}\right) \left((-1)_{[1]} (-1)_{[0]}\right) \left(H_{[0,1]} H_{[0,3]}\right) \left(H_{[1,2]} H_{[0,1]}\right) \left((-1)_{[4]} X_{[2,3]}\right)$$

$$\overset{\text{Corollary A.10}}{\approx} \quad \left(H_{[0,2]} H_{[1,3]}\right) \left((-1)_{[1]} (-1)_{[0]}\right) \left(H_{[0,1]} H_{[0,2]}\right) \left(H_{[1,3]} H_{[0,1]}\right)$$

*Proof of Equation* (80):

$$(H_{[0,1]}H_{[0,2]}) (H_{[1,3]}H_{[0,4]}) (H_{[1,5]}H_{[0,1]}) ((-1)_{[1]}(-1)_{[0]}) (H_{[0,4]}H_{[1,5]}) (H_{[0,2]}H_{[1,3]})$$

Corollary A.5
$$\approx ((-1)_{[6]}X_{[2,3]}) ((-1)_{[2]}(-1)_{[3]}) ((-1)_{[2]}(-1)_{[3]}) ((-1)_{[6]}X_{[2,3]}) (H_{[0,1]}H_{[0,2]}) (H_{[1,3]}H_{[0,4]}) (H_{[1,5]}H_{[0,1]}) ((-1)_{[1]}(-1)_{[0]}) (H_{[0,4]}H_{[1,5]}) (H_{[0,2]}H_{[1,3]})$$

Corollary A.10
$$\approx ((-1)_{[6]}X_{[2,3]}) ((-1)_{[2]}(-1)_{[3]}) (H_{[0,1]}H_{[0,3]}) ((-1)_{[2]}X_{[0,3]}) ((-1)_{[6]}X_{[2,3]}) (H_{[1,3]}H_{[0,4]}) (H_{[1,5]}H_{[0,1]}) ((-1)_{[1]}(-1)_{[0]}) (H_{[0,4]}H_{[1,5]}) (H_{[0,2]}H_{[1,3]})$$

Corollary A.10
$$\approx ((-1)_{[6]}X_{[2,3]}) ((-1)_{[2]}(-1)_{[3]}) (H_{[0,1]}H_{[0,3]}) (H_{[1,2]}H_{[3,4]}) (X_{[1,2]}X_{[0,3]}) ((-1)_{[6]}X_{[2,3]}) (H_{[1,5]}H_{[0,1]}) ((-1)_{[1]}(-1)_{[0]}) (H_{[0,4]}H_{[1,5]}) (H_{[0,2]}H_{[1,3]})$$

Corollary A.10
$$\approx ((-1)_{[6]}X_{[2,3]}) ((-1)_{[2]}(-1)_{[3]}) (H_{[0,1]}H_{[0,3]}) (H_{[1,2]}H_{[3,4]}) (H_{[2,5]}H_{[3,2]}) (X_{[1,2]}X_{[0,3]}) ((-1)_{[6]}X_{[2,3]}) ((-1)_{[1]}(-1)_{[0]}) (H_{[0,4]}H_{[1,5]}) (H_{[0,2]}H_{[1,3]})$$

Corollary A.5
$$\approx ((-1)_{[6]}X_{[2,3]}) ((-1)_{[2]}(-1)_{[3]}) (H_{[0,1]}H_{[0,3]}) (H_{[1,2]}H_{[3,4]}) (H_{[2,5]}H_{[3,2]}) ((-1)_{[2]}(-1)_{[3]}) (X_{[1,2]}X_{[0,3]}) ((-1)_{[6]}X_{[2,3]}) (H_{[0,4]}H_{[1,5]}) (H_{[0,2]}H_{[1,3]})$$

Corollary A.10
$$\approx ((-1)_{[6]}X_{[2,3]}) ((-1)_{[2]}(-1)_{[3]}) (H_{[0,1]}H_{[0,3]}) (H_{[1,2]}H_{[3,4]}) (H_{[2,5]}H_{[3,2]}) ((-1)_{[2]}(-1)_{[3]}) (H_{[3,4]}H_{[2,5]}) (X_{[1,2]}X_{[0,3]}) ((-1)_{[6]}X_{[2,3]}) (H_{[0,2]}H_{[1,3]})$$

Corollary A.10
$$\approx ((-1)_{[6]}X_{[2,3]}) ((-1)_{[2]}(-1)_{[3]}) (H_{[0,1]}H_{[0,3]}) (H_{[1,2]}H_{[3,4]}) (H_{[2,5]}H_{[3,2]}) ((-1)_{[2]}(-1)_{[3]}) (H_{[3,4]}H_{[2,5]}) (H_{[0,3]}H_{[1,2]}) ((-1)_{[2]}(-1)_{[3]}) ((-1)_{[6]}X_{[2,3]})$$

(70)(73)
$$\approx ((-1)_{[6]}X_{[2,3]}) ((-1)_{[2]}(-1)_{[3]}) (H_{[0,1]}H_{[0,3]}) (H_{[1,2]}H_{[3,4]}) (H_{[2,5]}H_{[3,2]}) (H_{[7,6]}H_{[4,5]}) (H_{[7,6]}H_{[4,5]}) ((-1)_{[2]}(-1)_{[3]}) (H_{[3,4]}H_{[2,5]}) (H_{[0,3]}H_{[1,2]}) ((-1)_{[2]}(-1)_{[3]}) ((-1)_{[6]}X_{[2,3]})$$

(73)
$$\approx ((-1)_{[6]}X_{[2,3]}) ((-1)_{[2]}(-1)_{[3]}) (H_{[0,1]}H_{[0,3]}) (H_{[1,2]}H_{[3,4]}) (H_{[2,5]}H_{[7,6]}) (H_{[3,2]}H_{[4,5]}) (H_{[7,6]}H_{[4,5]}) ((-1)_{[2]}(-1)_{[3]}) (H_{[3,4]}H_{[2,5]}) (H_{[0,3]}H_{[1,2]}) ((-1)_{[2]}(-1)_{[3]}) ((-1)_{[6]}X_{[2,3]})$$

(66)
$$\approx ((-1)_{[6]}X_{[2,3]}) ((-1)_{[2]}(-1)_{[3]}) (H_{[0,1]}H_{[0,3]}) (H_{[1,2]}H_{[3,4]}) (H_{[7,6]}H_{[2,5]}) (H_{[3,2]}H_{[4,5]}) (H_{[7,6]}H_{[4,5]}) ((-1)_{[2]}(-1)_{[3]}) (H_{[3,4]}H_{[2,5]}) (H_{[0,3]}H_{[1,2]}) ((-1)_{[2]}(-1)_{[3]}) ((-1)_{[6]}X_{[2,3]})$$

(73)
$$\approx ((-1)_{[6]}X_{[2,3]}) ((-1)_{[2]}(-1)_{[3]}) (H_{[0,1]}H_{[0,3]}) (H_{[1,2]}H_{[7,6]}) (H_{[3,4]}H_{[2,5]}) (H_{[3,2]}H_{[4,5]}) (H_{[7,6]}H_{[4,5]}) ((-1)_{[2]}(-1)_{[3]}) (H_{[3,4]}H_{[2,5]}) (H_{[0,3]}H_{[1,2]}) ((-1)_{[2]}(-1)_{[3]}) ((-1)_{[6]}X_{[2,3]})$$

(66)
$$\approx ((-1)_{[6]}X_{[2,3]}) ((-1)_{[2]}(-1)_{[3]}) (H_{[0,1]}H_{[0,3]}) (H_{[7,6]}H_{[1,2]}) (H_{[3,4]}H_{[2,5]}) (H_{[3,2]}H_{[4,5]}) (H_{[7,6]}H_{[4,5]}) ((-1)_{[2]}(-1)_{[3]}) (H_{[3,4]}H_{[2,5]}) (H_{[0,3]}H_{[1,2]}) ((-1)_{[2]}(-1)_{[3]}) ((-1)_{[6]}X_{[2,3]})$$

(73)
$$\approx ((-1)_{[6]}X_{[2,3]}) ((-1)_{[2]}(-1)_{[3]}) (H_{[0,1]}H_{[7,6]}) (H_{[0,3]}H_{[1,2]}) (H_{[3,4]}H_{[2,5]}) (H_{[3,2]}H_{[4,5]}) (H_{[7,6]}H_{[4,5]}) ((-1)_{[2]}(-1)_{[3]}) (H_{[3,4]}H_{[2,5]}) (H_{[0,3]}H_{[1,2]}) ((-1)_{[2]}(-1)_{[3]}) ((-1)_{[6]}X_{[2,3]})$$

(46)
$$\approx ((-1)_{[6]}X_{[2,3]}) ((-1)_{[2]}(-1)_{[3]}) (H_{[0,3]}H_{[1,2]}) (H_{[3,4]}H_{[2,5]}) ((-1)_{[2]}(-1)_{[3]}) (H_{[7,6]}H_{[4,5]}) (H_{[3,2]}H_{[4,5]}) (H_{[3,4]}H_{[2,5]}) (H_{[0,3]}H_{[1,2]}) (H_{[0,1]}H_{[7,6]}) ((-1)_{[2]}(-1)_{[3]}) ((-1)_{[6]}X_{[2,3]})$$

(73)(70)(66)
$$\approx ((-1)_{[6]}X_{[2,3]}) ((-1)_{[2]}(-1)_{[3]}) (H_{[0,3]}H_{[1,2]}) (H_{[3,4]}H_{[2,5]}) ((-1)_{[2]}(-1)_{[3]}) (H_{[3,2]}H_{[7,6]}) (H_{[3,4]}H_{[2,5]}) (H_{[0,3]}H_{[1,2]}) (H_{[0,1]}H_{[7,6]}) ((-1)_{[2]}(-1)_{[3]}) ((-1)_{[6]}X_{[2,3]})$$

(73)
$$\approx ((-1)_{[6]}X_{[2,3]}) ((-1)_{[2]}(-1)_{[3]}) (H_{[0,3]}H_{[1,2]}) (H_{[3,4]}H_{[2,5]}) ((-1)_{[2]}(-1)_{[3]}) (H_{[3,2]}H_{[3,4]}) (H_{[7,6]}H_{[2,5]}) (H_{[0,3]}H_{[1,2]}) (H_{[0,1]}H_{[7,6]}) ((-1)_{[2]}(-1)_{[3]}) ((-1)_{[6]}X_{[2,3]})$$

(66)
$$\approx ((-1)_{[6]}X_{[2,3]}) ((-1)_{[2]}(-1)_{[3]}) (H_{[0,3]}H_{[1,2]}) (H_{[3,4]}H_{[2,5]}) ((-1)_{[2]}(-1)_{[3]}) (H_{[3,2]}H_{[3,4]}) (H_{[2,5]}H_{[7,6]}) (H_{[0,3]}H_{[1,2]}) (H_{[0,1]}H_{[7,6]}) ((-1)_{[2]}(-1)_{[3]}) ((-1)_{[6]}X_{[2,3]})$$

(73)
$$\approx ((-1)_{[6]}X_{[2,3]}) ((-1)_{[2]}(-1)_{[3]}) (H_{[0,3]}H_{[1,2]}) (H_{[3,4]}H_{[2,5]}) ((-1)_{[2]}(-1)_{[3]}) (H_{[3,2]}H_{[3,4]}) (H_{[2,5]}H_{[0,3]}) (H_{[7,6]}H_{[1,2]}) (H_{[0,1]}H_{[7,6]}) ((-1)_{[2]}(-1)_{[3]}) ((-1)_{[6]}X_{[2,3]})$$

(66)
$$\approx ((-1)_{[6]}X_{[2,3]}) ((-1)_{[2]}(-1)_{[3]}) (H_{[0,3]}H_{[1,2]}) (H_{[3,4]}H_{[2,5]}) ((-1)_{[2]}(-1)_{[3]}) (H_{[3,2]}H_{[3,4]}) (H_{[2,5]}H_{[0,3]}) (H_{[1,2]}H_{[7,6]}) (H_{[7,6]}H_{[0,1]}) ((-1)_{[2]}(-1)_{[3]}) ((-1)_{[6]}X_{[2,3]})$$

(71)
$$\approx ((-1)_{[6]}X_{[2,3]}) ((-1)_{[2]}(-1)_{[3]}) (H_{[0,3]}H_{[1,2]}) (H_{[3,4]}H_{[2,5]}) ((-1)_{[2]}(-1)_{[3]}) (H_{[3,2]}H_{[3,4]}) (H_{[2,5]}H_{[0,3]}) (H_{[1,2]}H_{[0,1]}) ((-1)_{[2]}(-1)_{[3]}) ((-1)_{[6]}X_{[2,3]})$$

Corollary A.10
$$\approx ((-1)_{[6]}X_{[2,3]}) ((-1)_{[2]}(-1)_{[3]}) (H_{[0,3]}H_{[1,2]}) (H_{[3,4]}H_{[2,5]}) ((-1)_{[2]}(-1)_{[3]}) (H_{[3,2]}H_{[3,4]}) (H_{[2,5]}H_{[0,3]}) ((-1)_{[3]}X_{[1,2]}) ((-1)_{[6]}X_{[2,3]}) (H_{[1,3]}H_{[0,1]})$$

Corollary A.10
$$\approx ((-1)_{[6]}X_{[2,3]}) ((-1)_{[2]}(-1)_{[3]}) (H_{[0,3]}H_{[1,2]}) (H_{[3,4]}H_{[2,5]}) ((-1)_{[2]}(-1)_{[3]}) (H_{[3,2]}H_{[3,4]}) (X_{[1,2]}X_{[0,3]}) ((-1)_{[6]}X_{[2,3]}) (H_{[1,5]}H_{[0,2]}) (H_{[1,3]}H_{[0,1]})$$

Corollary A.10
$$\approx ((-1)_{[6]}X_{[2,3]}) ((-1)_{[2]}(-1)_{[3]}) (H_{[0,3]}H_{[1,2]}) (H_{[3,4]}H_{[2,5]}) ((-1)_{[2]}(-1)_{[3]}) (X_{[1,2]}X_{[0,3]}) ((-1)_{[6]}X_{[2,3]}) (H_{[0,1]}H_{[0,4]}) (H_{[1,5]}H_{[0,2]}) (H_{[1,3]}H_{[0,1]})$$

Corollary A.5
$$\approx ((-1)_{[6]}X_{[2,3]}) ((-1)_{[2]}(-1)_{[3]}) (H_{[0,3]}H_{[1,2]}) (H_{[3,4]}H_{[2,5]}) (X_{[1,2]}X_{[0,3]}) ((-1)_{[6]}X_{[2,3]}) ((-1)_{[1]}(-1)_{[0]}) (H_{[0,1]}H_{[0,4]}) (H_{[1,5]}H_{[0,2]}) (H_{[1,3]}H_{[0,1]})$$

Corollary A.10
$$\approx ((-1)_{[6]}X_{[2,3]}) ((-1)_{[2]}(-1)_{[3]}) (H_{[0,3]}H_{[1,2]}) (X_{[1,2]}X_{[0,3]}) ((-1)_{[6]}X_{[2,3]}) (H_{[0,4]}H_{[1,5]}) ((-1)_{[1]}(-1)_{[0]}) (H_{[0,1]}H_{[0,4]}) (H_{[1,5]}H_{[0,2]}) (H_{[1,3]}H_{[0,1]})$$

Corollary A.10
$$\approx ((-1)_{[6]}X_{[2,3]}) ((-1)_{[2]}(-1)_{[3]}) ((-1)_{[2]}(-1)_{[3]}) ((-1)_{[6]}X_{[2,3]}) (H_{[0,2]}H_{[1,3]}) (H_{[0,4]}H_{[1,5]}) ((-1)_{[1]}(-1)_{[0]}) (H_{[0,1]}H_{[0,4]}) (H_{[1,5]}H_{[0,2]}) (H_{[1,3]}H_{[0,1]})$$

Corollary A.5
$$\approx (H_{[0,2]}H_{[1,3]}) (H_{[0,4]}H_{[1,5]}) ((-1)_{[1]}(-1)_{[0]}) (H_{[0,1]}H_{[0,4]}) (H_{[1,5]}H_{[0,2]}) (H_{[1,3]}H_{[0,1]})$$

$$\overset{(20)(71)}{\approx} \left((-1)_{[1]}X_{[0,1]}\right)\left(H_{[0,1]}H_{[0,2]}\right)\left((-1)_{[3]}X_{[1,3]}\right)\left(H_{[1,3]}H_{[0,1]}\right)\left((-1)_{[0]}X_{[0,1]}\right)\left((-1)_{[1]}(-1)_{[0]}\right)\left(H_{[0,2]}H_{[1,3]}\right)\left((-1)_{[1]}X_{[1,3]}\right)$$

$$\overset{\text{Corollary A.10}}{\approx} \left((-1)_{[1]}(-1)_{[4]}\right)\left((-1)_{[4]}X_{[0,1]}\right)\left(H_{[0,1]}H_{[0,2]}\right)\left((-1)_{[3]}X_{[1,3]}\right)\left(H_{[1,3]}H_{[0,1]}\right)\left((-1)_{[0]}X_{[0,1]}\right)\left((-1)_{[1]}(-1)_{[0]}\right)\left(H_{[0,2]}H_{[1,3]}\right)\left((-1)_{[1]}X_{[1,3]}\right)$$

$$\overset{\text{Corollary A.10}}{\approx} \left((-1)_{[1]}(-1)_{[4]}\right)\left(H_{[0,1]}H_{[0,2]}\right)\left((-1)_{[4]}(-1)_{[1]}\right)\left((-1)_{[3]}X_{[1,3]}\right)\left(H_{[1,3]}H_{[0,1]}\right)\left((-1)_{[0]}X_{[0,1]}\right)\left((-1)_{[1]}(-1)_{[0]}\right)\left(H_{[0,2]}H_{[1,3]}\right)\left((-1)_{[1]}X_{[1,3]}\right)$$

$$\overset{\text{Corollary A.10}}{\approx} \left((-1)_{[1]}(-1)_{[4]}\right)\left(H_{[0,1]}H_{[0,2]}\right)\left(H_{[1,3]}H_{[0,1]}\right)\left((-1)_{[1]}(-1)_{[0]}\right)\left((-1)_{[4]}(-1)_{[1]}\right)\left(H_{[0,2]}H_{[1,3]}\right)\left((-1)_{[1]}X_{[1,3]}\right)$$

$$\overset{\text{Corollary A.10}}{\approx} \left((-1)_{[1]}(-1)_{[4]}\right)\left(H_{[0,1]}H_{[0,2]}\right)\left(H_{[1,3]}H_{[0,1]}\right)\left((-1)_{[1]}(-1)_{[0]}\right)\left(H_{[0,2]}H_{[1,3]}\right)\left((-1)_{[4]}(-1)_{[1]}\right)$$

$$\overset{(79)}{\approx} \left((-1)_{[1]}(-1)_{[4]}\right)\left(H_{[0,2]}H_{[1,3]}\right)\left((-1)_{[1]}(-1)_{[0]}\right)\left(H_{[0,1]}H_{[0,2]}\right)\left(H_{[1,3]}H_{[0,1]}\right)\left((-1)_{[4]}(-1)_{[1]}\right)$$

$$\overset{\text{Corollary A.10}}{\approx} \left(H_{[0,2]}H_{[1,3]}\right)\left((-1)_{[1]}X_{[1,3]}\right)\left((-1)_{[1]}(-1)_{[4]}\right)\left((-1)_{[1]}(-1)_{[0]}\right)\left(H_{[0,1]}H_{[0,2]}\right)\left(H_{[1,3]}H_{[0,1]}\right)\left((-1)_{[4]}(-1)_{[1]}\right)$$

$$\overset{\text{Corollary A.10}}{\approx} \left(H_{[0,2]}H_{[1,3]}\right)\left((-1)_{[1]}X_{[1,3]}\right)\left((-1)_{[1]}X_{[0,1]}\right)\left((-1)_{[1]}(-1)_{[0]}\right)\left(H_{[0,1]}H_{[0,2]}\right)\left((-1)_{[4]}(-1)_{[1]}\right)\left(H_{[1,3]}H_{[0,1]}\right)\left((-1)_{[4]}(-1)_{[1]}\right)$$

$$\overset{\text{Corollary A.10}}{\approx} \left(H_{[0,2]}H_{[1,3]}\right)\left((-1)_{[1]}X_{[1,3]}\right)\left((-1)_{[1]}X_{[0,1]}\right)\left((-1)_{[1]}(-1)_{[0]}\right)\left(H_{[0,1]}H_{[0,2]}\right)\left((-1)_{[3]}X_{[1,3]}\right)\left(H_{[1,3]}H_{[0,1]}\right)\left((-1)_{[0]}X_{[0,1]}\right)$$

$$\overset{(20)(70)(71)}{\approx} \left(H_{[0,2]}H_{[0,1]}\right)\left(H_{[0,1]}H_{[1,3]}\right)\left((-1)_{[1]}X_{[1,3]}\right)\left((-1)_{[1]}X_{[0,1]}\right)\left(H_{[0,1]}H_{[0,1]}\right)\left((-1)_{[1]}(-1)_{[0]}\right)\left((-1)_{[1]}(-1)_{[1]}\right)$$
$$\left(H_{[0,1]}H_{[0,2]}\right)\left((-1)_{[3]}X_{[1,3]}\right)\left(H_{[1,3]}H_{[0,1]}\right)\left(H_{[0,1]}H_{[0,1]}\right)\left((-1)_{[0]}X_{[0,1]}\right)$$

Proof of Equation (H-d3*):

$$\left(H_{[0,1]}H_{[0,1]}\right)\left(H_{[0,2]}H_{[0,1]}\right)\left(H_{[0,1]}H_{[1,3]}\right)\left(H_{[0,1]}H_{[0,1]}\right)$$
$$\left((-1)_{[1]}(-1)_{[0]}\right)\left((-1)_{[1]}(-1)_{[1]}\right)\left(H_{[0,1]}H_{[0,2]}\right)\left(H_{[1,3]}H_{[0,1]}\right)$$

$$\overset{(20)(70)(71)}{\approx} \left(H_{[0,2]}H_{[1,3]}\right)\left((-1)_{[1]}(-1)_{[0]}\right)\left(H_{[0,1]}H_{[0,2]}\right)\left(H_{[1,3]}H_{[0,1]}\right)$$

$$\overset{(79)}{\approx} \left(H_{[0,1]}H_{[0,2]}\right)\left(H_{[1,3]}H_{[0,1]}\right)\left((-1)_{[1]}(-1)_{[0]}\right)\left(H_{[0,2]}H_{[1,3]}\right)$$

$$\overset{(20)(70)(71)}{\approx} \left(H_{[0,1]}H_{[0,2]}\right)\left(H_{[1,3]}H_{[0,1]}\right)\left(H_{[0,1]}H_{[0,1]}\right)\left((-1)_{[1]}(-1)_{[0]}\right)\left((-1)_{[1]}(-1)_{[1]}\right)$$
$$\left(H_{[0,2]}H_{[0,1]}\right)\left(H_{[0,1]}H_{[1,3]}\right)\left(H_{[0,1]}H_{[0,1]}\right)$$

Proof of Equation (HZ-d3*):

$$(H_{[0,1]}H_{[0,1]})\ ((-1)_{[0]}X_{[0,1]})\ (H_{[0,2]}H_{[0,1]})\ (H_{[0,1]}H_{[1,3]})\ ((-1)_{[1]}X_{[0,1]})\ ((-1)_{[1]}(-1)_{[0,1]})\ (H_{[0,1]}H_{[0,1]})$$
$$((-1)_{[1]}(-1)_{[0]})\ ((-1)_{[1]}(-1)_{[1]})\ (H_{[0,1]}H_{[0,2]})\ ((-1)_{[3]}X_{[1,3]})\ (H_{[1,3]}H_{[0,1]})$$

$\overset{(20)(70)(71)}{\approx}$ $(H_{[0,1]}H_{[0,1]})\ ((-1)_{[0]}X_{[0,1]})\ (H_{[0,2]}H_{[0,1]})\ ((-1)_{[1]}X_{[1,3]})\ ((-1)_{[1]}X_{[0,1]})\ ((-1)_{[1]}(-1)_{[0]})\ (H_{[0,1]}H_{[0,2]})\ ((-1)_{[3]}X_{[1,3]})\ (H_{[1,3]}H_{[0,1]})$

$\overset{(71)(66)}{\approx}$ $(H_{[0,1]}H_{[4,5]})\ (H_{[0,1]}H_{[4,5]})\ ((-1)_{[0]}X_{[0,1]})\ (H_{[0,2]}H_{[1,3]})\ ((-1)_{[1]}X_{[1,3]})\ ((-1)_{[1]}X_{[0,1]})\ ((-1)_{[1]}(-1)_{[0]})\ (H_{[0,1]}H_{[0,2]})\ ((-1)_{[3]}X_{[1,3]})\ (H_{[1,3]}H_{[0,1]})$

$\overset{\text{Corollary A.10}}{\approx}$ $(H_{[0,1]}H_{[4,5]})\ ((-1)_{[1]}X_{[0,1]})\ (H_{[0,1]}H_{[4,5]})\ (H_{[0,2]}H_{[1,3]})\ ((-1)_{[1]}X_{[1,3]})\ ((-1)_{[1]}X_{[0,1]})\ ((-1)_{[1]}(-1)_{[0]})\ (H_{[0,1]}H_{[0,2]})\ ((-1)_{[3]}X_{[1,3]})\ (H_{[1,3]}H_{[0,1]})$

$\overset{(73)(66)}{\approx}$ $(H_{[0,1]}H_{[4,5]})\ ((-1)_{[1]}X_{[0,1]})\ (H_{[0,1]}H_{[0,2]})\ (H_{[1,3]}H_{[4,5]})\ ((-1)_{[1]}X_{[1,3]})\ ((-1)_{[1]}X_{[0,1]})\ ((-1)_{[1]}(-1)_{[0]})\ (H_{[0,1]}H_{[0,2]})\ ((-1)_{[3]}X_{[1,3]})\ (H_{[1,3]}H_{[0,1]})$

$\overset{\text{Corollary A.10}}{\approx}$ $(H_{[0,1]}H_{[4,5]})\ ((-1)_{[1]}X_{[0,1]})\ (H_{[0,1]}H_{[0,2]})\ ((-1)_{[3]}X_{[1,3]})\ (H_{[1,3]}H_{[4,5]})\ ((-1)_{[1]}X_{[0,1]})\ ((-1)_{[1]}(-1)_{[0]})\ (H_{[0,1]}H_{[0,2]})\ ((-1)_{[3]}X_{[1,3]})\ (H_{[1,3]}H_{[0,1]})$

$\overset{(71)}{\approx}$ $(H_{[0,1]}H_{[4,5]})\ ((-1)_{[1]}X_{[0,1]})\ (H_{[0,1]}H_{[0,2]})\ ((-1)_{[3]}X_{[1,3]})\ (H_{[1,3]}H_{[4,5]})\ ((-1)_{[1]}X_{[0,1]})\ ((-1)_{[1]}(-1)_{[0]})\ (H_{[0,1]}H_{[4,5]})\ (H_{[4,5]}H_{[0,2]})\ ((-1)_{[3]}X_{[1,3]})\ (H_{[1,3]}H_{[0,1]})$

$\overset{\text{Corollary A.10}}{\approx}$ $(H_{[0,1]}H_{[4,5]})\ ((-1)_{[1]}X_{[0,1]})\ (H_{[0,1]}H_{[0,2]})\ ((-1)_{[3]}X_{[1,3]})\ (H_{[1,3]}H_{[4,5]})\ (H_{[4,5]}H_{[0,1]})\ ((-1)_{[0]}X_{[0,1]})\ ((-1)_{[1]}(-1)_{[0]})\ (H_{[4,5]}H_{[0,2]})\ ((-1)_{[3]}X_{[1,3]})\ (H_{[1,3]}H_{[0,1]})$

$\overset{(71)}{\approx}$ $(H_{[0,1]}H_{[4,5]})\ ((-1)_{[1]}X_{[0,1]})\ (H_{[0,1]}H_{[0,2]})\ ((-1)_{[3]}X_{[1,3]})\ (H_{[1,3]}H_{[0,1]})\ ((-1)_{[0]}X_{[0,1]})\ ((-1)_{[1]}(-1)_{[0]})\ (H_{[4,5]}H_{[0,2]})\ ((-1)_{[3]}X_{[1,3]})\ (H_{[1,3]}H_{[0,1]})$

$\overset{\text{Corollary A.10}}{\approx}$ $(H_{[0,1]}H_{[4,5]})\ ((-1)_{[1]}X_{[0,1]})\ (H_{[0,1]}H_{[0,2]})\ ((-1)_{[3]}X_{[1,3]})\ (H_{[1,3]}H_{[0,1]})\ ((-1)_{[0]}X_{[0,1]})\ ((-1)_{[1]}(-1)_{[0]})\ ((-1)_{[3]}X_{[1,3]})\ (H_{[0,2]}H_{[4,5]})\ (H_{[1,3]}H_{[0,1]})$

$\overset{(73)}{\approx}$ $(H_{[0,1]}H_{[4,5]})\ ((-1)_{[1]}X_{[0,1]})\ (H_{[0,1]}H_{[0,2]})\ ((-1)_{[3]}X_{[1,3]})\ (H_{[1,3]}H_{[0,1]})\ ((-1)_{[0]}X_{[0,1]})\ ((-1)_{[1]}(-1)_{[0]})\ ((-1)_{[3]}X_{[1,3]})\ (H_{[0,2]}H_{[1,3]})\ (H_{[4,5]}H_{[0,1]})$

$\overset{\text{Corollary A.10}}{\approx}$ $(H_{[0,1]}H_{[4,5]})\ ((-1)_{[1]}X_{[0,1]})\ (H_{[0,1]}H_{[0,2]})\ ((-1)_{[3]}X_{[1,3]})\ (H_{[1,3]}H_{[0,1]})\ ((-1)_{[0]}X_{[0,1]})\ ((-1)_{[1]}(-1)_{[0]})\ (H_{[0,2]}H_{[1,3]})\ ((-1)_{[1]}X_{[1,3]})\ (H_{[4,5]}H_{[0,1]})$

$\overset{(20)(71)}{\approx}$ $(H_{[0,1]}H_{[4,5]})\ ((-1)_{[1]}X_{[0,1]})\ (H_{[0,1]}H_{[0,1]})\ (H_{[0,1]}H_{[0,2]})\ ((-1)_{[3]}X_{[1,3]})\ (H_{[1,3]}H_{[4,5]})\ (H_{[4,5]}H_{[0,1]})\ ((-1)_{[0]}X_{[0,1]})$
$((-1)_{[1]}(-1)_{[0]})\ ((-1)_{[1]}(-1)_{[1]})\ (H_{[0,2]}H_{[0,1]})\ ((-1)_{[3]}X_{[1,3]})\ (H_{[1,3]}H_{[0,1]})\ ((-1)_{[1]}X_{[0,1]})\ (H_{[4,5]}H_{[0,1]})$

$\overset{(Z\text{-d3*})}{\approx}$ $(H_{[0,1]}H_{[4,5]})\ (H_{[0,2]}H_{[0,1]})\ (H_{[0,1]}H_{[1,3]})\ ((-1)_{[1]}X_{[1,3]})\ ((-1)_{[1]}X_{[0,1]})\ (H_{[0,1]}H_{[0,1]})\ ((-1)_{[1]}(-1)_{[0]})\ ((-1)_{[1]}(-1)_{[1]})$
$(H_{[0,1]}H_{[0,2]})\ ((-1)_{[3]}X_{[1,3]})\ (H_{[1,3]}H_{[0,1]})\ (H_{[0,1]}H_{[0,1]})\ ((-1)_{[0]}X_{[0,1]})\ (H_{[4,5]}H_{[0,1]})$

$\overset{(20)(70)(71)}{\approx}$ $(H_{[0,1]}H_{[4,5]})\ (H_{[0,2]}H_{[1,3]})\ ((-1)_{[1]}X_{[1,3]})\ ((-1)_{[1]}X_{[0,1]})\ ((-1)_{[1]}(-1)_{[0]})\ (H_{[0,1]}H_{[0,2]})\ ((-1)_{[3]}X_{[1,3]})\ (H_{[1,3]}H_{[0,1]})\ ((-1)_{[0]}X_{[0,1]})\ (H_{[4,5]}H_{[0,1]})$

$\overset{\text{Corollary A.10}}{\approx}$ $(H_{[0,1]}H_{[4,5]})\ (H_{[0,2]}H_{[1,3]})\ ((-1)_{[1]}X_{[1,3]})\ ((-1)_{[1]}X_{[0,1]})\ ((-1)_{[1]}(-1)_{[0]})\ (H_{[0,1]}H_{[0,2]})\ (H_{[4,5]}H_{[1,3]})\ ((-1)_{[1]}X_{[1,3]})\ ((-1)_{[1]}X_{[0,1]})$

$\overset{(73)(66)}{\approx}$ $(H_{[0,1]}H_{[4,5]})\ (H_{[0,2]}H_{[1,3]})\ ((-1)_{[1]}X_{[1,3]})\ ((-1)_{[1]}X_{[0,1]})\ ((-1)_{[1]}(-1)_{[0]})\ (H_{[4,5]}H_{[0,1]})\ (H_{[0,2]}H_{[1,3]})\ ((-1)_{[1]}X_{[1,3]})\ ((-1)_{[1]}X_{[0,1]})$

$\overset{\text{Corollary A.10}}{\approx}$ $(H_{[0,1]}H_{[4,5]})\ (H_{[0,2]}H_{[1,3]})\ ((-1)_{[1]}X_{[1,3]})\ (H_{[4,5]}H_{[0,1]})\ ((-1)_{[0]}X_{[0,1]})\ ((-1)_{[1]}(-1)_{[0]})\ (H_{[0,2]}H_{[1,3]})\ ((-1)_{[1]}X_{[1,3]})\ ((-1)_{[1]}X_{[0,1]})$

$\overset{\text{Corollary A.10}}{\approx}$ $(H_{[0,1]}H_{[4,5]})\ ((-1)_{[3]}X_{[1,3]})\ (H_{[0,2]}H_{[1,3]})\ (H_{[4,5]}H_{[0,1]})\ ((-1)_{[0]}X_{[0,1]})\ ((-1)_{[1]}(-1)_{[0]})\ (H_{[0,2]}H_{[1,3]})\ ((-1)_{[1]}X_{[1,3]})\ ((-1)_{[1]}X_{[0,1]})$

$\overset{(73)(66)}{\approx}$ $(H_{[0,1]}H_{[4,5]})\ ((-1)_{[3]}X_{[1,3]})\ (H_{[4,5]}H_{[0,2]})\ (H_{[1,3]}H_{[0,1]})\ ((-1)_{[0]}X_{[0,1]})\ ((-1)_{[1]}(-1)_{[0]})\ (H_{[0,2]}H_{[1,3]})\ ((-1)_{[1]}X_{[1,3]})\ ((-1)_{[1]}X_{[0,1]})$

$\overset{\text{Corollary A.10}}{\approx}$ $(H_{[0,1]}H_{[0,2]})\ ((-1)_{[3]}X_{[1,3]})\ (H_{[1,3]}H_{[0,1]})\ ((-1)_{[0]}X_{[0,1]})\ ((-1)_{[1]}(-1)_{[0]})\ (H_{[0,2]}H_{[1,3]})\ ((-1)_{[1]}X_{[0,1]})\ ((-1)_{[1]}X_{[0,1]})$

$\overset{(20)(70)(71)}{\approx}$ $(H_{[0,2]}H_{[0,1]})\ ((-1)_{[3]}X_{[1,3]})\ (H_{[1,3]}H_{[0,1]})\ (H_{[0,1]}H_{[0,1]})\ ((-1)_{[0]}X_{[0,1]})\ ((-1)_{[1]}(-1)_{[0]})\ ((-1)_{[1]}(-1)_{[1]})$
$(H_{[0,2]}H_{[0,1]})\ (H_{[0,1]}H_{[1,3]})\ ((-1)_{[1]}X_{[0,1]})\ ((-1)_{[1]}X_{[0,1]})\ (H_{[0,1]}H_{[0,1]})$

Proof of Equation (ε-d4*):

$$(H_{[0,1]}H_{[0,1]})\ (H_{[0,1]}H_{[0,2]})\ (H_{[1,3]}H_{[0,1]})\ (H_{[0,1]}H_{[0,4]})\ (H_{[1,5]}H_{[0,1]})\ (H_{[0,1]}H_{[0,1]})$$
$$((-1)_{[1]}(-1)_{[0]})\ ((-1)_{[1]}(-1)_{[1]})\ (H_{[0,4]}H_{[0,1]})\ (H_{[0,1]}H_{[1,5]})\ (H_{[0,2]}H_{[0,1]})\ (H_{[0,1]}H_{[1,3]})$$

$\overset{(20)(71)}{\approx}$ $(H_{[0,1]}H_{[0,2]})\ (H_{[1,3]}H_{[0,4]})\ (H_{[1,5]}H_{[0,1]})\ ((-1)_{[1]}(-1)_{[0]})\ (H_{[0,4]}H_{[1,5]})\ (H_{[0,2]}H_{[1,3]})$

$\overset{(80)}{\approx}$ $(H_{[0,2]}H_{[1,3]})\ (H_{[0,4]}H_{[1,5]})\ ((-1)_{[1]}(-1)_{[0]})\ (H_{[0,1]}H_{[0,4]})\ (H_{[1,5]}H_{[0,2]})\ (H_{[1,3]}H_{[0,1]})$

$\overset{(20)(70)(71)}{\approx}$ $(H_{[0,2]}H_{[0,1]})\ (H_{[0,1]}H_{[1,3]})\ (H_{[0,4]}H_{[0,1]})\ (H_{[0,1]}H_{[1,5]})\ (H_{[0,1]}H_{[0,1]})\ ((-1)_{[1]}(-1)_{[0]})\ ((-1)_{[1]}(-1)_{[1]})$
$(H_{[0,1]}H_{[0,4]})\ (H_{[1,5]}H_{[0,1]})\ (H_{[0,1]}H_{[0,2]})\ (H_{[1,3]}H_{[0,1]})\ (H_{[0,1]}H_{[0,1]})$

Proof of Equation (Z-d4*):

$$((-1)_{[1]}X_{[0,1]}) \, (H_{[0,1]}H_{[0,1]}) \, (H_{[0,1]}H_{[0,2]}) \, ((-1)_{[3]}X_{[1,3]}) \, (H_{[1,3]}H_{[0,1]}) \, (H_{[0,1]}H_{[0,4]}) \, ((-1)_{[5]}X_{[1,5]}) \, (H_{[1,5]}H_{[0,1]}) \, (H_{[0,1]}H_{[0,1]}) \, ((-1)_{[0]}X_{[0,1]})$$
$$((-1)_{[1]}(-1)_{[0]}) \, ((-1)_{[1]}(-1)_{[1]}) \, (H_{[0,4]}H_{[1,5]}) \, (H_{[0,1]}H_{[1,5]}) \, ((-1)_{[1]}X_{[1,5]}) \, (H_{[0,2]}H_{[0,1]}) \, (H_{[0,1]}H_{[1,3]}) \, ((-1)_{[1]}X_{[1,3]})$$

$\overset{(20)(71)}{\approx}$ $\quad ((-1)_{[1]}X_{[0,1]}) \, (H_{[0,1]}H_{[0,2]}) \, ((-1)_{[3]}X_{[1,3]}) \, (H_{[1,3]}H_{[0,4]}) \, ((-1)_{[5]}X_{[1,5]}) \, (H_{[1,5]}H_{[0,1]}) \, ((-1)_{[0]}X_{[0,1]})$
$\quad\quad ((-1)_{[1]}(-1)_{[0]}) \, (H_{[0,4]}H_{[1,5]}) \, ((-1)_{[1]}X_{[1,5]}) \, (H_{[0,2]}H_{[1,3]}) \, ((-1)_{[1]}X_{[1,3]})$

$\overset{\text{Corollary A.10}}{\approx}$ $\quad ((-1)_{[1]}(-1)_{[6]}) \, ((-1)_{[6]}X_{[0,1]}) \, (H_{[0,1]}H_{[0,2]}) \, ((-1)_{[3]}X_{[1,3]}) \, (H_{[1,3]}H_{[0,4]}) \, ((-1)_{[5]}X_{[1,5]}) \, (H_{[1,5]}H_{[0,1]}) \, ((-1)_{[0]}X_{[0,1]})$
$\quad\quad ((-1)_{[1]}(-1)_{[0]}) \, (H_{[0,4]}H_{[1,5]}) \, ((-1)_{[1]}X_{[1,5]}) \, (H_{[0,2]}H_{[1,3]}) \, ((-1)_{[1]}X_{[1,3]})$

$\overset{\text{Corollary A.10}}{\approx}$ $\quad ((-1)_{[1]}(-1)_{[6]}) \, (H_{[0,1]}H_{[0,2]}) \, ((-1)_{[6]}(-1)_{[1]}) \, ((-1)_{[3]}X_{[1,3]}) \, (H_{[1,3]}H_{[0,4]}) \, ((-1)_{[5]}X_{[1,5]}) \, (H_{[1,5]}H_{[0,1]}) \, ((-1)_{[0]}X_{[0,1]})$
$\quad\quad ((-1)_{[1]}(-1)_{[0]}) \, (H_{[0,4]}H_{[1,5]}) \, ((-1)_{[1]}X_{[1,5]}) \, (H_{[0,2]}H_{[1,3]}) \, ((-1)_{[1]}X_{[1,3]})$

$\overset{\text{Corollary A.10}}{\approx}$ $\quad ((-1)_{[1]}(-1)_{[6]}) \, (H_{[0,1]}H_{[0,2]}) \, (H_{[1,3]}H_{[0,4]}) \, ((-1)_{[6]}(-1)_{[1]}) \, ((-1)_{[5]}X_{[1,5]}) \, (H_{[1,5]}H_{[0,1]}) \, ((-1)_{[0]}X_{[0,1]})$
$\quad\quad ((-1)_{[1]}(-1)_{[0]}) \, (H_{[0,4]}H_{[1,5]}) \, ((-1)_{[1]}X_{[1,5]}) \, (H_{[0,2]}H_{[1,3]}) \, ((-1)_{[1]}X_{[1,3]})$

$\overset{\text{Corollary A.10}}{\approx}$ $\quad ((-1)_{[1]}(-1)_{[6]}) \, (H_{[0,1]}H_{[0,2]}) \, (H_{[1,3]}H_{[0,4]}) \, (H_{[1,5]}H_{[0,1]}) \, ((-1)_{[6]}(-1)_{[1]})$
$\quad\quad ((-1)_{[1]}(-1)_{[0]}) \, (H_{[0,4]}H_{[1,5]}) \, ((-1)_{[1]}X_{[1,5]}) \, (H_{[0,2]}H_{[1,3]}) \, ((-1)_{[1]}X_{[1,3]})$

$\overset{\text{Corollary A.10}}{\approx}$ $\quad ((-1)_{[1]}(-1)_{[6]}) \, (H_{[0,1]}H_{[0,2]}) \, (H_{[1,3]}H_{[0,4]}) \, (H_{[1,5]}H_{[0,1]})$
$\quad\quad ((-1)_{[1]}(-1)_{[0]}) \, (H_{[0,4]}H_{[1,5]}) \, ((-1)_{[1]}X_{[1,5]}) \, (H_{[0,2]}H_{[1,3]}) \, ((-1)_{[1]}X_{[1,3]})$

$\overset{\text{Corollary A.10}}{\approx}$ $\quad ((-1)_{[1]}(-1)_{[6]}) \, (H_{[0,1]}H_{[0,2]}) \, (H_{[1,3]}H_{[0,4]}) \, (H_{[1,5]}H_{[0,1]})$
$\quad\quad ((-1)_{[1]}(-1)_{[0]}) \, (H_{[0,4]}H_{[1,5]}) \, (H_{[0,2]}H_{[1,3]}) \, ((-1)_{[6]}(-1)_{[1]})$

$\overset{(80)}{\approx}$ $\quad ((-1)_{[1]}(-1)_{[6]}) \, (H_{[0,2]}H_{[1,3]}) \, (H_{[0,4]}H_{[1,5]}) \, ((-1)_{[1]}(-1)_{[0]})$
$\quad\quad (H_{[0,1]}H_{[0,4]}) \, (H_{[1,5]}H_{[0,2]}) \, (H_{[1,3]}H_{[0,1]}) \, ((-1)_{[6]}(-1)_{[1]})$

$\overset{\text{Corollary A.10}}{\approx}$ $\quad (H_{[0,2]}H_{[1,3]}) \, ((-1)_{[1]}X_{[1,3]}) \, ((-1)_{[1]}(-1)_{[6]}) \, (H_{[0,4]}H_{[1,5]}) \, ((-1)_{[1]}(-1)_{[0]})$
$\quad\quad (H_{[0,1]}H_{[0,4]}) \, (H_{[1,5]}H_{[0,2]}) \, (H_{[1,3]}H_{[0,1]}) \, ((-1)_{[6]}(-1)_{[1]})$

$\overset{\text{Corollary A.10}}{\approx}$ $\quad (H_{[0,2]}H_{[1,3]}) \, ((-1)_{[1]}X_{[1,3]}) \, (H_{[0,4]}H_{[1,5]}) \, ((-1)_{[1]}X_{[1,5]}) \, ((-1)_{[1]}(-1)_{[6]}) \, ((-1)_{[1]}(-1)_{[0]})$
$\quad\quad (H_{[0,1]}H_{[0,4]}) \, (H_{[1,5]}H_{[0,2]}) \, (H_{[1,3]}H_{[0,1]}) \, ((-1)_{[6]}(-1)_{[1]})$

$\overset{\text{Corollary A.10}}{\approx}$ $\quad (H_{[0,2]}H_{[1,3]}) \, ((-1)_{[1]}X_{[1,3]}) \, (H_{[0,4]}H_{[1,5]}) \, ((-1)_{[1]}X_{[1,5]}) \, ((-1)_{[1]}X_{[0,1]}) \, ((-1)_{[1]}(-1)_{[0]})$
$\quad\quad (H_{[0,1]}H_{[0,4]}) \, ((-1)_{[1]}(-1)_{[6]}) \, (H_{[1,5]}H_{[0,2]}) \, (H_{[1,3]}H_{[0,1]}) \, ((-1)_{[6]}(-1)_{[1]})$

$\overset{\text{Corollary A.10}}{\approx}$ $\quad (H_{[0,2]}H_{[1,3]}) \, ((-1)_{[1]}X_{[1,3]}) \, (H_{[0,4]}H_{[1,5]}) \, ((-1)_{[1]}X_{[1,5]}) \, ((-1)_{[1]}X_{[0,1]}) \, ((-1)_{[1]}(-1)_{[0]})$
$\quad\quad (H_{[0,1]}H_{[0,4]}) \, ((-1)_{[5]}X_{[1,5]}) \, (H_{[1,5]}H_{[0,2]}) \, ((-1)_{[1]}(-1)_{[6]}) \, (H_{[1,3]}H_{[0,1]}) \, ((-1)_{[6]}(-1)_{[1]})$

$\overset{\text{Corollary A.10}}{\approx}$ $\quad (H_{[0,2]}H_{[1,3]}) \, ((-1)_{[1]}X_{[1,3]}) \, (H_{[0,4]}H_{[1,5]}) \, ((-1)_{[1]}X_{[1,5]}) \, ((-1)_{[1]}X_{[0,1]}) \, ((-1)_{[1]}(-1)_{[0]})$
$\quad\quad (H_{[0,1]}H_{[0,4]}) \, ((-1)_{[5]}X_{[1,5]}) \, (H_{[1,5]}H_{[0,2]}) \, ((-1)_{[3]}X_{[1,3]}) \, (H_{[1,3]}H_{[0,1]}) \, ((-1)_{[0]}X_{[0,1]})$

$\overset{(20)(70)(71)}{\approx}$ $\quad (H_{[0,2]}H_{[0,1]}) \, (H_{[0,1]}H_{[1,3]}) \, ((-1)_{[1]}X_{[1,3]}) \, (H_{[0,4]}H_{[0,1]}) \, (H_{[0,1]}H_{[1,5]}) \, ((-1)_{[1]}X_{[1,5]}) \, ((-1)_{[1]}X_{[0,1]}) \, (H_{[0,1]}H_{[0,1]}) \, ((-1)_{[1]}(-1)_{[0]}) \, ((-1)_{[1]}(-1)_{[1]})$
$\quad\quad (H_{[0,1]}H_{[0,4]}) \, ((-1)_{[5]}X_{[1,5]}) \, (H_{[1,5]}H_{[0,1]}) \, (H_{[0,1]}H_{[0,2]}) \, ((-1)_{[3]}X_{[1,3]}) \, (H_{[1,3]}H_{[0,1]}) \, (H_{[0,1]}H_{[0,1]}) \, ((-1)_{[0]}X_{[0,1]})$

Proof of Equation (H-d4*):

$$(H_{[0,1]}H_{[0,1]}) \, (H_{[0,2]}H_{[0,1]}) \, (H_{[0,1]}H_{[1,3]}) \, (H_{[0,4]}H_{[0,1]}) \, (H_{[0,1]}H_{[1,5]}) \, (H_{[0,1]}H_{[0,1]})$$
$$((-1)_{[1]}(-1)_{[0]}) \, ((-1)_{[1]}(-1)_{[1]}) \, (H_{[0,1]}H_{[0,4]}) \, (H_{[1,5]}H_{[0,1]}) \, (H_{[0,1]}H_{[0,2]}) \, (H_{[1,3]}H_{[0,1]})$$

$$\overset{(20)(70)(71)}{\approx} \quad \left(H_{[0,2]}H_{[1,3]}\right)\left(H_{[0,4]}H_{[1,5]}\right)\left((-1)_{[1]}(-1)_{[0]}\right)\left(H_{[0,1]}H_{[0,4]}\right)\left(H_{[1,5]}H_{[0,2]}\right)\left(H_{[1,3]}H_{[0,1]}\right)$$

$$\overset{(80)}{\approx} \quad \left(H_{[0,1]}H_{[0,2]}\right)\left(H_{[1,3]}H_{[0,4]}\right)\left(H_{[1,5]}H_{[0,1]}\right)\left((-1)_{[1]}(-1)_{[0]}\right)\left(H_{[0,4]}H_{[1,5]}\right)\left(H_{[0,2]}H_{[1,3]}\right)$$

$$\overset{(20)(70)(71)}{\approx} \quad \left(H_{[0,1]}H_{[0,2]}\right)\left(H_{[1,3]}H_{[0,1]}\right)\left(H_{[0,1]}H_{[0,4]}\right)\left(H_{[1,5]}H_{[0,1]}\right)\left(H_{[0,1]}H_{[0,1]}\right)\left((-1)_{[1]}(-1)_{[0]}\right)\left((-1)_{[1]}(-1)_{[1]}\right)$$
$$\left(H_{[0,4]}H_{[0,1]}\right)\left(H_{[0,1]}H_{[1,5]}\right)\left(H_{[0,2]}H_{[0,1]}\right)\left(H_{[0,1]}H_{[1,3]}\right)\left(H_{[0,1]}H_{[0,1]}\right)$$

**Proof of Equation (HZ-d4\*):**

$$\left(H_{[0,1]}H_{[0,1]}\right)\left((-1)_{[0]}X_{[0,1]}\right)\left(H_{[0,2]}H_{[0,1]}\right)\left(H_{[0,1]}H_{[1,3]}\right)\left((-1)_{[1]}X_{[1,3]}\right)\left(H_{[0,4]}H_{[0,1]}\right)\left(H_{[0,1]}H_{[1,5]}\right)\left((-1)_{[1]}X_{[1,5]}\right)\left((-1)_{[1]}X_{[0,1]}\right)\left(H_{[0,1]}H_{[0,1]}\right)$$
$$\left((-1)_{[1]}(-1)_{[0]}\right)\left((-1)_{[1]}(-1)_{[1]}\right)\left(H_{[0,1]}H_{[0,4]}\right)\left((-1)_{[5]}X_{[1,5]}\right)\left(H_{[1,5]}H_{[0,1]}\right)\left(H_{[0,1]}H_{[0,2]}\right)\left((-1)_{[3]}X_{[1,3]}\right)\left(H_{[1,3]}H_{[0,1]}\right)$$

$$\overset{(20)(71)}{\approx} \quad \left(H_{[0,1]}H_{[0,1]}\right)\left((-1)_{[0]}X_{[0,1]}\right)\left(H_{[0,2]}H_{[1,3]}\right)\left((-1)_{[1]}X_{[1,3]}\right)\left(H_{[0,4]}H_{[1,5]}\right)\left((-1)_{[1]}X_{[1,5]}\right)\left((-1)_{[1]}X_{[0,1]}\right)$$
$$\left((-1)_{[1]}(-1)_{[0]}\right)\left(H_{[0,1]}H_{[0,4]}\right)\left((-1)_{[5]}X_{[1,5]}\right)\left(H_{[1,5]}H_{[0,2]}\right)\left((-1)_{[3]}X_{[1,3]}\right)\left(H_{[1,3]}H_{[0,1]}\right)$$

$$\overset{(71)(66)}{\approx} \quad \left(H_{[0,1]}H_{[6,7]}\right)\left(H_{[0,1]}H_{[6,7]}\right)\left((-1)_{[0]}X_{[0,1]}\right)\left(H_{[0,2]}H_{[1,3]}\right)\left((-1)_{[1]}X_{[1,3]}\right)\left(H_{[0,4]}H_{[1,5]}\right)\left((-1)_{[1]}X_{[1,5]}\right)\left((-1)_{[1]}X_{[0,1]}\right)$$
$$\left((-1)_{[1]}(-1)_{[0]}\right)\left(H_{[0,1]}H_{[0,4]}\right)\left((-1)_{[5]}X_{[1,5]}\right)\left(H_{[1,5]}H_{[0,2]}\right)\left((-1)_{[3]}X_{[1,3]}\right)\left(H_{[1,3]}H_{[0,1]}\right)$$

$$\overset{\text{Corollary A.10}}{\approx} \quad \left(H_{[0,1]}H_{[6,7]}\right)\left((-1)_{[1]}X_{[0,1]}\right)\left(H_{[0,1]}H_{[6,7]}\right)\left(H_{[0,2]}H_{[1,3]}\right)\left((-1)_{[1]}X_{[1,3]}\right)\left(H_{[0,4]}H_{[1,5]}\right)\left((-1)_{[1]}X_{[1,5]}\right)\left((-1)_{[1]}X_{[0,1]}\right)$$
$$\left((-1)_{[1]}(-1)_{[0]}\right)\left(H_{[0,1]}H_{[0,4]}\right)\left((-1)_{[5]}X_{[1,5]}\right)\left(H_{[1,5]}H_{[0,2]}\right)\left((-1)_{[3]}X_{[1,3]}\right)\left(H_{[1,3]}H_{[0,1]}\right)$$

$$\overset{(73)(66)}{\approx} \quad \left(H_{[0,1]}H_{[6,7]}\right)\left((-1)_{[1]}X_{[0,1]}\right)\left(H_{[0,1]}H_{[0,2]}\right)\left(H_{[1,3]}H_{[6,7]}\right)\left((-1)_{[1]}X_{[1,3]}\right)\left(H_{[0,4]}H_{[1,5]}\right)\left((-1)_{[1]}X_{[1,5]}\right)\left((-1)_{[1]}X_{[0,1]}\right)$$
$$\left((-1)_{[1]}(-1)_{[0]}\right)\left(H_{[0,1]}H_{[0,4]}\right)\left((-1)_{[5]}X_{[1,5]}\right)\left(H_{[1,5]}H_{[0,2]}\right)\left((-1)_{[3]}X_{[1,3]}\right)\left(H_{[1,3]}H_{[0,1]}\right)$$

$$\overset{\text{Corollary A.10}}{\approx} \quad \left(H_{[0,1]}H_{[6,7]}\right)\left((-1)_{[1]}X_{[0,1]}\right)\left(H_{[0,1]}H_{[0,2]}\right)\left((-1)_{[3]}X_{[1,3]}\right)\left(H_{[1,3]}H_{[6,7]}\right)\left(H_{[0,4]}H_{[1,5]}\right)\left((-1)_{[1]}X_{[1,5]}\right)\left((-1)_{[1]}X_{[0,1]}\right)$$
$$\left((-1)_{[1]}(-1)_{[0]}\right)\left(H_{[0,1]}H_{[0,4]}\right)\left((-1)_{[5]}X_{[1,5]}\right)\left(H_{[1,5]}H_{[0,2]}\right)\left((-1)_{[3]}X_{[1,3]}\right)\left(H_{[1,3]}H_{[0,1]}\right)$$

$$\overset{(73)(66)}{\approx} \quad \left(H_{[0,1]}H_{[6,7]}\right)\left((-1)_{[1]}X_{[0,1]}\right)\left(H_{[0,1]}H_{[0,2]}\right)\left((-1)_{[3]}X_{[1,3]}\right)\left(H_{[1,3]}H_{[0,4]}\right)\left(H_{[1,5]}H_{[6,7]}\right)\left((-1)_{[1]}X_{[1,5]}\right)\left((-1)_{[1]}X_{[0,1]}\right)$$
$$\left((-1)_{[1]}(-1)_{[0]}\right)\left(H_{[0,1]}H_{[0,4]}\right)\left((-1)_{[5]}X_{[1,5]}\right)\left(H_{[1,5]}H_{[0,2]}\right)\left((-1)_{[3]}X_{[1,3]}\right)\left(H_{[1,3]}H_{[0,1]}\right)$$

$$\overset{\text{Corollary A.10}}{\approx} \quad \left(H_{[0,1]}H_{[6,7]}\right)\left((-1)_{[1]}X_{[0,1]}\right)\left(H_{[0,1]}H_{[0,2]}\right)\left((-1)_{[3]}X_{[1,3]}\right)\left(H_{[1,3]}H_{[0,4]}\right)\left((-1)_{[5]}X_{[1,5]}\right)\left(H_{[1,5]}H_{[6,7]}\right)\left((-1)_{[1]}X_{[0,1]}\right)$$
$$\left((-1)_{[1]}(-1)_{[0]}\right)\left(H_{[0,1]}H_{[0,4]}\right)\left((-1)_{[5]}X_{[1,5]}\right)\left(H_{[1,5]}H_{[0,2]}\right)\left((-1)_{[3]}X_{[1,3]}\right)\left(H_{[1,3]}H_{[0,1]}\right)$$

$$\overset{(71)}{\approx} \quad \left(H_{[0,1]}H_{[6,7]}\right)\left((-1)_{[1]}X_{[0,1]}\right)\left(H_{[0,1]}H_{[0,2]}\right)\left((-1)_{[3]}X_{[1,3]}\right)\left(H_{[1,3]}H_{[0,4]}\right)\left((-1)_{[5]}X_{[1,5]}\right)\left(H_{[1,5]}H_{[6,7]}\right)\left((-1)_{[1]}X_{[0,1]}\right)$$
$$\left((-1)_{[1]}(-1)_{[0]}\right)\left(H_{[0,1]}H_{[6,7]}\right)\left(H_{[6,7]}H_{[0,4]}\right)\left((-1)_{[5]}X_{[1,5]}\right)\left(H_{[1,5]}H_{[0,2]}\right)\left((-1)_{[3]}X_{[1,3]}\right)\left(H_{[1,3]}H_{[0,1]}\right)$$

$$\overset{\text{Corollary A.10}}{\approx} \quad \left(H_{[0,1]}H_{[6,7]}\right)\left((-1)_{[1]}X_{[0,1]}\right)\left(H_{[0,1]}H_{[0,2]}\right)\left((-1)_{[3]}X_{[1,3]}\right)\left(H_{[1,3]}H_{[0,4]}\right)\left((-1)_{[5]}X_{[1,5]}\right)\left(H_{[1,5]}H_{[0,1]}\right)\left((-1)_{[0]}X_{[0,1]}\right)$$
$$\left((-1)_{[1]}(-1)_{[0]}\right)\left(H_{[6,7]}H_{[0,4]}\right)\left((-1)_{[5]}X_{[1,5]}\right)\left(H_{[1,5]}H_{[0,2]}\right)\left((-1)_{[3]}X_{[1,3]}\right)\left(H_{[1,3]}H_{[0,1]}\right)$$

$$\overset{(71)}{\approx} \quad \left(H_{[0,1]}H_{[6,7]}\right)\left((-1)_{[1]}X_{[0,1]}\right)\left(H_{[0,1]}H_{[0,2]}\right)\left((-1)_{[3]}X_{[1,3]}\right)\left(H_{[1,3]}H_{[0,4]}\right)\left((-1)_{[5]}X_{[1,5]}\right)\left(H_{[1,5]}H_{[0,1]}\right)\left((-1)_{[0]}X_{[0,1]}\right)$$
$$\left((-1)_{[1]}(-1)_{[0]}\right)\left(H_{[6,7]}H_{[0,4]}\right)\left((-1)_{[5]}X_{[1,5]}\right)\left(H_{[1,5]}H_{[6,7]}\right)\left(H_{[6,7]}H_{[0,2]}\right)\left((-1)_{[3]}X_{[1,3]}\right)\left(H_{[1,3]}H_{[0,1]}\right)$$

$$\overset{\text{Corollary A.10}}{\approx} \quad \left(H_{[0,1]}H_{[6,7]}\right)\left((-1)_{[1]}X_{[0,1]}\right)\left(H_{[0,1]}H_{[0,2]}\right)\left((-1)_{[3]}X_{[1,3]}\right)\left(H_{[1,3]}H_{[0,4]}\right)\left((-1)_{[5]}X_{[1,5]}\right)\left(H_{[1,5]}H_{[0,1]}\right)\left((-1)_{[0]}X_{[0,1]}\right)$$
$$\left((-1)_{[1]}(-1)_{[0]}\right)\left(H_{[0,4]}H_{[1,5]}\right)\left((-1)_{[1]}X_{[1,5]}\right)\left(H_{[6,7]}H_{[0,2]}\right)\left((-1)_{[3]}X_{[1,3]}\right)\left(H_{[1,3]}H_{[0,1]}\right)$$

$$\overset{(71)}{\approx} \quad \left(H_{[0,1]}H_{[6,7]}\right)\left((-1)_{[1]}X_{[0,1]}\right)\left(H_{[0,1]}H_{[0,2]}\right)\left((-1)_{[3]}X_{[1,3]}\right)\left(H_{[1,3]}H_{[0,4]}\right)\left((-1)_{[5]}X_{[1,5]}\right)\left(H_{[1,5]}H_{[0,1]}\right)\left((-1)_{[0]}X_{[0,1]}\right)$$
$$\left((-1)_{[1]}(-1)_{[0]}\right)\left(H_{[0,4]}H_{[1,5]}\right)\left((-1)_{[1]}X_{[1,5]}\right)\left(H_{[6,7]}H_{[0,2]}\right)\left((-1)_{[3]}X_{[1,3]}\right)\left(H_{[1,3]}H_{[6,7]}\right)\left(H_{[6,7]}H_{[0,1]}\right)$$

$$\overset{\text{Corollary A.10}}{\approx} \quad \left(H_{[0,1]}H_{[6,7]}\right)\left((-1)_{[1]}X_{[0,1]}\right)\left(H_{[0,1]}H_{[0,2]}\right)\left((-1)_{[3]}X_{[1,3]}\right)\left(H_{[1,3]}H_{[0,4]}\right)\left((-1)_{[5]}X_{[1,5]}\right)\left(H_{[1,5]}H_{[0,1]}\right)\left((-1)_{[0]}X_{[0,1]}\right)$$
$$\left((-1)_{[1]}(-1)_{[0]}\right)\left(H_{[0,4]}H_{[1,5]}\right)\left((-1)_{[1]}X_{[1,5]}\right)\left(H_{[0,2]}H_{[1,3]}\right)\left((-1)_{[1]}X_{[1,3]}\right)\left(H_{[6,7]}H_{[0,1]}\right)$$

$\overset{(20)(70)(71)}{\approx}$ $(H_{[0,1]}H_{[6,7]})\;((-1)_{[1]}X_{[0,1]})\;(H_{[0,1]}H_{[0,1]})\;(H_{[0,1]}H_{[0,2]})\;((-1)_{[3]}X_{[1,3]})\;(H_{[1,3]}H_{[0,1]})\;(H_{[0,1]}H_{[0,4]})\;((-1)_{[5]}X_{[1,5]})\;(H_{[1,5]}H_{[0,1]})\;(H_{[0,1]}H_{[0,1]})\;((-1)_{[0]}X_{[0,1]})$
$((-1)_{[1]}(-1)_{[0]})\;((-1)_{[1]}(-1)_{[1]})\;(H_{[0,4]}H_{[0,1]})\;(H_{[0,1]}H_{[1,5]})\;((-1)_{[1]}X_{[1,5]})\;(H_{[0,2]}H_{[0,1]})\;(H_{[0,1]}H_{[1,3]})\;((-1)_{[1]}X_{[1,3]})\;(H_{[6,7]}H_{[0,1]})$

$\overset{(Z\text{-d4*})}{\approx}$ $(H_{[0,1]}H_{[6,7]})\;(H_{[0,2]}H_{[0,1]})\;(H_{[0,1]}H_{[1,3]})\;((-1)_{[1]}X_{[1,3]})\;(H_{[0,4]}H_{[0,1]})\;(H_{[0,1]}H_{[0,2]})\;((-1)_{[1]}X_{[1,5]})\;((-1)_{[1]}X_{[0,1]})\;(H_{[0,1]}H_{[0,1]})\;((-1)_{[1]}(-1)_{[0]})\;((-1)_{[1]}(-1)_{[1]})$
$(H_{[0,1]}H_{[0,4]})\;((-1)_{[5]}X_{[1,5]})\;(H_{[1,5]}H_{[0,1]})\;(H_{[0,1]}H_{[0,2]})\;((-1)_{[3]}X_{[1,3]})\;(H_{[1,3]}H_{[0,1]})\;(H_{[0,1]}H_{[0,1]})\;((-1)_{[0]}X_{[0,1]})\;(H_{[6,7]}H_{[0,1]})$

$\overset{(20)(70)(71)}{\approx}$ $(H_{[0,1]}H_{[6,7]})\;(H_{[0,2]}H_{[1,3]})\;((-1)_{[1]}X_{[1,3]})\;(H_{[0,4]}H_{[1,5]})\;((-1)_{[1]}X_{[1,5]})\;((-1)_{[1]}X_{[0,1]})\;((-1)_{[1]}(-1)_{[0]})$
$(H_{[0,1]}H_{[0,4]})\;((-1)_{[5]}X_{[1,5]})\;(H_{[1,5]}H_{[0,2]})\;((-1)_{[3]}X_{[1,3]})\;(H_{[1,3]}H_{[0,1]})\;((-1)_{[0]}X_{[0,1]})\;(H_{[6,7]}H_{[0,1]})$

$\overset{\text{Corollary A.10}}{\approx}$ $(H_{[0,1]}H_{[6,7]})\;(H_{[0,2]}H_{[1,3]})\;((-1)_{[1]}X_{[1,3]})\;(H_{[0,4]}H_{[1,5]})\;((-1)_{[1]}X_{[1,5]})\;((-1)_{[1]}X_{[0,1]})\;((-1)_{[1]}(-1)_{[0]})$
$(H_{[0,1]}H_{[0,4]})\;((-1)_{[5]}X_{[1,5]})\;(H_{[1,5]}H_{[0,2]})\;((-1)_{[3]}X_{[1,3]})\;(H_{[1,3]}H_{[0,1]})\;(H_{[6,7]}H_{[0,1]})\;((-1)_{[1]}X_{[0,1]})$

$\overset{\text{Corollary A.10}}{\approx}$ $(H_{[0,1]}H_{[6,7]})\;(H_{[0,2]}H_{[1,3]})\;((-1)_{[1]}X_{[1,3]})\;(H_{[0,4]}H_{[1,5]})\;((-1)_{[1]}X_{[1,5]})\;((-1)_{[1]}X_{[0,1]})\;((-1)_{[1]}(-1)_{[0]})$
$(H_{[0,1]}H_{[0,4]})\;((-1)_{[5]}X_{[1,5]})\;(H_{[1,5]}H_{[0,2]})\;(H_{[6,7]}H_{[1,3]})\;((-1)_{[1]}X_{[1,3]})\;((-1)_{[1]}X_{[0,1]})$

$\overset{(73)(66)}{\approx}$ $(H_{[0,1]}H_{[6,7]})\;(H_{[0,2]}H_{[1,3]})\;((-1)_{[1]}X_{[1,3]})\;(H_{[0,4]}H_{[1,5]})\;((-1)_{[1]}X_{[1,5]})\;((-1)_{[1]}X_{[0,1]})\;((-1)_{[1]}(-1)_{[0]})$
$(H_{[0,1]}H_{[0,4]})\;((-1)_{[5]}X_{[1,5]})\;(H_{[6,7]}H_{[1,5]})\;(H_{[0,2]}H_{[1,3]})\;((-1)_{[1]}X_{[1,3]})\;((-1)_{[1]}X_{[0,1]})$

$\overset{\text{Corollary A.10}}{\approx}$ $(H_{[0,1]}H_{[6,7]})\;(H_{[0,2]}H_{[1,3]})\;((-1)_{[1]}X_{[1,3]})\;(H_{[0,4]}H_{[1,5]})\;((-1)_{[1]}X_{[1,5]})\;((-1)_{[1]}X_{[0,1]})\;((-1)_{[1]}(-1)_{[0]})$
$(H_{[0,1]}H_{[0,4]})\;(H_{[6,7]}H_{[1,5]})\;((-1)_{[1]}X_{[1,5]})\;(H_{[0,2]}H_{[1,3]})\;((-1)_{[1]}X_{[1,3]})\;((-1)_{[1]}X_{[0,1]})$

$\overset{(73)(66)}{\approx}$ $(H_{[0,1]}H_{[6,7]})\;(H_{[0,2]}H_{[1,3]})\;((-1)_{[1]}X_{[1,3]})\;(H_{[0,4]}H_{[1,5]})\;((-1)_{[1]}X_{[1,5]})\;((-1)_{[1]}X_{[0,1]})\;((-1)_{[1]}(-1)_{[0]})$
$(H_{[6,7]}H_{[0,1]})\;(H_{[0,4]}H_{[1,5]})\;((-1)_{[1]}X_{[1,5]})\;(H_{[0,2]}H_{[1,3]})\;((-1)_{[1]}X_{[1,3]})\;((-1)_{[1]}X_{[0,1]})$

$\overset{\text{Corollary A.10}}{\approx}$ $(H_{[0,1]}H_{[6,7]})\;(H_{[0,2]}H_{[1,3]})\;((-1)_{[1]}X_{[1,3]})\;(H_{[0,4]}H_{[1,5]})\;((-1)_{[1]}X_{[1,5]})\;(H_{[6,7]}H_{[0,1]})\;((-1)_{[0]}X_{[0,1]})\;((-1)_{[1]}(-1)_{[0]})$
$(H_{[0,4]}H_{[1,5]})\;((-1)_{[1]}X_{[1,5]})\;(H_{[0,2]}H_{[1,3]})\;((-1)_{[1]}X_{[1,3]})\;((-1)_{[1]}X_{[0,1]})$

$\overset{(73)(66)}{\approx}$ $(H_{[0,1]}H_{[0,2]})\;(H_{[1,3]}H_{[6,7]})\;((-1)_{[1]}X_{[1,3]})\;(H_{[0,4]}H_{[1,5]})\;((-1)_{[1]}X_{[1,5]})\;(H_{[6,7]}H_{[0,1]})\;((-1)_{[0]}X_{[0,1]})\;((-1)_{[1]}(-1)_{[0]})$
$(H_{[0,4]}H_{[1,5]})\;((-1)_{[1]}X_{[1,5]})\;(H_{[0,2]}H_{[1,3]})\;((-1)_{[1]}X_{[1,3]})\;((-1)_{[1]}X_{[0,1]})$

$\overset{\text{Corollary A.10}}{\approx}$ $(H_{[0,1]}H_{[0,2]})\;((-1)_{[3]}X_{[1,3]})\;(H_{[1,3]}H_{[6,7]})\;(H_{[0,4]}H_{[1,5]})\;((-1)_{[1]}X_{[1,5]})\;(H_{[6,7]}H_{[0,1]})\;((-1)_{[0]}X_{[0,1]})\;((-1)_{[1]}(-1)_{[0]})$
$(H_{[0,4]}H_{[1,5]})\;((-1)_{[1]}X_{[1,5]})\;(H_{[0,2]}H_{[1,3]})\;((-1)_{[1]}X_{[1,3]})\;((-1)_{[1]}X_{[0,1]})$

$\overset{(73)(66)}{\approx}$ $(H_{[0,1]}H_{[0,2]})\;((-1)_{[3]}X_{[1,3]})\;(H_{[1,3]}H_{[0,4]})\;(H_{[1,5]}H_{[6,7]})\;((-1)_{[1]}X_{[1,5]})\;(H_{[6,7]}H_{[0,1]})\;((-1)_{[0]}X_{[0,1]})\;((-1)_{[1]}(-1)_{[0]})$
$(H_{[0,4]}H_{[1,5]})\;((-1)_{[1]}X_{[1,5]})\;(H_{[0,2]}H_{[1,3]})\;((-1)_{[1]}X_{[1,3]})\;((-1)_{[1]}X_{[0,1]})$

$\overset{\text{Corollary A.10}}{\approx}$ $(H_{[0,1]}H_{[0,2]})\;((-1)_{[3]}X_{[1,3]})\;(H_{[1,3]}H_{[0,4]})\;((-1)_{[5]}X_{[1,5]})\;(H_{[1,5]}H_{[0,1]})\;((-1)_{[0]}X_{[0,1]})\;((-1)_{[1]}(-1)_{[0]})$
$(H_{[0,4]}H_{[1,5]})\;((-1)_{[1]}X_{[1,5]})\;(H_{[0,2]}H_{[1,3]})\;((-1)_{[1]}X_{[1,3]})\;((-1)_{[1]}X_{[0,1]})$

$\overset{(20)(70)(71)}{\approx}$ $(H_{[0,1]}H_{[0,2]})\;((-1)_{[3]}X_{[1,3]})\;(H_{[1,3]}H_{[0,1]})\;(H_{[0,1]}H_{[0,4]})\;((-1)_{[5]}X_{[1,5]})\;(H_{[1,5]}H_{[0,1]})\;(H_{[0,1]}H_{[0,1]})\;((-1)_{[0]}X_{[0,1]})\;((-1)_{[1]}(-1)_{[0]})\;((-1)_{[1]}(-1)_{[1]})$
$(H_{[0,4]}H_{[0,1]})\;(H_{[0,1]}H_{[1,5]})\;((-1)_{[1]}X_{[1,5]})\;(H_{[0,2]}H_{[0,1]})\;(H_{[0,1]}H_{[1,3]})\;((-1)_{[1]}X_{[1,3]})\;((-1)_{[1]}X_{[0,1]})\;(H_{[0,1]}H_{[0,1]})$

$\square$

# B Some auxiliary Equations and Useful Properties

## B.1 A Few Useful General Properties of Circuits

**Definition B.1.** *The mirrored version of a circuit $C$, denoted by $C^\dagger$, is inductively defined as:*

- *for $g \in \left\{ \cdots, \;\text{—}, \;\diagdown\!\!\diagup, \;\boxed{H}, \;\text{•—}, \;\underset{\bullet}{\text{•}}, \;\boxed{H} \right\}$, $g^\dagger = g$*
- *$(C_1 \circ C_2)^\dagger = C_2^\dagger \circ C_1^\dagger$*

- $(C_1 \otimes C_2)^\dagger = C_1^\dagger \otimes C_2^\dagger$.

**Remark 7.** *Note that strictly speaking, the above definition is done on raw circuits. To show that it remains valid for circuits equipped with the PROP axioms, one has to show that if two raw circuits are equivalent according to the congruence generated by the rules of Figure 1, then this is also the case of their mirrored versions. This is the case since all the rules of Figure 1 are symmetric, except Equation (f) whose mirrored version can be derived (by using in particular Equation (g)).*

**Lemma B.2.** *For any $n$-qubit circuit $C$, one has $\mathrm{QC} \vdash C^\dagger \circ C = C \circ C^\dagger = (\!\text{———}\!)^{\otimes n}$.*

PROOF. Since $(C^\dagger)^\dagger = C$, it suffices to prove that $\mathrm{QC} \vdash C^\dagger \circ C = (\!\text{———}\!)^{\otimes n}$. We proceed by structural induction on $C$.[16] For $\raisebox{-2pt}{\text{⌣⌢}}$, ——— and $\raisebox{-2pt}{\text{⌣⌢}}$, the result is a consequence of the topological rules. For $\raisebox{-2pt}{\fbox{\scriptsize H}}$, —•—, $\raisebox{-2pt}{\text{⊤}}$ and $\raisebox{-2pt}{\text{⊥}}$, the result is a consequence of Equations (1), (2), (4) and (5). And for the induction steps, we have

$$(C_1 \circ C_2)^\dagger \circ (C_1 \circ C_2) = C_2^\dagger \circ C_1^\dagger \circ C_1 \circ C_2 \stackrel{\substack{\text{induction}\\\text{hypothesis}}}{=} C_2^\dagger \circ C_2 \stackrel{\substack{\text{induction}\\\text{hypothesis}}}{=} (\!\text{———}\!)^{\otimes n}$$

and

$$(C_1 \otimes C_2)^\dagger \circ (C_1 \otimes C_2) = (C_1^\dagger \circ C_1) \otimes (C_2^\dagger \circ C_2) \stackrel{\substack{\text{induction}\\\text{hypothesis}}}{=} (\!\text{———}\!)^{\otimes n}. \qquad \square$$

**Lemma B.3.** *For any $n$-qubit circuits $C_1, C_2$, if $\mathrm{QC} \vdash C_1 = C_2$ then $\mathrm{QC} \vdash C_1^\dagger = C_2^\dagger$.*

PROOF.

$$C_1^\dagger \stackrel{\text{Lemma B.2}}{=} C_2^\dagger \circ C_2 \circ C_1^\dagger \stackrel{\text{hypothesis}}{=} C_1^\dagger \circ C_1 \circ C_2^\dagger \stackrel{\text{Lemma B.2}}{=} C_2^\dagger \qquad \square$$

**Lemma B.4.** *For any $n$-qubit circuits $C, C'$, if $\mathrm{QC} \vdash C \circ C' = (\!\text{———}\!)^{\otimes n}$, then $\mathrm{QC} \vdash C' = C^\dagger$ and $\mathrm{QC} \vdash C' \circ C = (\!\text{———}\!)^{\otimes n}$.*

PROOF. To prove the first assertion, we have

$$C' \stackrel{\text{Lemma B.2}}{=} C^\dagger \circ C \circ C' \stackrel{\text{hypothesis}}{=} C^\dagger$$

and the second assertion follows by Lemma B.2. $\qquad \square$

One can further generalize Lemma B.3 in order to take into account blockwise circuits, with the dagger of each block replaced by an arbitrary circuit provably equivalent to it:

**Lemma B.5.** *Given four lists of $n$-qubit circuits $C_1, ..., C_k, D_1, ..., D_\ell, C'_1, ..., C'_k$, and $D'_1, ..., D'_\ell$, if for every $i$, one has $\mathrm{QC} \vdash C_i \circ C'_i = (\!\text{———}\!)^{\otimes n}$, and for every $j$, one has $\mathrm{QC} \vdash D_j \circ D'_j = (\!\text{———}\!)^{\otimes n}$, then*

$$\mathrm{QC} \vdash C_1 \circ \cdots \circ C_k = D_1 \circ \cdots \circ D_\ell \quad \text{if and only if} \quad \mathrm{QC} \vdash C'_k \circ \cdots \circ C'_1 = D'_\ell \circ \cdots \circ D'_1.$$

PROOF. By Lemma B.4, we also have $\mathrm{QC} \vdash C'_i \circ C_i = (\!\text{———}\!)^{\otimes n}$ and $\mathrm{QC} \vdash D'_j \circ D_j = (\!\text{———}\!)^{\otimes n}$, hence by symmetry it suffices to prove one half of the "if and only if".

If $\mathrm{QC} \vdash C_1 \circ \cdots \circ C_k = D_1 \circ \cdots \circ D_\ell$, then

$$C'_k \circ \cdots \circ C'_1 \stackrel{\text{hypothesis}}{=} C'_k \circ \cdots \circ C'_1 \circ D_1 \circ \cdots \circ D_\ell \circ D'_\ell \circ \cdots \circ D'_1 \stackrel{\text{assumption}}{=} C'_k \circ \cdots \circ C'_1 \circ C_1 \circ \cdots \circ C_k \circ D'_\ell \circ \cdots \circ D'_1 \stackrel{\text{Lemma B.4}}{=} D'_\ell \circ \cdots \circ D'_1. \square$$

**Lemma B.6.** *Given any three $n$-qubit circuits $C, D, D'$ such that $\mathrm{QC} \vdash D \circ D' = (\!\text{———}\!)^{\otimes n}$, one has*

$$\mathrm{QC} \vdash C \circ D = D \circ C \quad \text{if and only if} \quad \mathrm{QC} \vdash C \circ D' = D' \circ C.$$

PROOF. By Lemma B.4, we also have $\mathrm{QC} \vdash D' \circ D = (\!\text{———}\!)^{\otimes n}$, hence by symmetry it suffices to prove one half of the "if and only if".

If $\mathrm{QC} \vdash C \circ D = D \circ C$, then

$$C \circ D' \stackrel{\text{Lemma B.4}}{=} D' \circ D \circ C \circ D' \stackrel{\text{assumption}}{=} D' \circ C \circ D \circ D' \stackrel{\text{hypothesis}}{=} D' \circ C. \qquad \square$$

## B.2 Auxiliary Equations on 1 and 2 Qubits

**Lemma B.7.** *The equations of Figure 12 are consequences of those of Figure 4.*

PROOF.

*Proof of Equation (81):*

---

[16] More precisely, on an arbitrary underlying raw circuit.

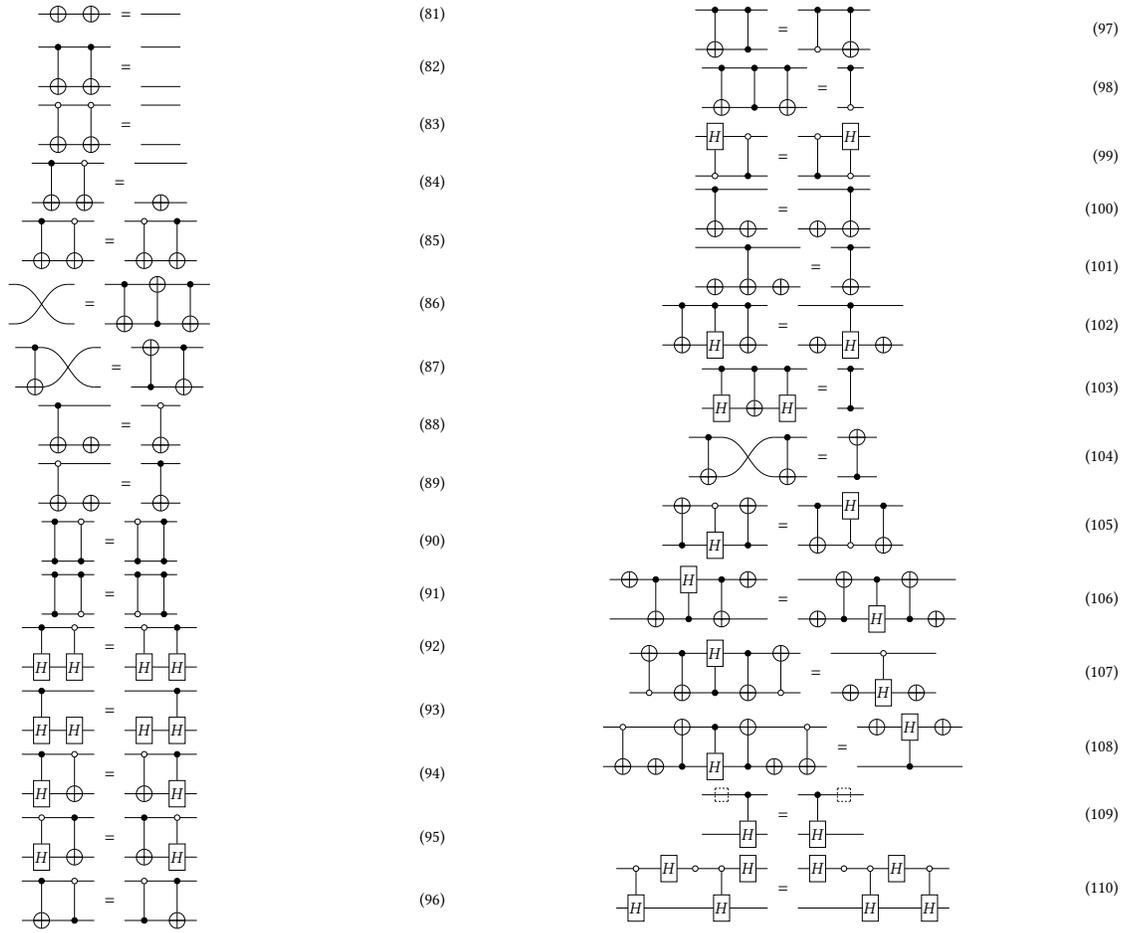

**Figure 12: Useful auxiliary equations.**

*Proof of Equation* (82):

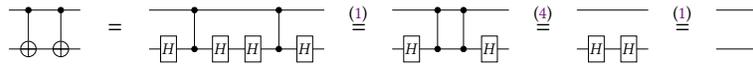

*Proof of Equation* (83):

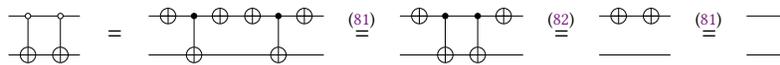

*Proof of Equation* (84):

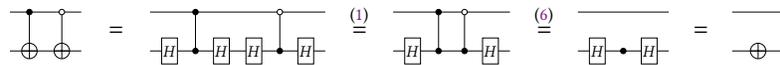

*Proof of Equation* (85):

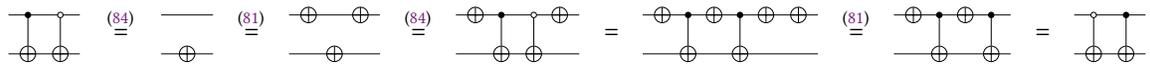

*Proof of Equation* (86):

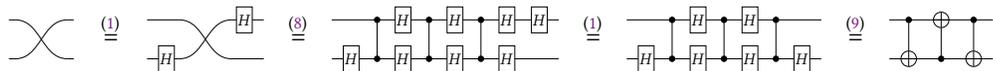

*Proof of Equation* (87):

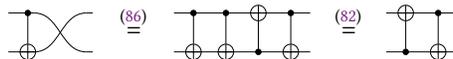

*Proof of Equation* (88):

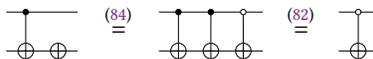

*Proof of Equation* (89):

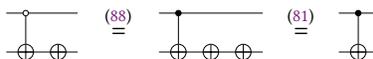

*Proof of Equation* (90):

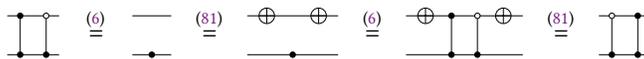

*Proof of Equation* (91):

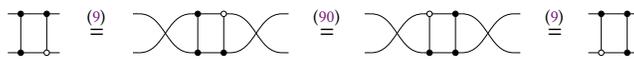

*Proof of Equation* (92):

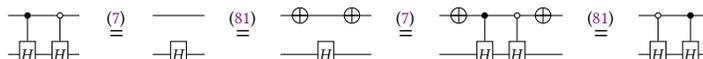

*Proof of Equation* (93):

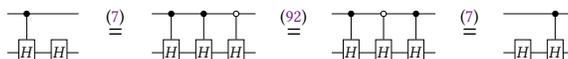

*Proof of Equation* (94):

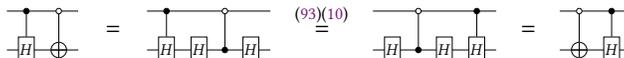

*Proof of Equation* (95):

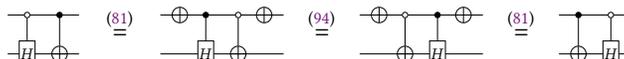

*Proof of Equation* (96):

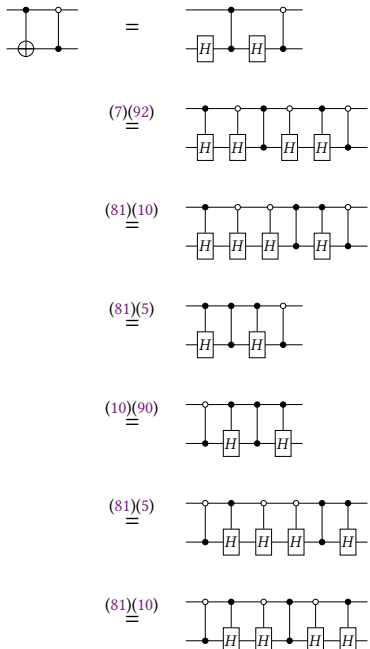

$$\overset{(92)(7)}{=}$$

$$=$$

*Proof of Equation* (97):

*Proof of Equation* (98):

*Proof of Equation* (99):

*Proof of Equation* (100):

*Proof of Equation* (101):

*Proof of Equation* (102):

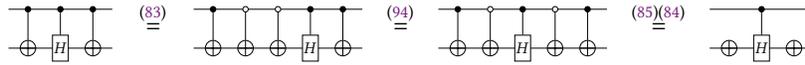

*Proof of Equation* (103):

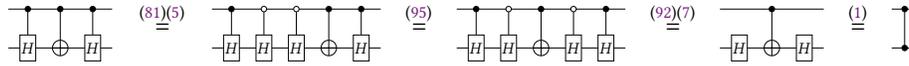

*Proof of Equation* (104):[17]

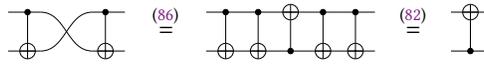

*Proof of Equation* (105):

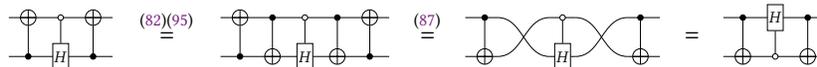

*Proof of Equation* (106):

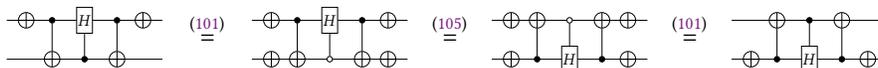

*Proof of Equation* (107):

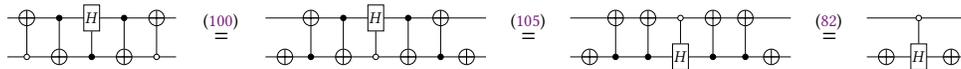

*Proof of Equation* (108):

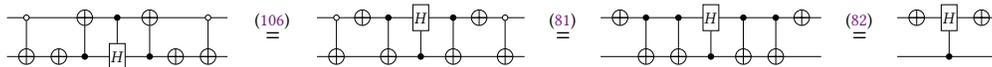

*Proof of Equation* (109):

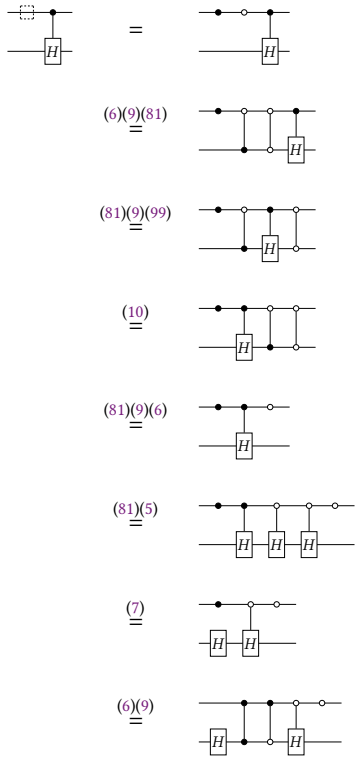

---

[17]Note that up to deformation, Equation (104) is just Equation (87) with the LHS and RHS exchanged and with an additional swap on the left, therefore we could also have proved it in one step by invoking Equation (87).

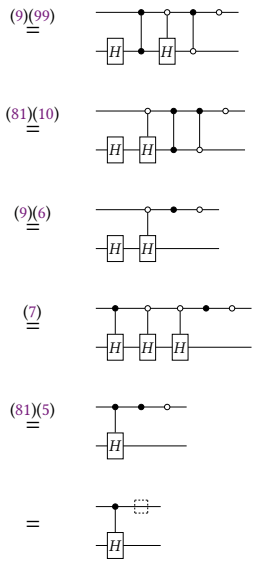

*Proof of Equation* (110):

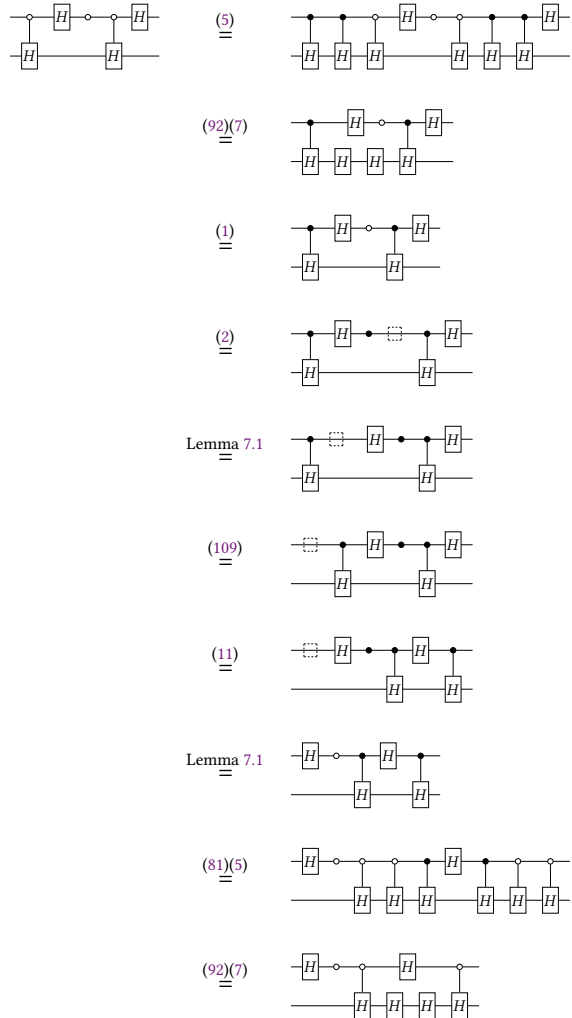

$$\overset{(1)}{=} \quad \boxed{\text{circuit diagram}}$$

□

## C  Two-Qubit Completeness

### C.1  Proof of Lemma 7.4

We are going to prove the slightly stronger result that for any $n$-qubit circuit $C \in \mathbf{QC}$, any $C' \in \mathbf{QC_{seq}}$ such that $|C'| = C$, and any $k, \ell$ such that $k + n + \ell = 2$,

$$\mathbf{QC} \vdash D(E_{k,\ell}(C')) = id_k \otimes C \otimes id_\ell.$$

More precisely, we are going to prove the equivalent statement that for any $n$-qubit raw circuit $C''$ such that $|||C''||| = C$, and any $k, \ell$, such that $k + n + \ell = 2$, one has

$$\mathbf{QC} \vdash D(E_{k,\ell}(||C''||)) = id_k \otimes C \otimes id_\ell$$

where $||C''||$ (resp. $|||C''|||$) denotes the image of $||C''||$ by the canonical projection onto $\mathbf{QC_{seq}}$ (resp. onto $\mathbf{QC}$). For simplicity, by abuse of notation we will omit the canonical projection $||\cdot||$ in the rest of the proof. We proceed by structural induction on $C''$.

For any two $n$-qubit raw circuits $C_1, C_2$ such that $k + n + \ell = 2$, one has

$$D(E_{k,\ell}(C_2 \circ C_1)) = D(E_{k,\ell}(C_2)) \circ D(E_{k,\ell}(C_1))$$

and for any $n$-qubit raw circuit $C_1$ and $m$-qubit raw circuit $C_3$ such that $k + n + m + \ell = 2$,

$$D(E_{k,\ell}(C_1 \otimes C_3)) = D(E_{k+n,\ell}(C_3)) \circ D(E_{k,\ell+m}(C_1)).$$

Hence, it remains the base cases:

$$D(E_{0,0}(\;\smile^\frown\;)) = D(X_{[1,2]}) = \;\smile^\frown\;,$$

$$D(E_{k,\ell}(\;\vdots\;)) = D(\epsilon) = \;\underline{\quad}\;,$$

$$D(E_{k,\ell}(\;\underline{\quad}\;)) = D(\epsilon) = \;\underline{\quad}\;,$$

$$D(E_{1,0}(-\boxed{H}-)) = D(H_{[0,1]}H_{[2,3]}) = \cdots \overset{(7)}{=} -\boxed{H}-,$$

$$D(E_{0,1}(-\boxed{H}-)) = D(H_{[0,2]}H_{[1,3]}) = \cdots \overset{(7)}{=} -\boxed{H}-,$$

$$D(E_{1,0}(-\bullet-)) = D((-1)_{[1]}(-1)_{[3]}) = \cdots \overset{(6)}{=} -\bullet-,$$

$$D(E_{0,1}(-\bullet-)) = D((-1)_{[2]}(-1)_{[3]}) = \cdots \overset{(9)(6)}{=} -\bullet-,$$

$$D(E_{0,0}(\;\begin{smallmatrix}\bullet\\\bullet\end{smallmatrix}\;)) = D((-1)_{[3]}) = \;\begin{smallmatrix}\bullet\\\bullet\end{smallmatrix}\;,$$

$$D(E_{0,0}(\;\begin{smallmatrix}H\\\bullet\end{smallmatrix}\;)) = D(H_{[2,3]}) = \boxed{H}.$$

### C.2  Proof of Lemma 7.5

By Proposition 4.8, it suffices to prove that for each rule of Figure 7, of the form $w_1 \approx w_2$ where $w_1, w_2 \in \mathcal{G}_4^*$, and for any two words $w_0, w_3 \in \mathcal{G}_4^*$, one has $\mathbf{QC} \vdash D(w_0 w_1 w_3) = D(w_0 w_2 w_3)$. Since $D$ is a monoid homomorphism, it suffices to prove that $\mathbf{QC} \vdash D(w_1) = D(w_2)$. Note that since we are only working in dimension 4, we do not need to treat Equation (d4*).

*Decoding Equation* (a1*):

$$D((-1)^2_{[0]}) \quad = \quad \boxed{\text{diagram}} \quad \overset{(81)(4)}{=} \quad \underline{\quad} \quad = \quad D(\epsilon)$$

*Decoding Equation* (a3): One has $D(H^2_{[j,k]}) = D(H_{[j,k]})^2$, and one can check that for any $j, k$, the circuit $D(H_{[j,k]})$ is symmetric (between left and right). Since according to Equations (1), (2), (4), (5) and (g), all generators are self-inverse, this implies that $D(H_{[j,k]})^2$ can be reduced to the identity.

*Decoding Equation* (a2): The argument is the same as for Equation (a3), after noting that $\begin{smallmatrix}\oplus\\\oplus\end{smallmatrix} = \begin{smallmatrix}\times\oplus\\\times\oplus\end{smallmatrix}$.

*Decoding Equation* (b1*):

$$D((-1)_{[0]}(-1)_{[1]}) \quad = \quad \boxed{\text{diagram}} \quad \overset{(81)(91)}{=} \quad \boxed{\text{diagram}} \quad = \quad D((-1)_{[1]}(-1)_{[0]})$$

*Decoding Equation* (b4\*):

$$D((-1)_{[1]} H_{[0,2]}) \quad = \quad \cdots \quad \overset{(99)}{=} \quad \cdots \quad = \quad D(H_{[0,2]} (-1)_{[1]})$$

*Decoding Equation* (b6\*):

$$D(H_{[0,1]} H_{[2,3]}) \quad = \quad \cdots \quad \overset{(92)}{=} \quad \cdots \quad = \quad D(H_{[2,3]} H_{[0,1]})$$

*Decoding Equation* (c1): If $(j,k) = (0,1)$, $(1,0)$, $(2,3)$, and $(3,2)$, then the result follows from Equation (97), together with Equation (81) and Lemma B.3. The cases $(j,k) = (0,2)$, $(2,0)$, $(1,3)$ and $(3,1)$ are similar but additionally use Equation (9).

The cases $(j,k) = (1,2)$ and $(2,1)$ follow from Equation (9) (together with the topological rules), and the cases $(j,k) = (0,3)$ and $(3,0)$ are similar but additionally use Equation (81).

*Decoding Equations* (c5) *and* (e2\*): In every case, we have to prove an equation of the form

$$D(X_{[j,k]}) \circ D(H_{[l,m]}) = D(H_{[l',m']}) \circ D(X_{[j,k]})$$

where either $m = j$ or $(l,m) = (j,k)$.

By Lemma B.2, this equation is equivalent to

$$D(X_{[j,k]}) \circ D(H_{[l,m]}) \circ D(X_{[j,k]})^{\dagger} = D(H_{[l',m']}).$$

Moreover, one can check that for any $j,k$, $D(X_{[j,k]})^{\dagger} = D(X_{[j,k]})$, so that this equation is equivalent to

$$D(X_{[j,k]}) \circ D(H_{[l,m]}) \circ D(X_{[j,k]}) = D(H_{[l',m']}).$$

By soundness, in order to derive this equation, it suffices to show that the left-hand side can be transformed into some $D(H_{[l',m']})$. Indeed, the indices $l', m'$ are uniquely determined by the semantics of $H_{[l',m']}$, and therefore by that of $D(H_{[l',m']})$.

We display in Table 1 the equations that we need to apply in order to transform $D(X_{[j,k]}) \circ D(H_{[l,m]}) \circ D(X_{[j,k]})$ into $D(H_{[l',m']})$ (the empty cells correspond to the cases that we do not need to address, namely those where we do not have $m = j$ or $(l,m) = (j,k)$ up to swapping $j$ and $k$ – note that since $D(X_{[k,j]}) = D(X_{[j,k]})$, we address both in the same column). In order to make it slightly easier to check the proof, we also depict the resulting $D(H_{[l',m']})$.

*Decoding Equation* (d2): We have to prove for every $j,k$ (with $j \neq k$) that the following equation is a consequence of those of Figure 4:

$$D(H_{[j,k]}) \circ D(X_{[j,k]}) = D((-1)_{[k]}) \circ D(H_{[j,k]}).$$

By Lemma B.2 (together with the fact that $D(H_{[j,k]})^{\dagger} = D(H_{[j,k]})$), this is equivalent to deriving the following equation:

$$D(H_{[j,k]}) \circ D(X_{[j,k]}) \circ D(H_{[j,k]}) = D((-1)_{[k]}).$$

For $(j,k) = (2,3)$, $(0,1)$, $(0,2)$, and $(1,3)$, one can easily check that it is a consequence of Equation (103), together with Equation (81). The cases $(j,k) = (3,2)$, $(1,0)$, $(2,0)$, and $(3,1)$ are similar, but with an additional use of Equation (101).

For $(j,k) = (2,1)$, $(1,2)$, $(0,3)$, and $(3,0)$, one has the following derivations:

$$D(H_{[2,1]}) \circ D(X_{[2,1]}) \circ D(H_{[2,1]}) \quad = \quad \cdots$$

$$\overset{(104)}{=} \quad \cdots$$

$$\overset{(103)}{=} \quad \cdots$$

$$\overset{(9)(98)}{=} \quad \cdots \quad = \quad D((-1)_{[1]})$$

$$D(H_{[1,2]}) \circ D(X_{[1,2]}) \circ D(H_{[1,2]}) \quad = \quad \cdots$$

$$\overset{(104)}{=} \quad \cdots$$

| $D(X_{[j,k]})$ → ／ $D(H_{[l,m]})$ ↓ | $D(X_{[2,3]})$ | $D(X_{[0,1]})$ | $D(X_{[0,2]})$ | $D(X_{[1,3]})$ | $D(X_{[1,2]})$ | $D(X_{[0,3]})$ |
|---|---|---|---|---|---|---|
| $D(H_{[2,3]})$ | (83)(94)(84) | | | none | | none |
| $D(H_{[0,1]})$ | | (82)(95)(84) | | none | none | |
| $D(H_{[0,2]})$ | (105) | | (82)(95)(84) | | none | |
| $D(H_{[1,3]})$ | none | | | (83)(94)(84) | | none |
| $D(H_{[3,2]})$ | (88)(94)(83) | | (81) | | none | |
| $D(H_{[1,0]})$ | | (89)(95)(82) | (81)(105)(101) | | | (81) |
| $D(H_{[2,0]})$ | | (81)(105)(101) | (89)(95)(82) | | | (81) |
| $D(H_{[3,1]})$ | | (81)(106) | | (88)(94)(83) | none | |
| $D(H_{[2,1]})$ | | (100)(105)(82) | | (82) | none | |
| $D(H_{[0,3]})$ | (105)(82) | | | (82) | | (101)(106) |
| $D(H_{[1,2]})$ | (82) | | (107) | | none | |
| $D(H_{[3,0]})$ | | (108) | (81)(82) | | | (81)(100)(105) |

**Table 1: Equations needed to transform $D(X_{[j,k]}) \circ D(H_{[l,m]}) \circ D(X_{[j,k]})$ into some $D(H_{[l',m']})$, together with the picture of the resulting $D(H_{[l',m']})$. We implicitly use Lemma B.3 when needed (as well as the topological rules and the shortcut definitions).**

$$\overset{(103)(9)}{=}$$

$$\overset{(98)}{=} \quad = \quad D((-1)_{[2]})$$

$$D(H_{[0,3]}) \circ D(X_{[0,3]}) \circ D(H_{[0,3]}) \quad = $$

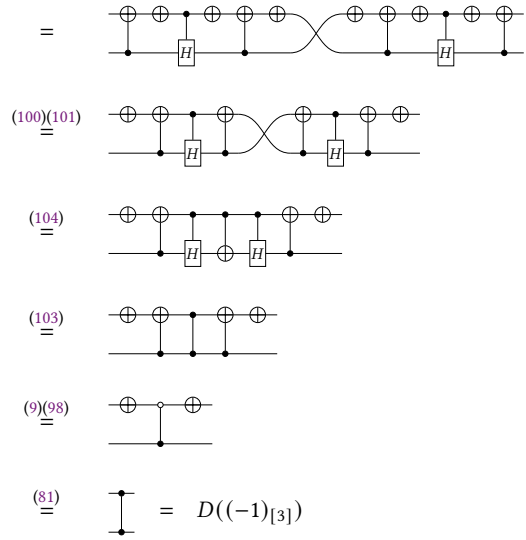

$$D(H_{[3,0]}) \circ D(X_{[3,0]}) \circ D(H_{[3,0]}) \quad =$$

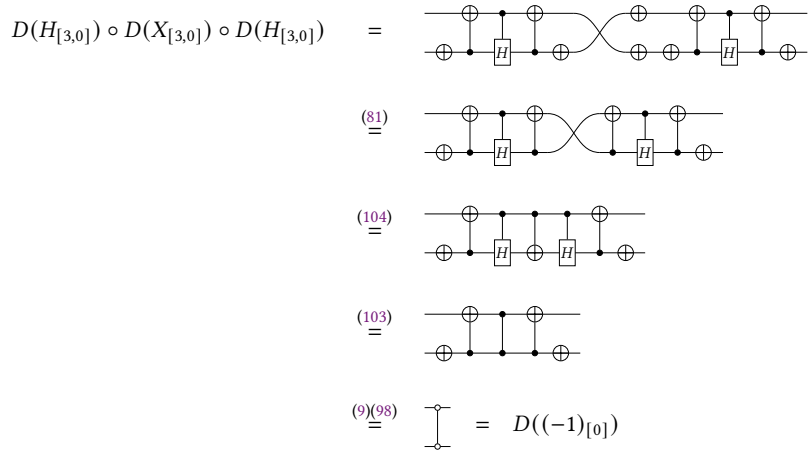

*Decoding Equation* (d3*):

$$D(H_{[0,1]}H_{[0,2]}H_{[1,3]}H_{[0,1]}(-1)_{[0]}(-1)_{[1]}H_{[0,2]}H_{[1,3]})$$

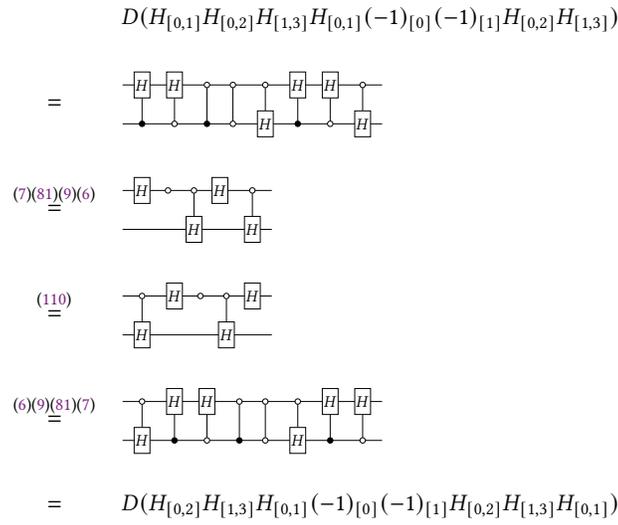

$$= \quad D(H_{[0,2]}H_{[1,3]}H_{[0,1]}(-1)_{[0]}(-1)_{[1]}H_{[0,2]}H_{[1,3]}H_{[0,1]})$$

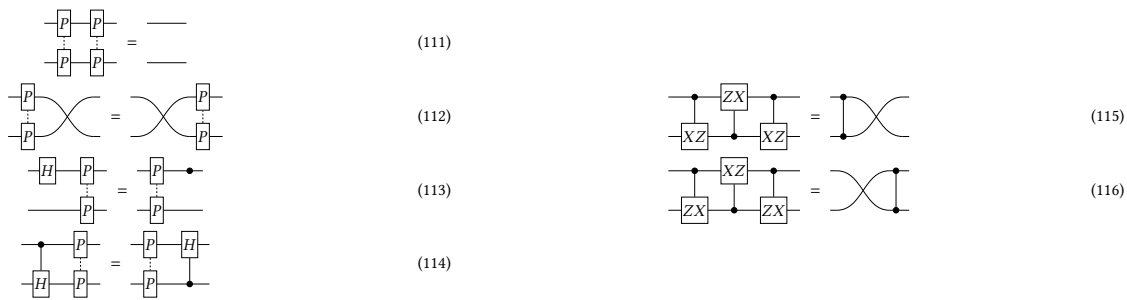

$$\text{(111)}$$

$$\text{(112)}$$

$$\text{(113)}$$

$$\text{(114)}$$

$$\text{(115)}$$

$$\text{(116)}$$

Figure 13: Other useful auxiliary equations on 2 qubits.

# D  Other Auxiliary Equations and Properties of Multi-Controlled gates

## D.1  Additional Auxiliary Equations on 2 Qubits

**Lemma D.1.** *The equations of Figure 13 are consequences of those of Figure 4.*

PROOF. This follows directly from Lemma 7.6.  □

## D.2  Auxiliary Equations on 3 Qubits

**Lemma D.2.** *The following equations are consequences of the equations of Figure 4:*

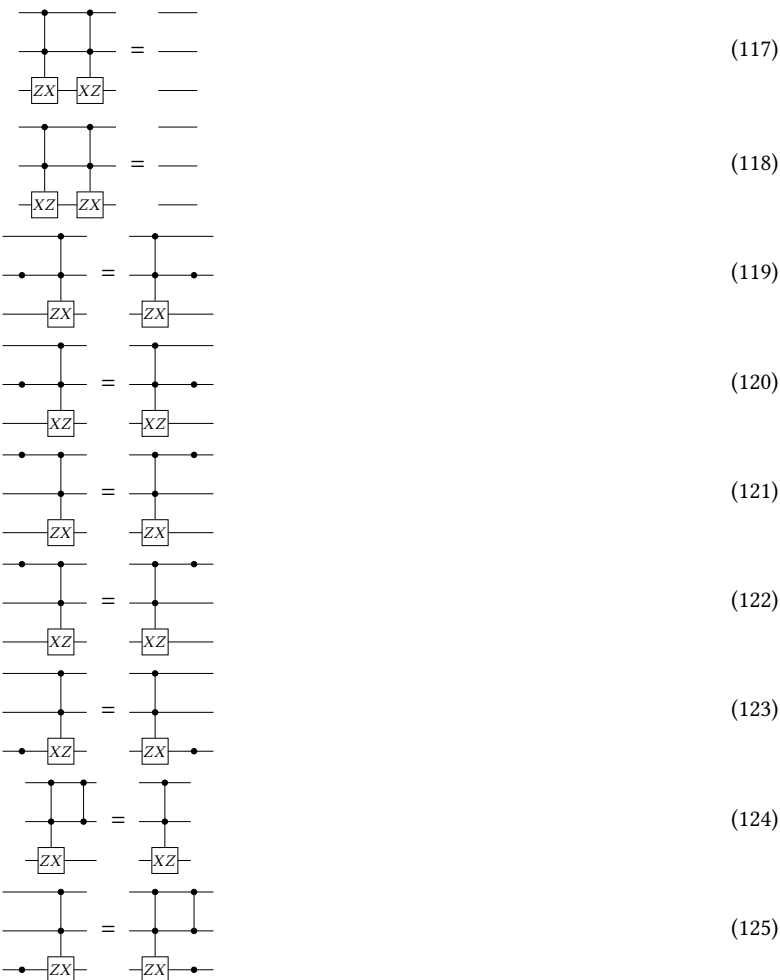

$$\text{(117)}$$

$$\text{(118)}$$

$$\text{(119)}$$

$$\text{(120)}$$

$$\text{(121)}$$

$$\text{(122)}$$

$$\text{(123)}$$

$$\text{(124)}$$

$$\text{(125)}$$

$$\tag{126}$$

$$\tag{127}$$

$$\tag{128}$$

$$\tag{129}$$

$$\tag{130}$$

$$\tag{131}$$

$$\tag{132}$$

$$\tag{133}$$

$$\tag{134}$$

$$\tag{135}$$

$$\tag{136}$$

$$\tag{137}$$

$$\tag{138}$$

$$\tag{139}$$

$$\tag{140}$$

$$\tag{141}$$

(142)

(143)

(144)

(145)

(146)

(147)

**PROOF.**

*Proof of Equation (117):* Equation (117) follows directly from Lemma B.2.

*Proof of Equation (118):* Equation (118) follows directly from Lemma B.2.

*Proof of Equation (119):*

*Proof of Equation (120):* Equation (120) follows directly from Lemma B.3 applied to Equation (119).

*Proof of Equation (121):*

*Proof of Equation (122):* Equation (122) follows directly from Lemma B.3 applied to Equation (121).

*Proof of Equation (123):*

*Proof of Equation (124):*

*Proof of Equation (125):*

*Proof of Equation (126):*

*Proof of Equation* (127):

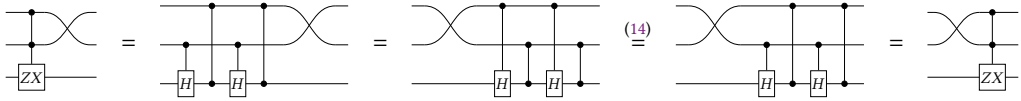

*Proof of Equation* (128):

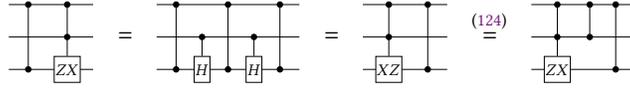

*Proof of Equation* (129):

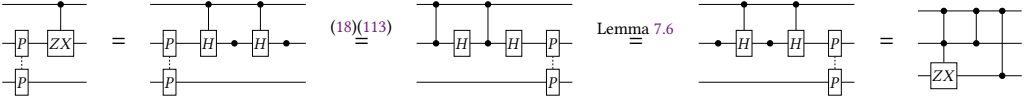

*Proof of Equation* (130):

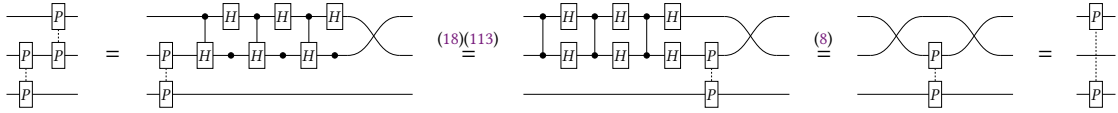

*Proof of Equation* (131):

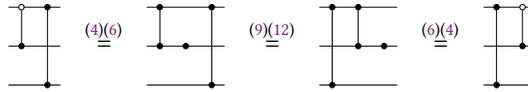

*Proof of Equation* (132):

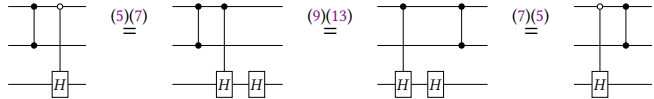

*Proof of Equation* (133):

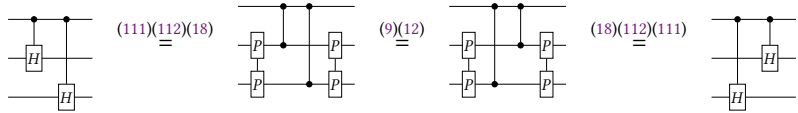

*Proof of Equation* (134):

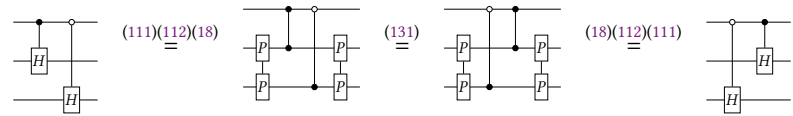

*Proof of Equation* (135):

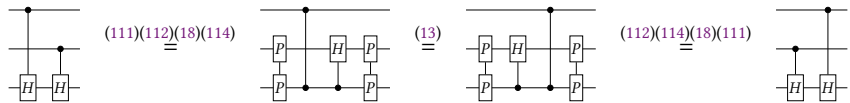

*Proof of Equation* (136):

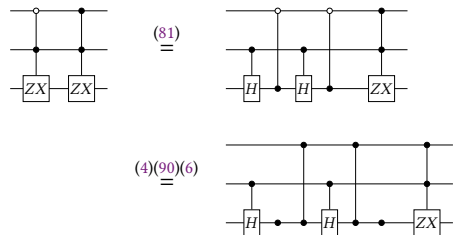

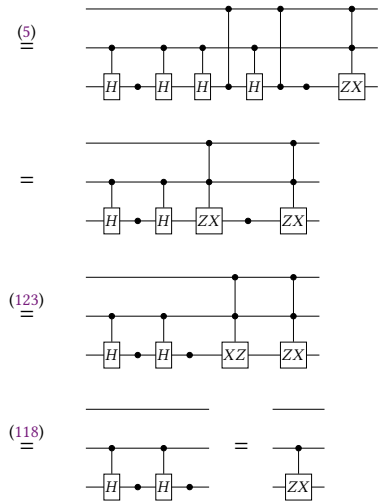

*Proof of Equation* (137): Equation (137) follows directly from Equation (136), together with Equation (81).

*Proof of Equation* (138): Equation (138) follows directly from Lemma B.3 applied to Equation (136).

*Proof of Equation* (139):

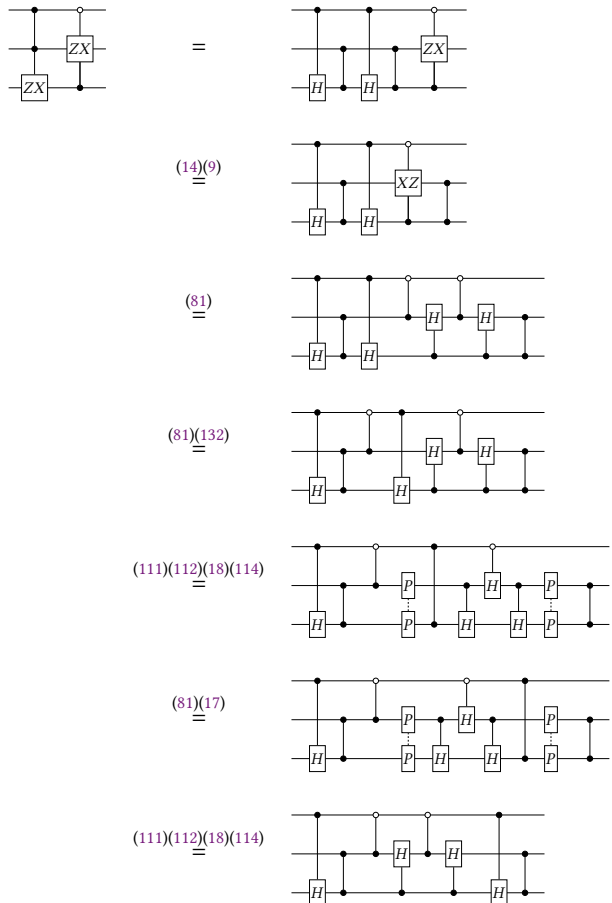

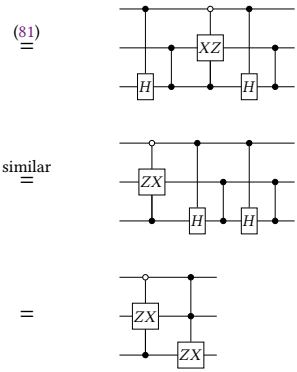

*Proof of Equation* (140):

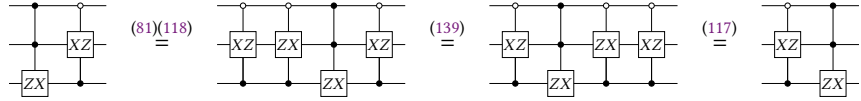

*Proof of Equation* (141):

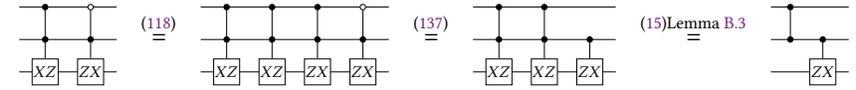

*Proof of Equation* (142):

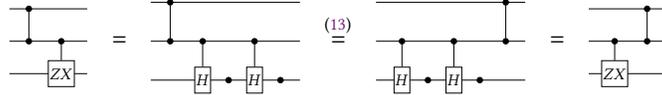

*Proof of Equation* (143):

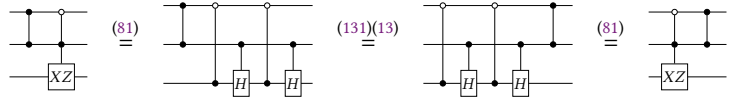

*Proof of Equation* (144):

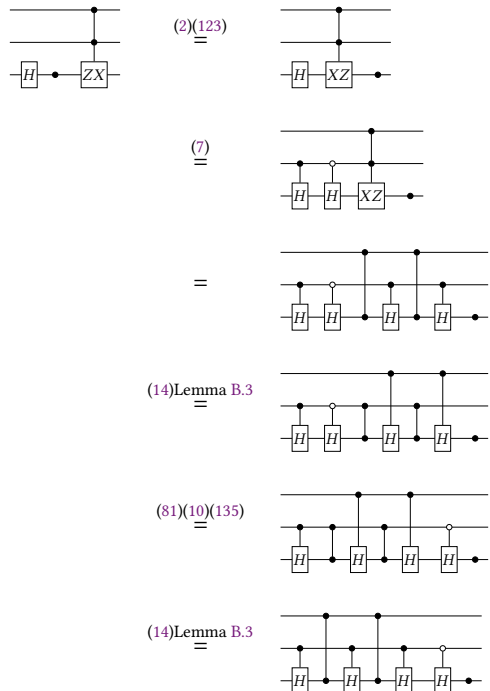

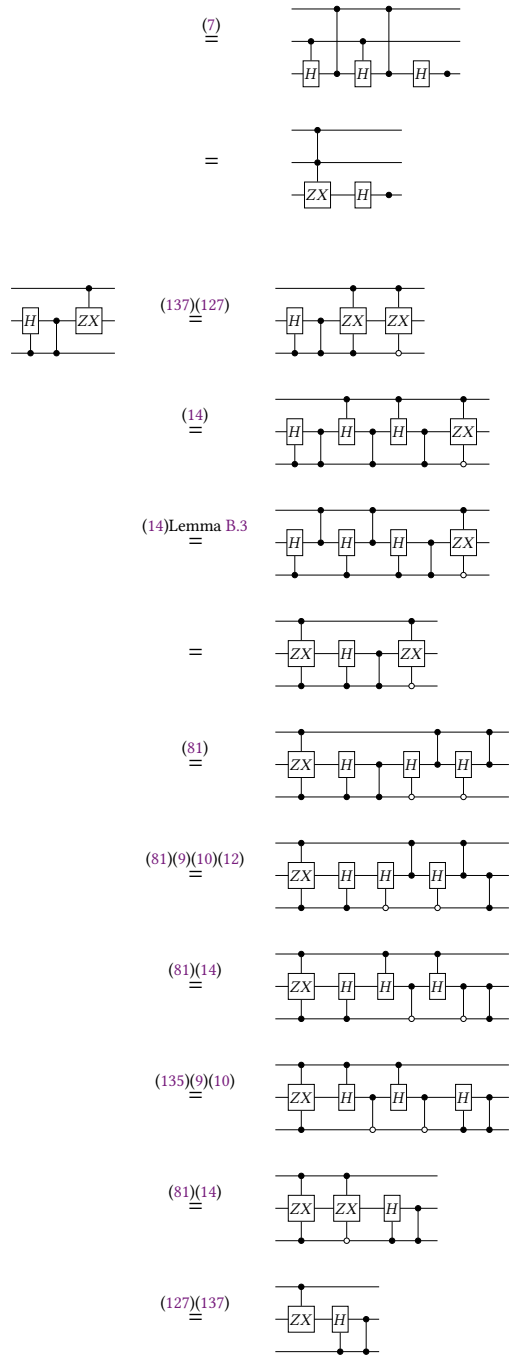

*Proof of Equation* (145):

*Proof of Equation* (146):

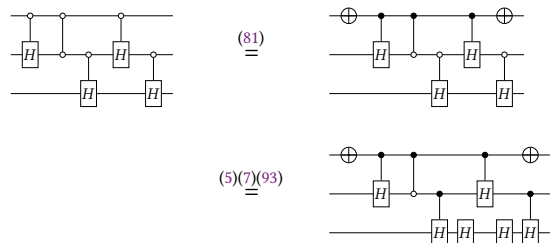

A Complete Equational Theory for Real-Clifford+CH Quantum Circuits

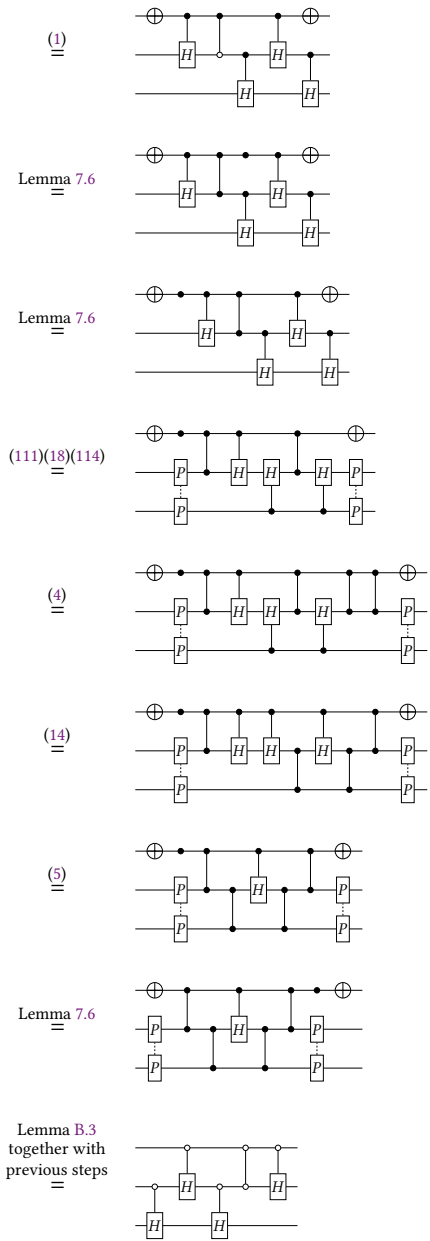

*Proof of Equation* (147):

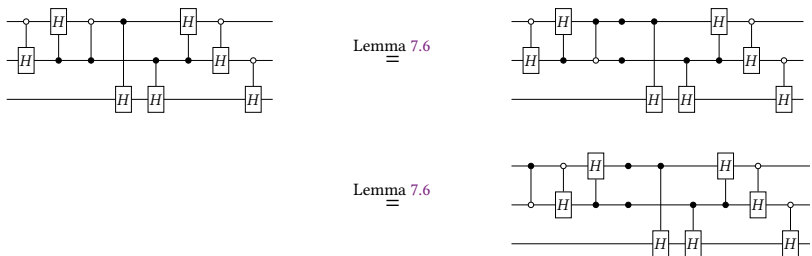

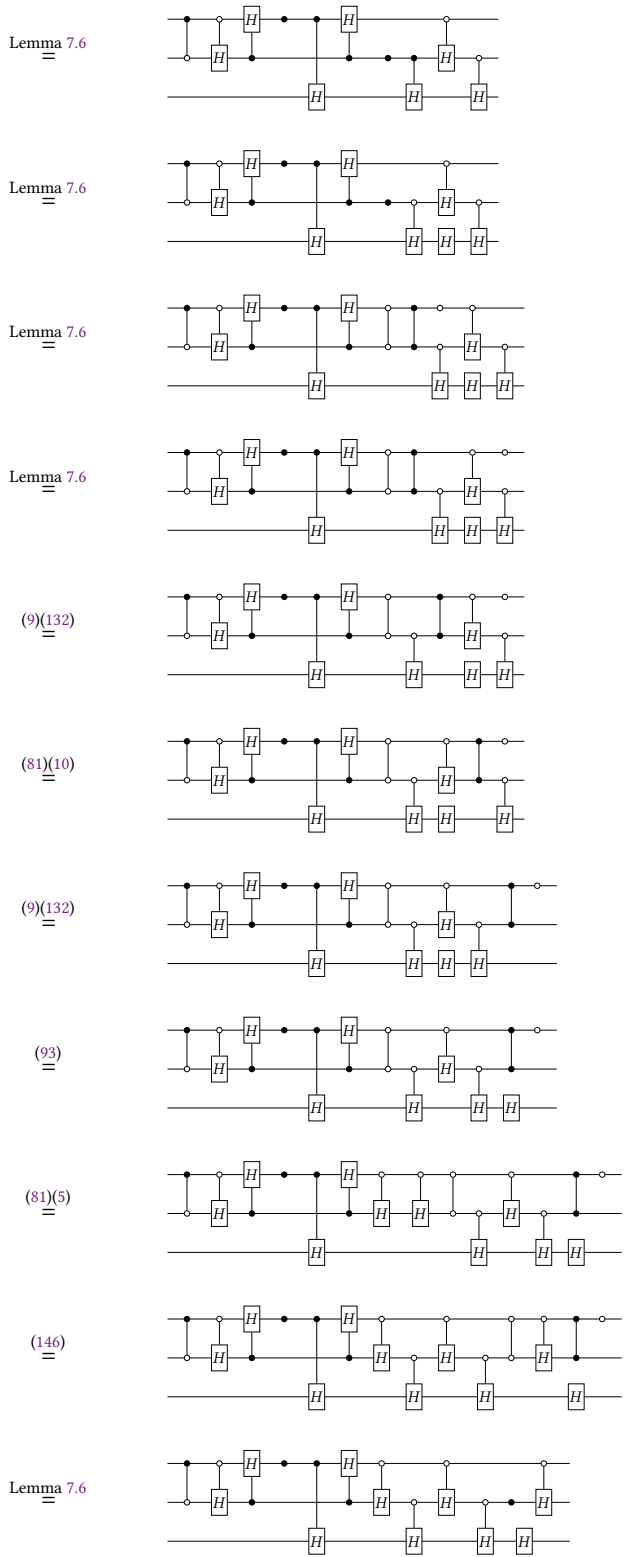

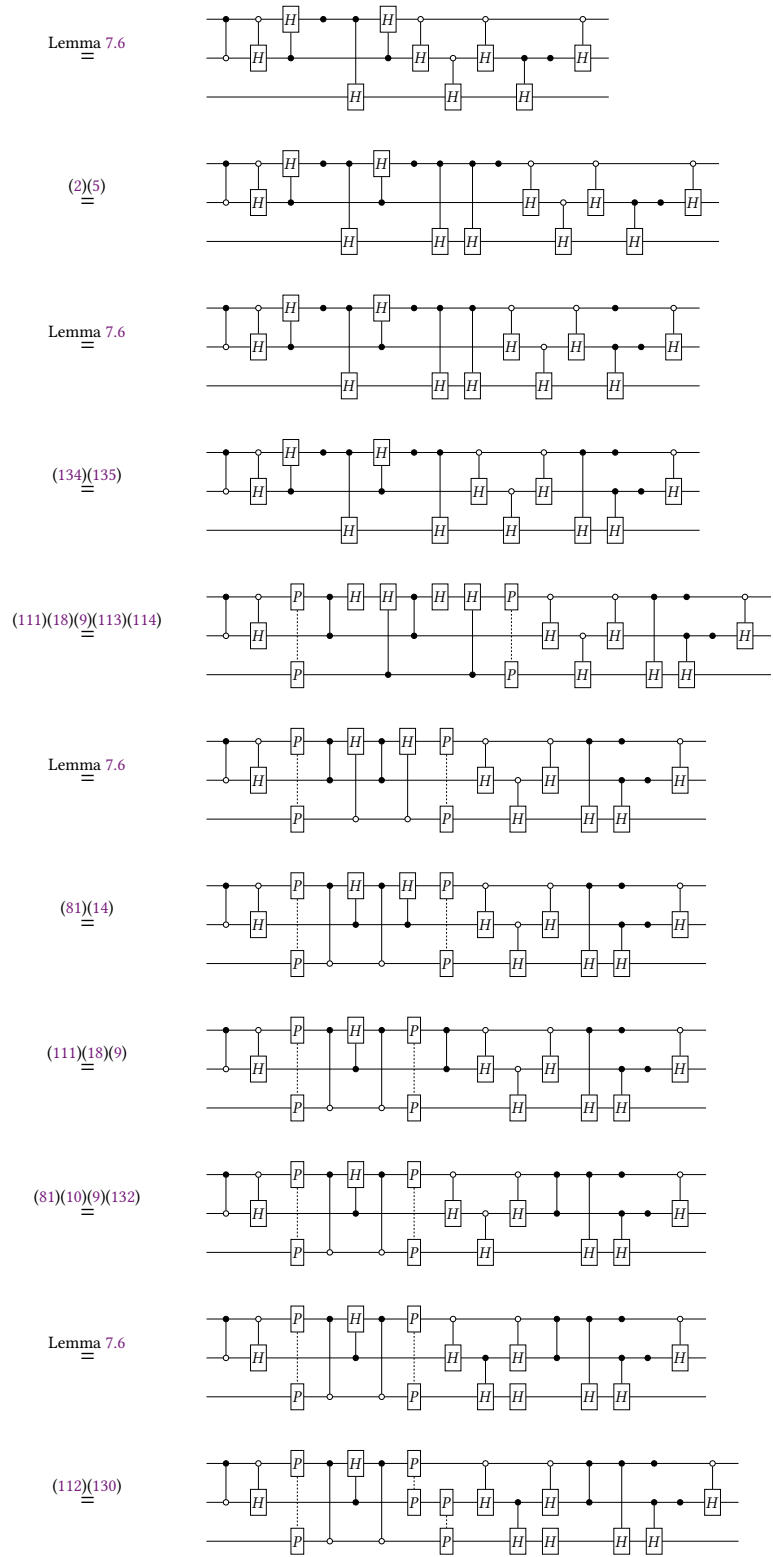

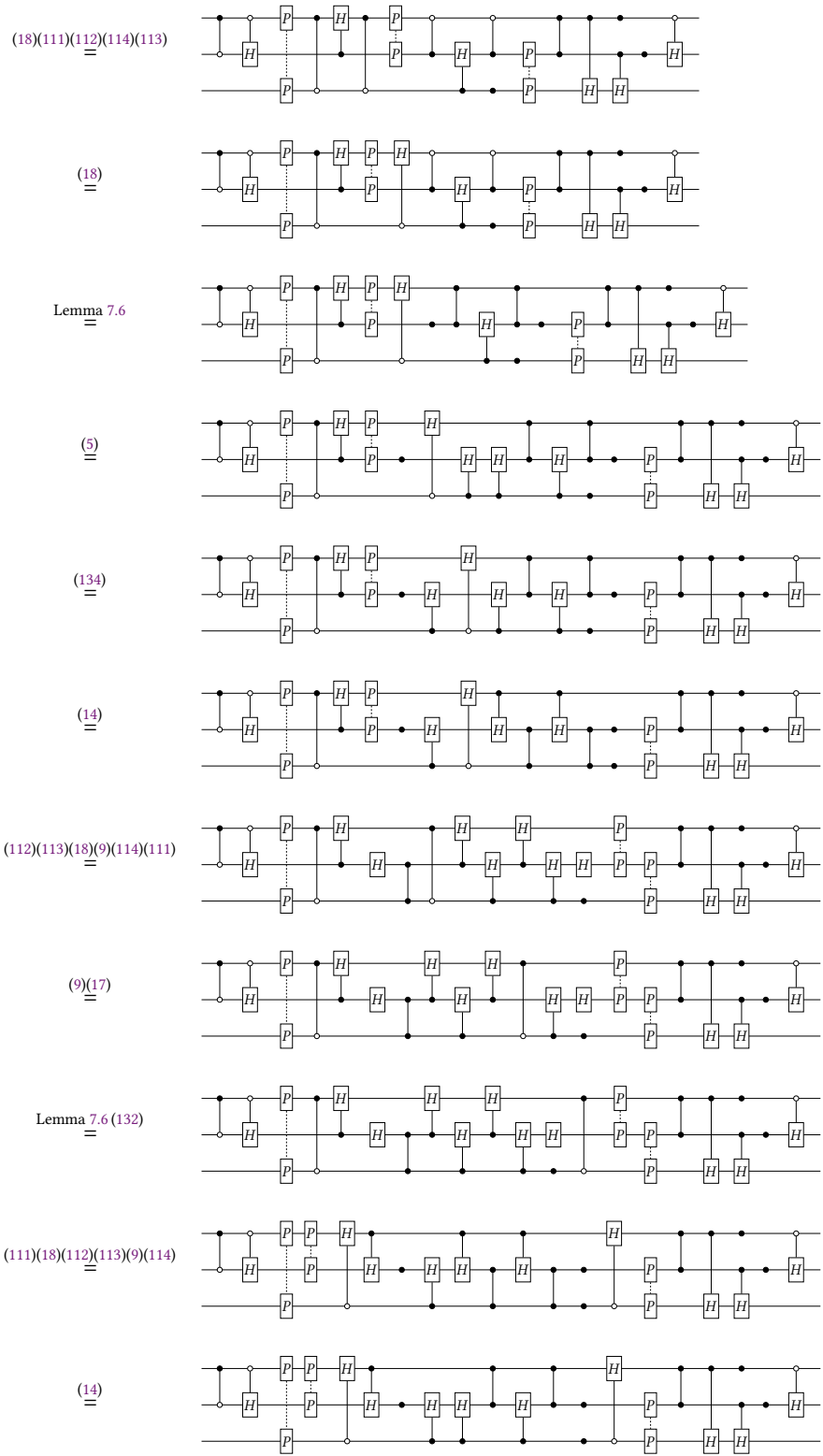

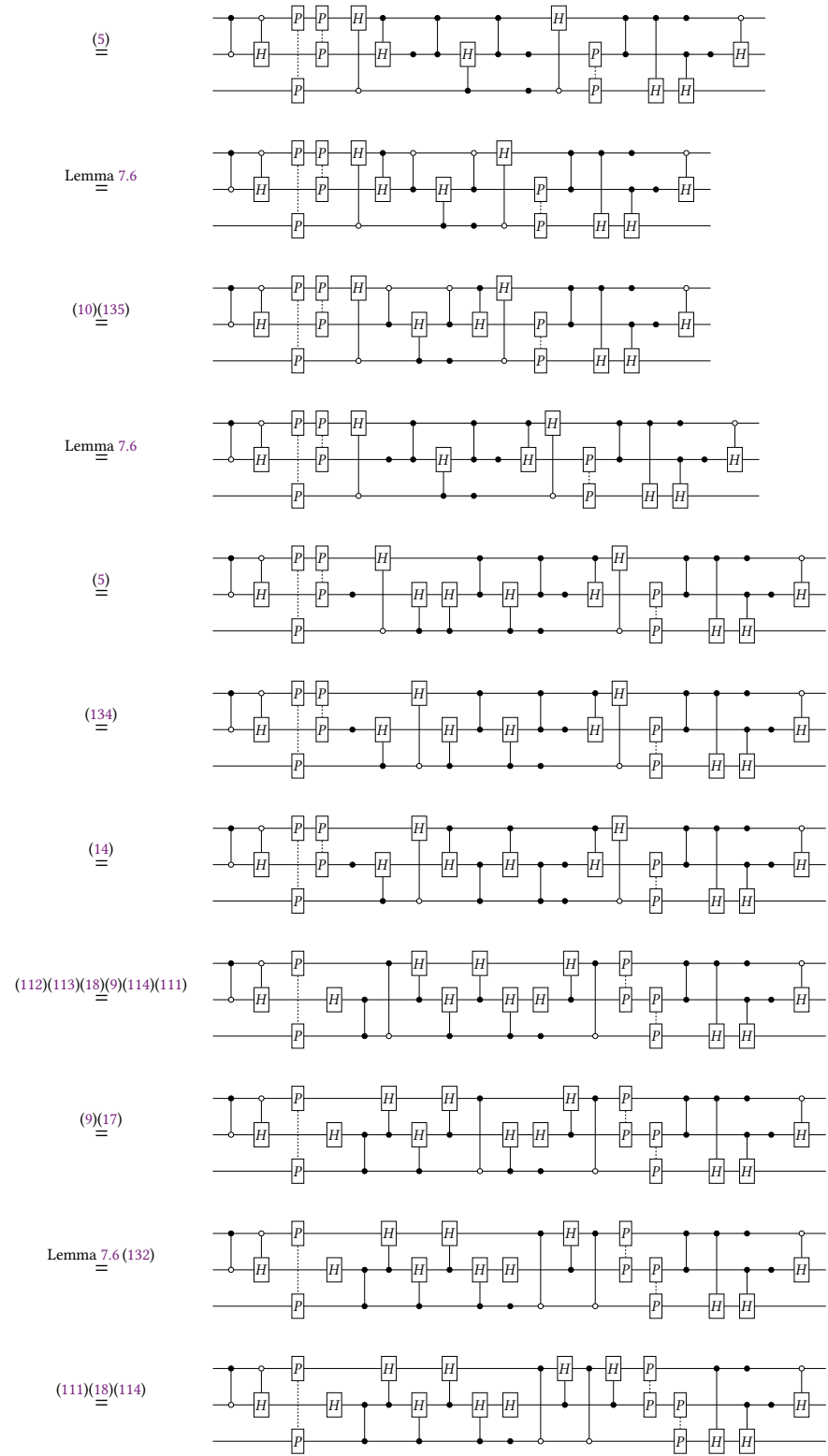

$\overset{(81)(14)}{=}$ 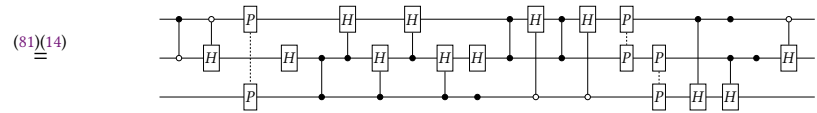

$\overset{(112)(130)(114)}{=}$ 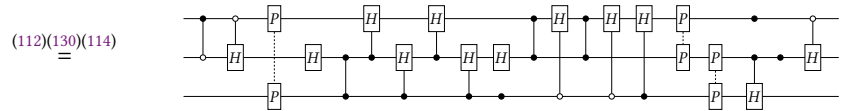

$\overset{(92)(7)}{=}$ 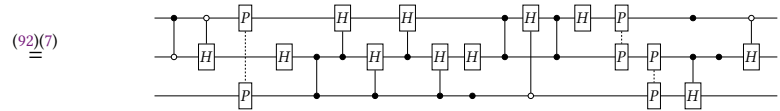

$\overset{(112)(130)}{=}$ 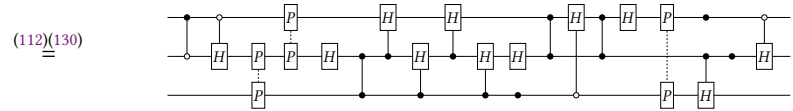

$\overset{(112)(111)(113)(18)(9)(114)}{=}$ 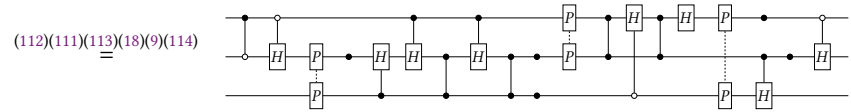

$\overset{(14)}{=}$ 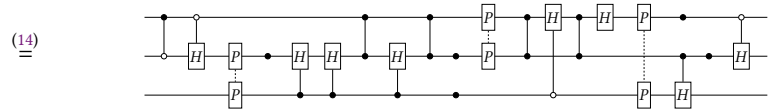

$\overset{(5)}{=}$ 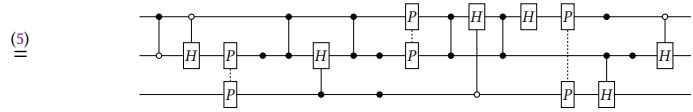

$\overset{(113)(111)(18)(114)(112)}{=}$ 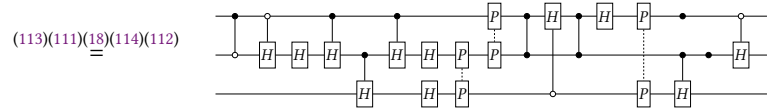

$\overset{(130)}{=}$ 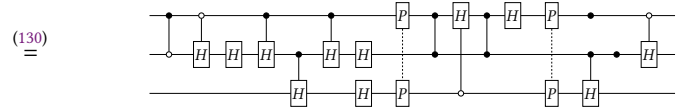

$\overset{\text{Lemma } 7.6}{=}$ 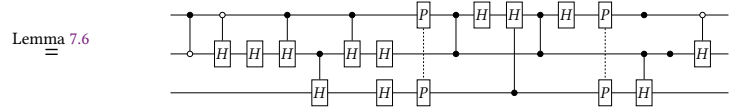

$\overset{\text{Lemma } 7.6}{=}$ 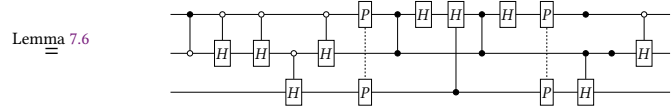

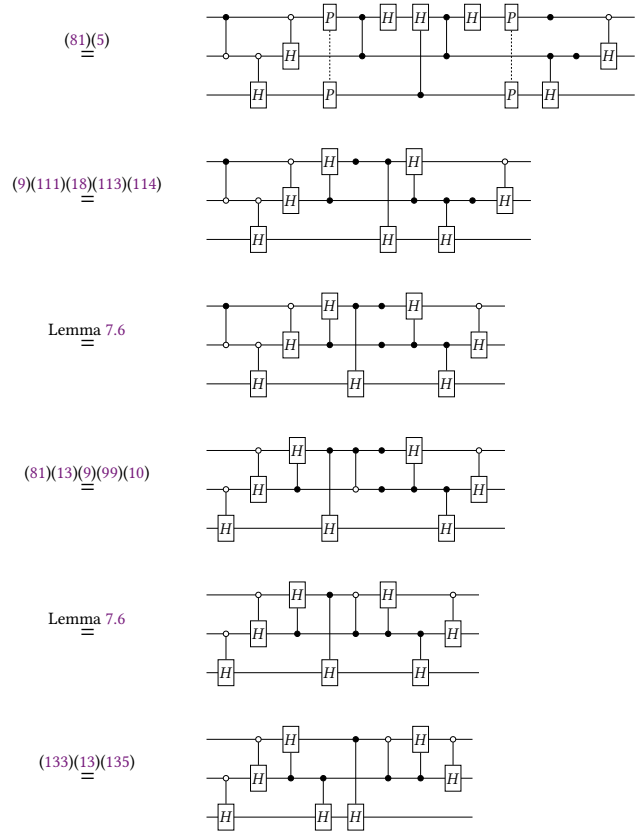

## D.3 Auxiliary Equations on 4 Qubits

In this section, we will use the 4-qubit case of Lemma 5.1:

**Lemma D.3 (Conditional 3-Qubit Completeness).** *For any 3-qubit quantum circuits $C_1, C_2$ over* $\left\{ -\boxed{H}-, -\bullet-, \begin{smallmatrix}\bullet\\\bullet\end{smallmatrix}, -\boxed{H}- \right\}$*, if* $[\![C_1]\!] = [\![C_2]\!]$ *then* $\text{QC} \vdash (-\!\!\!-\!\!\!- \otimes C_1) = (-\!\!\!-\!\!\!- \otimes C_2)$ *and* $\text{QC} \vdash (C_1 \otimes -\!\!\!-\!\!\!-) = (C_2 \otimes -\!\!\!-\!\!\!-)$.

In particular, we make the following two equations explicit in order to make the proof clearer where they are used:

**Corollary D.4.** *The following equations can be freely used in the proof of Lemma D.5, as if we had already proved that they are consequences of the equations of Figure 4:*

$$\text{(148)}$$

$$\text{(149)}$$

We now proceed to the proof of the useful 4-qubit auxiliary equations:

**Lemma D.5.** *The following equations are consequences of the equations of Figure 4:*

$$\text{(150)}$$

(151)

(152)

(153)

(154)

(155)

(156)

(157)

(158)

(159)

(160)

(161)

(162)

(163)

(164)

$$(165)$$

$$(166)$$

$$(167)$$

$$(168)$$

$$(169)$$

$$(170)$$

$$(171)$$

$$(172)$$

$$(173)$$

$$(174)$$

$$(175)$$

$$(176)$$

$$(177)$$

$$(178)$$

$$(179)$$

$$(180)$$

$$(181)$$

$$(182)$$

$$(183)$$

$$(184)$$

$$(185)$$

$$(186)$$

$$(187)$$

$$(188)$$

$$(189)$$

$$(190)$$

$$(191)$$

$$\text{(192)}$$

$$\text{(193)}$$

$$\text{(194)}$$

$$\text{(195)}$$

$$\text{(196)}$$

$$\text{(197)}$$

$$\text{(198)}$$

$$\text{(199)}$$

$$\text{(200)}$$

$$\text{(201)}$$

$$\text{(202)}$$

$$\text{(203)}$$

$$\text{(204)}$$

$$\text{(205)}$$

$$(206)$$

$$(207)$$

$$(208)$$

$$(209)$$

$$(210)$$

$$(211)$$

$$(212)$$

$$(213)$$

$$(214)$$

$$(215)$$

$$(216)$$

$$(217)$$

$$(218)$$

(219)

(220)

(221)

(222)

(223)

(224)

(225)

(226)

(227)

(228)

(229)

(230)

(231)

(232)

$$(233)$$

$$(234)$$

$$(235)$$

$$(236)$$

$$(237)$$

$$(238)$$

$$(239)$$

$$(240)$$

$$(241)$$

$$(242)$$

$$(243)$$

$$(244)$$

$$(245)$$

$$(246)$$

$$(247)$$

$$(248)$$

PROOF.

*Proof of Equation (150):*

*Proof of Equation (151):*

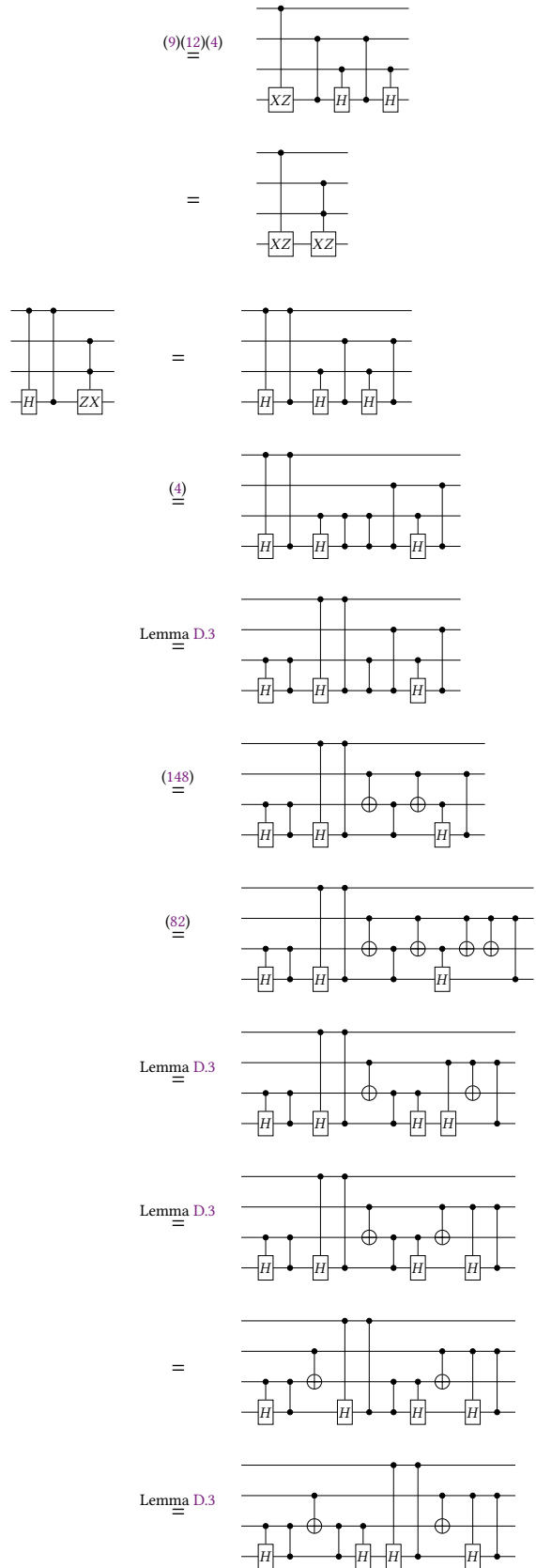

*Proof of Equation* (152):

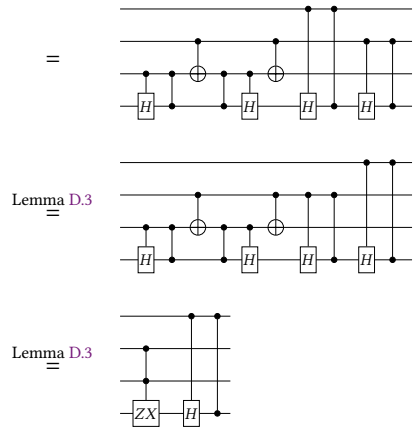

*Proof of Equation* (153):

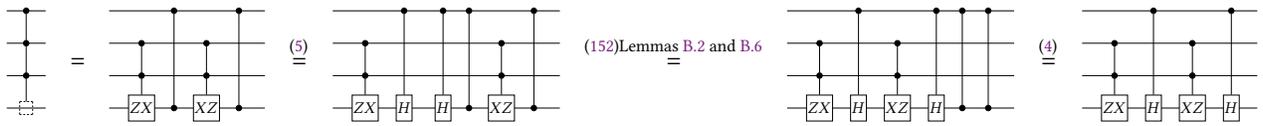

*Proof of Equation* (154):

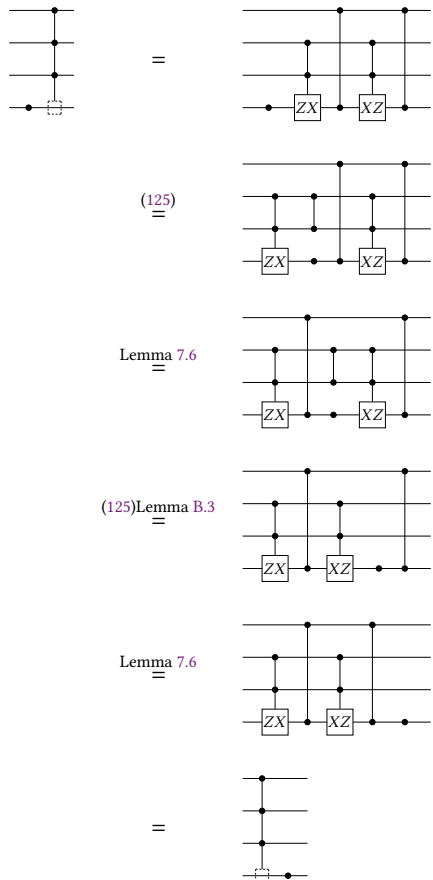

*Proof of Equation* (155):

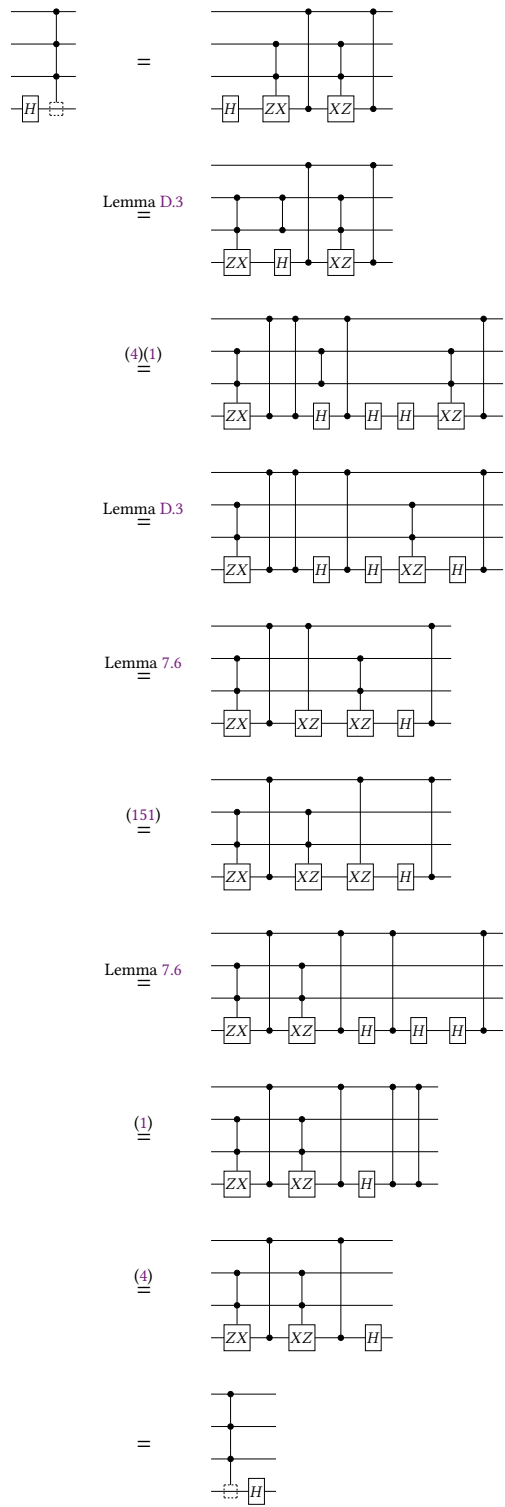

*Proof of Equation* (156): Equation (156) follows directly from Equations (154) and (155).

*Proof of Equation* (157):

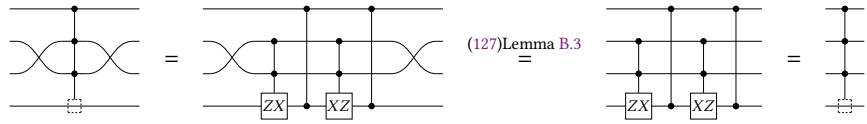

*Proof of Equation* (158):

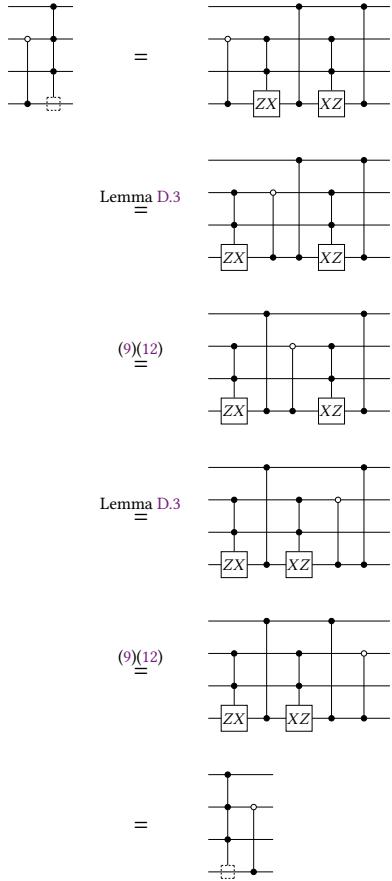

*Proof of Equation* (159):

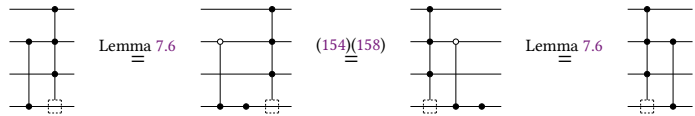

*Proof of Equation* (160): Equation (160) follows directly from Equations (157) and (159).

*Proof of Equation* (161): By Equation (150), we just need to prove the following equation:

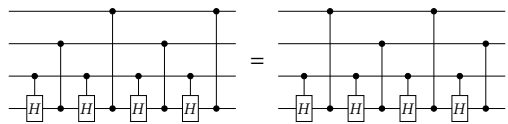

and by Lemma B.4, to do so it suffices to prove the following equation:

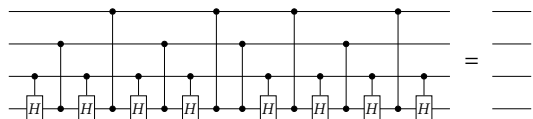

which we do as follows:

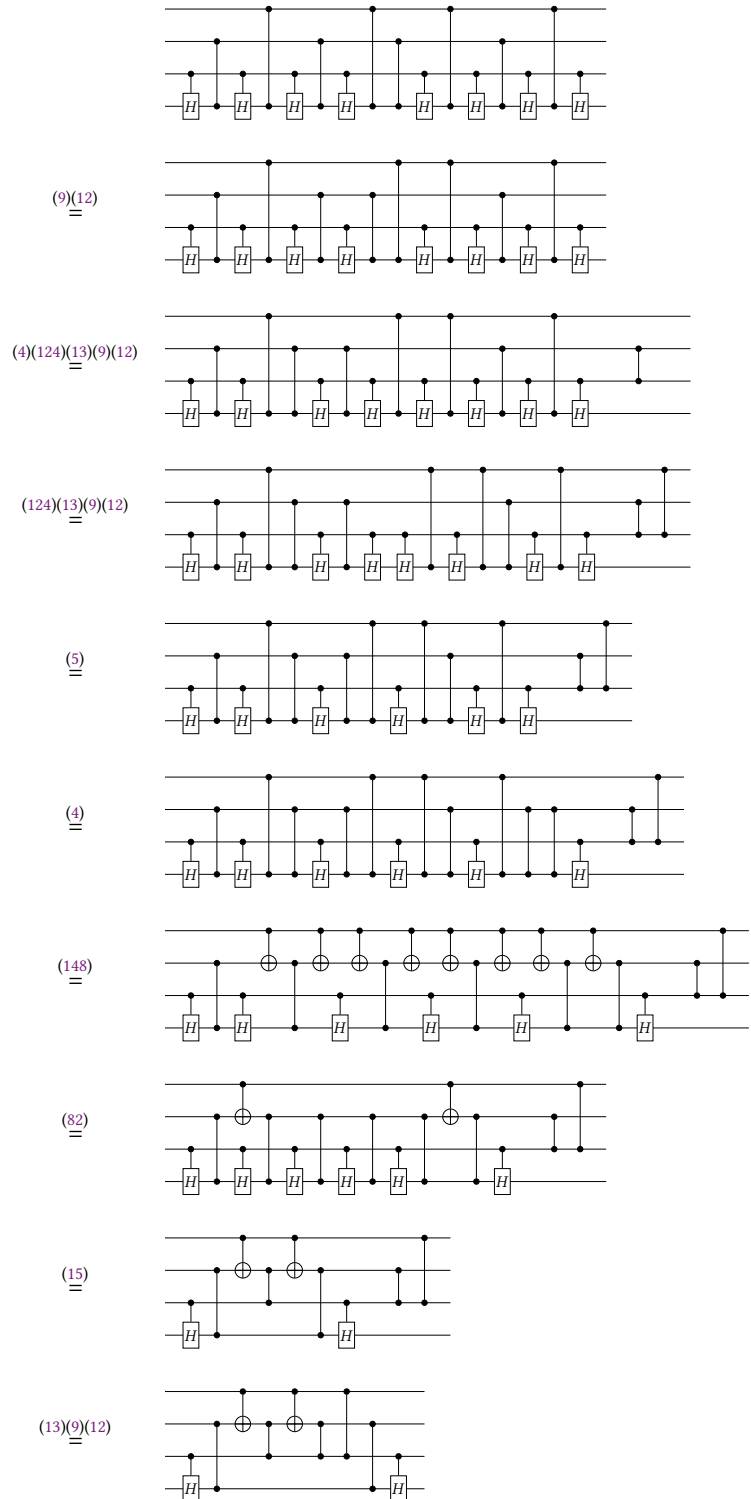

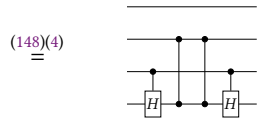

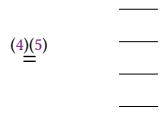

*Proof of Equation* (162):

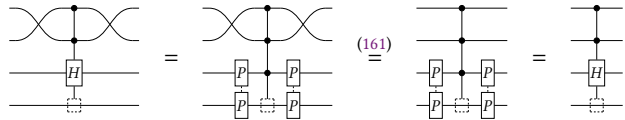

*Proof of Equation* (163):

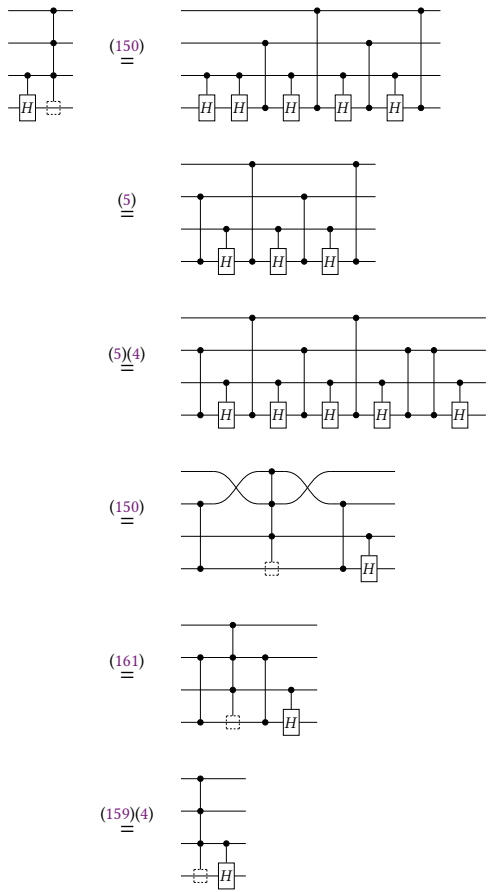

*Proof of Equation* (164):

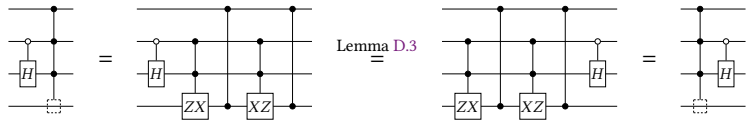

*Proof of Equation* (165):

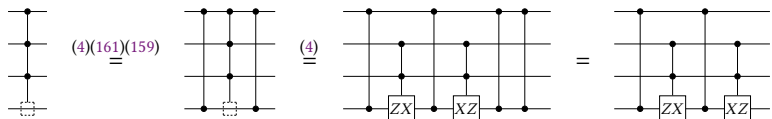

*Proof of Equation* (166):

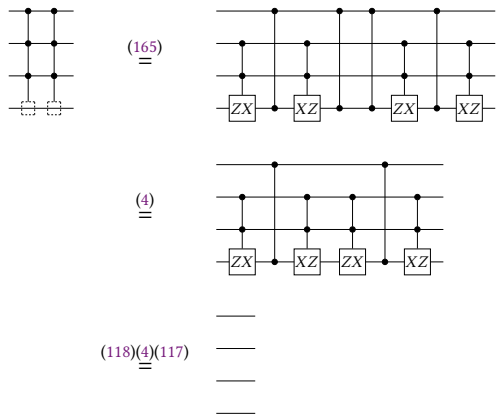

*Proof of Equation* (167):

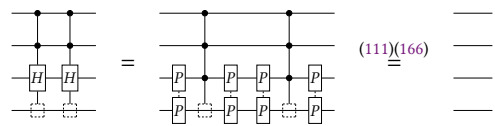

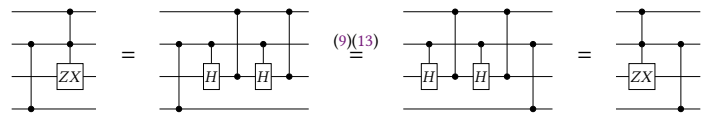

*Proof of Equation* (168):

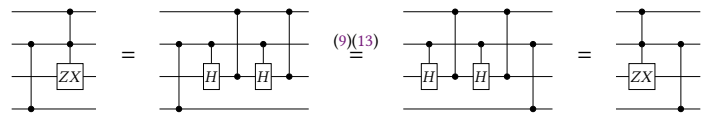

*Proof of Equation* (169):

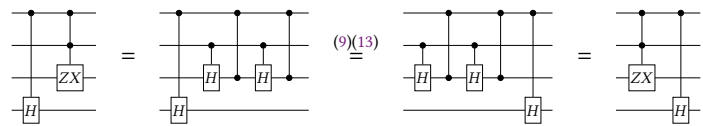

*Proof of Equation* (170): To prove the first equality, we have

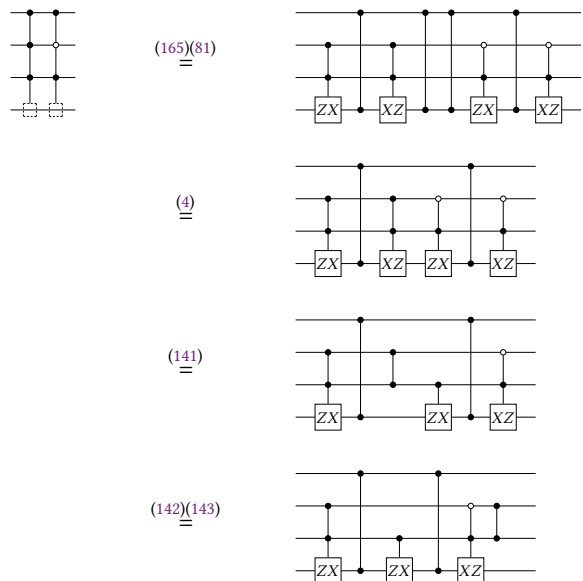

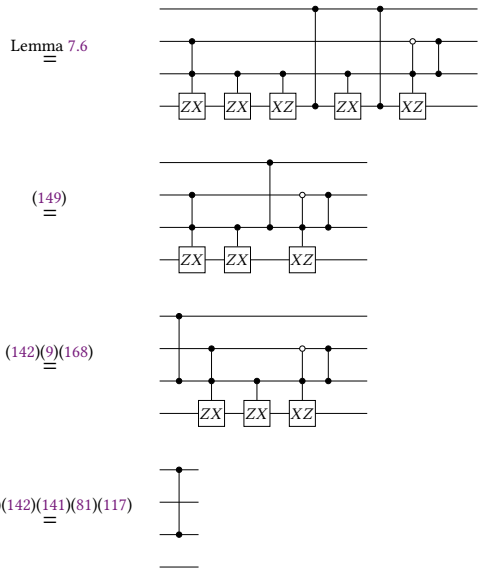

and to prove the second equality, we have

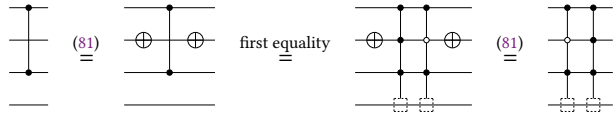

*Proof of Equation* (171):

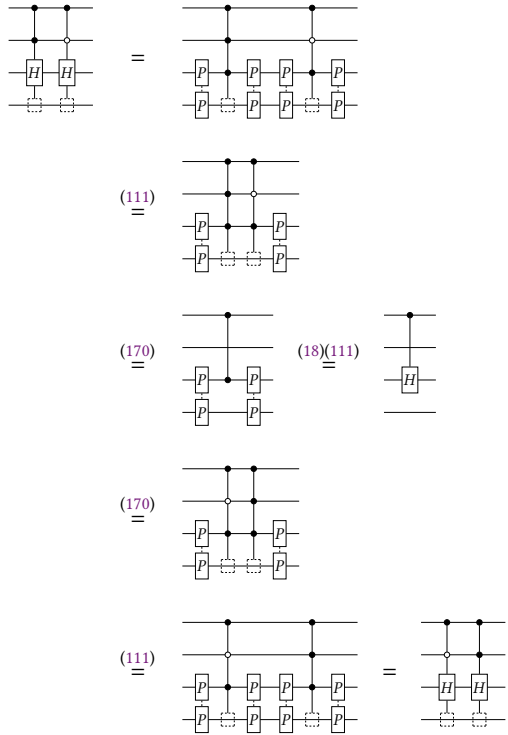

*Proof of Equation* (172):

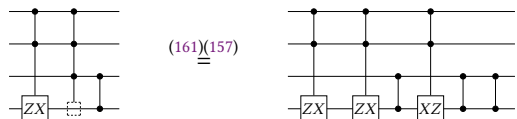

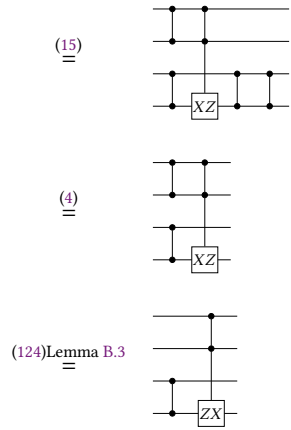

*Proof of Equation* (173): Given Equation (172), Equation (173) follows from Lemma B.5 used together with Equation (166).

*Proof of Equation* (174):

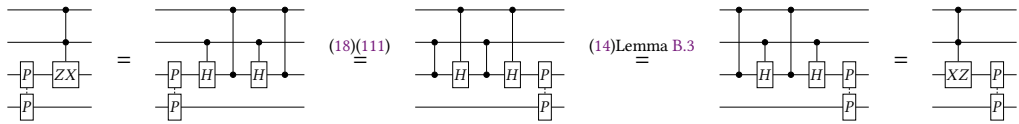

*Proof of Equation* (175):

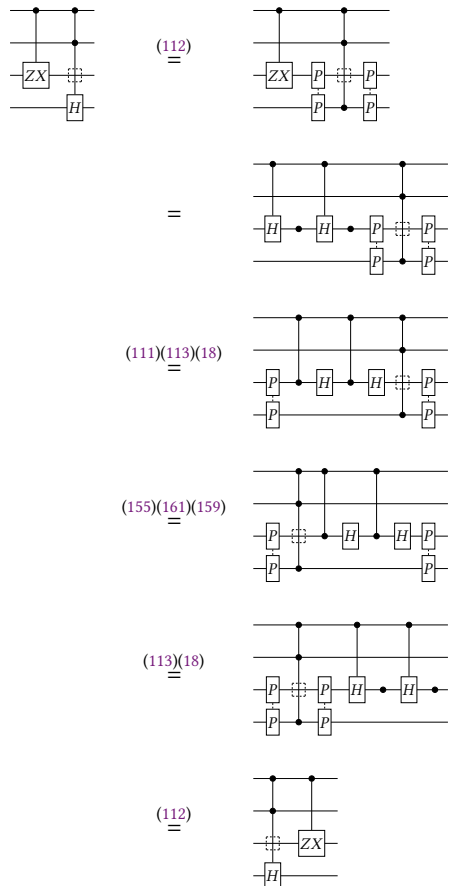

*Proof of Equation* (176):

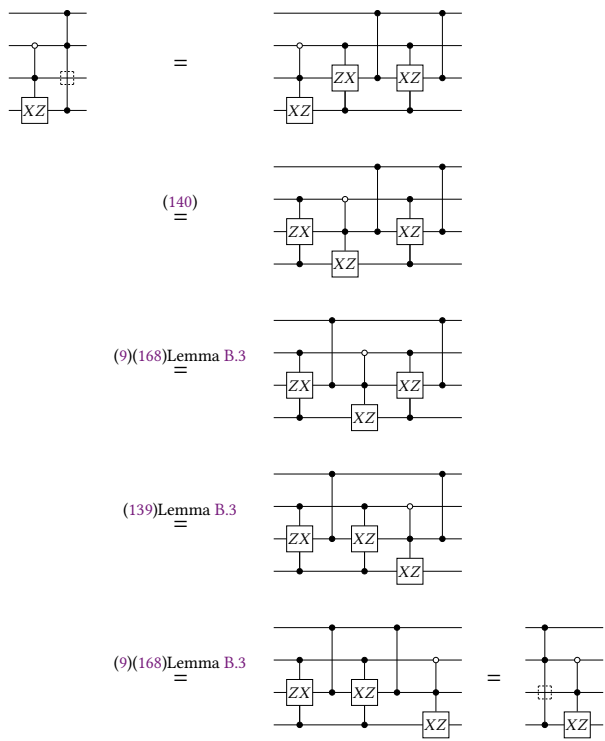

*Proof of Equation* (177):

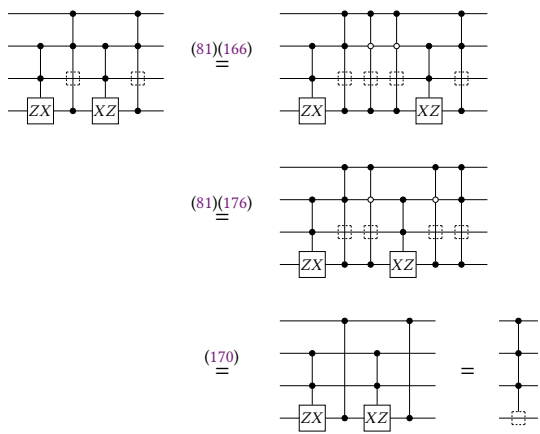

*Proof of Equation* (178):

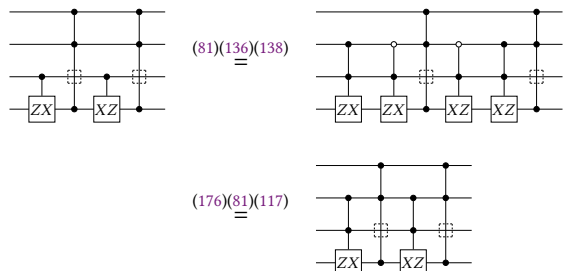

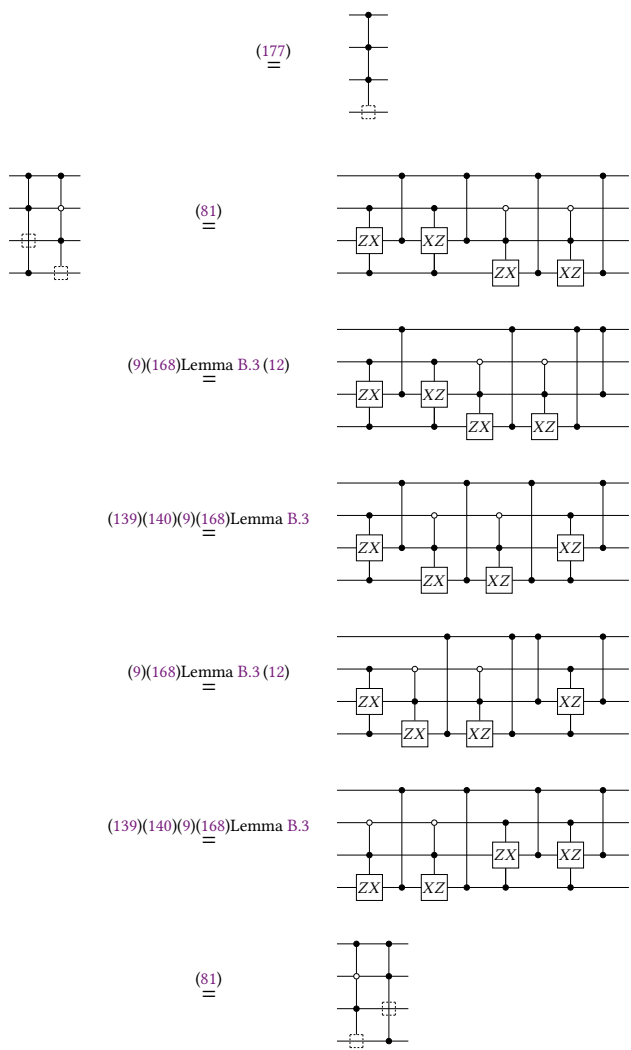

*Proof of Equation* (179):

*Proof of Equation* (180):

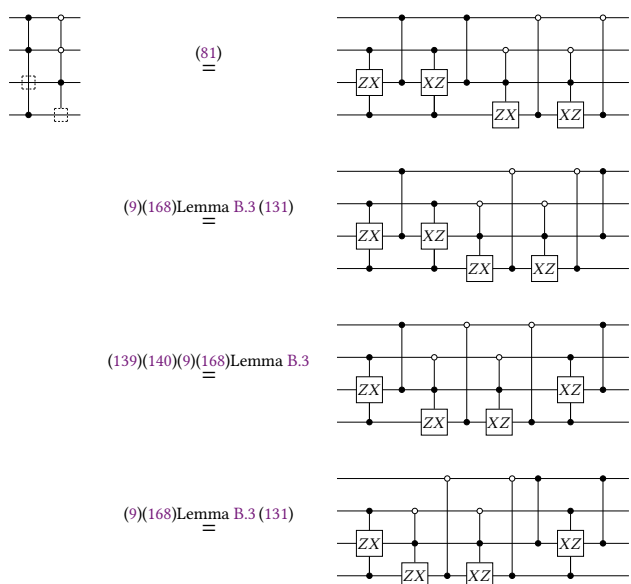

$$\overset{(139)(140)(9)(168)\text{Lemma B.3}}{=}$$

$$\overset{(81)}{=}$$

*Proof of Equation* (181):

$$\overset{(112)(157)(161)(153)(81)}{=}$$

$$\overset{(174)(114)(111)}{=}$$

$$\overset{(168)(127)(139)\text{Lemma B.3}(140)}{=}$$

$$\overset{\text{Lemma 7.6}}{=}$$

$$\overset{(2)}{=}$$

$$\overset{(81)}{=}$$

$$\overset{(9)(13)}{=}$$

$$\overset{(5)(9)(13)}{=}$$

$\overset{(17)}{=}$ 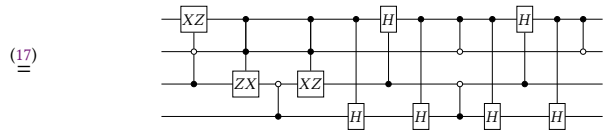

$\overset{(17)}{=}$ 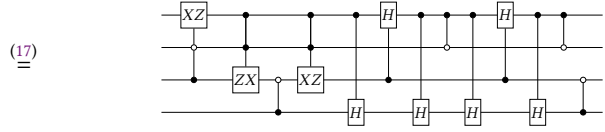

$\overset{(9)(13)(5)}{=}$ 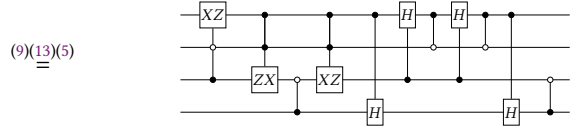

$\overset{(81)}{=}$ 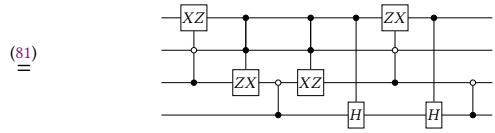

$\overset{(169)\text{Lemma B.3}}{=}$ 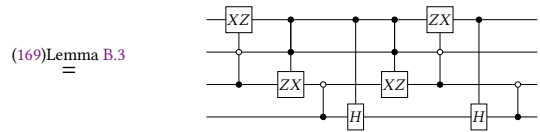

$\overset{(127)\text{Lemma B.3}(81)(140)}{=}$ 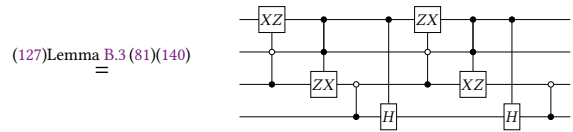

$\overset{(169)\text{Lemma B.3}}{=}$ 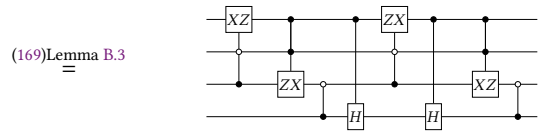

$\overset{(81)}{=}$ 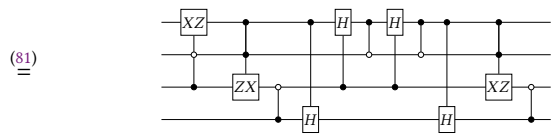

$\overset{(5)(9)(13)}{=}$ 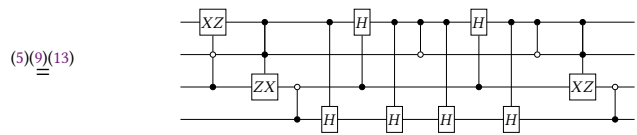

$\overset{(17)}{=}$ 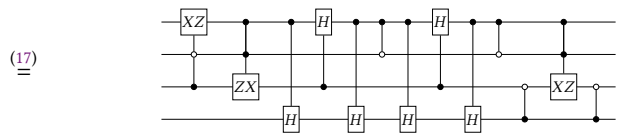

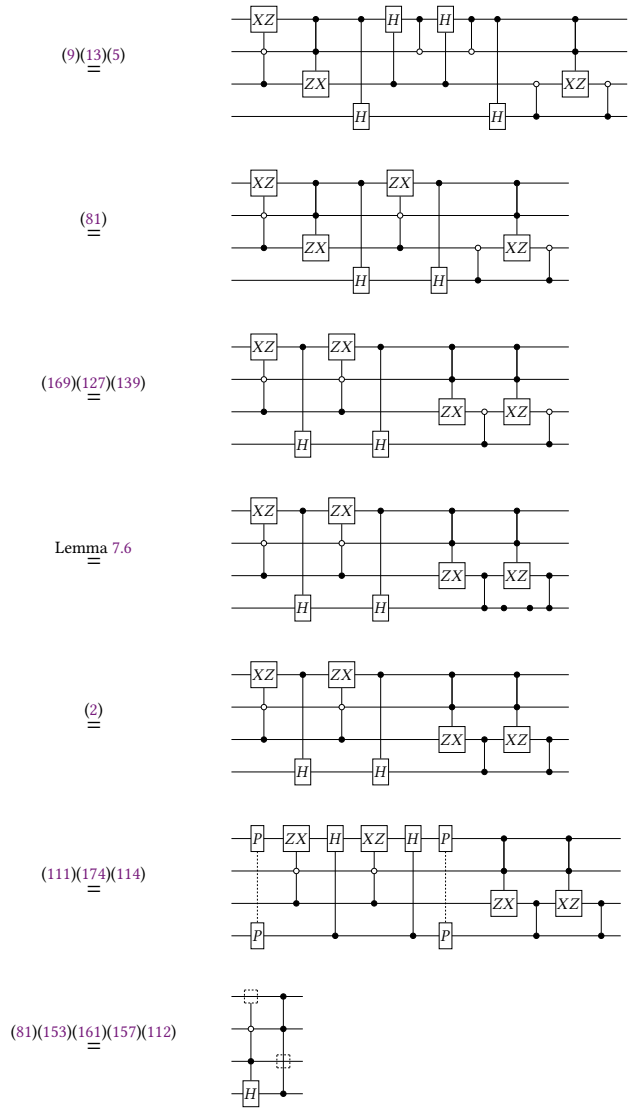

*Proof of Equations* (182) *to* (189): Equations (182) to (189) follow directly from Equation (181) together with Equations (81), (157), (161) and (162).

*Proof of Equation* (190):

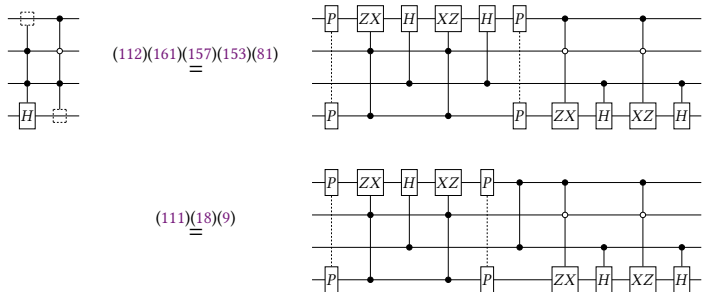

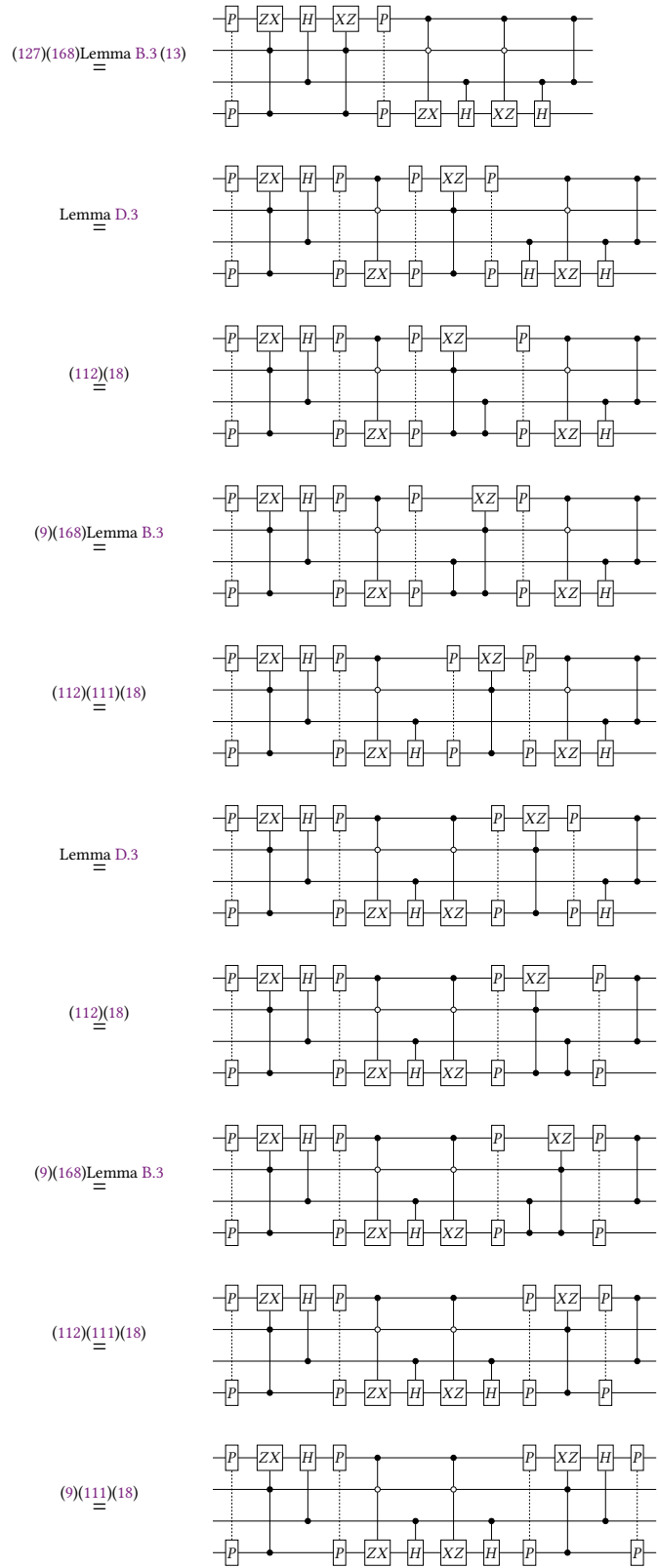

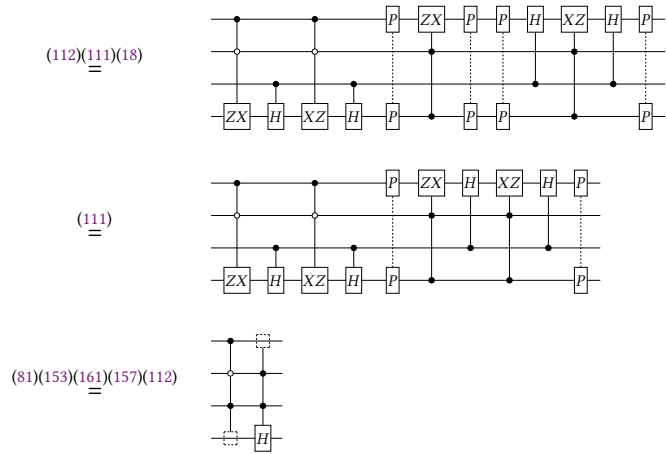

*Proof of Equations* (191) *to* (194): Equations (191) to (194) follow directly from Equation (190) together with Equations (81), (157), (161) and (162).

*Proof of Equation* (195):

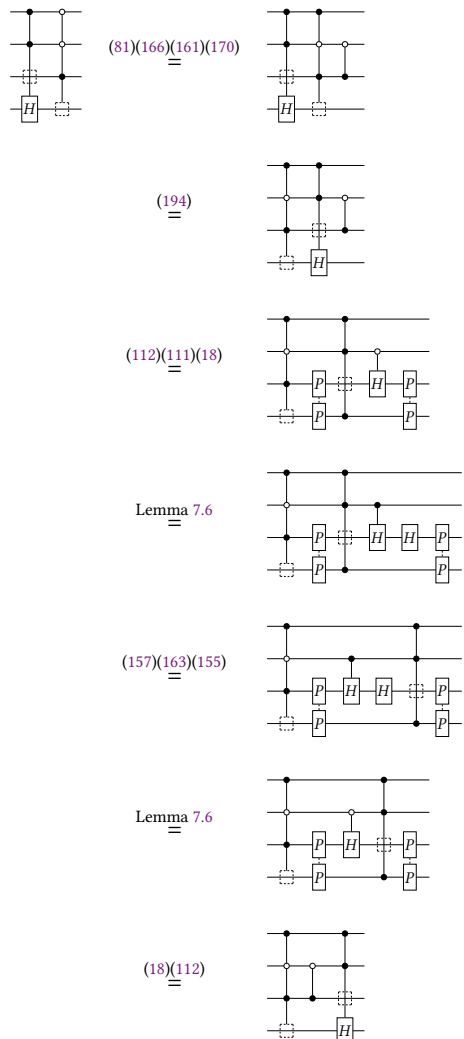

$$\overset{(170)(161)(81)(166)}{=}$$

*Proof of Equation* (196):

$$\overset{(161)(157)(178)(81)}{=}$$

$$\overset{(175)}{=}$$

$$\overset{(162)(157)(183)}{=}$$

$$\overset{(175)\text{Lemmas B.2 and B.6}}{=}$$

$$\overset{(162)(157)(183)}{=}$$

$$\overset{(81)(157)(178)(161)}{=}$$

*Proof of Equation* (197):

$$\overset{(81)(166)(161)(170)}{=}$$

$$\overset{(196)}{=}$$

$$\overset{(112)}{=}$$

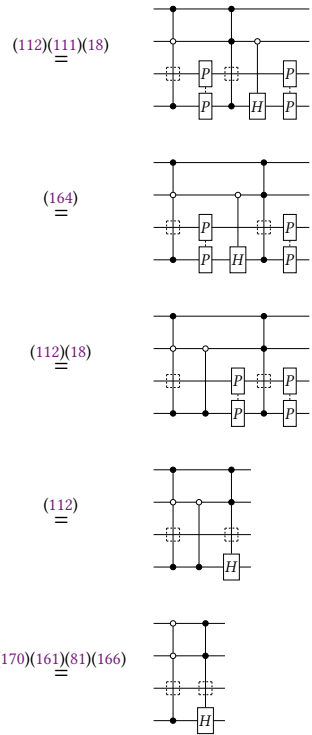

*Proof of Equations* (198) *to* (200): Equations (198) to (200) follow directly from Equation (196) together with Equations (81), (157), (161) and (162).

*Proof of Equation* (201): Equation (201) follows directly from Equations (162) and (171).

*Proof of Equation* (202):

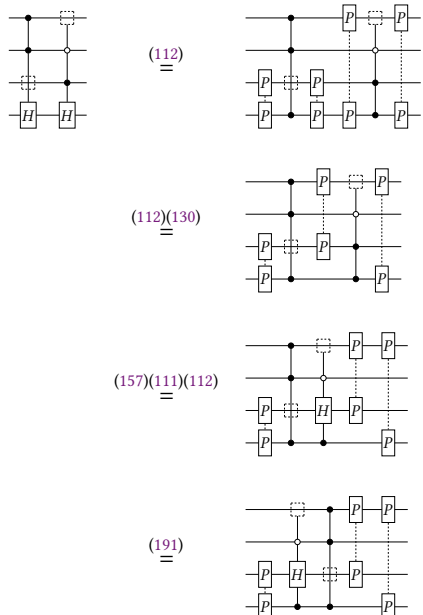

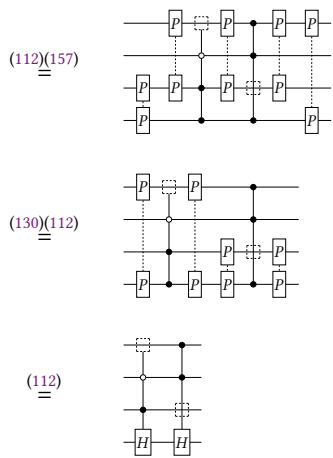

*Proof of Equation* (203):

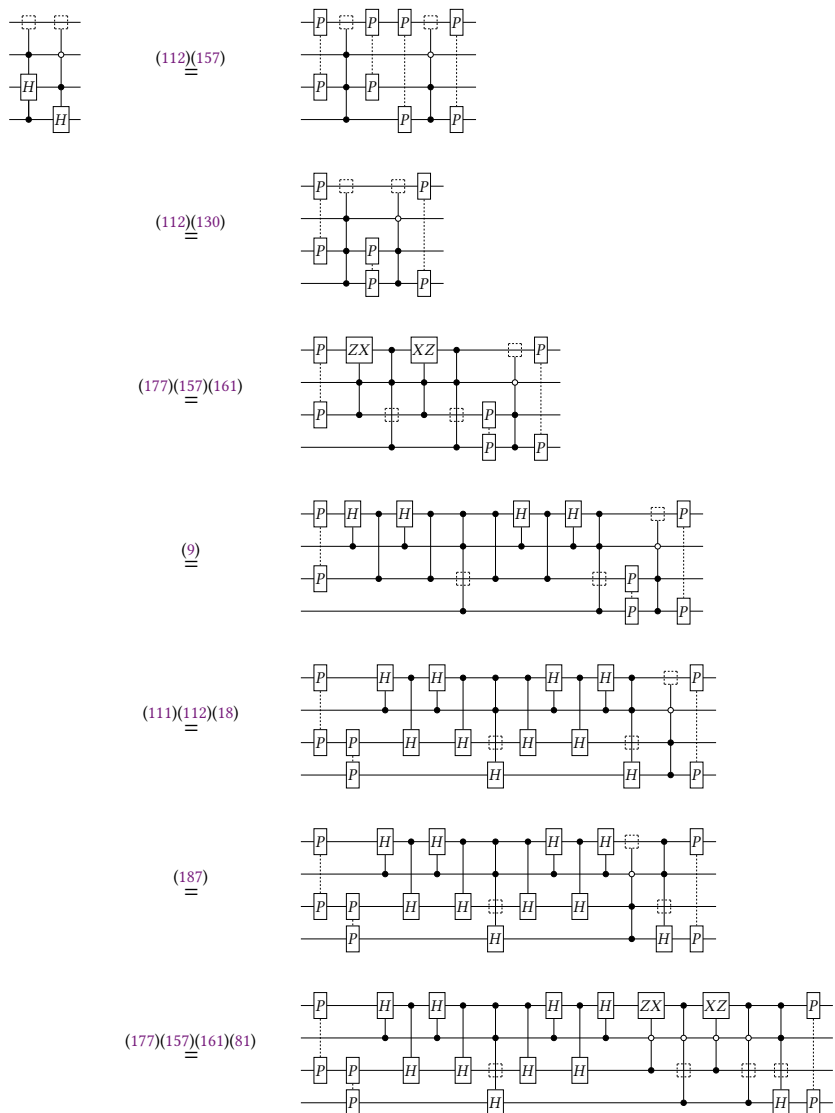

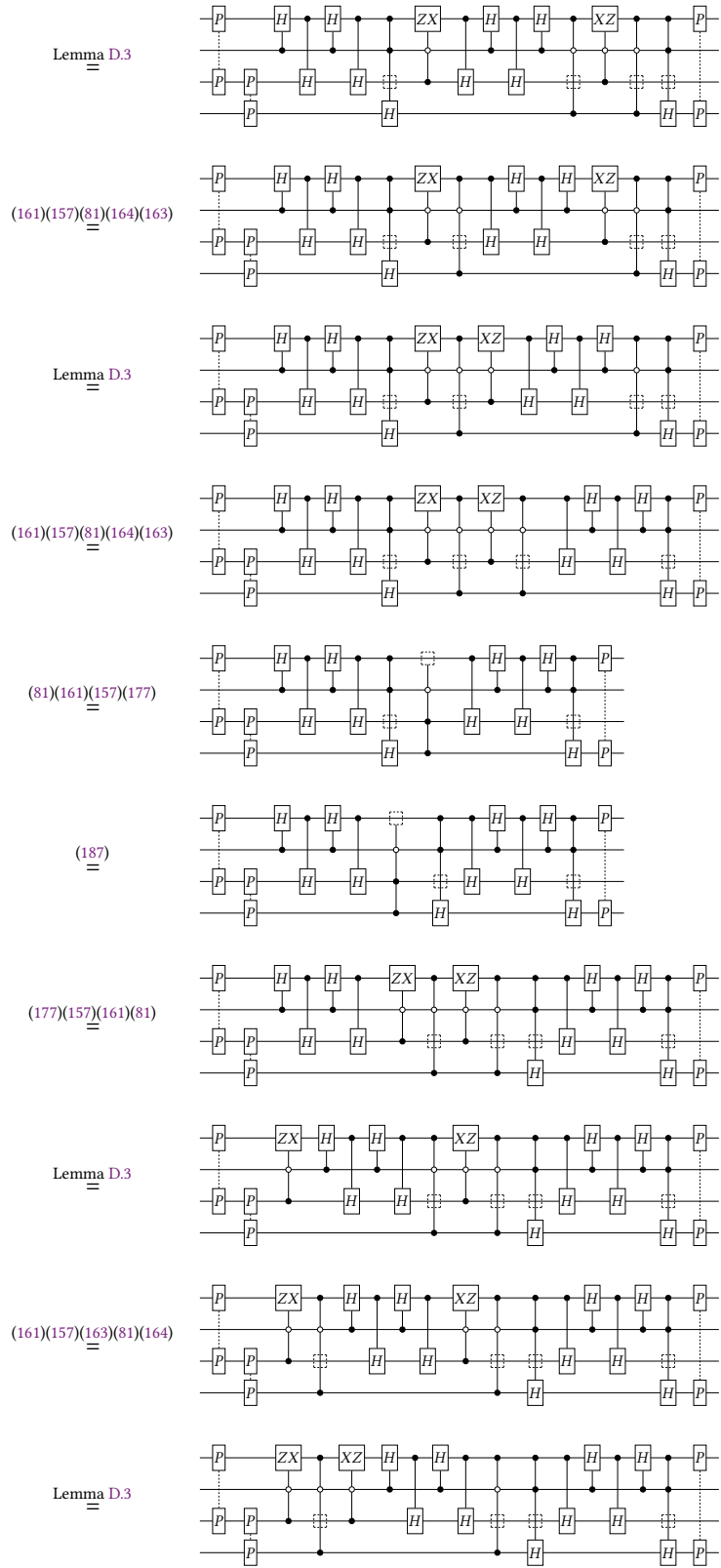

*Proof of Equation* (204):

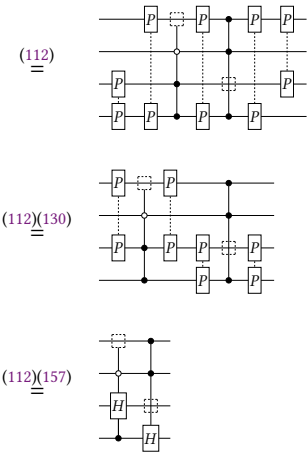

*Proof of Equation* (205):

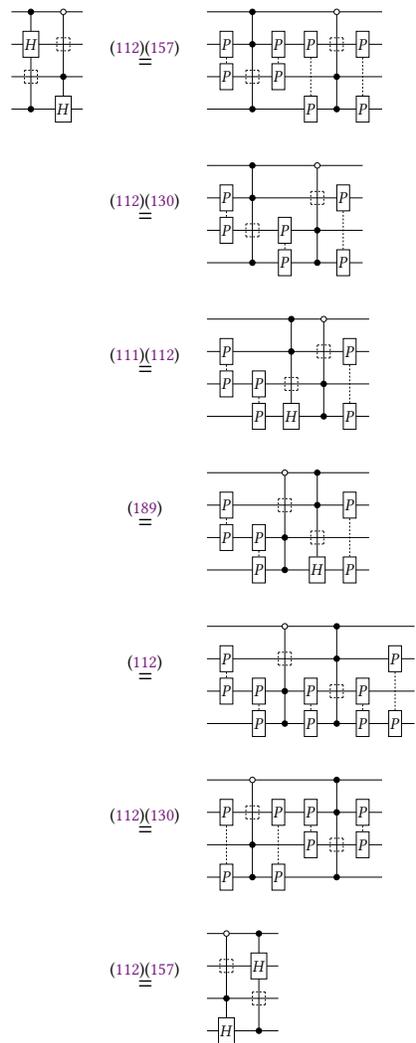

*Proof of Equation* (206):

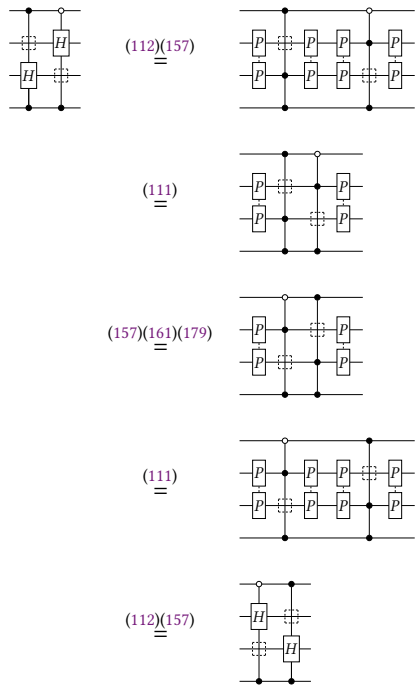

*Proof of Equation* (207):

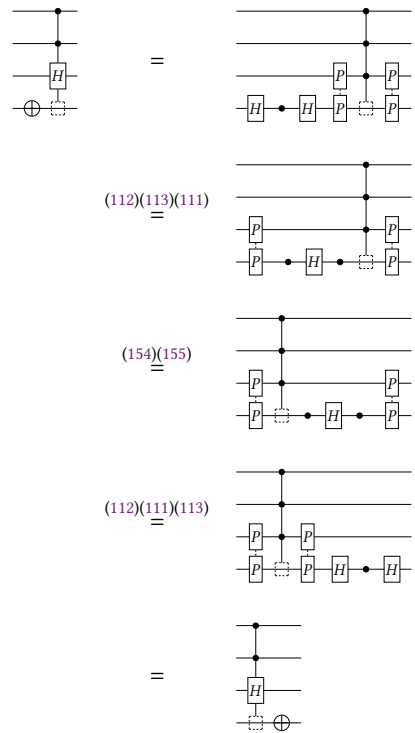

*Proof of Equation* (208):

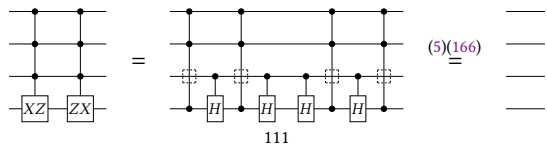

*Proof of Equation* (209): Equation (209) follows from Equations (81) and (208).

*Proof of Equation* (210):

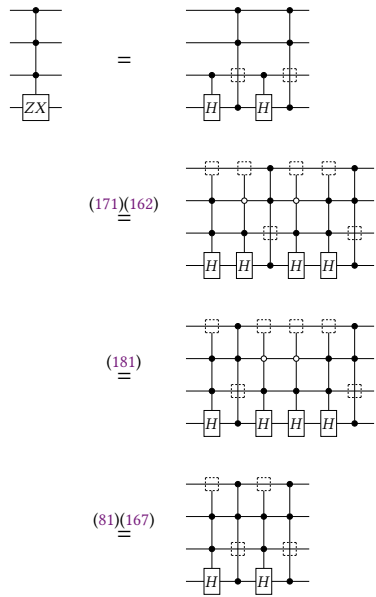

*Proof of Equation* (211): Given Equation (210), Equation (211) follows from Lemma B.5 used together with Equations (81), (166), (167) and (208).

*Proof of Equation* (212):

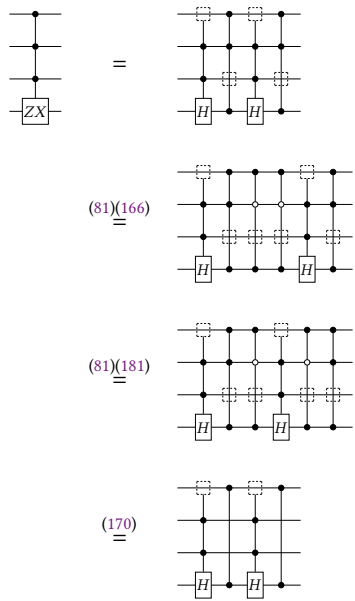

*Proof of Equation* (213): Given Equation (212), Equation (213) follows from Lemma B.5 used together with Equations (4), (81), (167) and (208).

*Proof of Equation* (214): If $a = 0$, then we have

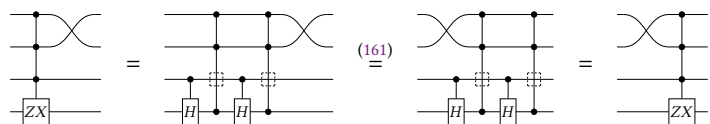

and if $a = 1$, we have

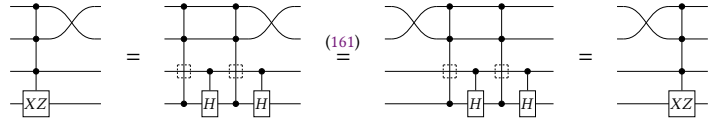

*Proof of Equation* (215): If $a = 0$, then we have

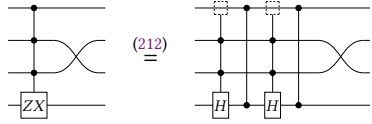

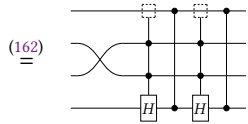

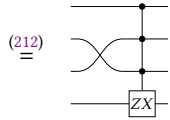

The case $a = 1$ follows by Lemma B.5 used together with Equation (208).

*Proof of Equation* (216):

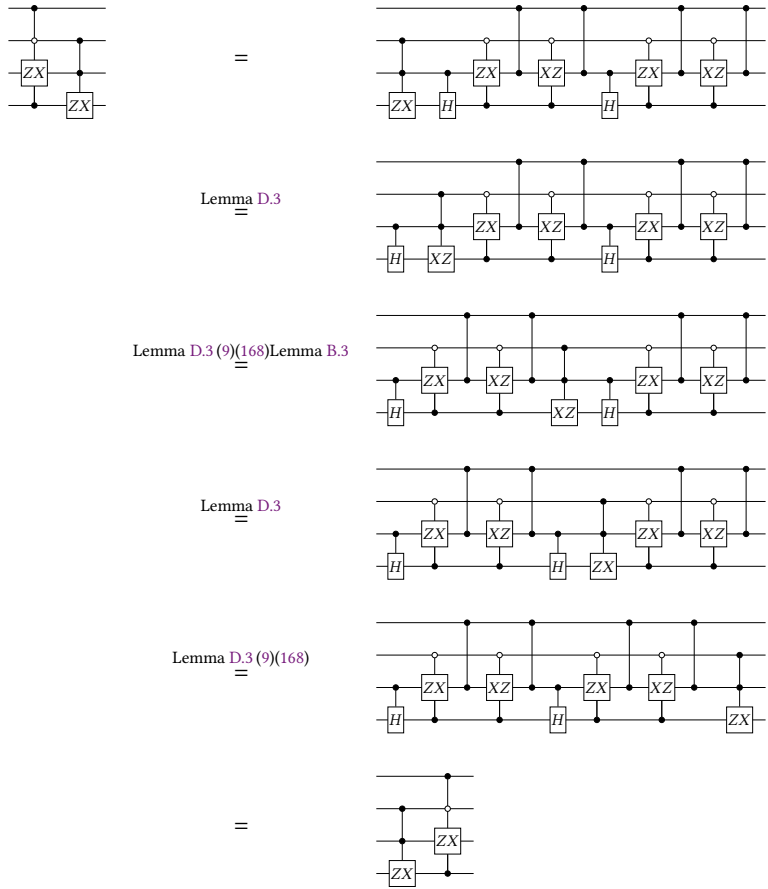

*Proof of Equation* (217): Given Equation (216), Equation (217) follows by Lemma B.6 used together with Equation (118).

*Proof of Equation* (218): Given Equation (216), Equation (218) follows by Lemma B.6 used together with Equations (81) and (208).

*Proof of Equation* (219):

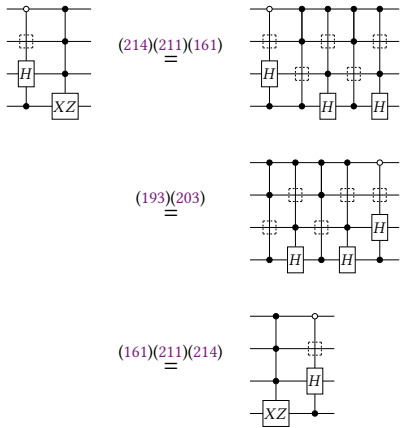

*Proof of Equation* (220):

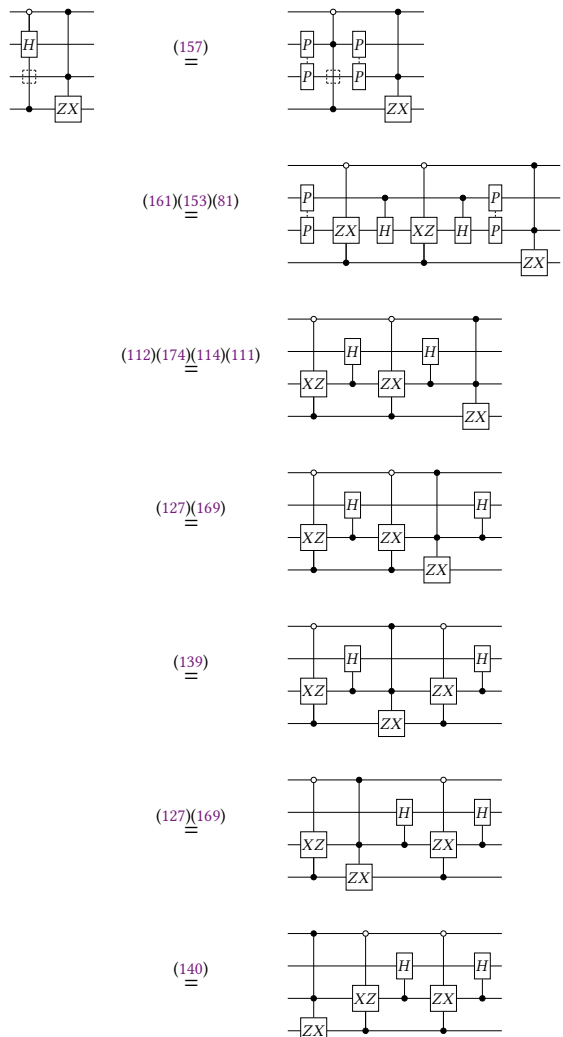

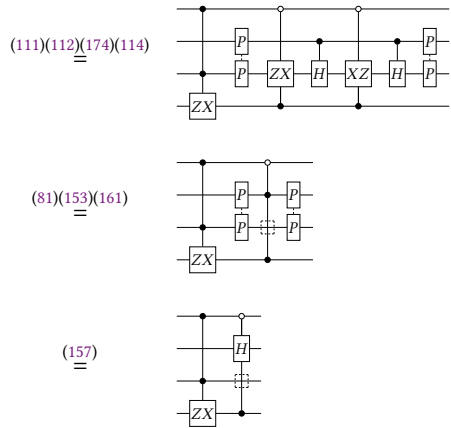

*Proof of Equation* (221):

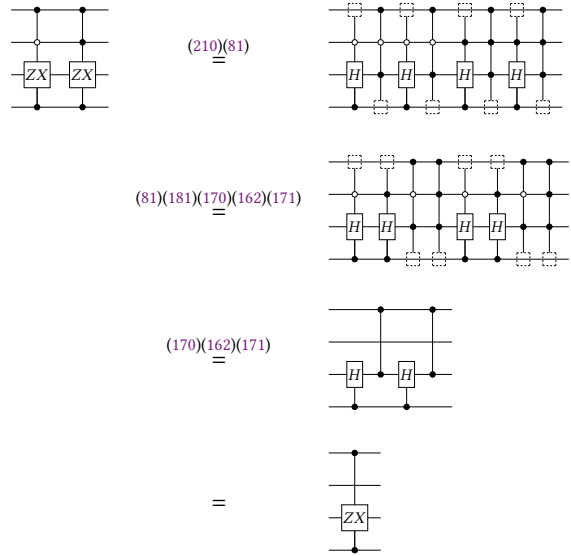

*Proof of Equation* (222): Given Equation (221), Equation (222) follows from Lemma B.5 used together with Equations (81), (118) and (208).

*Proof of Equation* (223): Equation (223) follows directly from Equation (221) together with Equations (81) and (214).

*Proof of Equation* (224): Equation (224) follows directly from Equation (222) together with Equations (81) and (214).

*Proof of Equation* (225):

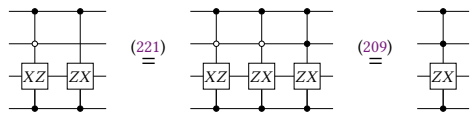

*Proof of Equation* (226):

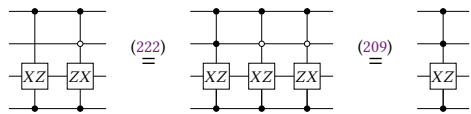

*Proof of Equation* (227):

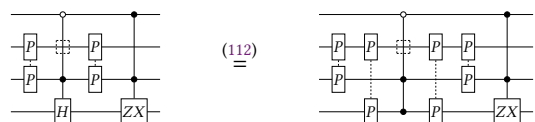

$$\overset{(112)(130)}{=}$$ 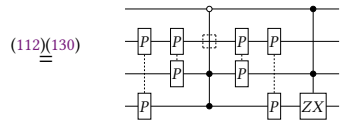

$$\overset{(112)(174)}{=}$$ 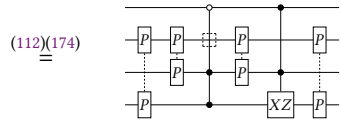

$$\overset{(157)(112)}{=}$$ 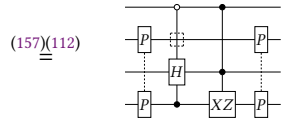

$$\overset{(222)}{=}$$ 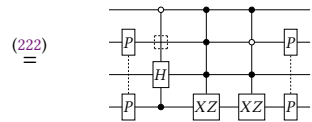

$$\overset{(219)}{=}$$ 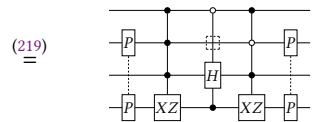

$$\overset{(207)(219)}{=}$$ 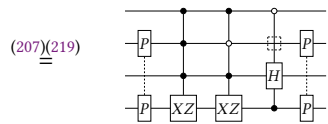

$$\overset{(222)}{=}$$ 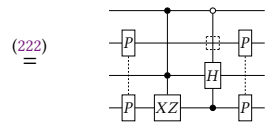

$$\overset{(112)(111)(174)}{=}$$ 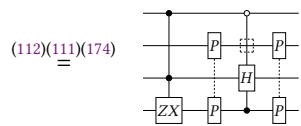

$$\overset{(112)(130)(157)}{=}$$ 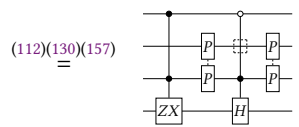

*Proof of Equation* (228)*:* If $\beta = 1$, then we have

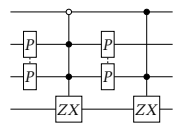 $$\overset{(214)(210)(161)(81)}{=}$$ 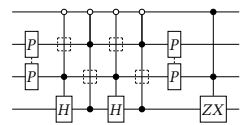

$$\overset{(111)(157)}{=}$$ 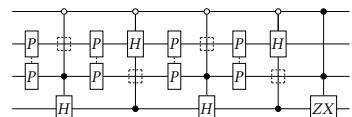

$\overset{(220)(227)}{=}$ 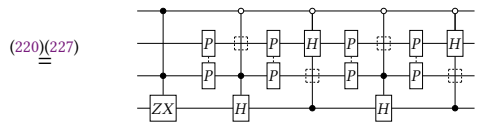

$\overset{(111)(157)}{=}$ 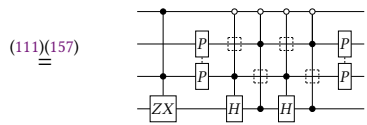

$\overset{(81)(161)(210)(214)}{=}$ 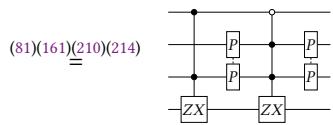

And if $\beta = 0$, we have

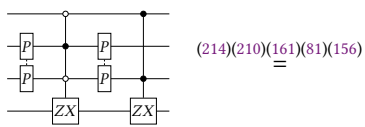 $\overset{(214)(210)(161)(81)(156)}{=}$ 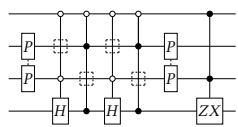

$\overset{(81)(167)(171)}{=}$ 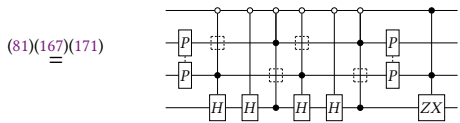

$\overset{(111)(157)}{=}$ 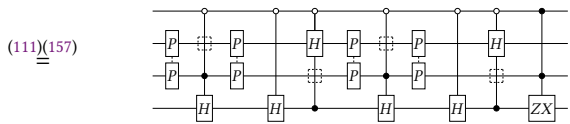

$\overset{(220)\text{Lemma D.3}(227)}{=}$ 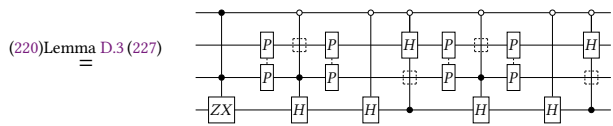

$\overset{(111)(157)}{=}$ 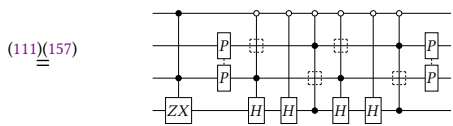

$\overset{(171)(81)(167)}{=}$ 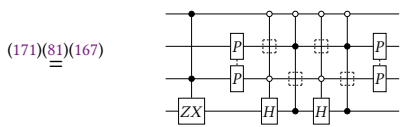

$\overset{(156)(81)(161)(210)(214)}{=}$ 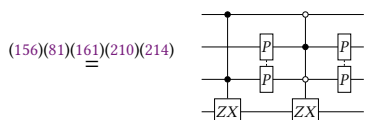

*Proof of Equation* (229): Up to using Lemma B.6 (together with Equations (81) and (208)), we can assume without loss of generality that $a = 0$. We have

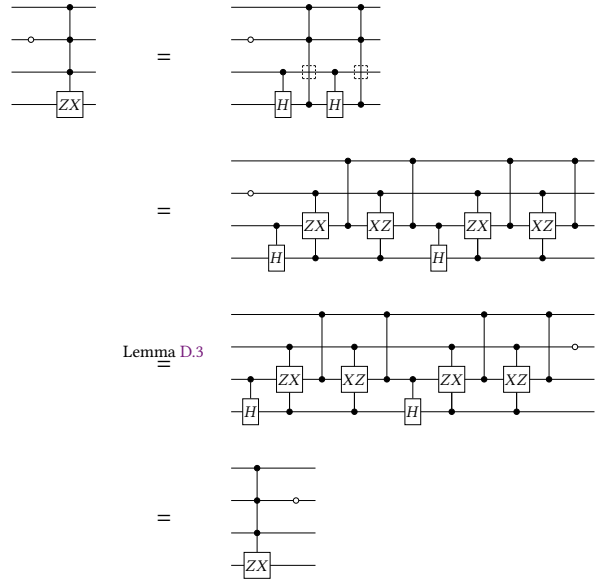

*Proof of Equation* (230):

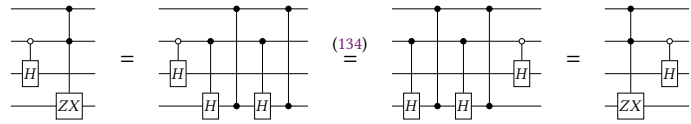

*Proof of Equation* (231):

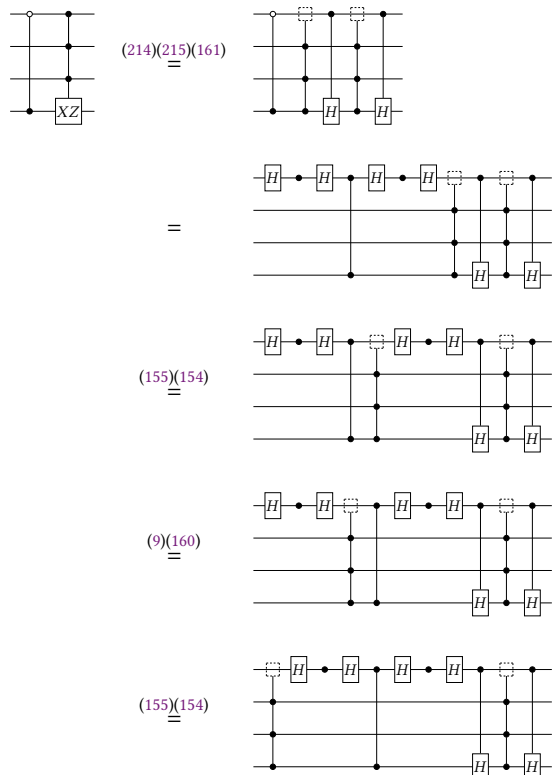

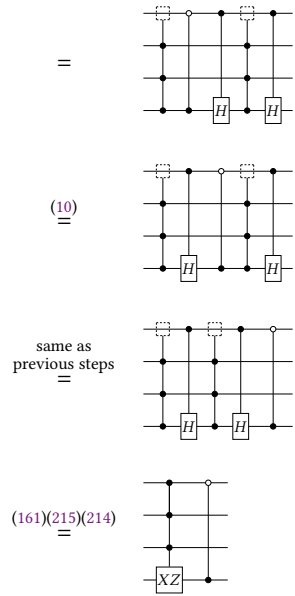

*Proof of Equation* (232):

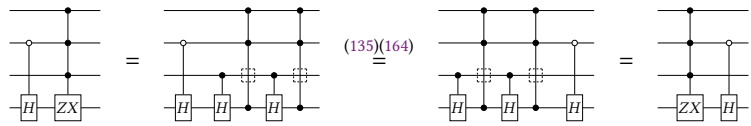

*Proof of Equation* (233): Given Equation (232), Equation (233) follows from Lemma B.5 used together with Equations (5), (81) and (208).

*Proof of Equation* (234):

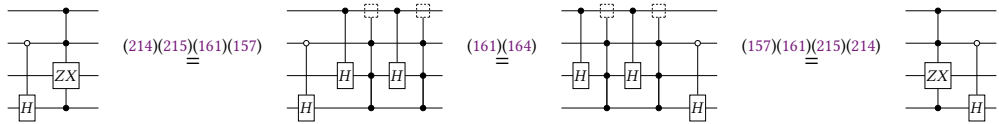

*Proof of Equation* (235): Given Equation (234), Equation (235) follows from Lemma B.5 used together with Equations (5), (81) and (208).

*Proof of Equation* (236):

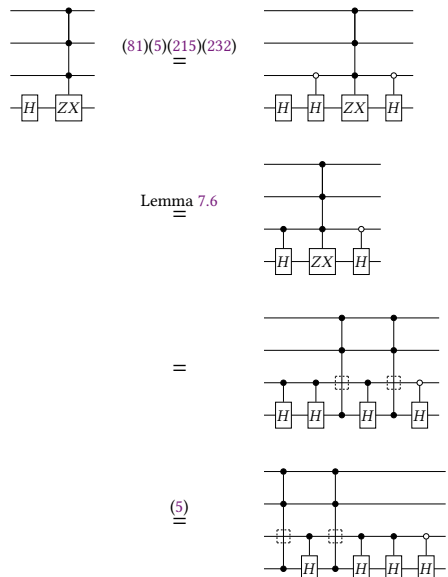

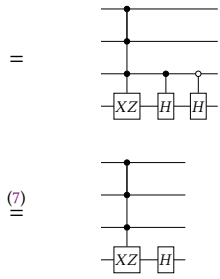

*Proof of Equation* (237):

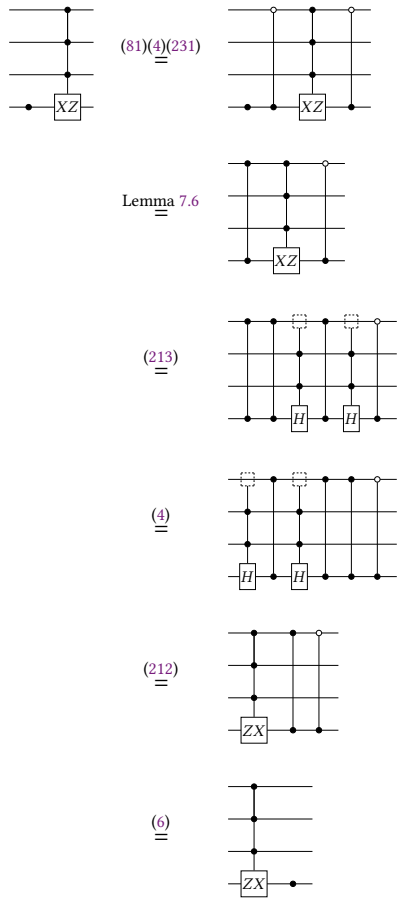

*Proof of Equation* (238):

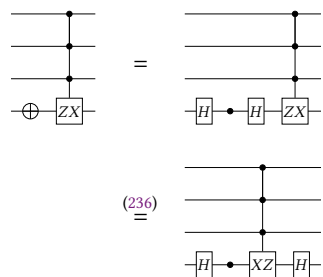

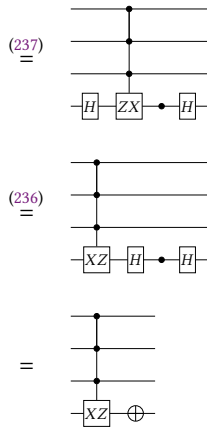

*Proof of Equation* (239): Up to using Lemma B.6 (together with Equations (81) and (118)), we can assume without loss of generality that $a = 1$. If $\gamma = 1$, then we have

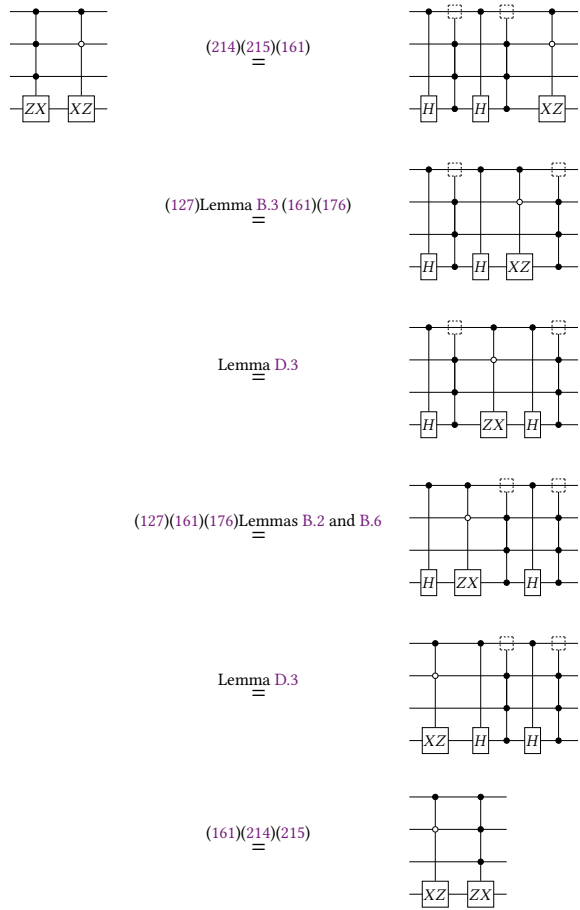

And if $\gamma = 0$, we have

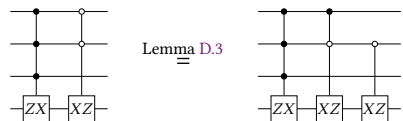

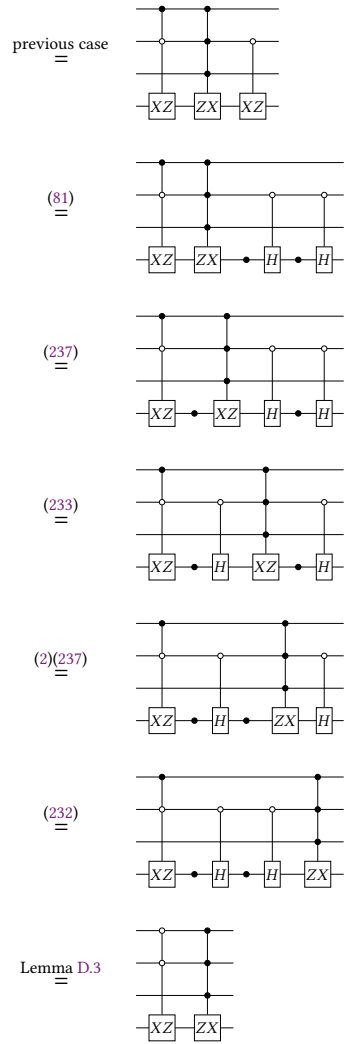

*Proof of Equation* (240): Up to using Equation (81), we can assume without loss of generality that $\alpha = 1$. Up to using Equations (81) and (238), we can assume without loss of generality that $\beta = 1$ and $\delta = 1$. And up to using Lemma B.6 (together with Equation (208)), we can assume without loss of generality that $a = 1$ and $b = 1$. That is, we just need to prove the following two equations:

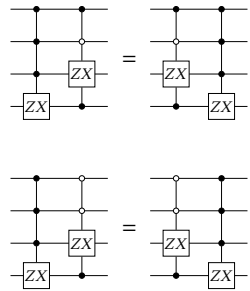

To prove the first one, we have

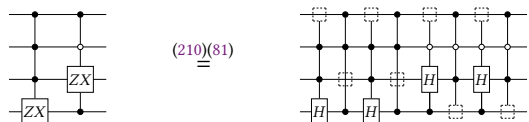

$$\overset{(191)(179)(203)(190)}{=}$$

$$\overset{(81)(210)}{=}$$

and to prove the second one, we have

$$\overset{(210)(81)(207)}{=}$$

$$\overset{(191)(180)(203)(207)(190)}{=}$$

$$\overset{(207)(81)(210)}{=}$$

*Proof of Equation* (241): Up to using Equation (81), we can assume without loss of generality that $\alpha = 1$ and $\beta = 1$. Up to using Equations (214) and (215), we can assume without loss of generality that $\gamma \leq \delta$. That is, we just need to prove the following three equations:

$$=\qquad\qquad\qquad (241a)$$

$$=\qquad\qquad\qquad (241b)$$

$$=\qquad\qquad\qquad (241c)$$

To prove Equation (241a), up to using Lemma B.6 (together with Equations (81) and (208)), we can assume without loss of generality that $a = 1$ and $b = 1$. We have

$$\overset{(81)(221)}{=}\qquad\overset{(221)}{=}$$

To prove Equation (241b), again up to using Lemma B.6 (together with Equations (81) and (208)), we can assume without loss of generality that $a = 1$ and $b = 1$. We have

$$\overset{(81)(208)(223)}{=}$$

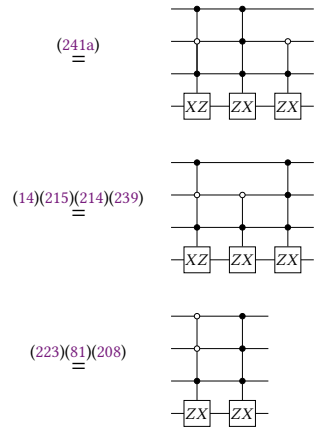

And to prove Equation (241c), again up to using Lemma B.6 (together with Equations (81) and (208)), we can assume without loss of generality that $a = 1$ and $b = 1$. We have

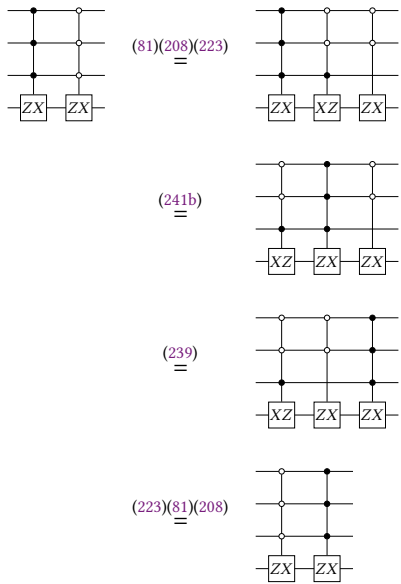

*Proof of Equation* (242): Up to using Equation (81), we can assume without loss of generality that $\alpha = 1$.

First, we treat the case where $\beta = \gamma = \delta = 1$. Up to using Lemma B.6 (together with Equations (81), (111) and (208)), we can assume without loss of generality that $a = 0$ and $b = 0$. We have

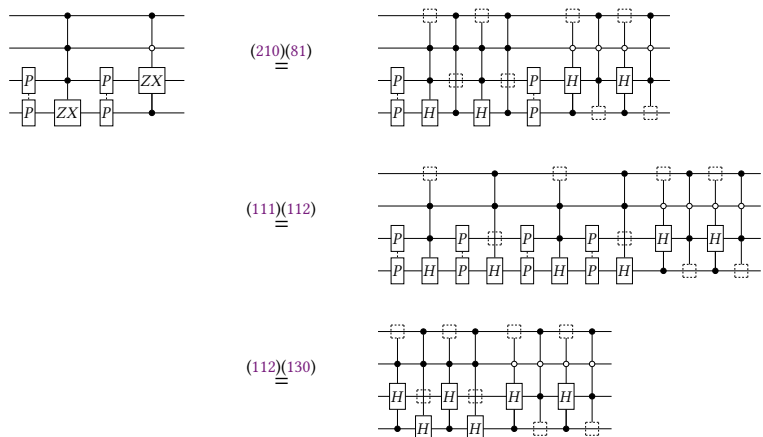

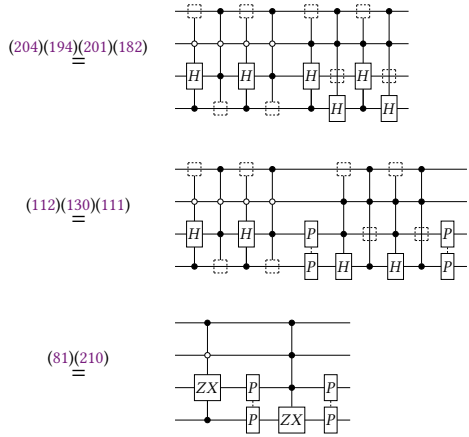

Then, we treat the case where $\beta = \delta = 1$ and $\gamma = 0$. Again, up to using Lemma B.6 (together with Equations (81), (111) and (208)), we can assume without loss of generality that $a = 0$ and $b = 0$. We have

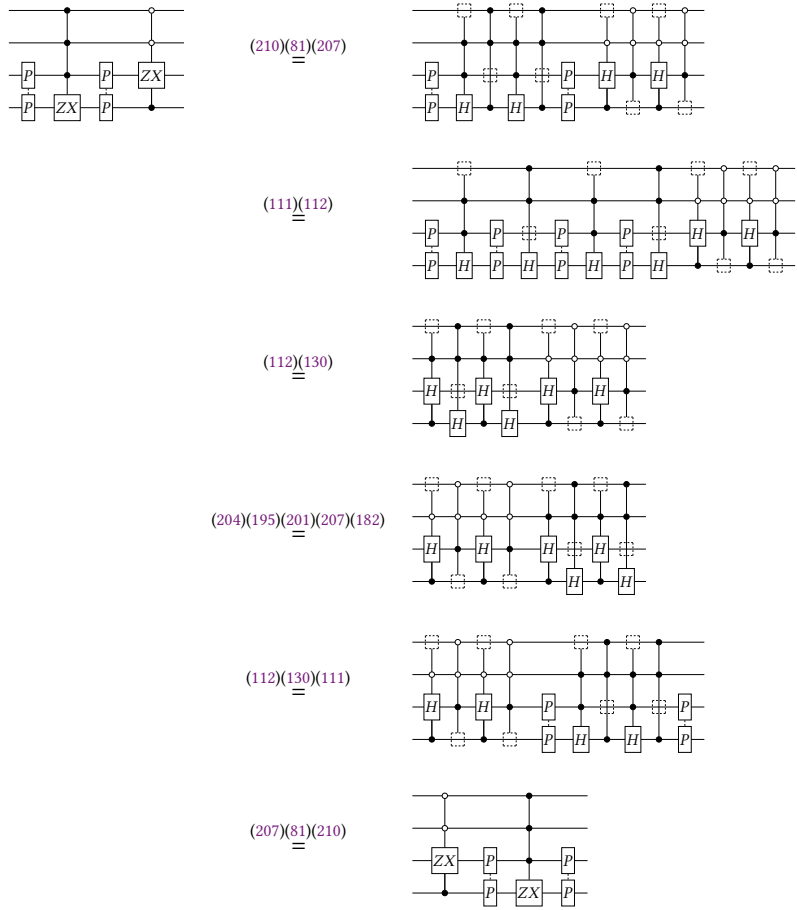

Then, to treat the cases where $\beta = 0$ and $\delta = 1$, we have

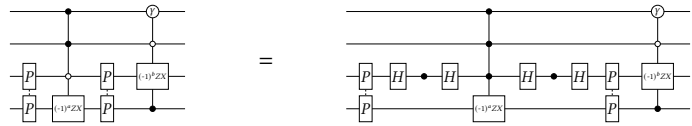

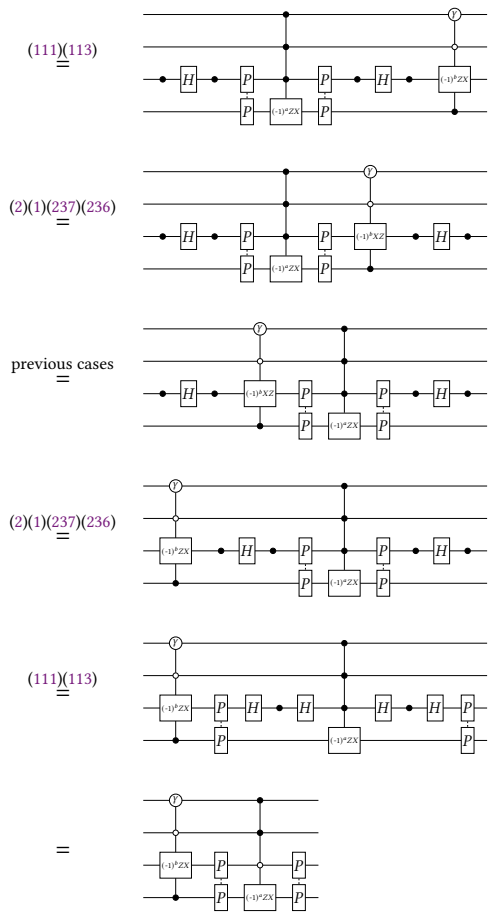

Finally, to treat the cases where $\delta = 0$, we have

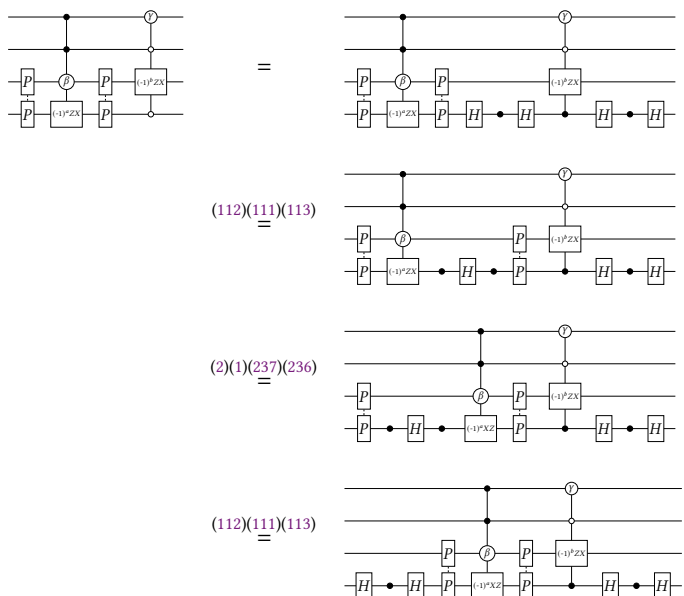

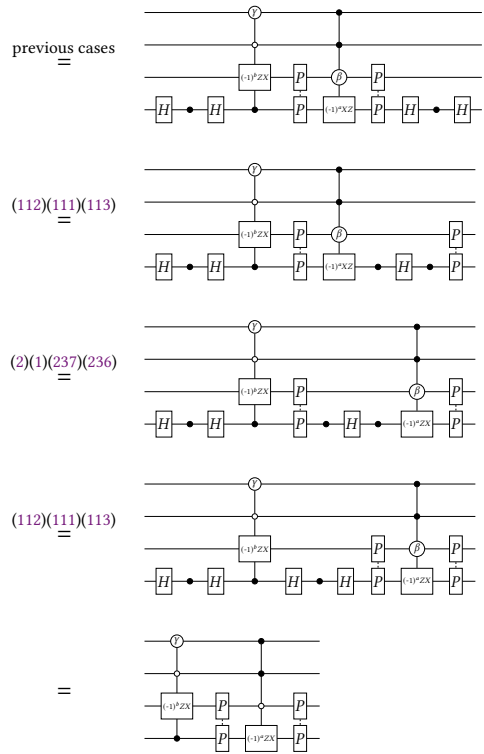

*Proof of Equation* (243): Up to using Equation (81), we can assume without loss of generality that $\alpha = 1$.

First, we treat the case where $\beta = \gamma = \delta = 1$. Up to using Lemma B.6 (together with Equations (81), (111) and (208)), we can assume without loss of generality that $a = 0$ and $b = 0$. We have

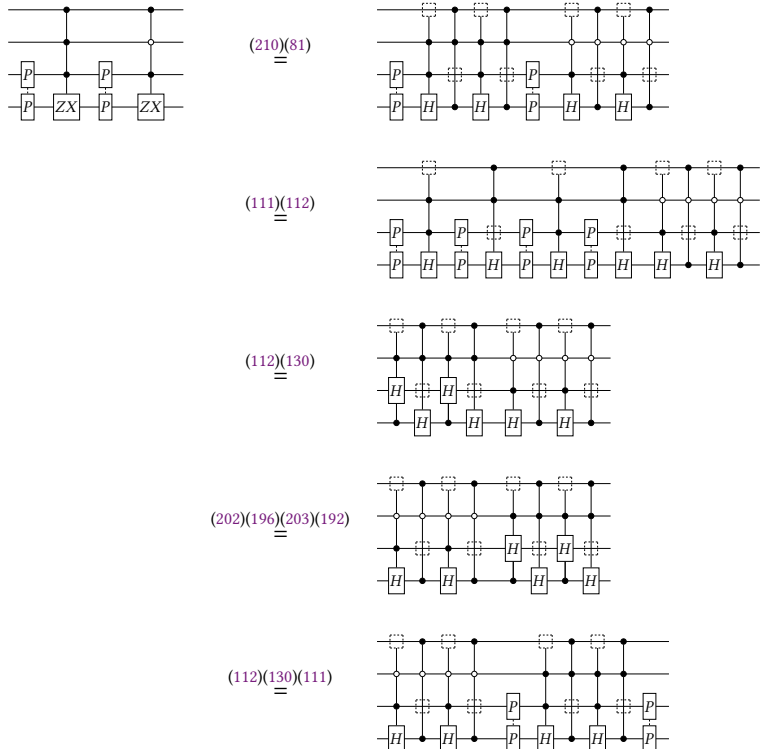

Then, we treat the case where $\beta = \delta = 1$ and $\gamma = 0$. Again, up to using Lemma B.6 (together with Equations (81), (111) and (208)), we can assume without loss of generality that $a = 0$ and $b = 0$. We have

Then, to treat the cases where $\beta = 0$ and $\delta = 1$, again up to using Lemma B.6 (together with Equations (81), (111) and (208)), we can assume without loss of generality that $a = 0$ and $b = 0$. We have

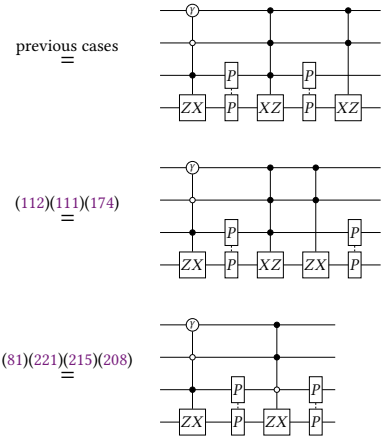

Finally, to treat the cases where $\delta = 0$, again up to using Lemma B.6 (together with Equations (81), (111) and (208)), we can assume without loss of generality that $a = 0$ and $b = 0$. We have

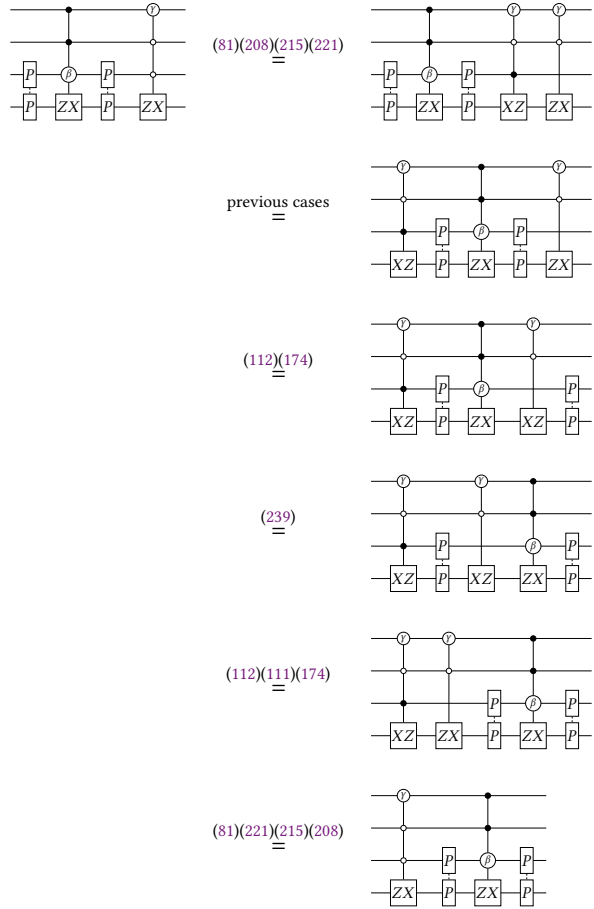

*Proof of Equation* (244): If $\alpha = \beta = 1$, then up to using Lemma B.6 (together with Equations (81), (111) and (208)), we can assume without loss of generality that $a = 0$ and $b = 0$. We have

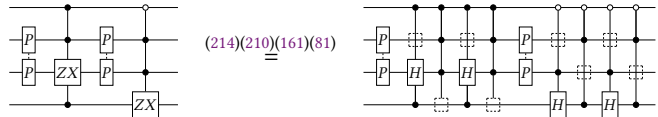

$(112)$
$=$

$(112)(130)$
$=$

$(111)(112)(130)$
$=$

$(183)(198)$
$=$

$(111)(112)(130)$
$=$

$(112)(130)$
$=$

$(157)(111)$
$=$

$(184)(206)$
$=$

$(157)$
$=$

$(112)(130)$
$=$

$(112)$
$=$

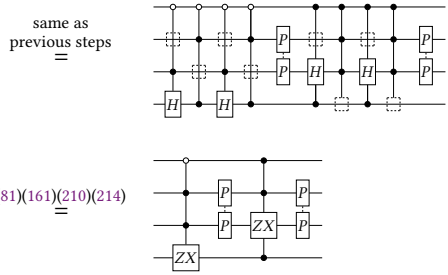

If $\beta = 1$ and $\alpha = 0$, then again, up to using Lemma B.6 (together with Equations (81), (111) and (208)), we can assume without loss of generality that $a = 0$ and $b = 0$. We have

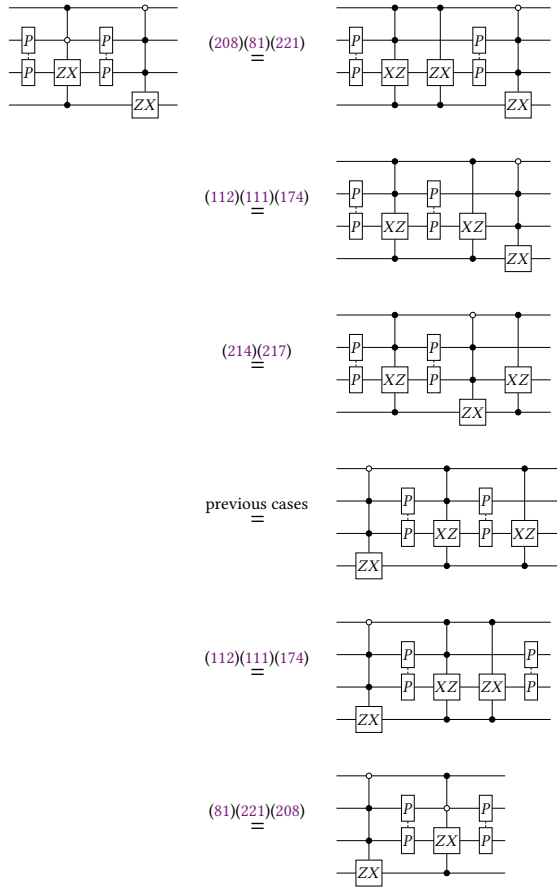

And if $\beta = 0$, then we have

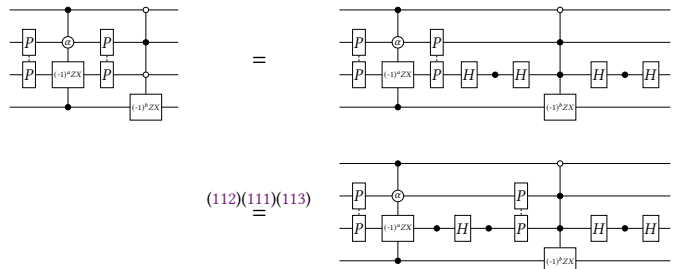

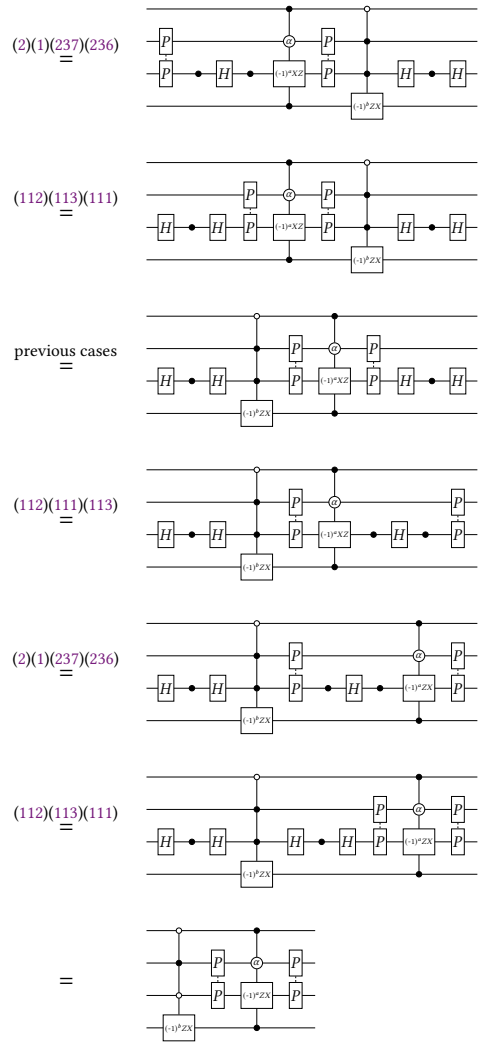

*Proof of Equation* (245): If $\alpha = \beta = 1$, then up to using Lemma B.6 (together with Equations (81), (111) and (208)), we can assume without loss of generality that $a = 0$ and $b = 0$. We have

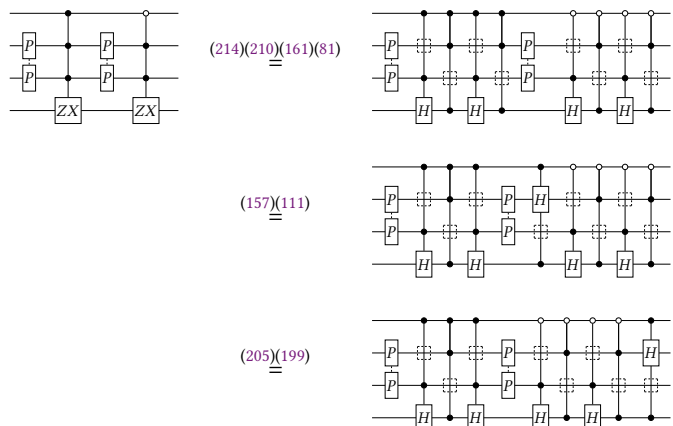

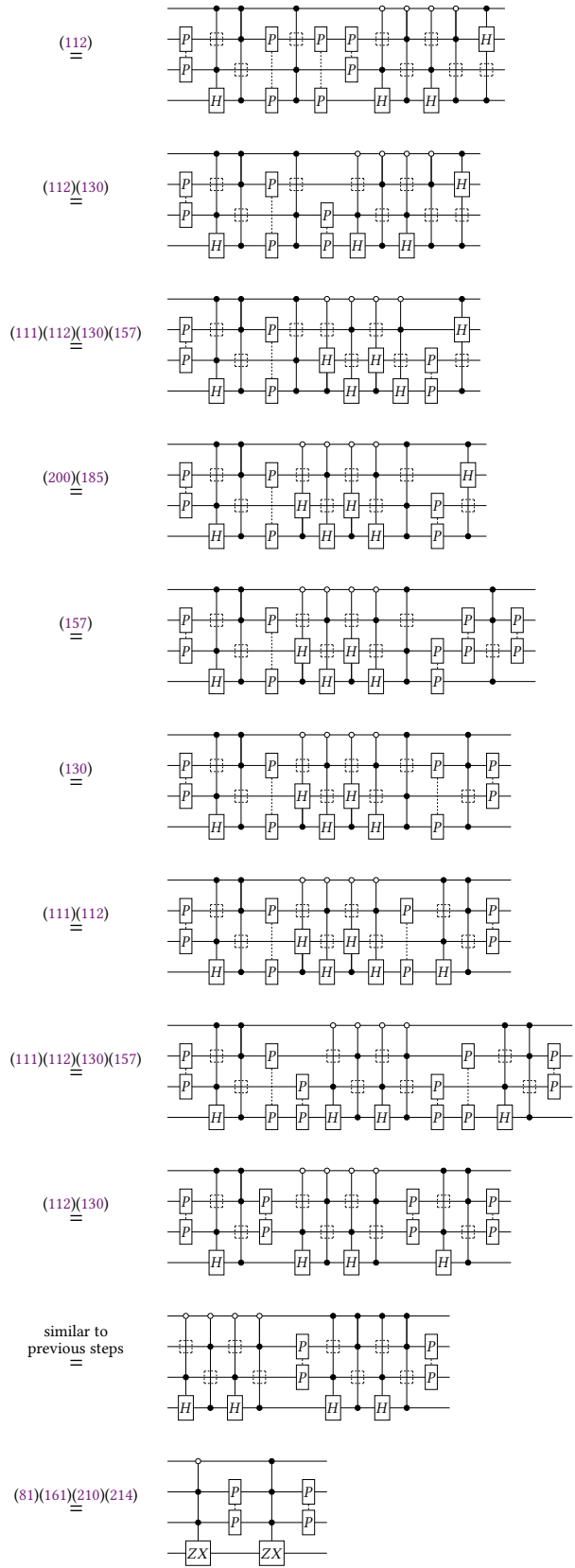

If $\alpha = 1$ and $\beta = 0$, then then again, up to using Lemma B.6 (together with Equations (81), (111) and (208)), we can assume without loss of generality that $a = 0$ and $b = 0$. We have

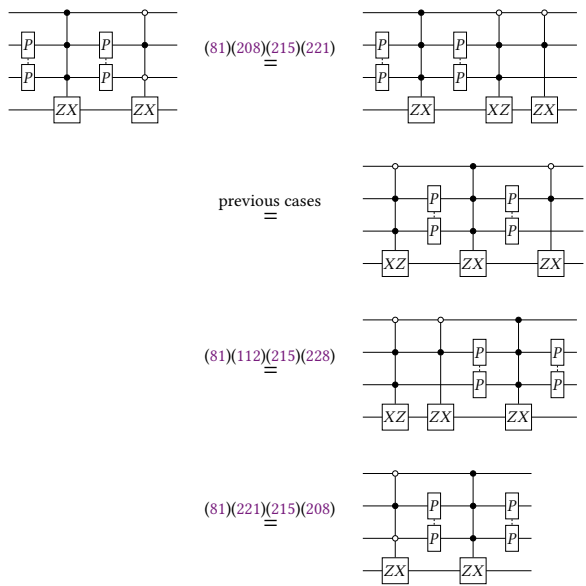

Finally, if $\alpha = 0$, then again, up to using Lemma B.6 (together with Equations (81), (111) and (208)), we can assume without loss of generality that $a = 0$ and $b = 0$. We have

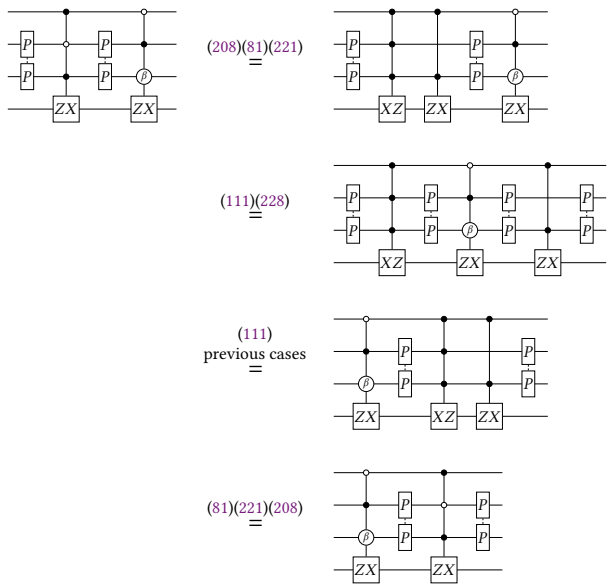

*Proof of Equation* (246)*:*

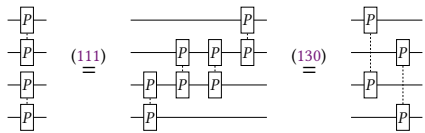

*Proof of Equation* (247)*:* If $\gamma = 1$, then we have

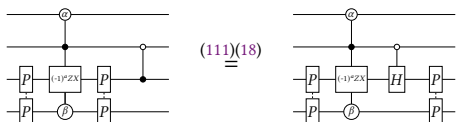

$$\stackrel{(232)(233)}{=}$$ 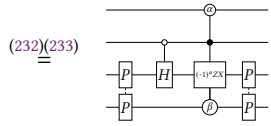

$$\stackrel{(18)}{=}$$ 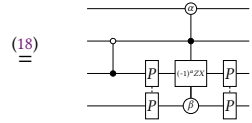

And if $\gamma = 0$, then we have

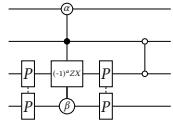 $\stackrel{\text{Lemma } 7.6}{=}$ 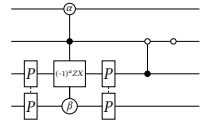

$$\stackrel{\text{previous case}}{=}$$ 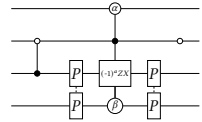

$$\stackrel{(229)}{=}$$ 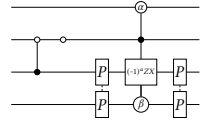

$$\stackrel{\text{Lemma } 7.6}{=}$$ 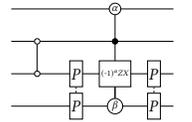

*Proof of Equation* (248): If $\gamma = 1$ and $\beta = 1$, then we have

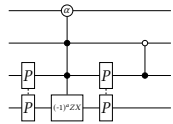 $\stackrel{(111)(18)}{=}$ 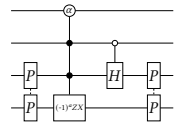

$$\stackrel{(234)(235)}{=}$$ 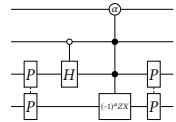

$$\stackrel{(18)}{=}$$ 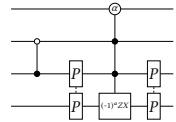

If $\gamma = 1$ and $\beta = 0$, then up to using Lemma B.6 (together with Equations (81), (111) and (208)), we can assume without loss of generality that $a = 0$. We have

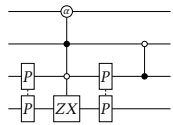 $\stackrel{(81)(208)(215)(221)}{=}$ 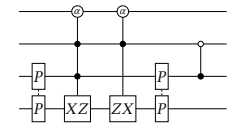

$$\underset{\equiv}{(111)(18)}$$ 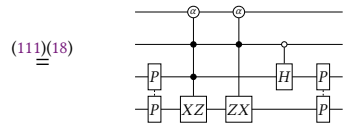

$$\underset{\equiv}{(230)(235)}$$ 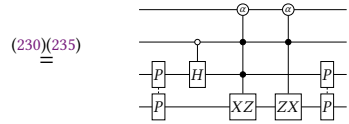

$$\underset{\equiv}{(18)}$$ 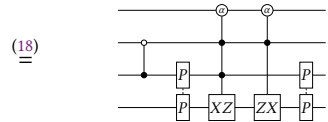

$$\underset{\equiv}{(81)(221)(215)(208)}$$ 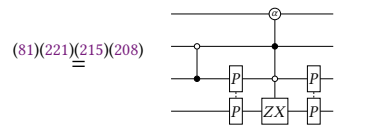

And if $\gamma = 0$, then we have

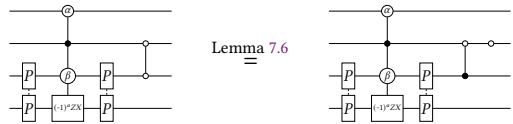

$$\underset{\equiv}{\text{Lemma 7.6}}$$ 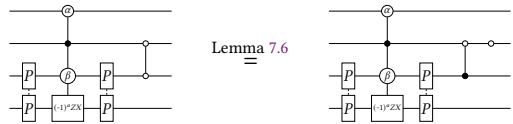

$$\underset{\equiv}{\text{previous cases}}$$ 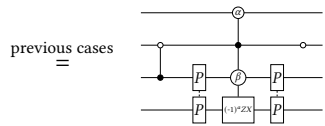

$$\underset{\equiv}{(229)}$$ 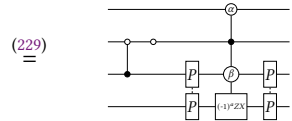

$$\underset{\equiv}{\text{Lemma 7.6}}$$ 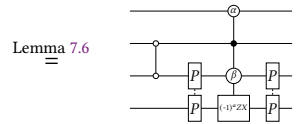

## D.4 Auxiliary Equations on 5 Qubits

Similarly to Appendix D.3, we will use the 4-qubit version of Lemma 5.1:

**Lemma D.6 (Conditional 4-qubit completeness).** *For any 4-qubit quantum circuits $C_1, C_2$ over $\left\{ \boxed{H}, \bullet, \vdots, \boxed{H} \right\}$, if $[\![C_1]\!] = [\![C_2]\!]$ then* $QC \vdash (\underline{\quad} \otimes C_1) = (\underline{\quad} \otimes C_2)$ *and* $QC \vdash (C_1 \ \underline{\quad}) = (C_2 \ \underline{\quad})$.

**Lemma D.7.** *The following equations are consequences of the equations of Figure 4:*

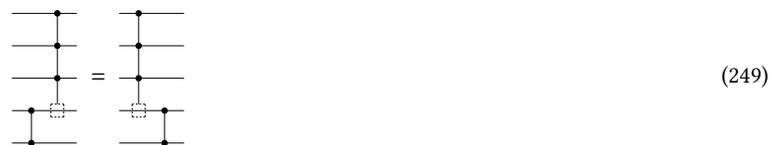

$$(249)$$

$$(250)$$

$$(251)$$

$$(252)$$

$$(253)$$

$$(254)$$

$$(255)$$

$$(256)$$

$$(257)$$

$$(258)$$

$$(259)$$

$$(260)$$

(261)

(262)

(263)

(264)

(265)

(266)

(267)

(268)

PROOF.

*Proof of Equation (249):*

(2)(118)(166)

(125)

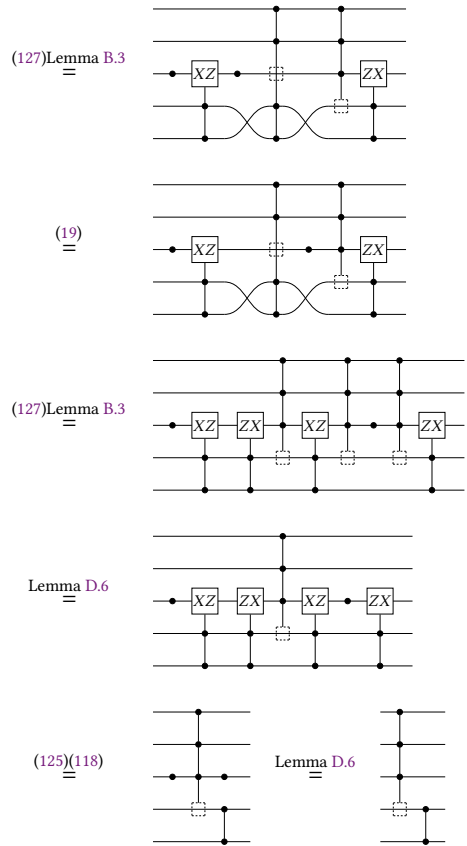

*Proof of Equation* (250):

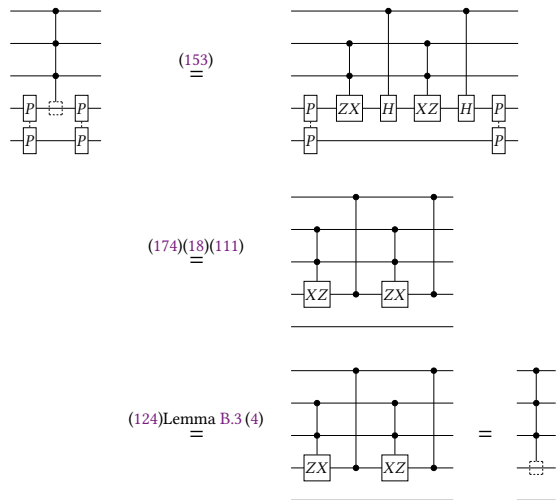

*Proof of Equation* (251):

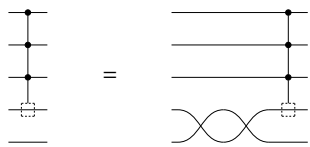

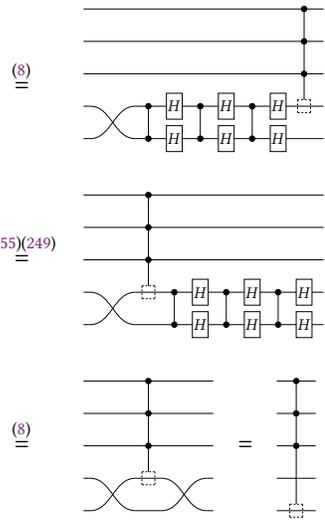

**Proof of Equation** (252):

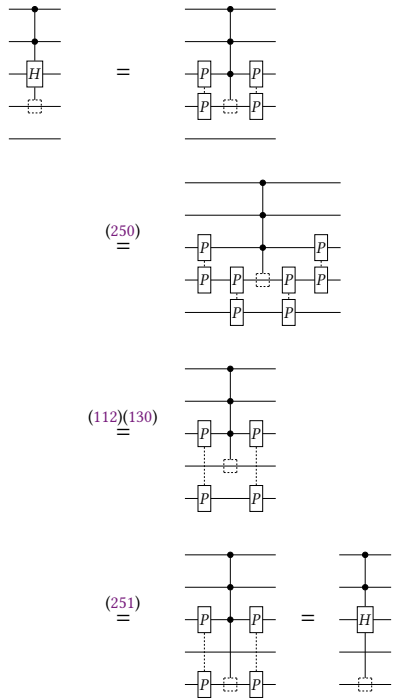

**Proof of Equation** (253):

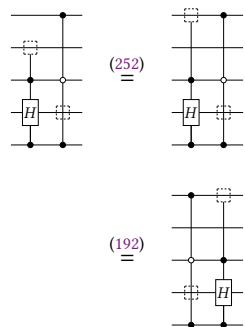

$\overset{(252)}{=}$ 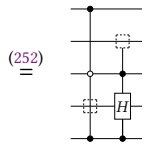

*Proof of Equation* (254):

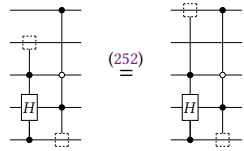

$\overset{(182)}{=}$ 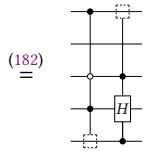

$\overset{(252)}{=}$ 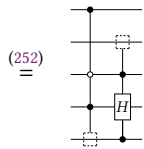

*Proof of Equation* (255):

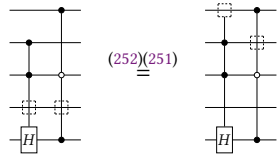

$\overset{(186)}{=}$ 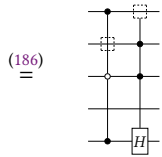

$\overset{(251)(252)}{=}$ 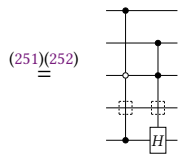

*Proof of Equation* (256):

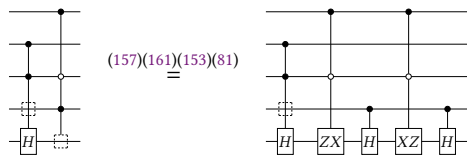

$\overset{(252)}{=}$ 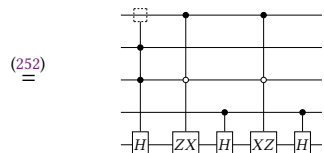

$\overset{\text{Lemma D.6}}{=}$ 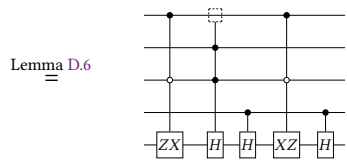

$\overset{(252)}{=}$ 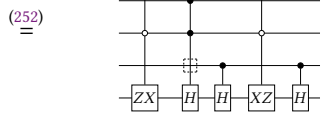

$\overset{\text{Lemma D.6}}{=}$ 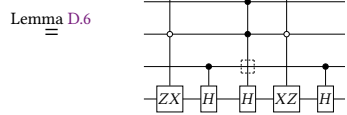

$\overset{(252)}{=}$ 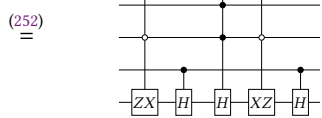

$\overset{\text{Lemma D.6}}{=}$ 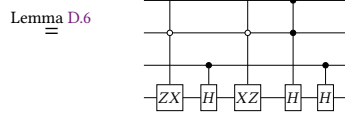

$\overset{(252)}{=}$ 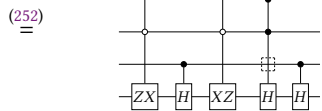

$\overset{\text{Lemma D.6}}{=}$ 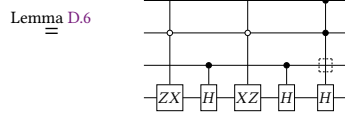

$\overset{(81)(153)(157)(161)}{=}$ 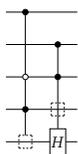

*Proof of Equation* (257):

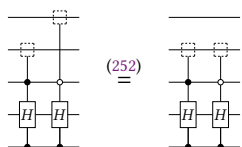

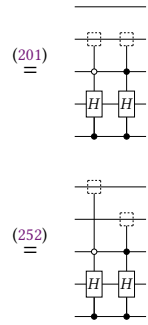

*Proof of Equation* (258):

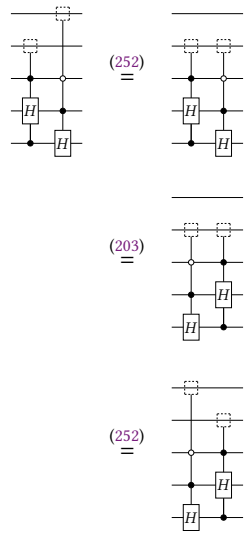

*Proof of Equation* (259):

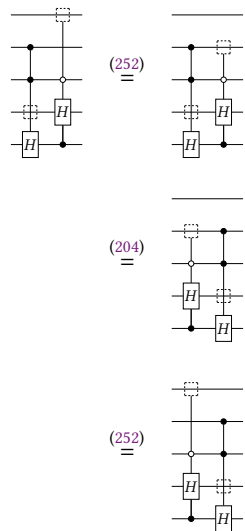

*Proof of Equation* (260):

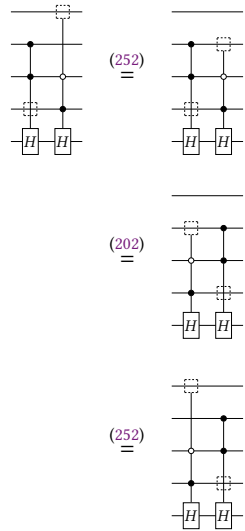

*Proof of Equation* (261):

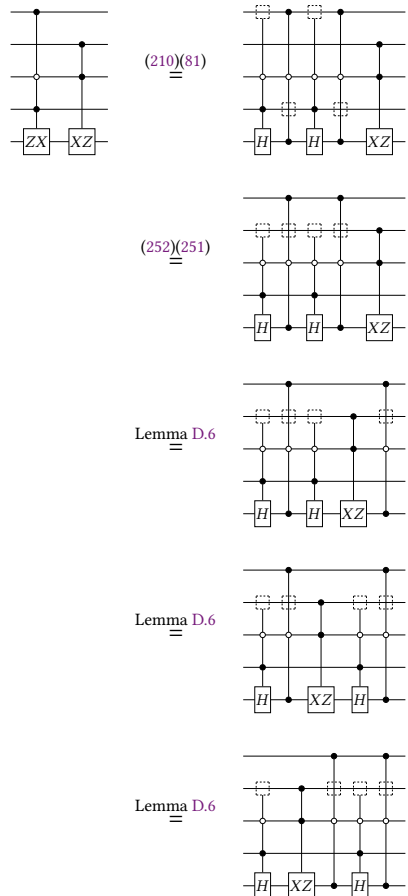

Lemma D.6
$=$

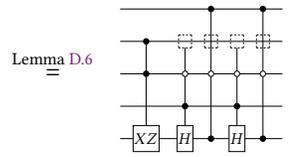

(252)(251)
$=$

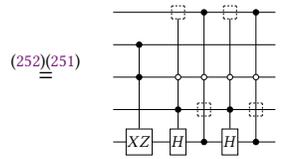

(81)(210)
$=$

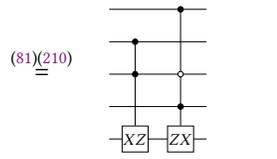

*Proof of Equation* (262):

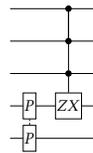
(210)
$=$
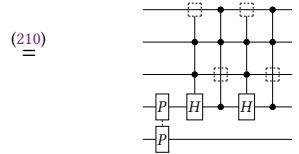

(252)(251)
$=$

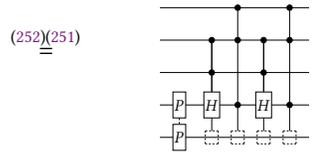

$=$

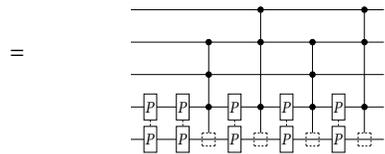

(111)
$=$

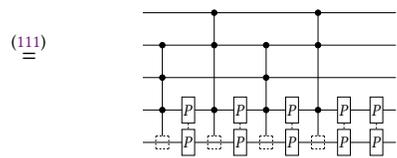

$=$

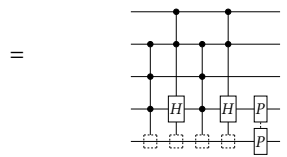

(251)(252)
$=$

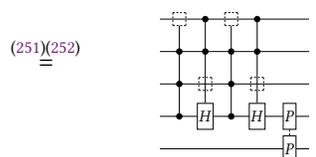

$$(161)(162)(211)(214)(215) \underset{=}{}$$

*Proof of Equation* (263):

$$(112)(111)(262) \underset{=}{}$$

$$(215)(112)(244) \underset{=}{}$$

$$(112)(111)(262) \underset{=}{}$$

*Proof of Equation* (264):

$$(112)(246) \underset{=}{}$$

$$(112)(111)(262) \underset{=}{}$$

$$(244) \underset{=}{}$$

$$(112)(111)(262) \underset{=}{}$$

$$(246)(112) \underset{=}{}$$

*Proof of Equation* (265):

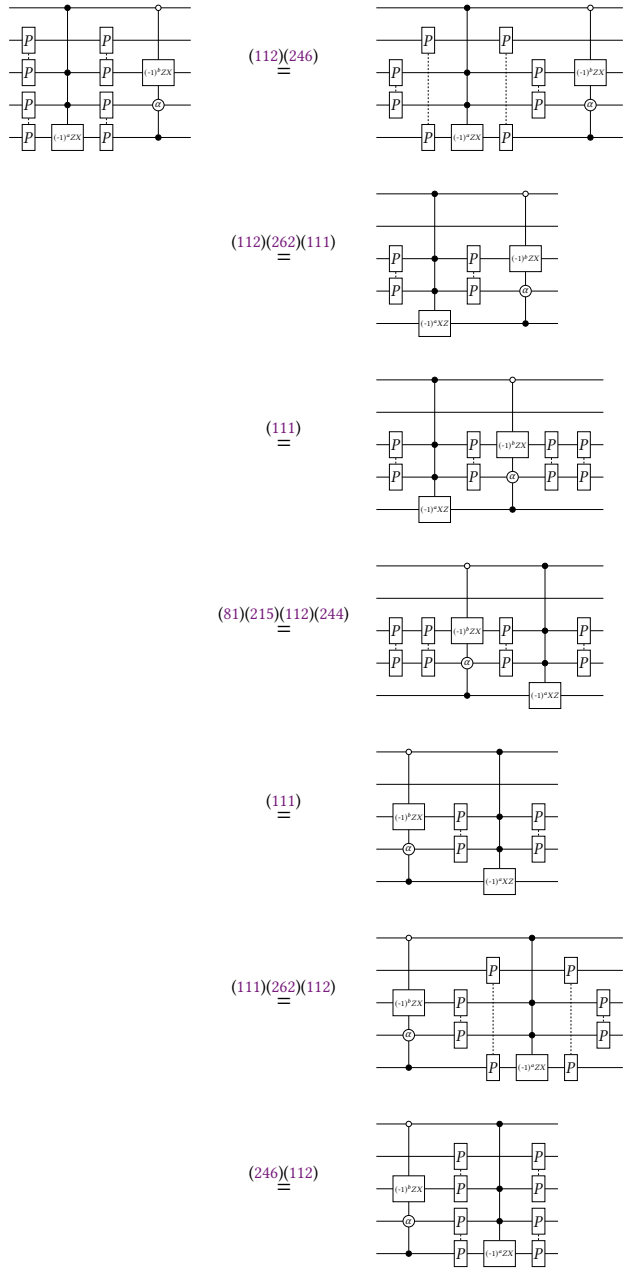

*Proof of Equation* (266):

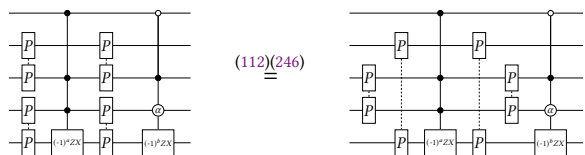

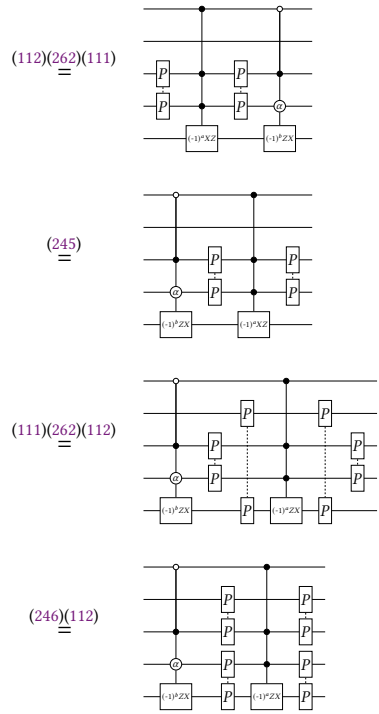

*Proof of Equation* (267): We can assume without loss of generality that $\alpha = 1$.

If $\beta = \gamma = 1$, then up to using Lemma B.6 (together with Equations (81), (111) and (208)), we can assume without loss of generality that $a = 0$ and $b = 1$. We have

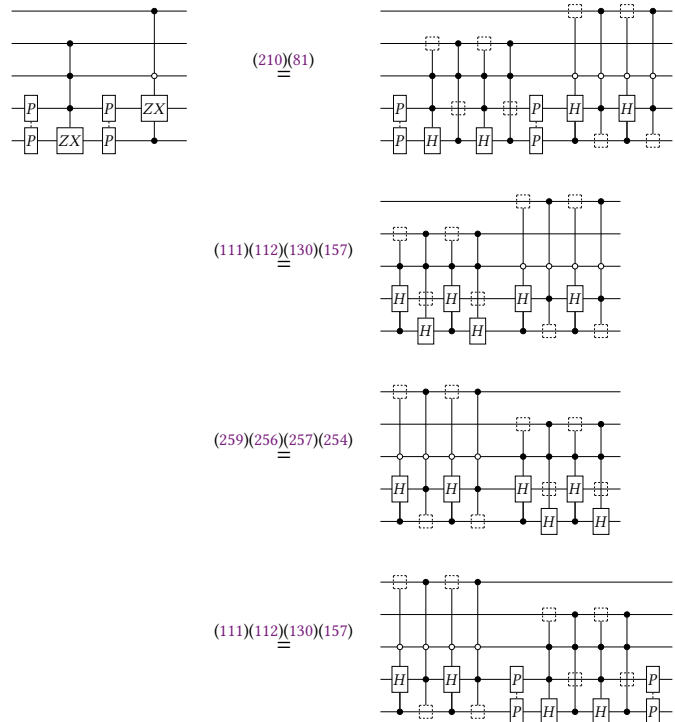

If $\gamma = 1$ and $\beta = 0$, then we have

And if $\gamma = 0$, then we have

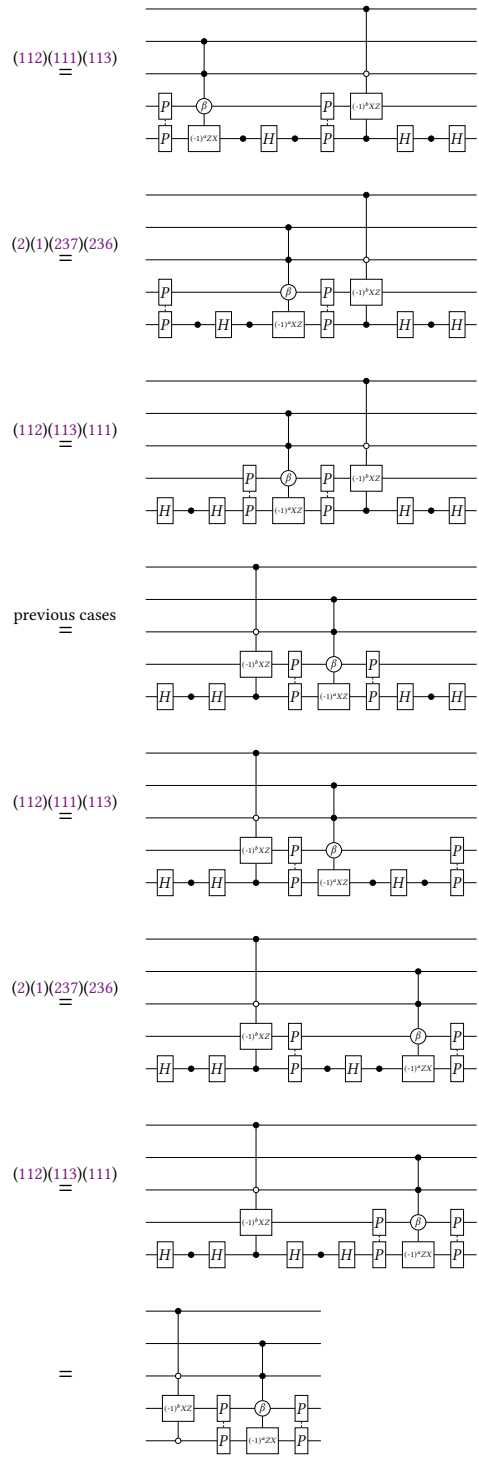

*Proof of Equation* (268)*:* We can assume without loss of generality that $\alpha = 1$.

If $\beta = \gamma = 1$, then up to using Lemma B.6 (together with Equations (81), (111) and (208)), we can assume without loss of generality that $a = 0$ and $b = 1$. We have

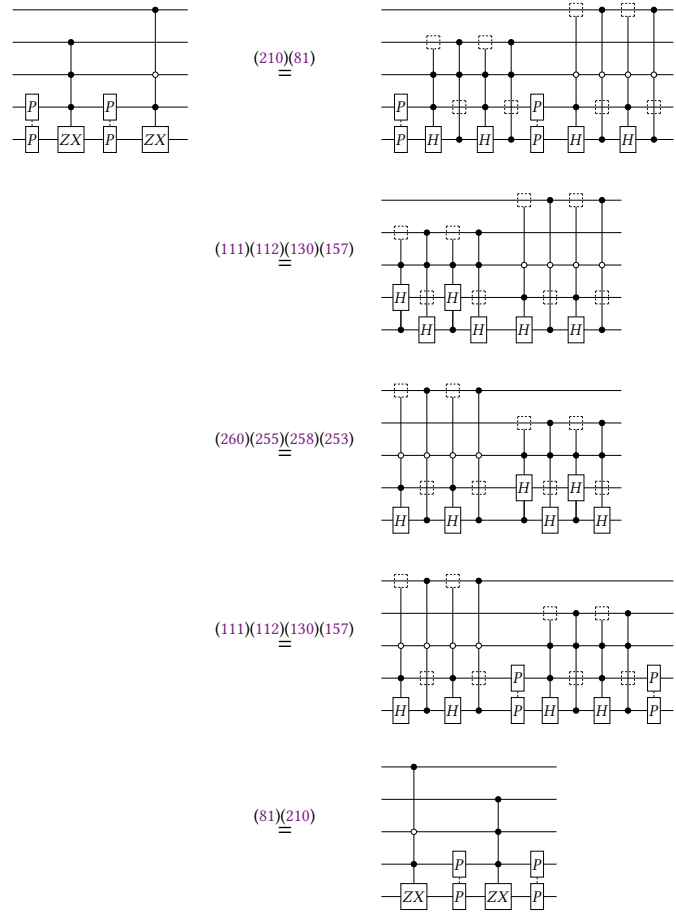

If $\gamma = 1$ and $\beta = 0$, then again, up to using Lemma B.6 (together with Equations (81), (111) and (208)), we can assume without loss of generality that $a = 0$ and $b = 1$. We have

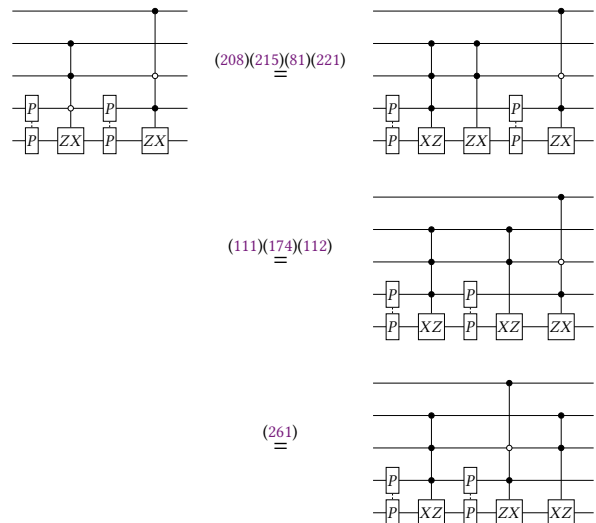

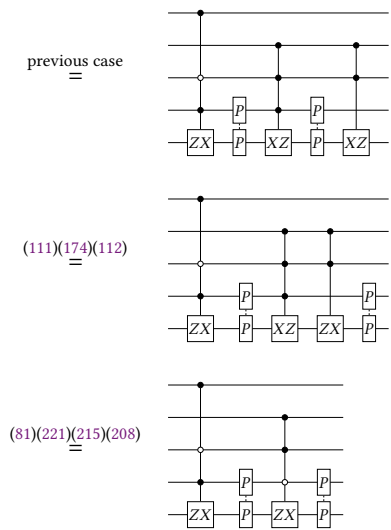

And if $\gamma = 0$, then again, up to using Lemma B.6 (together with Equations (81), (111) and (208)), we can assume without loss of generality that $a = 0$ and $b = 1$. We have

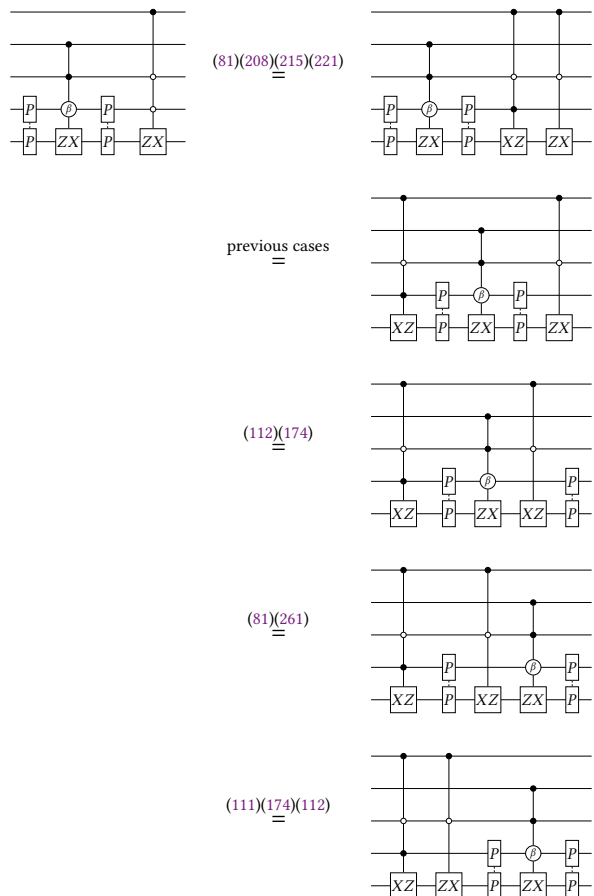

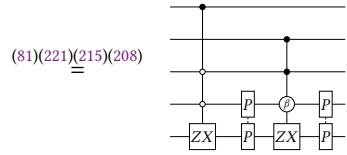

$$\overset{(81)(221)(215)(208)}{=}$$

□

## D.5 Basic Auxiliary Equations on $n$ Qubits and Properties of Multi-Controlled Gates

In the following, for convenience, we will use the notation $(XX)_k$ to refer to the $k$-qubit instance of a multi-qubit equation (schema) (XX). Note that in the derivations, we will sometimes invoke the $n{-}1$-qubit case of equations that are proved later: this should be understood as just a call to Lemma 5.1, the equation number being here only to be more explicit regarding the transformation performed. One can verify that the proof is well-founded by checking that all the references to $n$-qubit instances of equations, denoted by $(XX)_n$, refer to previously proven equations.

Additionally, we will sometimes refer to Lemmas D.3 and D.6 when making a 3- or 4-qubit transformation: this is also a call to Lemma 5.1 (which in particular requires the number of qubits to be strictly greater than 3, resp. 4).

**Lemma D.8.** *The following equations are consequences of the equations of Figure 4:*

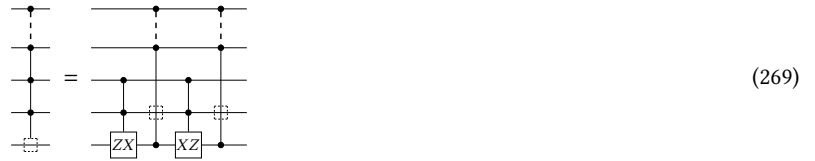

(269)

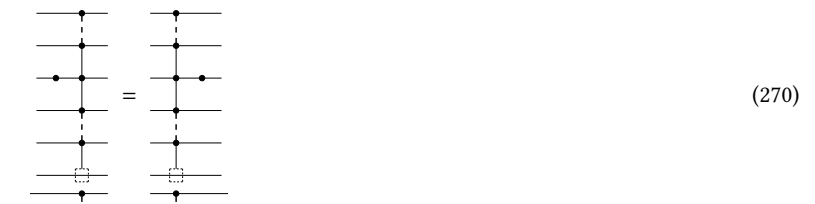

(270)

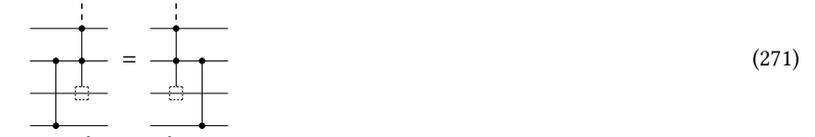

(271)

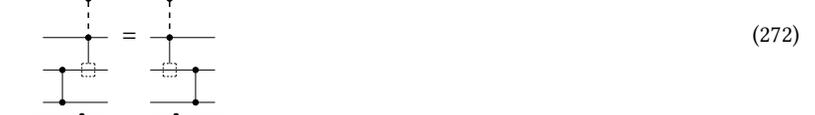

(272)

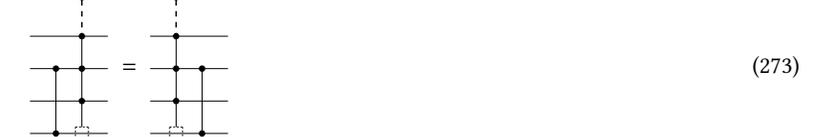

(273)

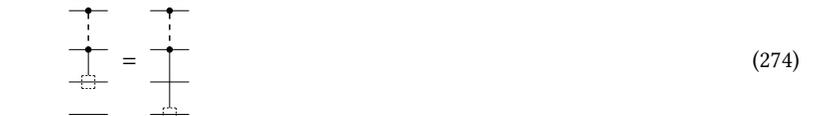

(274)

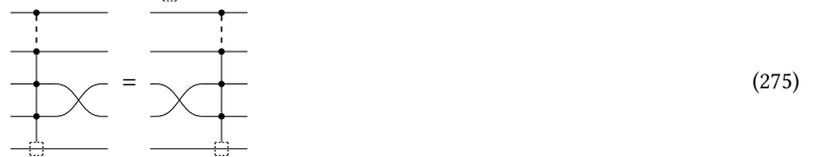

(275)

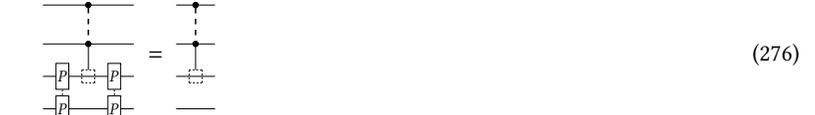

(276)

$$(277)$$

$$(278)$$

$$(279)$$

$$(280)$$

$$(281)$$

$$(282)$$

$$(283)$$

$$(284)$$

$$(285)$$

$$(286)$$

$$(287)$$

$$(288)$$

$$(289)$$

$$(290)$$

$$(291)$$

$$(292)$$

$$(293)$$

$$(294)$$

$$(295)$$

$$(296)$$

$$\tag{297}$$

$$\tag{298}$$

$$\tag{299}$$

$$\tag{300}$$

$$\tag{301}$$

$$\tag{302}$$

$$\tag{303}$$

$$\tag{304}$$

$$\tag{305}$$

$$\tag{306}$$

$$\tag{307}$$

PROOF.

*Proof of Equation* $(269)_n$: Equation $(269)_3$ is just Equation $(126)$, and for $n \geq 4$, Equation $(269)_n$ is just the definition of 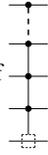.

*Proof of Equation* $(270)_n$: Equation $(270)_2$ is trivial. Equation $(270)_3$ follows from Lemma 7.6.

If $n \geq 4$, then we distinguish three cases depending on the vertical position of the $Z$ gate with respect to the multicontrolled-$Z$ gate:

If there are no controls below the $Z$ gate, then we have

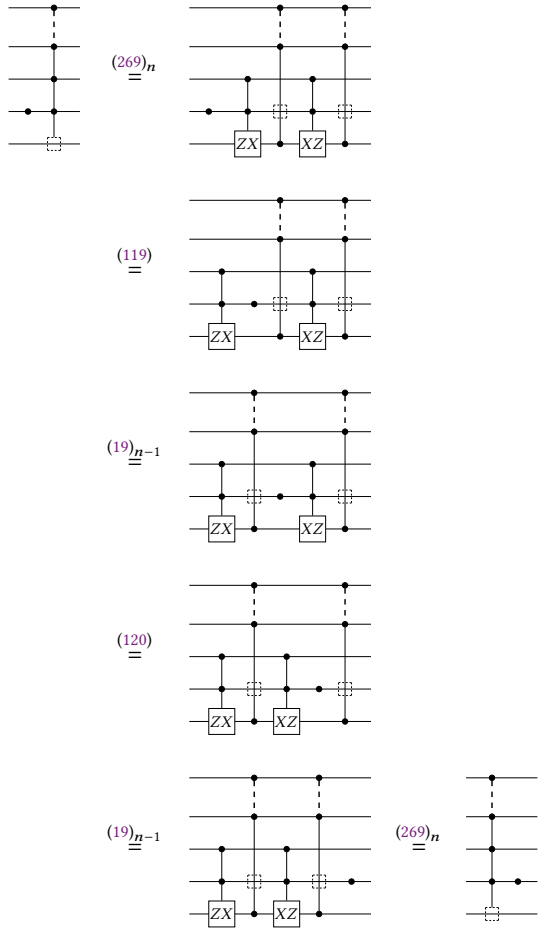

If there is exactly one control below it, then we have

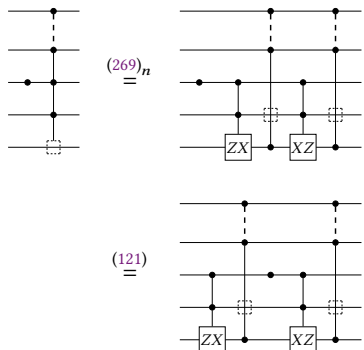

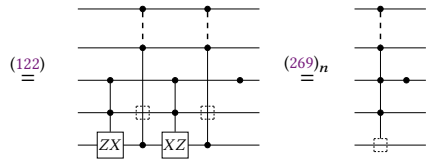

If there are two controls or more below it, then we have

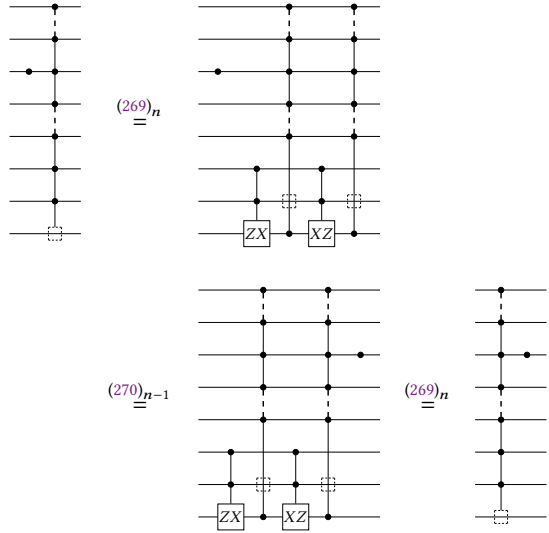

*Proof of Equation* $(271)_n$: Equation $(271)_3$ follows from Lemma 7.6. If $n \geq 4$, then we have

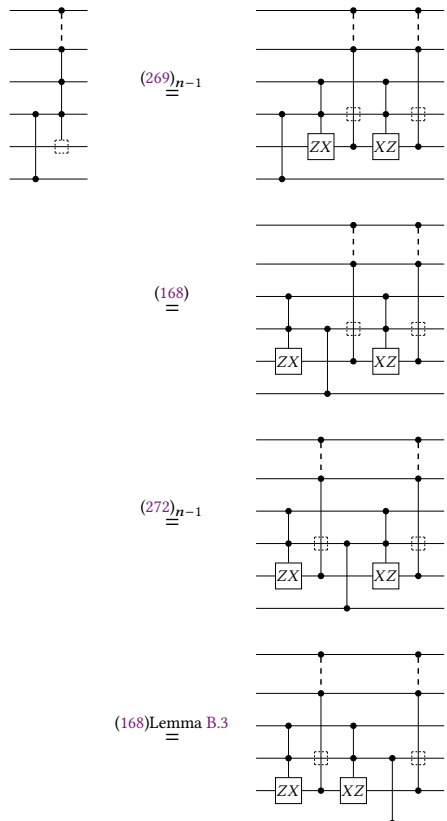

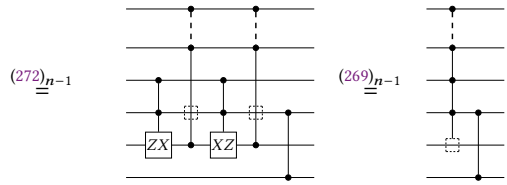

*Proof of Equation* $(272)_n$: Equation $(272)_2$ follows from Lemma 7.6. If $n \geq 3$, then we have

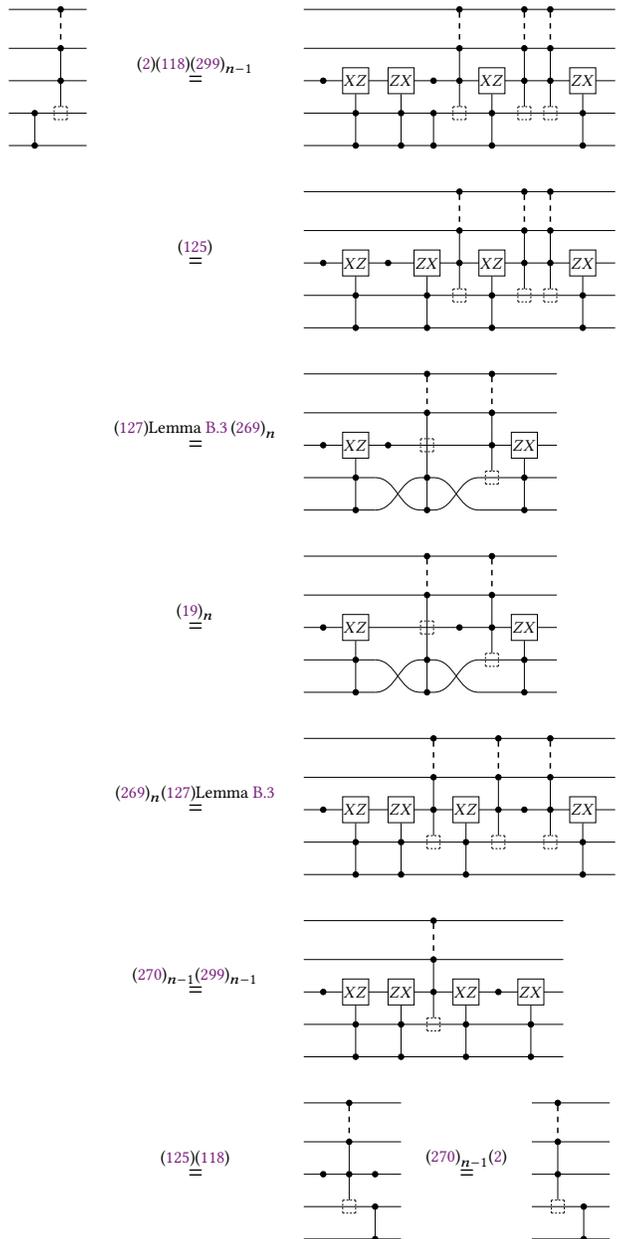

*Proof of Equation* (273)$_n$:

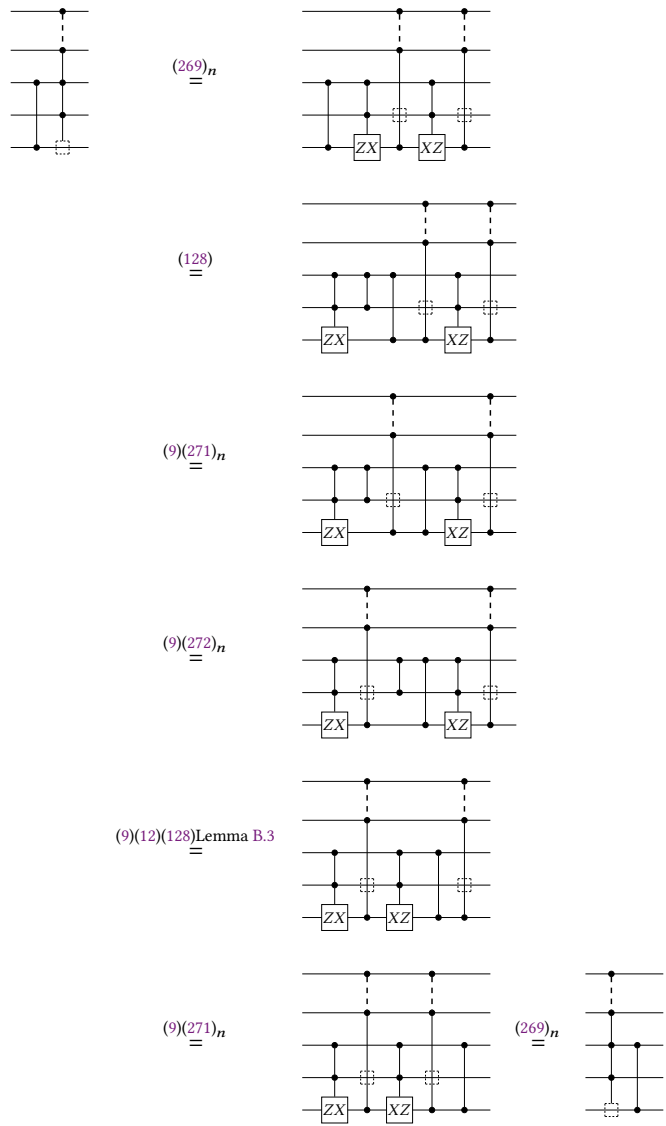

*Proof of Equation* (274)$_n$:

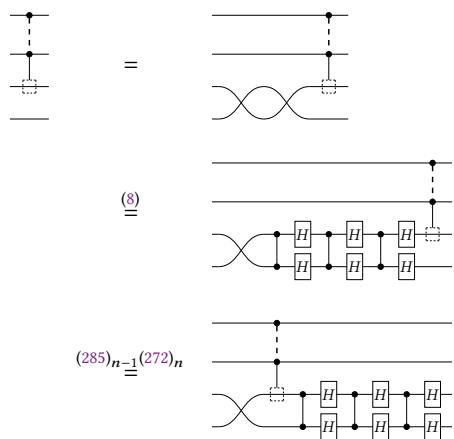

*Proof of Equation* (275)$_n$:

*Proof of Equation* (276)$_n$: Equation (276)$_2$ follows from Lemma 7.6. Equation (276)$_3$ follows from Equation (111). If $n \geq 4$, then we have

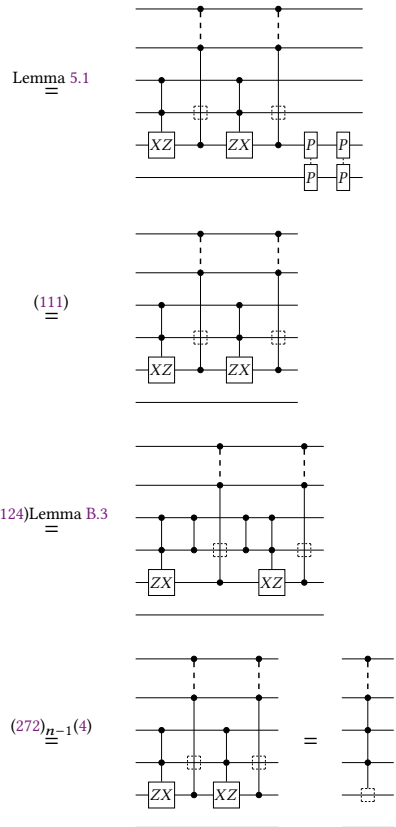

*Proof of Equation* $(277)_n$: Equation $(277)_3$ and Equation $(277)_4$ are trivial. If $n \geq 5$, then we have

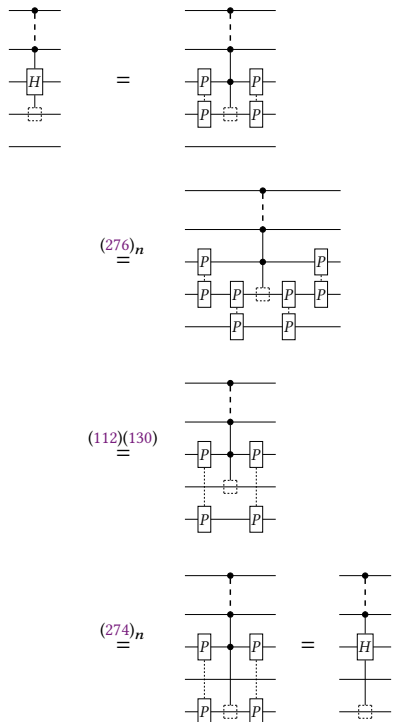

*Proof of Equation* $(278)_n$: Equation $(278)_2$ follows from Lemma 7.6. Equation $(278)_3$ is just Equation $(145)$. For $n \geq 4$, we have

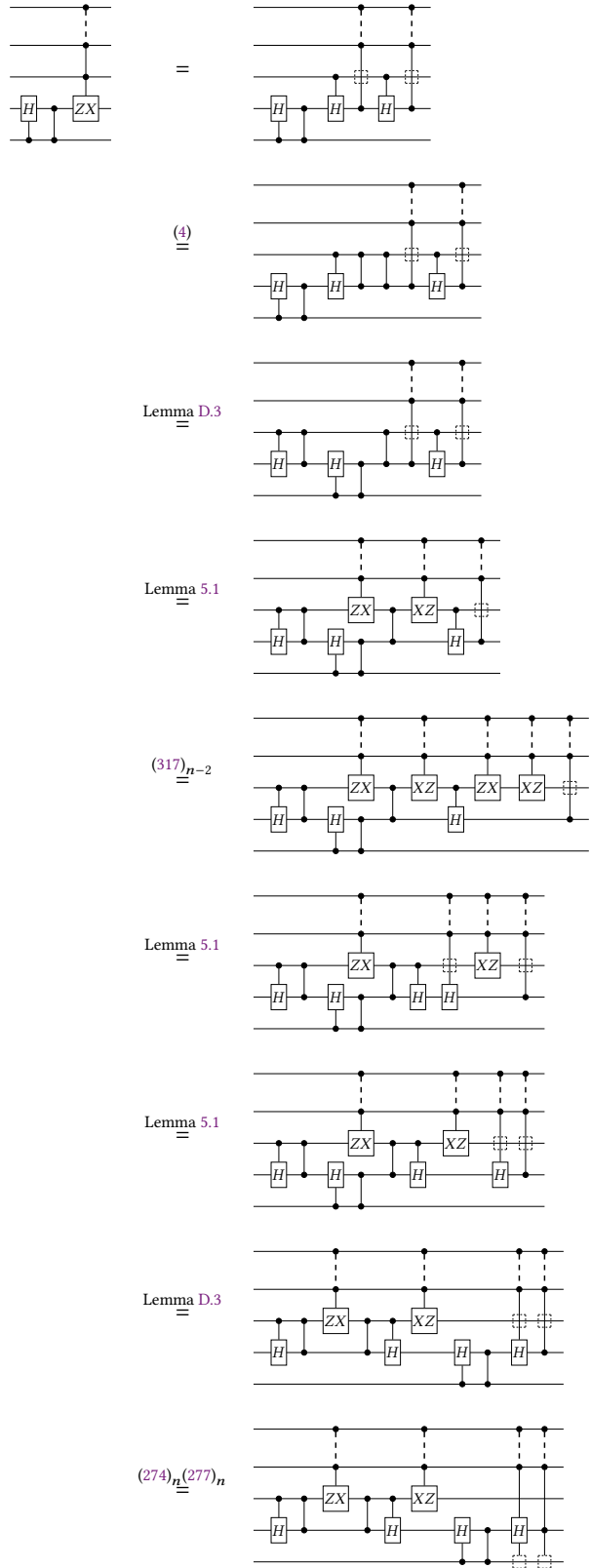

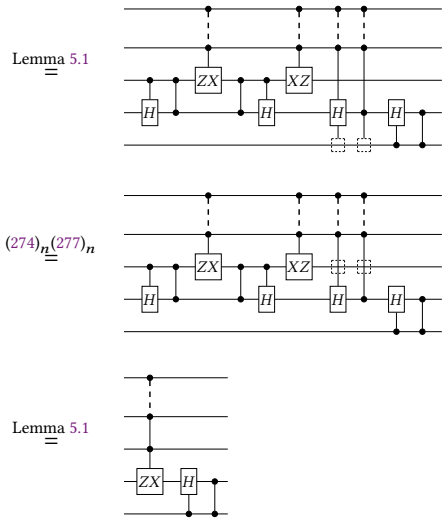

*Proof of Equation* (279)$_n$:

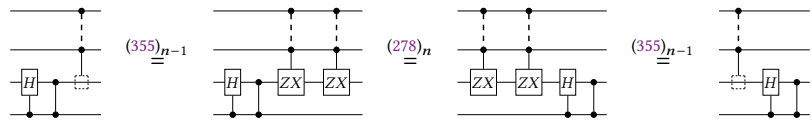

*Proof of Equation* (280)$_n$:

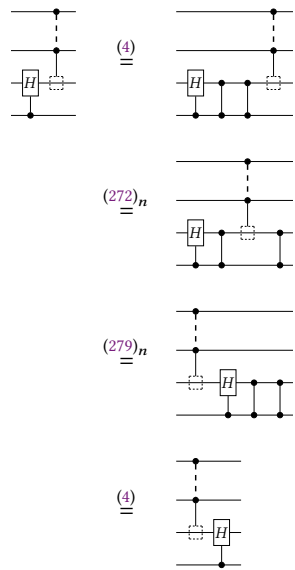

*Proof of Equation* (281)$_n$: Equation (281)$_2$ is trivial. Equation (281)$_3$ follows from Equation (93). If $n \geq 4$, then we have

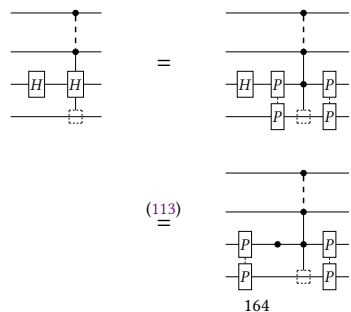

*Proof of Equation* $(282)_n$: Equation $(282)_3$ and Equation $(282)_4$ are trivial. If $n \geq 5$, then we have

*Proof of Equation* $(283)_n$: Equation $(283)_3$ is just Equation $(144)$. For $n \geq 4$, we have

$\overset{(148)}{=}$ 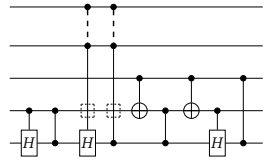

$\overset{(82)}{=}$ 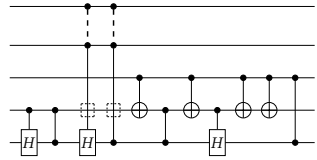

Lemma D.3 $=$ 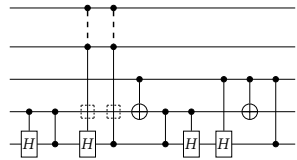

Lemma D.3 $=$ 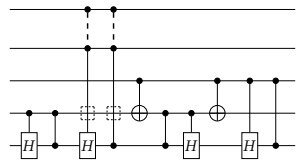

$(285)_{n-1}(303)_{n-1}\overset{(9)}{=}(272)_n(282)_n$ 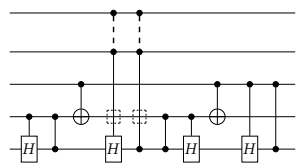

Lemma 5.1 $=$ 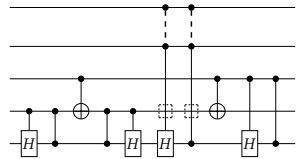

$(285)_{n-1}(303)_{n-1}\overset{(9)}{=}(272)_n(282)_n$ 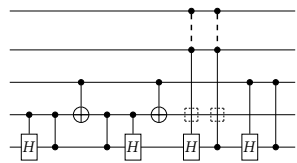

$(274)_n\overset{(277)_n}{=}$ 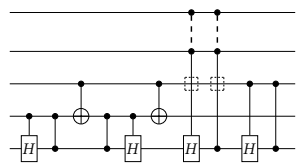

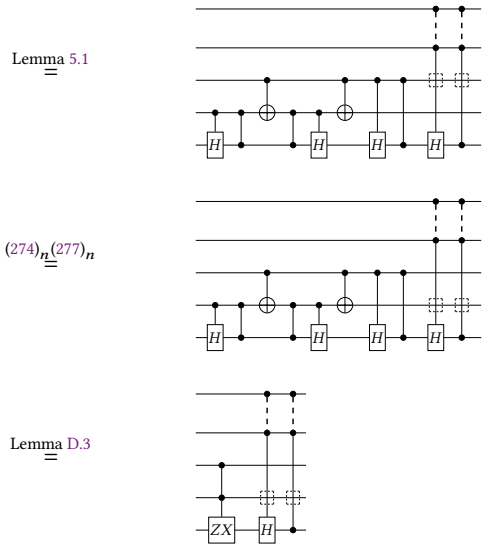

*Proof of Equation* $(284)_n$:

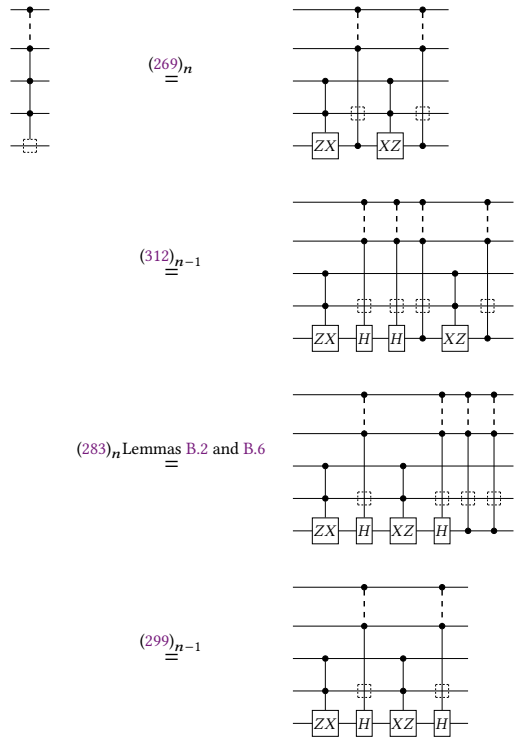

*Proof of Equation* $(285)_n$: Equation $(285)_1$ follows from Lemma 7.1. Equations $(285)_2$ and $(285)_3$ follow from the topological rules. For $n \geq 4$, we have

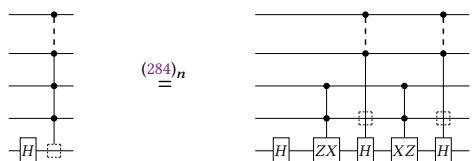

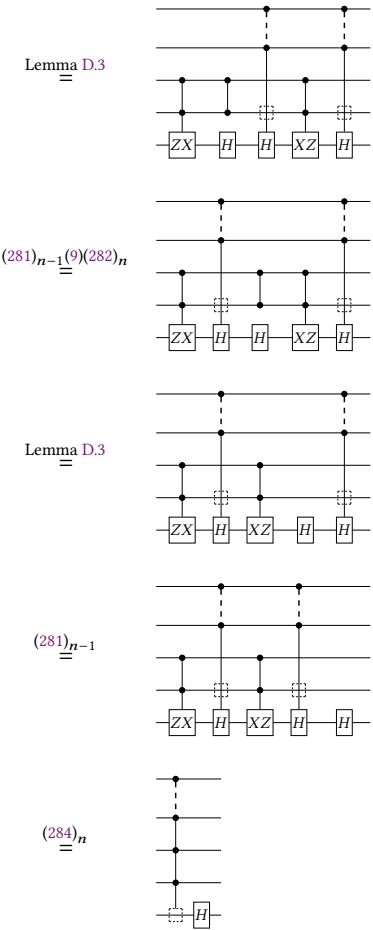

*Proof of Equation* $(286)_n$:

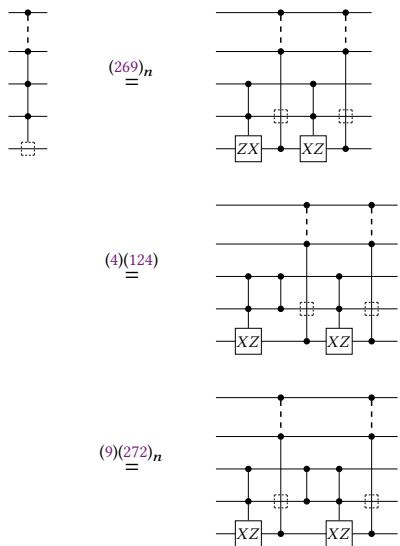

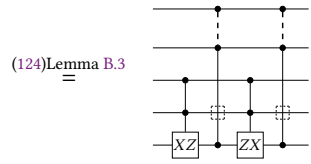

*Proof of Equation* (287)$_n$:

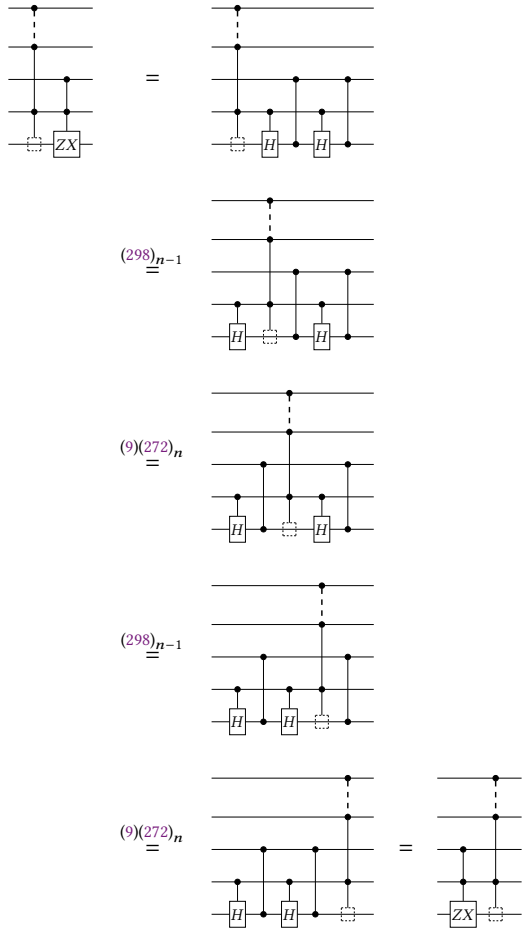

*Proof of Equation* (288)$_n$:

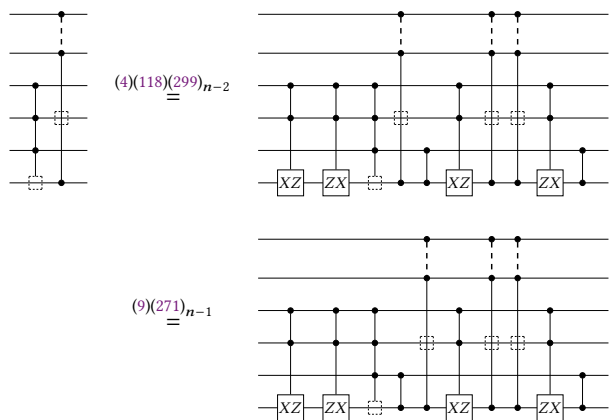

$\overset{(172)}{=}$ 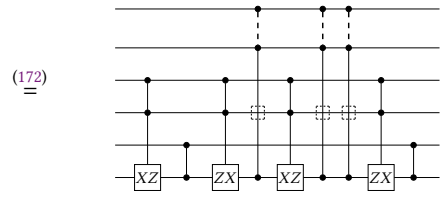

$\overset{(269)}{=}{}_{n-1}$ 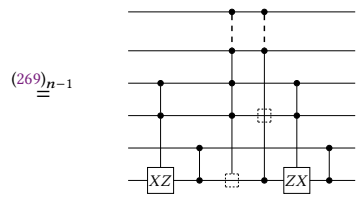

$\overset{(9)(272)}{=}{}_{n}$ 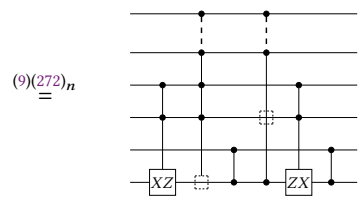

$\overset{(269)}{=}{}_{n-1}$ 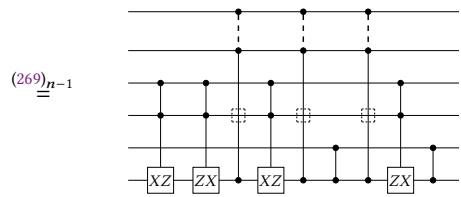

$\overset{(9)(271)}{=}{}_{n-1}$ 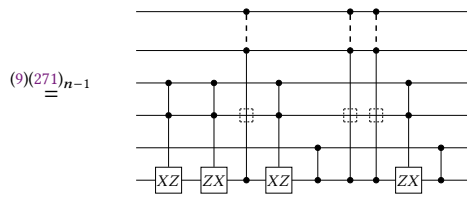

$\overset{(173)}{=}$ 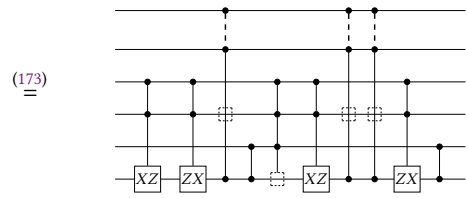

$\overset{(160)}{=}$ 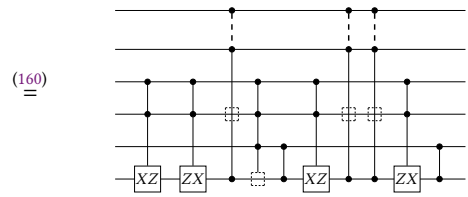

$$\overset{(299)_{n-2}(118)(4)}{=}$$ 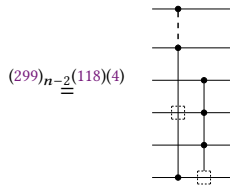

*Proof of Equation* $(289)_n$:

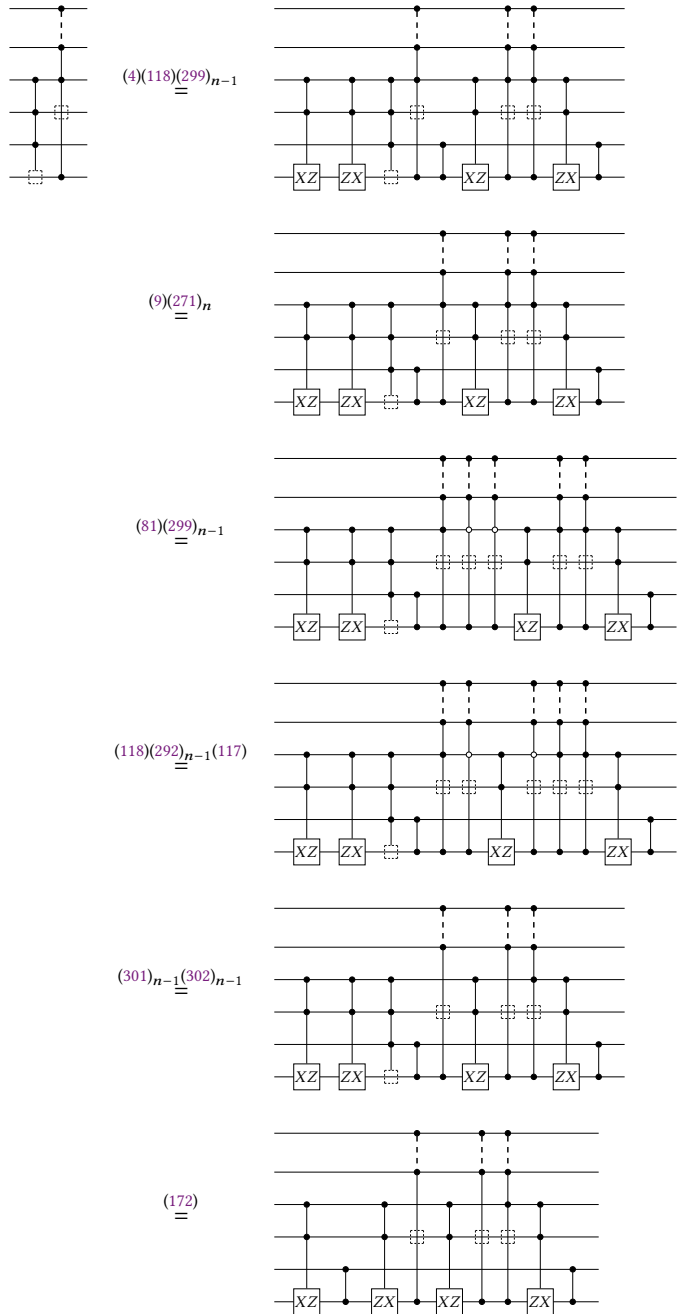

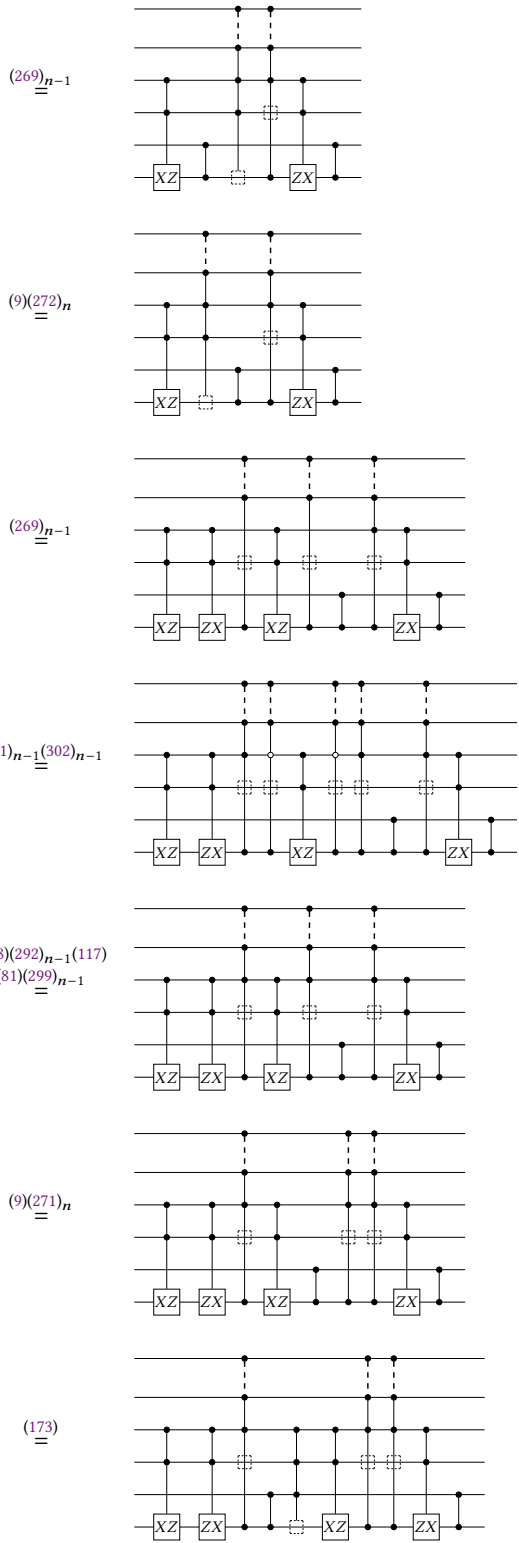

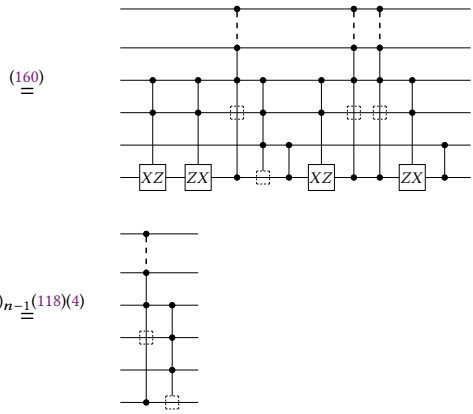

*Proof of Equation* $(290)_n$:

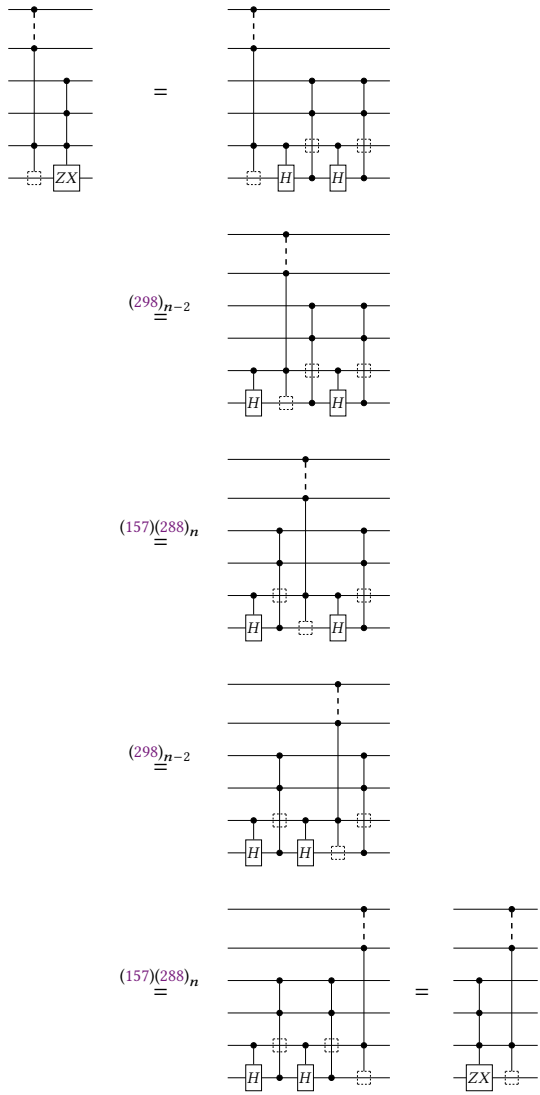

*Proof of Equation* $(291)_n$:

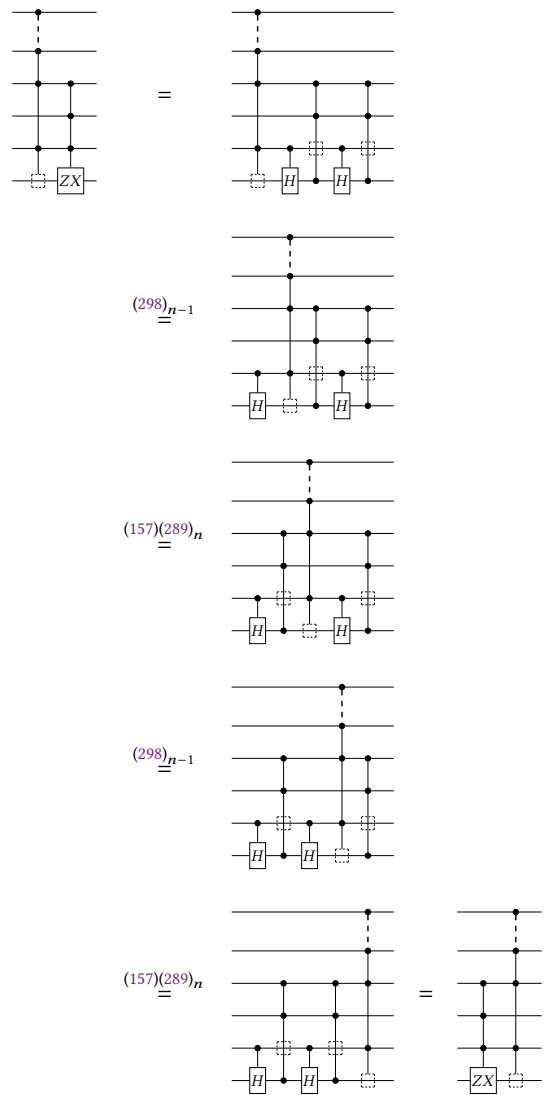

*Proof of Equation* $(292)_n$:

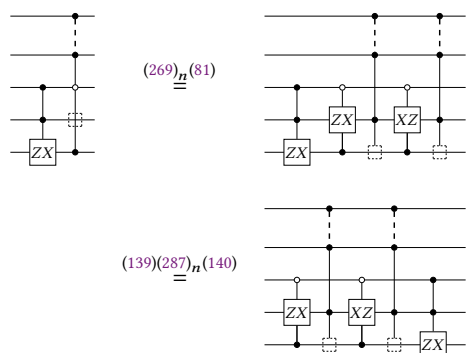

$$\overset{(81)(269)_n}{=}$$ 

*Proof of Equation* (293)$_n$:

$$\overset{(269)_n}{=}$$

$$\overset{(216)}{=}$$

$$\overset{(291)_n}{=}$$

$$\overset{(217)}{=}$$

$$\overset{(291)_n}{=} \qquad \overset{(269)_n}{=}$$

*Proof of Equation* (294)$_n$:

$$\overset{(269)_n}{=}$$

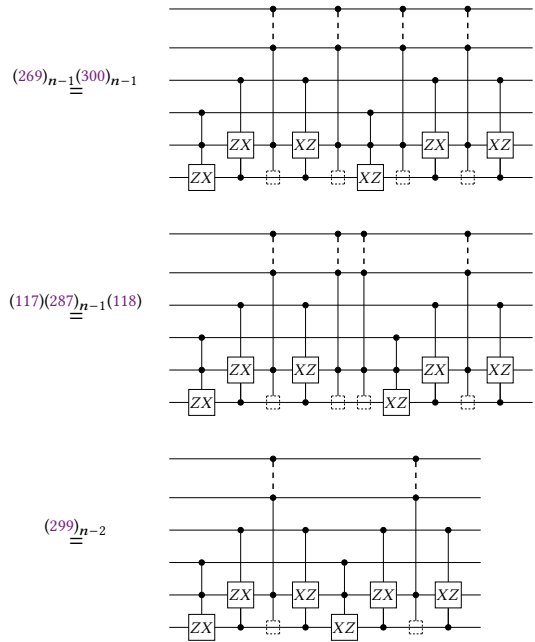

*Proof of Equation* $(295)_n$:

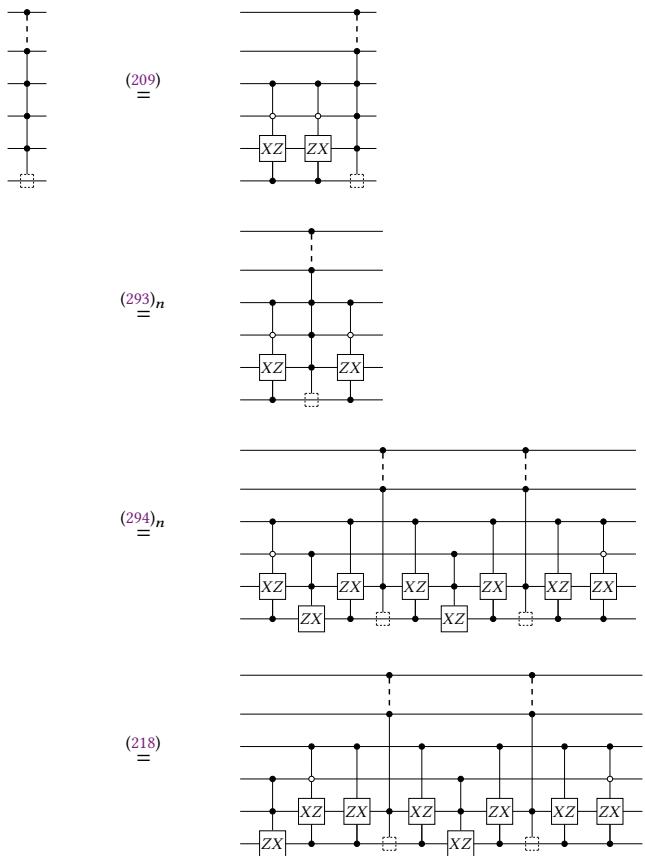

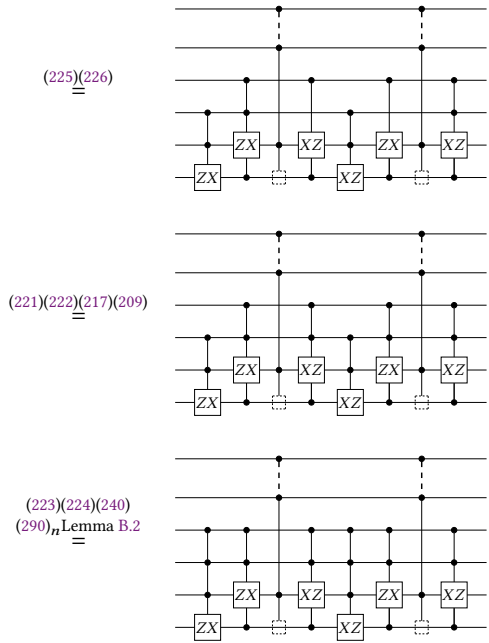

*Proof of Equation* $(296)_n$: Equation $(296)_3$ is just Equation $(132)$. If $n \geq 4$, then to prove Equation $(296)_n$, we have

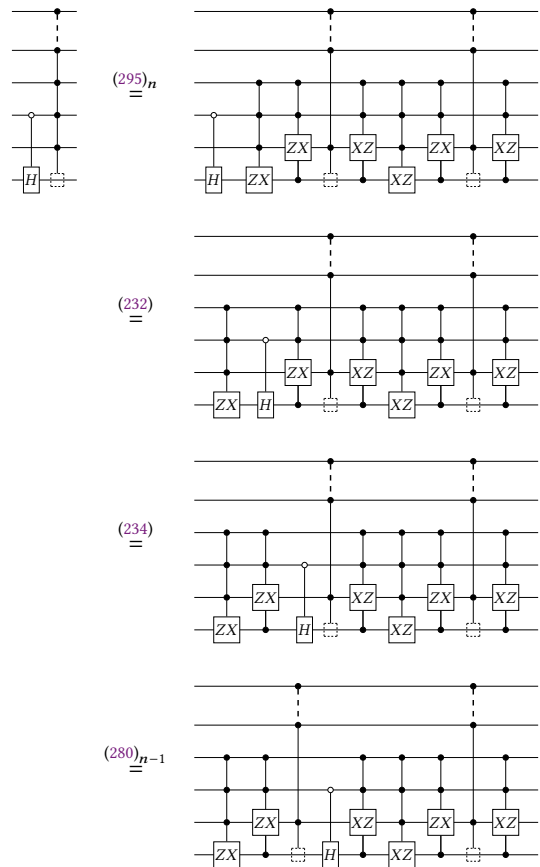

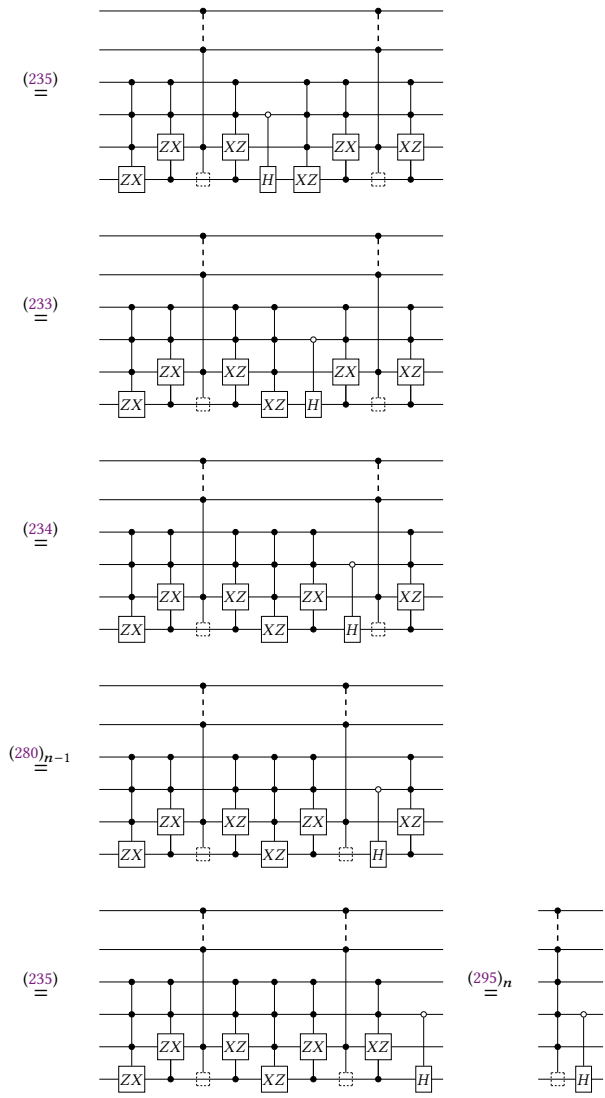

*Proof of Equation* $(297)_n$: Equation $(297)_2$ follows directly from Lemma 7.6, and for $n \geq 3$, Equation $(297)_n$ follows directly from Equations $(275)_n$ and $(296)_n$.

*Proof of Equation* $(298)_n$:

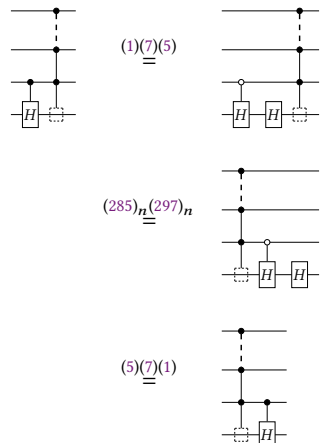

*Proof of Equation* $(299)_n$: Equations $(299)_1$ and $(299)_2$ follow directly from Lemma 7.1. Equation $(299)_3$ follows directly from Lemma 7.6. If $n \geq 4$, then to prove Equation $(299)_n$, we have

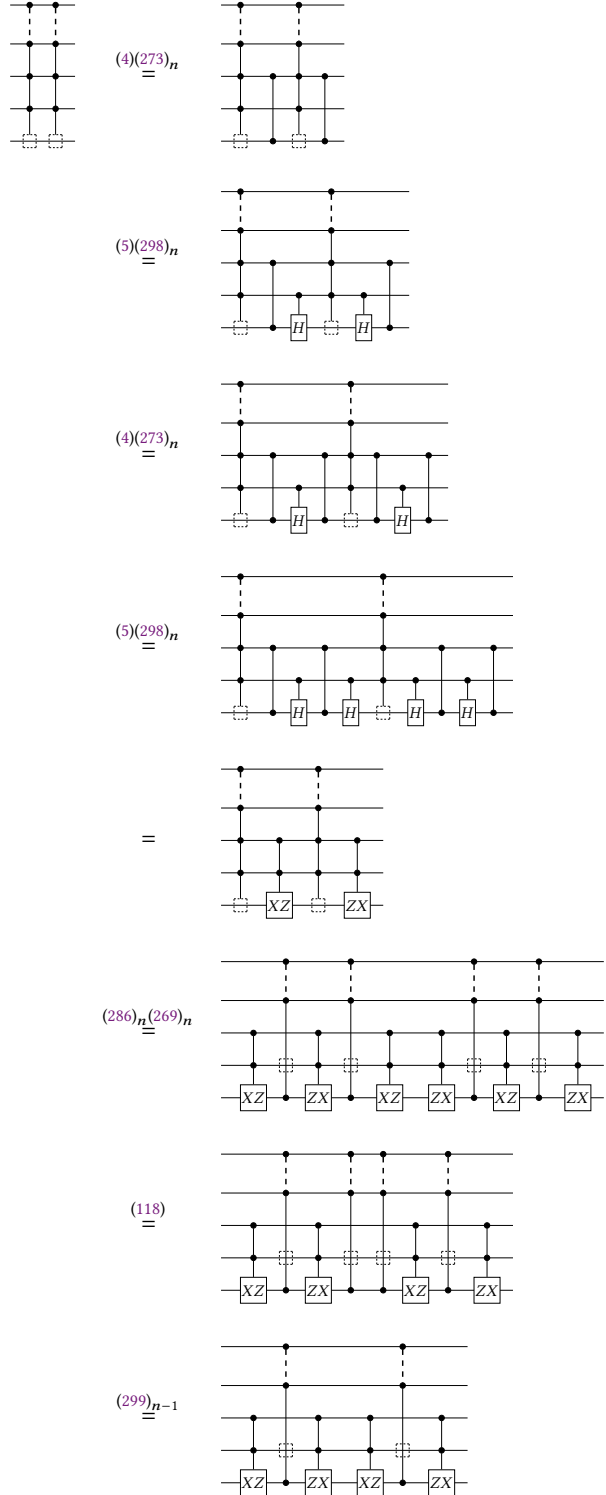

$\overset{(117)}{=}$ 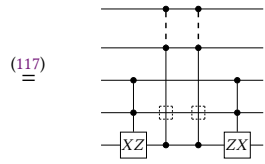

$\overset{(299)}{=}{}_{n-1}$ 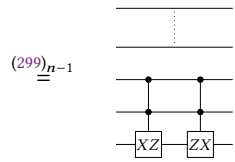

$\overset{(118)}{=}$ 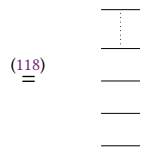

*Proof of Equation* $(300)_n$:

$\overset{(117)(299)_{n-1}(118)}{=}$ 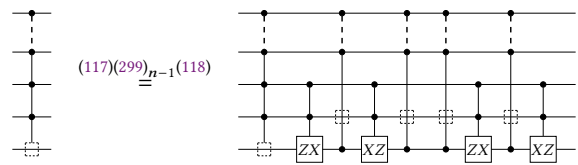

$\overset{(269)_n}{=}$ 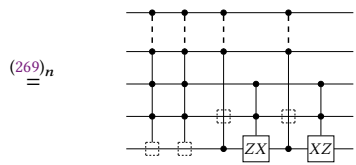

$\overset{(299)_n}{=}$ 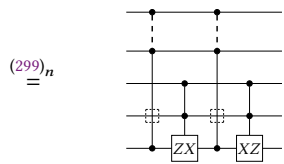

*Proof of Equation* $(301)_n$:

$\overset{(269)_n(300)_n(81)}{=}$ 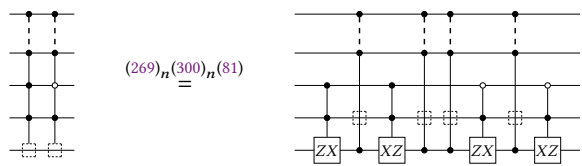

$\overset{(299)}{=}{}_{n-1}$ 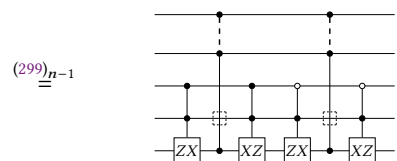

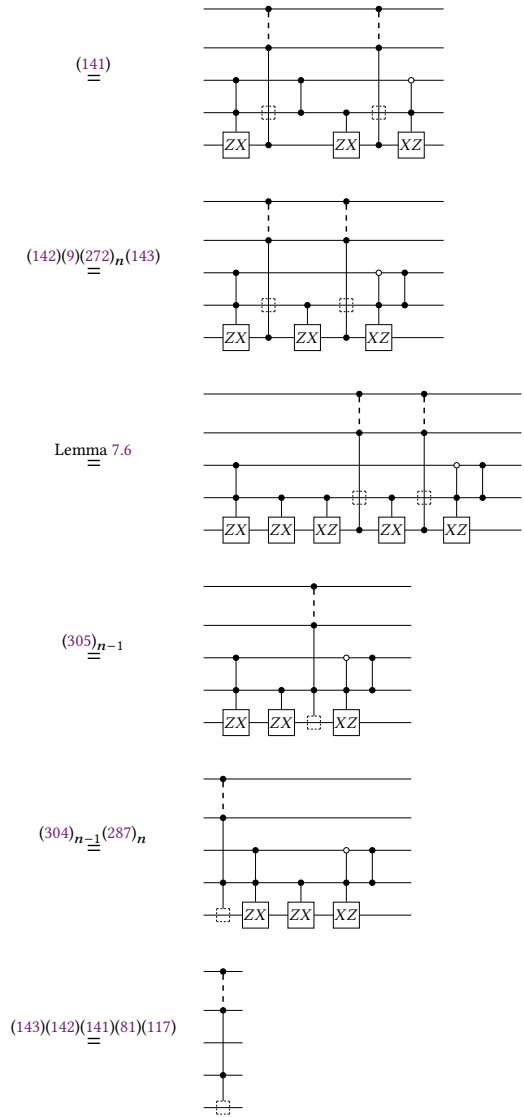

*Proof of Equation* (302)$_n$:

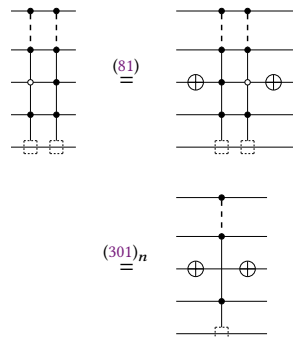

$$\overset{(81)}{=}$$

*Proof of Equation* $(303)_n$: Equation $(303)_2$ and Equation $(303)_3$ are trivial. If $n \geq 4$, then we have

$$\overset{(112)(113)}{=}$$

$$\overset{(19)_n}{=}$$

$$\overset{(111)(112)(113)}{=}$$

*Proof of Equation* $(304)_n$:

$$\overset{(298)_n}{=}$$

$$\overset{(19)_n}{=}$$

$$\overset{(298)_n}{=}$$

$$\overset{(19)_n}{=}$$

*Proof of Equation* (305)$_n$: Equation (305)$_2$ follows from Lemma 7.6. If $n \geq 3$, then we have

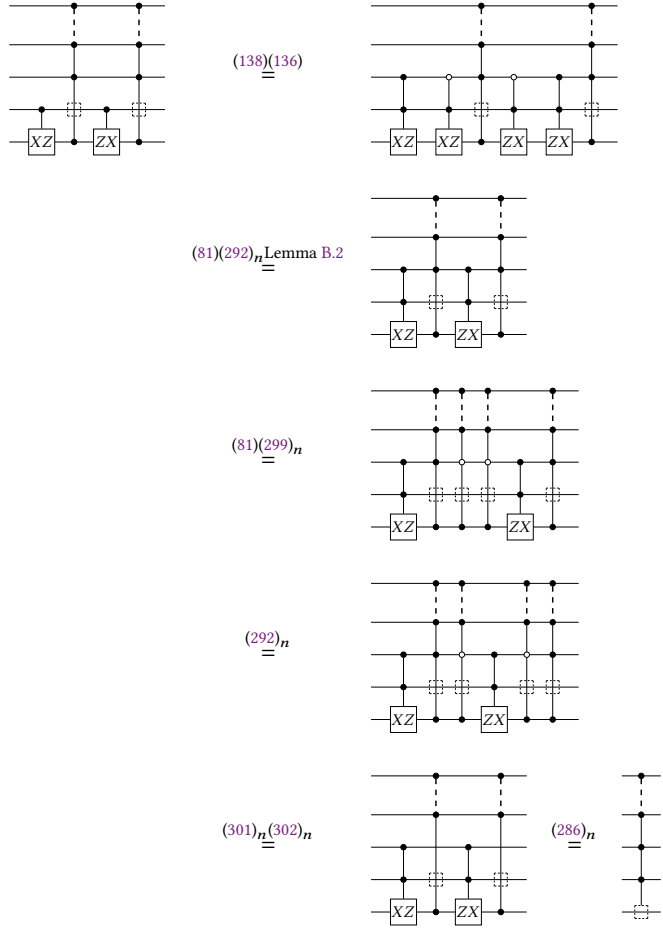

*Proof of Equation* (306)$_n$: Equation (306)$_n$ follows directly from Equation (295)$_n$ together with Equation (214).

*Proof of Equation* (307)$_n$: We proceed by case distinction on the number of controls strictly below the swap: If there are none, then Equation (307)$_n$ is just Equation (275)$_n$. If there is exactly one, then Equation (307)$_n$ is just Equation (306)$_n$. If there are more than one, then

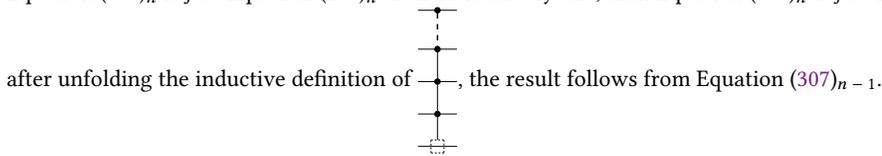

after unfolding the inductive definition of 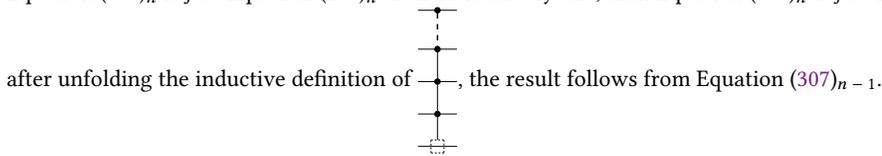, the result follows from Equation (307)$_{n-1}$. □

## D.6 Further Auxiliary Equations on $n$ Qubits and Properties of Multi-Controlled Gates

**Lemma D.9.** *For any circuit (over $\left\{ -\boxed{H}-, -\bullet-, \begin{array}{c}\bullet\\\bullet\end{array}, \begin{array}{c}\bullet\\\boxed{H}\end{array} \right\}$), the following equation is a consequence of the equations of Figure 4:*

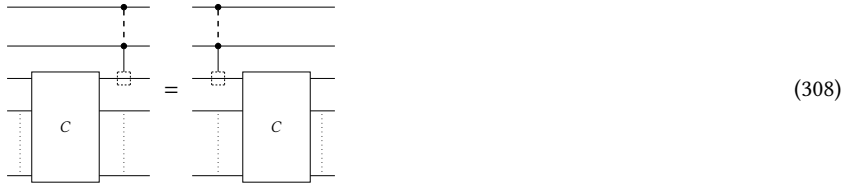

$$(308)$$

PROOF. We proceed by structural induction on $C$.[18]

If $C = $ ———, then the result is trivial. If $C = \diagdown\diagup$, $-\boxed{H}-$, $-\bullet-$, or $\begin{array}{c}\bullet\\\bullet\end{array}$, then the result follows from (274)$_n$, (19)$_n$, (285)$_n$, or (272)$_n$ respectively.

---

[18]More precisely, on an underlying raw circuit.

If $C = $ , then we have

If $C = C_2 \circ C_1$ (with $C_1, C_2$ different from the identity[19]), then we have

Finally, if $C = C_1 \otimes C_2$ (with $C_2 \neq$ ), then we have

$\square$

---

[19]Namely, different from ( —— )$^{\otimes n}$ if $C$ is an $n$-qubit circuit.

**Lemma D.10.** *The following equations are consequences of the equations of Figure 4:*

$$(309)$$

$$(310)$$

$$(311)$$

$$(312)$$

$$(313)$$

**Remark 8.** *The generalizations of Equations (277) and (313), where the dashed series of black controls are replaced by arbitrary $\Lambda^x$, are trivial to derive. However, we choose to keep these equations simple, because each time one would use their generalized versions in a derivation, this would in fact amount to using the all-black version, since $\Lambda^x$ is nothing but a series of black controls surrounded by some $-\oplus-$. The same remark holds for Equations (274), (307), (308) and (314).*

PROOF OF LEMMA D.10.

*Proof of Equation* (309)*:* By Equation (81), it suffices to prove the version of the equation where $x$ is the vector consisting only of 1's:

and this equation follows directly from Equations (301), (302) and (307).

*Proof of Equation* (310):

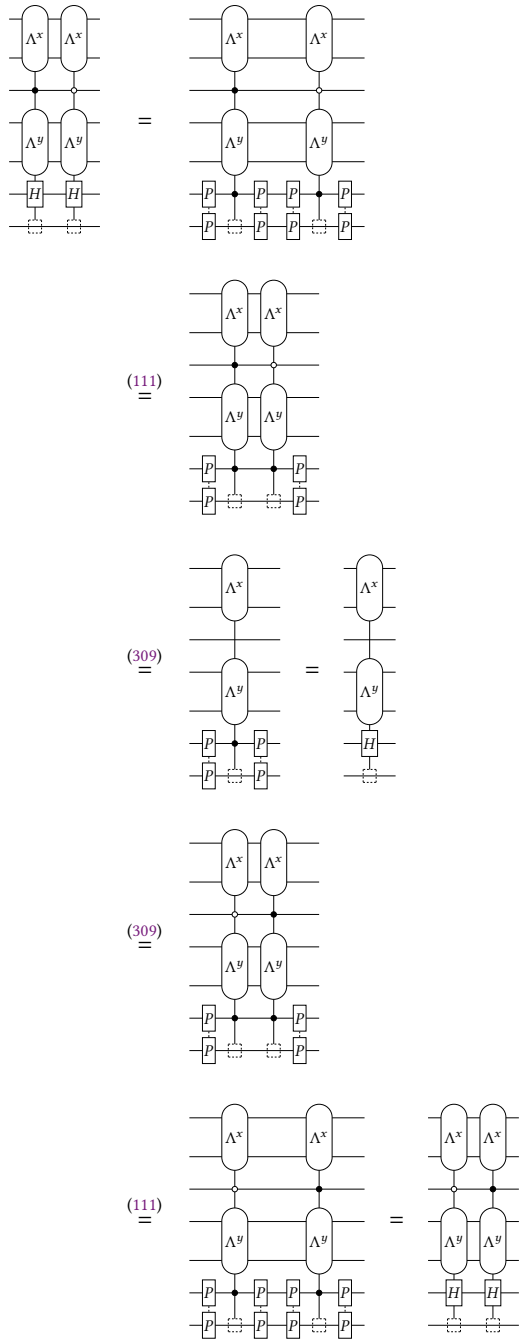

*Proof of Equation* (311): Equation (311) follows directly from Equations (81) and (299).

*Proof of Equation* (312):

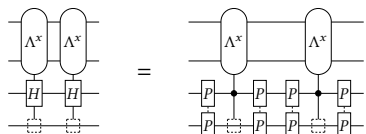

$$\overset{(111)}{=} \quad$$ 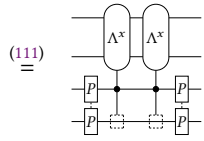

$$\overset{(311)}{=} \quad$$ 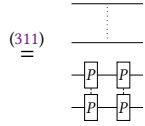

$$\overset{(111)}{=} \quad$$ 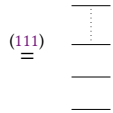

*Proof of Equation* (313):

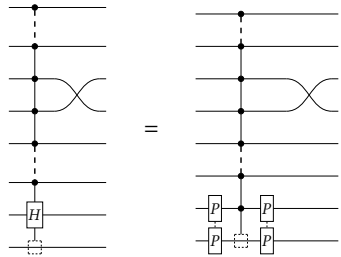

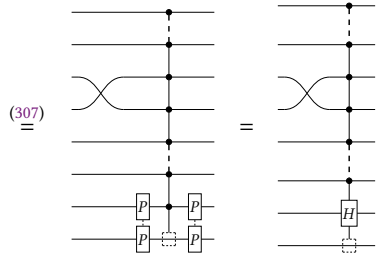

$\square$

**Lemma D.11.** *For any circuit (over* $\left\{ -\boxed{H}-, -\bullet-, \overset{\bullet}{\underset{\bullet}{\vert}}, \overset{\boxed{H}}{\underset{\boxed{H}}{\vert}} \right\}$ *), the following equation is a consequence of the equations of Figure* 4*:*

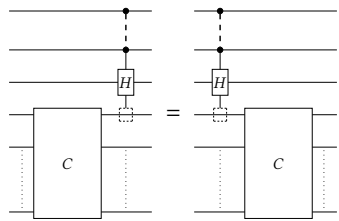

$$(314)$$

PROOF. We proceed by structural induction on $C$.[20]

If $C = \text{———}$, then the result is trivial. If $C = \times$, $-\boxed{H}-$ or $\overset{\bullet}{\underset{\bullet}{\vert}}$, then the result follows from Equation (277), (303) or (282) respectively.

If $C = -\bullet-$, then we have

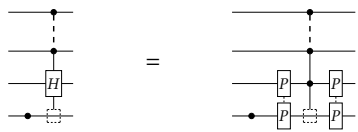

---

[20]More precisely, on an underlying raw circuit.

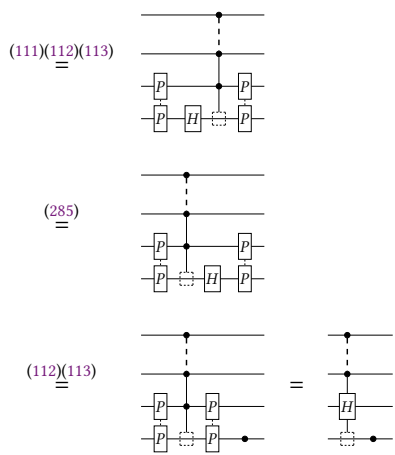

If $C = $ 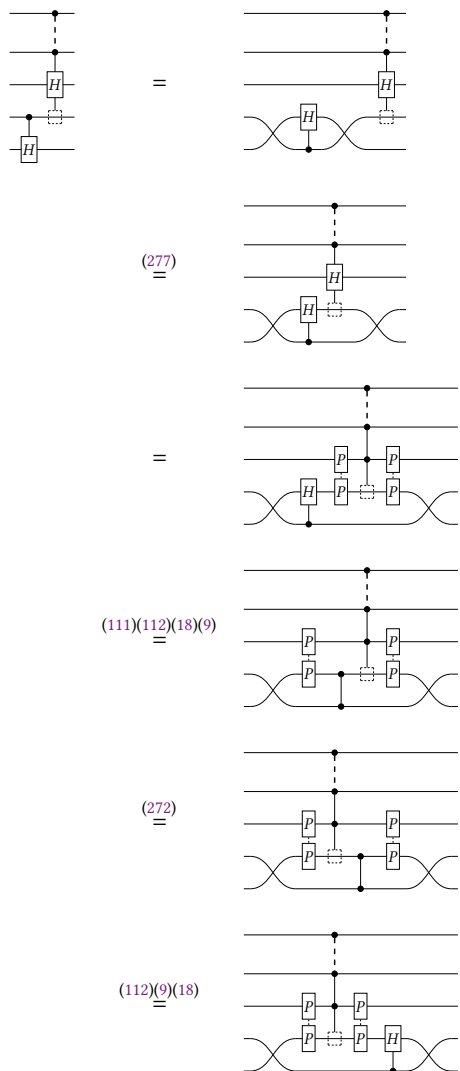, then we have

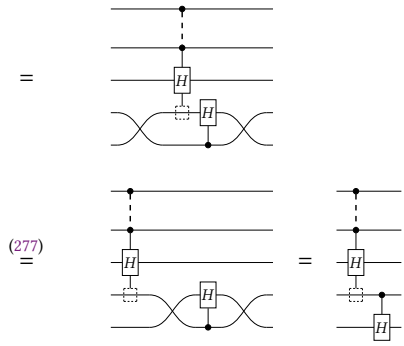

If $C = C_2 \circ C_1$ (with $C_1, C_2$ different from the identity), then we have

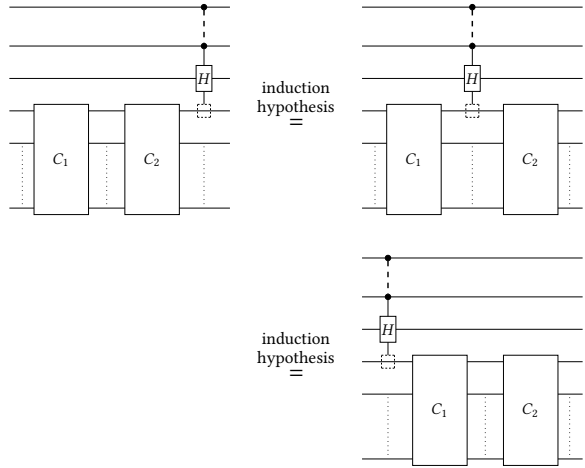

Finally, if $C = C_1 \otimes C_2$ (with $C_2 \neq \vdots$), then we have

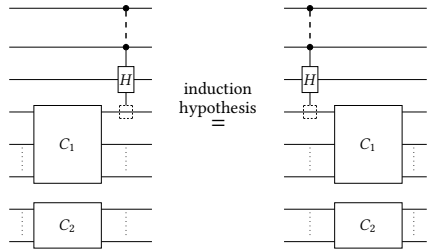

□

## D.7 Commutation Properties of Multi-Controlled Gates

**Lemma D.12.** *The following equations are consequences of the equations of Figure 4:*

$$\tag{315}$$

$$\tag{316}$$

$$\tag{317}$$

$$(318)$$

$$(319)$$

$$(320)$$

$$(321)$$

$$(322)$$

$$(323)$$

$$(324)$$

$$(325)$$

$$(326)$$

$$(327)$$

$$(328)$$

$$(329)$$

$$(330)$$

$$(331)$$

$$(332)$$

$$(333)$$

$$(334)$$

**Lemma D.13.** *The following equations are consequences of the equations of Figure 4, where* —•— *denotes either a black control or a white control:*

$$\forall x, y \in \{0,1\}^n, \qquad (335)$$

$$\forall x, y \in \{0,1\}^n, \qquad (336)$$

$$\forall x, y \in \{0,1\}^n \text{ with } x \neq y, \qquad (337)$$

$$\forall x, y \in \{0,1\}^n, \qquad (338)$$

$$\forall x, y \in \{0,1\}^n \text{ with } x \neq y, \qquad (339)$$

$$\forall x, y \in \{0,1\}^n, \qquad (340)$$

$$\forall x, y \in \{0,1\}^n, \qquad (341)$$

$$\forall x, y \in \{0,1\}^n \text{ with } x \neq y, \qquad (342)$$

$$\forall x, y \in \{0,1\}^n \text{ with } x \neq y, \qquad (343)$$

$$\forall x, y \in \{0,1\}^n \text{ with } x \neq y, \qquad (344)$$

$$\forall x, y \in \{0,1\}^n, \qquad (345)$$

$$\forall x, y \in \{0,1\}^n, \qquad (346)$$

$$\forall x, y \in \{0,1\}^n \text{ with } x \neq y, \qquad (347)$$

$$\forall x, y \in \{0,1\}^n \text{ with } x \neq y, \qquad (348)$$

$$\forall x, y \in \{0,1\}^n \text{ with } x \neq y, \qquad (349)$$

$$\forall x, y \in \{0,1\}^n \text{ with } x \neq y, \qquad (350)$$

$$\forall x, y \in \{0,1\}^n, \qquad (351)$$

$$\forall x, y \in \{0,1\}^n \text{ with } x \neq y, \qquad (352)$$

PROOF OF LEMMAS D.12 AND D.13. We prove all the equations in order, except Equations (333) and (334) which we prove only after having proved Equations (335) to (345).

*Proof of Equations* (315) *and* (316): Equations (315) and (316) are just particular cases of Equations (308) and (314) respectively. We make them explicit only to add more clarity to subsequent proofs.

*Proof of Equation* (317): Equation (317) follows directly from Equations (5), (81) and (299).

*Proof of Equation* (318): First, we prove the particular case of the equation where $x = \vec{1}$, $\alpha = \beta = 1$, and $a = 0$:

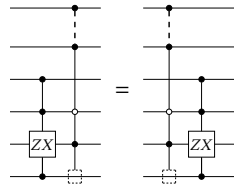

The derivation is as follows:

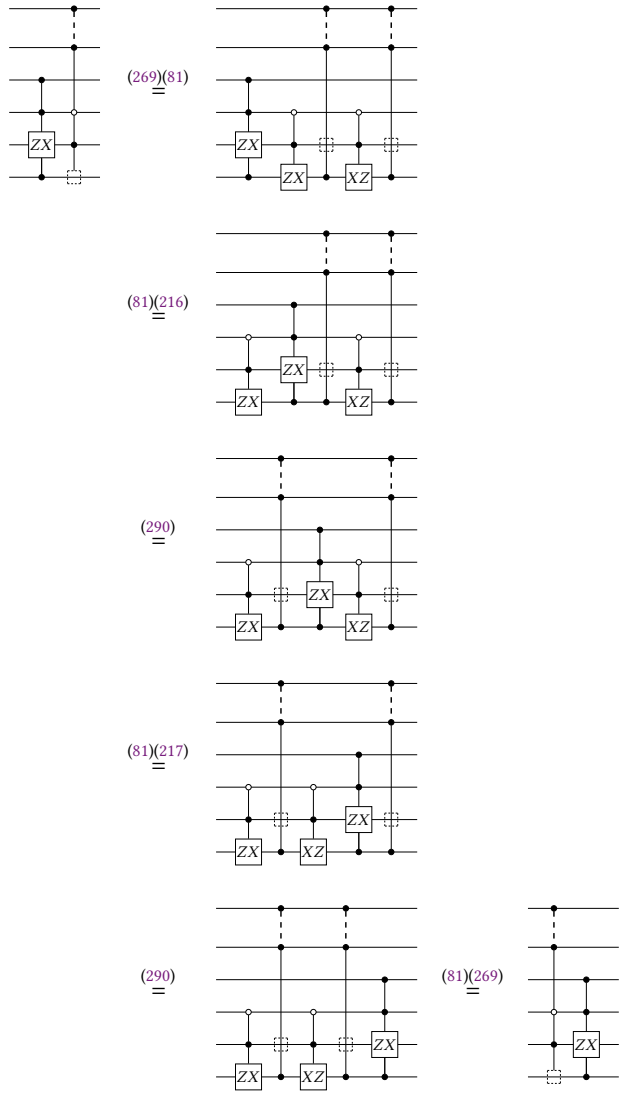

Then, the cases where $x \neq \vec{1}$ or $\alpha \neq 1$ follow trivially. The cases where $a = 1$ follow by Lemma B.5 (together with Equations (311) and (317)). And the cases where $\beta = 0$ follow by Equation (238) possibly together with Equation (81).

*Proof of Equation* (319): First, we prove the particular case of the equation where $x = \vec{1}$, $\alpha = \beta = 1$, and $a = 0$:

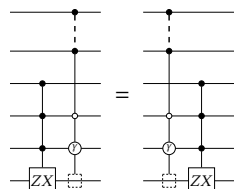

If $\gamma = 1$, then we have

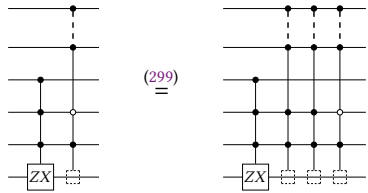

$\overset{(301)}{=}$

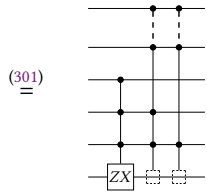

$\overset{(214)(291)}{=}$

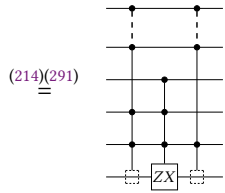

$\overset{(290)}{=}$

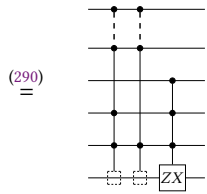

$\overset{(301)(299)}{=}$

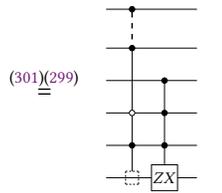

And if $\gamma = 0$, then we have

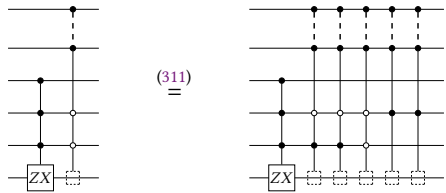

$\overset{(309)}{=}$

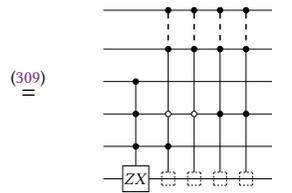

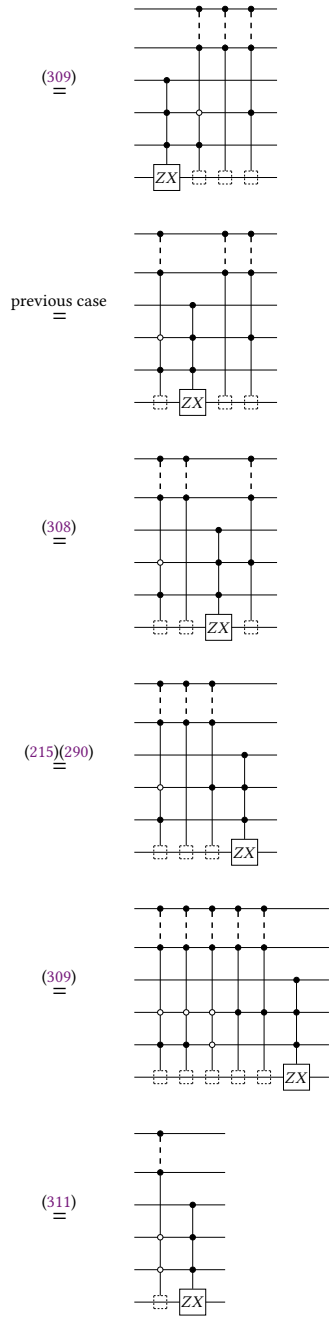

Finally, the cases where $x \neq \bar{1}$, $\alpha \neq 1$, or $\beta \neq 1$ follow trivially or by Equation (81), and the cases where $a = 1$ follow by Lemma B.5 (together with Equations (311) and (317)).

*Proof of Equation* (320):

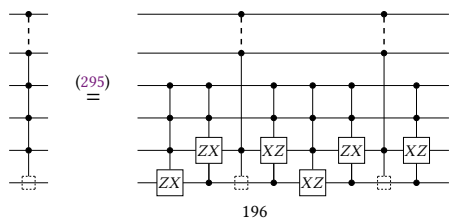

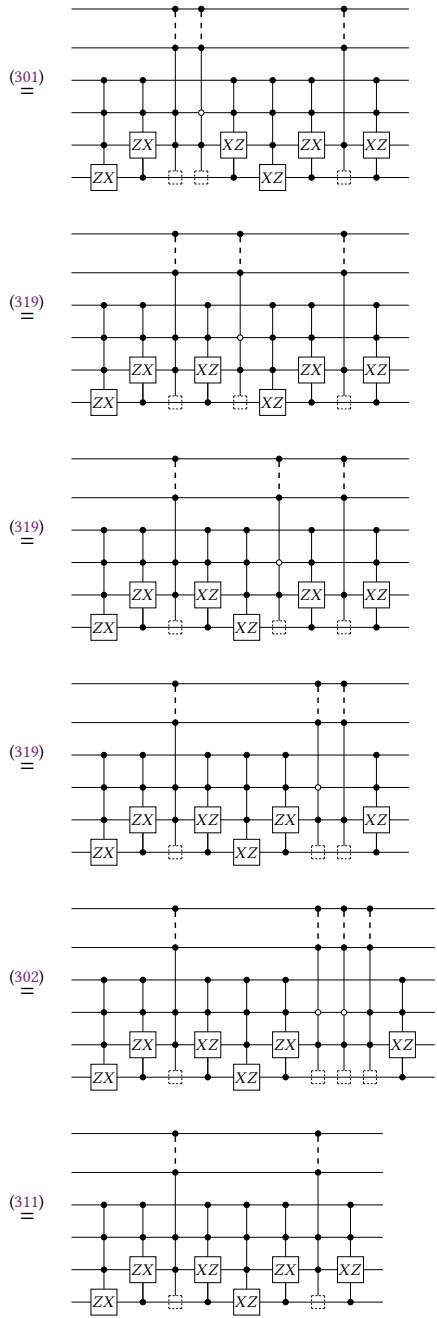

*Proof of Equation* (321): Given Equation (320), Equation (321) follows directly by Equations (81) and (238).

*Proof of Equation* (322): We proceed by induction on the length $n$ of $x$. If $n = 0$, then the result follows from Equation (247). If $n \geq 1$, then we can assume without loss of generality that the last bit of $x$ is 1, namely $x = z1$ with $z \in \{0,1\}^{n-1}$. Then we have

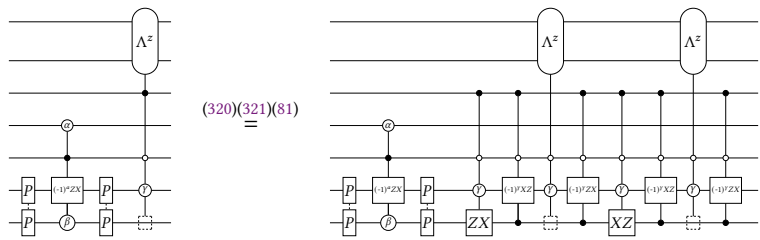

$$\overset{(112)(267)}{=}$$ 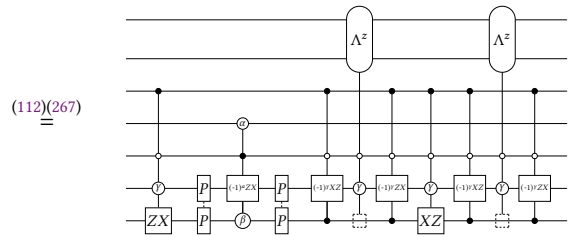

$$\overset{(112)(268)}{=}$$ 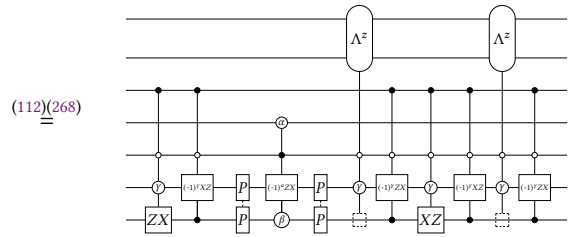

$$\overset{\text{induction}}{\underset{=}{\text{hypothesis}}}$$ 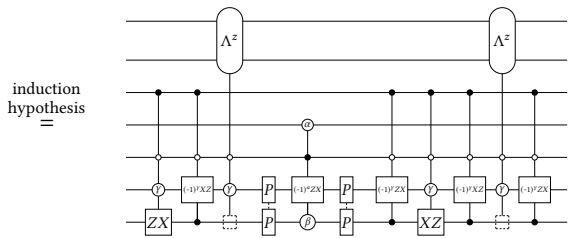

$$\overset{(112)(268)(267)}{\underset{=}{\text{induction}\atop\text{hypothesis}}}$$ 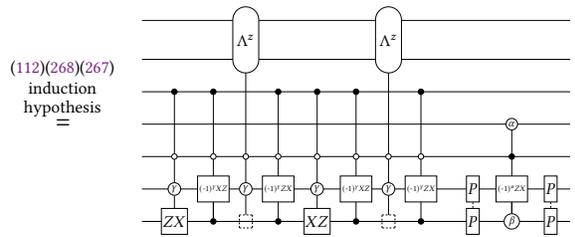

$$\overset{(81)(320)(321)}{=}$$ 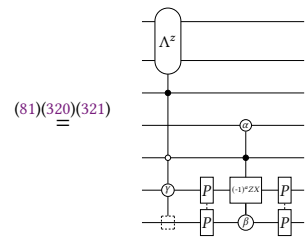

*Proof of Equation* (323): We proceed by induction on the length $n$ of $x$. If $n = 0$, then the result follows from Equation (248). If $n \geq 1$, then we can assume without loss of generality that the last bit of $x$ is 1, namely $x = z1$ with $z \in \{0, 1\}^{n-1}$. Then we have

$$\overset{(320)(321)(81)}{=}$$ 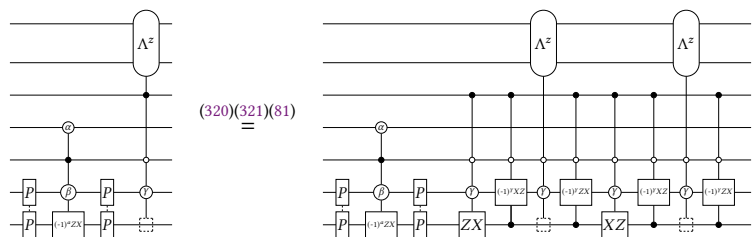

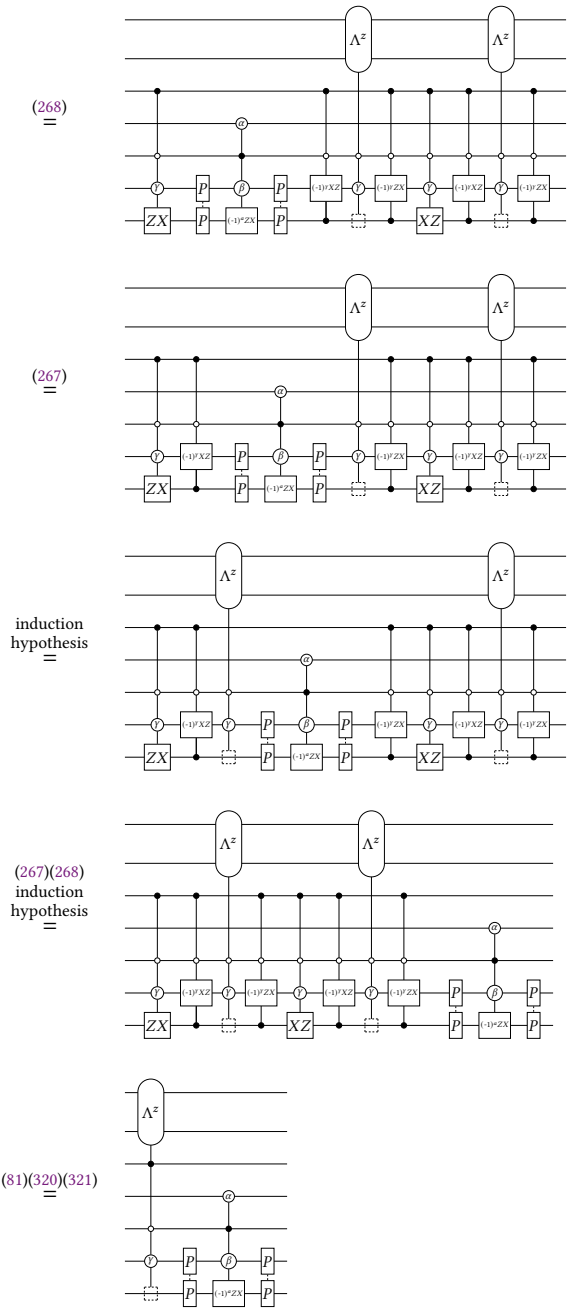

*Proof of Equation* (324):

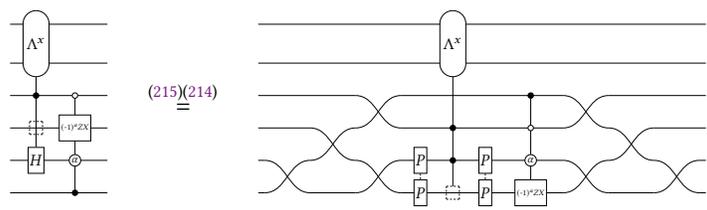

$\overset{(81)(111)(323)}{=}$ 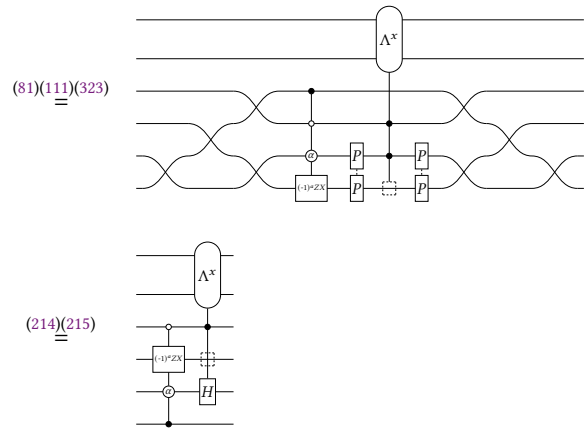

$\overset{(214)(215)}{=}$ 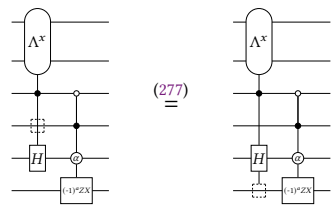

*Proof of Equation* (325):

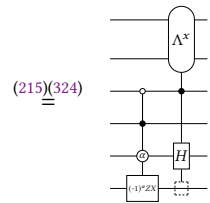

$\overset{(215)(324)}{=}$

$\overset{(277)}{=}$

$\overset{(277)}{=}$

*Proof of Equation* (326):

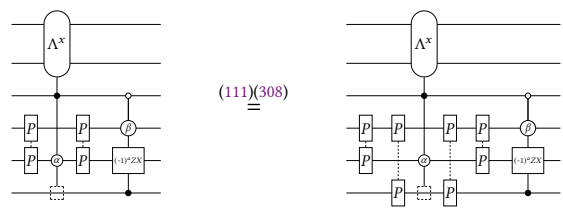

$\overset{(111)(308)}{=}$

$\overset{(112)(130)}{=}$ 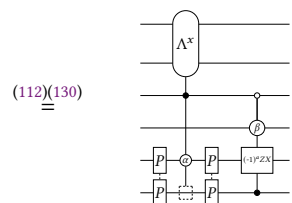

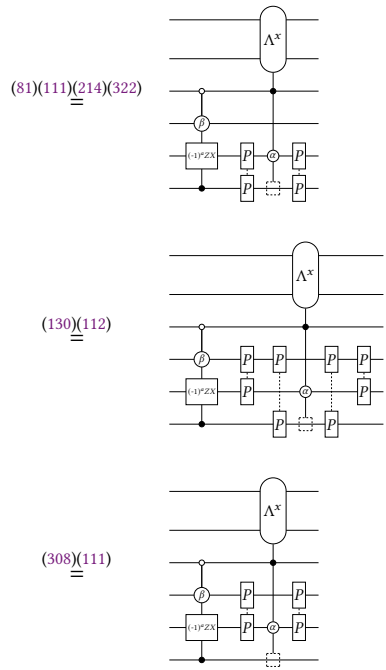

*Proof of Equation* (327)*:*

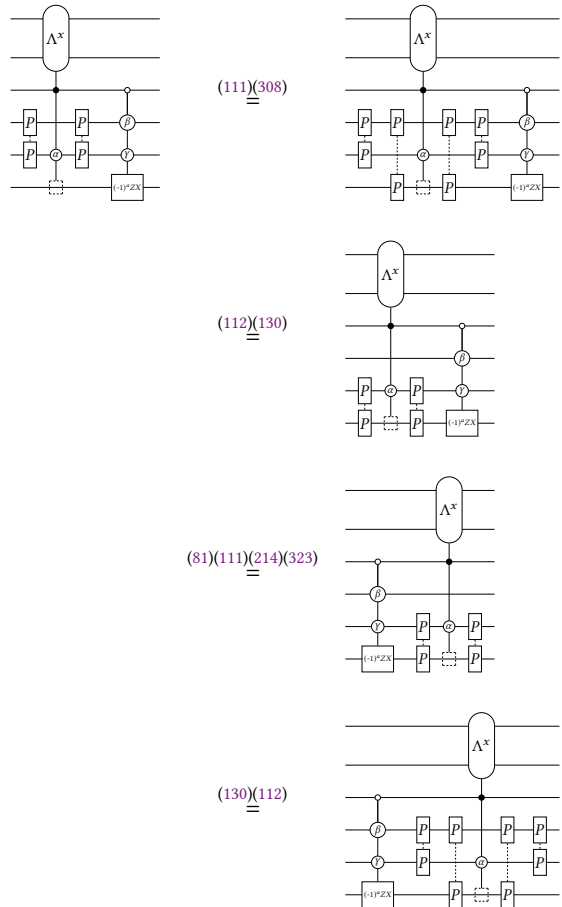

$$\overset{(308)(111)}{=}$$ 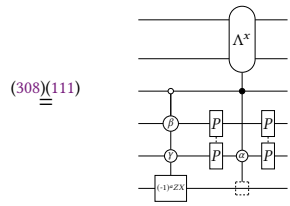

*Proof of Equation* (328):

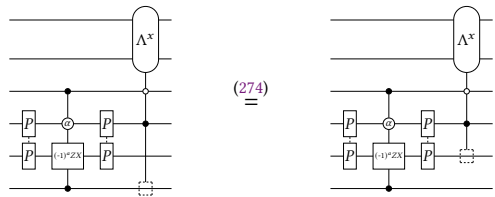

$$\overset{(215)(214)(323)}{=}$$ 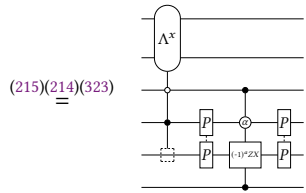

$$\overset{(274)}{=}$$ 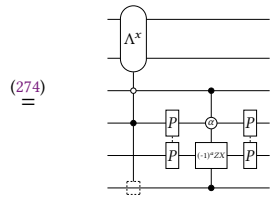

*Proof of Equation* (329):

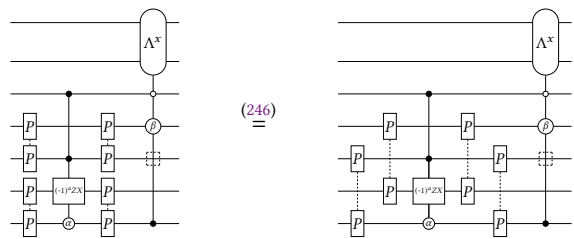

$$\overset{(112)(262)(111)}{=}$$ 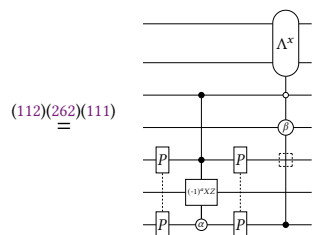

$$\overset{(111)(112)}{=}$$ 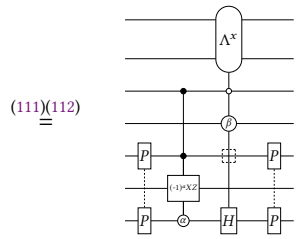

$$\overset{(307)(81)(325)}{=}$$ 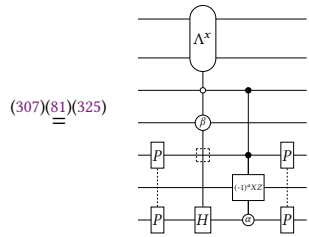

$$\overset{(112)(111)}{=}$$ 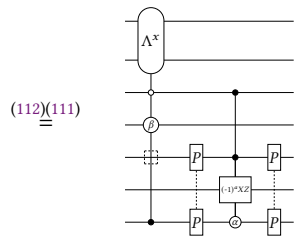

$$\overset{(111)(262)(112)}{=}$$ 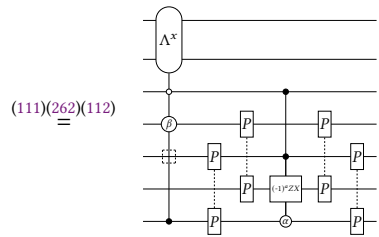

$$\overset{(246)}{=}$$ 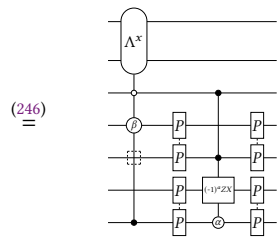

*Proof of Equation* (330):

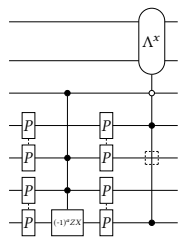 $$\overset{(112)(246)}{=}$$ 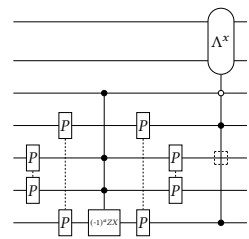

$(112)(262)(111)$
$=$

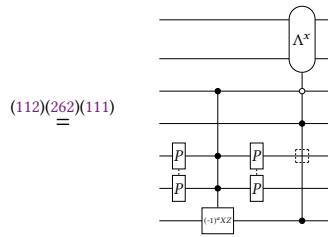

$(308)$
$=$

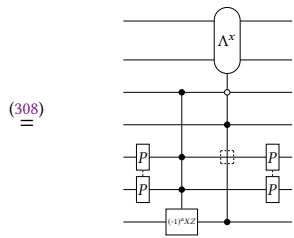

$(307)(215)(214)(318)$
$=$

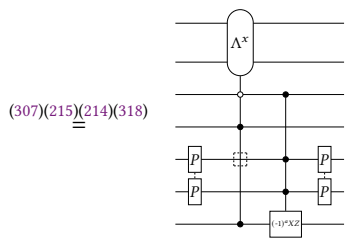

$(308)$
$=$

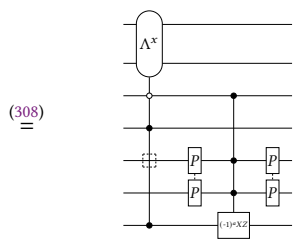

$(111)(262)(112)$
$=$

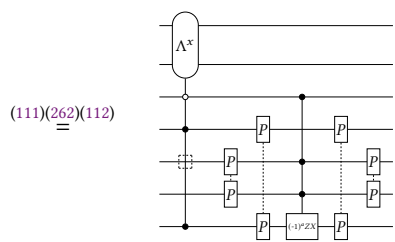

$(246)(112)$
$=$

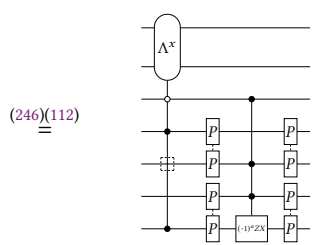

*Proof of Equation* (331):

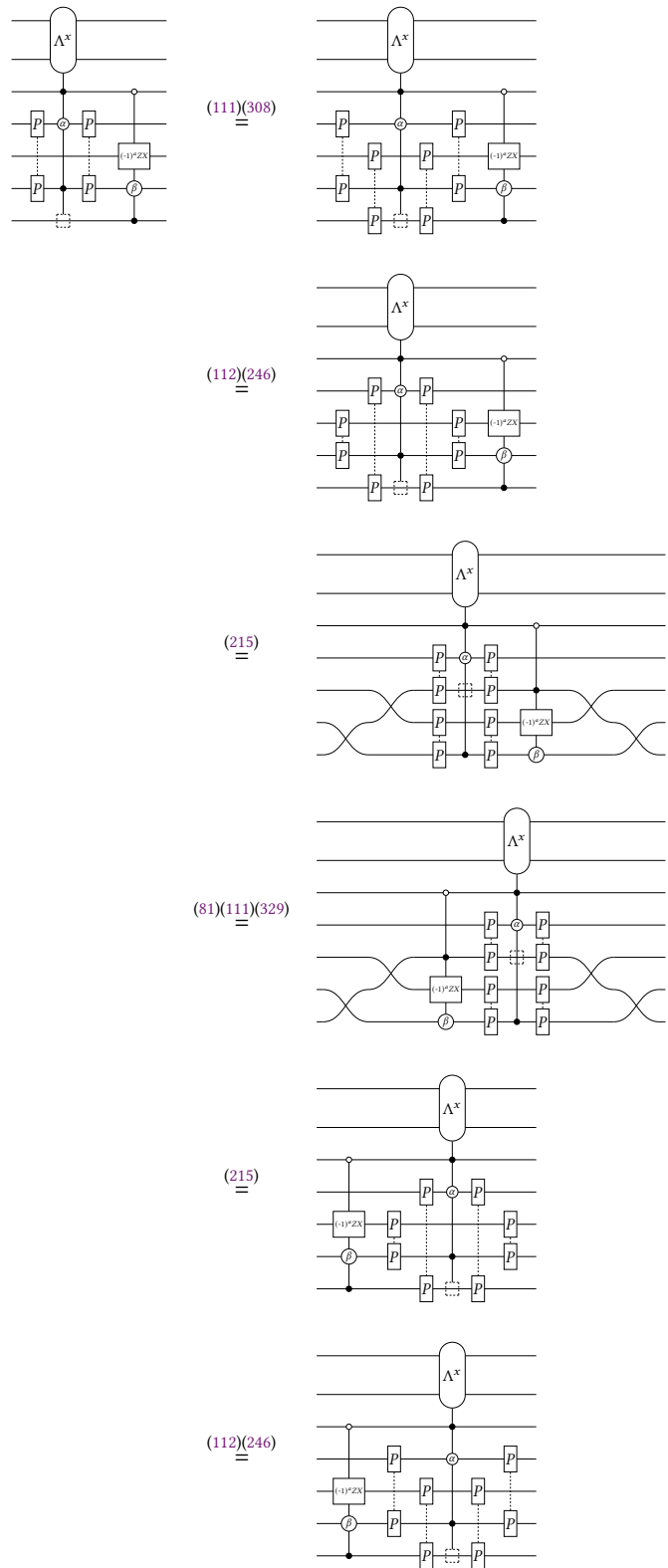

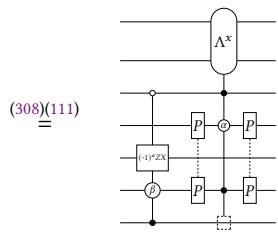

*Proof of Equation* (332):

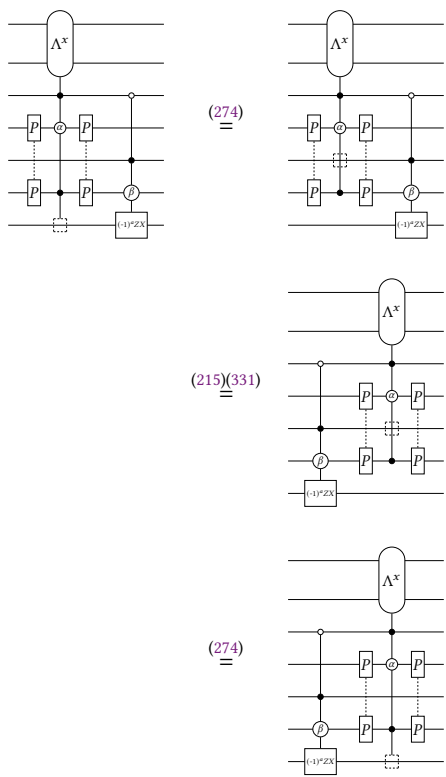

*Proof of Equation* (335): If $x = y$ then the equation is trivial.

If $x \neq y$, then $x$ and $y$ differ by at least one bit, which means that either a black control of $\Lambda^x$ is in front of a white control of $\Lambda^y$, or conversely. Since the equation is symmetric, we can assume without loss of generality that we are in the first case. Then we proceed by induction on $n$ (that is, the length of $x$). We cannot have $n = 0$ since this would imply $x = y$. If $n = 1$, then the result follows from Lemma 7.1. If $n = 2$, then the result follows from Lemma 7.6. If $n \geq 3$, then by Equation (307) we can assume without loss of generality that this is the penultimate bit that differs between $x$ and $y$, that is, $x = za1b$ and $y = tc0d$ where $z, t \in \{0, 1\}^{n-3}$ and $a, b, c, d \in \{0, 1\}$. Then we have

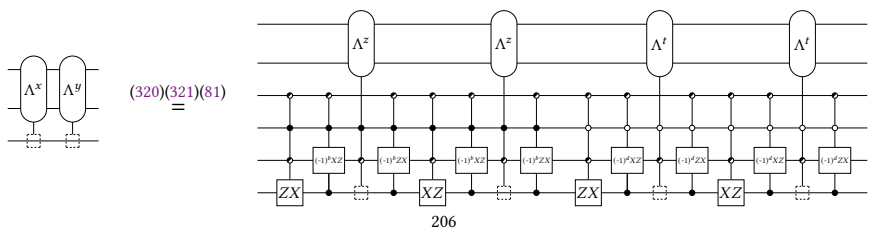

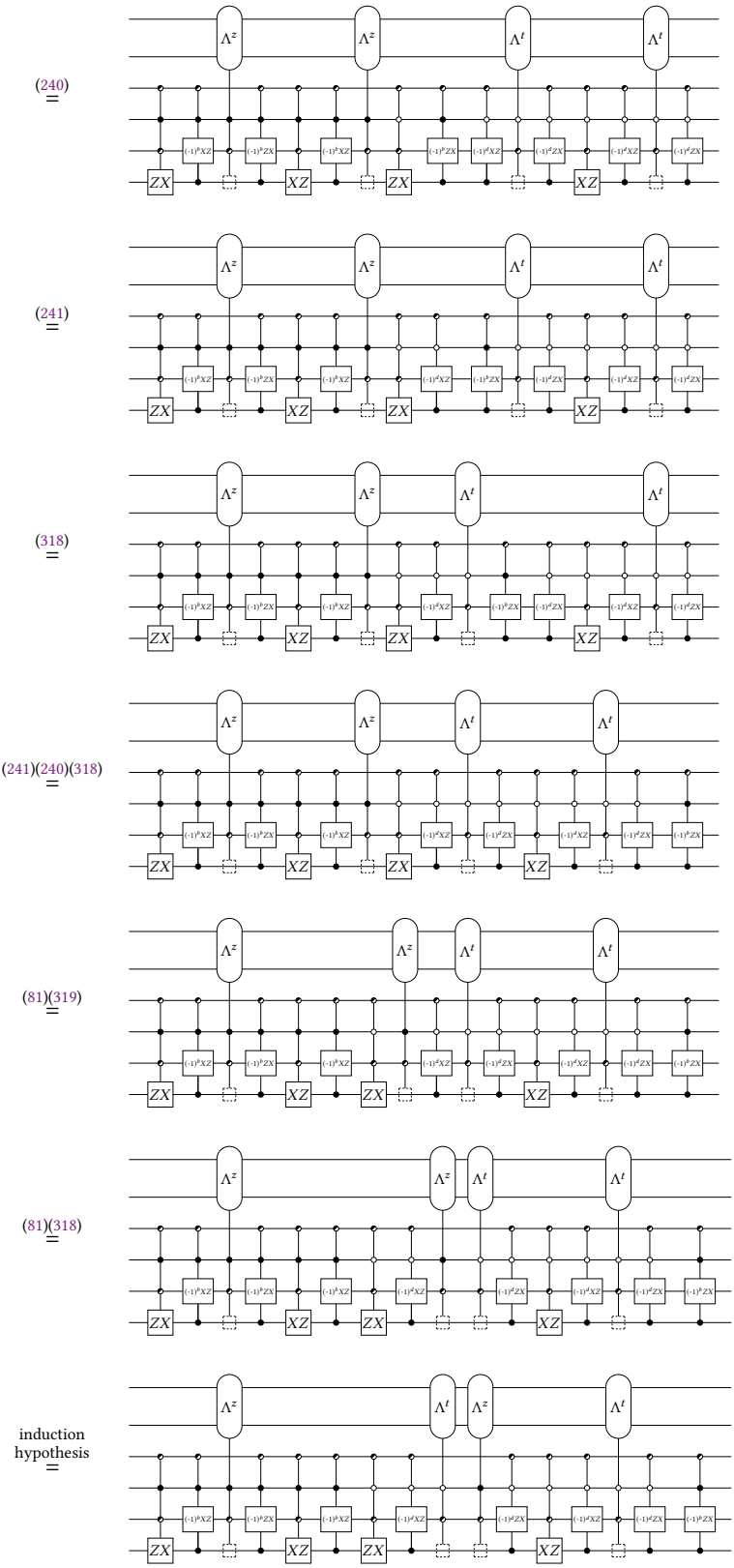

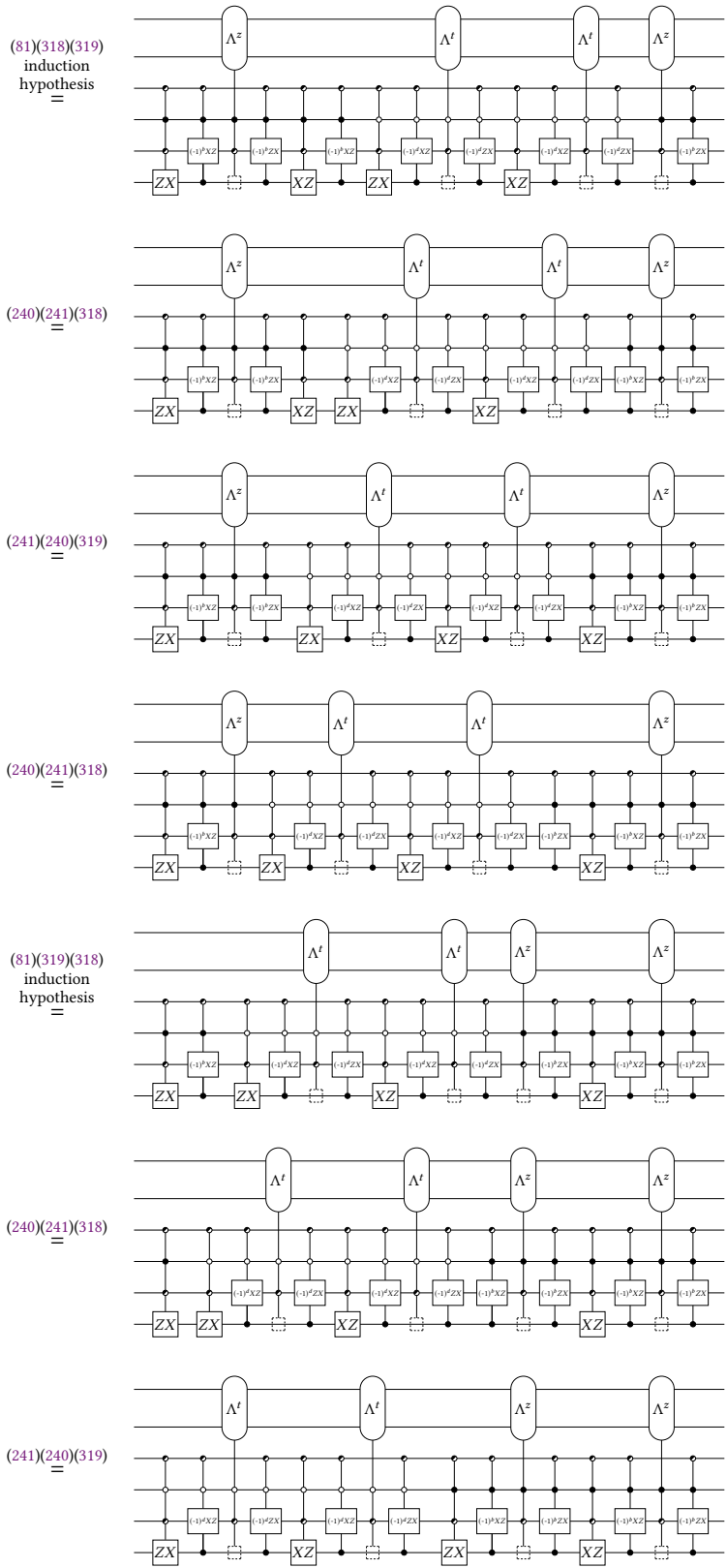

$$\overset{(81)(320)(321)}{=}$$

*Proof of Equation* (336): If $n = 0$, then the result follows from Lemma 7.6. If $x = y$ and $n \geq 1$, then by Equation (81), we can assume without loss of generality that $x$ (and therefore also $y$) is the vector consisting only of 1's. By Equation (315), we can also assume without loss of generality that the controls in front of the $\square$'s are black as well. Then we have

If $x \neq y$, then we proceed by induction on $n$. If $n = 1$, then the result follows from Equations (12) and (131) together with Equation (9). If $n \geq 2$, then by Equation (307) we can assume without loss of generality that this is the last bit that differs between $x$ and $y$, and by Equation (81) we can assume that that bit is 1 in $x$ and 0 in $y$. Namely, $x = za1$ and $y = tb0$ where $z, t \in \{0, 1\}^{n-2}$ and $a, b \in \{0, 1\}$. Then we

have

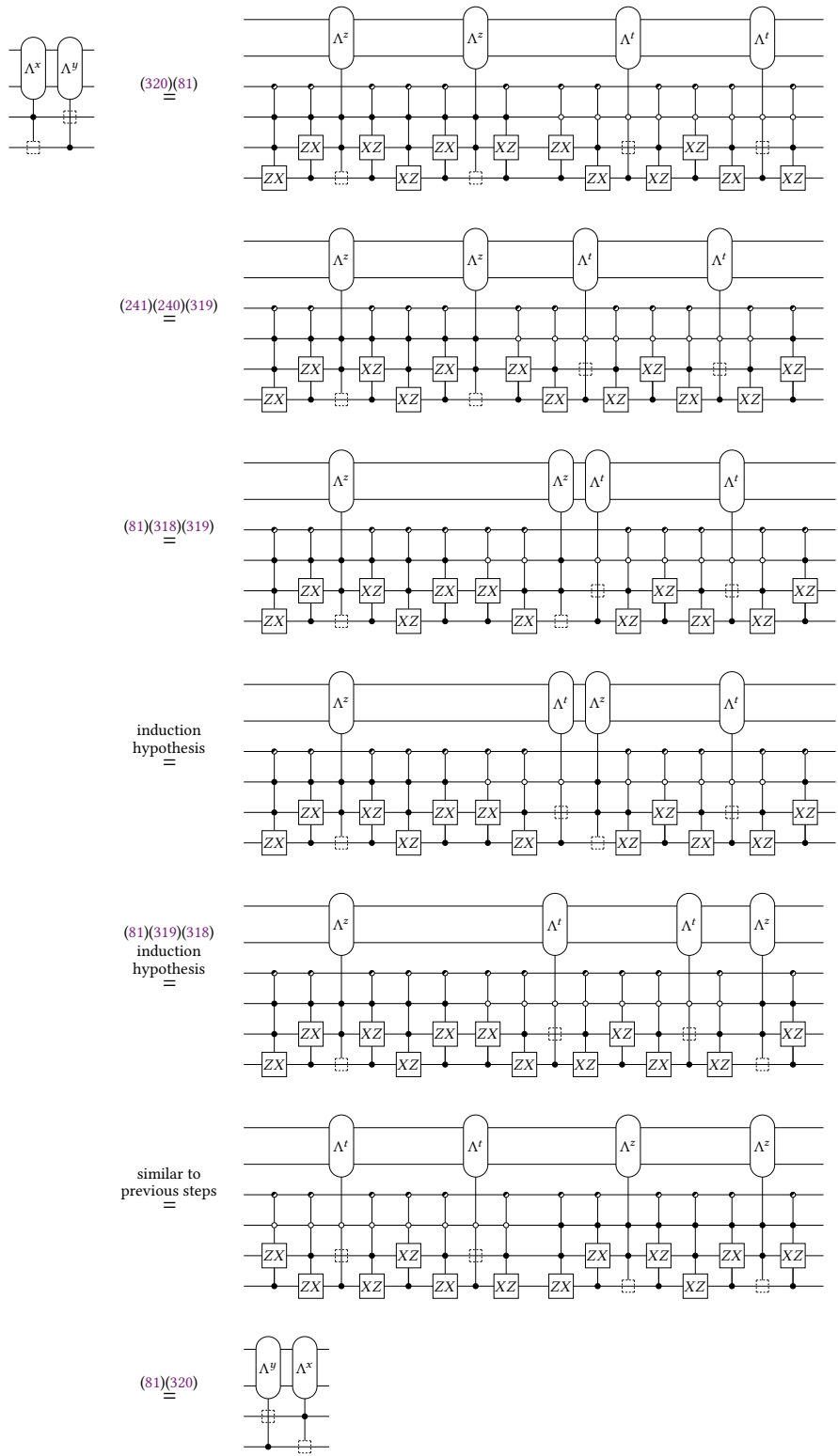

*Proof of Equation* (337): We proceed by induction on $n$. If $n = 1$, then the result follows from Lemma 7.6. If $n \geq 2$, then by Equations (307) and (313), we can assume without loss of generality that $x$ and $y$ differ by the last bit, and by Equation (81) we can assume that that bit is 1 in $x$ and 0 in $y$. Namely, $x = za1$ and $y = tb0$, where $z, t \in \{0,1\}^{n-2}$ and $a, b \in \{0,1\}$. Thus, we need to prove the following equation:

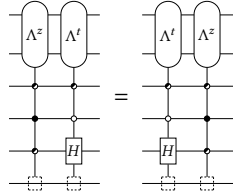

And we have (where $c$ denotes the bit corresponding to the control located in front of the $H$):

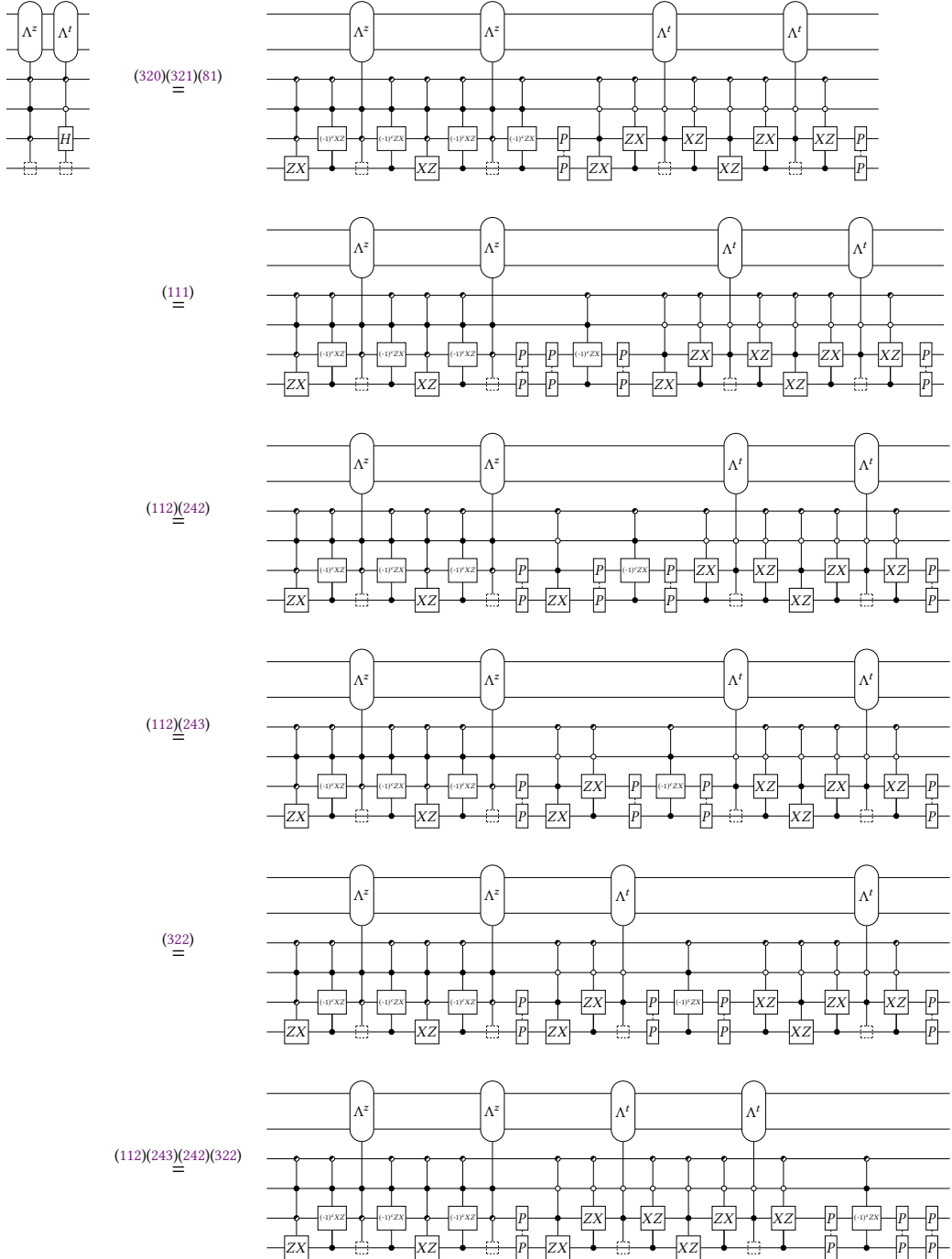

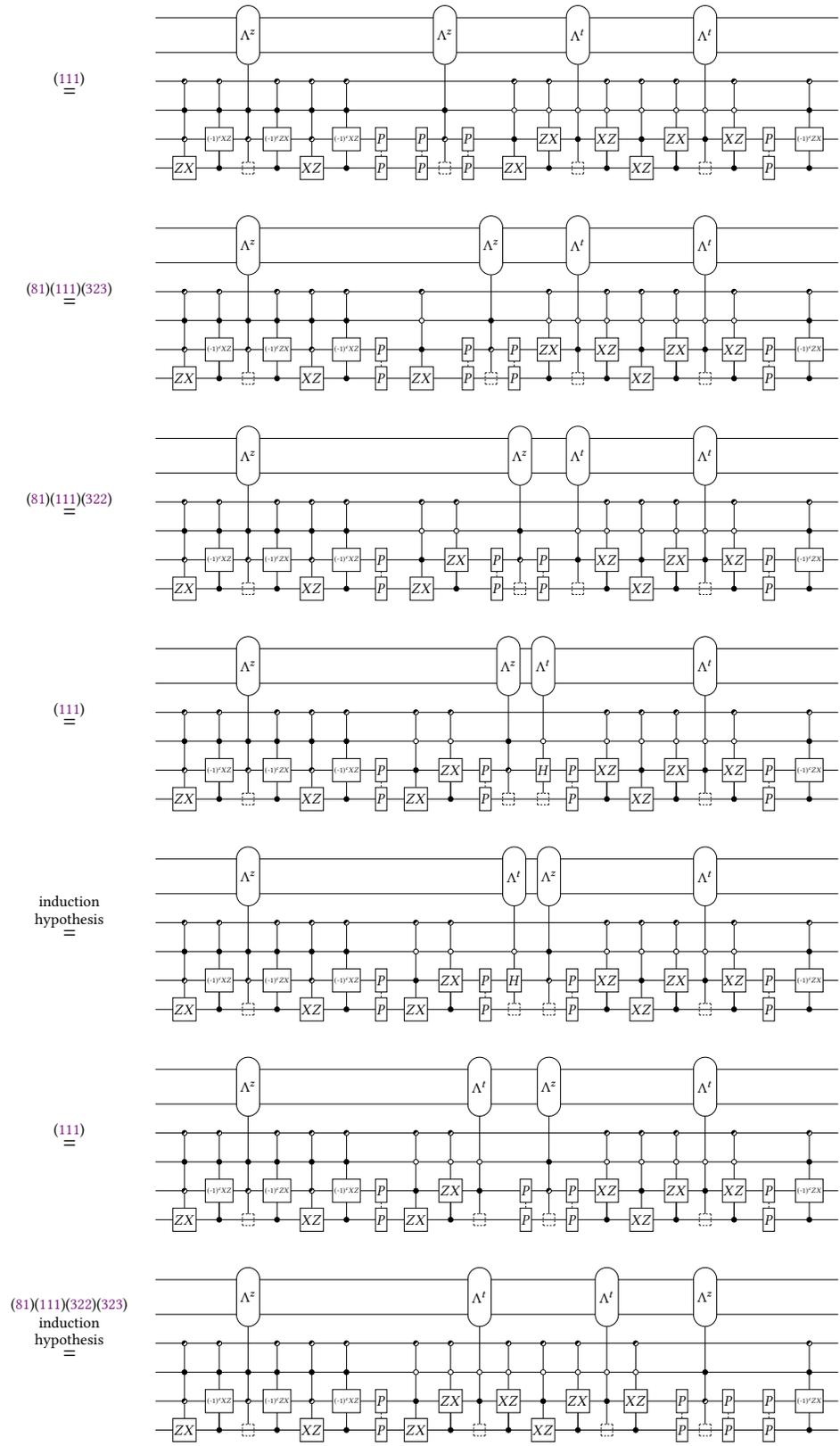

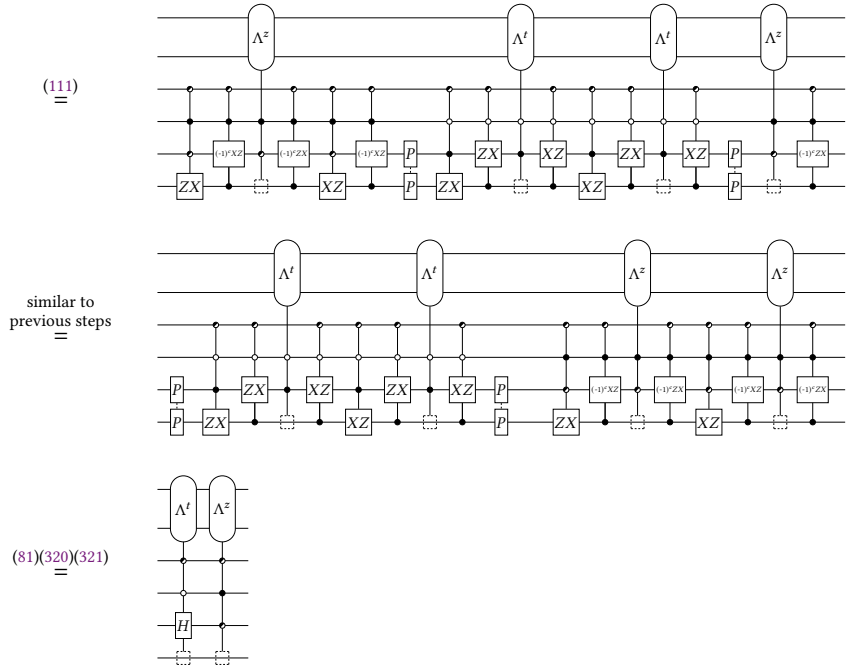

*Proof of Equation* (338): If $n = 0$, then the result is trivial.

If $n \geq 1$ and $x \neq y$, then we proceed by induction on $n$. If $n = 1$, then the result follows from Equations (13) and (132) together with Equations (9) and (81). If $n \geq 2$, then by Equations (307) and (313), we can assume without loss of generality that this is the last bit that differs between $x$ and $y$, and by Equation (81) we can assume without loss of generality that that bit is 1 in $x$ and 0 in $y$. Namely, $x = za1$ and $y = tb0$ where $z, t \in \{0, 1\}^{n-2}$ and $a, b \in \{0, 1\}$. Moreover, by Equation (316) we can assume without loss of generality that the control in front of the □ is black. Then we have

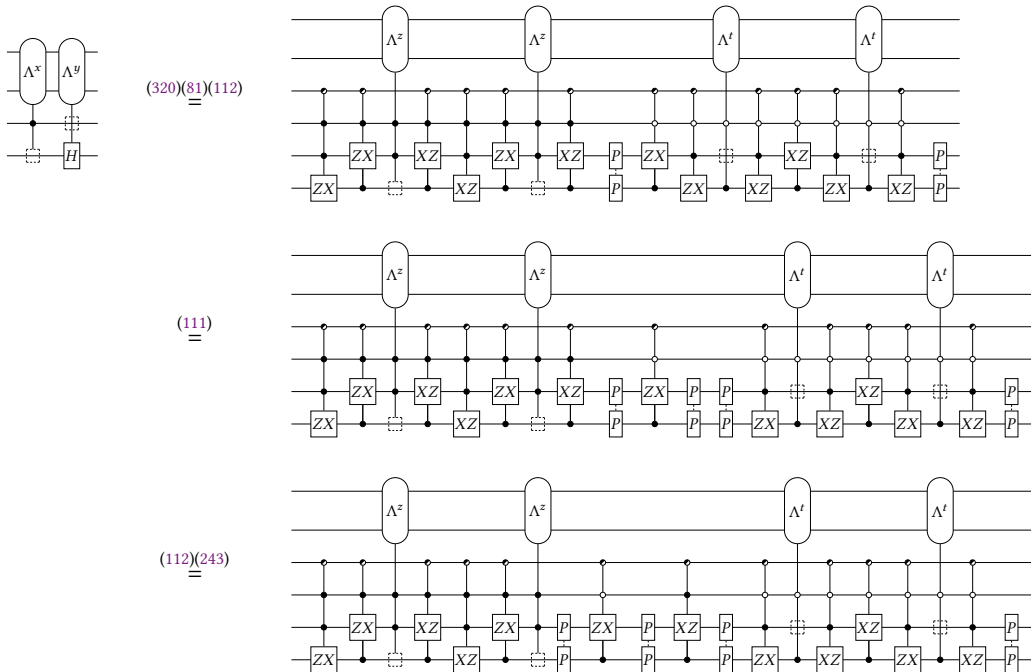

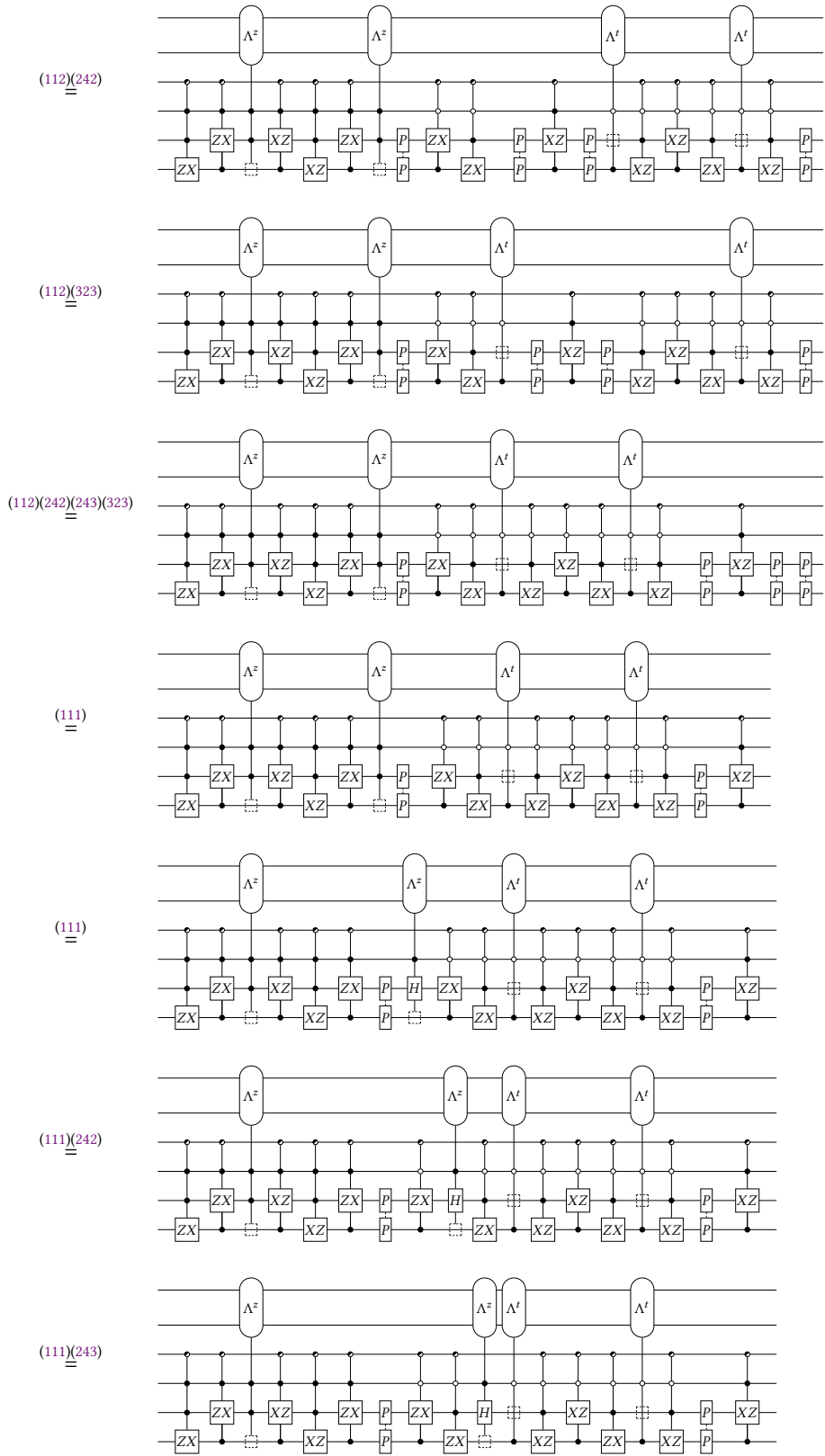

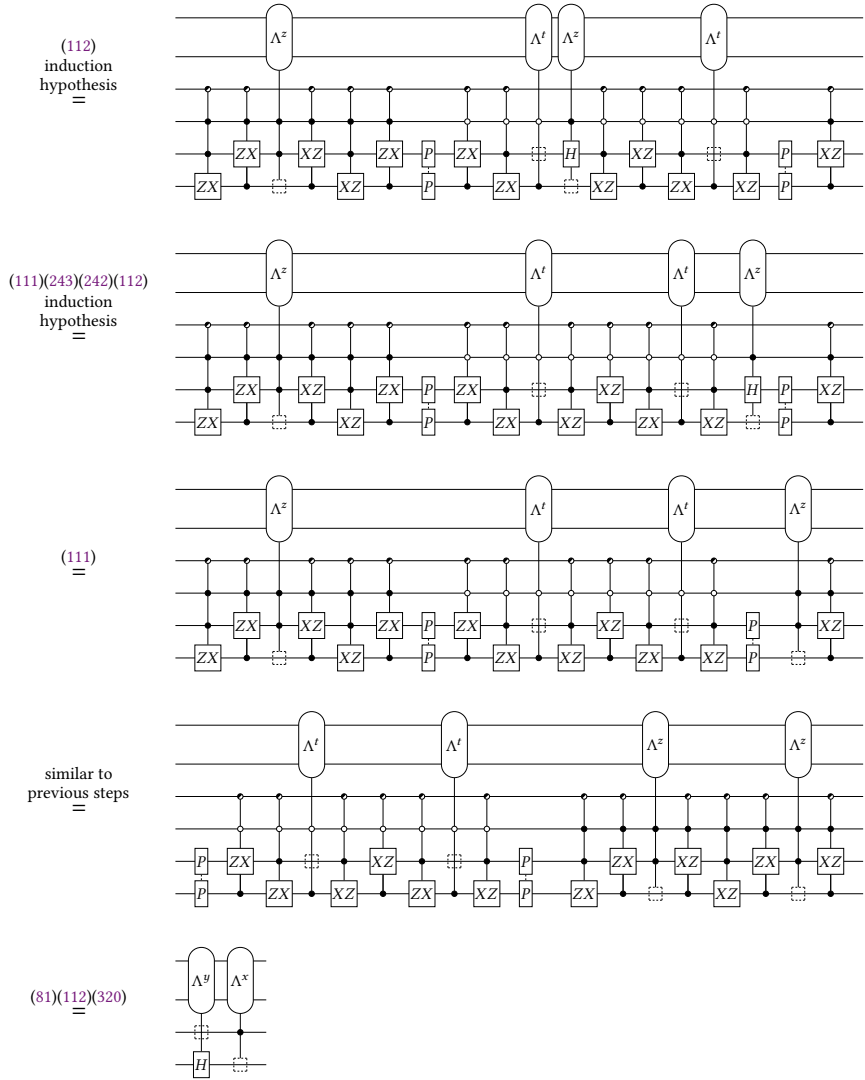

If $n \geq 1$ and $x = y$, then again by Equation (316) we can assume without loss of generality that the control in front of the □ is black. We have (where, similarly to [11], we use the notation

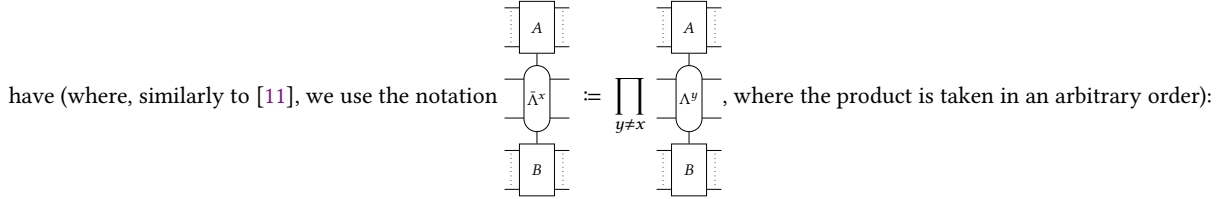

, where the product is taken in an arbitrary order):

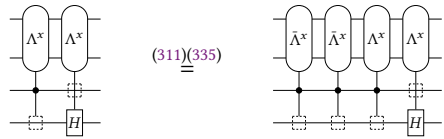

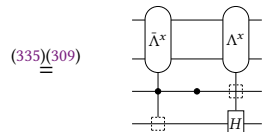

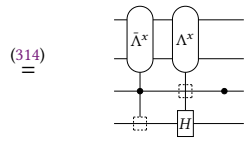

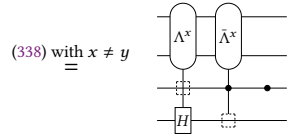

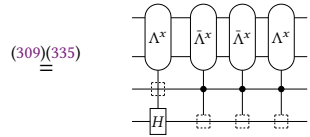

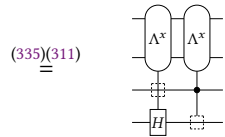

*Proof of Equation* (339)*:* By Equations (307) and (313), we can assume without loss of generality that $x$ and $y$ differ by the last bit, and by Equation (81) we can assume that that bit is 1 in $x$ and 0 in $y$. Moreover, by Equations (315) and (316) we can assume without loss of generality that the controls in front of the $\square$'s are black. Thus, it suffices to prove the following variant of the equation:

$$\forall x, y \in \{0,1\}^n,$$

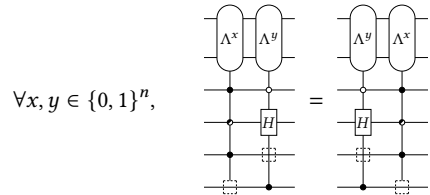

And we have

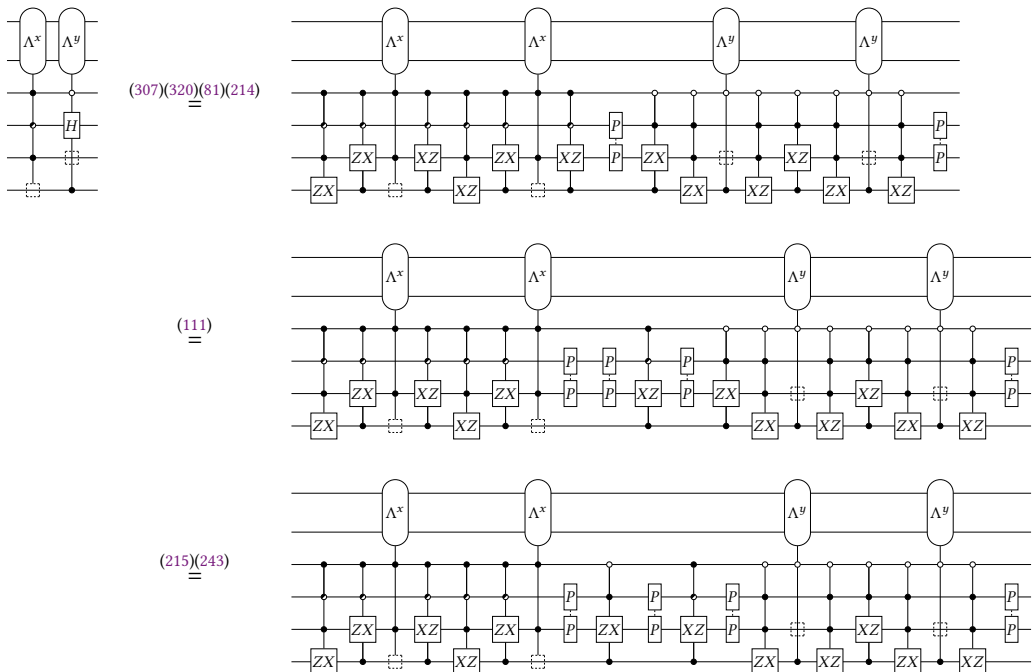

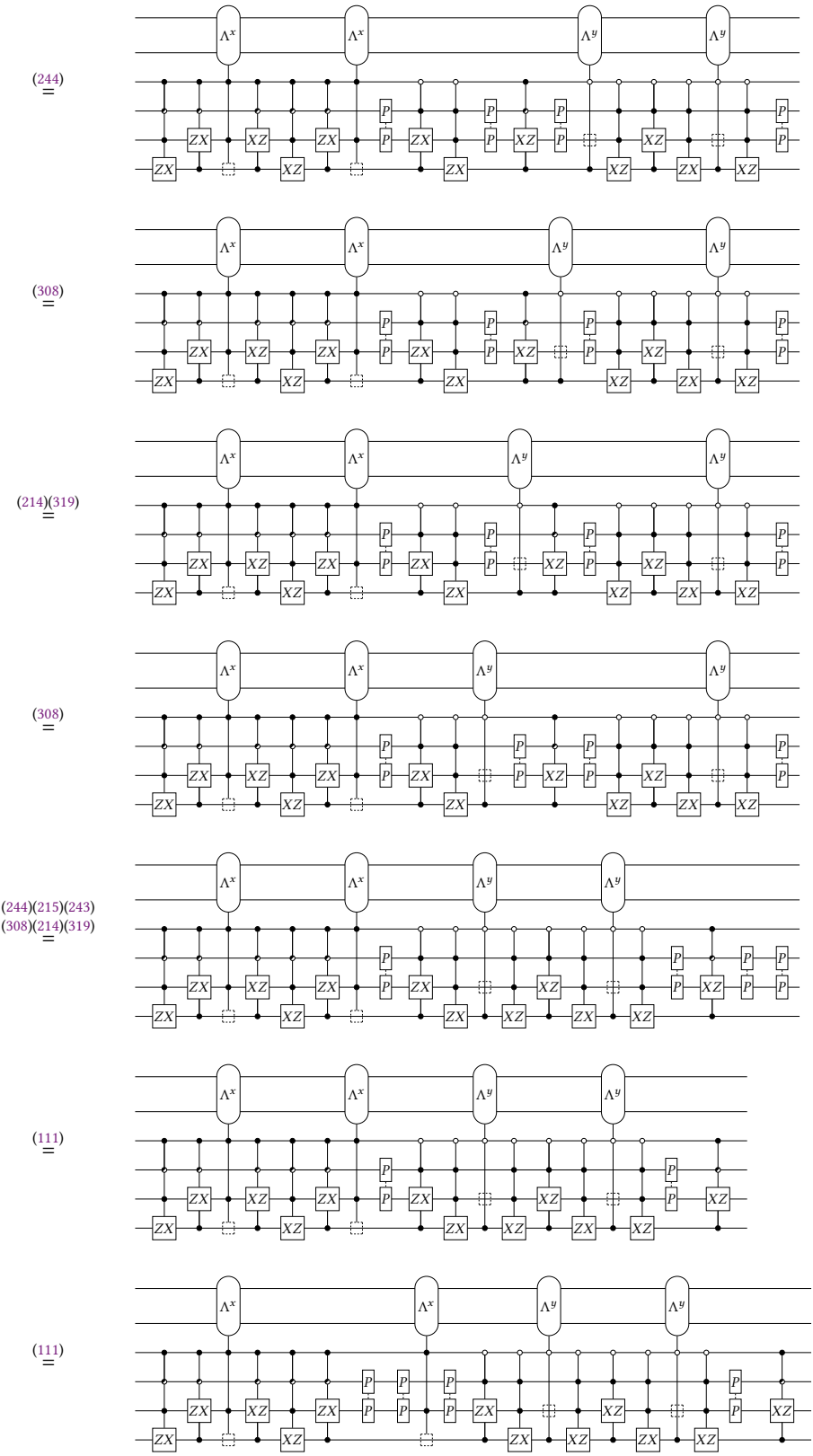

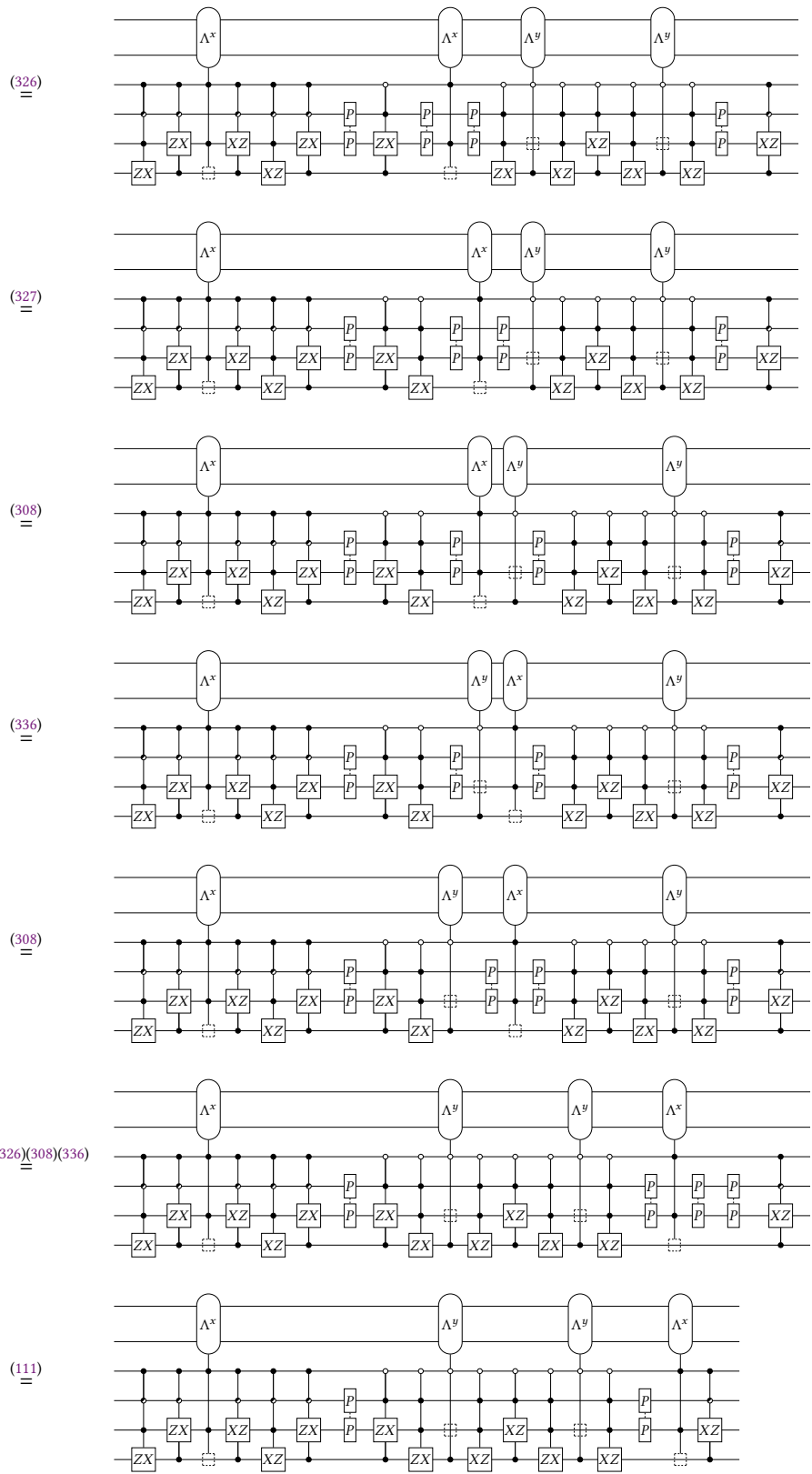

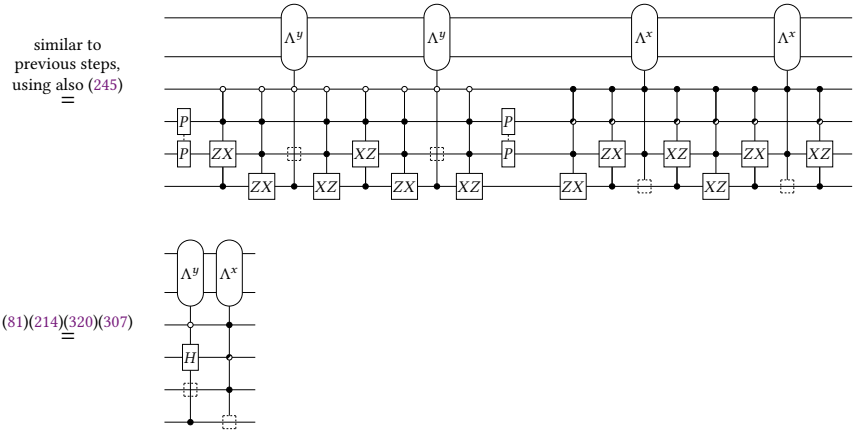

*Proof of Equation* (340):

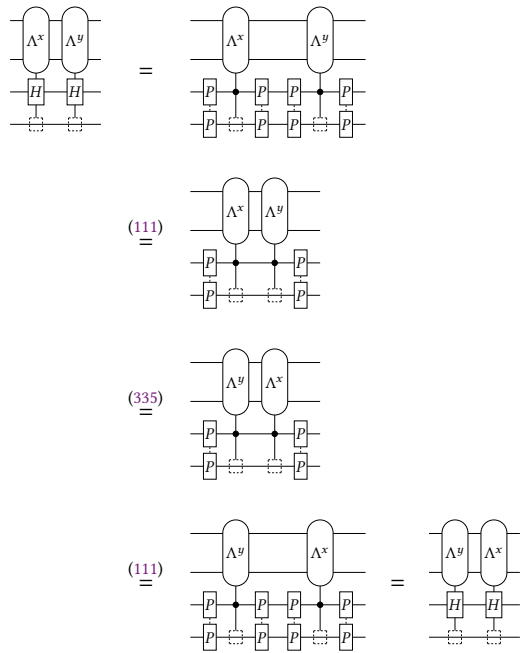

*Proof of Equation* (341): By Equation (316), we can assume without loss of generality that the controls in front of the $\square$'s are black. Then we have

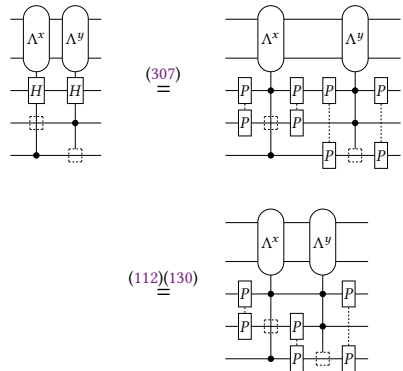

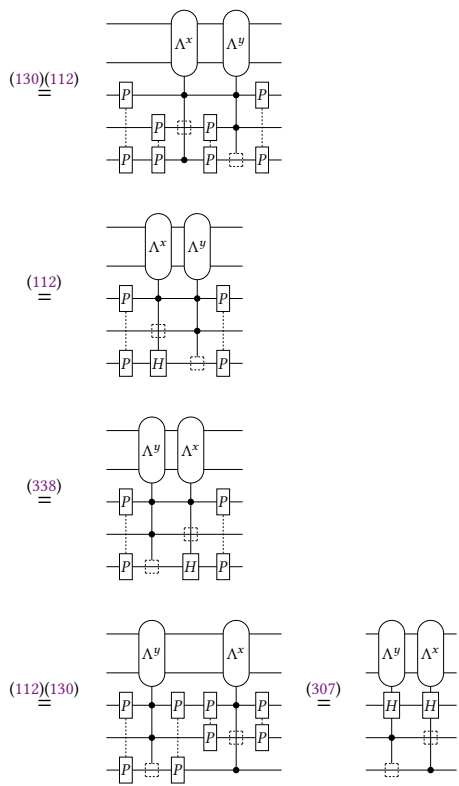

*Proof of Equation* (342): By Equations (307) and (313), we can assume without loss of generality that $x$ and $y$ differ by the last bit, and by Equation (81) we can assume without loss of generality that that bit is 1 in $x$ and 0 in $y$. Namely, it suffices to prove the following variant of the equation:

$$\forall x, y \in \{0, 1\}^n,$$

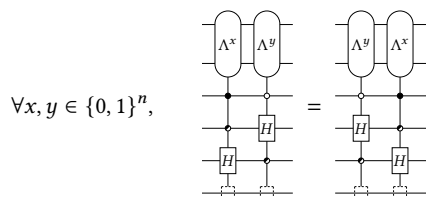

And we have

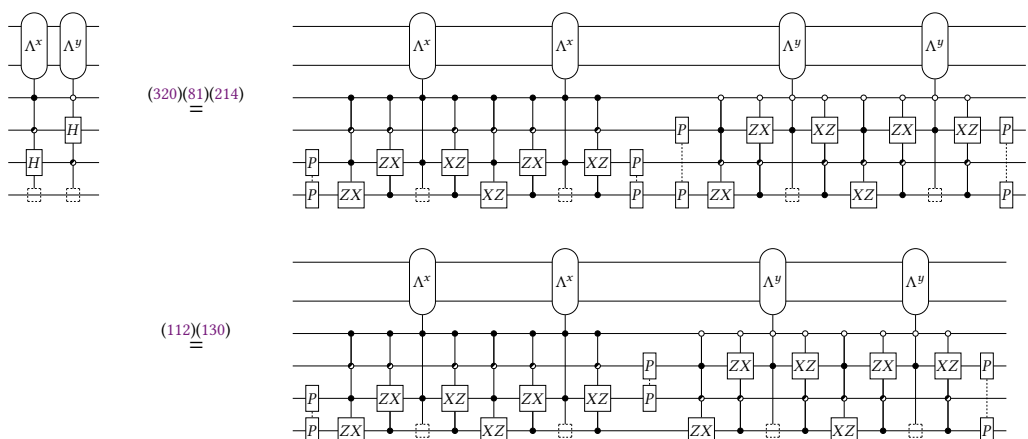

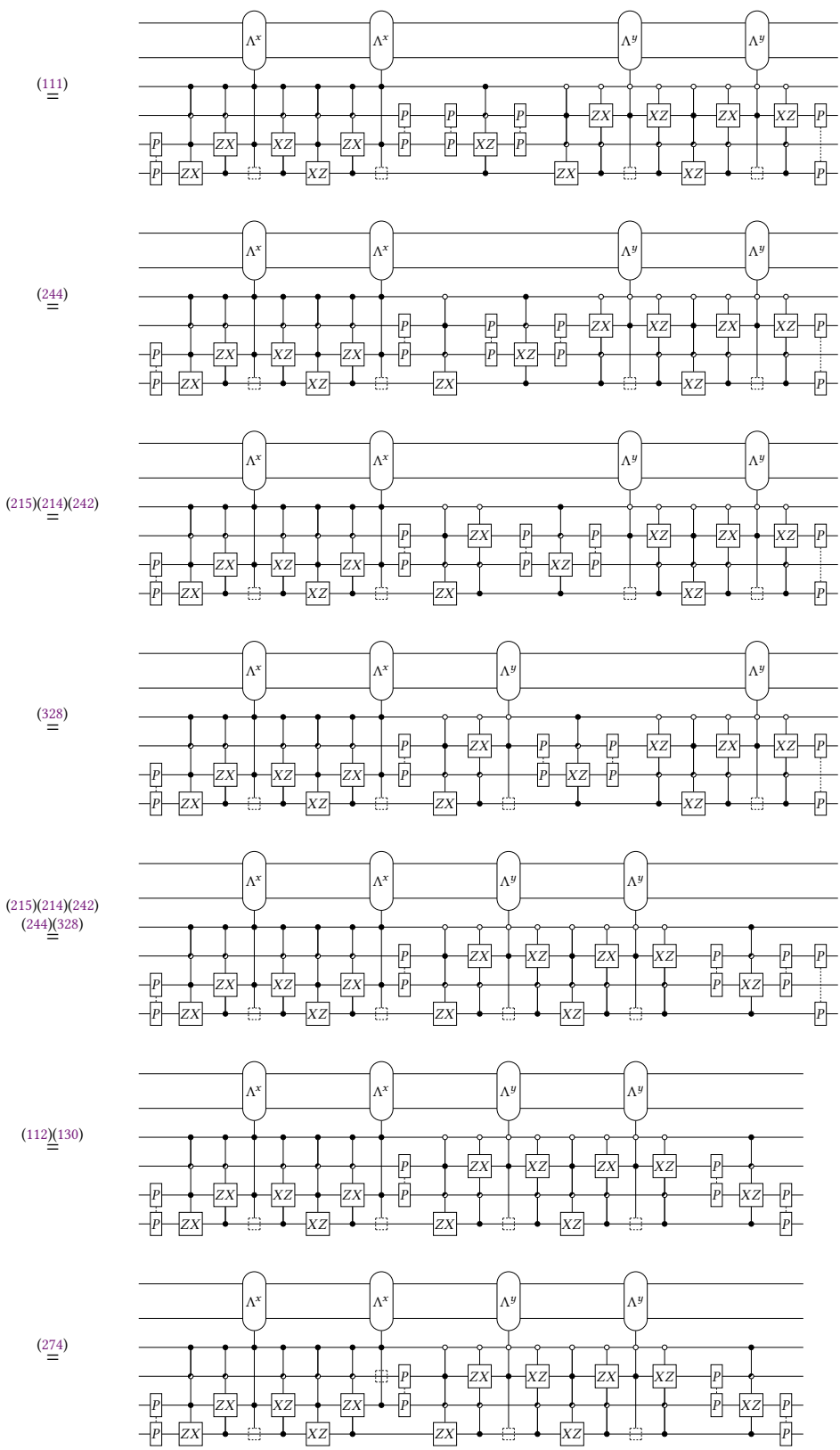

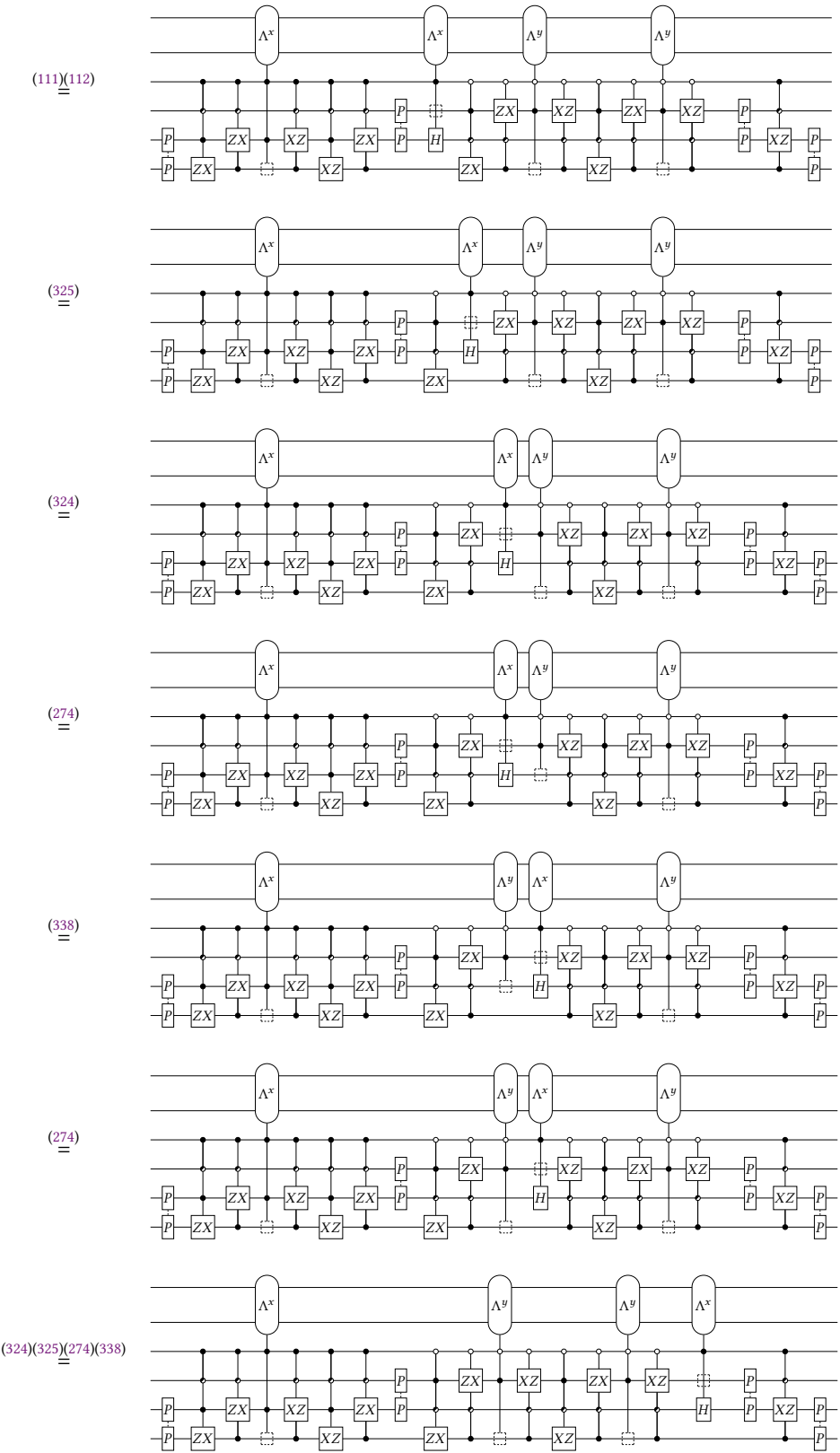

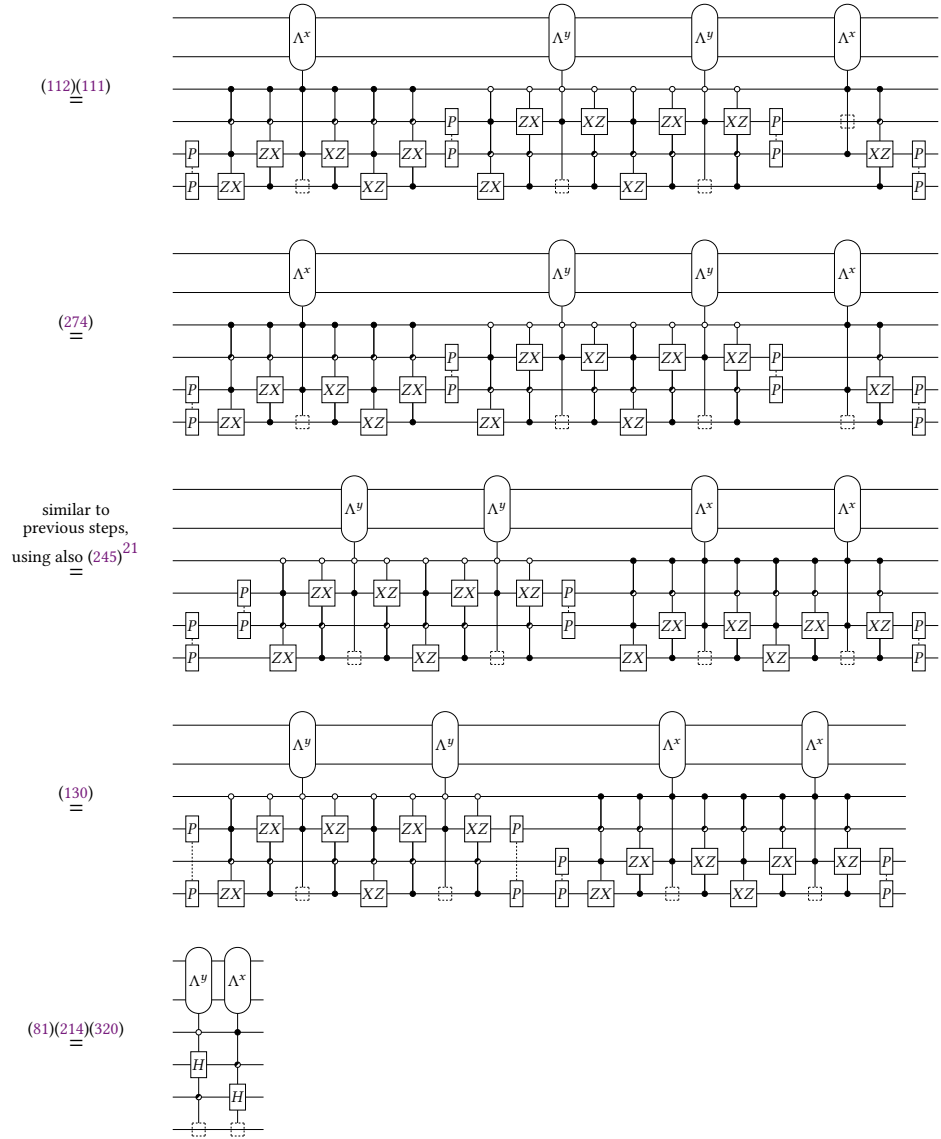

*Proof of Equation* (343): By Equations (307) and (313), we can assume without loss of generality that $x$ and $y$ differ by the last bit, and by Equation (81) we can assume without loss of generality that that bit is 1 in $x$ and 0 in $y$. Moreover, by Equation (316) we can assume without loss of generality that the controls in front of the $\square$'s are black. Thus, it suffices to prove the following variant of the equation:

$$\forall x, y \in \{0,1\}^n \text{ with } x \neq y,$$ 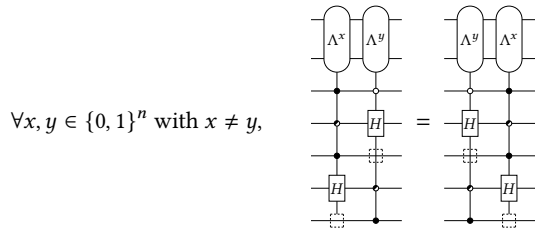

---

[21] Note that the $\Lambda^{0a1}ZX$ and $\Lambda^{0a1}XZ$ gates can be pushed through $\Lambda^{y01}(-I)$ in the same way that $\Lambda^{x11}(-I)$ was pushed through $\Lambda^{01b}ZX$.

And we have

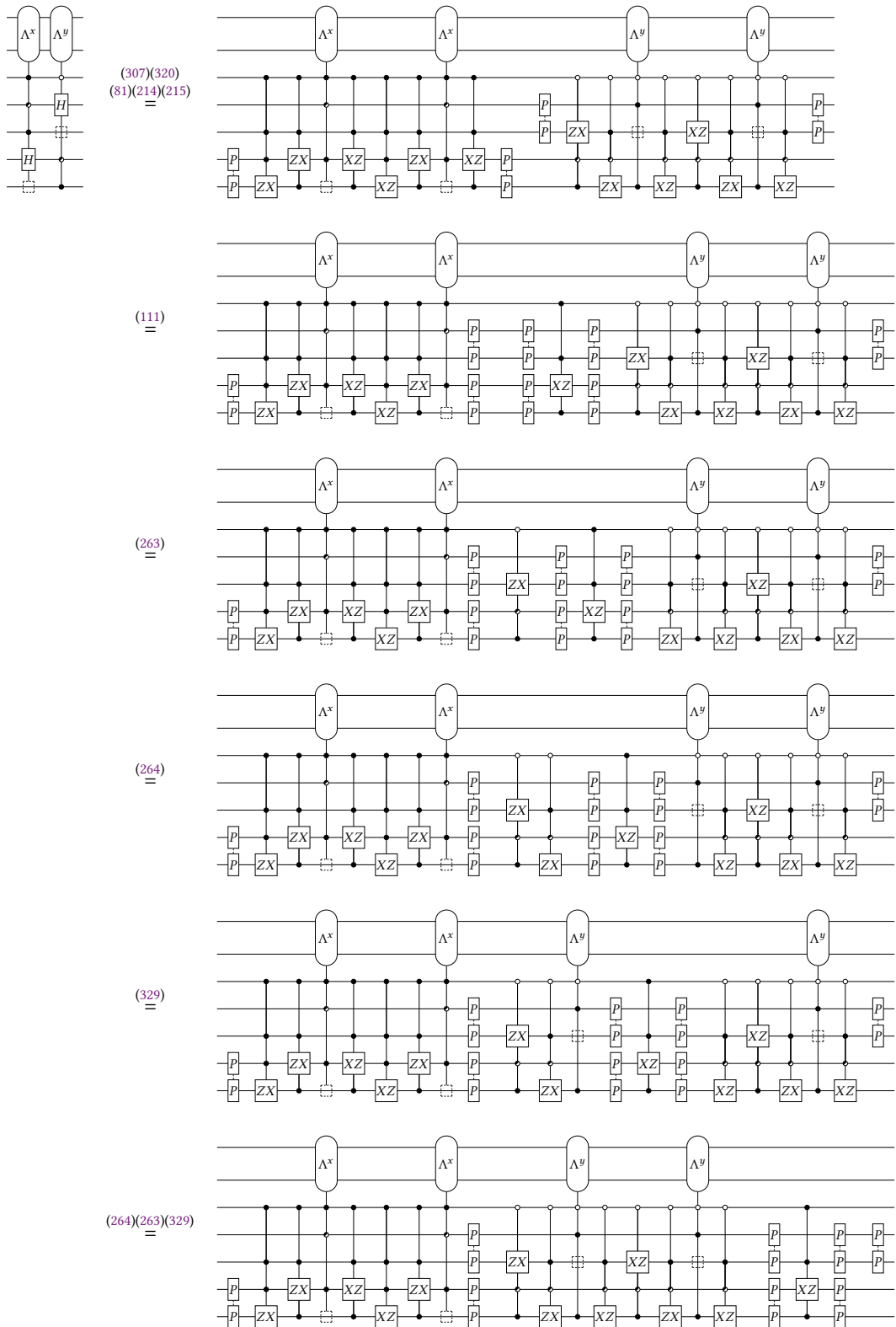

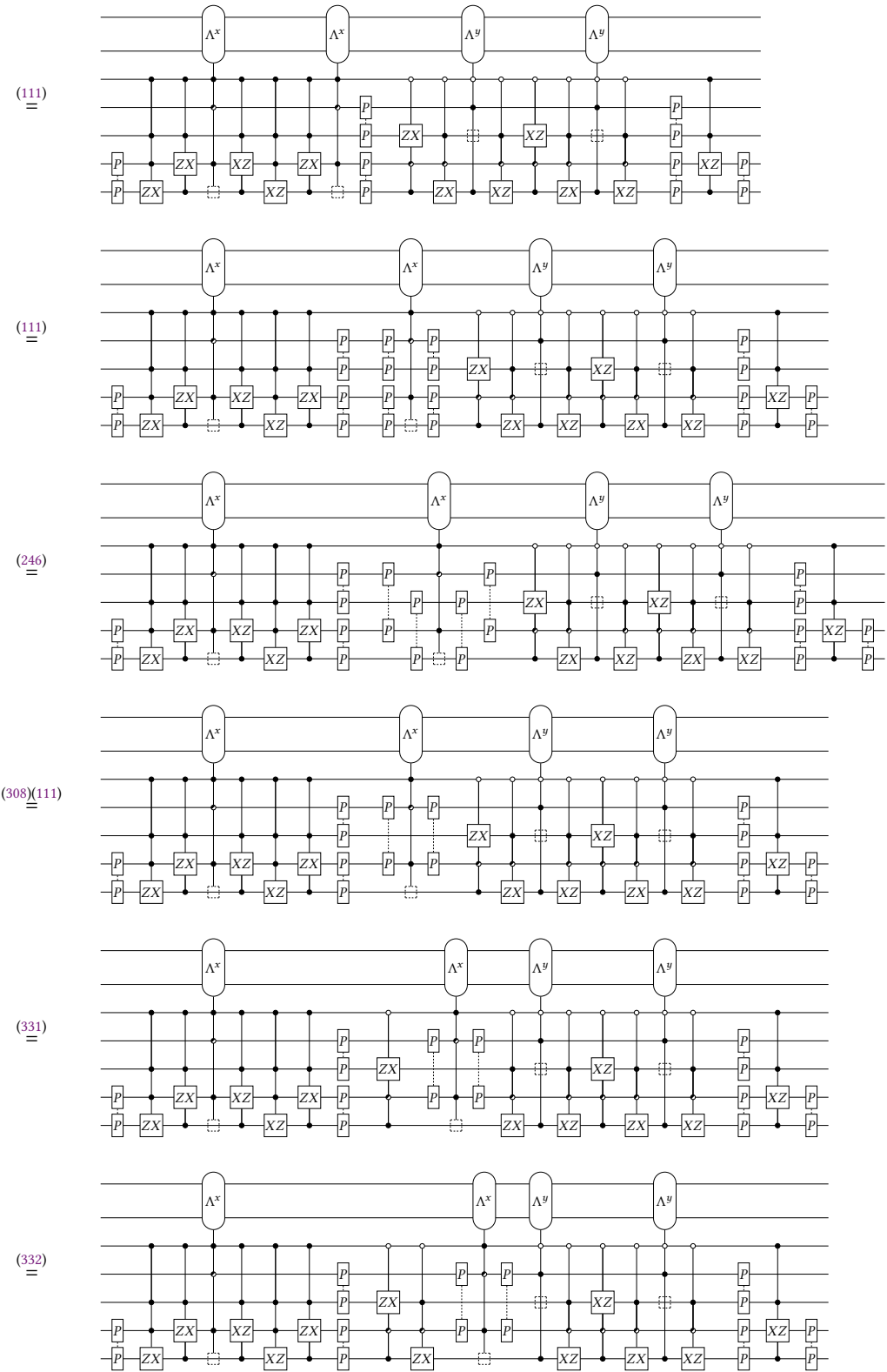

$\underset{=}{(274)}$ 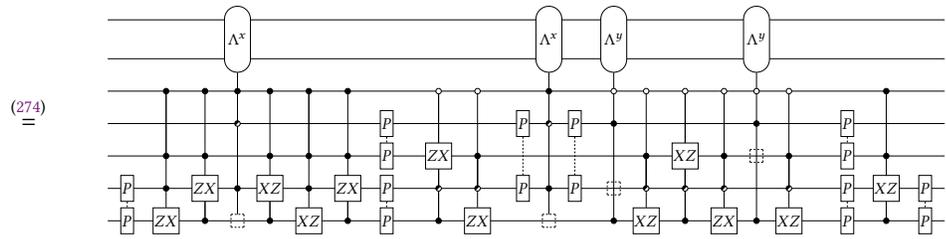

$\underset{=}{(307)(111)}$ 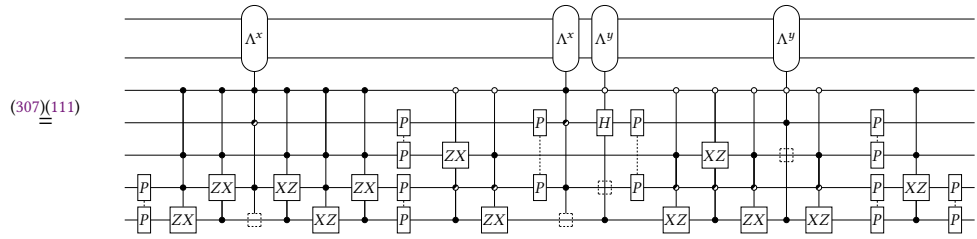

$\underset{=}{(339)}$ 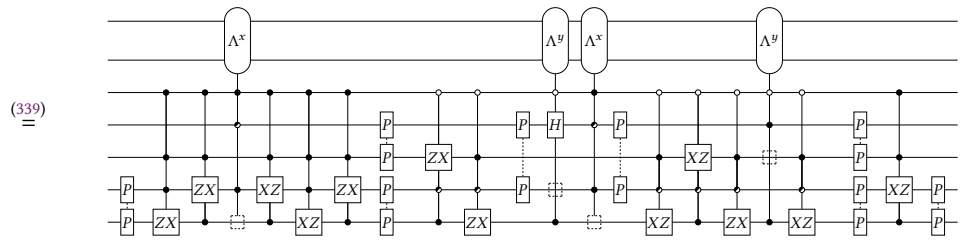

$\underset{=}{(111)(307)}$ 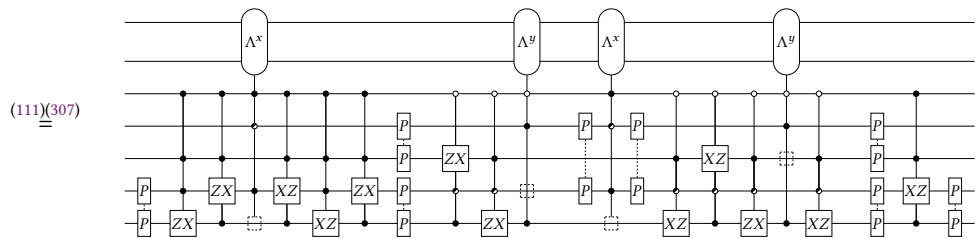

$\underset{=}{(274)}$ 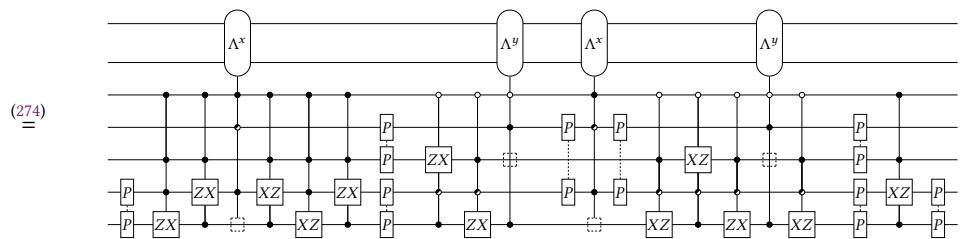

$\underset{=}{\begin{array}{l}(332)(331)(274)\\(307)(111)(339)\end{array}}$ 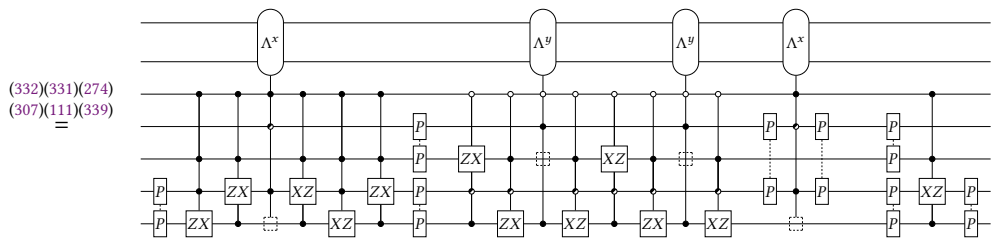

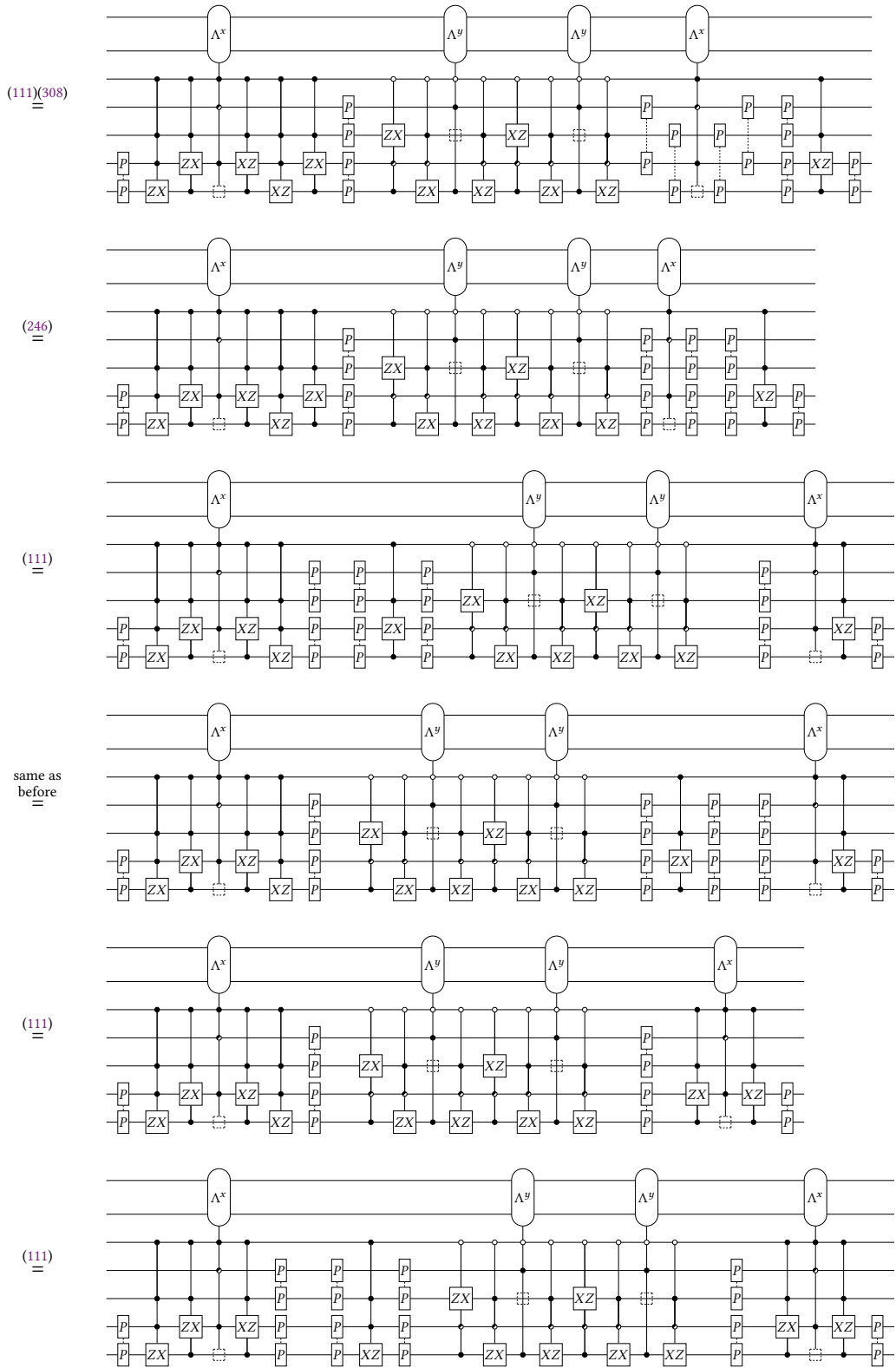

(265)
=

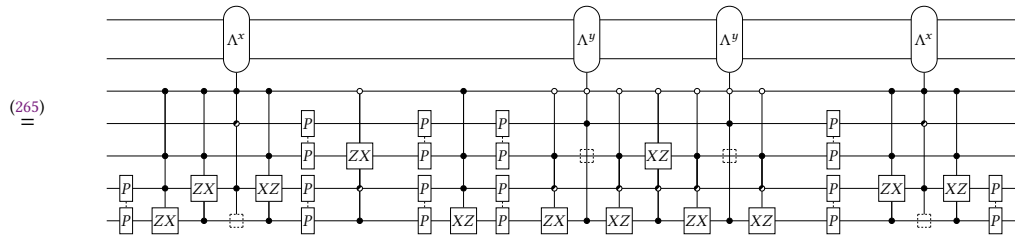

(266)
=

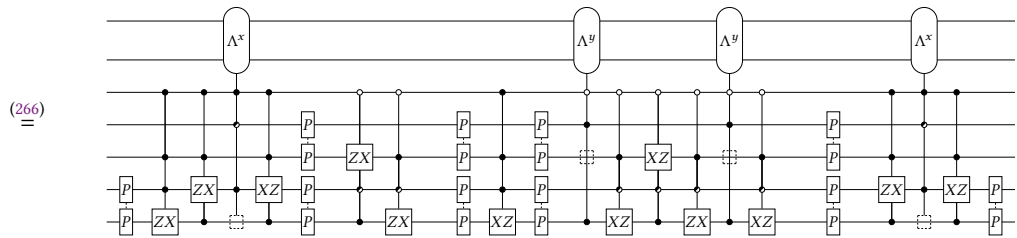

(330)
=

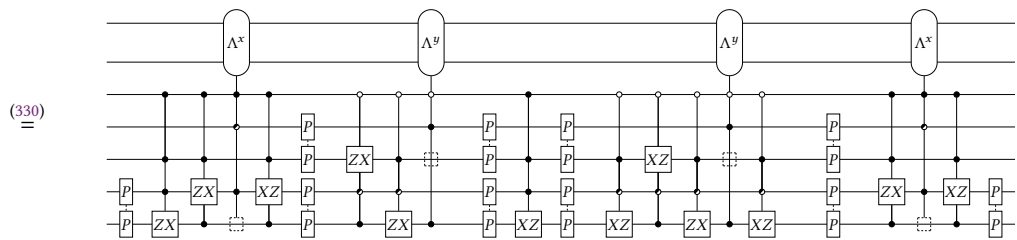

(266)(265)(330)
=

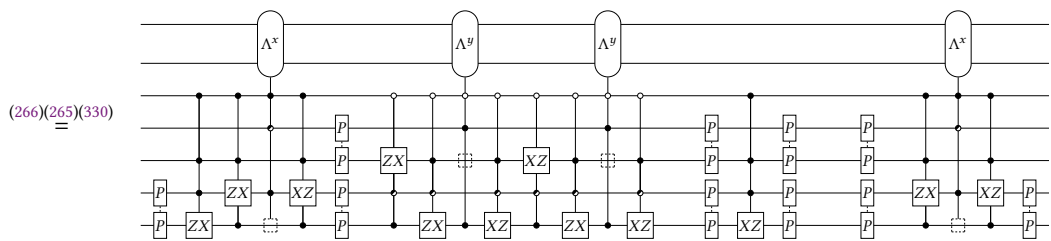

(111)
=

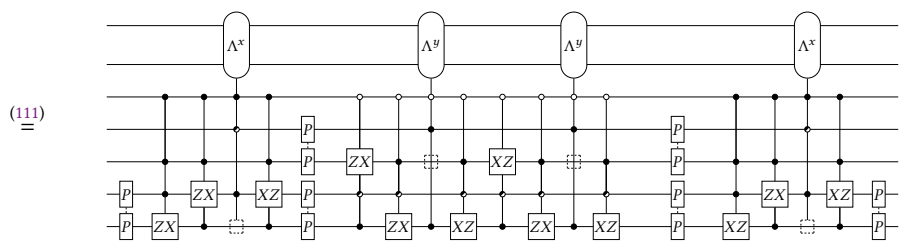

same as
previous steps
=

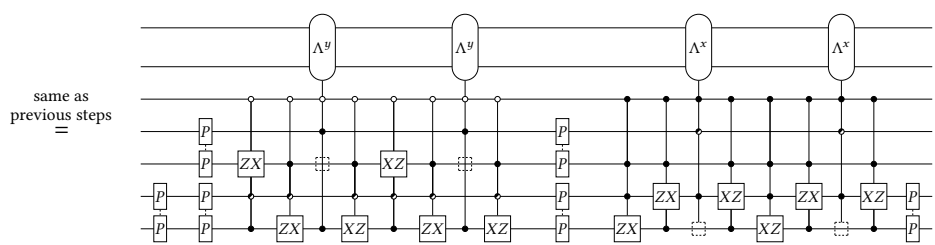

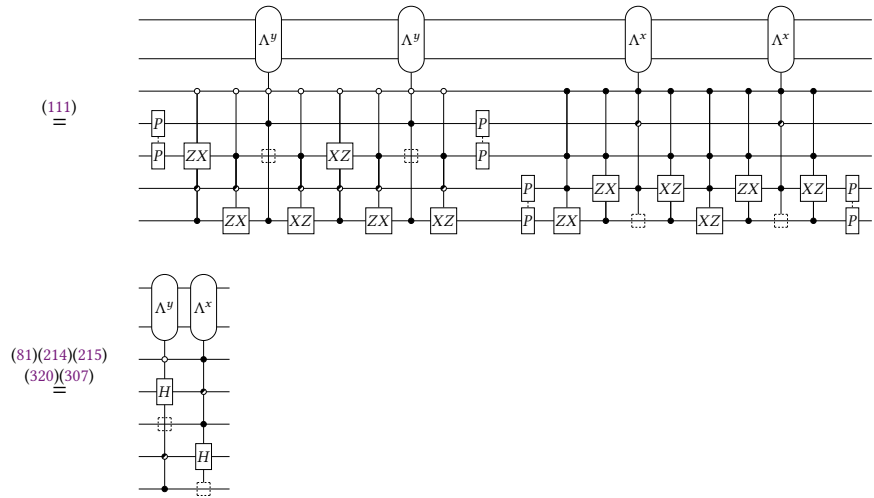

(111)
=

(81)(214)(215)
(320)(307)
=

*Proof of Equation* (344): By Equations (307) and (313), we can assume without loss of generality that $x$ and $y$ differ by the last bit, and by Equation (81) we can assume without loss of generality that that bit is 1 in $x$ and 0 in $y$. Moreover, by Equation (316) we can assume without loss of generality that the control in front of the $\square$ is black. Thus, it suffices to prove the following variant of the equation:

$\forall x, y \in \{0, 1\}^n$ with $x \neq y$,

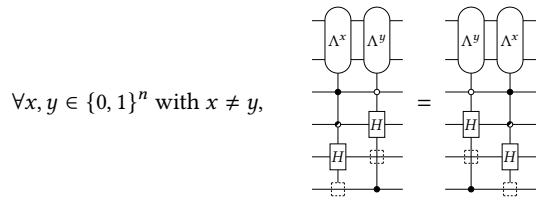

Then we distinguish whether the control in front of the $H$ is black or white.

If it is black, then we have

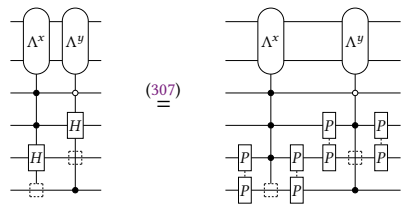

(307)
=

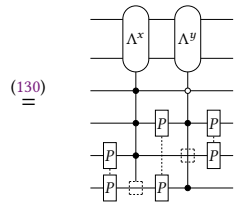

(130)
=

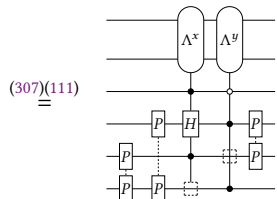

(307)(111)
=

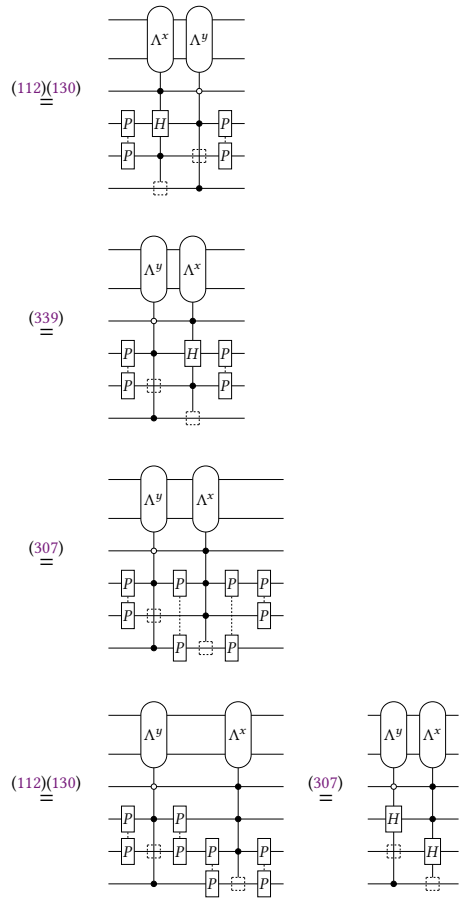

And if it is white, then we have

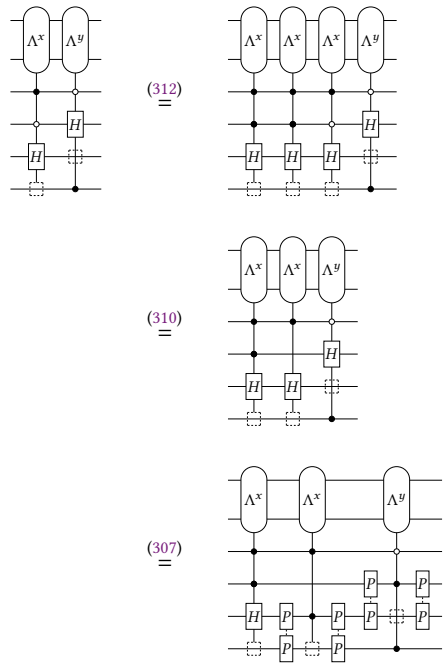

(130)
=
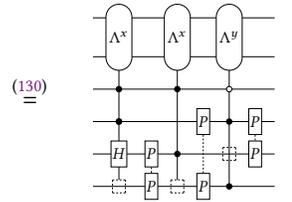

(308)
=
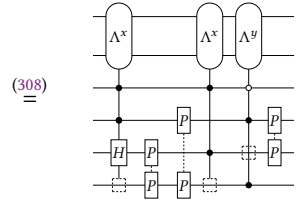

(112)(130)
=
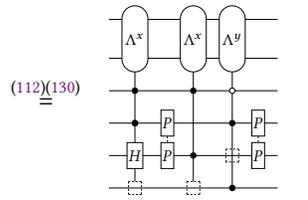

(309)
=
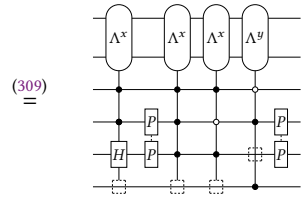

(336)
=
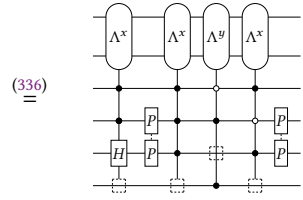

(336)
=
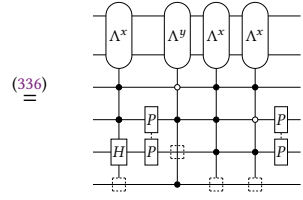

(111)(307)
=
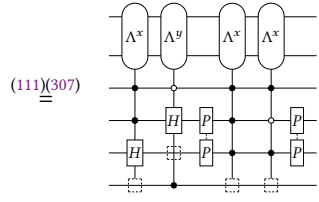

$\overset{(309)}{=}$ 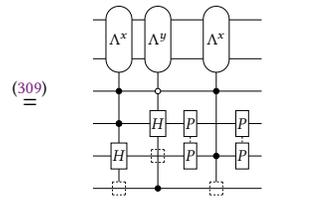

$\overset{(111)(308)}{=}$ 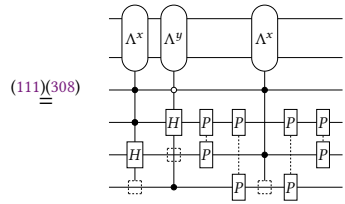

$\overset{(112)(130)}{=}$ 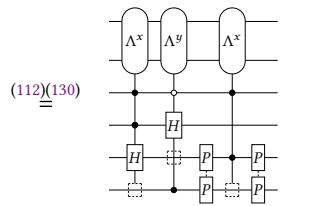

$=$ 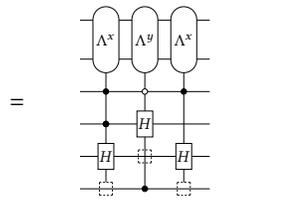

black case $=$ 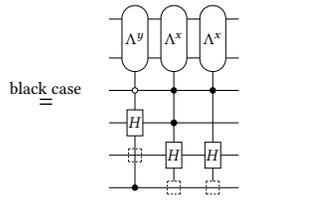

$\overset{(310)}{=}$ 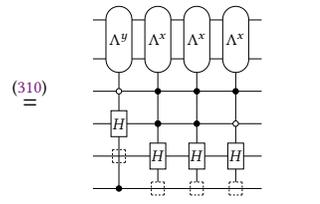

$\overset{(312)}{=}$ 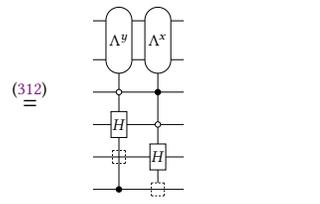

*Proof of Equation* (345):

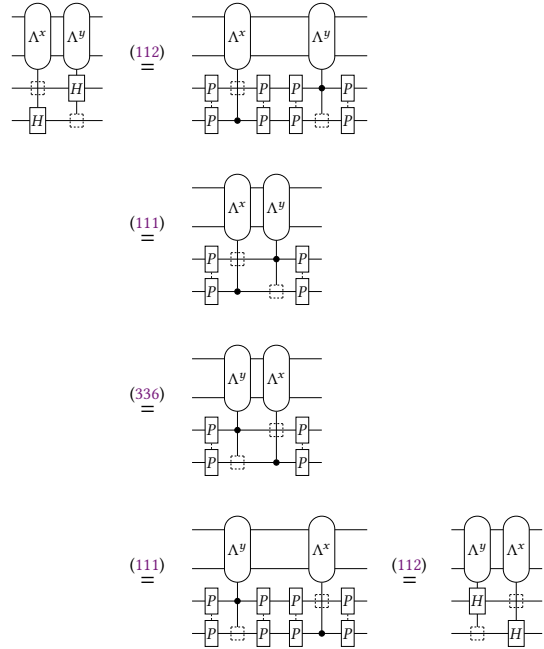

*Proof of Equation* (333):

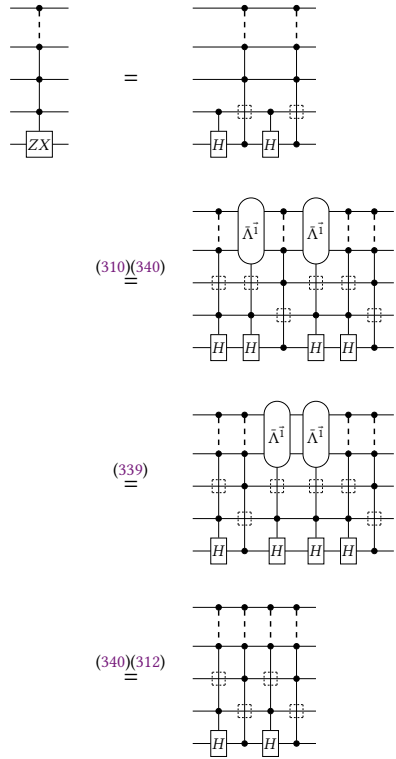

*Proof of Equation* (334): Equation (334) follows directly from Equation (333) and Lemma B.5 (used together with Equations (311), (312) and (317)).

*Proof of Equations* (346) *to* (352): The particular cases of these equations that are on one or two qubits follow directly from Lemmas 7.1 and 7.6,[22] therefore we only need to address the cases where there are at least three qubits.

By Equations (333) and (334) (together – if there are white controls – with Equation (81) and maybe with Equation (315) and/or (316)), any multi-controlled $ZX$ or $XZ$ gate with arbitrary combination of black and white controls can be written as a product of multi-controlled $(-I)$ and $H$ gates with a hole. Thus, we can deduce Equations (346) to (352) from Equations (335) to (345) whenever there is a black control in front of a white control, that is, in all cases except maybe in the particular case of Equations (346) and (351) where $x = y$. For Equation (351), this particular case is trivial if $a = b$, and follows directly from Equation (317) if $a \neq b$. Therefore it only remains to address it for Equation (346). By Equation (81), we can assume without loss of generality that $x$ (and therefore also $y$) is the vector consisting only of 1's; and by Lemma B.5 (used together with Equation (317)), we can assume without loss of generality that $a = 0$. Then we have

## D.8 Properties of Multi-Controlled $ZX$ and $XZ$ Gates

**Lemma D.14.** *The following equations are consequences of the equations of Figure 4:*

$$(353)$$

---

[22]One can notice that those cases are relevant only for Equations (346) and (351).

$$(354)$$

Proof.

*Proof of Equation* (353): By Lemma B.5 (used together with Equation (317)), we can assume without loss of generality that $a = 0$. The three-qubit case of the equation (where there are no controls above or below the swap) is just Equation (127).
If there is at least one control below the swap, then we have

Finally, if there is no control below the swap, but at least one control above, then we have

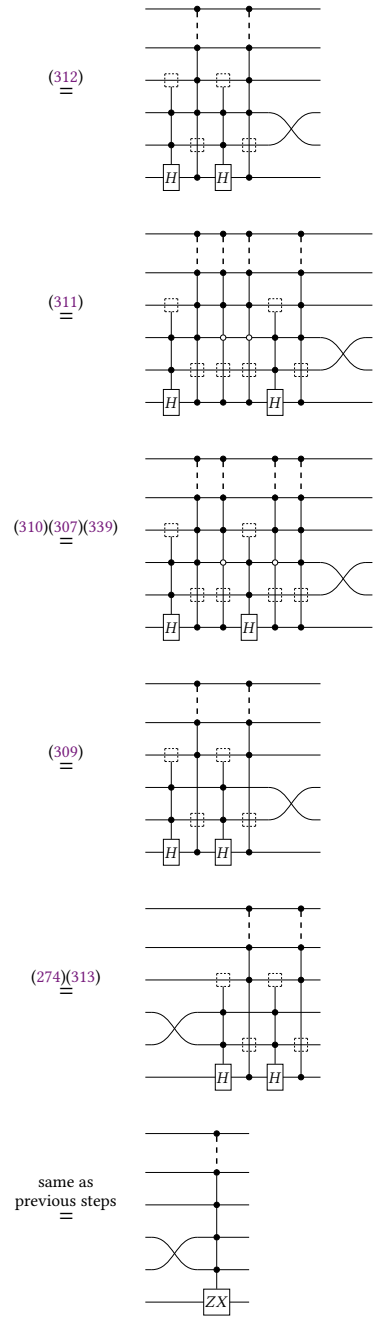

*Proof of Equation* (354): By Lemma B.5 (used together with Equation (317)), we can assume without loss of generality that $a = 0$, and by Equation (353), we can assume without loss of generality that $x = \epsilon$. If $y$ is of length 0, then the result follows from Lemma 7.6. If $y$ is of length 1, then the result follows from Equations (136) and (137) together with Equation (81). If $y$ is of length at least 2, then by Equation (81) we can assume without loss of generality that is the vector consisting only of 1's. Then we have

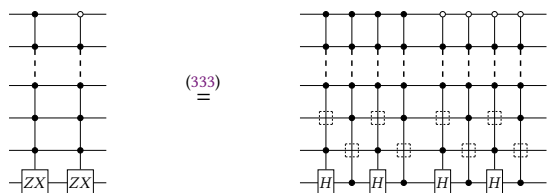

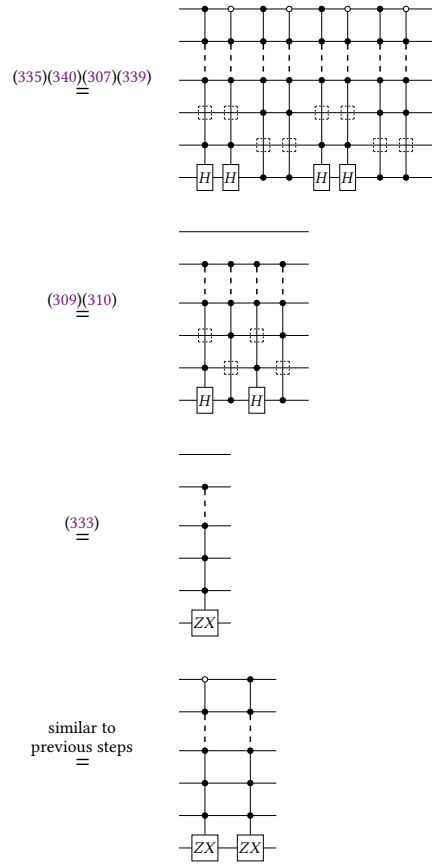

$$\square$$

## D.9   Additional Useful Auxiliary Equations on $n$ Qubits

**Lemma D.15.** *The following equations are consequences of the equations of Figure 4:*

$$(355)$$

$$(356)$$

$$(357)$$

$$(358)$$

$$(359)$$

$$(360)$$

Proof.

*Proof of Equation* (355)*:* If $x$ is of length 0, then the result follows from Lemma 7.1. If $x$ is of length at least 1, then by Equation (81), we can assume without loss of generality that $x$ is the vector consisting only of 1's, and by Lemma B.5 (used together with Equation (317)), we can assume without loss of generality that $a = 0$. Then we have

*Proof of Equation* (356)*:* The two-qubit case of the equation (where there are no controls above the two bottom qubits) follows from Lemma 7.6. In the other cases (where there is at least one control above), we have

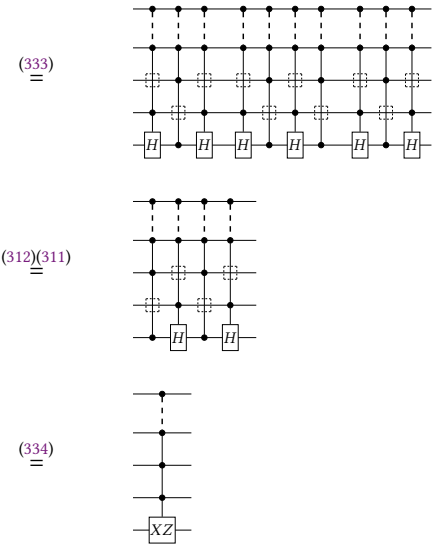

*Proof of Equation* (357): The two-qubit case of the equation (where there are no controls above the two bottom qubits) follows from Lemma 7.6. In the other cases (where there is at least one control above), we have

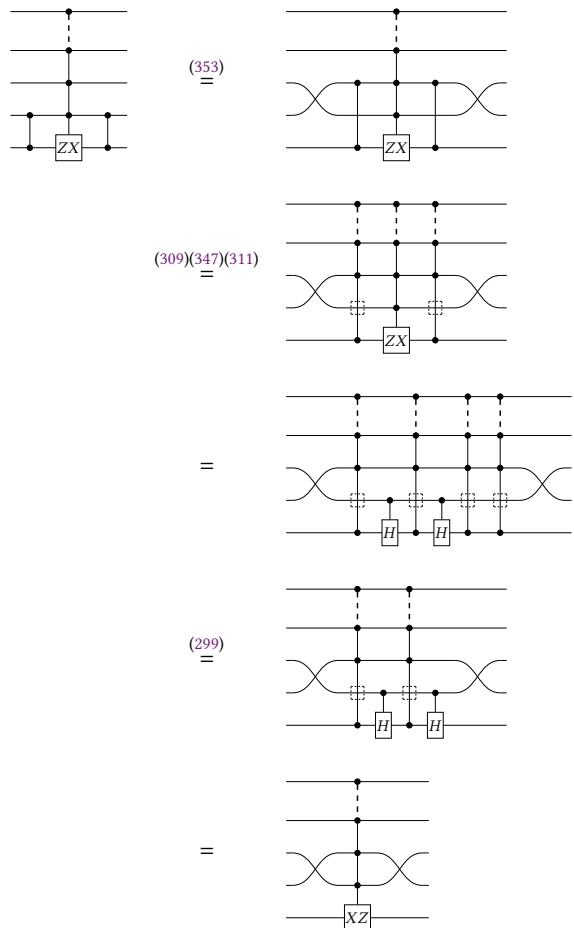

*Proof of Equation* (358):

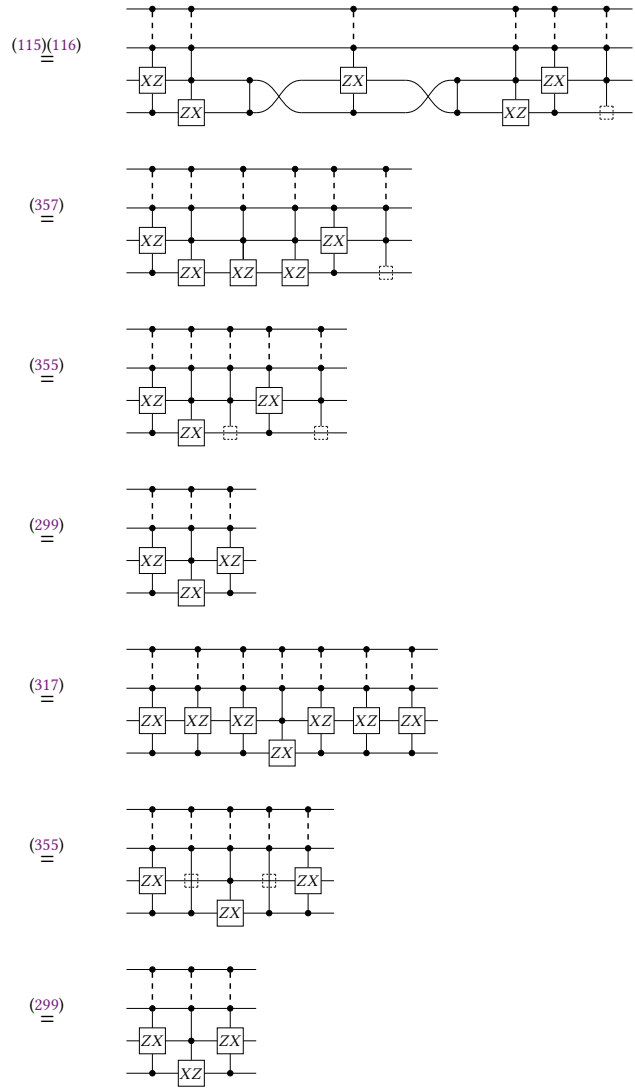

*Proof of Equation* (359):

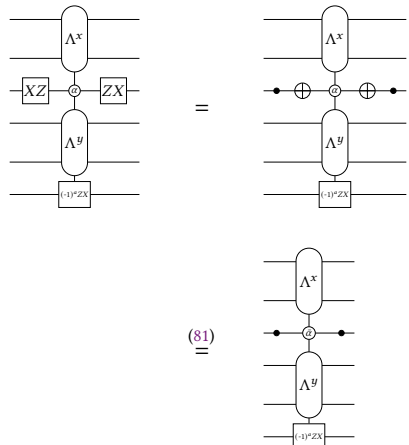

$$\overset{(309)(346)}{\equiv}$$

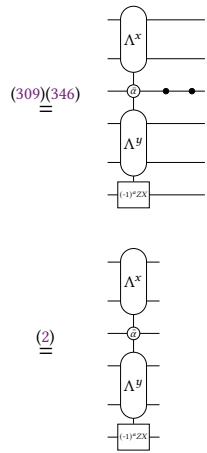

$$\overset{(2)}{\equiv}$$

and

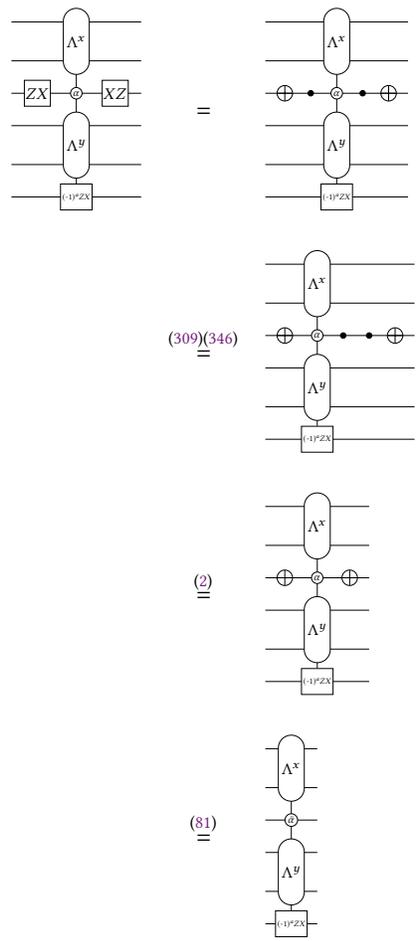

*Proof of Equation* (360): The two-qubit case of the equation (where the *ZX* gate has no controls) follows from Lemma 7.6. The three-qubit case of the equation (where the *ZX* gate has exactly one control) is just Equation (129). In the other cases, we have

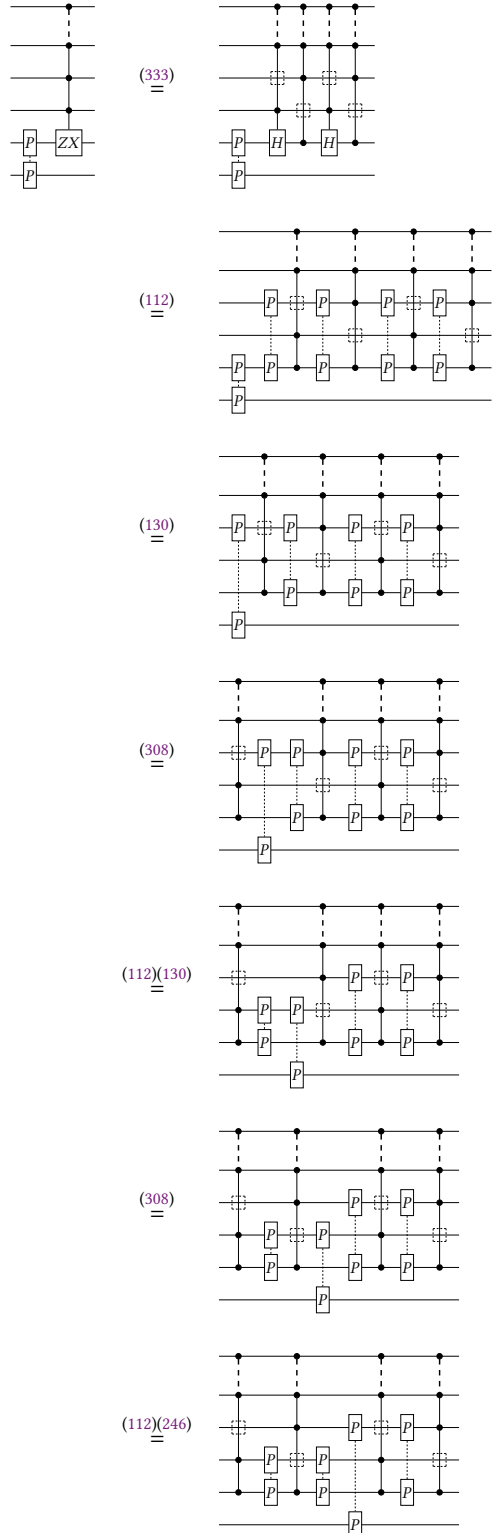

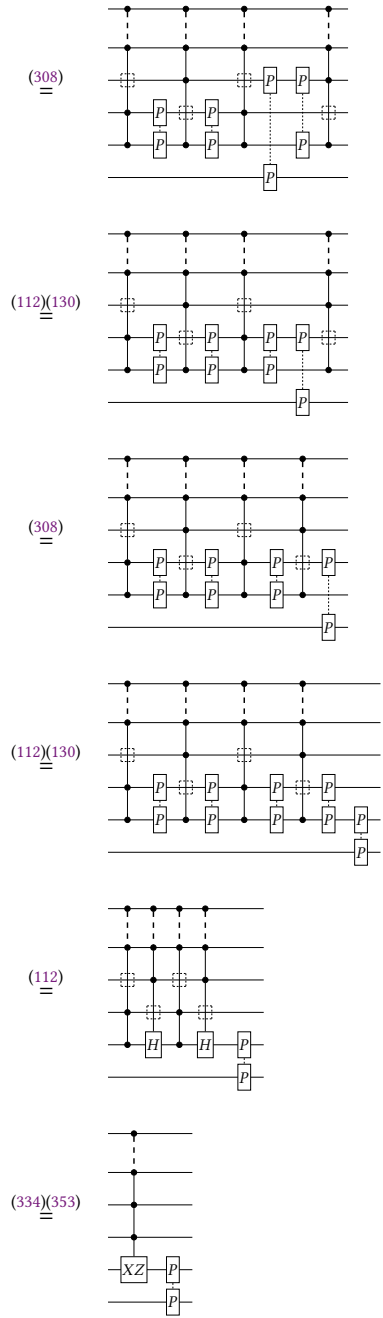

# E   Completeness in the General Case

## E.1   Definition of $\Sigma_{a,b,c,d}$ and $\Sigma'_{a,b,c,d}$

We need to provide two words $\Sigma_{a,b,c,d}$ and $\Sigma'_{a,b,c,d}$ over $\mathcal{P}_N$, not containing any generators of the form $\left(H_{[e,f]}H_{[g,h]}\right)$, such that $\left[\!\left[\Sigma'_{a,b,c,d}\right]\!\right] = \left[\!\left[\Sigma_{a,b,c,d}\right]\!\right]^{-1}$, and that there exists $\mu \notin \{0,1,3,2\}$ and a permutation matrix $\sigma$ mapping respectively $|0\rangle, |1\rangle, |3\rangle, |2\rangle$ to $|a\rangle, |b\rangle, |c\rangle, |d\rangle$, such that either $\left[\!\left[\Sigma_{a,b,c,d}\right]\!\right] = \sigma$ or $\left[\!\left[\Sigma_{a,b,c,d}\right]\!\right] = \sigma \circ \left[\!\left[(-1)_{[\mu]}\right]\!\right]$. The following definition fulfils these properties:

**Definition E.1.**
$$\Sigma_{a,b,c,d} = \left((-1)_{[m_0]}X_{[a,i_0]}\right)^{e_0}\left((-1)_{[m_1]}X_{[b,i_1]}\right)^{e_1}\left((-1)_{[m_2]}X_{[c,i_2]}\right)^{e_2}\left((-1)_{[m_3]}X_{[d,2]}\right)^{e_3}$$

$$\Sigma'_{a,b,c,d} = \left((-1)_{[4]}X_{[d,2]}\right)^{e_3}\left((-1)_{[m'_2]}X_{[c,i_2]}\right)^{e_2}\left((-1)_{[m'_1]}X_{[b,i_1]}\right)^{e_1}\left((-1)_{[m'_0]}X_{[a,i_0]}\right)^{e_0}$$

*where*

- $e_3 = \begin{cases} 1 & \text{if } d \neq 2 \\ 0 & \text{if } d = 2 \end{cases}$

- $m_3 = m'_2 = \tau^{e_3}_{d,2}(4)$

- $i_2 = \tau^{e_3}_{d,2}(3)$

- $e_2 = \begin{cases} 1 & \text{if } c \neq i_2 \\ 0 & \text{if } d = i_2 \end{cases}$

- $m_2 = m'_1 = \tau^{e_2}_{c,i_2}(\tau^{e_3}_{d,2}(4))$

- $i_1 = \tau^{e_2}_{c,i_2}(\tau^{e_3}_{d,2}(1))$

- $e_1 = \begin{cases} 1 & \text{if } b \neq i_1 \\ 0 & \text{if } b = i_1 \end{cases}$

- $m_1 = m'_0 = \tau^{e_1}_{b,i_1}(\tau^{e_2}_{c,i_2}(\tau^{e_3}_{d,2}(4)))$

- $i_0 = \tau^{e_1}_{b,i_1}(\tau^{e_2}_{c,i_2}(\tau^{e_3}_{d,2}(0)))$

- $e_0 = \begin{cases} 1 & \text{if } a \neq i_0 \\ 0 & \text{if } a = i_0 \end{cases}$

- $m_0 = \tau^{e_0}_{a,i_0}(\tau^{e_1}_{b,i_1}(\tau^{e_2}_{c,i_2}(\tau^{e_3}_{d,2}(4))))$

*where* $\tau_{k,i}: \{0, ..., 2^n - 1\} \to \{0, ..., 2^n - 1\}$ *denotes the transposition which exchanges* $k$ *and* $i$, *and where we allow* $k = i$ *in* $\left((-1)_{[m]}X_{[k,i]}\right)^e$ *whenever* $e = 0$ *(in which case this is still equal to the identity).*

## E.2    Proof of Lemma 8.4

We proceed by induction on $\left|G_n^{-1}(x0y) - G_n^{-1}(x1y)\right|$. Note that by an elementary property of the Gray code,[23] since $x0y$ and $x1y$ differ by an odd number of bits (namely 1), $\left|G_n^{-1}(x0y) - G_n^{-1}(x1y)\right|$ is necessarily odd.

If $\left|G_n^{-1}(x0y) - G_n^{-1}(x1y)\right| = 1$, then the result follows directly from Definition 8.3.

If $\left|G_n^{-1}(x0y) - G_n^{-1}(x1y)\right| \geq 3$, let $\alpha \in \{0, 1\}$ be such that $G_n^{-1}(x\alpha y) < G_n^{-1}(x\bar{\alpha}y)$. It follows from the properties of the Gray code that $G_n(G_n^{-1}(x\alpha y) + 1)$ differs from $x\alpha y$ by exactly one bit (which is either in $x$ or in $y$ since $G_n^{-1}(x\alpha y) + 1 \neq G_n^{-1}(x\bar{\alpha}y)$), and moreover, that $G_n(G_n^{-1}(x\bar{\alpha}y) - 1)$ differs from $x\bar{\alpha}y$ by the same bit. We assume that the changing bit is in $x$ (the proof being the same if it is in $y$). In other words, there exist two bitstrings $x_1 \in \{0,1\}^{k_1}, x_2 \in \{0,1\}^{k_2}$ and a bit $\beta \in \{0,1\}$ such that $x = x_1\beta x_2$, $G_n(G_n^{-1}(x\alpha y) + 1) = x_1\bar{\beta}x_2\alpha y$ and $G_n(G_n^{-1}(x\bar{\alpha}y) - 1) = x_1\bar{\beta}x_2\bar{\alpha}y$. Then we have

$$D\left((-1)_{[G_n^{-1}(x0y)]}(-1)_{[G_n^{-1}(x1y)]}\right)$$

$$= \prod_{k=G_n^{-1}(x\alpha y)}^{G_n^{-1}(x\bar{\alpha}y)-1} D\left((-1)_{[k]}(-1)_{[k+1]}\right)$$

$$= D\left((-1)_{[G_n^{-1}(x\bar{\alpha}y)-1]}(-1)_{[G_n^{-1}(x\bar{\alpha}y)]}\right) \circ \prod_{k=G_n^{-1}(x\alpha y)+1}^{G_n^{-1}(x\bar{\alpha}y)-2} D\left((-1)_{[k]}(-1)_{[k+1]}\right) \circ D\left((-1)_{[G_n^{-1}(x\alpha y)]}(-1)_{[G_n^{-1}(x\alpha y)+1]}\right)$$

$$= D\left((-1)_{[G_n^{-1}(x\bar{\alpha}y)-1]}(-1)_{[G_n^{-1}(x\bar{\alpha}y)]}\right) \circ D\left((-1)_{[G_n^{-1}(x\alpha y)+1]}(-1)_{[G_n^{-1}(x\bar{\alpha}y)-1]}\right) \circ D\left((-1)_{[G_n^{-1}(x\alpha y)]}(-1)_{[G_n^{-1}(x\alpha y)+1]}\right)$$

$$= D\left((-1)_{[G_n^{-1}(x_1\beta x_2\bar{\alpha}y)]}(-1)_{[G_n^{-1}(x_1\bar{\beta}x_2\bar{\alpha}y)]}\right) \circ D\left((-1)_{[G_n^{-1}(x_1\bar{\beta}x_2\alpha y)]}(-1)_{[G_n^{-1}(x_1\bar{\beta}x_2\bar{\alpha}y)]}\right) \circ D\left((-1)_{[G_n^{-1}(x_1\bar{\beta}x_2\alpha y)]}(-1)_{[G_n^{-1}(x_1\beta x_2\alpha y)]}\right)$$

---

[23] Namely, that the parity of a natural integer is always equal to the parity of the number of 1's in its Gray code.

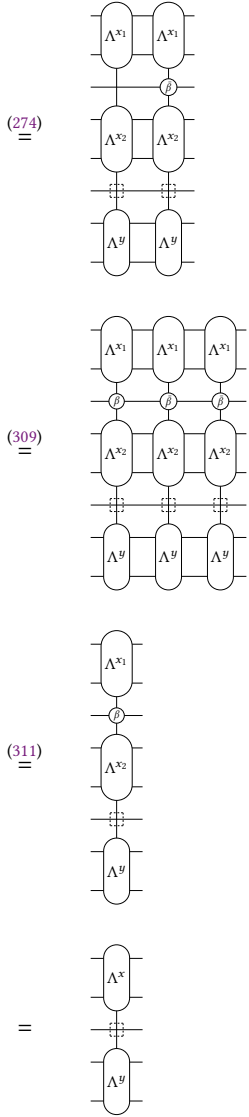

### E.3 Proof of Lemma 8.5

We proceed by induction on $\left|G_n^{-1}(x0y) - G_n^{-1}(x1y)\right|$. Note that for the same reason as in the proof of Lemma 8.4, $\left|G_n^{-1}(x0y) - G_n^{-1}(x1y)\right|$ is necessarily odd.

If $\left|G_n^{-1}(x0y) - G_n^{-1}(x1y)\right| = 1$, then the result follows directly from Definition 8.3.

If $\left|G_n^{-1}(x0y) - G_n^{-1}(x1y)\right| \geq 3$, let $\alpha \in \{0, 1\}$ be such that $G_n^{-1}(x\alpha y) < G_n^{-1}(x\bar{\alpha}y)$. Note that for any $a, b$, one has $D\left((-1)_{[a]}X_{[a,b]}\right) = D\left((-1)_{[a]}X_{[b,a]}\right)$ and $D\left((-1)_{[b]}X_{[a,b]}\right) = D\left((-1)_{[b]}X_{[b,a]}\right)$. Therefore the result that we want to prove is equivalent to

$$\text{QC} \vdash D\left((-1)_{[G_n^{-1}(x\alpha y)]}X_{[G_n^{-1}(x\alpha y),G_n^{-1}(x\bar{\alpha}y)]}\right) =$$

$$\text{QC} \vdash D\left((-1)_{[G_n^{-1}(x\alpha y)]} X_{[G_n^{-1}(x\alpha y), G_n^{-1}(x\breve{\alpha} y)]}\right) = \begin{array}{c} \boxed{\Lambda^x} \\ \boxed{(-1)^x X} \\ \boxed{\Lambda^y} \end{array}$$

Similarly as in the proof of Lemma 8.4, it follows from the properties of the Gray code that $G_n(G_n^{-1}(x\alpha y) + 1)$ differs from $x\alpha y$ by exactly one bit (which is either in $x$ or in $y$ since $G_n^{-1}(x\alpha y) + 1 \neq G_n^{-1}(x\breve{\alpha} y)$), and moreover, that $G_n(G_n^{-1}(x\breve{\alpha} y) - 1)$ differs from $x\breve{\alpha} y$ by the same bit. We assume again that the changing bit is in $x$ (the proof being the same if it is in $y$). In other words, there exist two bitstrings $x_1 \in \{0,1\}^{k_1}, x_2 \in \{0,1\}^{k_2}$ and a bit $\beta \in \{0,1\}$ such that $x = x_1 \beta x_2$, $G_n(G_n^{-1}(x\alpha y) + 1) = x_1 \breve{\beta} x_2 \alpha y$ and $G_n(G_n^{-1}(x\breve{\alpha} y) - 1) = x_1 \breve{\beta} x_2 \breve{\alpha} y$. Then we have

$$D\left((-1)_{[G_n^{-1}(x\alpha y)]} X_{[G_n^{-1}(x\alpha y), G_n^{-1}(x\breve{\alpha} y)]}\right)$$

$$= D\left((-1)_{[G_n^{-1}(x\alpha y)+1]} X_{[G_n^{-1}(x\alpha y), G_n^{-1}(x\alpha y)+1]}\right) \circ D\left((-1)_{[G_n^{-1}(x\breve{\alpha} y)-1]} X_{[G_n^{-1}(x\breve{\alpha} y)-1, G_n^{-1}(x\breve{\alpha} y)]}\right) \circ$$

$$D\left((-1)_{[G_n^{-1}(x\alpha y)+1]} X_{[G_n^{-1}(x\alpha y)+1, G_n^{-1}(x\breve{\alpha} y)-1]}\right) \circ$$

$$D\left((-1)_{[G_n^{-1}(x\breve{\alpha} y)]} X_{[G_n^{-1}(x\breve{\alpha} y)-1, G_n^{-1}(x\breve{\alpha} y)]}\right) \circ D\left((-1)_{[G_n^{-1}(x\alpha y)]} X_{[G_n^{-1}(x\alpha y), G_n^{-1}(x\alpha y)+1]}\right)$$

$$= D\left((-1)_{[G_n^{-1}(x_1 \breve{\beta} x_2 \alpha y)]} X_{[G_n^{-1}(x_1 \breve{\beta} x_2 \alpha y), G_n^{-1}(x_1 \breve{\beta} x_2 \alpha y)]}\right) \circ D\left((-1)_{[G_n^{-1}(x_1 \breve{\beta} x_2 \alpha y)]} X_{[G_n^{-1}(x_1 \breve{\beta} x_2 \alpha y), G_n^{-1}(x_1 \breve{\beta} x_2 \alpha y)]}\right) \circ$$

$$D\left((-1)_{[G_n^{-1}(x_1 \breve{\beta} x_2 \alpha y)]} X_{[G_n^{-1}(x_1 \breve{\beta} x_2 \alpha y), G_n^{-1}(x_1 \breve{\beta} x_2 \alpha y)]}\right) \circ$$

$$D\left((-1)_{[G_n^{-1}(x_1 \breve{\beta} x_2 \breve{\alpha} y)]} X_{[G_n^{-1}(x_1 \breve{\beta} x_2 \breve{\alpha} y), G_n^{-1}(x_1 \breve{\beta} x_2 \alpha y)]}\right) \circ D\left((-1)_{[G_n^{-1}(x_1 \breve{\beta} x_2 \alpha y)]} X_{[G_n^{-1}(x_1 \breve{\beta} x_2 \alpha y), G_n^{-1}(x_1 \breve{\beta} x_2 \alpha y)]}\right)$$

$$\overset{\text{induction}}{\underset{\text{hypothesis}}{=}} \quad$$ 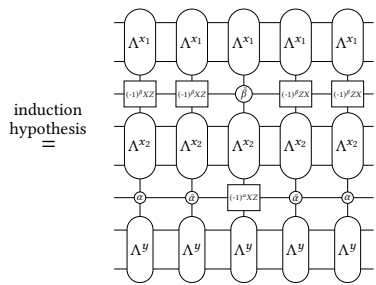

$$\overset{(354)}{=} \quad$$ 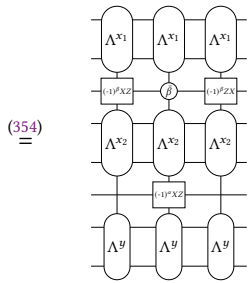

$$\overset{(317)(351)}{=}$$ 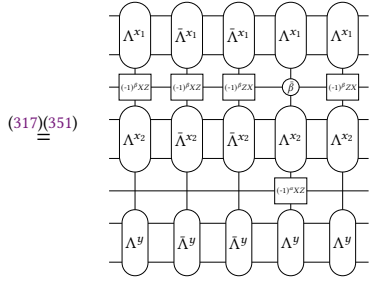

$$\overset{(352)}{=}$$ 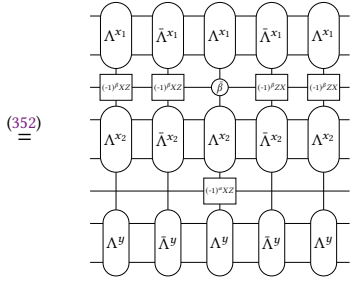

$$\overset{(351)(354)}{=}$$ 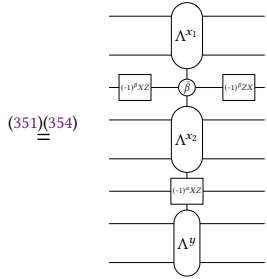

$$\overset{(359)}{=}$$ 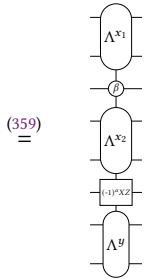

where  is an abuse of notation denoting $\prod\limits_{\substack{x_1' \in \{0,1\}^{k_1} \\ x_2' \in \{0,1\}^{k_2} \\ y' \in \{0,1\}^{f} \\ x_1' x_2' y' \neq x_1 x_2 y}}$ , where the order of the product is left unspecified.

We also have

$$D\left((-1)_{[G_n^{-1}(x\alpha y)]} X_{[G_n^{-1}(x\alpha y), G_n^{-1}(x\bar\alpha y)]}\right)$$

$$= \quad D\left((-1)_{[G_n^{-1}(x\alpha y)+1]} X_{[G_n^{-1}(x\alpha y)+1, G_n^{-1}(x\alpha y)+1]}\right) \circ D\left((-1)_{[G_n^{-1}(x\bar\alpha y)-1]} X_{[G_n^{-1}(x\bar\alpha y)-1, G_n^{-1}(x\bar\alpha y)]}\right) \circ$$

$$D\left((-1)_{[G_n^{-1}(x\bar{\alpha}y)-1]}X_{[G_n^{-1}(x\alpha y)+1,G_n^{-1}(x\bar{\alpha}y)-1]}\right)\circ$$

$$D\left((-1)_{[G_n^{-1}(x\bar{\alpha}y)]}X_{[G_n^{-1}(x\alpha y)-1,G_n^{-1}(x\alpha y)]}\right)\circ D\left((-1)_{[G_n^{-1}(x\alpha y)]}X_{[G_n^{-1}(x\alpha y),G_n^{-1}(x\alpha y)+1]}\right)$$

$$= \quad D\left((-1)_{[G_n^{-1}(x_1\bar{\beta}x_2\alpha y)]}X_{[G_n^{-1}(x_1\beta x_2\alpha y),G_n^{-1}(x_1\bar{\beta}x_2\alpha y)]}\right)\circ D\left((-1)_{[G_n^{-1}(x_1\beta x_2\alpha y)]}X_{[G_n^{-1}(x_1\beta x_2\alpha y),G_n^{-1}(x_1\beta x_2\alpha y)]}\right)\circ$$

$$D\left((-1)_{[G_n^{-1}(x_1\bar{\beta}x_2\alpha y)]}X_{[G_n^{-1}(x_1\beta x_2\alpha y),G_n^{-1}(x_1\bar{\beta}x_2\alpha y)]}\right)\circ$$

$$D\left((-1)_{[G_n^{-1}(x_1\bar{\beta}x_2\alpha y)]}X_{[G_n^{-1}(x_1\beta x_2\alpha y),G_n^{-1}(x_1\bar{\beta}x_2\alpha y)]}\right)\circ D\left((-1)_{[G_n^{-1}(x_1\beta x_2\alpha y)]}X_{[G_n^{-1}(x_1\beta x_2\alpha y),G_n^{-1}(x_1\beta x_2\alpha y)]}\right)$$

induction
hypothesis
$=$

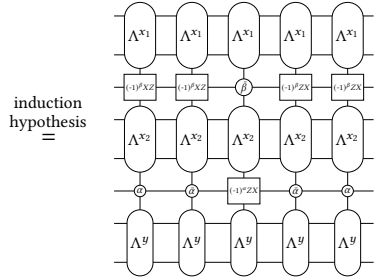

same as
above
$=$

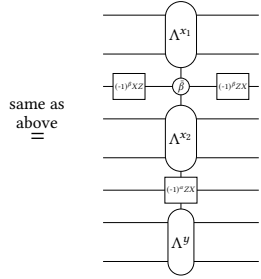

$\overset{(359)}{=}$

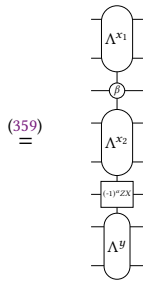

which completes the proof.

## E.4 Proof of Lemma 8.6

More precisely, we are going to prove the equivalent statement that for any raw circuit $C''$ such that $|||C''||| = C$, one has QC $\vdash$ $D(E_{k,\ell}(||C''||)) = id_k \otimes C \otimes id_\ell$, where $||C''||$ (resp. $|||C''|||$) denotes the image of $||C''||$ by the canonical projection onto $\mathbf{QC_{seq}}$ (resp. onto $\mathbf{QC}$). For simplicity, by abuse of notation we will omit the canonical projection $|| \cdot ||$ in the rest of the proof. We proceed by structural induction on $C''$.

For any two $n$-qubit raw circuits $C_1, C_2$ such that $k + n + \ell \geq 3$, one has

$$D(E_{k,\ell}(C_2 \circ C_1)) = D(E_{k,\ell}(C_2)) \circ D(E_{k,\ell}(C_1))$$

and for any $n$-qubit raw circuit $C_1$ and $m$-qubit raw circuit $C_3$ such that $k + n + m + \ell \geq 3$,

$$D(E_{k,\ell}(C_1 \otimes C_3)) = D(E_{k+n,\ell}(C_3)) \circ D(E_{k,\ell+m}(C_1)).$$

Hence, it remains the base cases:

*Case of* : $D(E_{k,\ell}($  $)) = D(\epsilon) = ($ ———— $)^{\otimes n}$

*Case of* ————: $D(E_{k,\ell}($ ———— $)) = D(\epsilon) = ($ ———— $)^{\otimes n}$

*Case of* •—•: If $\ell \geq 1$, then it follows from Definition 8.2 and Lemma 8.4 that

$$\text{QC} \vdash D(E_{k,\ell}(\,\bullet\!\!-\!\!\bullet\,)) = \prod_{\substack{x \in \{0,1\}^k \\ y \in \{0,1\}^{\ell-1}}} \text{}.$$

By Equation (335), one can put the factors in an appropriate order, so as to be able to merge them all together using Equation (309). One gets $id_k \otimes$ •—• $\otimes\, id_\ell$, which is the desired result.

If $\ell = 0$, then it follows from Definition 8.2 and Lemma 8.4 that

$$\text{QC} \vdash D(E_{k,0}(\,\bullet\!\!-\!\!\bullet\,)) = \prod_{x \in \{0,1\}^{k-1}} \text{}.$$

One can similarly merge all the factors together using Equations (335) and (309), so as to get $id_k \otimes$ •—•, which is the desired result.

*Case of* : If $\ell \geq 1$, then it follows from Definition 8.2 and Lemma 8.4 that

$$\text{QC} \vdash D(E_{k,\ell}(\text{})) = \prod_{\substack{x \in \{0,1\}^k \\ y \in \{0,1\}^{\ell-1}}} \text{}.$$

By Equation (335), one can put the factors in an appropriate order, so as to be able to merge them all together using Equation (309). One gets $id_k \otimes$  $\otimes\, id_\ell$, which is the desired result.

If $\ell = 0$, then it follows from Definition 8.2 and Lemma 8.4 that

$$\text{QC} \vdash D(E_{k,0}(\text{})) = \prod_{x \in \{0,1\}^{k-1}} \text{}.$$

One can similarly merge all the factors together using Equations (335) and (309), so as to get $id_k \otimes$ , which is the desired result.

*Case of* —[H]—: If $\ell \geq 1$, then it follows from Definitions 8.2 and 8.3 that

$$D(E_{k,\ell}(\text{—}[H]\text{—})) = \prod_{\substack{x \in \{0,1\}^k \\ y \in \{0,1\}^{\ell-1}}} \text{}.$$

By Equation (340), one can put the factors in an appropriate order, so as to be able to merge them all together using Equation (310). One gets $id_k \otimes$ —[H]— $\otimes\, id_\ell$, which is the desired result.

If $\ell = 0$, then it follows from Definitions 8.2 and 8.3 that

$$D(E_{k,0}(\text{-}\bullet\text{-})) = \prod_{x \in \{0,1\}^{k-1}} \Lambda^x .$$

One can similarly merge all the factors together using Equations (340) and (310), so as to get $id_k \otimes \text{-}\boxed{H}\text{-}$, which is the desired result.

*Case of* $\boxed{H}$ : If $\ell \geq 1$, then it follows from Definitions 8.2 and 8.3 that

$$D(E_{k,\ell}(\boxed{H})) = \prod_{\substack{x \in \{0,1\}^k \\ y \in \{0,1\}^{\ell-1}}} \Lambda^x \boxed{H} \Lambda^y .$$

By Equation (340), one can put the factors in an appropriate order, so as to be able to merge them all together using Equation (310). One gets $id_k \otimes \boxed{H} \otimes id_\ell$, which is the desired result.

If $\ell = 0$, then it follows from Definitions 8.2 and 8.3 that

$$D(E_{k,0}(\boxed{H})) = \prod_{x \in \{0,1\}^{k-1}} \Lambda^x \boxed{H} .$$

One can similarly merge all the factors together using Equations (340) and (310), so as to get $id_k \otimes \boxed{H}$, which is the desired result.

*Case of* $\asymp$ : It follows from Definition 8.2 and Lemma 8.5, and from the case of $\bullet$ above, that

$$QC \vdash D(E_{k,\ell}(\asymp)) = (id_k \otimes \bullet \otimes id_\ell) \circ \prod_{\substack{x \in \{0,1\}^k \\ y \in \{0,1\}^\ell}} \Lambda^x \boxed{ZX} \Lambda^y \circ \prod_{\substack{x \in \{0,1\}^k \\ y \in \{0,1\}^\ell}} \Lambda^x \boxed{XZ} \Lambda^y \circ \prod_{\substack{x \in \{0,1\}^k \\ y \in \{0,1\}^\ell}} \Lambda^x \boxed{ZX} \Lambda^y$$

In each product, by Equation (351), one can put the factors in an appropriate order, so as to be able to merge them all together using Equation (354). One gets

$$(id_k \otimes \bullet \otimes id_\ell) \circ (id_k \otimes \boxed{ZX} \otimes id_\ell) \circ (id_k \otimes \boxed{XZ} \otimes id_\ell) \circ (id_k \otimes \boxed{ZX} \otimes id_\ell)$$

$$= \quad id_k \otimes \boxed{XZ} \boxed{ZX} \boxed{ZX} \otimes id_\ell$$

$$\overset{(116)}{=} \quad id_k \otimes \asymp \otimes id_\ell$$

$$\overset{(4)}{=} \quad id_k \otimes \asymp \otimes id_\ell$$

## E.5 Proof of Lemma 8.8

To prove Lemma 8.8, it suffices to prove that for each rule of Figure 8, of the form $w_1 = w_2$ where $w_1, w_2 \in \mathcal{G}_{2n}^*$, and for any two words $w_0, w_3 \in \mathcal{G}_{2n}^*$, one has $QC \vdash D(w_0 w_1 w_3) = D(w_0 w_2 w_3)$. Since $D$ is a monoid homomorphism, it suffices to prove that $QC \vdash D(w_1) = D(w_2)$.

*Decoding Equation* ([20])*:* $D\left((-1)_{[a]}(-1)_{[a]}\right) = id_n = D(\epsilon)$.

*Decoding Equation* ([21])*:* The result follows from the definition of $D\left((-1)_{[a]}(-1)_{[b]}\right)$.

*Decoding Equation* ([22])*:* We distinguish depending on the respective order of $a, b, c$.

- If $a \leq b \leq c$, then we have

$$
\begin{aligned}
D\big(\left((-1)_{[a]}(-1)_{[b]}\right)\left((-1)_{[b]}(-1)_{[c]}\right)\big) &= \overset{b-1}{\underset{k=a}{\prod}} D\left((-1)_{[k]}(-1)_{[k+1]}\right) \circ \overset{c-1}{\underset{k=b}{\prod}} D\left((-1)_{[k]}(-1)_{[k+1]}\right) \\
&\overset{(335)(336)}{=} \overset{c-1}{\underset{k=b}{\prod}} D\left((-1)_{[k]}(-1)_{[k+1]}\right) \circ \overset{b-1}{\underset{k=a}{\prod}} D\left((-1)_{[k]}(-1)_{[k+1]}\right) \\
&= \overset{c-1}{\underset{k=a}{\prod}} D\left((-1)_{[k]}(-1)_{[k+1]}\right) \\
&= D\left((-1)_{[a]}(-1)_{[c]}\right)
\end{aligned}
$$

- If $a \leq c \leq b$, then we have

$$
\begin{aligned}
D\big(\left((-1)_{[a]}(-1)_{[b]}\right)\left((-1)_{[b]}(-1)_{[c]}\right)\big) &= \overset{b-1}{\underset{k=a}{\prod}} D\left((-1)_{[k]}(-1)_{[k+1]}\right) \circ \overset{b-1}{\underset{k=c}{\prod}} D\left((-1)_{[k]}(-1)_{[k+1]}\right) \\
&= \overset{b-1}{\underset{k=c}{\prod}} D\left((-1)_{[k]}(-1)_{[k+1]}\right) \circ \overset{c-1}{\underset{k=a}{\prod}} D\left((-1)_{[k]}(-1)_{[k+1]}\right) \circ \overset{b-1}{\underset{k=c}{\prod}} D\left((-1)_{[k]}(-1)_{[k+1]}\right) \\
&\overset{(335)(336)}{=} \overset{c-1}{\underset{k=a}{\prod}} D\left((-1)_{[k]}(-1)_{[k+1]}\right) \circ \overset{b-1}{\underset{k=c}{\prod}} \Big(D\left((-1)_{[k]}(-1)_{[k+1]}\right) \circ D\left((-1)_{[k]}(-1)_{[k+1]}\right)\Big) \\
&= \overset{c-1}{\underset{k=a}{\prod}} D\left((-1)_{[k]}(-1)_{[k+1]}\right) \\
&= D\left((-1)_{[a]}(-1)_{[c]}\right)
\end{aligned}
$$

- The other cases ($b \leq a \leq c$, $b \leq c \leq a$, $c \leq a \leq b$, and $c \leq b \leq a$) are similar.

*Decoding Equation* ([23])*:*

$$
D\left((-1)_{[a]}(-1)_{[a+1]}\right) = \;\cdots\; \overset{(355)}{=} \;\cdots\; = D\big(\left((-1)_{[a]}X_{[a,a+1]}\right)\left((-1)_{[a]}X_{[a,a+1]}\right)\big)
$$

where $G_n(a) = x\beta y$ and $G_n(a+1) = x\bar{\beta}y$, with $\beta \in \{0,1\}$, $x \in \{0,1\}^k$, and $y \in \{0,1\}^{n-k-1}$.

*Decoding Equation* ([24])*:* By the properties of the Gray code, there exist bitstrings $x, y, z$ (of combined length $n-2$) and bits $\alpha, \beta \in \{0,1\}$ such that $G_n(a) = x\alpha y\beta z$, either $G_n(a+1) = x\alpha y\bar{\beta}z$ or $G_n(a+1) = x\bar{\alpha}y\beta z$, and $G_n(a+2) = x\bar{\alpha}y\bar{\beta}z$. We treat only the case $G_n(a+1) = x\alpha y\bar{\beta}z$, the other case being similar. One has

$$D\left(\left((-1)_{[a]}X_{[a,a+1]}\right)\left((-1)_{[a+1]}X_{[a+1,a+2]}\right)\left((-1)_{[a]}X_{[a,a+1]}\right)\right) \;=\;$$ 

and

$$D\left(\left((-1)_{[a+1]}X_{[a+1,a+2]}\right)\left((-1)_{[a]}X_{[a,a+1]}\right)\left((-1)_{[a+1]}X_{[a+1,a+2]}\right)\right) \;=\;$$ .

To derive the equality between the two, up to using Equations (81) and (356), we can assume without loss of generality that $\alpha = 1$ and $\beta = 0$. Moreover, up to using Equation (81), we can assume without loss of generality that $x, y, z$ consist only of 1's, and up to using Equation (353), we can assume without loss of generality that $y$ and $z$ are of length zero. Thus, it suffices to prove the following equation:

which is exactly Equation (358).

*Decoding Equation* (25): The result follows directly from the definitions of $D\left((-1)_{[a]}X_{[a,c]}\right)$ and $D\left((-1)_{[a+1]}X_{[a+1,c]}\right)$.

*Decoding Equation* (26): The result follows directly from the definitions of $D\left((-1)_{[c]}X_{[a,c]}\right)$ and $D\left((-1)_{[c]}X_{[a+1,c]}\right)$.

*Decoding Equation* (27): The result follows directly from the definitions of $D\left((-1)_{[a]}X_{[a,c]}\right)$ and $D\left((-1)_{[a+1]}X_{[a+1,c]}\right)$.

*Decoding Equation* (28): The result follows directly from the definitions of $D\left((-1)_{[c]}X_{[a,c]}\right)$ and $D\left((-1)_{[c]}X_{[a+1,c]}\right)$.

*Decoding Equation* (29):

$$D\left(\left((-1)_{[a]}X_{[a,a+1]}\right)\left((-1)_{[a+1]}X_{[a,a+1]}\right)\right) \;=\; \text{} \;\overset{(317)}{=}\; id_n \;=\; D(\epsilon)$$

where $G_n(a) = x\beta y$ and $G_n(a+1) = x\bar{\beta}y$, with $\beta \in \{0,1\}$, $x \in \{0,1\}^k$, and $y \in \{0,1\}^{n-k-1}$.

*Decoding Equation* (30): The result follows directly from the definition of $D\left((-1)_{[c]}X_{[a,b]}\right)$, given that $c \neq a, b$.

*Decoding Equation* (31): We only treat the case $c > a+1$, the other case (where $c < a$) being similar. We have

$$D\left(\left((-1)_{[a+1]}(-1)_{[c]}\right)\left((-1)_{[a]}X_{[a,a+1]}\right)\right)$$

$$= \; \overset{\longleftarrow}{\prod_{k=a+1}^{c-1}} D\left((-1)_{[k]}(-1)_{[k+1]}\right) \circ D\left(\left((-1)_{[a]}X_{[a,a+1]}\right)\right)$$

$$= \prod_{k=a+2}^{c-1} D\left((-1)_{[k]}(-1)_{[k+1]}\right) \circ D\left((-1)_{[a+1]}(-1)_{[a+2]}\right) \circ D\left(\left((-1)_{[a]}X_{[a,a+1]}\right)\right)$$

By the properties of the Gray code, there exist bitstrings $x, y, z$ (of combined length $n - 2$) and bits $\alpha, \beta \in \{0, 1\}$ such that $G_n(a) = x\alpha y\beta z$, either $G_n(a + 1) = x\alpha y\bar{\beta} z$ or $G_n(a + 1) = x\bar{\alpha} y\beta z$, and $G_n(a + 2) = x\bar{\alpha} y\bar{\beta} z$. We treat only the case $G_n(a + 1) = x\alpha y\bar{\beta} z$, the other case being similar. We have

$$D\left((-1)_{[a+1]}(-1)_{[a+2]}\right) \circ D\left(\left((-1)_{[a]}X_{[a,a+1]}\right)\right) \ = \ $$ 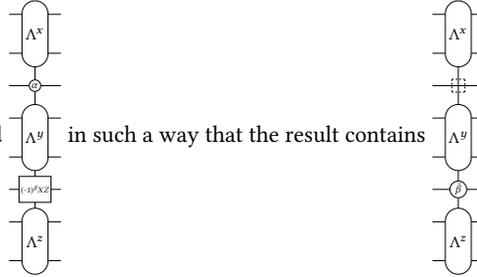 .

Up to using Equations (81) and (356), we can do as if the control parametrized by $\bar{\beta}$ was black, and up to using Equation (315), we can do as if the control parametrized by $\alpha$ was black as well. Then by using Equations (353), (307) and (313), we can swap the controls appropriately so that

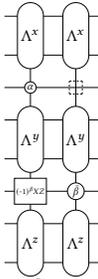

by using Equations (333) and (334) (and (81)), we unfold 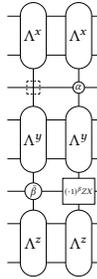 in such a way that the result contains . Then it becomes straightforward

to transform 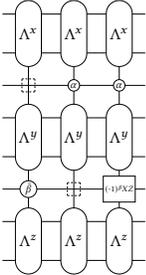 into , possibly with the help of Equation (317). By Equations (317) and (355), we can in turn transform it

into , which is equal to $D\left((-1)_{[a]}X_{[a,a+1]}\right) \circ D\left((-1)_{[a]}(-1)_{[a+1]}\right) \circ D\left((-1)_{[a+1]}(-1)_{[a+2]}\right)$. Thus, we get

$$\prod_{k=a+2}^{c-1} D\left((-1)_{[k]}(-1)_{[k+1]}\right) \circ \left((-1)_{[a]}X_{[a,a+1]}\right) \circ D\left((-1)_{[a]}(-1)_{[a+1]}\right) \circ D\left((-1)_{[a+1]}(-1)_{[a+2]}\right).$$

Finally, note that for any $k \notin \{a - 1, a, a + 1\}$, $D\left((-1)_{[k]}(-1)_{[k+1]}\right)$ commutes with $D\left((-1)_{[a]}X_{[a,a+1]}\right)$ by Equation (346) or (347) (and Equation (307)). Indeed, $G_n(k)$ differs from $G_n(k + 1)$ by exactly one bit, and $G_n(a)$ differs from $G_n(a + 1)$ by exactly one bit as well. Since $G_n(a), G_n(a + 1), G_n(k), G_n(k + 1)$ are all distinct, there must be at least one bit where $G_n(k)$ and $G_n(k + 1)$ agree, $G_n(a)$ and $G_n(a + 1)$ agree, but $G_n(k), G_n(k + 1)$ differ from $G_n(a), G_n(a + 1)$ (otherwise, we would be able to get two identical vectors by choosing appropriately between $G_n(k)$ and $G_n(k + 1)$ and between $G_n(a)$ and $G_n(a + 1)$, in contradiction with the fact that all four are distinct). This corresponds to a control of $D\left((-1)_{[k]}(-1)_{[k+1]}\right)$ being in front of a control of $D\left((-1)_{[a]}X_{[a,a+1]}\right)$ of the opposite color.

Thus, we get

$$D\left((-1)_{[a]}X_{[a,a+1]}\right) \circ \overleftarrow{\prod_{k=a+2}^{c-1}} D\left((-1)_{[k]}(-1)_{[k+1]}\right) \circ D\left((-1)_{[a]}(-1)_{[a+1]}\right) \circ D\left((-1)_{[a+1]}(-1)_{[a+2]}\right)$$

which by Equation (336) is equal to

$$D\left((-1)_{[a]}X_{[a,a+1]}\right) \circ \overleftarrow{\prod_{k=a+2}^{c-1}} D\left((-1)_{[k]}(-1)_{[k+1]}\right) \circ D\left((-1)_{[a+1]}(-1)_{[a+2]}\right) \circ D\left((-1)_{[a]}(-1)_{[a+1]}\right)$$

$$= \quad D\left((-1)_{[a]}X_{[a,a+1]}\right) \circ \overleftarrow{\prod_{k=a}^{c-1}} D\left((-1)_{[k]}(-1)_{[k+1]}\right)$$

$$= \quad D\left(((-1)_{[a]}X_{[a,a+1]})\,((-1)_{[a]}(-1)_{[c]})\right).$$

*Decoding Equation (32):* $G_n(a)$ differs from $G_n(a+1)$ by exactly one bit, and $G_n(b)$ differs from $G_n(b+1)$ by exactly one bit as well. Since $G_n(a), G_n(a+1), G_n(b), G_n(b+1)$ are all distinct, there must be at least one bit where $G_n(a)$ and $G_n(a+1)$ agree, $G_n(b)$ and $G_n(b+1)$ agree, but $G_n(a), G_n(a+1)$ differ from $G_n(b), G_n(b+1)$ (otherwise, we would be able to get two identical vectors by choosing appropriately between $G_n(a)$ and $G_n(a+1)$ and between $G_n(b)$ and $G_n(b+1)$, in contradiction with the fact that all four are distinct). This corresponds to a control of $D\left((-1)_{[a]}X_{[a,a+1]}\right)$ being in front of a control of $D\left((-1)_{[b]}X_{[b,b+1]}\right)$ of the opposite color, therefore the two commute by either Equation (351) or Equation (352) (together with Equation (353)).

*Decoding Equation (33):* The result follows directly from the definition of $D\left((-1)_{[a]}X_{[a,b]}\right)$ and $D\left((-1)_{[b]}X_{[a,b]}\right)$ in the case $a > b$.

*Decoding Equation (34):* The result follows directly from the definition of $D\left(X_{[a,b]}X_{[c,d]}\right)$.

*Decoding Equation (35):* The result follows directly from the definition of $D\left(H_{[a,b]}H_{[c,d]}\right)$ in the case where the tuple $(G_n(a), G_n(b), G_n(c), G_n(d))$ is neither of the form $(x0y0z, x1y0z, x0y1z, x1y1z)$ nor $(x0y0z, x0y1z, x1y0z, x1y1z)$.

*Decoding Equation (36):* On the one hand, by definition of $D\left(H_{[a,b]}H_{[c,d]}\right)$ in the case where the tuple $(G_n(a), G_n(b), G_n(c), G_n(d))$ is of the form $(x0y0z, x1y0z, x0y1z, x1y1z)$, the decoding of the left-hand side is $D\left(H_{[a,b]}H_{[c,d]}\right) = $ 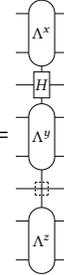 .

On the other hand, to decode the right-hand side, note first that $D\left(\mathcal{E}_{k,\ell}^{1-x}(-\oplus-)\right)$ is the identity if $x = 1$, and that if $x = 0$,

$$D\left(\mathcal{E}_{k,\ell}^{1-x}(-\oplus-)\right) \quad = \quad D(E_{k,\ell}(-\bullet-)) \circ \overrightarrow{\prod_{\substack{x\in\{0,1\}^k \\ y\in\{0,1\}^\ell}}} D\left((-1)_{[G_n^{-1}(x1y)]}X_{[G_n^{-1}(x0y),G_n^{-1}(x1y)]}\right)$$

$$\overset{\text{Lemmas 8.5 and 8.6}}{=} \quad (id_k \otimes -\bullet- \otimes id_\ell) \circ \overrightarrow{\prod_{\substack{x\in\{0,1\}^k \\ y\in\{0,1\}^\ell}}} \boxed{\begin{array}{c}\Lambda^x \\ \hline ZX \\ \hline \Lambda^y\end{array}}$$

$$\overset{(351)(354)}{=} \quad (id_k \otimes -\bullet- \otimes id_\ell) \circ (id_k \otimes \boxed{-ZX-} \otimes id_\ell)$$

$$= \quad id_k \otimes \oplus\!\!-\!\bullet\!-\!\bullet\!- \otimes id_\ell$$

$$\overset{(2)}{=} \quad id_k \otimes \ominus \oplus \ominus \otimes id_\ell$$

Together with Lemma 8.6 applied to ⤬, and with Equation (81), this implies that the decoding of the right-hand side of Equation (36) can

be transformed into 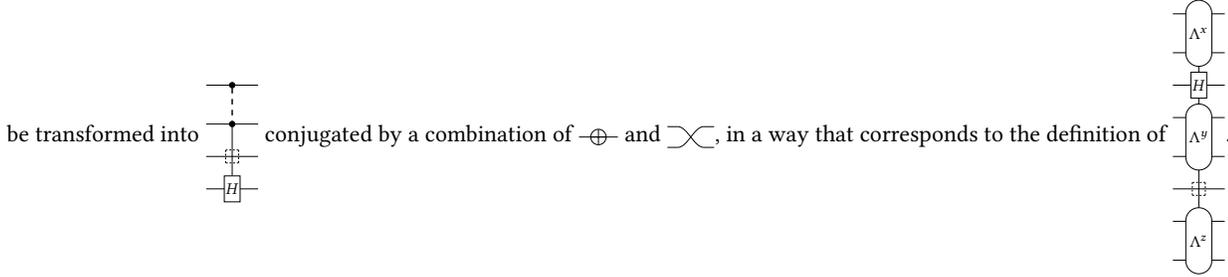 conjugated by a combination of $\ominus\oplus\ominus$ and ⤬, in a way that corresponds to the definition of 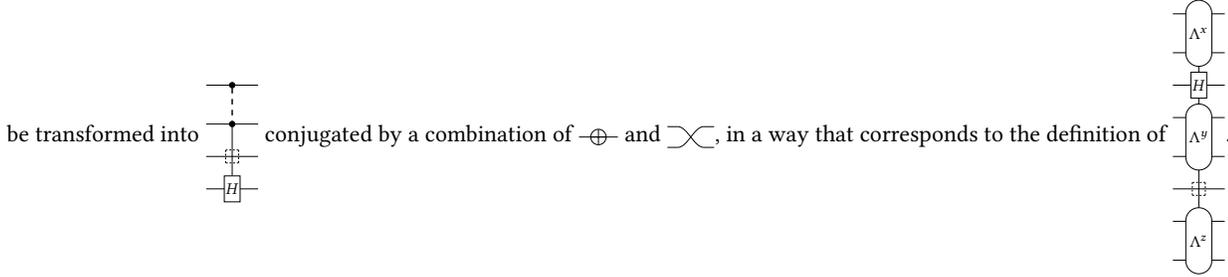.

*Decoding Equation* (37): On the one hand, by definition of $D\left(H_{[a,b]}H_{[c,d]}\right)$ in the case where the tuple $(G_n(a), G_n(b), G_n(c), G_n(d))$ is of the form $(x0y0z, x0y1z, x1y0z, x1y1z)$, the decoding of the left-hand side is $D\left(H_{[a,b]}H_{[c,d]}\right) = $ 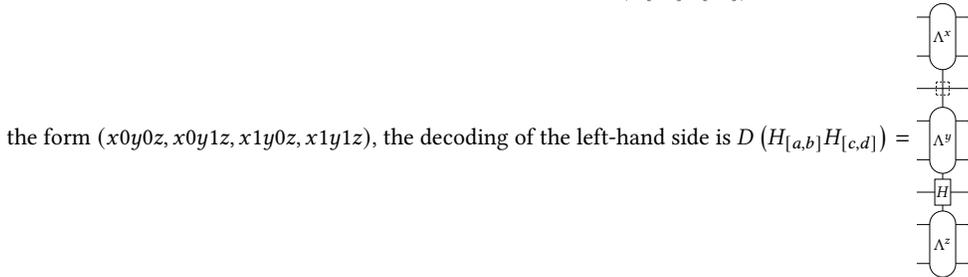.

On the other hand, with the same argument as for Equation (36), one can show that the decoding of the right-hand side of Equation (37)

can be transformed into 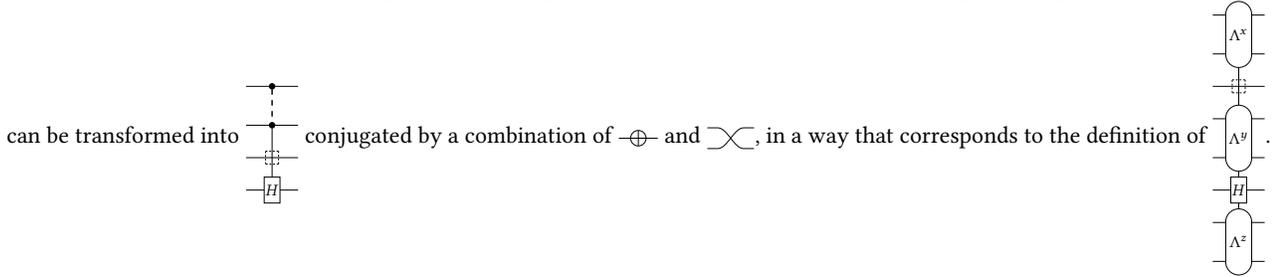 conjugated by a combination of $\ominus\oplus\ominus$ and ⤬, in a way that corresponds to the definition of 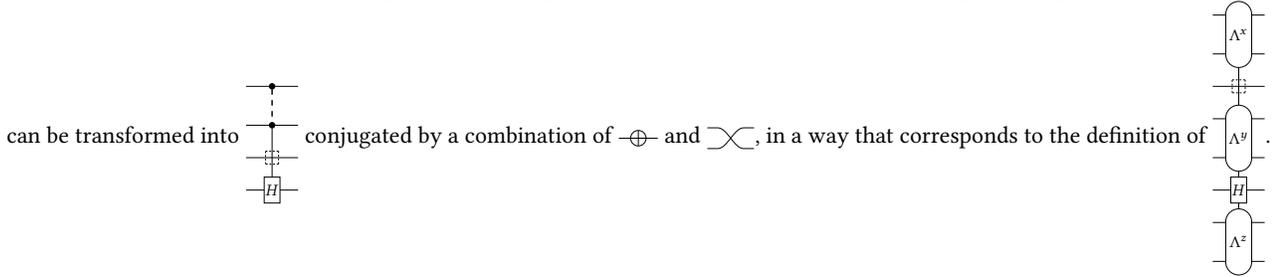.

*Decoding Equation* (38): On the one hand, $D\left(H_{[0,1]}H_{[3,2]}\right) = $ 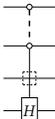.

On the other hand, for any $a \geq 4$, $G_n(a)$ and $G_n(a+1)$ necessarily have a common bit with value 1 which is not one of the last two. Indeed, otherwise, since they differ by only one bit, at least one of the two would be equal to either $0\ldots000$, $0\ldots001$, $0\ldots011$, or $0\ldots010$ (more precisely, if the changing bit was among the last two bits, both $G_n(a)$ and $G_n(a+1)$ would be equal to one of these codes, and if it was among the first $n-2$ bits, exactly one of $G_n(a)$ and $G_n(a+1)$ would be equal to one of these codes). But these are already the Gray codes of $0, 1, 2, 3$ respectively.

Consequently, $D\left((-1)_{[a]}X_{[a,a+1]}\right)$ has a black control in front of one of the white controls of 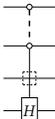, and therefore commutes with it by

Equation (348), (349) or (350).

*Decoding Equation* (39):

$$D\left(\left((-1)_{[0]}(-1)_{[1]}\right)\left(H_{[0,1]}H_{[3,2]}\right)\right) = \quad \overset{(338)}{=} \quad = D\left(\left(H_{[0,1]}H_{[3,2]}\right)\left((-1)_{[0]}(-1)_{[1]}\right)\right)$$

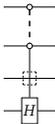

*Decoding Equation* (40):

$$D\Big(\big(H_{[0,1]}H_{[3,2]}\big)\big((-1)_{[3]}(-1)_{[4]}\big)\big((-1)_{[2]}X_{[2,3]}\big)\Big) \quad = \quad$$

$$\overset{(353)(333)(81)(315)}{=}$$

$$\overset{(309)}{=}$$

$$\overset{(274)(309)(338)}{=}$$

$$\overset{(312)}{=}$$

$$\overset{(309)}{=}$$

$$\overset{(311)}{=}$$

$$= \quad D\Big(\big((-1)_{[3]}(-1)_{[4]}\big)\big(H_{[0,1]}H_{[3,2]}\big)\Big)$$

*Decoding Equation* (41):

$$D\Big(\big(H_{[0,1]}H_{[3,2]}\big)\big(H_{[0,1]}H_{[3,2]}\big)\Big) = \quad$$  $$\overset{(312)}{=}$$  $$= D(\epsilon)$$

*Decoding Equation* (42): The result follows directly from the definition of $D\big(H_{[a,b]}H_{[c,d]}\big)$ in the case where $|\{a,b,c,d\}| \leq 3$.

*Decoding Equation* (43): First, we prove that the following auxiliary equation is a consequence of the equations of Figure 4:

$$D\left(X_{[0,3]}X_{[1,2]}\right) \;=\; \text{} \tag{361}$$

$$
\begin{aligned}
D\left(X_{[0,3]}X_{[1,2]}\right) \;&=\; D\Big(\big((-1)_{[0]}X_{[0,3]}\big)\,\big((-1)_{[3]}X_{[1,2]}\big)\Big) \\[2mm]
&=\; D\Big(\big((-1)_{[0]}X_{[0,3]}\big)\,\big((-1)_{[3]}(-1)_{[1]}\big)\,\big((-1)_{[1]}X_{[1,2]}\big)\Big) \\[2mm]
&=\; D\Big(\big((-1)_{[0]}X_{[0,3]}\big)\,\big((-1)_{[2]}(-1)_{[3]}\big)\,\big((-1)_{[1]}(-1)_{[2]}\big)\,\big((-1)_{[1]}X_{[1,2]}\big)\Big)
\end{aligned}
$$

$$\overset{\text{Lemma } 8.5}{=} \quad \text{}$$

$$\overset{(355)}{=} \quad \text{}$$

$$\overset{(317)}{=} \quad \text{}$$

$$\overset{(81)}{=} \quad \text{}$$

$$\overset{(311)}{=} \quad \text{}$$

$$\overset{(81)}{=} \quad \text{}$$

$$\overset{(354)}{=} \quad \text{}$$

We can then prove the result as follows:

$$D\Big(\big(H_{[0,1]}H_{[3,2]}\big)\big(H_{[3,2]}H_{[4,5]}\big)\Big) \quad = $$

$$\overset{(112)}{=}$$

$$\overset{(112)(130)}{=}$$

$$\overset{(112)(307)}{=}$$

$$\overset{(355)}{=}$$

$$\overset{(334)(81)(316)}{=}$$

$$\overset{(316)(81)(333)}{=}$$

$$\overset{(317)}{=}$$

$(353)(349) \atop =$ 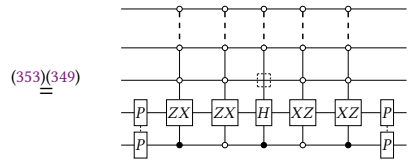

$(354) \atop =$ 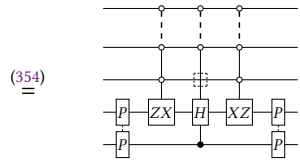

$(360) \atop =$ 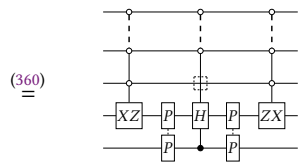

$(112)(307) \atop =$ 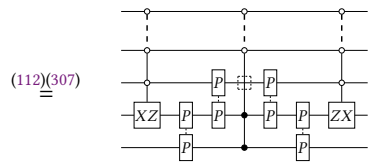

$(112)(130) \atop =$ 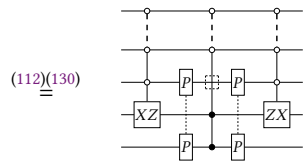

$(112) \atop =$ 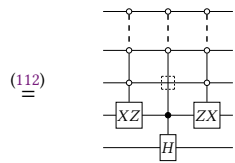

$(311)(355) \atop =$ 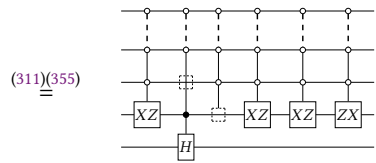

$(317) \atop =$ 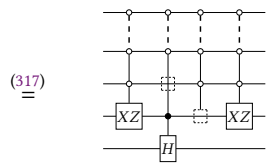

$\overset{(274)}{=}$ 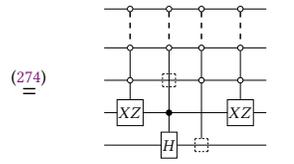

$\overset{(309)}{=}$ 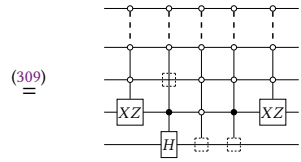

$\overset{(307)(338)}{=}$ 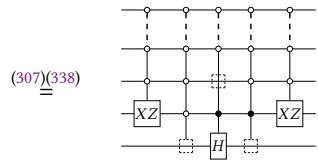

$\overset{(354)}{=}$ 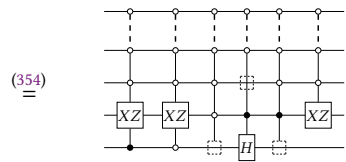

$\overset{(81)(315)}{=}$ 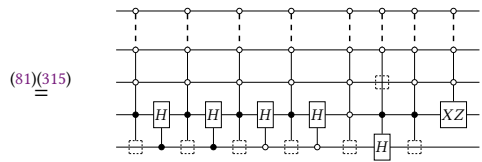

$\overset{(311)}{=}$ 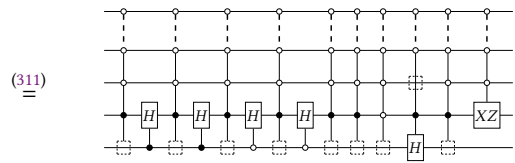

$\overset{(315)(81)}{=}$ 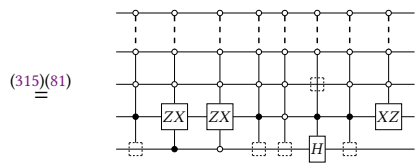

$\overset{(309)(354)}{=}$ 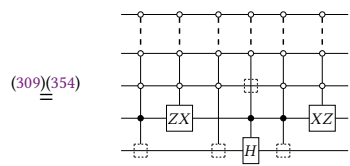

$\overset{(274)}{=}$

$\overset{(355)}{=}$

$\overset{(317)}{=}$

$\overset{(361)}{=} D\Big( \big( X_{[0,3]} X_{[1,2]} \big) \big( H_{[3,2]} H_{[4,5]} \big) \big( X_{[0,3]} X_{[1,2]} \big) \Big)$

*Decoding Equation* (44):

$D\Big( \big( X_{[0,3]} X_{[1,2]} \big) \big( H_{[0,1]} H_{[3,2]} \big) \Big) \overset{(361)}{=}$

$\overset{(338)}{=}$

$\overset{(112)}{=}$

$\overset{(111)(360)}{=}$

$\overset{(354)}{=}$

$\overset{(346)}{=}$

$$\overset{(354)}{=}$$

$$\overset{(360)}{=}$$

$$\overset{(112)}{=}$$

$$\overset{(361)}{=} D\Big(\big(H_{[0,1]}H_{[3,2]}\big)\big(X_{[0,3]}X_{[1,2]}\big)\Big)$$

*Decoding Equation* (45): The decoding of the left-hand side is

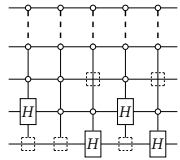

and the decoding of the left-hand side is

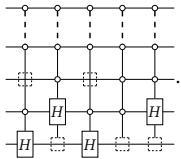

.

If $n = 3$, then the result is given by Equation (146).
If $n \geq 4$, then we have

$$\overset{(311)}{=}$$

$$\overset{(337)}{=}$$

$$\overset{(312)}{=}$$

(I)

$$\overset{(337)(344)}{=}$$

$$\overset{(309)}{=}$$

$$\overset{(310)}{=}$$

$$\overset{(277)}{=}$$

$$\overset{(274)}{=}$$

$$\overset{(112)}{=}$$

$$\overset{(308)}{=}$$

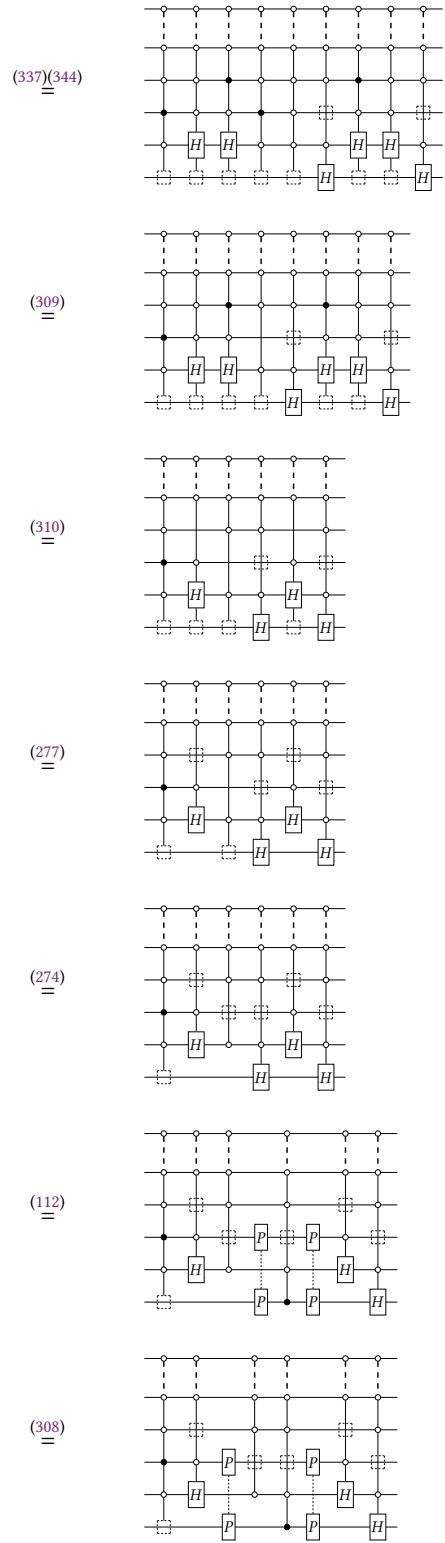

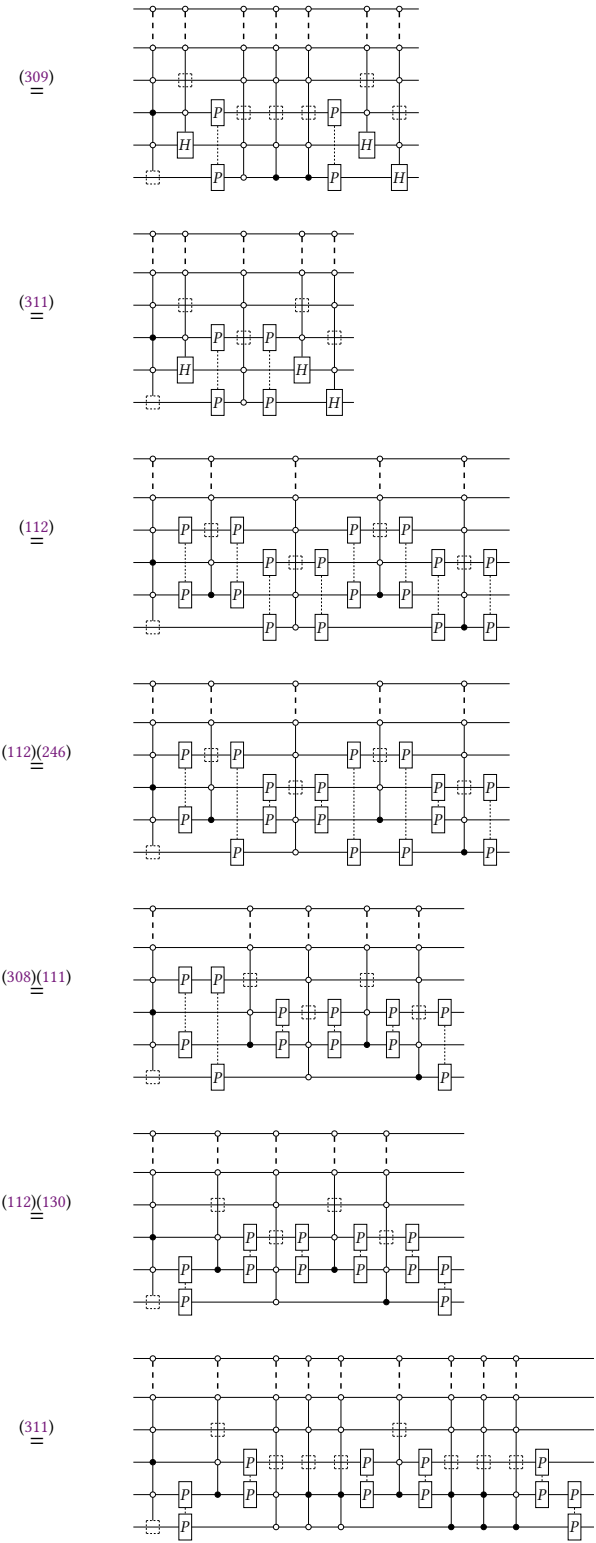

$\overset{(309)}{=}$

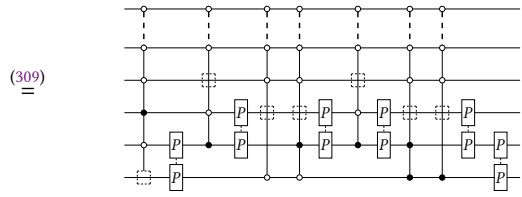

$\overset{(308)}{=}$

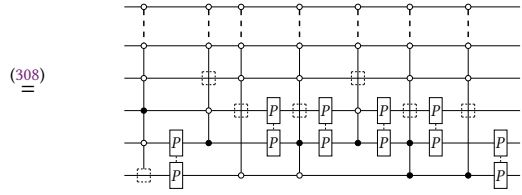

$\overset{(309)(307)(336)}{=}$

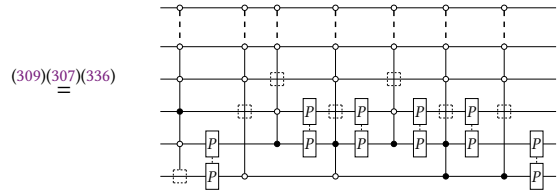

$\overset{(112)}{=}$

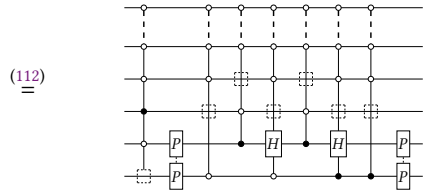

$\overset{(309)}{=}$

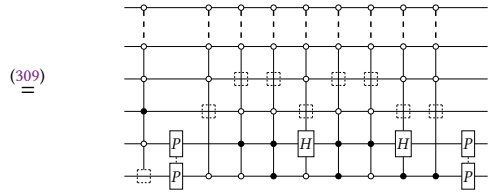

$\overset{(307)(313)(339)(311)}{=}$

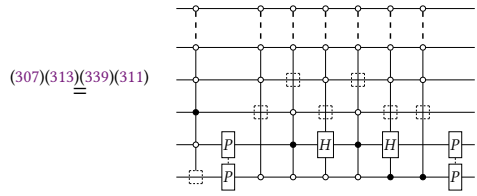

$\overset{(307)(313)(339)(340)}{=}$

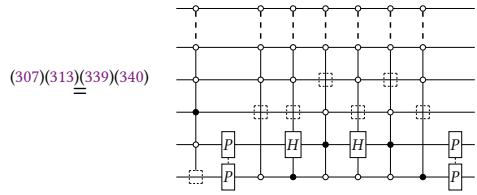

(II)

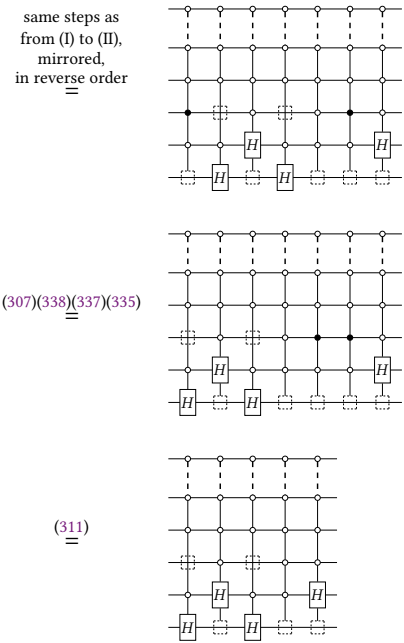

same steps as
from (I) to (II),
mirrored,
in reverse order
$\underset{=}{}$

$\underset{=}{(307)(338)(337)(335)}$

$\underset{=}{(311)}$

*Decoding Equation* (46): The decoding of the left-hand side is

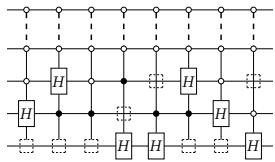

and the decoding of the left-hand side is

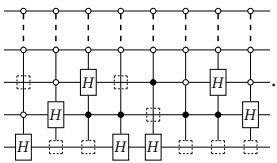

If $n = 3$, then the result is given by Equation (147).
If $n \geq 4$, then we have

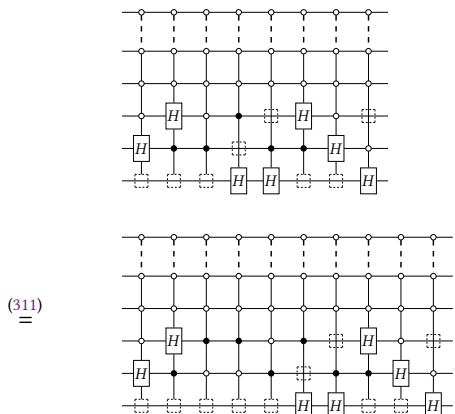

$\underset{=}{(311)}$

(III)

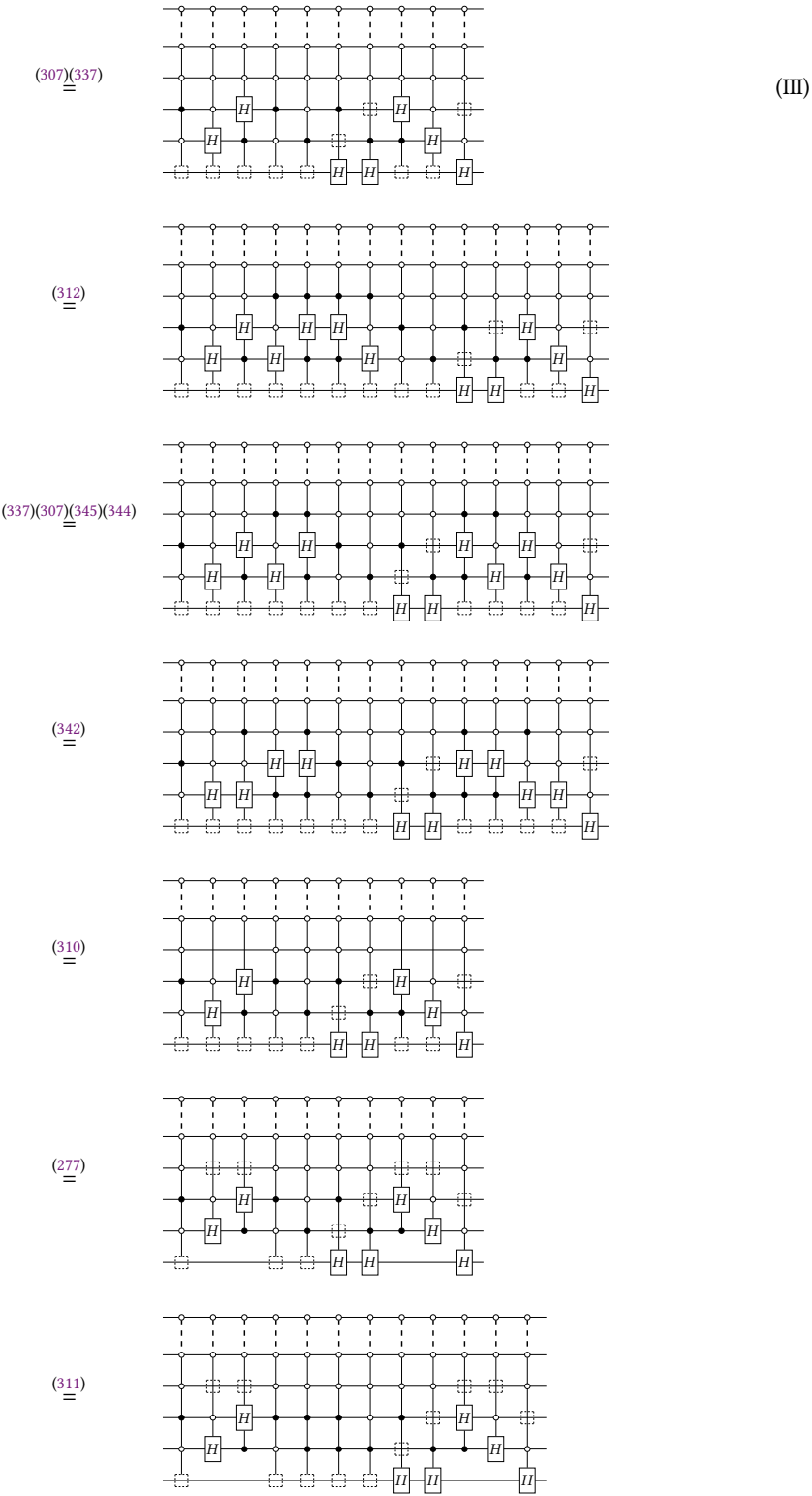

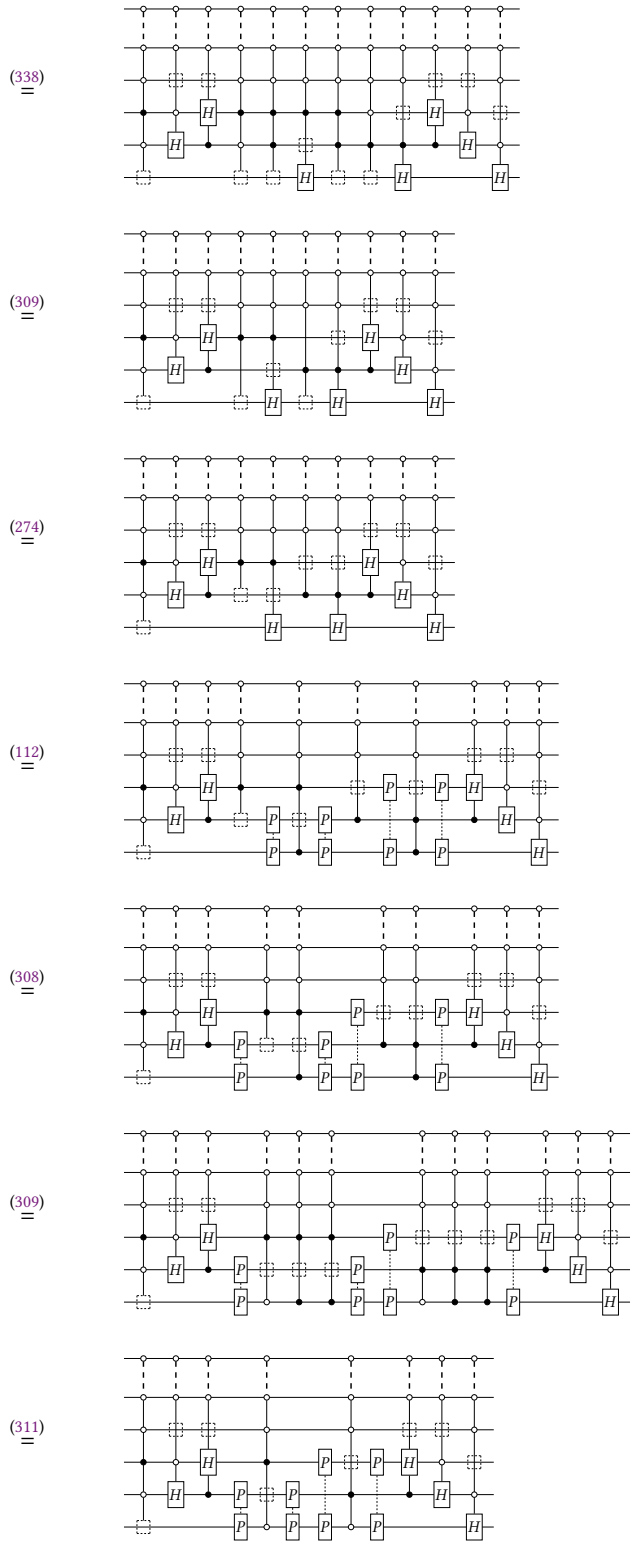

$\underset{=}{(112)(307)}$ 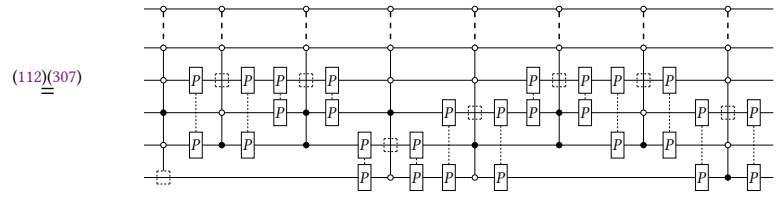

$\underset{=}{(112)(130)}$ 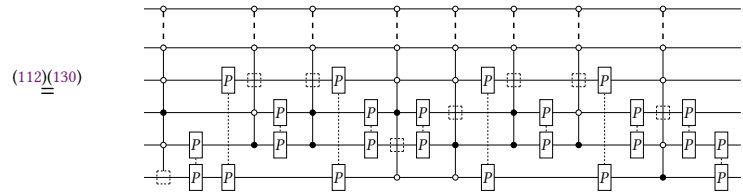

$\underset{=}{(308)(111)}$ 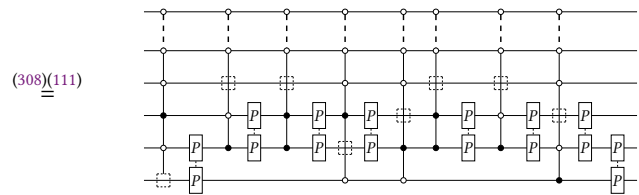

$\underset{=}{(307)}$ 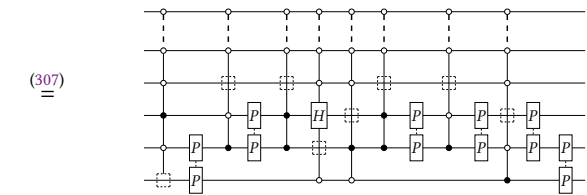

$\underset{=}{(309)}$ 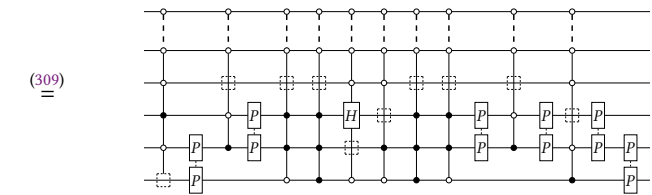

$\underset{=}{(307)(313)(339)(336)}$ 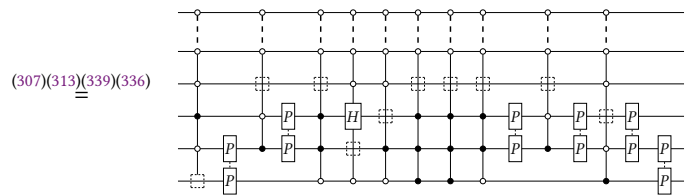

$\underset{=}{(311)}$ 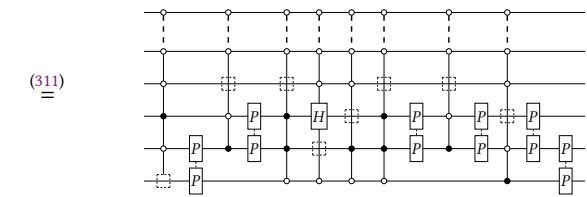

(309)
$\underset{=}{}$

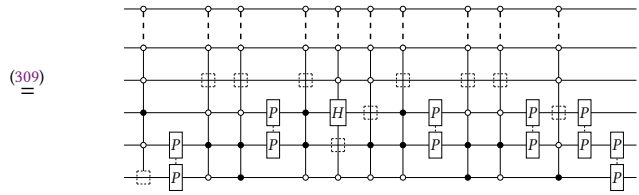

(130)(112)
$\underset{=}{}$

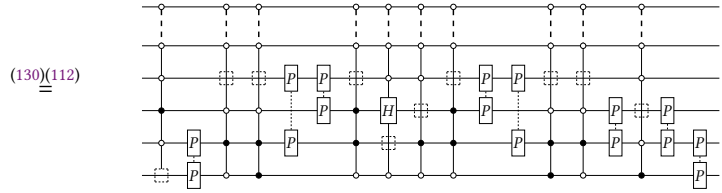

(111)(112)(307)
$\underset{=}{}$

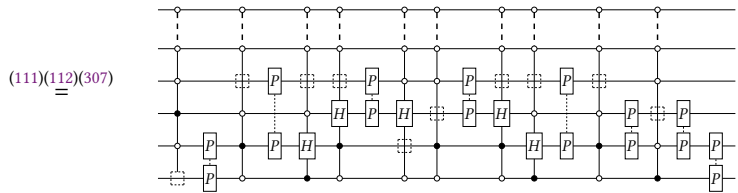

(313)(342)
$\underset{=}{}$

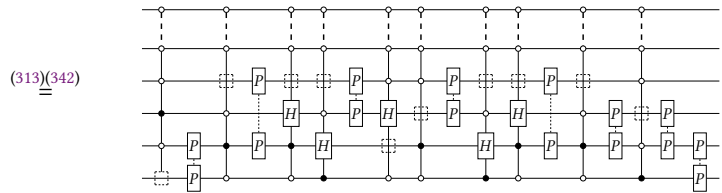

(112)(307)
$\underset{=}{}$

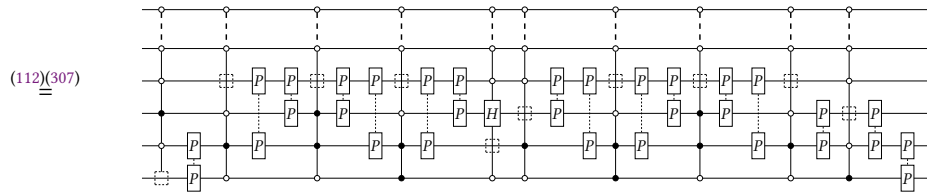

(112)(130)
$\underset{=}{}$

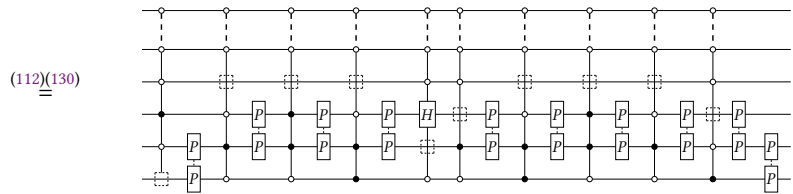

(111)(112)
$\underset{=}{}$

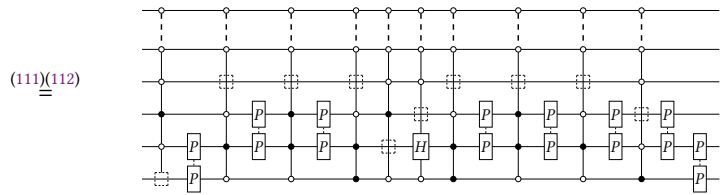

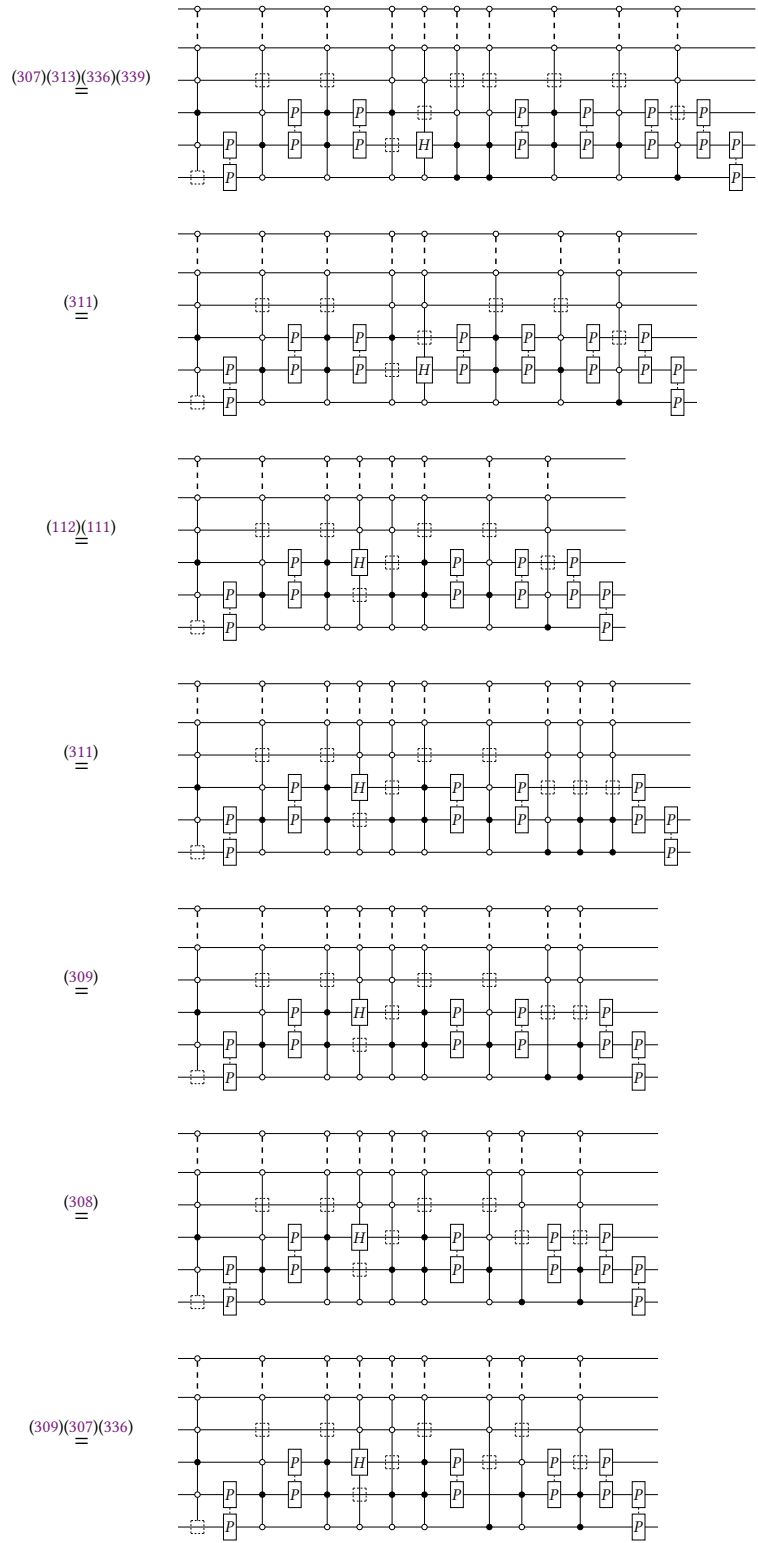

(IV)

(308)

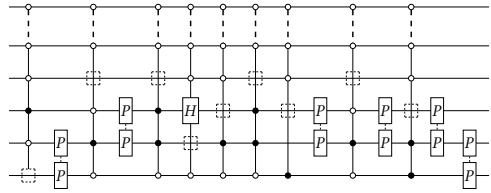

(309)(307)(336)

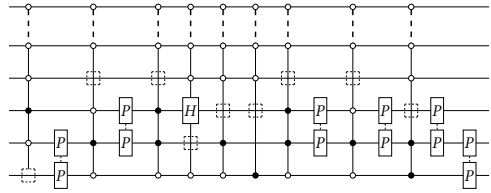

(309)(335)

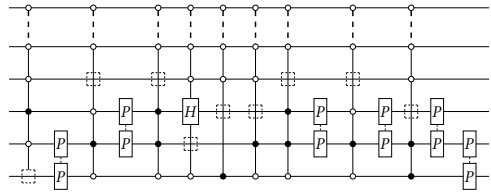

(309)(338)

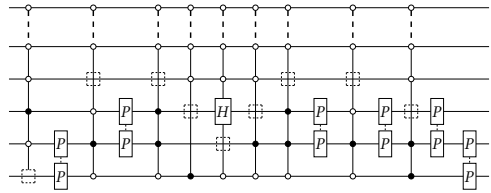

(309)(307)(336)

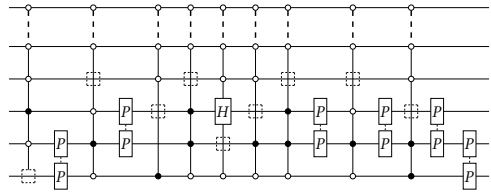

(308)

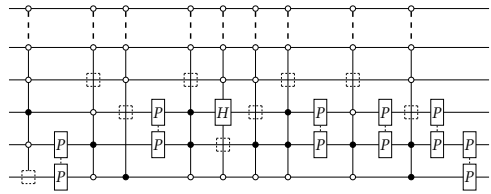

(309)(307)(336)

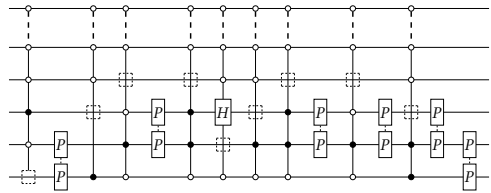

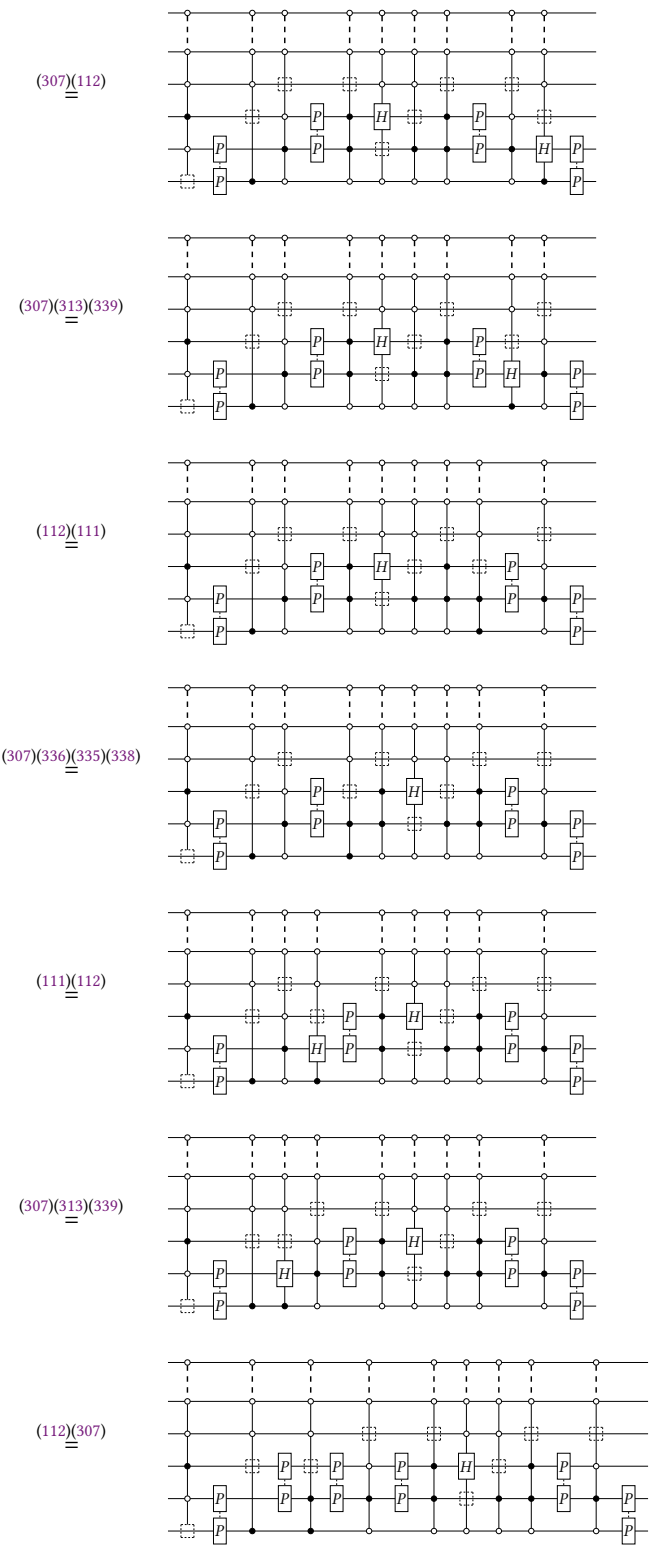

$(307)(112)$
$\equiv$

$(307)(313)(339)$
$\equiv$

$(112)(111)$
$\equiv$

$(307)(336)(335)(338)$
$\equiv$

$(111)(112)$
$\equiv$

$(307)(313)(339)$
$\equiv$

$(112)(307)$
$\equiv$

(308)
=

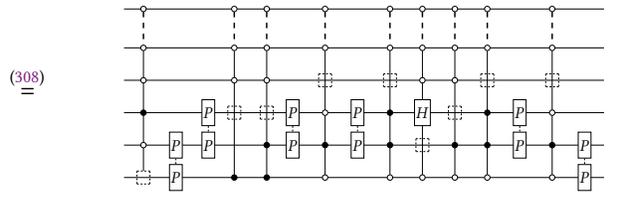

(309)
=

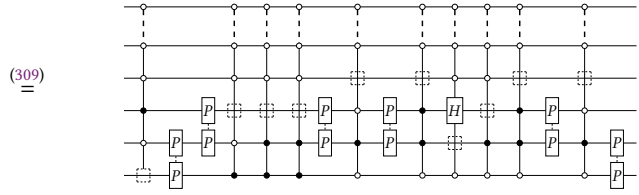

(311)
=

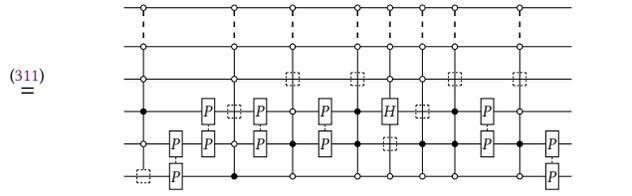

(307)(338)
=

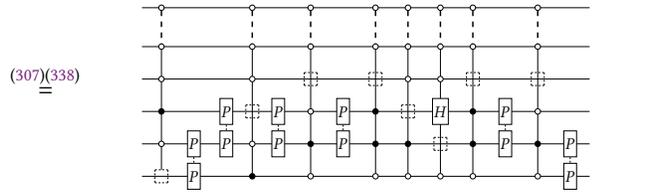

same steps as
from (III) to (IV),
mirrored,
in reverse order
=

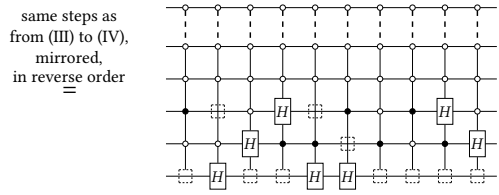

(307)(338)(337)(335)
=

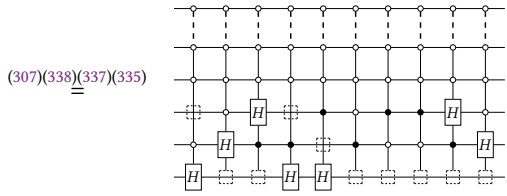

(311)
=

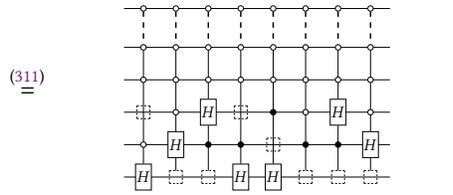